\pdfoutput=1
\documentclass{report}
\usepackage{macros2}

\usepackage{fbb}
\usepackage{newpxmath}

\usepackage{tikz-cd} 
\usepackage[utf8]{inputenc}
\usepackage[english]{babel}

\usepackage[style=numeric, sorting=none, sortcites=true, bibstyle=ieee, citestyle=numeric-comp, maxbibnames=99, giveninits=true]{biblatex}
\usepackage{hyperref}

\begin{document}

\begin{titlepage}
    \centering
    \vspace*{0.2in}
    
    {\Huge Chern-Simons Theories and Integrability \par}
    \vspace*{0.2in}
    \includegraphics[scale=0.1]{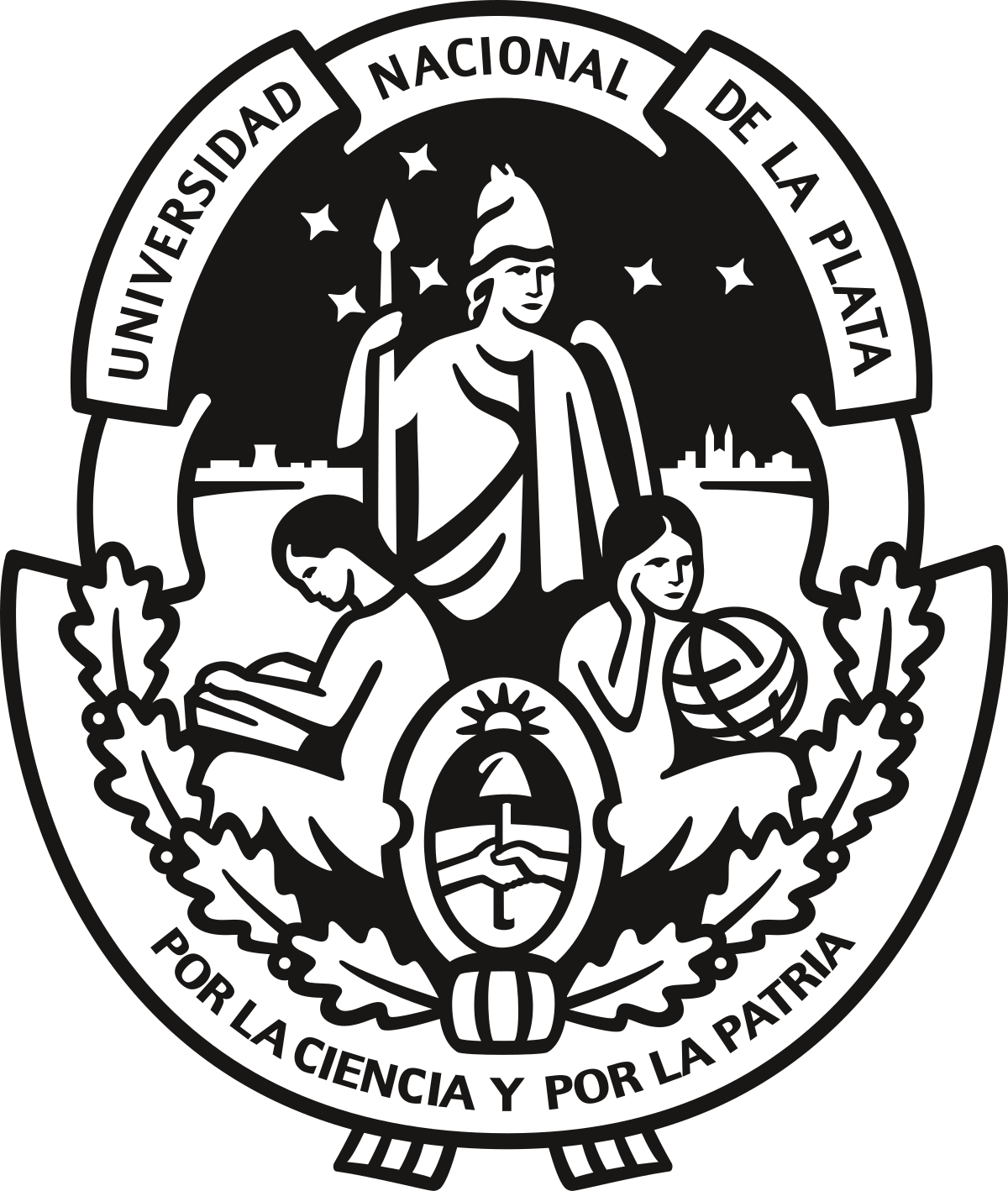}    
    \vspace*{0.2in}
    
    {\Large \textbf{Joaquin Liniado} \par}
    \vspace*{0.2in}
    
    {\Large Supervisor: Dr. Horacio Falomir \par}
    \vspace*{0.2in}
    
    {\large Departamento de Física \par}
    {\large Universidad Nacional de La Plata \par}
    
    \vspace{0.3in}

    {\large This dissertation is submitted for the degree of \par}
    {\large\itshape Doctor of Philosophy \par}
    
    \vspace*{0.1in}

    \begin{flushright}
        June 2025
    \end{flushright}
\end{titlepage}

\begin{abstract}

This thesis explores the correspondence between Chern-Simons theory and integrable field theories across different dimensions. It brings together all of my work in this area, including several distinct realizations of this correspondence. The thesis generalizes existing results on four-dimensional Chern-Simons theory and holomorphic Chern-Simons theory on twistor space. Additionally, it presents the discovery of a novel correspondence between five-dimensional holomorphic 2-Chern-Simons theory and integrable field theories in three dimensions.

\end{abstract}

\newpage
\section*{Agradecimientos}

A mi director, Horacio, por guiarme en mis primeros pasos como investigador, apoyarme en todas mis decisiones y acompañarme en mi osadía, siempre en la medida \emph{justa}, con debido estoicismo y profunda sabiduría.

A los amigos que me fui haciendo en la academia, desde los mas jóvenes a los mas viejos; por enseñarme, desafiarme y compartir esto que tanto nos gusta.  

A mis amigos de la vida, que festejan mis logros sin entenderlos, y que se enorgullecen de mi sin siquiera saber por qué. Eso es el amor desinteresado, y eso es todo lo que necesito. 

A mi compañera, Lucía, por darle el arte que le falta a mis cuentas. 

A mi familia, por amarme y enseñarme a amar, y así, a ser feliz.

A mi perro, Aquiles. Que conoció mi intimidad como nadie más lo hizo, y que aún así, eligió siempre estar a mi lado.

\tableofcontents

\newpage

\chapter{Introduction}

By the end of the 16th century, guided by the pioneering ideas of Galileo, it became increasingly evident that mathematics provided the ideal language for describing the physical world. These ideas were further developed by Descartes who introduced coordinate geometry, and by the end of the 17th century, Newton’s monumental work had firmly established mathematics as the foundation for understanding the laws of nature.

The introduction of differential calculus was the key to one of the most magical aspects of physics: the art of prediction. Once something that belonged to the realm of witches and wizards, suddenly came into the hands of scientists, with a level of precision that was simply unprecedented. Physicists suddenly found themselves able to predict the motion of planets, the paths of falling objects, and the simple harmonic motion of pendulums and springs: An unstoppable force had been set in motion.

But, as in every good tale ever told, power always comes at a price, and in this case, it lies in the dark land of differential equations. These equations, which correspond to different systems in nature, encode the information of all of their past, present, and future behavior (at least before we migrated away from determinism). However, to access this information, one must know how to solve them—and, unsurprisingly, they are almost always impossible to solve. I’m being a bit dramatic; the truth is we do know how to solve many differential equations. However, many of the most important ones remain beyond our reach. In general, solving differential equations is genuinely very difficult.

Yet, within this vast family, there is a special subset known as \emph{integrable equations}: these are equations that, due to very particular structural properties, are exactly solvable—equations for which we actually know how to find their solutions. Given the central role of differential equations in describing physical phenomena, and their notorious difficulty to solve them, it’s clear that identifying a class of equations that can be solved exactly is nothing short of remarkable.

Integrable equations appear in a wide range of areas in physics, from planetary dynamics and fluid motion to statistical mechanics, string theory, and quantum field theory. In this thesis we will focus on integrable equations within the context of quantum field theory. 

The equations governing field dynamics are partial differential equations (PDE), involving infinitely many coupled degrees of freedom. These are even more intricate than ordinary differential equations (ODE), which involve only a finite number of coupled degrees of freedom. In this regard, unlike the case of ordinary differential equations, the notion of integrability for partial differential equations is somewhat less ambitious and does not necessarily imply full solvability. In field theory, integrability usually means that the system is, only to some extent, solvable. This could mean that perhaps only a subsector of the system can be solved exactly, or that the system is constrained by an infinite number of conditions that reveal a wealth of information—even if the explicit solution remains out of reach.

One of the most intriguing aspects of integrable equations is that their origins have remained elusive for many years. While physicists and mathematicians have long identified conditions under which certain systems are considered integrable, there was, for decades, no real understanding of where integrable field theories actually came from.  

Recently, building on the ideas of Costello \cite{Costello:2013sla, Costello:2013zra} and further developed with Witten and Yamazaki, it has been proposed that the origin of integrable field theories can be traced back to a holomorphic variant of Chern-Simons theory \cite{Costello:2017dso, Costello:2018gyb, Costello:2019tri}. These ideas, once envisioned by Atiyah and Witten, have found a concrete realization in what is now known as four-dimensional Chern-Simons theory. At its core, the integrable structure descends from certain topological observables in the higher-dimensional gauge theory. 

This has led to a remarkable surge of work at the intersection of gauge theory and integrability, connecting diverse areas of mathematical, high-energy and condensed-matter physics. The  idea of constructing integrable structures from topological objects in higher-dimensional gauge theories has offered a revolutionary perspective on integrability in quantum field theory—perhaps the most far-reaching and cutting-edge understanding to date.

This PhD thesis presents a detailed exploration of the connection between Chern-Simons theory and various integrable structures, including Lax integrability in two-dimensional field theories, the Anti-Self-Dual Yang-Mills equations in four dimensions, and a categorical formulation of integrability in three dimensions.

\section{Outline}

The thesis is organized as follows:

In chapter \ref{chap:integrability}, we provide a pedagogical, guided, and thorough discussion of integrability across various dimensions, introducing all the mathematical background we consider essential. We focus on the concepts most relevant to our work, setting aside details that, while important in other contexts, would exceed the scope of this thesis. 

In chapter \ref{chap:hcstheory}, we introduce the different Chern-Simons theories that we use to construct integrable models in different dimensions, which forms the core content of the articles included in the subsequent chapters. 

Chapter \ref{chap:4dCS} presents my first paper in this area, written in collaboration with Benoit Vicedo. It extends some earlier work by Benoit and Sylvain Lacroix \cite{Lacroix:2021iit}, where we construct a broad family of integrable models in two dimensions—known as degenerate $\mathcal{E}$-models—starting from four-dimensional Chern-Simons theory. 

Chapter \ref{sec:intdeformations} is based on a paper in which we construct integrable deformations of the four-dimensional Wess-Zumino-Witten model, using six-dimensional holomorphic Chern-Simons theory. This work extends the results of Bittleston and Skinner \cite{Bittleston:2022cmz} to include new examples of integrable field theories in four dimensions.

Chapter \ref{sec:gaugingthediamond} also explores holomorphic Chern-Simons theory on twistor space, but focuses on the construction of gauged four-dimensional integrable models. This is a generalisation of Stedman's ideas \cite{Stedman:2021wrw} in four-dimensional Chern-Simons theory.   

Finally, chapter \ref{chap:5d2CS} introduces five-dimensional holomorphic 2-Chern-Simons theory—a categorical generalization of four-dimensional Chern-Simons theory—which leads to integrable models in three dimensions.

 \chapter{Integrability}
 \label{chap:integrability}

A Hamiltonian system with $n$ degrees of freedom has a $2n$-dimensional phase space. Solving such a system typically means determining the trajectories $(q_i(t),p_i(t))$ for $i=1,\dots,n$, which are often intractable without relying on symmetry arguments, extreme regimes or highly idealized scenarios. Indeed, these are the solutions to Hamilton's equations 
\begin{equation}
\label{ec:hamiltonseq}
    \frac{\mathrm{d} q_i}{\mathrm{d} t}=\frac{\partial H}{\partial p_i}\,,\quad \frac{\mathrm{d} p_i}{\mathrm{d} t} = -\frac{\partial H}{\partial q_i}\,,
\end{equation}
which are non-linear ordinary differential equations. Unlike linear systems, which are well understood, nonlinear systems can rarely be solved analytically. The absence of a universal method for solving them means that obtaining solutions usually depends on intuition, the clever identification of conserved quantities, or the ingenious discovery of symmetries.

Over time, and with considerable effort, people succeeded in solving certain problems exactly, such as the Kepler problem, the harmonic oscillator, or the Euler Top (describing rigid body rotation). However, it wasn’t until the nineteenth century that Liouville uncovered a unifying framework underlying all the known cases where Hamilton's equations could be solved exactly.

Liouville says that if you can find $n$ independent conserved quantities in involution, that is, with vanishing Poisson bracket, then the system can be solved by performing a finite number of definite integrals. This is fewer than one might expect to have to find, because in general $2n - 1$ constants are needed to integrate a flow in $2n$ dimensions. However, the independence and involutivity properties ensure that only $n$ conserved quantities are sufficient. Independence implies that the system is constrained to an $n$-dimensional manifold, effectively halving the number of dependent variables. Meanwhile, the conserved quantities generate Hamiltonian vector fields that describe the system's symmetries. Involutivity guarantees that these vector fields commute, forming an Abelian Lie group. This enables the construction of a special coordinate system, known as action-angle variables, in which the motion becomes linear in time, and thus can be solved by direct integration. These systems are said to be \emph{Liouville integrable}, and in principle, allow for an exact analytic solution.

While the framework of Liouville integrability provides a systematic method for solving finite-dimensional Hamiltonian systems, extending this notion to infinite-dimensional systems, such as field theories, introduces significant challenges. In plain old language, what would performing an infinite number of integrals even mean?

In this regard, many different attempts have been made in order to characterize integrability in the infinite dimensional context, but no single effective description covers all the known cases. One can, however, identify common features of the systems that are `integrable' in one accepted sense or another. First, the equations are to some degree soluble. In many cases, a large or even dense class of solutions can be
found explicitly and there exist general methods for constructing solutions. Second,
it is often possible to find nonlinear superpositions of solutions. Third,
ergodic behavior is ruled out by the existence of a large number of constants of motion, in other words, chaotic behavior is evidence of non-integrability.

The central goal of this thesis is to construct integrable field theories in a systematic way. We achieve this by exploring different realisations
of holomorphic Chern-Simons theory \cite{Costello:2013sla, Costello:2017dso, Costello:2018gyb, Costello:2019tri}, which, to date, provides the most general framework for systematically constructing integrable field theories. A particularly interesting aspect is that these models are integrable in distinct, yet well-established, senses. Indeed, the existence of multiple, mutually compatible definitions of integrability in field theory may reflect our incomplete understanding of the origins of integrability. In this regard, holomorphic Chern-Simons theory, which gives rise to models that are integrable in these various senses, may offer insight into a general, unified formulation of integrability.

In the remainder of this section we will discuss the different notions of integrability that will appear through this work, which are, Lax integrability, infinite-dimensional symmetry algebras and anti self-dual Yang Mills.

\section{Lax Integrability}

As mentioned in the introduction to this section, the key concept behind Liouville integrability is the existence of $n$ independent conserved quantities in involution. To generalize this idea to the infinite-dimensional setting, a natural requirement would be the existence of an infinite set of conserved charges in involution. While the existence of an infinite number of conserved quantities does not guarantee exact solvability of the system--in contrast to the finite-dimensional case--it remains one of the widely accepted definitions of integrability in field theories. We refer the reader to \cite{Torrielli:2016ufi, Retore:2021wwh, Babelon_Bernard_Talon_2003} for further discussion. 

Until now, we have overlooked the fact that, even in the finite-dimensional case, finding such a large number of conserved quantities can be a very challenging task. In the infinite-dimensional setting, this difficulty is only amplified. To address this, in the 1960s, Peter Lax introduced a formalism that reformulates the equations of motion as a compatibility condition for a pair of matrices, known as a Lax pair \cite{Lax:1968fm}. This elegant framework encodes the system's dynamics in terms of a matrix $L$ whose evolution is governed by another matrix $M$, and crucially, it provides a systematic approach for constructing conserved quantities. Furthermore, this formalism can be generalized to the field-theoretic context.

A Lax pair consists of two matrices $L$ and $M$, functions on the phase space of the system, such that Hamilton's equations \eqref{ec:hamiltonseq} are equivalent to
\begin{equation}
\label{ec:Lax1}
    \frac{\partial L}{\partial t} = [M,L]\,.
\end{equation}
The essence of the Lax formalism is that we can immediately construct $n$ conserved charges as follows. We consider the quantity $O_n = \mathrm{Tr}(L^n)$ such that
\begin{equation}
\begin{split}
    \frac{\partial O_n}{\partial t} 
    &= \sum_{k=0}^{n-1}\mathrm{Tr}\left(L^k \frac{\partial L}{\partial t}L^{n-1-k}\right) \\
    &=\sum_{k=0}^{n-1}\mathrm{Tr}\left(L^k [M,L]L^{n-1-k}\right)=0\,,
\end{split}
\end{equation}
due to the cyclicity of the trace. Thus, for every $n\in \mathbb{N}$, $O_n$ is conserved in time. Importantly, these quantities need not be independent nor in involution. Both conditions can be satisfied by imposing additional constraints on the matrices $L$ and $M$. To avoid diverting from our main argument, we will not discuss these conditions here. Further details can be found in \cite{Babelon_Bernard_Talon_2003}.

The Lax formalism becomes substantially more powerful when extended to include a complex parameter $\lambda \in \CP$, referred to as the spectral parameter. Indeed, in some cases we can find a family of Lax pairs, parameterised by $\lambda$, such that the equations of motion \eqref{ec:hamiltonseq} are equivalent to
\begin{equation}
\label{ec:Lax2}
    \frac{\partial L(\lambda)}{\partial t} = [M(\lambda),L(\lambda)]\,, \quad \forall \lambda \in \mathbb{C}P^1 \,.
\end{equation}

This has significant implications even in the finite-dimensional setting, as it leverages the interplay between group theory, arising through the Lie algebra commutator $[M,L]$, and complex analysis, which enters via the spectral parameter. However, for our purposes, the most important aspect of introducing the spectral parameter is the fact that it allows us to extend ideas of finite-dimensional integrability to the  infinite-dimensional case. 

\subsection{Two-Dimensional Field Theories}
\label{sec:PCM}

Let us consider, as a prototypical two-dimensional field theory, a sigma model on a 2-manifold $\Sigma$ with coordinates $t,x$ and target space some Lie group $G$ with Lie algebra 
$\mathfrak{g}=\mathrm{Lie}(G)$. The generalization of Lax integrability to this context requires the existence of two $\mathfrak{g}$-valued matrices $M,L$ such that the equations of motion of the theory are equivalent to
\begin{equation}
\label{ec:spectrallax}
    \frac{\partial L(\lambda)}{\partial t} - \frac{\partial M(\lambda)}{\partial x} = [M(\lambda), L(\lambda)]\,,\quad \forall \, \lambda \in \CP\,.
\end{equation}
We note that this is a zero-curvature equation for a connection defined by
\begin{equation}
    \mathcal{L}(x,t,\lambda)=M(x,t,\lambda) \mathrm{d}t + 
    L(x,t,\lambda) \mathrm{d}x \,,
\end{equation}
known as the \emph{Lax connection}. This object is central to two-dimensional integrability, as it enables the construction of an infinite number of independently conserved quantities as we now show.

\begin{figure}[h]
    \centering
    \begin{tikzpicture}
        \draw[thick] (0, 1) ellipse (2cm and 0.7cm);

        \draw[thick] (0, -1) ellipse (2cm and 0.7cm);

        \draw[thick, purple] (-2, -1) -- (-2, 1);
        \draw[thick, purple, ->] (-2, -1) -- (-2, 0);  

        \node at (2.5, 1) {$\gamma_1$};
        \node at (2.5, -1) {$\gamma_0$};
        \node[purple] at (-2.3, 0) {$\gamma_t$};

        \node at (-2.7, 1) {$(x_0,t_1)$};
        \node at (-2.7, -1) {$(x_0,t_0)$};

        \filldraw[purple] (-2, 1) circle (2pt);
        \filldraw[purple] (-2, -1) circle (2pt);
    \end{tikzpicture}
    \caption{Schematically, we may understand the path concatenation as follows. We start at $(x_0,t_1)$ we go back to $(x_0,t_0)$ with $\gamma_t^{-1}$, then we go around $(x_0,t_0)$ with $\gamma_0$ and then we go back to $(x_0,t_1)$ with $\gamma_t$. This defines a loop based at $(x_0,t_1)$.}
    \label{fig:pathconcintro}
\end{figure}
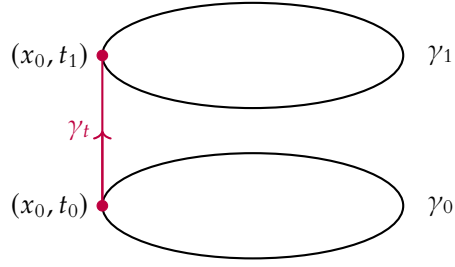

Given a smooth path $\gamma:[0,1]\to \Sigma$, we consider the differential equation 
\begin{equation}
    \frac{\mathrm{d}W}{\mathrm{d}\sigma} + (\gamma^\ast \mathcal{L})W =0
\end{equation}
with initial condition $W(0)=\mathrm{id}.$, where $\gamma^\ast \mathcal{L}$ is the pullback of $\mathcal{L}$ along $\gamma$. The solution to this equation is the parallel transport of $\mathcal{L}$ along $\gamma$ and is given by the holonomy
\begin{equation}
\label{ec:defholonomy}
    W_\gamma(\lambda)=P \mathrm{exp}\left(-\int_\gamma\mathcal{L}(\lambda)\right) \in G \,,
\end{equation}
where $P$ denotes the path-ordered exponential and we write the $\lambda$-dependence explicitly for later convenience. Given that the connection is flat, $W_\gamma$ is independent of the path, as long as the endpoints are kept fixed. We say that the holonomy solely depends on the homotopy class relative base points of the paths. Explicitly, if $\gamma_0,\gamma_1:[0,1]\to \Sigma$ are paths such that $\gamma_0(0)=\gamma_1(0)$ and $\gamma_0(1)=\gamma_1(1)$ then $W_{\gamma_{0}}=W_{\gamma_1}$. In particular if $\gamma$ is a contractible loop, then $W_\gamma=1$. We will use this invariance to construct conserved quantities. Let us consider at a time $t_0$ a loop $\gamma_0$ starting and ending at $x_0$, and at a later time $t_1$ a loop $\gamma_1$ starting and ending at $x_0$ as well. Note that since we are at two different time slices, $\gamma_0$ and $\gamma_1$ are not in the same homotopy class relative base point since they are loops based at $(x_0,t_0)$ and $(x_0,t_1)$ respectively.  Hence, we have $W_{\gamma_0}\neq W_{\gamma_1}$. However, we can  consider a path $\gamma_{t}$ connecting $(x_0,t_0)$ with $(x_0,t_1)$, such that the path concatenation $\gamma_t\ast\gamma_0\ast \gamma_t^{-1}$ is a loop based at $(x_0,t_1)$, see fig. \ref{fig:pathconcintro}.  

Given that $\gamma_t\ast\gamma_0\ast \gamma_t^{-1}$ and $\gamma_1$ are homotopical relative base point, flatness of the connection implies $W_{\gamma_t\ast\gamma_{0}\ast \gamma_t^{-1}} = W_{\gamma_1}$. Moreover, holonomies are multiplicative under path concatenation, namely, 
\begin{equation}
W_{\gamma_t\ast\gamma_{0}\ast \gamma_t^{-1}}=  W_{\gamma_t}W_{\gamma_0}W_{\gamma_t}^{-1}= W_{\gamma_1}\,.
\end{equation}
Hence, assuming that $G$ is a matrix Lie group, we find that 
\begin{equation}
    \mathrm{Tr} (W_{\gamma_0}) = \mathrm{Tr} (W_{\gamma_1})\,,
\end{equation}
and thus the quantity $\mathrm{Tr} (W_{\gamma_0})$ is conserved since $\gamma_0$ and $\gamma_1$ are paths on different time slices. We can now use this object to construct an infinite number of conserved quantities. We define the \emph{transfer matrix}
\begin{equation}
\label{ec:transfermatrix}
    \mathfrak{w}(\lambda)=\mathrm{Tr}(W_\gamma(\lambda)) \,,
\end{equation}
which satisfies $\partial_t \mathfrak{w}(\lambda)=0$. Moreover, it can be expanded in powers of $\lambda$ as
\begin{equation}
\label{ec:lambdaexp}
    \mathfrak{w}(\lambda)=\sum_{n \in \mathbb{Z}} Q_n \lambda^n\,,
\end{equation}
then, for every $n\in \mathbb{Z}$, we have that $\partial_t Q_n =0 $. It is worth mentioning that these charges are in general non-local. This non-locality makes it particularly challenging, and in some cases infeasible, to identify the symmetries of the theory corresponding to these conserved quantities. For this reason, they are often referred to as hidden symmetries. On the other hand, by construction, the charges $Q_n$ are independent, but once again, need not be in involution. Similarly to the finite-dimensional case, one can impose further constraints on $W_\gamma(\lambda)$ to guarantee that the charges are in involution, though again, we will not discuss these conditions here. Further details can be found in \cite{Babelon_Bernard_Talon_2003, Torrielli:2016ufi}. 

A two-dimensional field theory is said to be \textit{Lax integrable} or \textit{weakly integrable} if its equations of motion can be reformulated as a flatness condition for a Lax connection. \textit{Strong integrability}, on the other hand, requires that the conserved charges be in involution. Two-dimensional Lax integrable field theories are one of the central structures explored in this thesis. 

\begin{remark}
   The spectral parameter $\lambda$ not only allows for the construction of an infinite number of conserved quantities but it is also crucial in the construction of explicit solutions to the equations of motion. Specifically, the Lax equation \eqref{ec:spectrallax} can be reformulated as a Riemann-Hilbert factorization problem, whose solvability is guaranteed by the Riemann-Hilbert theorem \cite{Babelon_Bernard_Talon_2003}. While this theorem ensures the existence of solutions, it does not imply that explicit analytical solutions can always be found. Nonetheless, there are certain cases where solutions can be constructed; these are known as solitons and their generalizations.
\end{remark}


\subsubsection{The Principal Chiral Model}

The archetypal example of a two-dimensional integrable field theory is the integrable sigma model, also known as the principal chiral model (PCM). Its origins, or more precisely, the origin of its significance, can be traced back to string theory. Indeed, the dynamics of the bosonic string are governed by the Polyakov action, a two-dimensional field theory. This action is defined on a two-dimensional worldsheet $\Sigma$ with coordinate embeddings $\phi^I:\Sigma \to M$ into some target space $M$, given by \cite{POLYAKOV1981207}, 

\begin{equation}
    S_P = -\frac{1}{4\pi\alpha'} \int d^2 \sigma \sqrt{-\gamma} \, \gamma^{\alpha \beta} G_{IJ}(\phi) \partial_\alpha \phi^I \partial_\beta \phi^J\,,
\end{equation}
where \( \gamma_{\alpha \beta} \) is the worldsheet metric, \( G_{IJ}(\phi) \) is the target space metric and the coupling constant $\alpha'$ is related to the inverse of the string's tension. This action is Poincaré invariant, diffeomorphism invariant, and crucially, invariant under a two-dimensional Weyl transformation
\begin{equation}
    \gamma_{\alpha \beta}\mapsto e^{2\omega}\gamma_{\alpha \beta}
\end{equation}
for arbitrary $\omega:\Sigma \to \Sigma$. This invariance can be used to fix $\gamma$ to be the two-dimensional Minkowski metric, thus rendering the gauged-fixed Polyakov action 
\begin{equation}
    S_P=-\frac{1}{4\pi\alpha'}\int d^2 \sigma G_{IJ}(\phi) \partial_\alpha \phi^I \partial^{\alpha} \phi^J\,.
\end{equation}
The PCM action is obtained by taking the target space $M$ to be a Lie group $G$. There is a canonical metric on any Lie group induced from the Killing form on its Lie algebra. Denoting the latter by $\langle\cdot,\cdot\rangle:\fg \times \fg \to \bR$, we have that
\begin{equation}
   G_{IJ}(\phi) \partial_\alpha \phi^I \partial^{\alpha} \phi^J = \langle g^{-1}\partial_\alpha g, g^{-1}\partial^\alpha g\rangle\,,
\end{equation}
for a smooth embedding $g:\Sigma \to G$. Hence, we may write the PCM action in form notation as
\begin{equation}
    S_{\mathrm{PCM}}[g]=\frac{\mathscr{k}}{2}\int_\Sigma \langle g^{-1}\mathrm{d}g,\ast g^{-1}\mathrm{d}g\rangle \,,
\end{equation}
with $\mathscr{k}$ the inverse to the coupling constant of the model. If we take $\Sigma$ to be $(1+1)$-dimensional Minkowski space with light-cone coordinates $\sigma^\pm =\tfrac{1}{2}(t\pm \sigma)$, then the above can be written in components as
\begin{equation}
        S_{\mathrm{PCM}}[g]=\mathscr{k}\int_\Sigma \mathrm{d}\sigma^+\wedge \mathrm{d}\sigma^- \langle g^{-1}\partial_+ g, g^{-1}\partial_- g\rangle \,.
\end{equation}

The classical equation of motion, which can be computed by varying the action with respect to an infinitesimal variation of the field $g\mapsto g e^\epsilon \sim g (1+\epsilon)$, for an infinitesimal parameter $\epsilon$, is given by 
\begin{equation}
\label{ec;pcmeom}
    \mathrm{d}(\ast g^{-1}\mathrm{d}g)=\partial_+(g^{-1}\partial_- g)+\partial_-(g^{-1}\partial_+ g)=0\,.
\end{equation}
This characterizes the dynamics of the PCM, together with the Maurer-Cartan identity, 
\begin{equation}
    \mathrm{d}(g^{-1}\dr g) + \tfrac{1}{2}[g^{-1}\dr g,g^{-1}\dr g]=0\,.
\end{equation}
The PCM has a $G \times G$ global symmetry, meaning, the action is invariant under the transformation 
\begin{equation}
    g \mapsto  \ell ^{-1} g r \,,\quad \text{with}\quad (\ell,r)\in G\times G \,,
\end{equation}
which can be verified with a straightforward computation. The conserved current associated to the right acting global symmetry is precisely $\ast g^{-1}\dr g$, which can be readily obtained by realizing that the equation of motion \eqref{ec;pcmeom} corresponds to the variation of the action with respect to a right variation of the field $g$. If instead, we vary the action with respect to $\delta g = \epsilon g$, we find the conserved current associated to the left acting symmetry, given by $\ast \dr g g^{-1}$. 

Let us now discuss it's integrable structure. We look for a $\fg$-valued connection on $\Sigma$ depending on a spectral parameter $\lambda \in \mathbb{C}P^1$, such that the vanishing of it's curvature is equivalent to the equation of motion \eqref{ec;pcmeom}. This is given by  
\begin{equation}
\label{ec:2dLAXPCMintro}
    \mathcal{L}=\frac{g^{-1}\partial_+g}{1-\lambda}\dr \sigma^+ + \frac{g^{-1}\partial_-g}{1+\lambda}\dr \sigma^-\,.
\end{equation}
Note in particular, its meromorphic dependence on the spectral parameter $\lambda$. Its curvature $F=\mathrm{d}\cL + \cL \wedge \cL$ is then 
\begin{equation}
    F = \frac{1}{1-\lambda^2}\big(\mathrm{d}(g^{-1}\dr g) + \tfrac{1}{2}[g^{-1}\dr g,g^{-1}\dr g]\big) - \frac{\lambda}{1-\lambda^2} \dr(\ast g^{-1}\dr g)\,.
\end{equation}
In order for $F =0$ for every value of $\lambda \in \CP$ then both terms must vanish independently. The first term is the Maurer-Cartan identity, which is identically vanishing, whereas the second is zero due to the equation of motion \eqref{ec;pcmeom}. Thus, we conclude that the flatness of the Lax connection for every value of $\lambda$ is equivalent to the equation of motion of the PCM, which proves its Lax integrability. 

Let us now construct the holonomy \eqref{ec:defholonomy}. We consider a fixed time $t_0$ and a curve $\gamma:[0,1] \to \Sigma$ given by $\gamma(s)=(t_0,\sigma(s))$ with $\sigma(s) \in (-\infty,\infty)$. The holonomy is then
\begin{equation}
    W_\gamma(\lambda) = P\exp \left(-\int_{-\infty}^\infty \dr\sigma \mathcal{L}_\sigma(\lambda)\right) = P\exp \left(-\int_{-\infty}^\infty \dr\sigma \, \frac{g^{-1}\partial_\sigma g + \lambda g^{-1}\partial_t g}{1-\lambda^2}\right)
\end{equation}
where $\mathcal{L}_\sigma = \tfrac{1}{2}(\cL_+-\cL_-)$. We can look at conserved quantities by expanding in powers of $\lambda$ as
\begin{equation}
    W_\gamma(\lambda)=1+\frac{1}{\lambda}\int_{-\infty}^\infty \mathrm{d}\sigma j_t(t,\sigma)+\frac{1}{\lambda^2}\left(\int_{-\infty}^\infty \mathrm{d}\sigma j_\sigma(t,\sigma) + \int_{-\infty}^\infty \mathrm{d}\sigma\int_{-\infty}^\sigma \mathrm{d}\sigma'j_t(t,\sigma)j_t(t,\sigma')\right)+\dots 
\end{equation}
where we have used shorthand notation $j_i = g^{-1}\partial_i g$ for the Maurer-Cartan left forms. We find at the first non-trivial order in $\lambda$ the Noether charge associated to the right acting global symmetry. At higher orders, we observe that the conserved charges are non-local.

The principal chiral model is a fundamental example of an integrable field theory. In recent years, numerous two-dimensional integrable field theories have been constructed, with applications spanning a wide range of areas. Many of these models are deformations of the non-linear sigma model,  that preserve integrability. We refer the reader to \cite{Hoare:2021dix} for a recent review.

\section{Infinite-Dimensional Symmetry Algebras}

\label{sec:currentalgebras}

In the previous section, we discussed Lax integrability, where the integrable structure was tied to the existence of an infinite number of conserved quantities in involution, constructed from the Lax connection. These charges, however, are usually non-local, posing difficulties in uncovering the symmetries they correspond to. In this section, we introduce a distinct but related notion, which we will also regard as a form of integrability. We will focus on two-dimensional field theories with infinite-dimensional symmetry algebras \cite{DiFrancesco:1997nk, liniado2021}.

The archetypal example of an infinite-dimensional symmetry in two dimensions is conformal symmetry. While in higher dimensions the group of globally defined conformal transformations is finite-dimensional, in two dimensions something remarkable happens: the number of locally defined conformal transformations is infinite. This follows from  Liouville's rigidity theorem \cite{Harley} which states that, unlike in higher dimensions, not every local conformal transformation can be extended globally on the plane. In particular the number of parameters specifying such local transformations is infinite, because any locally analytic function provides a locally well defined conformal mapping on the complex plane. 


This infinite-dimensional symmetry imposes strong constraints on the system and leads to an infinite number of conserved quantities. However, these charges do not generally satisfy the involution condition required for Liouville integrability, meaning these models are not necessarily integrable in the Liouville sense. Nevertheless, the stringent structure imposed by these symmetries often allows for the exact computation of observables, making these theories exactly solvable. In this broader sense, we regard these models as possessing an integrable structure. A notable example which we will discuss in detail is the Wess–Zumino–Witten (WZW) model, where conformal symmetry is further enhanced by an affine Kac–Moody symmetry. 
\subsection{Conformal Symmetry}

Let $M$ be a $d$-dimensional semi-Riemannian manifold with with signature $(p,q)$. We say that a coordinate tranformation $x\mapsto x'$ is conformal if it leaves the metric tensor invariant up to scale
\begin{equation}
\label{ec:conformalcondition}
    g_{\mu \nu}'(x')= \frac{\partial x^\rho}{\partial x'^\mu}\frac{\partial x^\sigma}{\partial x'^\nu}g_{\rho\sigma}(x)=\Lambda^2(x)g_{\mu \nu}(x) \,.
\end{equation}
The set of globally defined invertible transformations which are conformal form the global conformal group which is isomorphic to $SO(d+1,1)$. This group is of dimension $\frac{(d+1)(d+2)}{2}$ and includes Poincaré transformations, dilations and special conformal transformations. When $d=1$, any transformation is conformal, since the metric is just a number. When $d\geq 3$,  Liouville's rigidity theorem states that every local conformal transformation can be extended to a globally defined one. However, when $d=2$, 
Liouville's theorem does not hold and we can treat global and local transformations separately.

Let us consider the two-dimensional Euclidean plane with coordinates $(x^0,x^1)$ and a coordinate transformation $(x^0,x^1)\mapsto (w^0(x),w^1(x))$. The condition \eqref{ec:conformalcondition} for the transformation to be conformal becomes
\begin{align*}
      \left(\frac{\partial w^0}{\partial x^0}\right)^2 + \left(\frac{\partial w^0}{\partial x^1}\right)^2 &= \left(\frac{\partial w^1}{\partial x^0}\right)^2 + \left(\frac{\partial w^1}{\partial x^1}\right)^2 \\
      \\
     \frac{\partial w^0}{\partial x^0}\frac{\partial w^1}{\partial x^0} &+ \frac{\partial w^0}{\partial x^1}\frac{\partial w^1}{\partial x^1}  = 0 \,.
\end{align*}
These conditions are equivalent either to:
\begin{equation}
\label{ec:holomorpic}
    \frac{\partial w^1}{\partial x^0}=\frac{\partial w^0}{\partial x^1} \quad \text{and} \quad \frac{\partial w^0}{\partial x^0}=-\frac{\partial w^1}{\partial x^1}
\end{equation}
or to:
\begin{equation}
\label{ec:antiholomorphic}
    \frac{\partial w^1}{\partial x^0}=-\frac{\partial w^0}{\partial x^1} \quad \text{and} \quad \frac{\partial w^0}{\partial x^0}=\frac{\partial w^1}{\partial x^1} \,, 
\end{equation}
from where we can immediately recognize the Cauchy-Riemann equations for holomorphic functions in equation \eqref{ec:holomorpic} and for anti-holomorphic functions in equation \eqref{ec:antiholomorphic}. This motivates the definition of complex coordinates
\begin{equation}
\label{ec:complexcoordinates1}
\begin{cases}
\begin{array}{l}
     z=x^0+ix^1  \\
     \partial_z
     =\frac{1}{2}(\partial_{0}-i\partial_{1})
\end{array}        
\end{cases}
      \quad \text{and}\qquad
\begin{cases}
 \begin{array}{l}
     \bar{z}=x^0-ix^1  \\
     \partial_{\bar{z}}=\frac{1}{2}(\partial_{0}+i\partial_{1}) 
\end{array}        
\end{cases}
\end{equation}
in terms of which the Cauchy-Riemann equations become
\begin{equation}
\label{ec:complexCR}
    \partial_{\bar{z}} w(z,\bar{z})= 0 \quad \text{and} \quad \partial_{z}\bar{w}(z,\bar{z})=0\,.
\end{equation}
Hence, a coordinate transformation $(z,\bar z)\mapsto (w(z),\bar w(\bar z))$ will be conformal as long as $w$ and $\bar w$ are holomorphic and anti-holomorphic mappings respectively. Crucially, the Cauchy-Riemann equations are inherently local, meaning that there can exist coordinate transformations that are holomorphic or anti-holomorphic in a local neighborhood but fail to be globally well-defined. This is precisely Liouville's rigidity theorem.

Having established the structure of conformal coordinate transformations on the Euclidean plane, we now explore their implications for field theories, focusing specifically on two-dimensional theories with conformal invariance. Since local conformal transformations are given by analytic functions, this naturally leads to an infinite number of conserved charges, as any analytic function admits a Laurent expansion on the complex plane. The key to constructing these conserved quantities lies in the stress–energy tensor of the theory. Indeed, in any conformally invariant theory, a conserved stress–energy tensor can always be constructed, and conformal symmetry further imposes the condition that this tensor is classically traceless. As a result, it has only two non-trivial components, $T_{zz}\coloneqq T$ and $T_{\bar z\bar z}\coloneqq\bar T$ whose conservation leads to
\begin{equation} \partial_{\bar z} T(z,\bar z) = 0 \quad \text{and} \quad \partial_{z} \bar{T}(z,\bar{z}) = 0. \end{equation}
In other words, $T(z)$ is a holomorphic function, whereas $\bar T(\bar z)$ is an anti-holomorphic function. This implies that 
we can Laurent expand
\begin{equation}
    T(z)=\sum_{n \in \mathbb{Z}}z^{-n-2} L_n \,,\quad L_{n}=\frac{1}{2\pi i}\oint \mathrm{d}z \, z^{n+1}T(z) \,,
\end{equation}
and similarly for the anti-holomorphic component. We can interpret the modes $L_n$ as conserved charges by introducing an Euclidean time coordinate $\tau = \ln |z|$, such that dilations in the complex plane correspond to translations in $\tau$. In this framework, contour integrals around the origin remain invariant under these transformations. Since the definition of $L_n$ depends only on the analytic structure of $T(z)$ and not on the specific choice of contour, the charges $L_n$ are independent of $\tau$ and thus conserved under radial time evolution. 

Crucially, these charges are not in involution, as they satisfy the Witt algebra\footnote{After quantisation, the Witt algebra is centrally extended to the Virasoro algebra due to the Weyl anomaly.}
\begin{equation}
    \{L_n,L_m\}=(m-n)L_{m+n} \,,
\end{equation}
and thus, they do not form an integrable structure in the Liouville sense. However, the stringent constraints imposed by the mere existence of this infinite dimensional symmetry algebra allows in many cases for the exact solvability of the the theory both classically and quantum mechanically. In this sense we will also consider conformal symmetry in two-dimensions as a form of integrability. 

\subsection{WZW Models}

\label{sec:infdimsymalg}

The Wess-Zumino-Witten (WZW) model is a two-dimensional conformal field theory describing a non-linear sigma model with additional topological structure. It plays a fundamental role in string theory, conformal field theory, and integrable systems. One of its key features is its invariance under an infinite-dimensional affine Lie algebra --the Kac-Moody algebra-- that extends the underlying conformal symmetry.

Let $\Sigma$ be a two-dimensional manifold, and let $G$ be a Lie group with Lie algebra $\fg$. The field of the model is a map $g:\Sigma \to G$ . The classical action of the WZW model is\footnote{If we were working in Minkowski spacetime the coefficient of the WZ term would simply be $\frac{\mathscr{k}}{6}$.}

\begin{equation}
S_{\rm WZW}[g] = \frac{\mathscr{k}}{2} \int_\Sigma \langle g^{-1}\dr g, \ast g^{-1}\dr g\rangle + \frac{i\mathscr{k}}{6} \int_{M} \langle \tilde{g}^{-1}\dr \tilde{g},[\tilde{g}^{-1}\dr \tilde{g},\tilde{g}^{-1}\dr \tilde{g}]\rangle.
\end{equation}

Here, $\mathscr{k}$ is known as the \emph{level}, and $M$ is a 3-manifold with boundary $\Sigma$. The function $\tilde g:M \to G$ is a smooth extension of $g$ into $M$. The first term corresponds to the PCM action introduced in \ref{sec:PCM}. The second term, known as the Wess-Zumino (WZ) term, despite its apparent three-dimensional form is effectively two-dimensional as we shall now discuss.

One way to see that the second term is effectively two-dimensional is to show that it is independent of the choice of extension $\tilde g: \Sigma \to G$ to $M$. If this holds—meaning that any two extensions $\tilde{g}$ and $\tilde{g}'$ yield the same contribution—then the WZ term depends solely on the two-dimensional field $g: \Sigma \to G$, ensuring its intrinsic two-dimensional nature. 

Let $B_3 = \tilde g(M)$ and $B_3'=\tilde{g}'(M)$ be the images of $M$ under $\tilde g $ and $\tilde g'$ respectively. Since $\tilde g$ and $\tilde g'$ agree on $\partial B_3 =\Sigma = \partial B_3'$, we can reverse the orientation of $B_3'$ and glue it to $B_3$ along $\Sigma$ to form a submanifold of $G$ diffeomorphic to $S^3\cong B_3 \cup_{\Sigma}(- B_3')$, where the minus sign means that the orientation is reversed. Then, denoting with $\omega = \langle \tilde{g}^{-1}\dr \tilde{g},[\tilde{g}^{-1}\dr \tilde{g},\tilde{g}^{-1}\dr \tilde{g}]\rangle$ we have\footnote{To integrate over $B_3$ and $B_3'$ we are pulling back $\omega$ by $\tilde g$ and $\tilde g'$ respectively.}
\begin{equation}
   \Delta = \int_{B_3} \omega - \int_{B_3'}\omega = \int_{S^3}\omega\,.
\end{equation}
The integral over $S^3$ of the $3$-form $\omega$ is a topological quantity which measures the winding number of a map from $S^3$ to the Lie group $G$. These maps are classified by the third homotopy group $\pi_3(G) \cong \mathbb{Z}$, which is isomorphic to the integers for any simple Lie group \cite{Witten:1983ar}. Hence, with a suitable normalisation of the bilinear form, we find that the change in the action given by choosing two different extensions of $g$ can be taken to be $\Delta \in 2\pi \mathbb{Z}$. Thus, the difference between any two extensions of $g$ will only introduce a phase factor to the path integral, leaving the physical observables unchanged.

The equations of motion of the WZW model can be found by considering a variation $g\mapsto g+ g \epsilon$ for some infinitesimal parameter $\epsilon$. The variation of the PCM action has been computed in \ref{sec:PCM}. For the WZ term, we have
\begin{equation}
    \delta \langle \tilde g^{-1}\dr \tilde g,[\tilde g^{-1}\dr \tilde g,\tilde g^{-1}\dr \tilde g]\rangle = 3\langle \mathrm{d}\epsilon + [\tilde g^{-1}\dr \tilde g,\epsilon],[\tilde g^{-1}\dr \tilde g,\tilde g^{-1}\dr \tilde g]\rangle \,.
\end{equation}
The second term vanishes due to the Jacobi identity. The first term, after an integration by parts and using Stokes theorem, contributes 
\begin{equation}
    \frac{i\mathscr{k}}{2}\big[g^{-1}\dr g ,g^{-1},\dr g\big]\,.
\end{equation}
Using the Maurer-Cartan equation for $g^{-1}\dr g $, we find that the equation of motion of the WZW model is 
\begin{equation}
\label{ec:eomwzw1}
    \dr \big(i g^{-1}\dr g +\ast g^{-1}\dr g \big)=0\,.
\end{equation}
To simplify the analysis, we take $\Sigma = \mathbb{R}^2$ with the usual complex structure. In complex coordinates we may write the above equation as
\begin{equation}
\label{ec:eomwzw2}
\partial_z(g^{-1}\partial_{\bar z }g)=0\,.
\end{equation}
By considering instead a left variation $g\mapsto g+\epsilon g$, the equation of motion can also be written  as\footnote{Equivalently, one may use the identity $\mathrm{Ad}_{g}\partial_{ z}(g^{-1}\partial_{\bar z} g )=\partial_{\bar z}(\partial_{ z}g g^{-1})$}
\begin{equation}
\partial_{\bar z}(\partial_{z}g g^{-1})=0\,,
\end{equation}
whose solutions are
\begin{equation}
    g(z,\bar z)=f(z)\bar f(\bar z)
\end{equation}
for arbitrary smooth functions $f(z)$ and $\bar f(\bar z)$. In other words, the WZW model is classically exactly solvable, which is one of the reasons we may consider it integrable. We will now show that it further possesses an infinite number of symmetries, both classically and quantum mechanically. These symmetries play a crucial role in the model's quantum integrability and, in certain cases, enable the exact computation of correlation functions. 

The equations of motion \eqref{ec:eomwzw1} and \eqref{ec:eomwzw2} imply that the current $J = \partial_z g g^{-1}$ is holomorphic, whereas $\bar J = g^{-1}\partial_{\bar z}g$ is anti-holomorphic. Indeed, these currents are associated to a semi-local symmetry of the WZW action given by
\begin{equation}
\label{ec:semiloctransf}
    g(z,\bar z) \mapsto \ell(z)^{-1}g(z,\bar z)r(\bar z)\,,
\end{equation}
with $\ell,r:\Sigma \to G$. The invariance of the action under the above transformation can be verified using (twice) the Polyakov-Wiegmmann identity \cite{Polyakov:1983tt}
\begin{equation}
S_{\text{WZW}}[\ell g] = S_{\text{WZW}}[\ell] + S_{\text{WZW}}[g] + \int \mathrm{d}^2z \, \langle \ell^{-1} \partial_z \ell , \partial_{\bar z} g g^{-1} \rangle.    
\end{equation}

We find that the global $G\times G$ invariance of the PCM discussed in \ref{sec:PCM} has been enhanced to a semi-local symmetry due to the presence of the WZ term. Similarly than in the case of local conformal symmetry, this will give rise to an infinite dimensional symmetry algebra. 

Let $\{t^a\}$ with $a=1,\dots \dim (\fg)$ be a basis of generators of the Lie algebra. We can write the currents $J(z)$ and $\bar J(\bar z)$ in this basis as
\begin{equation}
    J(z) = \sum_{a=1}^{\dim(\fg)}J^a(z) t^a \,,\quad \bar J (\bar z)=  \sum_{a=1}^{\dim(\fg)}\bar J^a(\bar z) t^a \,.
\end{equation}
Given that $J(z)$ is holomorphic, then $J^a(z)$ is holomorphic for every $a$ so that we can mode expand 
\begin{equation}
    J^a(z)=\sum_{n \in \bbZ}J^a_n z^{-n-1} \,, \quad J^a_n = \frac{1}{2\pi i}\oint \dr z \, z^{n}J^a(z) \,,
\end{equation}
and similarly for the anti-holomorphic current. With the same reasoning than with the modes of the stress-energy tensor, the $J_n^a$ will be conserved in time for every $n \in \mathbb{Z}$, as they are independent of contour deformations. We thus find an infinite number of conserved quantities, which again, are not in involution as their Poisson brackets are given by \cite{Witten:1983ar}
\begin{equation}
    \{J_n^a,J_m^b\}= f_c^{ab}J_{n+m}^c+\mathscr{k}n\delta^{ab}\delta_{m+n,0} \,,
\end{equation}
where $f_{c}^{ab}$ are the structure constants of the Lie algebra $\fg$. The modes $J_n^a$ satisfy the commutation relations of the affine Kac-Moody algebra at level $\mathscr{k}$. This algebra is often reffered to as a \emph{current algebra} and it is an infinite-dimensional symmetry algebra of the WZW model. 

At the beginning of this subsection we stated that the WZW model is a conformal field theory. Indeed, both its kinetic term and the WZ term are classically conformally invariant. Remarkably, this invariance persists at the quantum level. In fact, by applying the Sugawara construction \cite{Sugawara:1967rw, Sommerfield:1968pk}, one can build a conserved stress–energy tensor directly from the chiral currents $J$ and $\bar J$. This tensor is not only conserved but also traceless as an operator equation, ensuring the full quantum conformal symmetry of the model.

The presence of both the Kac-Moody and conformal algebras as symmetry algebras of the WZW model, implies that in certain cases, the theory can be exactly solved at the quantum level. More precisely, the Ward identities associated with the Kac-Moody symmetry give rise to the Knizhnik-Zamolodchikov  equation, which governs the correlation functions of primary fields \cite{KNIZHNIK198483}. In some cases, this equation can be solved exactly, yielding explicit expressions for these correlation functions. Since all other fields in the theory can be expressed in terms of primaries, this provides a complete solution for the correlation functions of the WZW model \cite{DiFrancesco:1997nk}. In this sense, the WZW model can be considered to be quantum integrable as well. 

\subsubsection{PCM + Wess-Zumino Term}

The invariance of the WZW model under the semi-local transformation \eqref{ec:semiloctransf} depends on the specific coefficient $\frac{i\mathscr{k}}{6}$ of the WZ term. This particular value ensures that the equations of motion take the form \eqref{ec:eomwzw1}. If instead we choose a different coefficient, say $\frac{i\mathscr{h}}{6}$, the equation of motion becomes 
\begin{equation}
\label{ec:eompcmpluswz}
    \dr \big(i\mathscr{h}g^{-1}\dr g +\mathscr{k}\ast g^{-1}\dr g)=0\,.
\end{equation}
This equation no longer yields holomorphic and anti-holomorphic conserved currents, as the Kac-Moody symmetry is explicitly broken. While the model remains classically conformally invariant, this symmetry does not persist at the quantum level. For this reason, the WZW model, which corresponds to the choice of coefficient $\frac{i\mathscr{k}}{6}$ is often called the \emph{conformal point}. 

Notably, the PCM + WZ term is integrable in the Liouville sense. Let us consider a Lax connection with components 
\begin{equation}
    \mathcal{L}_{z}=\left(1+ \frac{\mathscr{h}}{\mathscr{k}}\right)\frac{g^{-1}\partial_z g }{1+ \lambda}\,, \quad \mathcal{L}_{\bar z}=\left(1- \frac{\mathscr{h}}{\mathscr{k}}\right)\frac{g^{-1}\partial_{\bar z} g }{1- \lambda}
\end{equation}
where $\lambda \in \CP$ is the spectral parameter. With a straightforward computation it can be shown that the flatness of this Lax connection, is equivalent to the equations of motion \eqref{ec:eompcmpluswz} for every value of $\lambda \in \CP$, which proves Lax integrability.  

It is particularly interesting to examine the limit $\mathscr{h}\to \mathscr{k}$, corresponding to the conformal point. In this case, the $\mathcal{L}_{\bar z}$ component vanishes, leading to a one-component Lax connection for the WZW model.

\section{Higher-Categorical Integrability}
\label{sec:catintegrability}

This section is perhaps the most audacious in the chapter, as this notion of integrability is not yet established in the literature. Rather, it is a proposal that we will attempt to justify as a reasonable candidate for an integrable structure. To be fair, this idea is neither entirely new nor radical; it is simply a generalization of the previous two concepts—Lax integrability and infinite-dimensional symmetry algebras—to three dimensions. In this sense, as long as we can construct analogous structures to those introduced in the preceding sections, the approach should be well-founded. This is precisely what we aim to prove.

As the title of this section suggests, we will be working with higher categorical structures. However, this is by no means a thesis on higher category theory. Our goal is to introduce only the essential concepts needed to present a well-defined, self-contained narrative, ensuring that all necessary elements are in place for what follows. 

Before diving into the details, let us motivate why this could be useful approach in the first place. We will build on the  "categorical ladder = dimensional ladder" proposal \cite{Crane:1994ty, Baez:1995xq}. It states that higher-dimensional physics can be described by higher categorical structures, and that one can "climb" the dimensions by categorification. In brief, a category can be thought of as a collection of objects together with morphisms on them, such that certain "coherence conditions" are satisfied. Category theory itself can thus be understood as the study of \textit{structure} and the relations between them. \textit{Categorification} is then, abstractly, a way to impart relations between structures in a coherent manner. Applying this idea to categories themselves gives rise to the notion of higher categories, which consist of objects, relations between these objects, relations between these relations, and so on. Indeed, as one climbs this categorical ladder, the structures that appear are suited, in each step, for describing higher dimensional data.

We begin by introducing the algebraic structure that will play the role of a "Lie algebra" in our three-dimensional generalization of integrability: a Lie algebra crossed module. This is a pair of Lie algebras related by a differential and an action, satisfying certain compatibility conditions. As we will see, this structure is remarkably well suited to describe the three-dimensional analogues of the objects we are looking to generalize, namely Lax pairs and infinite-dimensional symmetry algebras.

\begin{definition}\label{lie2algintro}
Let $\mathbb{k}$ denote a field of characteristic zero (such as $\mathbb{R}$ or $\mathbb{C}$). A \textbf{Lie algebra crossed module} $\mathfrak{G}=(\fh\xrightarrow{\mu_1} \fg , \rhd)$ over $\mathbb{k}$ consist of two Lie algebras $\big(\fh,[\cdot,\cdot]_\fh\big)$ and $\big(\fg,[\cdot,\cdot]\big)$ over $\mathbb{k}$, a Lie algebra homomorphism $\mu_1:\mathfrak{h}\to \fg$ and an action $\rhd:\mathfrak{g}\to \mathrm{Der}(\fh)$ satisfying
\begin{enumerate}
    \item Equivariance condition
    \begin{equation}
    \mu_1(\sfx\rhd \sfy) =[\sfx,\mu_1(\sfy)]
\end{equation}
for every $\sfx\in \fg$ and $\sfy \in \fh$.
    \item Peiffer identity 
    \begin{equation}
    \mu_1(\sfy)\rhd \sfy'=[\sfy,\sfy']_{\fh}
\end{equation}
for every $\sfy,\sfy'\in \fh$.
\end{enumerate}
\end{definition}
A Lie algebra crossed module \( \fG=(\fh \xrightarrow{\mu_1} \fg, \rhd) \) can be equivalently described as a strict \( 2 \)-term \( L_\infty \) algebra. In this formulation, we consider a chain complex of vector spaces concentrated in degrees \(-1\) and \(0\):
\[
L_{-1} = \fh, \qquad L_0 = \fg\,,
\]
with a differential \( \mu_1: \fh \to \fg \) of degree \( +1 \). The higher structure is encoded in a bilinear bracket \( \mu_2 \) of degree zero, which plays the role of a graded Lie bracket:
\begin{itemize}
  \item \( \mu_2(\sfx,\sfx') = [\sfx,\sfx'] \) for \( \sfx,\sfx' \in \fg \), recovering the Lie bracket on \( \fg \),
  \item \( \mu_2(\sfx,\sfy) = \sfx \rhd \sfy \) for \( \sfx \in \fg, \sfy \in \fh \), giving the action of \( \fg \) on \( \fh \),
  \item \( \mu_2(\sfy,\sfy') = 0 \) for \( \sfy,\sfy' \in \fh \), since the \( L_\infty \) structure is strict.
\end{itemize}

The two descriptions---as a Lie algebra crossed module and as a strict 2-term \( L_\infty \) algebra---are completely equivalent, and we will use them interchangeably throughout this work. While the crossed module language is often more familiar in the context of higher gauge theory, the \( L_\infty \) perspective makes the underlying grading structure more transparent. This grading will play a crucial role in the construction of higher connections and curvatures in what's next. 



\subsection{Higher Lax Integrability}
\label{sec:highercatlax}

Based on the construction of conserved quantities in two-dimensional field theories using the holonomy, it is natural to ask how these ideas might extend to higher dimensions. Indeed, one may think of the Lax equation \eqref{ec:spectrallax} as a conservation law, in the sense that the holonomy along a closed (contractible) path is a constant independent of the path. This reflects the presence of a topological invariant whose invariance under deformations gives rise to a conserved quantity. We want to generalize this idea to three dimensions, by looking for an analogous invariant defined over a two-dimensional surface, associated to a rank-2 antisymmetric tensor. Along these lines, conserved quantities in 3d integrable field theories may be constructed using a generalized holonomy over surfaces.

These generalized holonomies should be such that when integrated over a surface, they are independent of deformations of the surface as long as the boundary is fixed. These are the kind of invariants we are looking for, which would allow for the construction of conserved quantities.  Naively, in a space-time of dimension 3, one would think of introducing just a rank 2 antisymmetric tensor, and define the counterpart of $W_\gamma$ in \eqref{ec:defholonomy} as a surface-ordered exponential of this tensor. 
However, this approach alone is insufficient. Unlike in the case of line integrals, where a path has a natural ordering, a surface can be scanned in many different ways, each leading to a potentially different definition of surface ordering. To make the construction well-defined and geometrically meaningful, we must ensure that the resulting surface holonomy depends only on the surface itself, and not on the particular way it is traversed.

Following \cite{Alvarez:1997ma}, we begin with a very geometric construction of surface-ordered exponentials. While this approach is compelling and intuitive, we will see that it runs into limitations when trying to describe systems whose underlying Lie algebras are non-Abelian. We will then argue that these limitations can be naturally resolved using higher categorical structures—structures that, in hindsight, were implicitly present all along.

\begin{figure}
    \centering
\includegraphics[width=0.6\linewidth]{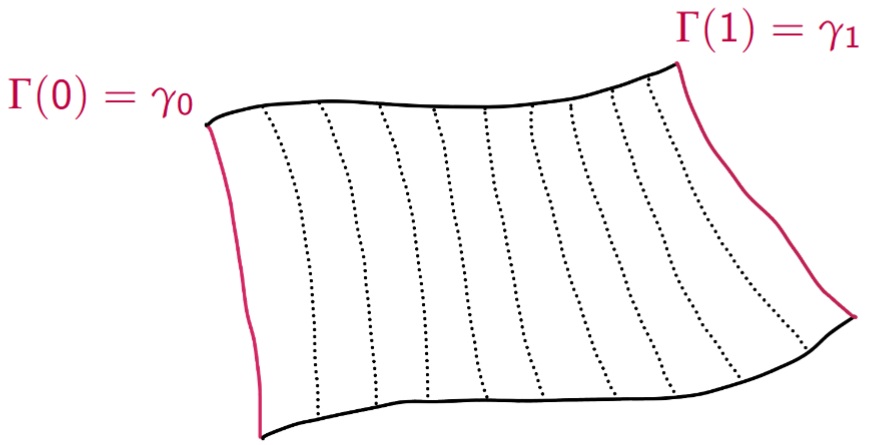}
    \caption{Surface swiped out by the path $\Gamma$ on $PY$}
    \label{fig:surfacespacetime}
\end{figure}

Recall that in the two-dimensional setting, we considered a principal $G$-bundle $P\to \Sigma$ over the surface $\Sigma$, equipped with a Lie algebra-valued connection $\cL$. The holonomy \eqref{ec:defholonomy} was then defined as the path-ordered exponential of $\cL$  along a curve $\gamma \subset \Sigma$. To generalize this construction to three dimensions, we consider a three-manifold $Y$ and seek a natural way to define a surface holonomy. A particularly elegant approach is to work with the \emph{space of paths} in $Y$, defined by
\begin{equation}
    PY \coloneqq \left\{ \gamma : [0,1] \to Y \,\middle|\, \gamma \in C^\infty([0,1], Y) \right\}\,.
\end{equation}
The key geometric idea is that a path in path space, \( \Gamma: [0,1] \to PY \), corresponds to a smooth family of paths \( \gamma_\tau = \Gamma(\tau) \) in \( Y \), which together sweep out a surface in $Y$, see fig \ref{fig:surfacespacetime}. This observation allows us to think of surface observables in \( Y \) as arising from path-ordered exponentials in \( PY \). To define such an object, we aim to construct a connection \( \mathcal{A} \) on a bundle over $PY$, whose path-ordered exponential over a path on $PY$ gives us a well defined notion of a surface ordered exponential over the corresponding surface in $Y$.

The natural question is then: how does one define a connection on path space, and what kind of geometric data is required for this? 
Providing a rigorous answer demands performing differential calculus on path space, which is an infinite-dimensional manifold and involves considerable technical subtleties. To keep the discussion accessible, we will instead offer a heuristic motivation that illustrates why the structures we introduce are reasonable. At a certain point, we will present the final expression for the connection $\cA$ on path space without going into the full formal derivation. Readers interested in the mathematical foundations are encouraged to consult \cite{Alvarez:1997ma, Baez:2004in, Schreiber:2004ma, Soncini:2014zra}.

In the usual setting, a connection on a bundle $E \to Y$ prescribes how to move within the total space $E$ as one moves through the base $Y$. Concretely, at each point in the base, the connection takes as input a tangent vector—i.e., a direction in which we're moving—and returns a way of lifting that motion to the bundle, effectively telling us how to "move" in the fibre.

Let us now consider a bundle over the path space $E\to PY$. Analogously, a connection $\cA$ here should describe how to move in the total space of the bundle as we move from one path $\gamma \in PY$ to a nearby path. Just as before, the connection should take as input a point in the base—a path $\gamma$—and a direction in which it is being deformed.

To unpack this geometrically, we shift perspective and think of the path $\gamma$ not as a point in the infinite-dimensional manifold $PY$, but rather as an extended object on $Y$. From this point of view, describing an infinitesimal deformation of $\gamma$ means analyzing how the entire path changes point by point in $Y$. At each point $\gamma(t)\in Y$, two pieces of data are relevant: first, the velocity vector $\dot \gamma(t) \in T_{\gamma(t)}Y$, which describes the direction of the path at the time $t$; and second, the deformation vector $\delta \gamma(t)  \in T_{\gamma(t)}Y$, which encodes how the path is being varied "transversely"—how the shape of the path changes as we move to a nearby path $\gamma' \in PY$. 

Hence, we look for an object that takes as input the pair of tangent vectors \((\dot{\gamma}(t), \delta\gamma(t))\) at each point \(\gamma(t) \in Y\) and returns an element of the Lie algebra. The natural candidate for such an object is a Lie algebra–valued 2-form \(B\) on \(Y\), which can be evaluated pointwise along the path. To compute the total contribution of \(B\) associated with a deformation of the path \(\gamma\), we integrate these local evaluations over the parameter \(t \in [0,1]\). However, the value of \(B\) at each point \(\gamma(t)\in Y\) lives in a different copy of the Lie algebra, so we cannot directly sum the contributions along the path. To make this possible, we use the parallel transport operator \( W_{\gamma(t)} \), defined by a connection $A$ on the principal bundle \( E \to Y \), to transport all values of \( B \) to a common reference point before integrating. 
We are thus led to the expression for the connection $\cA$ on the bundle over $PY$ given by
\begin{equation}
  \mathcal{A}[\gamma] = \int_0^1 \mathrm{d}t\, W_{\gamma(t)}^{-1} B_{\mu\nu}(\gamma(t)) W_{\gamma(t)} \, \dot \gamma(t)^\mu \, \delta \gamma^\nu(t)\,.
\end{equation}
It can be shown that $\cA$ has the correct transition laws and it therefore serves as a local $G$-connection on $PY$. 

Now remember that we want to use this connection over $PY$ to define a surface holonomy. Specifically, we define it as follows. Let $\Gamma:[0,1]\to PY$ be a path on path space. Let $M \subset Y$ be the surface defined by $M=\Gamma([0,1])$. We define the surface holonomy along $M$ as
\begin{equation}
    V_M \coloneqq P\mathrm{exp}\left(-\int_\Gamma \mathcal{A}\right)\,.
\end{equation}
This is a path-ordered exponential over path space $PY$, though geometrically, we may interpret it as surface-ordered exponential over the surface $M\subset Y$ swept out by $\Gamma$. The idea is to use \( V_M \) to construct conserved quantities, much like in the two-dimensional case. Recall that in two dimensions, the key condition for the path-ordered exponential to yield conserved quantities was the flatness of the connection \( \mathcal{L} \). In complete analogy, for \( V_M \) to be independent of the choice of \( \Gamma \) — as long as its endpoints are fixed — we require that the curvature $\mathcal{F}=\delta \cA + \tfrac{1}{2}[\cA,\cA]$ vanishes. The latter can be obtained by a direct though tedious computation and is given by \cite{Alvarez:1997ma}
\begin{multline}
        \mathcal{F}=\int_0^1 \mathrm{d}t \,\mathrm{Ad}^{-1}_{W(t)}(\mathrm{d}B+[A,B])_{\mu \nu \rho}\dot \gamma^\mu \delta \gamma^\nu \wedge \delta \gamma^\rho \\
         - \int_0^1 \mathrm{d}t\int_0^1 \dr t'[\mathrm{Ad}^{-1}_{W(t)}F_{\mu\rho}(\gamma(t)),\mathrm{Ad}^{-1}_{W(t')}B_{\nu \sigma}(\gamma(t'))]\frac{\mathrm d \gamma^\mu}{\dr t}\frac{\dr \gamma^\nu}{\dr t'}\delta \gamma^\rho\wedge\delta\gamma^\sigma \\
         +\int_0^1 \mathrm{d}t\int_0^1 \dr t'[\mathrm{Ad}^{-1}_{W(t)}B_{\mu\rho}(\gamma(t)),\mathrm{Ad}^{-1}_{W(t')}B_{\nu \sigma}(\gamma(t'))]\frac{\mathrm d \gamma^\mu}{\dr t}\frac{\dr \gamma^\nu}{\dr t'}\delta \gamma^\rho\wedge\delta\gamma^\sigma \,.
\end{multline}
In particular, we find that a set of conditions for $\cF = 0$ are 
\begin{equation}
    F = 0\,,\quad \mathrm{d}B+[A,B]=0\,,\quad [B,B]=0\,,
\end{equation}
that is, $A$ is a flat connection, the covariant derivative of $B$ with respect to $A$ vanishes, and $B$ is valued in some abelian subalgebra of $\fg$. While the first two conditions are geometrically quite natural, the requirement that $B$ takes values in an Abelian subalgebra is considerably more restrictive. In particular, it rules out the possibility of capturing genuinely non-Abelian features in the theory. To overcome this limitation, we turn to the framework of \emph{higher differential geometry} \cite{Baez:2002jn,Baez:2003fs, Baez:2004, Baez:2004in}, where, as we will see, the constraint $[B,B]=0$ can be  relaxed.

In regular differential geometry, the connection is modeled as a Lie algebra-valued 1-form, \( A \in \Omega^1(Y, \mathfrak{g}) \). The relevant geometric structure in this context is the tensor product \( \Omega^\bullet(Y) \otimes \mathfrak{g} \), where the total degree is simply the form degree. In the higher setting, we generalize this by considering instead $\Omega^\bullet(Y)\otimes \mathfrak{G}$ where $\mathfrak{G} = (\mathfrak{h} \xrightarrow{\mu_1} \mathfrak{g}, \rhd) $ is the Lie algebra crossed module we introduced earlier. In particular, when tensored with the de Rham complex, we may define the total degree as the sum of the form degree and the vector space grading degree. Thus, since $\deg(\fh)=-1$ and $\deg(\fg)=0$ we note that $A \in \Omega^1(Y)\otimes \fg$ and $B \in \Omega^2(Y)\otimes \fh$ have the same total degree equal to one. This motivates treating $(A,B)$ as a unique object which we shall call a $2$-connection. As we shall see, this 2-connection is precisely the appropriate structure for describing parallel transport in path space. Just as an ordinary connection has an associated curvature, we now seek to define the curvature associated to a 2-connection. This can be motivated from the theory of $L_\infty$ algebras (see, for instance, \cite{Jurco:2018sby} \S 4.1) and is defined by 
\begin{equation}
    \mathscr{F}(A,B)=(\underbrace{\mathrm{d}A+\tfrac{1}{2}[A,A]-\mu_1(B)}_{\in \Omega^1(Y)\otimes \fg},\underbrace{\mathrm{d}B+A \wedge^{\rhd}B}_{\in \Omega^2(Y)\otimes \fh})\,.
\end{equation}
The $\fg$-valued piece of the higher curvature, is usually referred to as the \textit{fake-curvature} because it differs from the usual curvature by the $\mu_1(B)$ piece. The second term, is known as the $2$-curvature. The object $A \wedge^\rhd B$ can be expressed in components as $(A_i \rhd B_{jk}) \mathrm{d}y^i\wedge \mathrm{d}y^j\wedge \mathrm{d}y^k$.

With the higher notions of connection and curvature in place, we are now ready to extend our earlier construction and define the corresponding connection on path space $PY$ in this generalized setting:
\begin{equation}
    \mathcal{A}[\gamma]=\int_0^1\mathrm{d}t\left(W^{-1}_{\gamma(t)}\rhd B_{\mu \nu}(\gamma(t))\right)\dot \gamma^\mu \delta \gamma^\nu \,.
\end{equation}
Note that $\mathcal{A}$ is now $\fh$ valued since $B$ is an $\fh$-valued $2$-form. In particular, in the same way than before, we may compute the curvature $\cF=\delta \cA + \tfrac{1}{2}[\cA,\cA]$ which now becomes
\begin{multline}
    \cF = \int_0^1\mathrm{d}t \, W_{\gamma(t)}^{-1}\rhd(\mathrm{d}B+A\wedge^\rhd B)_{\mu\nu\rho}\dot \gamma^\mu(t)\delta\gamma^\nu(t)\wedge \delta \gamma^\rho(t)\\
    -\int_0^1\mathrm{d}t\int_0^1\mathrm{d}t'\left(W_{\gamma(t)}^{-1}\left[F(A)-\mu_1(B)\right]W_{\gamma(t)}\right)_{\mu\rho}\rhd\left(W_{\gamma(t')}^{-1}\rhd B_{\nu \sigma}(t')\right)\dot \gamma^\mu(t)\dot \gamma^\nu(t')\delta\gamma^\rho(t)\wedge \delta  \gamma^{\sigma}(t') \,.
\end{multline}
We find that $\mathcal{F}=0$ if and only if the higher connection $(A,B)$ is flat. Namely, if 
\begin{equation}
\label{ec:flatnessss}
    F(A)-\mu_1(B)=0 \quad \text{and} \quad \mathrm{d}B+A\wedge^\rhd B=0\,.
\end{equation}
Crucially, this formulation overcomes the main limitation of the 
approach discussed earlier. There, in order to define a flat connection on path space, one was forced to impose the condition $[B,B]=0$ by hand. In the higher setting, by contrast, this term appears naturally as part of the fake curvature $F(A)-\mu_1(B)$, and its presence is fully dictated by the underlying $L_\infty$ structure. We thus see that this framework provides a natural setting for surface holonomies: the flatness of the connection on path space is precisely equivalent to the (higher) flatness of the higher connection $(A,B)$. 

Having constructed a flat connection on path space, we know that the associated path-ordered exponential $V_M$ is invariant under deformations of the path $\Gamma$ with fixed endpoints. Geometrically, this means that the surface holonomy is invariant under deformations of the surface traced out by $\Gamma$, as long as its boundary remains fixed. A rigorous proof of this statement is presented in \S\ref{sec:holonomies}.

This property allows us to define conserved quantities in direct analogy with the two-dimensional case discussed in \S \ref{sec:PCM}. Indeed, given a character $\chi$ of the Lie algebra $\fh$, under suitable conditions (see \S \ref{sec:holonomies} for details) the quantity 
\begin{equation}
    \mathfrak{v} = \chi (V_M)
\end{equation} 
will be conserved in time. In particular, as with the holonomy $W_\gamma$, if $V_M$ depends on a spectral parameter $\lambda \in \mathbb{C}P^1$, as in the case of the transfer matrix (see \eqref{ec:transfermatrix}), we may expand in powers of $\lambda$ to construct an infinite number of conserved quantities. 

We will say that a three-dimensional field theory is \textit{weakly integrable} if the equations of motion of the theory are equivalent to the flatness equations \eqref{ec:flatnessss} for some pair of spectral parameter dependent fields $(A,B)$ constructed from the fields of the theory. This definition is less established in the literature compared to the two-dimensional case. However, the core idea remains the same: the construction of an infinite number of independently conserved quantities. Ensuring involution, on the other hand, presents a different challenge. Unlike in the two-dimensional case, no analogous method for guaranteeing involutivity of the charges is currently known. Some initial investigations along these lines have been carried out in \cite{Chen:2023rxf}.

\begin{remark}
 In two dimensions, the existence of a Lax connection allows the equations of motion to be reformulated as a Riemann-Hilbert factorization problem, whose solvability is ensured by the Riemann-Hilbert theorem. In three dimensions, certain integrable models—such as the Ward model \cite{Ward:1988ie, Ward_1988}—also admit a Riemann-Hilbert formulation, with explicit methods for constructing soliton solutions \cite{Ward_1988, Ward:1988ie}. This raises the natural question of whether the Riemann-Hilbert problem in these cases is related to the higher Lax connection introduced above. 
\end{remark}

\subsection{Higher Kac-Moody Algebras}

\label{sec:introhigherkacmoody}

In \S\ref{sec:currentalgebras}, we discussed two-dimensional field theories with infinite-dimensional symmetry algebras, such as the Kac–Moody and conformal algebras, and argued that these highly constraining structures play a central role in enabling exact solutions at the quantum level. 

The existence of these algebras is closely tied to the structure of local conformal transformations in two dimensions, which correspond to analytic maps on the complex plane. Given the remarkable power of conformal symmetry in two dimensions, it is natural to ask whether the tools of conformal field theory can be extended to higher dimensions. However, one quickly encounters fundamental obstructions. Most notably, Liouville’s rigidity theorem implies that for \(d \geq 3\), all local conformal transformations extend to global ones, and the resulting conformal group is finite-dimensional. This rules out the possibility of local, infinite-dimensional conformal symmetry algebras in higher dimensions.

Thus, to realize infinite-dimensional symmetry in three-dimensional field theories, one must generalize the very notion of a chiral algebra into an infinite-dimensional structure suited to describe symmetries in three-dimensional dynamics. A natural candidate exhibiting this kind of symmetry structure are the so-known 3d topological-holomorphic field theories, which first appeared in the context of twisted 3d $\mathcal{N}=2$ supersymmetric field theories \cite{Aganagic:2017tvx, Costello:2020ndc}, and more recently in the contexts of Poisson vertex algebras \cite{Zeng:2021zef, Khan:2025rah} and twistorial field theories \cite{Garner:2023izn}. It is also the type of theory obtained in \cite{Chen:2024axr} from 5d 2-Chern-Simons theory. These are three-dimensional theories defined on $\bR\times \bC$ which are invariant under transformations along both the holomorphic and topological directions.

It turns out that the appropriate structure for studying topological-holomorphic symmetries in three dimensions is the so-called \emph{raviolo} \cite{Garner:2023zqn}. This space is constructed by gluing two formal disks along a shared punctured disk, \(\mathrm{Rav} = D \cup_{D^{\times}} D\), and serves as the three-dimensional analogue of \(\mathbb{C}^\times=\bC \setminus \{0\} \) in the algebraic formulation of three-dimensional current algebras. In particular, it provides the natural setting for defining the three-dimensional generalizations of vertex algebras, namely, raviolo vertex algebras. 
 
In \cite{Chen:2025ujx}, we show that the three-dimensional field theory constructed in \cite{Chen:2024axr}, which is the subject of Chapter~\ref{chap:5d2CS}, is a holomorphic–topological field theory whose symmetry algebra can be described using the raviolo structure introduced in \cite{Garner:2023zqn}, and defines a higher categorical analogue of the Kac-Moody algebra. As this work is still in preparation, we will limit ourselves here to outlining the main ideas.

Recall that in two-dimensions, we said that the infinite dimensional Kac-Moody symmetry was related to the existence of a holomorphic current whose modes were the generators of the Kac-Moody algebra. Notably, the Laurent expansion of the current $J(z)$ can be understood from a cohomological perspective. The condition $\bar \partial J = 0$ implies that
\begin{equation}
    J \in H^{(1,0)}_{\bar \partial}(\mathbb{C}^\times\otimes \fg) = \bC(\!(z)\!)\otimes \fg \,,
\end{equation}
where $\bC(\!(z)\!)$ denotes the space of formal Laurent polynomials with coefficients in $\bC$. In other words, the mode expansion of the current—central to the construction of the Kac–Moody algebra—arises from its realization as a nontrivial Dolbeault cohomology class.

To generalize this notion to \(\mathbb{R} \times \mathbb{C}\), one must choose the differential used to compute cohomology. In \cite{Garner:2023zqn}, the authors consider
\begin{equation}
    \dr' = \bar\partial + \dr_\tau,
\end{equation}
where \(\bar\partial\) is the Dolbeault differential on \(\mathbb{C}\), and \(\dr_\tau\) is the de Rham differential along the real line. In this setting, the generalization of holomorphicity is encoded in the condition
\begin{equation}
    \dr' J = 0\,,
\end{equation}
for some $J \in \Omega^\bullet(\bR\times \bC)$. Thus, one is naturally led to study the cohomology of the differential \(\dr'\), which is precisely the focus of \cite{Garner:2023zqn}. In particular, they show that
\begin{equation}
\label{ec:01cohomology}
    H_{\dr'}^{(1,0)}(\mathbb{R} \times \mathbb{C} \setminus \{0\}) = \mathbb{C}[\![z]\!]\,,
\end{equation}
where \(\mathbb{C}[\![z]\!]\) denotes the space of formal Taylor series in \(z\) with coefficients in \(\mathbb{C}\). Elements of this cohomology group are of the form \(L = L_z\, \dr z\), subject to the condition \(\dr' L = 0\). In particular, equation \eqref{ec:01cohomology} implies that any $L\in  H_{\dr'}^{(1,0)}(\mathbb{R} \times \mathbb{C} \setminus \{0\})$ admits a mode expansion of the form
\begin{equation}
    L = \sum_{n=0}^\infty L_n z^n \,.
\end{equation}

Since these field theories are built from higher categorical structures, it is natural to expect that conserved currents—like the connection introduced in the previous section—come in pairs. Indeed, we previously saw that the 2-connection takes the form \((A, B)\), with \(A \in \Omega^1(M, \mathfrak{g})\) and \(B \in \Omega^2(M, \mathfrak{h})\). In parallel, we now expect the conserved currents to also come in pairs \((L, H)\), where \(L \in \Omega^{(0,1)}(M, \mathfrak{g})\) and \(H \in \Omega^{(1,1)}(M, \mathfrak{h})\). Accordingly, we are led to compute the cohomology group of \((1,1)\)-forms—that is, objects of the form
\begin{equation}
\label{ec:rav2form}
   H = H_{z\bar z}\, \dr z \wedge \dr \bar z + H_{z\tau}\, \dr z \wedge \dr \tau\,. 
\end{equation}
Notably, in \cite{Garner:2023zqn} they show that
\begin{equation}
\label{ec:11cohomology}
    H_{\dr'}^{(1,1)}(\mathbb{R} \times \mathbb{C} \setminus \{0\}) = \operatorname{Span}_{\mathbb{C}}\{\Omega^0, \Omega^1, \Omega^2, \dots\} \,,
\end{equation}
where \(\Omega^m\) are the so-called degree-one \textit{raviolo differential forms}, given explicitly by
\begin{equation}
    \Omega^m = \frac{(-1)^m}{m!} \partial_z^m \omega, \qquad \omega = \frac{\tau\, \dr \bar{z} - \bar{z}\, \dr \tau}{\tau^2 + |z|^2}\,.
\end{equation}
Elements of  $H_{\dr'}^{(1,1)}(\mathbb{R} \times \mathbb{C} \setminus \{0\})$ are of the form \eqref{ec:rav2form} and satisfy $\dr'H =0$. In particular \eqref{ec:11cohomology} implies that any $H \in H_{\dr'}^{(1,1)}(\mathbb{R} \times \mathbb{C} \setminus \{0\})$ admits a mode expansion of the form 
\begin{equation}
    H = \sum_{m=0}^\infty H_m \Omega^m,
\end{equation}
where the \(\Omega^m\) are degree-one differential forms. The idea is that, in direct analogy with the two-dimensional case—where the modes \(J_n\) of the current \(J\) generate the Kac--Moody algebra—the modes \((L_n, H_m)\) of the higher current \((L, H)\) generate a higher Kac--Moody algebra. In principle, one can also construct the \emph{raviolo} vertex algebra associated with this higher Kac--Moody algebra. The ultimate goal is to use this structure to perform exact computations of correlation functions in the three-dimensional quantum field theory.

In this sense, we will also regard a three-dimensional theory as integrable if its symmetries are governed by an infinite-dimensional algebra, such as the higher Kac–Moody algebra.

\section{Anti Self-Dual Yang-Mills Equations}

\label{sec:ASDYMEQ}

Here, we introduce the final notion of integrability considered in this work: the anti-self-dual Yang-Mills (ASDYM) equations. In some respects, this definition differs from the previous ones, as it does not rely on the existence of an infinite number of conserved quantities. Instead, it is associated with the fact that ASDYM equations form a class of exactly solvable nonlinear partial differential equations \cite{ADHM}. 

There is, however, a strong connection between ASDYM equations and Lax integrability, as many Lax integrable field equations in two dimensions can be obtained from dimensional reductions of the ASDYM equations\footnote{We will not go into detail on this point in this introduction; for a complete treatment the reader may refer to \cite{Mason:1991rf}.}. Indeed, this led to the conjecture that ASDYM could serve as a unifying framework for two-dimensional integrable models, with every such field theory arising as a dimensional reduction of ASDYM. A central theme of this thesis is a generalization of this conjecture \cite{Kcostello, Bittleston:2020hfv}, proposing that both ASDYM and 2d Lax integrability belong to a broader framework, with holomorphic Chern-Simons theory serving as the unifying structure.

A natural starting point for discussing the ASDYM equations is Yang-Mills theory, the non-Abelian generalization of Maxwell's theory, which plays a central role in the Standard Model of particle physics. It also underpins key developments in topological field theory, string theory, and holography, particularly in the AdS/CFT correspondence. It is arguably the most important theory of modern theoretical physics.

Given a Lie algebra $\mathfrak{g}$ and a $\mathfrak{g}$-valued connection $A$, we define the Yang-Mills action 
\begin{equation}
    S_{\mathrm{YM}}[A]=-\frac{1}{4}\int_{\mathbb{R}^4} \langle F,\ast F\rangle \,,
\end{equation}
where $F=\dr A + \tfrac{1}{2}[A,A]$ is the curvature, and $\ast$ is the Hodge dual. The equations of motion of Yang-Mills theory can be obtained by varying the action with respect to $A$ and are given by
\begin{equation}
\label{ec:fullYMeq}
    \dr \ast F + [A,\ast F]=0 \,.
\end{equation}
These are the Yang-Mills equations, which are highly nonlinear and generally difficult to solve. However, there is a special subset of solutions which come from imposing a self-duality condition on the field strength. Recall that the Hodge dual in four-dimensions maps 2-forms to 2-forms and in Euclidean signature, it squares to the identity. Thus, we may consider gauge fields whose field strength satisfies
\begin{equation}
    F = \pm \ast F \,.
\end{equation}
The plus sign corresponds to self-duality, whereas the minus sign, to anti-self-duality. Notably, both self-dual and anti-self dual field strengths are solutions to the full Yang-Mills equations, as can be verified using the Bianchi identity
\begin{equation}
    \dr F + [A,F]=0\,.
\end{equation}
An exceptional feature of the ASDYM equations is that they admit exact solutions, known as instantons. Their solvability stems from the introduction of twistor theory by Penrose \cite{Penrose:1967wn,Penrose:1968me}, and from the subsequent development of the Penrose–Ward correspondence \cite{Ward:1977ta}, which relates solutions of the ASDYM equations to certain holomorphic vector bundles over twistor space. This framework was later leveraged by Atiyah, Drinfeld, Hitchin, and Manin to construct explicit solutions, leading to what is now known as the ADHM construction \cite{ADHM}. Since twistor theory and the Penrose-Ward correspondence play a central role in this thesis, we will now introduce these topics in detail.

\subsection{Twistor Space}

Twistor space was introduced by Roger Penrose in the late 1960s as an attempt to unify general relativity and quantum mechanics \cite{Penrose:1967wn, Penrose:1968me}. This approach was actively explored until the mid-1970s, when progress stalled due to technical problems. For the next twenty-five years, twistor methods found a natural application in the study of integrable equations, particularly in the context of self-dual field theories. During this period, notable developments included the Penrose-Ward transform \cite{Ward:1977ta} and the ADHM construction \cite{ADHM}. In the early 2000s, the subject was revitalized by the work of Witten \cite{Witten:2003nn}, who showed that twistor theory, when combined with string perturbation theory, provides an elegant framework for computing the tree-level S-matrix of Yang-Mills theory in four dimensions. Today, twistor methods play a central role in cutting-edge computations of scattering amplitudes. Their use builds on ideas originating from the Parke-Taylor formula \cite{Parke:1986gb}, which provides a compact expression for all MHV tree-level gluon scattering in Yang-Mills theory.

Here we will offer a custom-made introduction to twistors, which is greatly inspired by \cite{Cole:2024hyt}. Twistor space is a rather intricate mathematical structure, and depending on how it's presented, it can feel more or less motivated. The reality is that it's difficult to satisfy all tastes at once—different perspectives and interpretations of twistor space are useful for different purposes. In the context of this thesis, the most practical approach is to understand how to work with twistors in their spinorial form. However, from the author's perspective, introducing twistor space via spinors can feel abrupt and unmotivated. Perhaps the presentation here will still feel abstract or unmotivated. If the main interest is to learn how to use twistors in practical computations, particularly in their spinorial form, it is entirely reasonable to skip ahead to the relevant sections. The geometric intuition and background provided here can always be revisited later, as needed.

Twistor space is not an independent object but is always associated with an underlying space. For instance, Euclidean twistor space refers to the twistor space corresponding to four-dimensional spacetime with Euclidean signature. Similarly, one can define Lorentzian or hyperbolic twistor space, each adapted to different space-time signatures. Here we will focus on Euclidean twistor space, which we will denote with $\mathbb{PT}_{\mathbb{E}}$.

Various formulations of Euclidean twistor space appear in the literature, depending on the context and motivation. The most common are either 
\begin{equation}
\label{ec:twistordef1}
    \mathbb{PT}_{\mathbb{E}}=\mathbb{C}P^3\setminus \mathbb{C}P^1\,,
\end{equation}
or as the total space of the holomorphic vector bundle 
\begin{equation}
\label{ec:twistordef2}
    O(1)\oplus O(1)\to \mathbb{C}P^1\,.
\end{equation}
The aim of this section is to clarify the meaning of the definitions above and to explain why they are equivalent. We will initially adopt some nonstandard notation, as we believe it offers the clearest perspective on these constructions. As we proceed, we will gradually align our notation with the standard conventions.

Let us begin by briefly reviewing complex projective space, as it plays a central role in the constructions that follows. Projective $n$-space is the space of complex lines through the origin in $\mathbb{C}^{n+1}$, obtained as the quotient of nonzero elements of $\bC^{n+1}$ by complex scaling. More precisely, 
\begin{equation}
    \bC P^{n} = (\bC^{n+1}\setminus \{0\}) / \sim
\end{equation}
where $(\pi^0,\dots, \pi^n) \sim (\zeta^0,\dots ,\zeta^n)$ if there exist $r \in \bC^\ast$ such that 
\begin{equation}
    (\zeta^0,\dots ,\zeta^n) = r(\pi^0,\dots ,\pi^n) \,.
\end{equation}
In other words, in $\bC P^n$, two elements related by scaling belong to the same equivalence class, which we will denote by $[\pi^0: \dots: \pi^n] \in \bC P^n$. 

A twistor is an element of \( \mathbb{PT}_{\mathbb{E}}=\mathbb{C}P^3\setminus \mathbb{C}P^1 \), meaning it is of the form \( [\pi^0:\pi^1:\pi^2:\pi^3] \), where at least one \( \pi^i \) is nonzero. However, \( \mathbb{PT}_{\mathbb{E}} \) is \emph{not} exactly \( \mathbb{C}P^3 \), as it is defined by removing an embedded \( \mathbb{C}P^1 \) subspace. Specifically, the subspace we remove is given by 
\begin{equation}
 \{[0:0:\pi^2:\pi^3] \in \mathbb{C}P^3\} \subset \mathbb{C}P^3,
\end{equation}
which consists of all points in \( \mathbb{C}P^3 \) for which \( \pi^0 = \pi^1 = 0 \), leaving only the homogeneous coordinates \( [\pi^2:\pi^3] \), which parametrize a \( \mathbb{C}P^1 \). Twistor space is thus defined as \( \mathbb{C}P^3 \) with this copy of \( \mathbb{C}P^1 \) removed, explicitly given by 
\begin{equation}
    \mathbb{PT}_{\mathbb{E}} = \{[\pi^0:\pi^1:\pi^2:\pi^3] \in \mathbb{CP}^3 \mid (\pi^0, \pi^1) \neq (0,0)\}\,.
\end{equation}
A key consequence of \( \pi^0 \) and \( \pi^1 \) never vanishing simultaneously is that they define a \( \mathbb{C}P^1 \) on their own. This allows us to interpret \( \mathbb{PT}_{\mathbb{E}} \) as a bundle over \( \mathbb{C}P^1 \), with fibers given by the remaining two coordinates, which take values in \( \mathbb{C}^2 \). At first glance, one might expect that this implies that $\PTE$ is a trivial $\bC^2$ bundle over $\CP$, namely, \( \mathbb{PT}_{\mathbb{E}}  =  \mathbb{C}P^1 \times \mathbb{C}^2 \). However, as we shall see, this is not the case, and this  distinction is central to the nature of twistor space.

To analyze the structure of \( \mathbb{PT}_{\mathbb{E}} \) as a bundle over \( \mathbb{C}P^1 \), we introduce the natural projection \( p: \mathbb{PT}_{\mathbb{E}} \to \mathbb{C}P^1 \). Our goal is to show that \( \mathbb{PT}_{\mathbb{E}} \) is not a trivial \( \mathbb{C}^2 \) bundle over \( \mathbb{C}P^1 \). To do so, we compute the transition function between local frames and demonstrate that it is not globally trivializable. 

We cover \( \mathbb{CP}^1 \) with two standard charts, \( U_0 = \{ \pi^0 \neq 0 \} \) and \( U_1 = \{ \pi^1 \neq 0 \} \), and define local trivializations
\begin{equation}
\label{eq:loctriv0}
    \phi_0:U_0\times \mathbb{C}^2 \to p^{-1}(U_0), \quad \phi_0([\pi^0:\pi^1]\,, \pi^2, \pi^3) \mapsto [\pi^0:\pi^1: \pi^0\pi^2:\pi^0\pi^3]\,,
\end{equation}
\begin{equation}
\label{eq:loctriv1}
    \phi_1:U_1\times \mathbb{C}^2 \to p^{-1}(U_1), \quad \phi_1([\pi^0:\pi^1]\,,\pi^2, \pi^3) \mapsto [\pi^0:\pi^1: \pi^1\pi^2:\pi^1\pi^3]\,.
\end{equation}
The factors of \( \pi^0 \) and \( \pi^1 \) appearing in the fiber coordinates of \( \phi_0 \) and \( \phi_1 \) respectively, ensure that these maps are well-defined on \( \mathbb{C}P^1 \); namely, that they do not depend on the choice of representative of the equivalence class in $\CP$. 

Taking a point on the overlap region $[\pi^0:\pi^1]\in U_0\cap U_1$, the transition function $t_{[\pi^0:\pi^1]}:\bC^2 \to \bC^2$ must satisfy the compatibility condition  
\begin{equation}
    \phi_1\left([\pi^0:\pi^1]\,,\pi^2, \pi^3\right)=\phi_0\left([\pi^0:\pi^1],t_{[\pi^0:\pi^1]}(\pi^2,\pi^3)\right)\,.
\end{equation}
Using the explicit forms of $\phi_0$ and $\phi_1$ given in \eqref{eq:loctriv0} and \eqref{eq:loctriv1}, this condition becomes
\begin{equation}
    [\pi^0:\pi^1:\pi^1\pi^2:\pi^1\pi^3] = [\pi^0:\pi^1:\pi^0t_{[\pi^0:\pi^1]}(\pi^2,\pi^3)]\,,
\end{equation}
so that the transition function is given by the complex linear map
\begin{equation}
    t_{[\pi^0:\pi^1]} = \begin{pmatrix}
        \frac{\pi^1}{\pi^0} & 0 \\
        0 & \frac{\pi^1}{\pi^0}
    \end{pmatrix}\,.
\end{equation}
Since the transition function is not the identity, we conclude that \( \mathbb{PT}_{\mathbb{E}} \) is not a trivial \( \mathbb{C}^2 \) bundle over \( \mathbb{C}P^1 \). Instead, we observe that the transition function $t_{[\pi^0:\pi^1]}$ coincides with that of the standard holomorphic vector bundle $O(1)\oplus O(1)\to \mathbb{C}P^1$. We thus conclude that twistor space is the total space of this bundle, recovering the alternative definition \eqref{ec:twistordef2} commonly found in the literature

\subsubsection{Moduli Space of Complex Structures}

\begin{figure}[h]
    \centering
    \begin{tikzpicture}

    \draw[thick] (1,0,0) -- (2.75,0,0);

    \draw[dashed,->] (2.75,0,0) -- (2.75,2.5,0);
    \draw[dashed,->] (2.75,0,0) -- (4,-1,0);
    \draw[dashed,->] (5.25,0,0) -- (6,2.5,0); \draw[dashed,->] (5.25,0,0) -- (6.5,-1,0);   
    \draw[dashed,->] (8.5,0,0) -- (10.1,2.1,0); 
    \draw[dashed,->] (8.5,0,0) -- (9.6,-1.5,0); 
    
    \draw[thick] (3.5,0,0) -- (5.25,0,0);

    \draw[thick] (6.2,0,0) -- (8.5,0,0);

    \draw[thick] (9.65,0,0) -- (10.5,0,0);

    \draw[black] 
        (2,-1.5,0) coordinate (A3) -- 
        (2, 2,0) coordinate (B3) -- 
        (3.5, 1,0) coordinate (C3) -- 
        (3.5,-2.5,0) coordinate (D3) -- 
        cycle;

        \draw[black] 
        (4,-1.5,0) coordinate (A3) -- 
        (5, 2,0) coordinate (B3) -- 
        (6.5, 1,0) coordinate (C3) -- 
        (5.5,-2.5,0) coordinate (D3) -- 
        cycle; 

        \draw[black] 
        (7,-1,0) coordinate (A3) -- 
        (9, 2,0) coordinate (B3) -- 
        (10, 0.5,0) coordinate (C3) -- 
        (8,-2.5,0) coordinate (D3) -- 
        cycle;

    \node at (0.4,0) {\large \(\mathbb{C}P^1\)}; 
    \node at (1.6,2) {\large \(\mathbb{C}^2\)};
    \node at (4.5,2) {\large \(\mathbb{C}^2\)};
    \node at (8.5,2) {\large\(\mathbb{C}^2\)};
    
\end{tikzpicture}

    \caption{Twistor space as a non trivial $\mathbb{C}^2$ fibration over $\mathbb{C}P^1$. As one moves along $\CP$, the complex structure on the copies of $\bC^2$ changes.}
    \label{fig:twistorspace}
\end{figure}
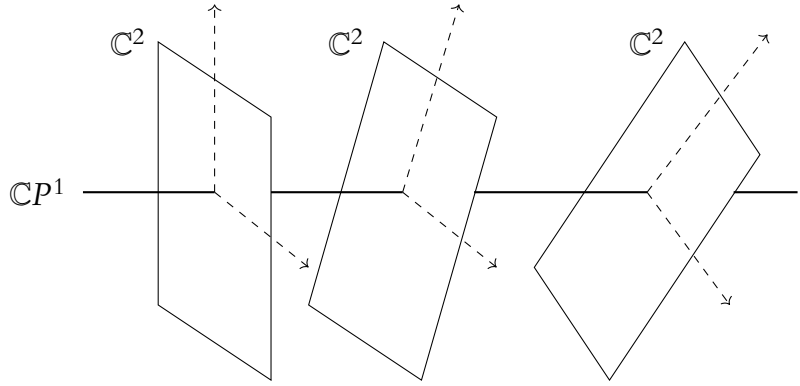

Having rigorously described twistor space as a mathematical object, we may now shift perspective and bring some geometric intuition into the picture. The key idea is that the non-triviality of the bundle \( O(1)\oplus O(1)\to \mathbb{CP}^1 \) reflects a simple but profound geometric fact: as one moves along \( \mathbb{C}P^1 \), the complex structure on the fibers \( \mathbb{C}^2 \) changes. In other words, as we move along \( \mathbb{C}P^1 \), the notion of which directions are ``holomorphic'' varies smoothly, as depicted in figure \ref{fig:twistorspace}.

Let us remind the reader what it means for a coordinate to be holomorphic on a manifold $M$. The key idea is to introduce an almost complex structure — a linear map \( J: TM \to TM \) on the tangent bundle satisfying \( J^2 = -1 \). This structure allows us to view the manifold locally as a complex one by specifying which tangent directions are holomorphic. Indeed, a coordinate \( z \) is said to be holomorphic if its associated vector field \( \partial/\partial z \) is an eigenvector of \( J \) with eigenvalue \( i \). Similarly, its complex conjugate \( \bar{z} \) is anti-holomorphic and corresponds to an eigenvector of \( J \) with eigenvalue \( -i \).

Notably, every even-dimensional real manifold can be equipped with an almost complex structure locally, so an important question is whether this structure extends globally. In order for $J$ to define a complex structure (as opposed to an almost complex structure) it must satisfy an integrability condition,
\begin{equation}
    N_J(X,Y)\coloneqq [X,Y]+J[JX,Y]+J[X,JY]-[JX,JY]=0
\end{equation}
for any two vector fields $X$ and $Y$. The object $N_J$ is known as the Nijenhuis tensor, and it measures a sort of curvature of the almost complex structure. If the Nijenhuis tensor vanishes, then one can consistently glue local $\mathbb{C}^n$ charts using holomorphic transition functions, which turns the manifold $M$ into a complex manifold. 

Returning to our case of interest, consider \(\mathbb{C}^2\) and its identification with \(\mathbb{R}^4\) as a real manifold. What distinguishes \(\mathbb{C}^2\) from \(\mathbb{R}^4\) is the fact that \(\mathbb{C}^2\) is endowed with a specific integrable complex structure. In this sense, \(\mathbb{C}^2\) can be seen as \(\mathbb{R}^4\) equipped with a particular choice of holomorphic and anti-holomorphic directions.

Crucially, this choice is not unique: there are infinitely many complex structures that one can place on \(\mathbb{R}^4\), each corresponding to a different way of identifying it with \(\mathbb{C}^2\). In the context of twistor theory, this freedom is used in a very specific way. To endow twistor space itself with a complex structure, one selects a particular complex structure on \(\mathbb{R}^4\) that depends smoothly on the coordinate of \(\mathbb{C}P^1\), the base of the twistor fibration. This varying choice of complex structure across \(\mathbb{C}P^1\) is what allows the identification of $\mathbb{PT}_\mathbb{E}$ with a complex three-manifold.

We can make this complex structure on \(\mathbb{R}^4\) explicit by expressing a set of holomorphic coordinates in terms of the standard real coordinates \(x^\mu\). In the patch \(U_0\), where \(\pi^0 \neq 0\), we introduce the inhomogeneous coordinate \(\zeta = \pi^1 / \pi^0\). The holomorphic coordinates on $\mathbb{R}^4$ are (locally) given by
\begin{equation}
\label{ec:cp1depcoord}
    v^1 = x^0+ ix^3 + \zeta( i x^1 + x^2)\,, \qquad
v^2 = ix^1 - x^2 + \zeta(x^0 - i x^3)\,.
\end{equation}
In the other patch \( U_1 \), where \( \pi^1 \neq 0 \) we introduce local coordinates \( (\tilde{\zeta}, \tilde{v}^1, \tilde{v}^2) \). Consistency of the coordinate charts on the overlap \( U_0 \cap U_1 \) requires that these be related to \( (\zeta, v^1, v^2) \) on \( U_0 \) by the transition functions
\begin{equation}
    \zeta = \frac{1}{\tilde{\zeta}} \,, \qquad v^1 = \frac{\tilde{v}^1}{\tilde{\zeta}} \,, \qquad v^2 = \frac{\tilde{v}^2}{\tilde{\zeta}} \,.
\end{equation}
The relation between the coordinates $v^i$, the real spacetime coordinates $x^\mu$, and the $\mathbb{C}P^1$ coordinate $\zeta$ is known as the incidence relations. Later on, we will re-express this relation in the more familiar spinorial form that commonly appears in the literature, and confirm its equivalence to the expression above.

Hence, for each \(\zeta \in \mathbb{C}P^1\), the coordinates $v^1$ and $v^2$ are holomorphic coordinates on $\mathbb{R}^4$. In other words, one can define a complex structure \(J_\zeta\) on \(\mathbb{R}^4\) such that the vector fields \(\partial / \partial v^1\) and \(\partial / \partial v^2\) span the \(+i\) eigenspace. When $J_\zeta$ is combined with the standard complex structure on \(\mathbb{C}P^1\), this yields a complex structure on the total space of the fibration, thereby endowing Euclidean twistor space with the structure of a complex three-manifold.

\subsubsection{The Correspondence Space}

Everything we have said so far pertains to Euclidean twistor space, which will remain our primary focus throughout this work. However, it is useful to place twistor space in a broader context that gives an alternative geometric origin. This viewpoint arises from the original ideas of Penrose and is most naturally formulated in complexified spacetime, namely, \(\mathbb{C}^4\) equipped with the Minkowski metric
\begin{equation}
    \mathrm{d}s^2 = (\dr x^0)^2 - (\dr x^1)^2 - (\dr x^2)^2 - (\dr x^3)^2\,,
\end{equation}
where \(x^\mu \in \mathbb{C}^4\). In this picture, twistor space can be seen as parametrizing the family of lightcones centered at points in $\mathbb{C}^4$, providing a natural arena for encoding the causal structure of the theory.

More precisely, a point in twistor space corresponds to  a totally null, self-dual 2-plane in \( \mathbb{C}^4 \). \emph{Totally null} means that every vector in the plane is null with respect to the complexified Minkowski metric, and any pair of vectors in the plane are orthogonal. \emph{Self-dual} refers to the property that any bivector formed by wedging two such vectors lies entirely within the self-dual subspace of \( \Lambda^2 \mathbb{C}^4 \). Planes satisfying both of these conditions are called $\alpha$-planes. It turns out (see, for instance \cite{Huggett_Tod_1994}) that the data of such an $\alpha$-plane can be succinctly captured by a triple of complex coordinates \( (\zeta, v^1, v^2) \), where \( \zeta \in \mathbb{C}P^1 \) specifies the direction of the plane and \( (v^1, v^2) \in \mathbb{C}^2 \) encodes its position. These coordinates parametrize complex twistor space $\mathbb{PT}_{\mathbb{C}}$ which is a three-dimensional complex manifold. In particular, the set of totally null, self dual 2-planes passing through a given point in $\mathbb{C}^4$ sweeps out the complex lightcone at that point.

On the other hand, we may ask the reverse question: what does a point in complexified spacetime correspond to in twistor space? The answer is that a point \( x \in \mathbb{C}^4 \) determines a family of totally null, self-dual 2-planes—namely, all the \(\alpha\)-planes that pass through \(x\). Each of these \(\alpha\)-planes corresponds to a point in twistor space, and together they trace out a complex projective line. In this way, a point in \( \mathbb{C}^4 \) defines a holomorphically embedded Riemann sphere \( \mathbb{C}P^1 \subset \mathbb{PT}_{\mathbb{C}} \).

The relationship between complexified spacetime \( \mathbb{C}^4 \) and twistor space \( \mathbb{PT}_{\mathbb{C}} \) is elegantly captured by the following diagram:
\begin{equation}
\label{ec:twistorcorrespondence}
\begin{tikzcd}[column sep=large, row sep=large]
& \mathbb{C}P^1 \times \mathbb{C}^4 \arrow[dl, "q"'] \arrow[dr, "p"] & \\
\mathbb{PT}_{\mathbb{C}} & & \mathbb{C}^4
\end{tikzcd}
\end{equation}
The correspondence space \( \mathbb{C}P^1 \times \mathbb{C}^4 \) maps to twistor space via the projection \( q \), and to spacetime via the projection \( p \). Explicitly, these maps act on a point \( (\zeta, x^\mu) \in \mathbb{C}P^1 \times \mathbb{C}^4 \) as
\begin{equation}
\label{ec:incidencerelation1}
    q(\zeta, x^\mu) = (\zeta, v^1, v^2)\,, \qquad p(\zeta, x^\mu) = x^\mu\,,
\end{equation}
where \( v^1 \) and \( v^2 \) are defined by the incidence relations given in~\eqref{ec:cp1depcoord}. 

This perspective makes it clear that twistor spaces for various spacetimes can be constructed by replacing the \( \mathbb{C}^4 \) in the bottom right corner of the diagram with other manifolds and defining the corresponding twistor correspondence. For instance, one can obtain the twistor space of a real spacetime by selecting an appropriate real slice of complexified spacetime, which corresponds to imposing specific reality conditions on the coordinates \( x^\mu \). Typical examples include:
\begin{align}
\label{ec:Euclideansignature}
    &\text{Euclidean:}\quad \mathbb{R}^4\,, && x^0 \in \mathbb{R} \,, \quad x^1,x^2,x^3 \in i\mathbb{R}\\
    &\text{Minkowski:}\quad \mathbb{M}^4\,, && x^\mu \in \mathbb{R}\\
    &\text{Split:}\quad \mathbb{R}^{2,2}\,, && x^0,x^2,x^3 \in \mathbb{R}\,, \quad x^1 \in i\mathbb{R}
\end{align}
with their corresponding twistor spaces given by
\begin{equation}
    \mathbb{PT}_{\mathbb{E}} = q(\mathbb{C}P^1 \times \mathbb{R}^4)\,, \quad 
    \mathbb{PT}_{\mathbb{M}} = q(\mathbb{C}P^1 \times \mathbb{M}^4)\,, \quad 
    \mathbb{PT}_{\mathbb{U}} = q(\mathbb{C}P^1 \times \mathbb{R}^{2,2})\,.
\end{equation}
These fixed-signature twistor spaces are all realized as subspaces of the complexified twistor space \( \mathbb{PT}_{\mathbb{C}} \), but they differ from one another in important ways.

It follows directly from the definition of \( q \) in~\eqref{ec:incidencerelation1} and the discussion of complex structures in the previous section that \( \mathbb{PT}_{\mathbb{E}} = q(\mathbb{C}P^1 \times \mathbb{R}^4) \) coincides with the total space of the vector bundle \( O(1) \oplus O(1) \to \mathbb{C}P^1 \). In the Euclidean case, the projection \( q \) is in fact a diffeomorphism, so that \( \mathbb{PT}_{\mathbb{E}} \cong \mathbb{C}P^1 \times \mathbb{R}^4 \) as real manifolds. This is consistent with our earlier observation: if we disregard the complex structures, Euclidean twistor space appears as a trivial \( \mathbb{R}^4 \)-bundle over \( \mathbb{C}P^1 \). However, the non-triviality of the complex geometry—namely, the twisting of the \( \mathbb{C}^2 \) fibers—is only apparent when we view twistor space as a complex manifold.

In the other signature cases, the projection $q$ is no longer a diffeomorphism; instead, the corresponding twistor spaces are lower-dimensional real manifolds. We will not explore these cases in detail here, but interested readers may consult~\cite{Huggett_Tod_1994, Adamo:2017qyl} for further discussion.

\subsection{Spinorial Form}

In the previous section, we introduced the twistor space of complexified spacetime \(\mathbb{C}^4\) as the general structure from which specific examples, such as the Minkowski, Euclidean, and split-signature twistor spaces, arise by considering different real slices of $\mathbb{C}^4$. A key advantage of working over complexified spacetime is that the relevant symmetry group is 
\begin{equation}
    SO(4,\mathbb{C}) \cong SL(2,\mathbb{C}) \times SL(2,\mathbb{C})/\mathbb{Z}_2 \,.
\end{equation}
This implies that we can replace the usual vector lorentz index $\mu=0,1,2,3$ with a pair of spinor indices $a=0,1$ and $\dot a=0,1$. More precisely, the fundamental (vector) representation of $SO(4,\bC)$ corresponds to the product representation $(\tfrac{1}{2},\tfrac{1}{2})$ of $SL(2,\bC)\times SL(2,\bC)$. In particular a spacetime vector $x^\mu$ can be mapped to a bispinor $x^{a\dot a}$ via
\begin{equation}
    x^{a \dot a} = \sigma_\mu^{a\dot a}x^\mu = \begin{pmatrix}
        x^0+x^3 & x^1-ix^2 \\
        x^1 + ix^2 & x^0-x^3
    \end{pmatrix}
\end{equation}
where $\sigma_\mu = (\mathbf{1},\sigma_i)$  is the set of Pauli matrices including the identity. In particular, this translation extends naturally to higher-rank tensors; for example, $T^{\mu \nu}\to T^{a\dot a b\dot b}$. Spinor indices are raised and lowered using the \(SL(2,\mathbb{C})\)-invariant Levi-Civita symbols, which serve as the spinor analog of the metric tensor. These are given by
\begin{equation}
    \epsilon_{ab} = \begin{pmatrix}
        0 & 1 \\
        -1 & 0
    \end{pmatrix} = \epsilon_{\dot{a}\dot{b}}\,,
\end{equation}
which in particular, are antisymmetric: \(\epsilon_{ab} = -\epsilon_{ba}\). Because of this antisymmetry, it is crucial to adopt a consistent convention for raising and lowering indices to avoid sign errors. In what follows, we use the mnemonic \emph{``raise from the right, lower from the left''}, meaning
\begin{equation}
    \alpha^a =\alpha_b \epsilon^{ba} \,, \qquad \alpha_a = \epsilon_{ab} \alpha^b\,,
\end{equation}
with analogous conventions for dotted indices. We can use the Levi-Civita symbols to define $SL(2,\mathbb{C})$-invariant inner products for both dotted and undotted spinor indices. We denote them by 
\begin{equation}
    \langle \alpha \beta \rangle = \alpha_a \beta^a = \epsilon_{ab} \alpha^b \beta^a \,,\quad [\kappa \mu]=\kappa_{\dot a} \mu^{\dot a}= \epsilon_{\dot a \dot b}\kappa^{\dot b}\mu^{\dot a} \,.
\end{equation}
Note that in particular, $\langle \alpha \beta \rangle = -\langle \beta \alpha \rangle$ and $\langle \alpha \alpha \rangle = 0$, as expected from the antisymmetry of the inner product. We will make frequent use of such expressions, especially when reformulating geometric structures in spinor terms. 

Let us now return to the twistor correspondence~\eqref{ec:twistorcorrespondence} and express it using spinor notation. The correspondence space \( \mathbb{CP}^1 \times \mathbb{C}^4 \) is often referred to as the \emph{projective spinor bundle}, where the \( \mathbb{CP}^1 \) fiber is parametrized by homogeneous spinor coordinates \( \pi_a \), and the base \( \mathbb{C}^4 \) is coordinatized by spacetime variables \( x^{a\dot a} \). In these terms, the projection map \( q \) defined in~\eqref{ec:incidencerelation1} takes the form
\begin{equation}
    q(\pi_a, x^{a\dot a}) = \left( \pi_a, \, \pi_a x^{a\dot a} \right) \,.
\end{equation}
The relation \( \mu^{\dot a} = \pi_a x^{a\dot a} \) is precisely the incidence relation~\eqref{ec:cp1depcoord}. 

As discussed in the previous section, all of this holds over complexified spacetime. Twistor spaces associated with different real spacetimes arise by choosing different real slices of \( \mathbb{C}^4 \). In what follows, we focus on the Euclidean case, and to do so, we must understand how to implement real slices directly in spinor notation. Recall from equation \eqref{ec:Euclideansignature} that taking spacetime coordinates $x^\mu$ with \( x^0 \in \mathbb{R} \) and \( x^1, x^2, x^3 \in i\mathbb{R} \) defines the Euclidean slice of \( \mathbb{C}^4 \). The question is how this condition translates to the bispinor form \( x^{a\dot a} \). In other words, we must define a conjugation map \( \hat{x}^{a\dot a} \) such that the condition \( \hat{x}^{a\dot a} = x^{a\dot a} \) encodes the correct Euclidean reality conditions on the underlying spacetime coordinates. This is given by 
\begin{equation}
    \hat{x}^{a\dot a} = \begin{pmatrix}
        \bar x^0-\bar x^3 & -\bar x^1+i\bar x^2 \\
        -\bar x^1-i \bar x^2 & \bar x^0+ \bar x^3
    \end{pmatrix} \,.
\end{equation}
In particular, the requirement that $\hat x^{a\dot a}=x^{a \dot a}$ is satisfied if and only if $x^0\in \mathbb{R}$ and $x^1,x^2,x^3 \in i\mathbb{R}$. Moreover, this operation induces a conjguation on spinors $\alpha^a$ and $\mu^{\dot a}$ given by 
\begin{equation}
    \hat \alpha^a = (-\bar \alpha^1, \bar \alpha^0) \,,\quad \hat \mu^{\dot a}=(-\bar \mu^{ 1}, \bar \mu^{0}) \,.
\end{equation}
Crucially, this operation is not an involution, applying it twice yields $\hat{\hat{\alpha}}^a = -\alpha^a$ rather than returning the original spinor. The same holds for dotted spinors. We therefore conclude that, in this context, a twistor of Euclidean twistor space is given by a pair $(\pi_a,\mu^{\dot a})$, where the spinor components satisfy the reality conditions discussed above.

Let us now pause for a moment and revisit the various representations of a twistor introduced so far, highlighting that they all describe the same underlying object:
\begin{equation}
\label{ec:inhomogcoord}
    [\pi^0 : \pi^1 : \pi^2 : \pi^3] \quad \simeq \quad (\zeta, v^1, v^2) \quad \simeq \quad (\pi_a, \mu^{\dot a}) \,.
\end{equation}
Here, the equivalence symbols indicate that these expressions are not strictly equal, but rather different representations of the same quantity.

The first form describes a twistor as a point in projective space \( \mathbb{CP}^3 \), expressed in homogeneous coordinates. Although it appears to involve four complex numbers, the overall rescaling symmetry reduces the number of independent complex degrees of freedom to three.

The second form, \( (\zeta, v^1, v^2) \), is a local coordinate expression valid on the patch \( \pi^0 \neq 0 \), where \( \zeta = \pi^1/\pi^0 \) and \( v^1, v^2 \) are fiber coordinates on the holomorphic vector bundle \( O(1) \oplus O(1) \to \mathbb{C}P^1 \).

Finally, the spinorial expression \( (\pi_a, \mu^{\dot a}) \) encapsulates the previous two by organizing the data in a way that makes the incidence relation \( \mu^{\dot a} = \pi_a x^{a\dot a} \) manifest. In this form, \( \mu^{\dot a} \) corresponds locally to the pair \( (v^1, v^2) \).

Throughout this work, we will primarily use the spinorial form \( (\pi_a, \mu^{\dot a}) \). However, in other contexts, authors often work in the inhomogeneous coordinate description \( (\zeta, v^1, v^2) \). The important point is that all of these are equivalent and describe the same object, simply presented in different coordinate systems or conventions.

\subsubsection{Calculus}

We conclude this section by introducing some of the basic differential geometry on twistor space, namely the vector fields and one-forms that will be used consistently throughout this work. While we will not go into the details behind certain choices made here, we refer the interested reader to the excellent discussions found in the PhD theses \cite{Cole:2024hyt, Sharma:2022arl}. Our aim in this section is simply to present the relevant expressions and highlight some of their key properties, particularly those that may appear unfamiliar. 

First, consider the spinor coordinates \((\pi_a, \mu^{\dot a}) = (\pi_a, \pi_a x^{a\dot a})\). There are several natural candidates for constructing a basis of 1-forms on twistor space, such as \(\mathrm{d}\pi_a\), \(\mathrm{d}\mu^{\dot a}\), and \(\mathrm{d}x^{a\dot a}\). While these are perfectly valid and perhaps the most immediate choices, they each present drawbacks in the twistor setting. For instance, \(\mathrm{d}x^{a\dot a}\) is not holomorphic on \(\mathbb{PT}_{\mathbb{E}}\), as the complex structure depends on the \(\mathbb{C}P^1\) fiber coordinate. Meanwhile, \(\mathrm{d}\pi_a\) fails to be homogeneous under projective rescalings: under \(\pi_a \mapsto r \pi_a\), it transforms as \(\mathrm{d}\pi_a \mapsto r\,\mathrm{d}\pi_a + \pi_a\,\mathrm{d}r\), acquiring an inhomogeneous term. Similarly, \(\mathrm{d}\mu^{\dot a} = \pi_a\,\mathrm{d}x^{a\dot a} + \mathrm{d}\pi_a\,x^{a\dot a}\) is not homogeneous due to the presence of the \(\mathrm{d}\pi_a\) term, which introduces the same inhomogeneous transformation behavior. To circumvent such issues, we adopt a different basis of 1-forms, originally introduced in~\cite{Boels:2006ir}, which is better adapted to the complex geometry of twistor space. These are given by:
\begin{equation}
\label{ec:1formbasis}
\begin{aligned}
    e^0 &= \langle \pi\, \mathrm{d}\pi \rangle \in \Omega^{1,0}(\mathbb{PT}_E) \otimes O(2)\,,\quad
    \bar{e}^0 = \frac{\langle \hat{\pi}\, \mathrm{d}\hat{\pi} \rangle}{\langle \pi \hat{\pi} \rangle^2} \in \Omega^{0,1}(\mathbb{PT}_E) \otimes O(-2)\,, \\
    e^{\dot{a}} &= \pi_a\, \mathrm{d}x^{a \dot{a}} \in \Omega^{1,0}(\mathbb{PT}_E) \otimes O(1)\,,\quad
    \bar{e}^{\dot{a}} = \frac{\hat{\pi}_a\, \mathrm{d}x^{a \dot{a}}}{\langle \pi \hat{\pi} \rangle} \in \Omega^{0,1}(\mathbb{PT}_E) \otimes O(-1)\,.
\end{aligned}
\end{equation}
Let us now see how these definitions address the issues discussed above. The 1-forms \(e^0\) and \(\bar{e}^0\) serve as replacements for \(\mathrm{d}\pi\) and \(\mathrm{d}\hat{\pi}\), providing well-defined 1-forms along the \(\mathbb{C}P^1\) base of the twistor fibration \(O(1) \oplus O(1) \to \mathbb{C}P^1\). The form \(e^0\) is holomorphic, as it depends only on the spinor \(\pi\), while \(\bar{e}^0\) is anti-holomorphic, depending only on \(\hat{\pi}\). Both \(e^0\) and \(\bar{e}^0\) are projectively well-defined: under the rescaling \(\pi \mapsto r \pi\), we have \(\langle \pi\, \mathrm{d}\pi \rangle \mapsto r^2 \langle \pi\, \mathrm{d}\pi \rangle\), so \(e^0\) transforms homogeneously and is valued in the line bundle \(O(2)\). Similarly, \(\bar{e}^0\) is valued in \(O(-2)\). The forms \(e^{\dot a}\) and \(\bar{e}^{\dot a}\), which are 1-forms along the \(\mathbb{C}^2\) fiber directions, are similarly well-behaved. They are projectively well-defined and, importantly, holomorphic and anti-holomorphic with respect to the \(\mathbb{C}P^1\)-dependent complex structure on \(\mathbb{PT}_{\mathbb{E}}\). 

In addressing the issues of holomorphicity and homogeneity, it becomes necessary to allow the 1-forms to take values in \(O(n)\) line bundles. This means that the forms transform nontrivially under projective rescalings. However, physical observables must remain invariant under such rescalings. For instance, when expressing a gauge field as \(\mathcal{A} = \bar e^0\, \mathcal{A}_0 + \bar e^{\dot a} \mathcal{A}_{\dot a}\), invariance under scaling requires that the components \(\mathcal{A}_0\) and \(\mathcal{A}_{\dot a}\) transform in the opposite way to the basis forms: specifically, \(\mathcal{A}_0\) must be a section of \(O(2)\), and \(\mathcal{A}_{\dot a}\) must be valued in \(O(1)\), so that the full 1-form \(\mathcal{A}\) is projectively well-defined, i.e., a section of \(O(0)\). In any case, this will be the primary choice of 1-forms used throughout this work.

The dual basis of vector fields is given by
\begin{equation}
\label{ec:twistorvecfields}
\begin{aligned}
    \partial_0 &= \frac{\hat{\pi}_a}{\langle \pi \hat{\pi} \rangle} \frac{\partial}{\partial \pi_a} \in \mathfrak{X}^{1,0}(\mathbb{PT}_E) \otimes O(-2)\,, \quad 
    \partial_{\dot a} = -\frac{\hat{\pi}^a}{\langle \pi \hat{\pi} \rangle} \partial_{a \dot a} \in \mathfrak{X}^{1,0}(\mathbb{PT}_E) \otimes O(-1)\,, \\
    \bar{\partial}_0 &= - \langle \pi \hat{\pi} \rangle \pi_a \frac{\partial}{\partial \hat{\pi}_a} \in \mathfrak{X}^{0,1}(\mathbb{PT}_E) \otimes O(2)\,,\quad
    \bar{\partial}_{\dot a} = \pi^a \partial_{a \dot a} 
    \in \mathfrak{X}^{0,1}(\mathbb{PT}_E) \otimes O(1)\,.
\end{aligned}
\end{equation}
An important and somewhat unusual feature of this choice of vector fields is that they do not commute. More precisely, their commutation relations are given by
\[
[\partial_0, \bar{\partial}_{\dot a}] = -\partial_{\dot a}, \qquad [\bar{\partial}_0, \partial_{\dot a}] = \bar{\partial}_{\dot a}.
\]
This behavior contrasts sharply with the more familiar case of coordinate vector fields forming a commuting basis of the tangent bundle.

\subsection{The Penrose-Ward Correspondence}

\label{sec:penroseward}

We are now ready to introduce and prove the Penrose–Ward correspondence, which asserts that holomorphic vector bundles over twistor space \(\mathbb{PT}_{\mathbb{E}}\) of rank \(k\), whose restriction to each \(\mathbb{C}P^1\) is holomorphically trivial, correspond bijectively to gauge equivalence classes of anti-self-dual Yang--Mills connections on \(\mathbb{R}^4\) with structure group \(GL(k, \mathbb{C})\).

At the beginning of this section, we emphasized that this correspondence is crucial because constructing holomorphic vector bundles over twistor space is often significantly easier than solving the ASDYM equations directly. This was the key insight behind the ADHM construction, where a large class of solutions to the ASDYM equations was obtained precisely by constructing holomorphic vector bundles over twistor space which are trivial when restricted to $\mathbb{C}P^1$\cite{ADHM}.

Before proceeding with the proof, we clarify what is meant by a holomorphic vector bundle of rank $k$ over twistor space. As the name suggests, it is a vector bundle \( E \to \mathbb{PT}_{\mathbb{E}} \) with typical fibre \(\mathbb{C}^k\) such that the local trivialisations are biholomorphic and the transition functions are holomorphic.  
These conditions can equivalently be formulated by requiring the existence of an anti-holomorphic covariant derivative $\nabla^{(0,1)} =  \bar{\partial} + \bar{\mathcal{A}}$ on twistor space that is integrable, meaning that $(\nabla^{(0,1)})^2 = 0$. We may express the latter condition in components using the basis of vector fields introduced in \eqref{ec:twistorvecfields} as
\begin{equation}
\label{ec:intconditionspw}
[\bar{\partial}_0 + \bar{\mathcal{A}}_0, \bar{\partial}_{\dot{a}} + \bar{\mathcal{A}}_{\dot{a}}] = 0\,,
\qquad
[\bar{\partial}_{\dot{a}} + \bar{\mathcal{A}}_{\dot{a}}, \bar{\partial}_{\dot{b}} + \bar{\mathcal{A}}_{\dot{b}}] = 0\,.
\end{equation}

Next, we recall how the ASDYM equations are naturally expressed in spinor language, which will allow us to relate them to holomorphic structures on twistor space. First, a basis for self-dual and anti-self-dual 2-forms on $\mathbb{R}^4$ are given by
\begin{equation}
\Sigma^{ab} = \epsilon_{\dot{a}\dot{b}}\, \mathrm{d}x^{a\dot{a}} \wedge \mathrm{d}x^{b\dot{b}},
\qquad
\tilde{\Sigma}^{\dot{a}\dot{b}} = \epsilon_{ab}\, \mathrm{d}x^{a\dot{a}} \wedge \mathrm{d}x^{b\dot{b}},
\end{equation}
where \(\Sigma^{ab}\) is symmetric in \(a,b\) and \(\tilde{\Sigma}^{\dot{a}\dot{b}}\) is symmetric in \(\dot{a},\dot{b}\). In this decomposition, we are making use of the fact that in four dimensions the group \(SO(4,\mathbb{R})\) is locally isomorphic to \(SL(2,\mathbb{R}) \times SL(2,\mathbb{R})\), and that the Hodge star distinguishes between these two factors: it acts as \(+1\) on the self-dual sector and as \(-1\) on the anti-self-dual sector.  
Explicitly, $\star \Sigma^{ab} = +\Sigma^{ab}$ and $\star \tilde{\Sigma}^{\dot{a}\dot{b}} = -\tilde{\Sigma}^{\dot{a}\dot{b}}$, 
so that \(\Sigma^{ab}\) spans the self-dual subspace and \(\tilde{\Sigma}^{\dot{a}\dot{b}}\) spans the anti-self-dual subspace. Using this basis, the field strength \(F\) decomposes as
\begin{equation}
F= \tilde{f}_{ab}\, \tilde{\Sigma}^{ab}
+f_{\dot{a}\dot{b}}\, \Sigma^{\dot{a}\dot{b}}   =(\tilde f_{ab}\epsilon_{\dot a \dot b}+f_{\dot a \dot b} \epsilon_{ab})\dr x^{a\dot a}\wedge \dr x^{b\dot b},
\end{equation}
where \(f_{\dot{a}\dot{b}}\) encodes the anti-self-dual part and \(\tilde{f}_{ab}\) the self-dual part. Thus, the anti-self-duality condition \(\star F = -F\) is equivalent to setting the self-dual component to zero, that is, $\tilde f_{ab} = 0$. In particular, using the anti-symmetry of the Levi-Civitta symbols, this condition can be written as
\begin{equation}
\label{ec:asdymeqpi}
    \pi^a\pi^b (\tilde f_{ab}\epsilon_{\dot a \dot b}+f_{\dot a \dot b} \epsilon_{ab})= 0\,,\quad \forall \pi^a \in \mathbb{C}P^1\,.
\end{equation}
With these preparations in place, we are now ready to begin the proof of the Penrose--Ward correspondence.  
We start with a holomorphic vector bundle over \(\mathbb{PT}_{\mathbb{E}}\), which, equivalently, is specified by an integrable partial connection \(\mathcal{A} \in \Omega^{(0,1)}(\mathbb{PT}_{\mathbb{E}}) \otimes \mathfrak{g}\), whose restriction to each \(\mathbb{C}P^1\) fibre is holomorphically trivial.  
Locally, we can expand \(\mathcal{A}\) in the basis of \((0,1)\)-forms introduced in \eqref{ec:1formbasis} as
\begin{equation}
    \mathcal{A} = \mathcal{A}_0\, \bar{e}^0 + \mathcal{A}_{\dot{a}}\, \bar{e}^{\dot{a}},
\end{equation}
where \(\mathcal{A}_0\) and \(\mathcal{A}_{\dot{a}}\) are Lie algebra valued sections of the line bundles \(O(2)\) and \(O(1)\), respectively.  
This follows from the requirement that \(\mathcal{A}\) must be homogeneous of weight zero, while the basis one-forms \(\bar{e}^0\) and \(\bar{e}^{\dot{a}}\) carry nontrivial weights. Moreover, the condition that the bundle is holomorphically trivial when restricted to each \(\mathbb{C}P^1\) imposes that the partial connection along the \(\bar{e}^0\) direction vanishes, that is, $\mathcal{A}_0 = 0$. Under these conditions, the first integrability equation \eqref{ec:intconditionspw} reduces to $[\bar\partial_0, \bar\partial_{\dot{a}} + \mathcal{A}_{\dot{a}}] = 0$, 
which implies that
\begin{equation}
    \bar\partial_0 \mathcal{A}_{\dot{a}} = 0\,.
\end{equation}
Thus, \(\mathcal{A}_{\dot{a}}\) is holomorphic along each \(\mathbb{C}P^1\) fibre.  
Since it is also a section of \(O(1)\), it must necessarily be linear in \(\pi^a\), and hence can be written as
\begin{equation}
    \mathcal{A}_{\dot{a}} = \pi^a A_{a\dot{a}},
\end{equation}
for some functions \(A_{a\dot{a}}\) on \(\mathbb{R}^4\), independent of \(\pi\). Defining $A = A_{a\dot{a}}\, \mathrm{d}x^{a\dot{a}} \in \Omega^1(\mathbb{R}^4) \otimes \mathfrak{g}$, we claim that \(A\) is an anti-self-dual connection.  
Indeed, the second of the integrability conditions \eqref{ec:intconditionspw} for \(\mathcal{A}\) reads
\begin{equation}
    0 = [\bar{\partial}_{\dot{a}} + \mathcal{A}_{\dot{a}},\, \bar{\partial}_{\dot{b}} + \mathcal{A}_{\dot{b}}]
    = \pi^a \pi^b [\partial_{a\dot{a}} + A_{a\dot{a}},\, \partial_{b\dot{b}} + A_{b\dot{b}}]\,,
\end{equation}
which is precisely the spinor form of the anti-self-dual Yang--Mills equation \eqref{ec:asdymeqpi}.

Conversely, given an ASDYM connection \(A = A_{a\dot{a}}\, \mathrm{d}x^{a\dot{a}}\) on \(\mathbb{R}^4\), we can construct an integrable partial connection \(\mathcal{A}\) over \(\mathbb{PT}_{\mathbb{E}}\) by setting
\begin{equation}
    \mathcal{A} = \pi^a A_{a\dot{a}}\, \bar{e}^{\dot{a}} \in \Omega^{(0,1)}(\mathbb{PT}_{\mathbb{E}}) \otimes \mathfrak{g}.
\end{equation}
This connection satisfies \(\mathcal{A}_0 = 0\), ensuring that it is holomorphically trivial when restricted to each \(\mathbb{C}P^1\) fibre.  
The integrability condition
\begin{equation}
    [\bar{\partial}_0 + \mathcal{A}_0,\, \bar{\partial}_{\dot{a}} + \mathcal{A}_{\dot{a}}] = 0
\end{equation}
is then satisfied given that $\mathcal{A}_{\dot a}$ is holomorphic by construction, while the second integrability condition reduces precisely to the ASDYM equations for \(A\), ensuring that \((\nabla^{(0,1)})^2 = 0\).

This completes the proof of the correspondence.

\chapter{Holomorphic Chern-Simons Theory}

\label{chap:hcstheory}

In the previous sections, we explored various notions of integrability across different dimensions, each grounded in distinct principles. Some were characterized by the existence of an infinite number of independently conserved quantities, leading to exact solvability, while others exhibited exact solvability without manifesting an underlying infinite-dimensional symmetry algebra. The idea that these seemingly disparate notions might be unified within a common framework is compelling—not merely for aesthetic reasons, but because such unifying structures often enable the transfer of techniques and insights across dual formulations. Notably, while integrable models are idealized and do not directly describe physical reality, the models discussed thus far often arise as subsectors or controlled deformations of more realistic theories relevant to our world. Hence, a unifying framework could provide a more systematic understanding of integrable structures themselves, while also clarifying the role of integrable models as building blocks within broader, non-integrable theories that contain integrable subsectors.

A major step toward such a unifying framework came from the work of Costello \cite{Costello:2013sla, Costello:2013zra}. These ideas were later developed and refined in collaboration with Witten and Yamazaki, culminating in the seminal series Gauge Theory and Integrability I, II, and III \cite{Costello:2017dso, Costello:2018zrm, Costello:2019tri}. In this work, the authors introduced a four-dimensional holomorphic variant of Chern-Simons theory that provides a gauge-theoretic origin for integrability. The first two papers focus on integrable lattice models in statistical mechanics and explain the origin of their quantum integrability from this perspective. The third paper, which is most relevant to our interests, turns to two-dimensional integrable field theories and provides the first attempt at explaining the appearance of the Lax connection, whose origin had remained somewhat mysterious until then.

Since the appearance of these articles, a great deal of work has gone into developing the four-dimensional framework further. These studies have led to the systematic construction of a broad class of two-dimensional integrable field theories, encompassing many well-known examples as well as a number of novel models that had not previously appeared in the literature \cite{Liniado:2023uoo}. It is yet unclear if this framework allows for the construction of every integrable field theory in two dimensions. 

Notably, these ideas sparked generalizations in other directions as well, in particular, to higher dimensional integrability. The first development came shortly after the original set of papers, once again inspired by ideas of Costello \cite{Kcostello}, and carried out independently by Bittleston and Skinner \cite{Bittleston:2020hfv} and by Penna \cite{Penna:2020uky}. Their key insight was that the anti-self-dual Yang–Mills (ASDYM) equations could be derived from a holomorphic Chern–Simons action on twistor space. This construction provided a concrete realization of the Penrose–Ward correspondence at the level of an action principle, and, in particular, provided an action-based origin of the ASDYM equations. 

Along a different line, higher-dimensional integrable models have been constructed using ideas from higher category theory. Building on the “categorical ladder = dimensional ladder” proposal, the author—together with Hank Chen, and independently Benoit Vicedo and Alexander Schenkel—proposed a five-dimensional holomorphic extension of higher Chern–Simons theory, from which one could construct integrable field theories in three dimensions through a mechanism analogous to that in \cite{Costello:2019tri}. Here, the three dimensional field theories are integrable in the sense described in \S \ref{sec:catintegrability}.

There are several connections between these holomorphic Chern–Simons theories, yet it remains unclear whether a unifying relation among them truly exists. Each of the notions of integrability described in the previous section finds a natural origin within one of these gauge-theoretic frameworks. However, despite their apparent similarities, it is still an open question whether these notions represent different facets of a single underlying structure. This remains an active area of investigation, and answering these questions may ultimately lead to a complete classification of integrable systems in arbitrary dimensions.

This chapter is organized as follows. We begin with an overview of three-dimensional Chern–Simons theory, focusing on its historical significance and the conventions we will adopt. We then introduce four-dimensional Chern–Simons theory, aiming to explain conceptually why the Lax connection arises naturally within this framework. To make this concrete, we work through a detailed example of how a two-dimensional integrable field theory emerges from the 4d theory. This example is presented thoroughly and accessibly, as it effectively captures the structure of all other constructions discussed in the thesis. Next, we discuss six-dimensional holomorphic Chern–Simons theory, emphasizing its role as an action-based realization of the Penrose–Ward correspondence. Finally, we turn to holomorphic 2-Chern–Simons theory, addressing how it provides a natural gauge-theoretic origin for higher categorical integrability.

\section{3d Chern-Simons Theory}

We consider a three-dimensional manifold $M$ and a compact, simple Lie group $G$. We pick a $G$-bundle $E\to M$ over $M$, and on $E$ we put a connection $A$ which we view as a Lie-algebra valued $1$-form $A \in \Omega^{1}(M)\otimes \fg$. The Chern-Simons action is given by 
\begin{equation}
    S_{\mathrm{CS}}[A] =\frac{k}{4\pi}\int_M \langle A,\mathrm{dA}+\tfrac{2}{3}A\wedge A\rangle\,.
\end{equation}
Here we use the notation $\langle \cdot, \cdot \rangle : \fg \times \fg \to \mathbb{R}$ for a symmetric, ad-invariant bilinear form on $\fg$. If $G$ were a matrix Lie group, then we can take $\langle \cdot, \cdot \rangle$ to be the trace. Chern-Simons theory is a topological field theory, in the sense that it does not depend on the metric of the manifold, rather, only on its topology. This is not only true classically, but crucially, it is also true at the quantum level. The understanding of this fact led Witten to uncover the connection between quantum Chern–Simons theory and the Jones polynomial \cite{Witten:1988hf}, a breakthrough that earned him the Fields Medal in mathematics. 

The equations of motion of Chern-Simons can be found by considering a variation of the action with respect to $\delta A$ and are given by 
\begin{equation}
    F(A)= \mathrm{d}A + A\wedge A = 0\,,
\end{equation}
namely, flatness of $A$. This is the first indication that Chern–Simons theory may be related to Lax integrability, as its dynamics are governed by the flatness of a connection. We will return to this later. Next, we examine how the action transforms under gauge transformations. For a transformation of the form
\begin{equation}
    A^g = g^{-1}Ag + g^{-1}\mathrm{d}g\,,
\end{equation}
we find that the action transforms as
\begin{equation}
\label{ec:introCSgt}
    S_{\mathrm{CS}}[A^g]=S_{\mathrm{CS}}[A]+\frac{k}{4\pi}\int_M \mathrm{d}\langle A,\dr g g^{-1}\rangle +\frac{k}{12\pi}\int_M\langle g^{-1}\dr g, g^{-1}\dr g\wedge g^{-1}\dr g\rangle \,.
\end{equation}
The first term is the original Chern–Simons action, while the second is a boundary term that vanishes when the manifold has no boundary. The third term, however, is non-zero and matches precisely the Wess–Zumino term encountered in \S\ref{sec:infdimsymalg}. This provides a second indication of a connection between Chern–Simons theory and integrable field theories, particularly those exhibiting infinite-dimensional symmetry algebras. Recall that the integral of the three-form
$ \omega = \langle g^{-1} \mathrm{d}g,\, g^{-1} \mathrm{d}g \wedge g^{-1} \mathrm{d}g \rangle$
measures the winding number of the map \( g \) around \( S^3 \). As a result, under gauge transformations with non-zero winding number, the Chern--Simons action is not strictly gauge invariant. However, in quantum field theory, it is the path integral of \( \exp(iS) \) that must be well defined, not the action \( S \) itself. Since the winding number is an integer, consistency requires that \( \exp(iS) \) remain invariant under gauge transformations, which holds if 
$\frac{k}{12\pi} \times (\text{winding number}) \in 2\pi \mathbb{Z}$. 
This leads to the condition that \( k \in \mathbb{Z} \), assuming a suitable normalization of the invariant bilinear form.

As mentioned in the introduction to this chapter, the first two papers relate holomorphic Chern–Simons theory to discrete integrable models. This connection can be motivated by observing that the Reidemeister moves in knot theory bear a strong resemblance to the Yang–Baxter equation. There is an insightful lecture by Witten that develops this perspective in detail, and we will not attempt to improve upon it here; instead, we refer the reader to \cite{Witten:2016spx}. Our aim in these final couple of paragraphs is to motivate the relevance of Chern–Simons theory from the standpoint of the Lax formalism, by clarifying both what the gauge-theoretic approach is able to capture and what essential features it leaves out, pointing to the need for a higher-dimensional framework.

As noted earlier, the equations of motion of Chern--Simons theory are given by \( F(A) = 0 \), i.e., the connection is flat. This mirrors the key condition for Lax integrability in two-dimensional field theories, where the equations of motion are equivalent to the flatness of a connection constructed from the fields of the theory. However, in that case, the connection must also depend on an additional parameter, the spectral parameter \( \lambda \in \mathbb{C}P^1 \), which is essential for generating the infinite set of conserved quantities. Let us now attempt to proceed along these lines and examine where the approach breaks down.

We consider a gauge field \( A \in \Omega^1(M) \otimes \mathfrak{g} \), where \( M \) is a three-dimensional manifold with coordinates \( x^1, x^2, x^3 \), so that we can write
\begin{equation}
   A = A_i(x^1, x^2, x^3)\, \mathrm{d}x^i\,. 
\end{equation}
As is standard in gauge theory, part of the gauge freedom can be used to fix one component of the gauge field; for example, we set \( A_1 = 0 \). Suppose now that we label the remaining coordinates as \( x^2 = t \) and \( x^3 = \sigma \), so that the gauge field becomes
\begin{equation}
  A = A_t(x^1, t, \sigma)\, \mathrm{d}t + A_\sigma(x^1, t, \sigma)\, \mathrm{d}\sigma  
\end{equation}
This setup looks quite promising for a Lax connection. However, recall that we need the connection to depend on a complex spectral parameter \( \lambda \in \mathbb{CP}^1 \), which should not live on the spacetime of the field theory, but rather on an auxiliary complex curve. In our case, the only remaining coordinate is \( x^1 \), which is real and cannot play this role directly. This suggests that we should complexify \( x^1 \) for it to become the spectral parameter, yet this leads to a second issue. By complexifying one of the real directions of the manifold, the gauge field would acquire an additional component along the new direction. Since we have already used gauge freedom to fix \( A_1 = 0 \), we no longer have the freedom to eliminate this new component.

It turns out that complexifying the remaining direction is indeed the right thing to do, but it must be done very carefully. This is precisely the insight behind the work of Witten, Costello, and Yamazaki, to which we now turn.

\section{4d Chern-Simons Theory}

Let us take the three-dimensional manifold \( M \) to be of the form \( M = \mathbb{R} \times \Sigma \), and complexify and compactify the \( \mathbb{R} \) factor into \( \mathbb{C}P^1 \). We define the resulting four-dimensional space as \( X = \mathbb{CP}^1 \times \Sigma \), with local coordinates \( (z, \bar{z}, t, \sigma) \). Our goal is to construct an action using the Chern--Simons 3-form. However, since \( X \) is now four-dimensional, integrating a 3-form alone does not yield a well-defined action. To obtain a valid integrand on a four-manifold, we must wedge the Chern--Simons 3-form with a 1-form. We can therefore write:
\begin{equation}
    S_{\mathrm{CS}_4}[A]=\int_X \omega \wedge \langle A,\mathrm{d}A+\tfrac{2}{3}A\wedge A\rangle\,.
\end{equation}
In principle, \( \omega \) could be any 1-form on \( \mathbb{C}P^1 \). However, the particular choice of \( \omega \) plays a crucial role. Following the construction in~\cite{Costello:2019tri}, and as we will assume throughout this thesis, \( \omega \) is taken to be a meromorphic 1-form along the \( \mathrm{d}z \) direction, meaning that it depends solely on \( z \), and not on \( \bar{z} \). Explicitly, we write
\begin{equation}
\label{ec:introomega1}
    \omega = \varphi(z)\, \mathrm{d}z
\end{equation}
for some meromorphic function \( \varphi(z) \). This is not, in fact, the most general choice that yields sensible results. As shown in~\cite{Cole:2024noh}, one can also consider choices of \( \omega \) that depend on \( \Sigma \), leading to integrable models that arise in the context of gravity. 

Before specifying an explicit form for \( \varphi(z) \), let us see what can already be deduced from the structure of equation~\eqref{ec:introomega1}. First, since \( \omega \sim \mathrm{d}z \), any term in the Chern--Simons 3-form that also contains a \( \mathrm{d}z \) will not contribute to the action. In particular, there is a symmetry \( A \to A + \chi \), with \( \chi \in \Omega^{(1,0)}(\mathbb{CP}^1) \otimes \mathfrak{g} \), which allows us to fix \( A_z = 0 \). Note that this is not the usual gauge symmetry of the theory, but rather an additional symmetry that emerges due to the insertion of \( \omega \) in the action.

On the other hand, the Riemann--Roch theorem for 1-forms on \( \mathbb{C}P^1 \) tells us that
\begin{equation}
    \text{number of poles} - \text{number of zeroes} = 2\,,
\end{equation}
counted with multiplicity. For readers unfamiliar with this result, one can simply write down a few meromorphic 1-forms and convince themselves that it holds. In particular, the 1-form \( \mathrm{d}z \) has a double pole at infinity, which can be seen by performing the change of variables \( z \to \frac{1}{\zeta} \).

Since we always have at least two poles (or a double pole), it is natural to ask what physical role these play. We can think of the poles as introducing pointlike punctures in \( \mathbb{C}P^1 \), effectively generating a boundary in the manifold. Given that the total space is \( X = \mathbb{C}P^1 \times \Sigma \), the boundary becomes a codimension-two submanifold of the form \( \{\text{pole}\} \times \Sigma \). 
In gauge theory, the presence of a boundary typically breaks the gauge symmetry—at least partially—at the boundary’s location. Now the role of gauge transformations is to identify physically indistinguishable
field configurations, by killing would-be degrees of freedom. So
restricting the kind of gauge transformations we allow by breaking part of the symmetry will resurrect some of
these degrees of freedom from the dead \cite{Tong:2016kpv}. Said differently, breaking the gauge symmetry of the theory induces dynamics precisely where the symmetry is broken, namely, on the boundary. 

Crucially, the dynamics of the boundary theory are inherited from those of the bulk, whose equations of motion impose the flatness of the connection. This immediately suggests—before performing any explicit computation—that the resulting two-dimensional theory on the boundary should be governed by the flatness of a connection, which is precisely the hallmark of Lax integrability. The goal of the remainder of this section is to make this idea precise.

\subsubsection{The Principal Chiral Model}

To make the argument as concrete as possible, we will carry it out for a specific choice of $\omega$, namely
\begin{equation}
\omega = \frac{1-z^2}{z^2}\mathrm{d}z \,.
\end{equation}
This example captures the essential features of the general argument while remaining technically straightforward, avoiding the need for additional machinery. Of course, this is neither the only possible choice nor the most general. In fact, Chapter \ref{chap:4dCS} is devoted to studying the most general admissible forms of $\omega$, and to analyzing the corresponding two-dimensional integrable field theories that arise from them.

Let us begin by considering a variation of the action with respect to a field variation
\begin{equation}
    \delta S_{\mathrm{CS}_4} = \frac{i}{2\pi}\int_X \omega \wedge \langle \delta A, F(A)\rangle + \frac{i}{4 \pi}\int_X \omega \wedge \mathrm{d}\langle \delta A, A\rangle \,,
\end{equation}
where $F(A)=\mathrm{d}A+\tfrac{1}{2}[A,A]$ is the curvature. The first term leads to the bulk equations of motion, $F(A)=0$, whereas the second term is a boundary term. In the three-dimensional setting, if $X$ has no boundary, this term would simply vanish. However, the presence of $\omega$ will now lead to a non trivial contribution. The fact that this will become a non trivial boundary term is attributed to the action of the exterior derivative on $\omega$. Indeed, using the identity\footnote{See Appendix A in \cite{Lacroix:2021iit} for a proof of this identity.}
\begin{equation}
     \quad \partial_{\bar z}\left(\frac{1}{(z-y)^{p+1}}\right)=\frac{(-1)^{p+1}2\pi i}{p!} \partial_z^p\delta^{(2)}(z-y)
\end{equation}
where the complex delta distribution satisfies
\begin{equation}
    \int_{\CP}\mathrm{d}z\wedge \mathrm{d}\bar z \, \delta^{(2)}(z-y)f(z,\bar z)=f(y,\bar y)\,,
\end{equation}
one finds, upon integration by parts 
\begin{equation}
       \frac{i}{4 \pi}\int_X \mathrm{d}\omega \wedge \langle \delta A, A\rangle 
       = \frac{1}{2}\int_\Sigma \langle \delta A, A\rangle|_{z=0} - \langle \delta A, A\rangle|_{z=\infty} \,.
\end{equation}

Hence, to ensure that the action principle is well-defined—that is, that \( \delta S_{\mathrm{CS}_4} = 0 \) on shell—we must impose boundary conditions on the gauge field at the locations of the poles of \( \omega \). We choose
\begin{equation}
    A|_{z=0} = 0\,, \quad A|_{z=\infty} = 0\,,
\end{equation}
and restrict to field variations that satisfy the same conditions. This ensures that the variation of the action vanishes on shell.

Let us now examine the bulk equations of motion. These arise from the requirement that the bulk variation term vanishes. Note, however, that the 1-form \( \omega \) vanishes at \( z = \pm 1 \), and therefore the equations of motion \( F(A) = 0 \) hold only away from those points. 

Recall, as mentioned earlier, that the presence of \( \omega \) induces an additional symmetry of the theory, namely \( A \to A + \chi \), where \( \chi \in \Omega^{(1,0)}(\mathbb{C}P^1) \otimes \mathfrak{g} \). This allows us to fix the holomorphic component \( A_z \) to zero. In this gauge, the gauge field takes the form
\begin{equation}
    A = A_{\bar z}(z, \bar z, t, \sigma)\, \mathrm{d}\bar z + A_t(z, \bar z, t, \sigma)\, \mathrm{d}t + A_\sigma(z, \bar z, t, \sigma)\, \mathrm{d}\sigma \,,
\end{equation}
and the equations of motion become
\begin{align}
    & \partial_{\bar z} A_t - \partial_t A_{\bar z} + [A_{\bar z}, A_t] = 0\,, \\
    & \partial_{\bar z} A_\sigma - \partial_\sigma A_{\bar z} + [A_{\bar z}, A_\sigma] = 0\,, \\
    & \partial_t A_\sigma - \partial_\sigma A_t + [A_t, A_\sigma] = 0\,.
\end{align}
Next, we turn to gauge transformations. From equation \eqref{ec:introCSgt}, we find that under a gauge transformation the action transforms as
\begin{equation}
\label{ec:introCS4gt}
    S_{\mathrm{CS}_4}[A^g] = S_{\mathrm{CS}_4}[A] + \frac{ik}{4\pi} \int_X \omega \wedge \mathrm{d} \langle A, \mathrm{d}g\, g^{-1} \rangle + \frac{ik}{12\pi} \int_X \omega \wedge \langle g^{-1} \mathrm{d}g, g^{-1} \mathrm{d}g \wedge g^{-1} \mathrm{d}g \rangle \,.
\end{equation}
The presence of $\omega$ implies that gauge invariance of the action will only hold for transformations that satisfy specific boundary conditions. Let us analyze the second and third terms separately. Integrating by parts, and letting the exterior derivative act on \( \omega \), we find
\begin{equation}
\begin{split}
    \frac{ik}{4\pi} \int_X \omega \wedge \mathrm{d} \langle A, \mathrm{d}g\, g^{-1} \rangle 
    &= \frac{k}{2}\int_\Sigma \partial_z \langle A, \mathrm{d}g\, g^{-1} \rangle \big|_{z=0} - \partial_z \langle A, \mathrm{d}g\, g^{-1} \rangle \big|_{z=\infty} \\
    &= \frac{k}{2}\int_\Sigma \langle \partial_z A, \mathrm{d}g\, g^{-1} \rangle \big|_{z=0} - \langle \partial_z A, \mathrm{d}g\, g^{-1} \rangle \big|_{z=\infty} \,.
\end{split}
\end{equation}
In deriving the second line, we used the boundary conditions \( A|_{z=0} = A|_{z=\infty} = 0 \). Therefore, for this term to vanish, we must impose \( \mathrm{d}g\, g^{-1} |_{z=0} = 0 \) and \( \mathrm{d}g\, g^{-1} |_{z=\infty} = 0 \), which can be achieved by requiring\footnote{These are not the most general gauge transformations that preserve the boundary conditions. One could, for instance, allow \( g|_{z = 0,\infty} \) to be a constant element of the group \( G \). The fact that we make this choice will result in an additional global symmetry of the resulting two-dimensional theory.}
\begin{equation}
\label{ec:gbcintro}
    g|_{z=0} = \mathrm{id}\,, \quad g|_{z=\infty} = \mathrm{id}\,.
\end{equation}
These boundary conditions also ensure that the third term in \eqref{ec:introCS4gt} vanishes. Indeed, we can apply a standard trick: extend the manifold \( X \) to \( X \times [0,1] \), and extend the gauge transformation \( g \) to a map \( \tilde{g}: X \times [0,1] \to G \) such that \( \tilde{g}|_{X \times \{0\}} = g \) and \( \tilde{g}|_{X \times \{1\}} = \mathrm{id} \). Since the Wess–Zumino 3-form is closed,
\[
\mathrm{d} \langle g^{-1} \mathrm{d}g, g^{-1} \mathrm{d}g \wedge g^{-1} \mathrm{d}g \rangle = 0\,,
\]
we can apply Stokes' theorem to write:
\begin{equation}
\begin{split}
    \frac{ik}{12\pi} \int_X \omega \wedge \langle g^{-1} \mathrm{d}g, g^{-1} \mathrm{d}g \wedge g^{-1} \mathrm{d}g \rangle 
    &= \frac{ik}{12\pi} \int_{X \times [0,1]} \mathrm{d} \left( \omega \wedge \langle \tilde{g}^{-1} \mathrm{d} \tilde{g}, \tilde{g}^{-1} \mathrm{d} \tilde{g} \wedge \tilde{g}^{-1} \mathrm{d} \tilde{g} \rangle \right) \\
    &= \frac{ik}{12\pi} \int_{X \times [0,1]} \mathrm{d}\omega \wedge \langle \tilde{g}^{-1} \mathrm{d} \tilde{g}, \tilde{g}^{-1} \mathrm{d} \tilde{g} \wedge \tilde{g}^{-1} \mathrm{d} \tilde{g} \rangle \,.
\end{split}
\end{equation}

As before, the exterior derivative \( \mathrm{d}\omega \) localizes the integral to the poles, and the boundary conditions \eqref{ec:gbcintro} ensure that the third term vanishes as well.\footnote{Some care is required here. Since we extended \( g \) to a map \( \tilde{g} \) on \( X \times [0,1] \), the integral is effectively supported on \( \Sigma \times [0,1] \). We assume that the boundary conditions imposed on \( \Sigma \) are sufficient to guarantee that the full integrand vanishes upon integration.} This makes explicit how the presence of the boundary breaks gauge invariance: not all gauge transformations leave the action invariant, but only those that satisfy the appropriate boundary conditions.

Armed with all these preliminaries, we are now ready to perform the localization of the four-dimensional action to a two-dimensional one. The guiding principle is to construct a two-dimensional integrable field theory characterized by a Lax connection. Our goal is to reinterpret the gauge field $A$ as this Lax connection. However, the presence of a non-vanishing $A_{\bar z}$ component obstructs this interpretation. Notably, this component can be gauged away everywhere except at the poles of $\omega$, where, as we have shown, the gauge symmetry is broken. At these points, eliminating $A_{\bar z}$ induces nontrivial dynamics, corresponding to a two-dimensional integrable field theory. What follows is the precise mathematical formulation of this argument.

Let us begin by introducing new field variables \( A' \in \Omega^1(X) \otimes \mathfrak{g} \) and \( h' \in C^\infty(X) \otimes G \), which reparametrize the original gauge field \( A \) as
\begin{equation}
    A = (h')^{-1} A' h' + (h')^{-1} \mathrm{d}h'\,.
\end{equation}
This transformation resembles a gauge transformation, and indeed, if \( h' \) were to satisfy \( h'|_{z=0} = h'|_{z=\infty} = \mathrm{id} \), then \( A \) and \( A' \) would be gauge equivalent, and the action would satisfy \( S_{\mathrm{CS}_4}[A] = S_{\mathrm{CS}_4}[A'] \). However, we will \emph{not} impose this boundary condition on \( h' \), since \( h' \) will instead encode the boundary degrees of freedom.

It is important to note that this reparametrization introduces redundancy: we can simultaneously transform both \( A' \) and \( h' \) via a new field \( u \in C^\infty(X) \otimes G \) as
\begin{equation}
    A' \mapsto u^{-1} A' u + u^{-1} \mathrm{d}u \,, \quad h' \mapsto u^{-1} h' \,,
\end{equation}
without changing \( A \). We will now use this redundancy to simplify the form of \( A' \) by eliminating its \( \bar{z} \)-component. This is achieved by performing a gauge transformation with \( u_1 \) satisfying
\begin{equation}
   \partial_{\bar z} u_1\, u_1^{-1} = - A'_{\bar z} \,.
\end{equation}
We denote the resulting field by \( L' \), which satisfies $L'_{\bar z} = 0$. To preserve \( A \), we must simultaneously act on \( h' \), so we define \( h'_1 = u_1^{-1} h' \), and then \( A \) becomes
\begin{equation}
    A = (h'_1)^{-1} L' h'_1 + (h'_1)^{-1} \mathrm{d}h'_1 \,.
\end{equation}
Having partially fixed the redundancy using \( u_1 \), we ask whether any symmetry remains that preserves the condition \( L'_{\bar z} = 0 \). The answer is yes: we can still transform \( L' \) and \( h'_1 \) by another map \( u_2 \), provided that \( u_2 \) is independent of \( \mathbb{C}P^1 \). We can now use this remaining freedom to fix the value of \( h'_1 \) at one point on \( \Sigma \). Although \( u_2 \) must be constant on \( \mathbb{C}P^1 \), we have complete freedom in the \( \Sigma \) directions. We choose \( u_2 = h'_1|_{z=\infty} \), by which we mean that \( u_2 \) is constant in \( \mathbb{C}P^1 \) but varies over \( \Sigma \), and takes the same value as \( h'_1 \) at \( z = \infty \). Under this transformation, we get
\begin{equation}
    L = u_2^{-1} L' u_2 + u_2^{-1} \mathrm{d}u_2 \,, \quad h_2 = u_2^{-1} h_1 \,.
\end{equation}
By construction, this ensures that \( h_2|_{z = \infty} = \mathrm{id} \). The conclusion of this argument is that we can always bring the gauge field into the form
\begin{equation}
    A = h_2^{-1} L h_2 + h_2^{-1} \mathrm{d}h_2 \,,
\end{equation}
with \( L_{\bar z} = 0 \) and \( h_2|_{z = \infty} = \mathrm{id} \). There is one final step remaining in our construction. So far, all our manipulations have involved redefining the auxiliary fields \( A' \) and \( h' \), while leaving the original gauge field \( A \) untouched. However, the theory still enjoys a residual gauge symmetry: we are allowed to perform gauge transformations on \( A \) by any group-valued function \( g \) that preserves the boundary conditions \ref{ec:gbcintro}.

This raises the question of how such a transformation acts on the reparametrization fields. Under
\begin{equation}
A \mapsto g^{-1} A g + g^{-1} \mathrm{d}g\,,    
\end{equation}
one finds it corresponds to the following action on the reparametrization fields:
\[
L \mapsto L\,, \quad h_2 \mapsto h_2 g\,,
\]
as can be verified by direct substitution. We now use this residual gauge freedom to impose the condition \( \partial_z h|_{z=0} = 0 \), where \( h = h_2 g \) for a suitable choice of \( g \). Taking \( g \) such that \( g|_{z=0} = \mathrm{id} \), we have
\begin{equation}
\left.\partial_z h\right|_{z=0} = \left.\partial_z h_2\right|_{z=0} + h_2|_{z=0} \cdot \left.\partial_z g\right|_{z=0}\,.
\end{equation}
Then, \( \partial_z h|_{z=0} = 0 \) will be satisfied provided we take
\begin{equation}
  \left.\partial_z g\right|_{z=0} = - (h_2|_{z=0})^{-1} \left.\partial_z h_2\right|_{z=0}\,.  
\end{equation}
This is a first-order constraint on \( g \), which can always be satisfied since we are free to choose \( g \) arbitrarily near \( z = 0 \), as long as it reduces to the identity at \( z = 0 \). This brings us to the final expression for $A$, given by
\begin{equation}
\label{ec:lastALH}
    A= h^{-1}Lh + h^{-1}\mathrm{d}h
\end{equation}
with $L_{\bar z}=0$, $h|_{z=\infty}=\mathrm{id}.$, and $\partial_zh|_{z=0}=0$. 

With this form of $A$ in hand, we can now substitute it into the action. Using equation \eqref{ec:introCS4gt}, we find:
\begin{equation}
\label{ec:CSactionforL}
    S_{\mathrm{CS}_4}[A] = S_{\mathrm{CS}_4}[L] + \frac{ik}{4\pi} \int_X \omega \wedge \mathrm{d} \langle L, \mathrm{d}h\, h^{-1} \rangle + \frac{ik}{12\pi} \int_X \omega \wedge \langle h^{-1} \mathrm{d}h, h^{-1} \mathrm{d}h \wedge h^{-1} \mathrm{d}h \rangle \,.
\end{equation}
The first term is the four-dimensional Chern--Simons action for \( L \), which governs the bulk dynamics. The second term will give rise to a two-dimensional theory localized on the boundary, while the third term will vanish, as we now show.

We begin by addressing this third term. As in previous steps, we extend \( h \) to a field \( \tilde{h} \) defined on \( X \times [0,1] \). Using the closure of the Wess--Zumino--Witten 3-form, we can rewrite the third term as
\begin{multline}
    \frac{ik}{12\pi} \int_{X \times [0,1]} \mathrm{d}\omega \wedge \left\langle \tilde{h}^{-1} \mathrm{d} \tilde{h}, \tilde{h}^{-1} \mathrm{d} \tilde{h} \wedge \tilde{h}^{-1} \mathrm{d} \tilde{h} \right\rangle \\
    = \frac{k}{6} \int_{\Sigma \times [0,1]} \partial_z \left\langle \tilde{h}^{-1} \mathrm{d} \tilde{h}, \tilde{h}^{-1} \mathrm{d} \tilde{h} \wedge \tilde{h}^{-1} \mathrm{d} \tilde{h} \right\rangle \Big|_{z=0}
    - \frac{k}{6} \int_{\Sigma \times [0,1]} \partial_z \left\langle \tilde{h}^{-1} \mathrm{d} \tilde{h}, \tilde{h}^{-1} \mathrm{d} \tilde{h} \wedge \tilde{h}^{-1} \mathrm{d} \tilde{h} \right\rangle \Big|_{z=\infty}.
\end{multline}
Assuming that, like \( h \), its extension \( \tilde{h} \) satisfies \( \tilde{h}|_{z=\infty} = \mathrm{id} \) and \( \partial_z \tilde{h}|_{z=0} = 0 \), both contributions vanish, and hence this term does not contribute to the action\footnote{For a more explicit version of this argument, the reader may consult \S 7.4 of \cite{Lacroix:2021iit}.
}.

We now turn to the bulk dynamics, captured by the four-dimensional Chern–Simons action for $L$. Varying $L$ in the bulk leads to the equations of motion, which in light-cone coordinates $x^\pm = t\pm \sigma$ take the form:
\begin{align}
    & \partial_{\bar z} L_+ = 0\,, \\
    & \partial_{\bar z} L_-= 0\,, \\
    & \partial_+ L_- - \partial_- L_+ + [L_+, L_-] = 0\,.
\end{align}
The first two equations imply that $L_+$ and $L_-$ are holomorphic away from the zeroes of $\omega$. The third equation takes the familiar form of a Lax equation. 

We now partially go on-shell\footnote{This step is somewhat unconventional. At the classical level, imposing some or all of the equations of motion poses no issue. However, in the quantum theory, one is required to integrate over all gauge-inequivalent field configurations, not just those that satisfy the classical equations of motion. Nonetheless, since our analysis is purely classical, we will not concern ourselves with this subtlety here.} by imposing the first two equations of motion. As a result, the first term in \eqref{ec:CSactionforL} vanishes, since it is proportional to \( \omega \wedge \langle \partial_{\bar z} L, L \rangle \).

The most general solutions for $\partial_{\bar z}L_\pm=0$ on $\Sigma \times (\mathbb{C}P^1 \setminus \{\pm 1\})$, namely, away from the zeroes of $\omega$ are 
\begin{equation}
\label{ec:introgensol}
    L_\pm = \frac{U^{(1)}_\pm}{z+1} + \frac{U_\pm^{(-1)}}{z-1} + V_{\pm}\,,
\end{equation}
where $U^{\pm 1}_\pm$ and $V_\pm$ are $\CP$-independent functions. In practice, one can always choose coordinates on \( \Sigma \) such that each pole appears in only one of the two components of \( L \) along \( \Sigma \). Thus, without loss of generality, we may place one pole along \( \mathrm{d}x^+ \) and the other along \( \mathrm{d}x^- \), and write
\begin{equation}
    L_\pm = \frac{U_\pm}{z\mp 1} + V_\pm\,.
\end{equation}
The values of \( U_\pm \) and \( V_\pm \) are fixed by imposing the boundary conditions on \( A \). Recall that, to ensure a well-defined variational principle, we required that \( A|_{z=0} = A|_{z=\infty} = 0 \). Using $A=h^{-1}Lh + h^{-1}\dr h$, we evaluate at \( z = \infty \) to find
\begin{equation}
    A_\pm|_{z=\infty} = 0 = V_\pm\,.
\end{equation}
Similarly, at \( z = 0 \), we have
\begin{equation}
  A_\pm|_{z=0} = 0 = h^{-1} (\mp U_\pm) h + h^{-1} \partial_\pm h\,,  
\end{equation}
where, with a slight abuse of notation, we have denoted \( h = h|_{z=0} \). Solving for \( U_\pm \), we obtain the final expression
\begin{equation}
 L_\pm = \pm\frac{\partial_\pm h\, h^{-1}}{z \mp 1}\,.   
\end{equation}
We now return to the second term in the action. Integrating by parts and letting the exterior derivative act on $\omega$, we get:
\begin{equation}
         \frac{ik}{4\pi} \int_X \omega \wedge \mathrm{d} \langle L, \mathrm{d}h\, h^{-1} \rangle 
         =\frac{k}{2}\int_\Sigma \langle \partial_z L, \mathrm{d}hh^{-1}\rangle|_{z=0}
\end{equation}
where we used that $h|_{z=\infty}=\mathrm{id.}$, and that $\partial_z h|_{z=0}=0$. Substituting the form of $L_\pm$, we obtain the two-dimensional action:
\begin{equation}
    S[h]=k\int_\Sigma \langle \partial_+hh^{-1},\partial_- hh^{-1}\rangle \dr x^+\wedge \dr x^{-}
\end{equation}
which is the action of the principal chiral model. Moreover, the Lax connection is given by
\begin{equation}
    L=\frac{\partial_+h h^{-1}}{z-1}\dr x^+ - \frac{\partial_-h h^{-1}}{z+1}\dr x^-\,,
\end{equation}
which coincides with the expression in \eqref{ec:2dLAXPCMintro}, up to an adjoint action by $h^{-1}$, which we are always free to apply to the Lax connection.

\subsubsection{State of The Art}

The example we have described is perhaps the simplest application of a much broader framework that allows for the construction of a wide class of two-dimensional integrable field theories. Since its introduction in \cite{Costello:2019tri}, numerous variants and generalizations have been explored. Many new integrable models have been constructed—though not always with clear physical interpretation—and most known two-dimensional integrable field theories have been recovered within this framework \cite{Vicedo:2019dej, Delduc:2019whp, Schmidtt:2019otc, Fukushima:2020dcp, Tian:2020ryu, Tian:2020pub, Benini:2020skc, Ashwinkumar:2020krt, Lacroix:2020flf, Fukushima:2020tqv, Fukushima:2021eni, Stedman:2021wrw, Fukushima:2021ako, Py:2022hoa, Levin:2022dnq, Levine:2023wvt, Lacroix:2025ias}. However, it remains an open question what is the most general admissible choice of meromorphic $1$-form and corresponding boundary conditions, and, in particular, whether every two-dimensional integrable field theory can be obtained from four-dimensional Chern–Simons theory.

\section{6d Chern-Simons Theory on Twistor Space}

Six dimensional Chern-Simons theory  has a fairly long history, certainly predating the work of CWY. It first appeared in \cite{Witten:1992fb} as the cubic open string field theory action for
the type B topological string. The action is given by 
\begin{equation}
\label{ec:6dCSactionintro}
    S_{\mathrm{CS}_6}[\bar \cA]=\frac{1}{4\pi i}\int_X \Omega \wedge \langle \bar \cA, \bar \partial\bar \cA + \tfrac{2}{3} \bar\cA \wedge \bar\cA\rangle 
\end{equation}
where $X$ is a three-dimensional complex Calabi-Yau manifold, $\Omega$ a holomorphic $(3,0)$-form and $\bar\cA$ a $(0,1)$-form gauge field. The equations of motion impose flatness of the partial connection,
\begin{equation}
\label{ec:PTEflatcon}
    \bar \partial \bar \cA+\bar \cA \wedge \bar \cA=0 \,.
\end{equation}
Since Euclidean twistor space is also a three-dimensional complex manifold, it would be very appealing to interpret $\bar \cA$ as a $(0,1)$-connection on a vector bundle over twistor space. In that case, as discussed in \ref{sec:penroseward}, flatness would endow the bundle with a holomorphic structure, which is precisely one of the two key ingredients in the Penrose–Ward correspondence. 

The problem is that \( \PTE \) is not a Calabi--Yau manifold and, as such, does not admit a globally defined holomorphic \((3,0)\)-form. In the spirit of the four-dimensional Chern--Simons construction, Costello proposed in \cite{Kcostello} a way to circumvent this obstruction: namely, to allow \( \Omega \) to be a \emph{meromorphic} \((3,0)\)-form rather than a holomorphic one. Schematically, this can be viewed as defining a non-compact Calabi--Yau threefold by excising the poles of \( \Omega \) from \( \PTE \).

The proposal was subsequently realized by Bittleston and Skinner \cite{Bittleston:2020hfv}, and independently by Penna \cite{Penna:2020uky}. The idea follows precisely the reasoning outlined in the previous section: by viewing \( \mathbb{P}\mathbb{T}_\mathbb{E} \cong \mathbb{C}P^1 \times \mathbb{R}^4 \) as a real manifold, the poles of the meromorphic \((3,0)\)-form \( \Omega \) effectively introduce a boundary in \( \mathbb{P}\mathbb{T}_\mathbb{E} \), partially breaking the gauge symmetry of the theory. This symmetry breaking induces a boundary theory on \( \mathbb{R}^4 \), whose dynamics are inherited from the original six-dimensional theory. In particular, the flatness condition \eqref{ec:PTEflatcon} for the connection $\mathcal{A}$ on $\PTE$ reduces to the anti-self-dual Yang–Mills equations for a four-dimensional gauge field $A$ on $\mathbb{R}^4$. This construction thus offers a concrete realization of the Penrose–Ward correspondence.

\begin{figure}[h]
\centering
\begin{tikzpicture}[scale=0.8, >=Latex, node distance=3cm and 5cm]
\label{fig:diamondintro1}
\node (top) at (0,3) {CS$_6$ on $\PTE$};
\node (left) at (-4,0) {CS$_4$ on  \( \mathbb{C}P^1 \times \mathbb{R}^2\)};
\node (right) at (4,0) {ASDYM on \( \mathbb{R}^4 \)};
\node (bottom) at (0,-3) {2d IFT on \( \mathbb{R}^2 \)};

\draw[->] (top) -- (left) node[midway, above left] {\footnotesize dimensional reduction};
\draw[->] (top) -- (right) node[midway, above right] {\footnotesize integrating along $\CP$};
\draw[->] (left) -- (bottom) node[midway, below left] {\footnotesize integrating along $\CP$};
\draw[->] (right) -- (bottom) node[midway, below right] {\footnotesize dimensional reduction};

\end{tikzpicture}
\caption{Diamond correspondence of theories.}
\end{figure}
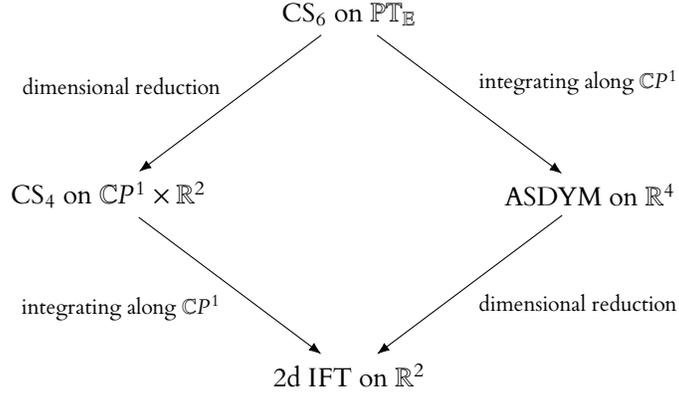

Notably, as mentioned at the beginning of \S\ref{sec:ASDYMEQ}, there is a deep connection between ASDYM theory and Lax integrable systems, as many integrable field equations in two dimensions arise as dimensional reductions of the ASDYM equations. This naturally raises the question of whether these various structures can be brought together within a unified framework. The realisation of this idea was indeed one of the key insights in the work of Bittleston and Skinner, where they showed that one can construct a diamond correspondence of theories, as depicted in Figure~\ref{fig:diamondintro1}.

Starting from holomorphic Chern--Simons theory on twistor space, one can follow the right-hand path of the diamond: integrating over \( \mathbb{C}P^1 \) yields a four-dimensional theory on \( \mathbb{R}^4 \) whose equations of motion are equivalent to ASDYM equations. From there, applying the well-known dimensional reduction of ASDYM leads to a two-dimensional integrable field theory equipped with a Lax connection. Alternatively, one can begin by reducing the six-dimensional holomorphic Chern--Simons theory on \( \mathbb{P}\mathbb{T}_\mathbb{E} \) to obtain four-dimensional Chern--Simons theory on \( \mathbb{C}P^1 \times \mathbb{R}^2 \). Integrating over \( \mathbb{C}P^1 \) as described in the previous section, yields a two-dimensional integrable field theory. Crucially, both routes lead to the same two-dimensional integrable model.

Here we briefly summarize the main results of one of the key examples, illustrating how the diamond correspondence is realized for a specific choice of $(3,0)$-form. A detailed treatment of this example is presented in \S\ref{sec:gaugingthediamond}.

We consider the action \eqref{ec:6dCSactionintro} with $X= \PTE$ and the meromorphic $(3,0)$-form 
\begin{equation}
    \Omega = \frac{1}{2}\frac{e^0 \wedge e^{\dot a} \wedge e_{\dot a}}{\langle\pi \alpha\rangle^2\langle \pi \beta \rangle^2} = \frac{1}{2} \frac{\dr \zeta  \wedge e^{\dot a} \wedge e_{\dot a}}{(\zeta-\alpha)^2(\zeta-\beta)^2} \,.
\end{equation}
The expression in inhomogeneous coordinates is valid on the southern patch, where \( \pi^0 \neq 0 \), so that \( \zeta = \pi^1 / \pi^0 \), \( \alpha = \alpha^1 / \alpha^0 \), and \( \beta = \beta^1 / \beta^0 \). Working in homogeneous coordinates avoids the need to specify coordinate patches and transition functions, providing a globally defined description on \( \mathbb{C}P^1 \). Regardless of the coordinate system, there are two key features to emphasize about the chosen \((3,0)\)-form. First, it has support only along the holomorphic directions of \( \mathbb{P}\mathbb{T}_{\mathbb{E}} \). Second, it exhibits two double poles on \( \mathbb{C}P^1 \), located at \( \zeta = \alpha \) and \( \zeta = \beta \), or, in homogeneous coordinates, $\pi^a = \alpha^a$ and $\pi^a=\beta^a$.


Following the same procedure than in the previous section (see \ref{sec:gaugingthediamond} for details), integration along $\CP$ renders the four-dimensional action
\begin{equation}
\label{ec:4dWZWmodelintro}
    S[g] = \int_{\bR^4}\langle g^{-1}\dr g ,\ast g^{-1} \dr g \rangle + \frac{1}{3} \int_{\bR^4\times [0,1]}\omega_{\alpha,\beta} \wedge \langle \tilde g^{-1}\dr \tilde g, \tilde g^{-1}\dr \tilde g\wedge \tilde g^{-1}\dr \tilde g\rangle \,,
\end{equation}
where $\omega_{\alpha,\beta}$ is a $2$-form defined by 
\begin{equation}
    \omega_{\alpha,\beta} = \frac{1}{\langle \alpha \beta \rangle}\alpha_a \beta_b \epsilon_{\dot a\dot b}\dr x^{a\dot a}\wedge \dr x^{b\dot b} \,,
\end{equation}
and $\tilde g$ is a suitable extension of $g$ from $\bR^4$ to $\bR^4\times [0,1]$. The action \eqref{ec:4dWZWmodelintro}, originally proposed by Donaldson \cite{Donaldson:1985zz} and studied in detail in \cite{Losev:1995cr}, defines a four-dimensional generalization of the WZW model. Varying the action with respect to \( g \), one obtains the equation of motion
\begin{equation}
\label{ec:Yangeqintro}
    \epsilon^{ab} \alpha^{\dot a} \beta^{\dot b} \partial_{b\dot b}(\partial_{a\dot a} g\, g^{-1})=0\,,
\end{equation}
which is known as Yang's equation \cite{Yang:1977zf}. In particular, the four-dimensional connection \( A_{a\dot a} \) arising from the six-dimensional theory (see \S \ref{sec:gaugingthediamond} for details) is given by
\begin{equation}
    A_{a\dot a} = - \beta_{\dot a} \alpha^{\dot b} \partial_{a\dot b} g\, g^{-1}\,.
\end{equation}
It is straightforward to verify that the ASDYM equations for \( A_{a\dot a} \) are equivalent to Yang's equation \eqref{ec:Yangeqintro}. 

One can then dimensionally reduce the action \eqref{ec:4dWZWmodelintro} to obtain the PCM with a WZ term. Alternatively, following the left-hand path of the diamond, one may begin by reducing the six-dimensional action. This yields four-dimensional Chern--Simons theory with a meromorphic 1-form
\begin{equation}
    \omega = \frac{(\zeta - 1)(\zeta + 1)}{(\zeta - \alpha)^2(\zeta - \beta)^2} \, \mathrm{d}\zeta\,,
\end{equation}
which can then be localized to the PCM with WZ term by integrating over \( \mathbb{C}P^1 \), as described in the previous section. Further details can be found in \S \ref{sec:gaugingthediamond} and \cite{Bittleston:2020hfv, Cole:2024hyt}.

Hence, we conclude that both paths lead to the same two-dimensional integrable field theory. In particular, six-dimensional holomorphic Chern--Simons theory provides an action principle for the Penrose--Ward correspondence: it relates flat connections on \( \mathbb{P}\mathbb{T}_{\mathbb{E}} \) that are trivial when restricted to \( \mathbb{C}P^1 \) (i.e., elements of the moduli space of solutions to the equations of motion) with solutions to the ASDYM equations, which are precisely encoded in the four-dimensional gauge field \( A_{a\dot a} \) obtained by integrating along \( \mathbb{C}P^1 \).

\subsubsection{State of the Art}

Following the original works of Bittleston and Skinner \cite{Bittleston:2020hfv} and Penna \cite{Penna:2020uky}, an initial attempt to generalize their construction to other choices of meromorphic \((3,0)\)-forms was made in \cite{He:2021xoo}, inspired by the generalization principle of four-dimensional Chern--Simons theory. However, this first attempt was only partially successful, as it overlooked several essential aspects of the complex geometry of \( \mathbb{P}\mathbb{T}_{\mathbb{E}} \).

In this context, the author, together with Lewis Cole, Ryan Cullinan, Ben Hoare, and Daniel Thompson, extended the framework to include meromorphic $(3,0)$-forms with simple poles. This led to the realization of the diamond \ref{fig:diamondintro1} corresponding to a four-dimensional deformation of the WZW model \cite{Cole:2024ess}, which is the focus of Chapter~\ref{sec:intdeformations}. The same collaboration subsequently considered a generalized six-dimensional holomorphic Chern--Simons theory involving two gauge fields and an interaction term, which resulted in the construction of the diamond corresponding to four-dimensional coset models of the WZW type. This is the focus of Chapter~\ref{sec:gaugingthediamond}.
 
The development of holomorphic Chern–Simons theory also sparked significant progress in celestial holography. In \cite{Costello:2021bah}, Costello showed that certain four-dimensional quantum field theories arising from local holomorphic actions on twistor space—known as twistorial field theories—can be quantized in a systematic way. These theories are characterized by rational correlators and operator product expansions, and although they are anomalous on twistor space, the anomalies can be canceled by coupling to suitable background fields, yielding consistent four-dimensional QFTs \cite{Costello:2015xsa, Costello:2021bah}.

A close connection between twistorial QFTs and celestial holography was established in \cite{Costello:2022jpg, Costello:2022upu}. In particular, it was shown that local operators in these four-dimensional theories correspond to conformal blocks of a two-dimensional chiral algebra closely related to the algebra of asymptotic symmetries appearing in celestial holography. 

The concrete realization of this correspondence was achieved through a top-down duality construction relating two distinct theories originating from string theory. This was carried out in \cite{Costello:2022jpg, Costello:2023hmi}, where a duality was established between a simple two-dimensional conformal field theory and a self-dual field theory defined on a four-dimensional manifold known as Burns space. This was further generalized in \cite{Bittleston:2024efo}, where a broad class of top-down dualities of this type was constructed. Each duality relates an explicit two-dimensional system, in the large-$N$ limit, to self-dual QCD on a non-trivial self-dual background. 

\section{5d 2-Chern-Simons Theory}

We now introduce the final example of a holomorphic Chern–Simons theory considered in this work. It was developed independently by the author in collaboration with Hank Chen \cite{Chen:2024axr}, and by Benoît Vicedo and Alexander Schenkel \cite{Schenkel:2024dcd}. The central idea is to construct integrable systems— in the sense described in \S\ref{sec:catintegrability}—from a higher-dimensional gauge theory. In this section we will give a brief overview as the full development of this is the content of chapter \ref{chap:5d2CS}. 

Recall that we defined a three-dimensional integrable field theory to be \emph{weakly integrable} if its equations of motion are equivalent to the flatness conditions of a spectral-parameter-dependent 2-connection \((A, B)\), constructed from the fields of the theory. In the original formulation of four-dimensional Chern--Simons theory, the motivation was the fact that its equations of motion correspond to the flatness condition \( F(A) = 0 \) for the connection. This naturally raises the question of whether an analogous theory exists for the higher connection \((A, B)\)---that is, whether there exists an action functional depending on \((A, B)\) whose equations of motion are precisely the higher flatness conditions
\begin{equation}
    \mathscr{F}(A,B)=(\mathrm{d}A+\tfrac{1}{2}[A,A]-\mu_1(B),\mathrm{d}B+A \wedge^{\rhd}B) = 0\,.
\end{equation}
Indeed, such theories exist and are known as \emph{homotopy Maurer--Cartan theories} \cite{Zwiebach:1992ie}, which are defined for general Lie \( n \)-algebras. Each theory is characterized by an action functional whose equations of motion correspond to the vanishing of an associated \( n \)-curvature. When \( n = 1 \), one recovers three-dimensional Chern--Simons theory. The case \( n = 2 \) leads to four-dimensional 2-Chern--Simons theory, which can be defined as follows: Let \( \mathfrak{G} = (\fh \xrightarrow{\mu_1} \fg, \rhd) \) be a Lie 2-algebra, and let \( M \) be a four-dimensional manifold. Consider a 1-form \( A \in \Omega^1(M, \fg) \) and a 2-form \( B \in \Omega^2(M, \fh) \). The 2-Chern--Simons theory is defined by the action
\begin{equation}
    S_{2\mathrm{CS}}[A, B] = \int_M \langle F(A) - \mu_1(B), B \rangle\,,
\end{equation}
where \( \langle \cdot, \cdot \rangle : \fg \times \fh \to \mathbb{R} \) is a graded-symmetric, ad-invariant bilinear form. The equations of motion are obtained by varying the action with respect to the fields $A$ and $B$, and are given by:
\begin{align}
    &\delta B:  \mathrm{d}A + \tfrac{1}{2}[A, A] - \mu_1(B) = 0\,, \\
    &\delta A: \mathrm{d}B + \mu_2(A, B) = 0\,,
\end{align}
where we recongize the fake-flatness and the 2-flatness equations. We proceed in the same way that as \cite{Costello:2019tri}; we assume $M=\bR \times Y$ for some $3$-manifold $Y$, we complexify and compactify $\bR$ to $\CP$ and define the five-dimensional holomorphic 2-CS action \cite{Chen:2024axr, Schenkel:2024dcd}
\begin{equation}
    S_{\mathrm{h2CS}_5}[A,B]=\int_{\CP \times Y} \omega \wedge \langle F(A) - \mu_1(B), B \rangle\,,
\end{equation}
where $\omega$ is a meromorhpic $(1,0)$-form as in the 4d CS case. The idea is now the same as in four-dimensional Chern--Simons theory and six-dimensional holomorphic Chern--Simons theory on twistor space: the  poles in $\omega$ introduce a boundary in the manifold. This partially breaks the gauge symmetry and, as a result, induces non-trivial dynamics on the three-dimensional boundary \( \{\text{poles}\} \times Y \). The dynamics of the boundary theory are inherited from the bulk, whose equations of motion impose the fake-flatness and 2-flatness conditions.

We will not work through an explicit example in this section, as the main one is developed in detail in Chapter~\ref{chap:5d2CS}. However, we will offer some remarks on the example presented there and in \cite{Schenkel:2024dcd}, as they differ significantly and illustrate the two distinct notions of integrability discussed in \S\ref{sec:catintegrability}. These differences are both important and instructive, and stem primarily from the choice of meromorphic 1-form. Specifically, we have:
\begin{align*}
   & \text{Chen--Liniado:} \quad \omega = \frac{\mathrm{d}z}{z} \\
   & \text{Schenkel--Vicedo:} \quad \omega = \frac{z \prod_{i=1}^3 (z - a_i)}{(z - r)^2 (z - s)^2} \, \mathrm{d}z
\end{align*}
The Chen--Liniado choice has simple poles at \( z = 0 \) and \( z = \infty \), and is nowhere vanishing. In contrast, the Schenkel--Vicedo form has double poles at \( z = r \), \( z = s \), and \( z = \infty \), along with four simple zeroes at \( z = 0 \) and \( z = a_i \) for \( i = 1, 2, 3 \). 

\subsubsection{Chen-Liniado}

The three-dimensional model constructed in \cite{Chen:2024axr}, with the meromorphic \(1\)-form \(\omega = \tfrac{\dr z}{z}\), is integrable in the sense described in \S\ref{sec:introhigherkacmoody}; that is, it exhibits an infinite-dimensional symmetry algebra. This model is discussed in detail in \S\ref{chap:5d2CS}, so we will not elaborate further here. However, we briefly mention some recent developments on this topic, which will be presented in a forthcoming work \cite{Chen:2025ujx}.

Taking the 3d manifold $Y=\bR \times \bC$ with local coordinates $(\tau,z,\bar z)$, the three-dimensional action we construct (see Chapter \ref{chap:5d2CS} for the explicit construction) is given by 
\begin{multline}
\label{ec:3dtopholactionintro}
       S[g,\Theta] = \int_Y \operatorname{vol}_Y \Big(\langle \partial_z (\partial_{\bar z}gg^{-1}),g\rhd\Theta_\tau\rangle - \langle \partial_z (\partial_\tau gg^{-1}),g\rhd\Theta_{\bar z}\rangle \\
       +\langle \mu_1\big(\partial_z (g\rhd\Theta_{\bar z})\big),g\rhd\Theta_\tau\rangle\Big)\,,
\end{multline}
where $g\in C^{\infty}(M,G)$ is a $G$-valued smooth field, $\Theta \in \Omega^{1}(M,\fh)$ an $\fh$-valued $1$-form, and $\langle\cdot,\cdot\rangle:\fg \times \fh \to \mathbb{R}$ a degree $-1$, non degenerate, invariant bilinear form. The appearance of the 1-form field \(\Theta\), not as a gauge connection but rather as a matter field playing a role analogous to that of \(g\), is closely tied to the higher categorical nature of the action.

In complete analogy with the WZW model, the action \eqref{ec:3dtopholactionintro} is invariant under both left- and right-acting symmetries. A direct computation shows that it is invariant under the transformations
\begin{equation}
\label{ec:3dleftsym}
    g \mapsto g(1 + \tilde\alpha) \,, \quad \Theta \mapsto \Theta-\tilde\alpha \rhd \Theta + \tilde\Gamma \,,
\end{equation}
where the infinitesimal parameters \(\tilde\alpha \in C^{\infty}(M, \fg)\) and \(\tilde\Gamma \in \Omega^{1}(M, \fh)\) satisfy the constraints
\begin{equation}
    \partial_{\bar z} \tilde\alpha + \mu_1(\tilde\Gamma_{\bar z}) = 0\,, \quad 
    \partial_{\tau} \tilde\alpha + \mu_1(\tilde\Gamma_{\tau}) = 0\,, \quad 
    \partial_\tau \tilde\Gamma_{\bar z} - \partial_{\bar z} \tilde\Gamma_\tau = 0\,.
\end{equation}
These define the right-acting symmetries of the model. The left-acting symmetries take the form
\begin{equation}
\label{ec:3drightsym}
    g \mapsto (1 - \alpha)g \,, \quad \Theta \mapsto \Theta - g^{-1}(1 + \alpha) \rhd \Gamma \,,
\end{equation}
where the infinitesimal parameters \(\alpha \in C^{\infty}(M, \fg)\) and \(\Gamma \in \Omega^{1}(M, \fh)\) satisfy
\begin{equation}
    \partial_z \alpha = 0\,, \quad \partial_z \Gamma_\tau = 0\,, \quad \partial_z \Gamma_{\bar z} = 0\,.
\end{equation}
The symmetries \eqref{ec:3dleftsym} and \eqref{ec:3drightsym} give rise to associated conserved currents. In line with the higher categorical structure of the theory, these currents will also be a pair, in this case, a \(\fg\)-valued 1-form and an \(\fh\)-valued 2-form. This structure is consistent with the higher-form symmetries of the action. With this in mind, these currents can be derived using \emph{Noether's trick} as in the two-dimensional case. Specifically, for the right-acting symmetry, one considers a variation of the action under $g \mapsto g(1 + \tilde\alpha) $ and $\Theta \mapsto \Theta - \tilde\alpha \rhd \Theta + \tilde\Gamma$ with arbitrary infinitesimal parameters \( \tilde\alpha \in C^\infty(M, \fg) \) and \( \tilde\Gamma \in \Omega^1(M, \fh) \). Requiring the action to be stationary under such variations, i.e., \( \delta S = 0 \), leads to 
\begin{equation}
\label{ec:conservationeq1}
     \partial_{\bar z}\tilde L_z -\mu_1\tilde H_{{\bar z}z}=0,\qquad \partial_{\tau}\tilde L_z-\mu_1\tilde H_{\tau z}=0,\qquad \partial_\tau H_{\bar zz} + \partial_{\bar z}H_{\tau z}=0\,.
\end{equation}
where we have introduced the currents
\begin{equation}
    \tilde L_z = g^{-1}\partial_z g \,,\quad \tilde H_{z{\bar z}} = \partial_z\Theta_{\bar z} + \mu_2(L_z,\Theta_{\bar z}),\quad \tilde H_{z\tau} =\partial_z\Theta_\tau+\mu_2(L_z,\Theta_\tau) \,.
\end{equation}
In a completely analogous manner, we can compute the currents associated to the left symmetry, given by
\begin{equation}
    \begin{aligned}
      &L_{\bar z} = -\partial_{\bar z}gg^{-1}-\mu_1(g\rhd \Theta_{\bar z})\,,\quad L_\tau = -\partial_\tau gg^{-1}-\mu_1(g\rhd \Theta_\tau)
      \,,\\
      &\qquad H_{{\bar z}\tau} = g\rhd (\partial_{\bar z}\Theta_\tau-\partial_\tau\Theta_{\bar z}-[\Theta_{\bar z},\Theta_\tau]) \,,
\end{aligned}
\end{equation}
which satisfy the conservation equations
\begin{equation}
\label{ec:conservationeq2}
    \partial_z L_{\bar z} = 0\,,\quad \partial_zL_\tau=0\,,\quad \partial_zH_{\bar z\tau} = 0\,.
\end{equation}
The expressions we have obtained suggest a natural splitting of coordinates, treating \((\bar z, \tau)\) and \(z\) separately. More precisely, we can introduce the differential
\begin{equation}
    \mathrm{d}' = \bar{\partial} + \mathrm{d}_\tau\,,
\end{equation}
and decompose differential forms into components along \(\mathrm{d}'\) and along \(\partial\). With this in mind, the currents associated with the right-acting symmetry can be written as
\begin{equation}
\label{ec:currents1}
    \tilde{L} = g^{-1} \partial g\,, \quad \tilde{H} = \partial \Theta' + \mu_2(g^{-1} \partial g, \Theta')\,,
\end{equation}
and those associated with the left-acting symmetry as
\begin{equation}
\label{ec:currents2}
    L = -\mathrm{d}'g\, g^{-1} - \mu_1(\Theta')\,, \quad 
    H = g \rhd \left( \mathrm{d}'\Theta - \tfrac{1}{2}[\Theta', \Theta'] \right)\,.
\end{equation}
In terms of this decomposition, the conservation equations \eqref{ec:conservationeq1} and \eqref{ec:conservationeq2} take the form
\begin{equation}
\label{ec:introdiff0}
    \mathrm{d}'\tilde{L} - \mu_1(\tilde{H}) = 0\,, \quad \mathrm{d}'\tilde{H} = 0\,,
\end{equation}
and
\begin{equation}
    \partial L = 0\,, \quad \partial H = 0\,.
\end{equation} 
Note that the conditions in \eqref{ec:introdiff0} closely resemble those discussed in Section~\ref{sec:introhigherkacmoody}, with the exception of the presence of \(\mu_1\), which introduces an additional subtlety. This allows us to expand the fields \(\tilde{L}\) and \(\tilde{H}\) in modes as
\begin{equation}
    \tilde{L} = \sum_{n=0}^\infty \tilde{L}_n z^n \,,\quad 
    \tilde{H} = \sum_{m=0}^\infty \tilde{H}_m \Omega^m\,.
\end{equation}
Using the action \eqref{ec:3dtopholactionintro}, one can compute the Poisson brackets among these modes and show that they generate a centrally extended affine Lie 2-algebra, which serves as a higher-dimensional analogue of the Kac–Moody algebra. One can also construct the corresponding \emph{raviolo} vertex algebra, which is a direct generalization of a vertex algebra to the topological-holomorphic setting. We expect that this structure may serve as the foundation for a three-dimensional analogue of the conformal field theory framework, largely because many of the important theorems of vertex operator algebras extend naturally to the raviolo setting. The details of these computations are presented in~\cite{Chen:2025ujx}.

\subsubsection{Schenkel-Vicedo}

For the choice of meromorphic \(1\)-form used in \cite{Schenkel:2024dcd},
\begin{equation}
    \omega = \frac{z \prod_{i=1}^3 (z - a_i)}{(z - r)^2 (z - s)^2} \, \dr z,
\end{equation}
something qualitatively different occurs: the \(2\)-connection \((A, B)\), when reduced to the three-dimensional manifold, becomes meromorphic with poles located at the zeroes of \(\omega\). This is fully analogous to the four-dimensional case, where the Lax connection develops poles as in \eqref{ec:introgensol}. In principle, this feature allows for the construction of an infinite number of conserved quantities, as discussed in \S\ref{sec:highercatlax}, by series-expanding the conserved surface holonomies.

However, the physical interpretation of these charges remains unclear, as does their potential role in constructing explicit solutions—even at the classical level. It would be particularly interesting to clarify the relationship between their three-dimensional model and the Ward model \cite{Ward_1988, Ward:1988ie}, and to investigate whether the conserved quantities arising from surface holonomies are related to the Riemann–Hilbert formulation of such systems.

\chapter{Degenerate E-Models from 4d CS Theory}
\label{chap:4dCS}

This work was done in collaboration with Benoit Vicedo \cite{Liniado:2023uoo}.

\section{Introduction}

Given the action for a $2$d field theory, it is a difficult problem to decide whether or not it is classically integrable. Indeed, proving a theory is integrable requires showing, in particular, that the field equations of motion are equivalent to the flatness equation for a $2$d connection $\cL$ valued in some finite-dimensional Lie algebra $\fg$ and depending meromorphically on some auxiliary complex parameter. Unfortunately there is no systematic procedure for finding such a Lax connection when it exists.

In recent years, however, there has been tremendous progress towards the problem of classifying $2$d integrable field theories. In particular, Costello and Yamazaki proposed a general approach \cite{Costello:2019tri}, based on earlier work in the context of integrable spin chains \cite{Costello:2013sla, Costello:2013zra, Costello:2017dso, Costello:2018gyb}, for constructing $2$d integrable field theories starting from $4$d Chern-Simons theory. This approach, being rooted in the Lagrangian formalism, provides an elegant way of computing the $2$d action for the integrable field theory as well as its Lax connection. It has been extensively applied to reproduce many existing $2$d integrable field theories and also construct a wide variety of new ones \cite{Delduc:2019whp, Bykov:2019vkf, Bykov:2020nal, Bykov:2020tao, Lacroix:2020flf, Costello:2020lpi, Fukushima:2020dcp, Fukushima:2020kta, Schmidtt:2019otc, Schmidtt:2020dbf, Tian:2020ryu, Tian:2020pub, Caudrelier:2020xtn, Fukushima:2020tqv, Derryberry:2021rne}. See also \cite{Benini:2020skc, Ashwinkumar:2020krt, Bittleston:2020hfv, Penna:2020uky, Bittleston:2022cmz, Khan:2022vrx, Costello:2021zcl, Stedman:2021wrw, Levin:2022dnq, Zhou:2022mzb} for further developments in relation to $4$d Chern-Simons theory.

It is important to note that establishing the complete integrability of a $2$d field theory also requires showing that the integrals of motion constructed out of the Lax connection Poisson commute with one another, which necessitates moving to the Hamiltonian formalism. A general approach for constructing classical $2$d integrable field theories directly in the Hamiltonian formalism was proposed in \cite{Vicedo:2017cge}, and further developed in \cite{Delduc:2019bcl, Lacroix:2019xeh}, by starting from affine Gaudin models. It was shown in \cite{Vicedo:2019dej} by performing a Hamiltonian analysis of $4$d Chern-Simons theory (see also \cite{Schmidtt:2020dbf} for the $\bbZ_2$- and $\bbZ_4$-equivariant settings in the context of the $\lambda$-model) that the formalisms of \cite{Costello:2019tri} and \cite{Vicedo:2017cge} are intimately related. In particular, all the $2$d integrable field theories constructed from $4$d Chern-Simons theory are integrable in this stronger sense.

\medskip

The action of $4$d Chern-Simons theory is specified by a choice of meromorphic $1$-form $\omega$ on $\CP$ (one could also consider higher genus Riemann surfaces \cite{Costello:2019tri} but in this article we focus on the Riemann sphere). 
Different $2$d integrable field theories arise from different choices of $\omega$ and various other data to be reviewed in \S\ref{sec: review}.
The case when $\omega$ has at most double poles was studied in detail in \cite{Delduc:2019whp}, where a very simple `unifying' $2$d action was derived for all integrable field theories belonging to this class of meromorphic $1$-forms. A generalisation of this `unifying' $2$d action for arbitrary $\omega$ was then obtained in \cite{Benini:2020skc}, where a different perspective on the passage from $4$d Chern-Simons theory to $2$d integrable field theory was also advocated.

In order to write the `unifying' $2$d actions of \cite{Delduc:2019whp} or \cite{Benini:2020skc} in terms of the field of the $2$d integrable field theory alone, one needs to solve a certain constraint relating this field to the Lax connection. A very general class of solutions to this constraint was constructed in \cite{Lacroix:2020flf} when $\omega$ has arbitrary poles and zeroes in the complex plane but a double pole at infinity. This technical assumption was required in order to fix some of the gauge invariance of $4$d Chern-Simons theory and as a result remove any constant term from the Lax connection, as we recall at the start of \S\ref{sec:reductionto2d}. It was also shown in \cite{Lacroix:2020flf} that all of the $2$d integrable field theories arising from such solutions of the constraint are described by \emph{integrable} non-degenerate $\cE$-models.

\medskip

Non-degenerate $\cE$-models were introduced by Klim\v{c}\'\i{}k and \v{S}evera in \cite{Klimcik:1995dy, Klimcik:1995ux, Klimcik:1996np} as $\sigma$-models providing a natural setting for describing a non-abelian generalisation of $T$-duality, known as Poisson-Lie $T$-duality. Even though the generic non-degenerate $\cE$-model is not integrable, it turns out that many interesting examples were found to be integrable \cite{Klimcik:2015gba, Klimcik:2017ken, Klimcik:2020fhs, Hoare:2020mpv}. A simple condition was also formulated on the data of the non-degenerate $\cE$-model which ensures it is integrable \cite{Severa:2017kcs}. Of course, the data of the integrable non-degenerate $\cE$-models constructed in \cite{Lacroix:2020flf} all satisfy this condition.

A generalisation of non-degenerate $\cE$-models, known as degenerate $\cE$-models or dressing cosets, was also introduced by Klim\v{c}\'\i{}k and \v{S}evera in \cite{Klimcik:1996np}. Whereas the non-degenerate $\cE$-model can be described as a $\sigma$-model on a certain quotient space $K \setminus D$ where $K$ is an maximal isotropic subgroup of a Lie group $D$, the degenerate $\cE$-model is a $\sigma$-model on a double quotient $K \setminus D \,/\, F$ where $F$ is another isotropic subgroup of $D$. It was recently shown in \cite{Klimcik:2019kkf} that a certain integrable $\sigma$-model that was constructed in \cite{Delduc:2017fib} provides an example of an \emph{integrable} degenerate $\cE$-model. Very recently, conditions on the degenerate $\cE$-model data ensuring its integrability, in the Hamiltonian sense recalled above, were also given in \cite{Klimcik:2021bqm} and a family of new integrable degenerate $\cE$-models were constructed, including the pseudo dual chiral model and its multifield generalisations.

The purpose of this article is to extend the construction of \cite{Lacroix:2020flf} to the most general setting of an arbitrary meromorphic $1$-form $\omega$. In particular, we drop the technical assumption made in \cite{Lacroix:2020flf} that $\omega$ is required to have a double pole at infinity. We show that the solutions of the constraint equation from \cite{Benini:2020skc} which we construct by generalising the approach of \cite{Lacroix:2020flf} all give rise to \emph{integrable} degenerate $\cE$-models. Just as in \cite{Lacroix:2020flf}, the Lie group $D$ is determined by the pole structure of $\omega$ and the maximal isotropic subgroup $K \subset D$ is determined by the choice of boundary condition imposed on the Chern-Simons field at the collection of poles of $\omega$. On the other hand, the isotropic subgroup $F$, which is specific to the present case, is a remnant of the gauge symmetry of the $4$d Chern-Simons theory under the Lie group $G$ and is given by the image of the diagonal embedding $G \hookrightarrow D$.

\medskip

The plan of the paper is as follows.

In \S\ref{sec: review} we review the alternative, less conventional approach of \cite{Benini:2020skc} for extracting the action of a $2$d integrable field theory from the $4$d Chern-Simons theory of Costello and Yamazaki \cite{Costello:2019tri}. One advantage of this approach is that it makes the passage from $4$d to $2$d more direct, with the field $h$ of the $2$d theory being introduced along surface defects in the $4$d theory to ensure gauge invariance.

In \S\ref{sec:reductionto2d} we generalise the approach of \cite{Lacroix:2020flf} for solving a constraint between the $2$d field $h$ and the $4$d gauge field appearing in the construction of \cite{Benini:2020skc}. More precisely, we do away with the technical requirement in \cite{Lacroix:2020flf} that the $1$-form $\omega$ in the $4$d Chern-Simons action should have a double pole at infinity. Starting from a general meromorphic $1$-form $\omega$, the resulting $2$d integrable field theories are degenerate $\cE$-models.

In \S\ref{sec: examples} we give two detailed examples of the construction from \S\ref{sec:reductionto2d}. Namely, we apply the general formalism to recover the pseudo dual of the principal chiral model, or pseudo-chiral model for short, of Zakharov and Mikhailov \cite{Zakharov:1973pp} and the bi-Yang-Baxter $\sigma$-model proposed by Klim\v{c}\'\i{}k in \cite{Klimcik:2008eq}.

\section{A review on 4d Chern-Simons and 2d IFT} \label{sec: review}

In this section we begin by reviewing the correspondence between $4$d Chern-Simons theory and $2$d integrable field theories, proposed by Costello and Yamazaki in \cite{Costello:2019tri}. We will, however, follow the approach advocated in \cite{Benini:2020skc} which puts special emphasis on the principle of gauge invariance. In this approach, the $4$d Chern-Simons field $A$ is coupled to additional degrees of freedom, the so-called \emph{edge modes}, living on certain surface defects. This is to ensure the full gauge invariance of the theory. The $2$d integrable field theory is then seen to emerge in a particular gauge by going partly on-shell. Although the main ideas of \cite{Benini:2020skc} are intrinsically physical, the constructions rely heavily on methods of homotopical analysis and the theory of groupoids. So the purpose of this section is to review the key steps of the approach of \cite{Benini:2020skc} using a language more familiar to theoretical physicists.

\subsection{4d Chern-Simons theory} \label{sec: 4d CS action}

Let $X \coloneqq \Sigma\times \mathbb{C}P^1$ where $\Sigma$ denotes a $2$d manifold which will eventually correspond to the space-time of the $2$d integrable field theory. We take $\Sigma=\mathbb{R}^2$ or $\mathbb{R}\times S^1$ with coordinates $(\tau,\sigma)$.

Let $G$ be a real, simply connected Lie group with Lie algebra $\fg$. Let $\fg^\bC \coloneqq \fg \otimes_\bR \bC$ be the complexification of $\fg$ and let $G^\bC$ denote the corresponding complex Lie group. We fix a non-degenerate, symmetric and ad-invariant bilinear form $\langle\cdot,\cdot\rangle:\fg \times \fg \to \bR$ and denote by $\langle\cdot,\cdot\rangle:\fg^\bC \times \fg^\bC \to \bC$ its complex linear extension to $\fg^\bC$.

\subsubsection{The meromorphic 1-form} \label{sec: omega}

The key ingredient entering the definition of the $4$d variant of Chern-Simons theory \cite{Costello:2019tri} is a choice of meromorphic 1-form $\omega$ on $\mathbb{C}P^1$.

We denote by $\bZ \subset \mathbb{C}P^1$ the set of poles of $\omega$ and by $n_x$ the order of the pole $x \in \bZ$. Let $\bZ' \coloneqq \bZ \setminus \{\infty\}$ be the subset of finite poles of $\omega$. The reason for this notation will be justified shortly. Fixing a coordinate $z$ on $\mathbb{C} \subset \mathbb{C}P^1$ we can write $\omega$ explicitly as
\begin{equation} \label{omega def}
\omega=\left(\sum_{x\in \bZ' }\sum_{p=0}^{n_x-1}\frac{\ell^x_p}{(z-x)^{p+1}}-\sum_{p=1}^{n_{\infty}-1}\ell^{\infty}_p z^{p-1}\right)\dr z \eqqcolon \varphi(z)\dr z\,,
\end{equation}
for some $\ell^x_p\in \mathbb{C}$ which we call \emph{levels}. We impose reality conditions on each $x\in \bZ$ and its corresponding levels by requiring that $\overline{\varphi(z)} = \varphi(\bar{z})$. In particular, introducing the subset of real poles $\bz_{\mathrm{r}} \coloneqq \bz'_{\mathrm{r}} \,\sqcup\, \{\infty\}$, where $\bz'_{\mathrm{r}} \coloneqq \bZ' \,\cap\, \mathbb{R}$, the associated levels are real, i.e. $\ell^x_p \in \mathbb{R}$ for $x\in \bz_{\mathrm{r}}$. The remaining poles come in complex conjugate pairs and we define $\bz_{\mathrm{c}}\coloneqq \{x \in \bZ \,|\, \Im x > 0\}$ so that $\bZ = \bz_{\mathrm{r}}\,\sqcup \,\bz_{\mathrm c}\,\sqcup\, \bar{\bz}_{\mathrm c}$. For every $x \in \bz_{\mathrm{c}} \,\sqcup\, \bar{\bz}_{\mathrm c}$ we have $n_{\bar{x}} = n_x$ and $\overline{\ell^x_p}=\ell^{\bar{x}}_p$ for $p = 0,\dots, n_{x}-1$. It is convenient to introduce the subset $\bz \coloneqq  \bz_{\mathrm{r}} \,\sqcup\, \bz_{\mathrm{c}}$ of \emph{independent} poles, namely which are independent under complex conjugation. Finally, we introduce the subset $\bz'\coloneqq  \bz'_{\mathrm{r}}\,\sqcup\, \bz_{\mathrm{c}}\subset \bz$ of finite independent poles in $\bz$. We let $\Pi \coloneqq \{ \mathrm{id}, \ms t \}$ denote the group $\bbZ_2$ generated by the element $\ms t$ acting by complex conjugation on $\CP$. The notation $\bZ$ (resp. $\bZ'$) introduced above then corresponds to the orbit of the set $\bz$ (resp. $\bz'$) under $\Pi$.

We will also be interested in the set of zeroes of $\omega$ which can be similarly decomposed as $\bzeta_{\mathrm{r}} \, \sqcup \, \bzeta_{\mathrm{c}} \, \sqcup \, \bar{\bzeta}_{\mathrm{c}}$ with $\bzeta_{\mathrm{r}}\subset \mathbb{R}$ the subset of real zeroes and $\bzeta_{\mathrm{c}}\subset  \{y \in \mathbb{C} \,|\, \Im y > 0\}$ the subset of complex zeroes with positive imaginary part. Moreover, we introduce the
set $\bzeta \coloneqq  \bzeta_{\mathrm{r}} \,\sqcup \, \bzeta_{\mathrm{c}}$ of independent zeroes and let $m_y \in \mathbb{Z}_{\geq 1}$ denote the order of the zero $y \in \bzeta$. For $y \in \bzeta_{\mathrm{c}}$, $\omega$ also has a zero of order $m_{\bar{y}}\coloneqq  m_y$ at $\overline{y} \in \overline{\bzeta}_{\mathrm{c}}$. The set of all zeroes of $\omega$ is $\Pi \bzeta$ and the set of all finite zeroes is $\Pi \bzeta'$.

The total number of poles of $\omega$ (counting multiplicities) is related to the total number of zeroes of $\omega$ (counting multiplicities) by
\begin{equation} \label{poles vs zeroes}
\sum_{x \in \bZ} n_x = \sum_{y \in \Pi \bzeta} m_y + 2.
\end{equation}
We will assume that the total number of poles of $\omega$ (counting multiplicities) is even, so that the total number of zeroes of $\omega$ (counting multiplicities) also is by \eqref{poles vs zeroes}.

We summarize the notations for the different subsets of poles and zeroes of $\omega$ introduced above in the tables below.
\begin{center}
\begin{tabular}{||c | c||}
 \hline
 \multicolumn{2}{|c|}{Subsets of poles of $\omega$} \\ [0.5ex] 
 \hline\hline
 $\bZ$ & all \\
 \hline
 $\bZ'$ & finite \\ 
 \hline
 $\bz_{\mathrm{r}}$ & real \\
 \hline
 $\bz_{\mathrm{r}}'$ & finite and real \\
 \hline 
 $\bz_{\mathrm{c}}$ & positive imaginary part \\
 \hline
 $\bz$ & independent\\ 
 \hline
 $\bz'$ & finite and independent\\
 \hline
\end{tabular}
\qquad
\begin{tabular}{||c | c||}
 \hline
 \multicolumn{2}{|c|}{Subsets of zeroes of $\omega$} \\ [0.5ex] 
 \hline\hline
 $\Pi \bzeta$ & all \\
 \hline
 $\Pi \bzeta'$ & finite \\
 \hline
 $\bzeta_{\mathrm{r}}$ & real \\
 \hline
 $\bzeta'_{\mathrm{r}}$ & finite and real \\
 \hline 
 $\bzeta_{\mathrm{c}}$ & positive imaginary part \\
 \hline
 $\bzeta$ & independent\\ 
 \hline
 $\bzeta'$ & finite and independent\\ 
 \hline
\end{tabular}
\end{center}

Following \cite{Lacroix:2020flf}, to account for the fact that poles and zeroes of $\omega$ have multiplicities, it will sometimes be convenient to use the notation $\bsb{\bz}$ for the set of pairs $\up{x, p}$ with $x \in \bz$ and $p = 0, \ldots, n_x - 1$, and similarly $\brb{\bzeta}$ for the set of pairs $\uz{y, q}$ with $y \in \bzeta$ and $q=0,\ldots, m_y-1$. We will use other similar notations, such as $\bsb{\bZ}$ and $\brb{\Pi\bzeta}$ for the set of all poles and all zeroes of $\omega$ with multiplicities included, respectively.

Finally, we shall also often make use of the local coordinates $\xi_x \coloneqq z-x$ at any finite pole $x \in \bz'$ or finite zero $x \in \bzeta'$ and the local coordinate at infinity $\xi_\infty \coloneqq z^{-1}$ if $\infty \in \bz$ or $\infty \in \bzeta$. The expansion of the meromorphic $1$-form at each pole $x \in \bz$ can then be written uniformly as
\begin{equation} \label{omega expansion}
\iota_x \omega = \sum_{p=0}^{n_x-1} \ell^x_p \xi_x^{-p-1} d\xi_x.
\end{equation}
For the point at infinity we have used the fact that $z^{p-1} \dr z = - \xi_\infty^{-p-1} \dr \xi_\infty$ for any $p \in \bbZ$ and introduced the additional level $\ell^\infty_0 \coloneqq - \sum_{x\in \bZ'} \ell^x_0$ for convenience.

\subsubsection{The 4d action and space of fields} \label{sec: 4d action and field}

Given a choice of meromorphic $1$-form $\omega$ as described in \S\ref{sec: omega}, the $4$d Chern-Simons action for a $\fg$-valued $1$-form $A$ on $X$ is given by \cite{Costello:2019tri}
\begin{equation} \label{ec:4dCSaction}
    S_{4d}(A)=\frac{i}{4\pi}\int_{X}\omega\wedge \mathrm{CS}(A),
\end{equation}
where $\mathrm{CS}(A) \coloneqq \langle A, \dr A + \tfrac{1}{3}[A,A]\rangle$ denotes the Chern-Simons $3$-form.

Strictly speaking, the action \eqref{ec:4dCSaction} only makes sense when $\omega$ has at most simple poles, i.e. when $n_x = 1$ for all $x \in \bz$. Indeed, when $\omega$ has higher order poles, i.e. $n_x > 1$ for some $x \in \bz$, the $4$-form $\omega \wedge \mathrm{CS}(A)$ is not locally integrable near the \emph{surface defects} $\Sigma_x \coloneqq \Sigma \times \{ x \}$ with $n_x > 1$ and therefore needs to be suitably regularised \cite[\S 3.1]{Benini:2020skc}. In what follows we will not need the precise form of the regularised action, only its variation under gauge transformations to be discussed shortly, so we refer to \cite[\S 3.1]{Benini:2020skc}, see also \cite{Li:2020ljm}, for details of this regularisation procedure. We only point out that the regularisation is `local' in the sense that it consists in modifying the $4$-form $\omega \wedge \mathrm{CS}(A)$ only locally in small neighbourhoods of the surface defects $\Sigma \times \{ x \}$ for each pole $x \in \bz$ of $\omega$. We will keep denoting the action as $S_{4d}(A)$ in the presence of higher order poles in $\omega$.

Note that the $\dr z$-component of $A$ drops out from the action \eqref{ec:4dCSaction} due to the presence of the meromorphic $1$-form $\omega = \varphi(z) \dr z$. Another way to say this is that \eqref{ec:4dCSaction} is trivially invariant under translations $A \mapsto A + \chi \dr z$ for any $\chi \in C^\infty(X, \fg)$ and we can fix this invariance by simply focusing on gauge fields with no $\dr z$-component. This remains true when $\omega$ has higher order poles \cite{Benini:2020skc}. From now on we will therefore always focus on fields of the form
\begin{equation} \label{no dz component}
A = A_\tau \dr \tau + A_\sigma \dr \sigma + A_{\bar z} \dr \bar z.
\end{equation}

On the other hand, it is too restrictive to consider only smooth $\fg$-valued $1$-forms $A$, namely \eqref{no dz component} where the components $A_\tau$, $A_\sigma$ and $A_{\bar z}$ are smooth $\fg$-valued functions on $X$. Indeed, one can allow these component functions to be singular at the zeroes of $\omega$ provided that the Lagrangian $\omega \wedge \mathrm{CS}(A)$, or its regularised version in the case when $\omega$ has higher order poles, remains locally integrable there. More precisely, let us fix a partition $\bzeta = \bzeta_+ \sqcup \bzeta_-$ of the independent zeroes of $\omega$ such that $\sum_{y \in \bzeta_+} m_y = \sum_{y \in \bzeta_-} m_y$. In particular, if all the zeroes of $\omega$ are simple, which will be the case in all our examples, then the latter condition means $|\Pi \bzeta_+| = |\Pi \bzeta_-|$.
We will take the space of fields of $4$d Chern-Simons theory to consist of $\fg$-valued $1$-forms as in \eqref{no dz component} such that:
\begin{itemize}
  \item[$(i)$] $A_\pm \coloneqq A_\tau \pm A_\sigma$ has singularities at $\Sigma \times \bzeta_\pm$ and is smooth elsewhere,
  \item[$(ii)$] $A_{\bar z}$ is smooth everywhere,
  \item[$(iii)$] the $4$-form $\omega \wedge \mathrm{CS}(A)$ is locally integrable near $\Sigma \times \bzeta$.
\end{itemize}
Condition $(iii)$ puts constraints on the type of singularities allowed in condition $(i)$ so that the action is well defined. On the other hand, condition $(ii)$ is consistent with the gauge fixing condition $A_{\bar z} = 0$ which we shall impose later on in \S\ref{sec: 2d IFT review} in order to describe integrable field theories, once we have established the gauge invariance of $4$d Chern-Simons theory in the next section.

\subsection{Gauge invariance}

Having defined the $4$d Chern-Simons action in \S\ref{sec: 4d CS action}, the next step in the approach of \cite{Benini:2020skc} is to study its gauge invariance. We therefore consider the variation of the $4$d Chern-Simons action under gauge transformations
\begin{equation} \label{gauge transf def}
A \longmapsto {}^g A \coloneqq gAg^{-1} - \dr g g^{-1}
\end{equation}
for $g \in C^{\infty}(X,G)$ an arbitrary smooth $G$-valued function. Notice once again, as in \S\ref{sec: 4d action and field}, that the $\dr z$-component of the term $\dr g g^{-1}$ will automatically drop out from the action due to the presence of the $1$-form $\omega$, so that the gauge transformation \eqref{gauge transf def} effectively acts on connections of the form \eqref{no dz component}.

In the case when $\omega$ has only simple poles, and the action takes the form \eqref{ec:4dCSaction}, we easily see that
\begin{align} \label{ec:gaugetransformation}
S_{4d}({}^g A) &= S_{4d}(A)+\frac{i}{4\pi}\int_X \omega\, \wedge\, \dr\langle g^{-1}\dr g,A\rangle \notag\\
&\qquad\qquad\qquad + \frac{i}{4\pi}\int_X \omega\, \wedge \,\langle g^{-1}\dr g,[g^{-1}\dr g,g^{-1}\dr g]\rangle \,.
\end{align}
The $4$d Chern-Simons action is thus manifestly \emph{not} gauge invariant. It is instructive to compare \eqref{ec:gaugetransformation} with the gauge variation of the usual $3$d Chern-Simons action. In particular, the first additional term generated on the right hand side of \eqref{ec:gaugetransformation} is not the integral of an exact differential precisely due to the presence of the $1$-form $\omega$. In fact, neither of the two additional terms in \eqref{ec:gaugetransformation} will vanish in general. Therefore, obtaining a better understanding of these two terms is key to being able to promote $4$d Chern-Simons theory to a gauge invariant theory.

Before stating the general result from \cite{Benini:2020skc}, it is helpful to first explain the result in the simplest case when $\omega$ has only simple poles. It can be shown, see \cite{Costello:2019tri, Delduc:2019whp}, that the first additional term on the right hand side of \eqref{ec:gaugetransformation} localises on the surface defects $\Sigma_x = \Sigma \times \{ x \}$. Explicitly, if we suppose for simplicity that all the poles are real, i.e. $\bz = \bz_{\rm r}$, then we have
\begin{equation*}
\frac{i}{4\pi}\int_X \omega\, \wedge\, \dr\langle  g^{-1}\dr g,A\rangle
= - \frac{1}{2} \sum_{x \in \bz} \ell^x_0 \int_{\Sigma_x} \langle g^{-1}\dr g, A \rangle|_{\Sigma_x}.
\end{equation*}
In turn, the right hand side of the above can be rewritten as a single integral over $\Sigma$ as follows. The collection $(A|_{\Sigma_x})_{x \in \bz}$ of the restrictions of $A \in \Omega^1(X, \fg)$ to each surface defect $\Sigma_x$, $x \in \bz$ defines a $\fg$-valued $1$-form on the disjoint union of surface defects $\sqcup_{x \in \bz} \Sigma_x$. Alternatively, this can also be thought of as defining a $1$-form on $\Sigma$ but valued in the direct product of Lie algebras $\fd = \prod_{x \in \bz} \fg$. Moreover, we have a map $\jb^\ast : \Omega^1(X,\fg) \to \Omega^1(\Sigma, \fd)$ given by $\jb^\ast A = (A|_{\Sigma_x})_{x \in \bz}$. Likewise, the collection $(g|_{\Sigma_x})_{x \in \bz}$ of the restrictions of $g \in C^\infty(X, G)$ to the surface defects defines a smooth function on $\Sigma$ valued in the direct product Lie group $D = \prod_{x \in \bz} G$, and we have a map $\jb^\ast : C^\infty(X,G) \to  C^\infty(\Sigma, D)$ given by $\jb^\ast g = (g|_{\Sigma_x})_{x \in \bz}$. Defining the bilinear form $\langle\!\langle \cdot, \cdot \rangle\!\rangle_{\fd} : \fd \times \fd \to \bR$ as $\langle\!\langle (\ms u_x)_{x \in \bz}, (\ms v_x)_{x \in \bz} \rangle\!\rangle_{\fd} = \sum_{x \in \bz} \ell^x_0 \langle \ms u_x, \ms v_x \rangle$ we may finally rewrite the first additional term on the right hand side of \eqref{ec:gaugetransformation} as
\begin{equation} \label{second term gt}
\frac{i}{4\pi}\int_X \omega\, \wedge\, \dr\langle  g^{-1}\dr g,A\rangle
= - \frac{1}{2} \int_{\Sigma} \langle\!\langle (\jb^\ast g)^{-1}\dr (\jb^\ast g), \jb^\ast A \rangle\!\rangle_{\fd}.
\end{equation}
The second additional term in \eqref{ec:gaugetransformation} may similarly be rewritten as a WZ-term for an extension of the $D$-valued field $\jb^\ast g \in C^\infty(\Sigma, D)$ to $\Sigma \times [0,1]$, see \cite{Costello:2019tri, Delduc:2019whp}.

\subsubsection{Defect Lie algebra and Lie group} \label{sec: defect Lie alg}

When $\omega$ has higher order poles, it was shown in \cite{Benini:2020skc} that the above rewriting of the gauge variation \eqref{ec:gaugetransformation} of the $4$d Chern-Simons action goes through with the obvious modifications. In particular, instead of just restricting the $\fg$-valued $1$-form $A$ on $X$ to each surface defect $\Sigma_x$ one should keep the first $n_x - 1$ orders in the Taylor expansion of $A$ near $\Sigma_x$. Correspondingly, the direct product Lie algebra $\fd$ and Lie group $D$ need to be replaced by the defect Lie algebra and Lie group \cite{Benini:2020skc,Lacroix:2020flf}.

Let $\cT_x \coloneqq \bR[\varepsilon_x]/ (\varepsilon_x^{n_x})$ for each real pole $x \in \bz_{\rm r}$ and $\cT_x \coloneqq \bC[\varepsilon_x]/ (\varepsilon_x^{n_x})$ for each complex pole $x \in \bz_{\rm c}$. We define the \emph{defect Lie algebra} as the real Lie algebra
\begin{equation} \label{defect alg}
\fd \coloneqq \prod_{x \in \bz_{\rm r}} \fg \otimes_\bR \cT_x \times \prod_{x \in \bz_{\rm c}} \fg^\bC \otimes_\bC \cT_x,
\end{equation}
where $\fg^\bC \otimes_\bC \cT_x$ is regarded as a Lie algebra over $\bR$.
The Lie algebra relations of $\fd$ are given explicitly as
\begin{equation*}
\big[ \ms u \otimes \varepsilon^p_x, \ms v \otimes \varepsilon^q_y \big] = \delta_{xy} [\ms u, \ms v] \otimes \varepsilon^{p+q}_x
\end{equation*}
where $\varepsilon_x^{p+q}=0$ for $p+q\geq n_x$. The truncated polynomial Lie algebras $\fg \otimes_\bR \cT_x$ and $\fg^\bC \otimes_\bC \cT_x$ are sometimes referred to as Takiff algebras and we will refer to the integer $p$ as the \emph{Takiff} degree of the element $\ms u \otimes \varepsilon^p_x$. Note that
\begin{equation} \label{dim poles}
\dim \fd = \dim \fg \sum_{x \in \bz_{\rm r}} n_x + 2 \dim \fg \sum_{x \in \bz_{\rm c}} n_x = \dim \fg \sum_{x \in \bZ} n_x.
\end{equation}
Recall from \S\ref{sec: omega} that we are assuming the total number of poles of $\omega$ (counting multiplicities) to be even, which implies by \eqref{dim poles} that $\dim \fd$ is even.

We define a non-degenerate invariant symmetric bilinear form
\begin{subequations} \label{form on gz rc}
\begin{equation}
\langle\!\langle \cdot, \cdot \rangle\!\rangle_{\fd} : \fd \times \fd \longrightarrow \bR
\end{equation}
with respect to which all the factors in \eqref{defect alg} are orthogonal, for any $x,y \in \bz_{\rm r}$ we set
\begin{equation} \label{form on gz 0}
\big\langle{\mkern-4mu}\big\langle \ms u \otimes \varepsilon^p_x,\ms v \otimes \varepsilon^q_y \big\rangle{\mkern-4mu}\big\rangle_{\fd} \coloneqq \delta_{xy} \, \ell^x_{p+q} \langle \ms u, \ms v \rangle,
\end{equation}
where $\ell^x_p = 0$ for all $p \geq n_x$, and for any $x, y \in \bz_{\rm c}$ we set
\begin{equation} \label{form on gz 1}
\big\langle{\mkern-4mu}\big\langle \ms u \otimes \varepsilon^p_x, \ms v \otimes \varepsilon^q_y \big\rangle{\mkern-4mu}\big\rangle_{\fd} \coloneqq \delta_{xy} \, \big( \ell^x_{p+q} \langle \ms u, \ms v \rangle + \ell^{\bar x}_{p+q} \langle \tau \ms u, \tau \ms v \rangle \big).
\end{equation}
\end{subequations}
In the case when all poles are real and simple, i.e. $\bz = \bz_{\rm r}$ and $n_x = 1$ for all $x \in \bz$, the defect Lie algebra $\fd$ reduces to the direct product Lie algebra $\prod_{x \in \bz} \fg$ considered above in the motivating example.

An important subalgebra of the defect Lie algebra $\fd$ which will play a central role in the description of \emph{degenerate} $\cE$-models is the diagonal subalgebra. To introduce it, we define the diagonal map
\begin{equation} \label{diagonal map}
\Delta : \fg \longrightarrow \fg^{\oplus|\bz|} \subset \fd, \qquad
\ms u \longmapsto (\ms u \otimes \varepsilon^0_x)_{x \in \bz}.
\end{equation}
Note that at complex points $x \in \bz_{\rm c}$ the defect Lie algebra contains a copy of the real Lie algebra $\fg$ in Takiff degree $0$ so that $\fg^{\oplus |\bz|}$ is indeed a subalgebra of $\fd$. The image of \eqref{diagonal map} then defines the \emph{diagonal subalgebra} $\ff \coloneqq \mathrm{im} \Delta \subset \fd$.

One can also introduce a real Lie group with Lie algebra $\fd$ which we will call the \emph{defect Lie group} and denote by $D$. As a set this is given by the direct product
\begin{equation}
\label{ec:defectliegroup}
D = \prod_{x \in \bz_{\rm r}} \big( G \times (\fg \otimes_\bR \cT'_x) \big) \times \prod_{x \in \bz_{\rm c}} \big( G^\bC \times (\fg^\bC \otimes_\bC \cT'_x) \big),
\end{equation}
where $\cT'_x \coloneqq \varepsilon_x \bR[\varepsilon_x]/ (\varepsilon_x^{n_x})$ for $x \in \bz_{\rm r}$ and $\cT'_x \coloneqq \varepsilon_x \bC[\varepsilon_x]/ (\varepsilon_x^{n_x})$ for $x \in \bz_{\rm c}$.
However, the general definition of the Lie group structure on $D$, which can be found in \cite{Vizman}, is quite involved so we will not include it here to avoid clutter. In practice, we will only require the group law on $D$ when discussing specific examples in \S\ref{sec: examples}, where the corresponding expressions will be explicitly stated. Finally, note that we have the diagonal embedding $\Delta : G \to G^{\times |\bz|} \subset D$ corresponding to \eqref{diagonal map} at the group level.

Just as in the above motivating example, the purpose of introducing the defect Lie algebra is that we then have a map \cite{Benini:2020skc,Lacroix:2020flf}
\begin{equation} \label{ec:jstaralg}
    \jb^*:\Omega^1(X,\fg) \longrightarrow \Omega^1 (\Sigma, \fd), \qquad
    A \longmapsto \bigg(\sum_{p=0}^{n_x-1}\frac{1}{p!} (\partial_{\xi_x}^p A)|_{\Sigma_x}\otimes \varepsilon_x^p\bigg)_{x\in \bz}
\end{equation}
where $(\partial_{\xi_x}^p A)|_{\Sigma_x} \in \Omega^{1}(\Sigma,\fg)$ denotes the pullback of $\partial_{\xi_x}^p A$ to each surface defect $\Sigma_x$. In other words, $\jb^*$ sends a $\fg$-valued $1$-form $A$ on $X$ to the first $n_x$ terms in its Taylor expansion at each surface defect $\Sigma_x$ for $x \in \bz$. Similarly, for smooth $G$-valued functions on $X$ we have a map
\begin{align} 
\label{ec:jstargroup}
\jb^*:C^{\infty}(X,G)&\longrightarrow C^{\infty}(\Sigma,D) \notag\\
g & \longmapsto \jb^*g=\bigg(g|_{\Sigma_x}, \sum_{p=1}^{n_x-1}\frac{1}{p!} \big(\partial_{\xi_x}^{p-1}(\partial_{\xi_x} g g^{-1}) \big)\big|_{\Sigma_x} \otimes \varepsilon_x^p\bigg)_{x\in \bz} \,.
\end{align}
This definition differs from the one given in \cite{Benini:2020skc} which applies to matrix Lie groups.

With the above definitions in place, we are now in a position to state one of the main results of \cite{Benini:2020skc}. Namely, for an arbitrary meromorphic $1$-form $\omega$ as in \eqref{omega def}, the variation \eqref{ec:gaugetransformation} of the (regularised) $4$d Chern-Simons action under an arbitrary gauge transformation \eqref{gauge transf def} can be expressed as
\begin{equation}
\label{ec:defectgaugetransf}
    S_{4d}({}^g A)=S_{4d}(A)-\frac{1}{2}\int_{\Sigma}\langle\!\langle (\jb^* g)^{-1}\dr (\jb^*g),\jb^* A \rangle \!\rangle_{\fd}-\frac{1}{2}I_{\fd}^{\rm WZ}[\jb^* g] \,,
\end{equation}
where we have introduced the standard WZ-term for a field $h\in C^{\infty}\left(\Sigma, D\right)$, namely
\begin{equation}
\label{ec:wesszumino}
  I_{\fd}^{\rm WZ}[h]= -\frac{1}{6}\int_{\Sigma \times I}\langle\!\langle \widehat{h}^{-1}\dr \widehat{h},[\widehat{h}^{-1}\dr \widehat{h},\widehat{h}^{-1}\dr \widehat{h}]\rangle\!\rangle_{\fd}
\end{equation}
where $I\coloneqq [0,1]$ and $\widehat{h}\in C^{\infty}\left(\Sigma\times I, D \right)$ is any smooth extension of $h$ to $\Sigma \times I$ with the property that $\widehat{h}=h$ near $\Sigma \times \{0\} \subset \Sigma \times I$ and $\widehat{h}=\mathrm{id}$ near $\Sigma \times \{1\} \subset \Sigma \times I$. Of course, the second term on the right hand side in \eqref{ec:defectgaugetransf} coincides with \eqref{second term gt} when $\omega$ has only simple poles. The virtue of the result \eqref{ec:defectgaugetransf} is that it holds for any meromorphic $1$-form $\omega$ with poles of arbitrary order.

\subsubsection{Isotropy and edge-modes}
\label{sec:isotropy}

As already anticipated, it is now clear from \eqref{ec:defectgaugetransf} that the 4d Chern-Simons action is not gauge invariant. However, gauge invariance of the theory may still be achieved upon imposing boundary conditions on both $\jb^*A$ and $\jb^*g$ at the surface defects, in order for the two additional terms appearing in \eqref{ec:defectgaugetransf} to vanish.

Recall that a Lie subalgebra $\fk \subset \fd$ is said to be \emph{isotropic} with respect to \eqref{form on gz rc} if $\langle\!\langle \ms x,\ms y \rangle\!\rangle_\fd=0$ for every $\ms x, \ms y \in \fk$. Given a subgroup $K\subset D$ whose Lie algebra $\fk\subset \fd$ is isotropic with respect to $\langle\!\langle \cdot,\cdot\rangle\!\rangle_\fd$, we can impose the boundary conditions
\begin{equation} \label{strict BCs}
\jb^*A \in \Omega^{1}(\Sigma,\fk) \qquad \text{and} \qquad \jb^*g \in C^{\infty}(\Sigma,K)
\end{equation}
on both the field $A$ and the gauge transformation parameter $g$, so that in particular $(\jb^* g)^{-1}\dr (\jb^*g)\in \Omega^{1}(\Sigma,\fk)$. The action then becomes manifestly gauge invariant since the last two terms on the right hand side of \eqref{ec:defectgaugetransf} vanish due to isotropy.

There are, however, two important related issues with the boundary conditions in \eqref{strict BCs}. Firstly, the condition imposed on $A$ is a \emph{strict} boundary condition which equates $\jb^\ast A$, the restriction of $A$ to the surface defects, with a $\fk$-valued gauge field on $\Sigma$. But in a gauge theory one should only compare gauge fields via gauge transformations and not via equalities. Secondly, the condition imposed on $g$ restricts the set of allowed gauge transformations, thereby partially breaking the gauge invariance we are trying to achieve. In particular, the strict boundary condition imposed on $A$ is preserved only by these restricted gauge transformations. Therefore, strictly speaking, even upon imposing boundary conditions, $4$d Chern-Simons theory is not a fully gauge invariant theory.

Now the role of gauge transformations is to identify physically indistinguishable field configurations, by killing \emph{would-be} degrees of freedom. So restricting the kind of gauge transformations we allow will resurrect some of these degrees of freedom from the dead \cite{Tong:2016kpv}. In particular, if we insist on establishing a fully gauge invariant theory then these \emph{resurrected} degrees of freedom must be included somehow. 

This brings us to the second main result of \cite{Benini:2020skc}. Both issues with the boundary condition \eqref{strict BCs} can be resolved by introducing a new degree of freedom living on the surface defects, namely a smooth $D$-valued field $h\in C^{\infty}(\Sigma,D)$ called the \emph{edge mode}. It was shown in \cite{Benini:2020skc} that $4$d Chern-Simons theory with the boundary conditions \eqref{strict BCs} is equivalent to $4$d Chern-Simons theory coupled to the edge mode by introducing the extended action
\begin{equation}
\label{ec:edgemodeaction}
    S_{4d}^{\rm ext}(A,h)=S_{4d}(A)-\frac{1}{2}\int_{\Sigma}\langle\!\langle h^{-1}\dr h,\jb^* A \rangle \!\rangle_\fd - \frac{1}{2}I_{\fd}^{\rm WZ}[h] \,,
\end{equation}
together with the alternate boundary condition 
\begin{equation}
\label{ec:edgemodecondition}
{}^h (\jb^*A) \in \Omega^1(\Sigma,\fk)\,.    
\end{equation}
That is, instead of imposing boundary conditions on $A$ and $g$ as in \eqref{strict BCs}, we only impose the boundary condition on $A$ and only up to a gauge transformation by $h$.

One can verify, using \eqref{ec:defectgaugetransf}, the Polyakov-Wiegmann identity \cite{POLYAKOV1983121} and the invariance of the bilinear form, that both the extended action \eqref{ec:edgemodeaction} and the constraint \eqref{ec:edgemodecondition} are invariant under the gauge transformation
\begin{equation}
\label{ec:Gsymmetry}
    A\longmapsto {}^g A\,, \qquad h \longmapsto h (\jb^* g)^{-1}
\end{equation}
with arbitrary $g\in C^{\infty}(X,G)$. Thus, we have defined a fully gauge invariant theory, at the price of adding a new field.

Observe that if we restrict the edge mode $h$ to take values in $K$ then we recover the original $4$d Chern-Simons action together with the original boundary conditions \eqref{strict BCs}. More precisely, the extended action \eqref{ec:edgemodeaction} enjoys the additional symmetry
\begin{equation}
\label{ec:gaugek}
    h \longmapsto k h
\end{equation}
for arbitrary $k\in C^{\infty}(\Sigma,K)$. The invariance of \eqref{ec:edgemodeaction} under \eqref{ec:gaugek}, can be verified using the Polyakov-Wiegmann identity, the constraint \eqref{ec:edgemodecondition} and the isotropy of $\fk$.
In other words, the degrees of freedom added can be described by a smooth field on $\Sigma$ valued in the quotient $K \setminus D$.

\subsection{2d integrable field theories} \label{sec: 2d IFT review}

Although $4$d Chern-Simons theory coupled to the edge mode as described above is equivalent to the original $4$d Chern-Simons theory with boundary conditions \eqref{strict BCs}, the advantage of the former is that it leads more naturally to $2$d integrable field theories. In particular, the field content of the latter will correspond precisely to the edge mode degrees of freedom living on the defect. Moreover, the Lax connection of the $2$d integrable field theory will come directly from the gauge field $A$ of the $4$d Chern-Simons theory in \eqref{no dz component}.

There are, however, two glaring issues with interpreting the gauge field \eqref{no dz component} as a Lax connection. The first is that $A$ has a component along the $\dr \bar z$ direction whereas a Lax connection should be a $1$-form along $\Sigma$. The second is that as it stands $A$ is not meromorphic in the $z$-coordinate which we would like to interpret as the spectral parameter of the Lax connection.

The first issue is easily resolved. Indeed, we can partially fix the gauge invariance of \eqref{ec:edgemodeaction} using the gauge fixing condition $A_{\bar z} = 0$ in order to get rid of the undesired $\dr \bar{z}$-component. This is analogous to the axial gauge in electrodynamics and Yang-Mills theories, where one of the component of the gauge field is set to vanish.
We will suggestively denote the gauge field $A$ in this gauge by the letter $\mathcal{L}$. Note that there is a residual gauge symmetry \eqref{gauge transf def} by $g\in C^{\infty}(X,G)$ satisfying $\partial_{\bar{z}}g g^{-1}=0$.

The second issue is more problematic. In order to resolve it we will have to go partly on-shell, which we turn to next.

\subsubsection{Solving the bulk equations of motion} \label{sec: solving bulk eom}

The action \eqref{ec:edgemodeaction} defines a $4$-dimensional theory due to the presence of the `bulk' term $S_{4d}(\cL)$ for the `bulk' field $\cL = \cL_\tau \dr \tau + \cL_\sigma \dr \sigma$. To obtain a $2$-dimensional theory we will therefore restrict to solutions of the bulk equations of motion. Specifically, varying the action \eqref{ec:edgemodeaction} with respect to both $\cL$ and $h$, subject to the constraint \eqref{ec:edgemodecondition}, we find the bulk and boundary field equations of motion \cite{Benini:2020skc}
\begin{subequations}
\begin{align}
\label{ec:bulkeom} \partial_{\bar z} \mathcal{L} &= 0 \quad \text{on} \quad \Sigma \times (\CP \setminus \bzeta),\\
\label{ec:boundaryeom} \dr_{\Sigma}(\jb^*\cL)+\tfrac{1}{2}\left[\jb^*\cL,\jb^*\cL\right] &= 0 \quad \text{on} \quad \Sigma,
\end{align}
\end{subequations}
where $\dr_\Sigma$ denotes the de Rham differential on $\Sigma$.

The bulk equation of motion \eqref{ec:bulkeom} expresses the fact that the components of $\cL = \cL_\tau \dr \tau + \cL_\sigma \dr \sigma$ are holomorphic along $\CP$ away from the set $\bzeta$ of zeroes of $\omega$. However, recall from condition $(i)$ in \S\ref{sec: 4d action and field} that we allow the light-cone components $\cL_\pm = \cL_\tau \pm \cL_\sigma$ of the gauge field to have singularities at the subset of poles $\bzeta_\pm$, so long as the Lagrangian $\omega \wedge \mathrm{CS}(\cL)$ remained integrable along each surface $\Sigma \times \{ y \}$ for $y \in \bzeta$. We can therefore take the components of $\cL$ to be of the form
\begin{subequations} \label{L rational form}
\begin{equation} \label{ec:polestructureadm}
    \cL_\mu = \sum_{y \in \bzeta'}\sum_{q=0}^{m_y-1}\frac{\cL_\mu^{(y,q)}}{(z-y)^{q+1}}+\sum_{q=0}^{m_\infty-1}\cL_\mu^{(\infty,q)}z^{q+1}+\cL_\mu^{\rm c}
\end{equation}
for $\mu=\tau,\sigma$, where the coefficient functions $\cL_\mu^{\rm c}\in C^{\infty}(\Sigma,\fg)$, $\cL_{\mu}^{\uz{y,q}} \in C^{\infty}(\Sigma,\fg)$ for $\uz{y, q} \in \brb{\bzeta_{\rm r}}$ and $\cL_{\mu}^{\uz{y,q}} \in C^{\infty}(\Sigma,\fg^\bC)$ for $\uz{y, q} \in \brb{\bzeta_{\rm c}}$ are related by
\begin{equation} \label{E-model condition}
\cL_\tau^{\uz{y,q}} = \epsilon_y \cL_\sigma^{\uz{y,q}}
\end{equation}
\end{subequations}
with $\epsilon_y = \pm 1$ for $y \in \bzeta_\pm$. Note that \eqref{E-model condition} ensures that the light-cone component $\cL_\pm$ only has poles in $\bzeta_\pm$, and not in $\bzeta_\mp$, as required by condition $(i)$ from \S\ref{sec: 4d action and field}.
To see why singularities of the form \eqref{L rational form} are allowed by condition $(iii)$, consider the case when $\omega$ has only simple poles so that the action takes the form \eqref{ec:4dCSaction}. The cubic term in the Chern-Simons $3$-form drops out since $\cL$ only has legs along $\dr \sigma$ and $\dr \tau$ so that the bulk Lagrangian reads
\begin{subequations} \label{admissibility quadr cubic}
\begin{equation} \label{admissibility quadr}
\omega \wedge \mathrm{CS}(\cL) = \omega \wedge \big( \langle \cL_+, \bar \partial \cL_- \rangle - \langle \cL_-, \bar \partial \cL_+ \rangle\big) \wedge \dr \sigma^+ \wedge \dr \sigma^-.
\end{equation}
It is then easy to see that this $4$-form is locally finite near the surface $\Sigma \times \{y\}$ for each $y \in \bzeta$. Notice, in particular, that the condition \eqref{E-model condition} is used here to guarantee that the set of poles of $\cL_\pm$ are disjoint so that, for instance, the $\delta$-functions at the set $\bzeta_-$ arising from $\bar \partial \cL_-$ are multiplied by poles in $\bzeta_+$ coming from $\cL_+$. Moreover, note that the singularities of the form \eqref{ec:polestructureadm} are also consistent with the cubic term in the Lagrangian before moving to the gauge $A_{\bar z} = 0$, i.e. when $A_{\bar z}$ is non-zero but smooth as in condition $(ii)$ of \ref{sec: 4d action and field}, namely
\begin{equation} \label{admissibility cubic}
\omega \wedge \langle A, \tfrac{1}{3}[A,A]\rangle = -2 \omega \wedge \langle A_{\Bar{z}},[A_+,A_-]\rangle \wedge \dr \bar z \wedge \dr \sigma^+ \wedge \dr \sigma^-.
\end{equation}
\end{subequations}
The above remains true also when $\omega$ has higher order poles since the regularisation procedure, to make sense of the action in that case, only modifies the Lagrangian locally near the surface defects $\Sigma \times \{ x \}$ for each $x \in \bz$.

Gauge fields $\cL$ of the form \eqref{ec:polestructureadm} satisfying the condition \eqref{E-model condition} were referred to in \cite{Benini:2020skc, Lacroix:2020flf} as being \emph{admissible}. More precisely, \eqref{E-model condition} is the most natural solution of the admissibility conditions given there, cf. \cite[Example 5.4]{Benini:2020skc} and \cite[\S 3.5 \& \S 4.2]{Lacroix:2020flf}. The observation we made above is that the admissibility condition can be traced back to the $4$d Chern-Simons Lagrangian as the requirement that it be locally integrable.
In fact, this new perspective on admissible solutions of the bulk equations of motion \eqref{ec:bulkeom} leads to the following observation which will be useful later.

\begin{remark} \label{rem: stronger poles in L}
The rational expression \eqref{ec:polestructureadm}, with poles of order $m_y$ at each $y \in \bzeta$, is not the most general one for which the bulk action is locally integrable. Indeed, one could take poles of order $m_y + 1$ at each $y \in \bzeta$ while maintaining the local integrability of the expression \eqref{admissibility quadr} and also of \eqref{admissibility cubic} before fixing the gauge $A_{\bar z} = 0$. This is because both of the top forms in \eqref{admissibility quadr cubic} are locally integrable near a \emph{simple} pole of the component \cite[Lemma 2.1]{Benini:2020skc}. Note that this is precisely why the $4$d Chern-Simons action \eqref{ec:4dCSaction} was well defined in the case when $\omega$ has only simple poles. The reason we have kept the strength of the poles in \eqref{ec:polestructureadm} as they are is to ensure that the $2$d action we end up with is integrable, as we will see shortly.
\end{remark}

Upon restricting the gauge field $\cL$ to be a solution of the bulk equation of motion \eqref{ec:bulkeom} as in \eqref{ec:polestructureadm}, the bulk $4$-dimensional term in the action \eqref{ec:edgemodeaction} disappears and we are left with the $2$-dimensional action
\begin{equation}
\label{ec:edgemodele1}
    S_{2d}(\mathcal{L},h)=-\frac{1}{2}\int_{\Sigma}\langle\!\langle h^{-1}\dr h,\jb^* \mathcal{L} \rangle \!\rangle_\fd-\frac{1}{2}I_\fd^{\rm WZ}[h] \,.
\end{equation}
The equation of motion of this action is the boundary equation of motion \eqref{ec:boundaryeom}. It was shown in \cite[Proposition 5.6]{Benini:2020skc} that for admissible solutions of the bulk equations of motion, the flatness equation \eqref{ec:boundaryeom} for $\jb^*\cL$ lifts to a flatness equation for $\cL$ itself, namely
\begin{equation} \label{flatness of Lc}
    \dr \cL +\tfrac{1}{2}[\cL,\cL]=0 \quad \text{on }\Sigma\,.
\end{equation}
Here we can make use of the observation in Remark \ref{rem: stronger poles in L} by noting that the argument in the proof of \cite[Proposition 5.6]{Benini:2020skc} still applies if we increase the order of \emph{one} of the poles of $\cL$ with components \eqref{ec:polestructureadm} by $1$. In other words, although the requirement that the action be well defined allows us to increase the order of \emph{all} the poles in $\cL$ by $1$, the requirement that the boundary equations of motion \eqref{ec:boundaryeom} lift to the flatness of $\cL$ in \eqref{flatness of Lc}, which ultimately ensures integrability, only enables us to increase the order of \emph{one} of the poles in $\cL$ by $1$ while keeping the strength of all the other poles the same. We will see another proof of this later in \S\ref{sec: j hat iso}.

Finally, recall that after removing the $\dr \bar z$-component of the gauge field the gauge symmetry \eqref{ec:Gsymmetry} was restricted to those $g \in C^\infty(X, G)$ such that $\partial_{\bar z} g g^{-1} = 0$. And to preserve the pole structure of the meromorphic gauge field $\cL$ in \eqref{ec:polestructureadm} we can restrict to $g \in C^\infty(\Sigma, G)$ which are independent of the coordinate on $\CP$. It follows from \eqref{ec:jstargroup} that for such gauge transformation parameters we have $\jb^\ast g = \Delta(g)$, where recall that $\Delta : G \to G^{\times |\bz|} \subset D$ is the diagonal embedding.

\subsubsection{The 2d action} \label{sec: 2d action}

The $2$-dimensional action \eqref{ec:edgemodele1} which we obtained from \eqref{ec:edgemodeaction} by solving the bulk equations of motion in \S\ref{sec: solving bulk eom} should, of course, be supplemented by the boundary condition given in \eqref{ec:edgemodecondition}. In other words, the fields $\cL$ and $h$ on which the action \eqref{ec:edgemodele1} depends are not independent but instead are related by the constraint
\begin{equation} \label{constraint L h}
{}^h (\jb^*\cL) \in \Omega^{1}(\Sigma,\fk).
\end{equation}
The final step for obtaining a $2$-dimensional integrable field theory is therefore to solve the constraint \eqref{constraint L h} to find an expression for $\cL$ in terms of $h$.

Indeed, \emph{suppose} that we can find a unique solution $\cL=\cL(h)$ to the constraint \eqref{constraint L h}. In order to respect the gauge invariance \eqref{ec:Gsymmetry} and \eqref{ec:gaugek} we further assume that this solution is such that
\begin{equation}
\label{ec:equivariance}
    \jb^*\cL \big( kh\Delta(g)^{-1} \big) = {}^{\Delta(g)} \big( \jb^*\cL(h) \big)
\end{equation}
for every $g \in C^{\infty}(\Sigma, G)$ and $k\in C^{\infty}(\Sigma, K)$. Note that the existence and uniqueness of such a solution depends on the choice of Lagrangian subalgebra $\fk \subset \fd$. One of the main results of \cite{Lacroix:2020flf} was to explicitly construct such solutions. The resulting models were shown to coincide with integrable non-degenerate $\cE$-models. In the remainder of this article we will generalise the construction of \cite{Lacroix:2020flf} to obtain a more general class of solutions to \eqref{constraint L h} leading to the class of integrable \emph{degenerate} $\cE$-models.

Given any solution of the boundary condition \eqref{constraint L h} satisfying the equivariance property \eqref{ec:equivariance}, the action \eqref{ec:edgemodele1} reduces to a $2$-dimensional action for the edge mode field $h \in C^{\infty}(\Sigma,D)$ \emph{alone} given by
\begin{equation}
\label{ec:2daction}
S_{2d}(h)=-\frac{1}{2}\int_\Sigma \langle \!\langle h^{-1}\dr h, \jb^*\mathcal{L}(h)\rangle\!\rangle_\fd-\frac{1}{2}I_\fd^{\rm WZ}[h] \,.
\end{equation}
By virtue of the property \eqref{ec:equivariance}, this action is invariant under the transformations
\begin{equation}
\label{ec:gaugesymmetry2daction}
    h \longrightarrow kh\Delta(g)^{-1}
\end{equation}
for any $k\in C^{\infty}(\Sigma,K)$ and $g\in C^{\infty}(\Sigma,G)$. Moreover, the equations of motion \eqref{flatness of Lc} which arose from the boundary equations of motion of the original $4$d Chern-Simons action now read
\begin{equation} \label{flatness of Lh}
    \dr \cL(h) + \tfrac{1}{2}[\cL(h),\cL(h)]=0\,.
\end{equation}
In other words, the $2$-dimensional action \eqref{ec:2daction} has an associated Lax connection $\mathcal{L}(h)$ and therefore describes a $2$-dimensional integrable field theory.

\section{Obtaining degenerate \texorpdfstring{$\cE$}{E}-models} \label{sec:reductionto2d}

The purpose of this section is to complete the passage from $4$d Chern-Simons theory to $2$d integrable field theories. In particular, we will show how to obtain integrable degenerate $\cE$-models. Since the details of this section are quite technical, the reader interested in applying the construction may wish, on first read, to skip to \S\ref{sec: examples} where we present various examples of the procedure in detail. They may then refer back to the present section for further details of the construction. Before presenting these details, and in order to facilitate the reading of this section, we begin by giving a brief outline of the main strategy.

\medskip

As recalled in \S\ref{sec: 2d action}, the very last step in the approach of \cite{Benini:2020skc} for passing from $4$d Chern-Simons theory to $2$d integrable field theories consists in finding a solution of the constraint equation \eqref{constraint L h} which satisfies the transformation property \eqref{ec:equivariance}. Such solutions were constructed in \cite{Lacroix:2020flf} under the assumption that $\omega$ has a double pole at infinity. This technical assumption was used to fix the gauge symmetry under $F$. Specifically, under the assumption that $n_\infty = 2$, the component of the edge mode $h \in C^\infty(\Sigma, D)$ at infinity is a field on $\Sigma$ valued in the semi-direct product $G \ltimes \fg$. The latter can be brought to the identity by using the $F$ symmetry and the component of the $K$ symmetry associated with the point at infinity, see \cite[\S 3.6]{Lacroix:2020flf} for details. With the gauge symmetry under $F$ fixed in this way, the component at infinity of the constraint \eqref{constraint L h} forces the constant term in the Lax connection to vanish (note that since $\omega$ initially has a pole at infinity, the Lax connection cannot have poles there). Therefore, the Lax connections considered in \cite{Lacroix:2020flf} are of the special form
\begin{equation} \label{L from LV}
\cL_\mu = \sum_{y \in \bzeta'}\sum_{q=0}^{m_y-1}\frac{\cL_\mu^{(y,q)}}{(z-y)^{q+1}}.
\end{equation}
Moreover, the property \eqref{ec:equivariance} that the solution $\cL = \cL(h)$ is required to satisfy boils down to $\jb^*\cL (kh) = \jb^*\cL(h)$. And indeed, the solutions constructed in \cite{Lacroix:2020flf} were shown to have this property and the resulting $2$d integrable field theories were shown to coincide with integrable non-degenerate $\cE$-models.

The main purpose of this section is to generalise the results of \cite{Lacroix:2020flf} to the case of a generic $1$-form $\omega$, as defined in \S\ref{sec: omega}. The key idea behind the approach of \cite{Lacroix:2020flf} for solving \eqref{constraint L h} is to construct an involution $\cE : \fd \SimTo \fd$ on the defect Lie algebra with the property that
\begin{equation} \label{non-deg E-model condition}
\jb_\bz \cL_\tau = \cE (\jb_\bz \cL_\sigma),
\end{equation}
where we wrote $\jb^\ast \cL = \jb_z \cL_\tau \dr \tau +\jb_z \cL_\sigma \dr \sigma$ in components.
More precisely, the property \eqref{non-deg E-model condition} was the one imposed in \cite{Lacroix:2020flf} but it will have to be adapted in the present case, see \eqref{deg E-model condition} below.
The property \eqref{non-deg E-model condition} was then used in \cite{Lacroix:2020flf} as the starting point for solving the constraint \eqref{constraint L h}.

In order to build such an involution $\cE : \fd \SimTo \fd$ satisfying \eqref{non-deg E-model condition}, observe that the relationship between $\cL_\tau$ and $\cL_\sigma$ is in fact very simple to describe in terms of the coefficients of these rational functions \eqref{L from LV}. Indeed, recall that these are related by \eqref{E-model condition}, namely $\cL_\tau^{(y,q)} = \epsilon_y \cL_\sigma^{(y,q)}$ where $\epsilon_y = \pm 1$ for each $y \in \bzeta$. The idea of \cite{Lacroix:2020flf} is then to build two isomorphisms
\begin{equation} \label{two isos}
\begin{tikzcd}
\fd & R'_{\Pi \bzeta'}(\fg^\bC)^\Pi \arrow[l, "\jb_\bz"', "\cong"] \arrow[r, "\boldsymbol{\pi_{\bzeta'}}", "\cong"'] & \fg^{\brb{\bzeta'}}
\end{tikzcd}
\end{equation}
from a certain space of rational functions $R'_{\Pi \bzeta'}(\fg^\bC)^\Pi$, where the components \eqref{L from LV} of the Lax connection live, to the defect Lie algebra $\fd$ by Taylor expanding at each $x \in \bz$ and to a vector space $\fg^{\brb{\bzeta'}}$ by extracting the coefficients at each pole $y \in \bzeta'$.

If we do not fix the gauge symmetry by $F$, as was done in \cite{Lacroix:2020flf}, then the components of the Lax connection still have a constant term compared to \eqref{L from LV} and can in general also have a pole at infinity, cf. \eqref{L rational form}. In this section we will adapt the construction of \cite{Lacroix:2020flf} summarised above to this case, in particular defining suitable generalisations of the above isomorphisms \eqref{two isos} in \S\ref{sec: pi zeta iso} and \S\ref{sec: j hat iso}. These will then be used in \S\ref{sec: E-operator} to build an involution $\cE :\fd \SimTo \fd$ which is symmetric with respect to the bilinear form on $\fd$ introduced in \S\ref{sec: defect Lie alg}. Finally, we will use the latter in \S\ref{sec: E-model} to construct solutions of the constraint \eqref{constraint L h} satisfying \eqref{ec:equivariance} and thereby obtain the action of integrable degenerate $\cE$-models.

\subsection{The real vector space \texorpdfstring{$R_{\Pi\bzeta}(\fg^\bC)^\Pi$}{Rg}} \label{sec: vs Rg}

Given a complex vector space $V$ we let $R_{\Pi\bzeta}(V)$ denote the space of $V$-valued rational functions with poles at each $y \in \Pi \bzeta$ of order at most $m_y$, the order of the zero $y$ of $\omega$. It will also be useful to define the subspace $R'_{\Pi\bzeta}(V) \subset R_{\Pi\bzeta}(V)$ of $V$-valued rational functions without constant term.

If $V$ is equipped with an anti-linear involution $\tau : V \to V$ then we can define an action of $\Pi$ on $V$ by letting $\ms t \in \Pi$ act as $\tau$. This then also lifts to an action of $\Pi$ on $R_{\Pi\bzeta}(V)$. We can also define an action of $\Pi$ on $R_{\Pi\bzeta}(V)$ by letting $\ms t \in \Pi$ act as the pullback by complex conjugation $\mu_{\ms t} : z \mapsto \bar z$. We let $R_{\Pi\bzeta}(V)^\Pi$ denote the real vector space of rational functions in $R_{\Pi\bzeta}(V)$ on which these two actions coincide. We will also make use of the subspace $R'_{\Pi\bzeta}(V)^\Pi \subset R_{\Pi\bzeta}(V)^\Pi$ of such rational functions without constant term.

In what follows we will either take $V = \fg^\bC$ or $V = C^\infty(\Sigma, \fg^\bC)$, where the action of $\Pi$ on the latter is induced from the action of $\Pi$ on $\fg^\bC$.
Explicitly, an element $f \in R_{\Pi\bzeta}(\fg^\bC)^\Pi$ is a $\Pi$-equivariant $\fg$-valued rational function of the form
\begin{equation} \label{uz def}
f(z) = \sum_{y \in \Pi \bzeta'} \sum_{q=0}^{m_y-1} \frac{f^{\uz{y, q}}}{(z-y)^{q+1}} + \sum_{q=0}^{m_\infty-1} f^{(\infty, q)} z^{q+1} + f^{\rm c}
\end{equation}
where $f^{\rm c} \in \fg$, $f^{\uz{y, q}} \in \fg$ for all $\uz{y, q} \in \brb{\bzeta_{\rm r}}$ with $q = 0, \ldots, m_y - 1$, and $f^{\uz{y, q}} \in \fg^\bC$ for all $\uz{y, q} \in \brb{\bzeta_{\rm c}}$ with $q = 0, \ldots, m_y - 1$ and $f^{\uz{\bar y, q}} = \tau f^{\uz{y, q}}$.

The dimension of the real vector space $R_{\Pi\bzeta}(\fg^\bC)^\Pi$ is given by
\begin{equation} \label{dim Rg}
\dim R_{\Pi\bzeta}(\fg^\bC)^\Pi = \dim \fg \bigg( \sum_{y \in \Pi \bzeta} m_y + 1 \bigg).
\end{equation}
The term $\dim \fg \sum_{y \in \Pi \bzeta} m_y$ comes from counting the degrees of freedom in the pole parts at each $y \in \Pi \bzeta$ of a generic rational function \eqref{uz def}, see in particular the second equality in \eqref{dim zeroes} later. The additional $\dim \fg$ comes from the constant term $f^{\rm c}$ in \eqref{uz def}. An alternative way of counting the dimension of $R_{\Pi\bzeta}(\fg^\bC)^\Pi$ is to consider instead the isomorphic space $R_{\Pi\bzeta}(\fg^\bC)^\Pi \omega$ which consists of $\Pi$-equivariant $\fg^\bC$-valued meromorphic $1$-forms with poles in $\bZ$. Its dimension is then given by
\begin{equation} \label{dim Rg 2}
\dim R_{\Pi\bzeta}(\fg^\bC)^\Pi = \dim \fg \bigg( \sum_{x \in \bZ} n_x - 1 \bigg),
\end{equation}
where the term $\dim \fg \sum_{x \in \bZ} n_x$ comes from counting the degrees of freedom in the pole parts of $f \omega$ for $f \in R_{\Pi\bzeta}(\fg^\bC)^\Pi$ at each $x \in \bZ$ and the additional $- \dim \fg$ accounts for the fact that the sum of the residues of a meromorphic $1$-form vanishes.
Of course, the two expressions \eqref{dim Rg} and \eqref{dim Rg 2} coincide by virtue of \eqref{poles vs zeroes}.

A $\Pi$-equivariant $\fg$-valued rational function $f \in R'_{\Pi\bzeta}(\fg^\bC)^\Pi$ takes the form
\begin{equation} \label{upz def}
f(z) = \sum_{y \in \Pi \bzeta'} \sum_{q=0}^{m_y-1} \frac{f^{\uz{y, q}}}{(z-y)^{q+1}} + \sum_{q=0}^{m_\infty-1} f^{(\infty, q)} z^{q+1}.
\end{equation}
Note that the only difference with \eqref{uz def} is that the `constant' term $f^{\rm c} \in \fg$ is missing in \eqref{upz def}. The constant rational functions in $R_{\Pi\bzeta}(\fg^\bC)^\Pi$ form a subspace isomorphic to $\fg$ and we have a direct sum decomposition
\begin{equation} \label{rat func split off constant}
R_{\Pi\bzeta}(\fg^\bC)^\Pi = \fg \dotplus R'_{\Pi\bzeta}(\fg^\bC)^\Pi
\end{equation}
given explicitly by writing a function $f \in R_{\Pi \bzeta}(\fg^\bC)^\Pi$ as in \eqref{uz def}, with the first two sums defining the component in $R'_{\Pi\bzeta}(\fg^\bC)^\Pi$, cf. \eqref{upz def}, and the constant term $f^{\rm c} \in \fg$ corresponding to the component in $\fg$.

It is useful to adjoin another copy of $\fg$ to the space $R_{\Pi\bzeta}(\fg^\bC)^\Pi$ by considering the direct sum $R_{\Pi\bzeta}(\fg^\bC)^\Pi \oplus \fg$. It is important to note that this additional copy of $\fg$ is distinct from the copy of $\fg$ already present in \eqref{rat func split off constant}, representing the constant term in the rational function. It follows from comparing \eqref{dim poles} with \eqref{dim Rg 2} that
\begin{equation} \label{dim R plus g}
\dim \big( R_{\Pi\bzeta}(\fg^\bC)^\Pi \oplus \fg \big) = \dim \fd.
\end{equation}

We define the symmetric bilinear form
\begin{subequations} \label{bilinear R zeta}
\begin{equation} \label{bilinear R zeta a}
\langle\!\langle \cdot, \cdot \rangle\!\rangle_\omega : \big( R_{\Pi\bzeta}(\fg^\bC)^\Pi \oplus \fg \big) \times \big( R_{\Pi\bzeta}(\fg^\bC)^\Pi \oplus \fg \big) \longrightarrow \bR,
\end{equation}
given for any $f, g \in R_{\Pi\bzeta}(\fg^\bC)^\Pi$ and $\ms u, \ms v \in \fg$ by
\begin{equation} \label{bilinear R zeta b}
\langle\!\langle (f, \ms u), (g, \ms v) \rangle\!\rangle_\omega \coloneqq \sum_{x \in \bZ} \mathrm{res}_x \langle f, g \rangle \omega + \langle \ms u, g^{\rm c} \rangle + \langle f^{\rm c}, \ms v \rangle.
\end{equation}
\end{subequations}
By a slight abuse of notation, we will often denote the restriction of \eqref{bilinear R zeta} to the subspace $R_{\Pi\bzeta}(\fg^\bC)^\Pi$ also as $\langle\!\langle \cdot, \cdot \rangle\!\rangle_\omega$. That is, we will write $\langle\!\langle f, g \rangle\!\rangle_\omega \coloneqq \langle\!\langle (f, 0), (g, 0) \rangle\!\rangle_\omega$ for any $f, g \in R_{\Pi\bzeta}(\fg^\bC)^\Pi$.

\begin{lemma}
The bilinear form \eqref{bilinear R zeta} is non-degenerate. Its restriction to $R_{\Pi\bzeta}(\fg^\bC)^\Pi$ is degenerate and its restriction to $R'_{\Pi\bzeta}(\fg^\bC)^\Pi$ is also non-degenerate.
\begin{proof}
Let us first show that the restriction to $R_{\Pi\bzeta}(\fg^\bC)^\Pi$ is degenerate. Consider a constant function $f = f^{\rm c} \in R_{\Pi\bzeta}(\fg^\bC)^\Pi$. Then for any $g \in R_{\Pi\bzeta}(\fg^\bC)^\Pi$ the poles of the $1$-form $\langle f, g \rangle \omega$ are contained in $\bZ$ so it follows that $\langle\!\langle f, g \rangle\!\rangle_\omega = 0$.

Consider now the restriction of the bilinear form \eqref{bilinear R zeta} to $R'_{\Pi\bzeta}(\fg^\bC)^\Pi$. We show that this is non-degenerate.
For every $f, g \in R'_{\Pi\bzeta}(\fg^\bC)^\Pi$ we have
\begin{equation*}
\langle\!\langle f, g \rangle\!\rangle_\omega = \sum_{x \in \bZ} \mathrm{res}_x \langle f, g \rangle \omega.
\end{equation*}
Since $f$ is not constant, it has poles at some of the zeroes of $\omega$. By choosing $g$ to also have a pole at one of these same zeroes, we can ensure that some of the poles of the $1$-form $\langle f, g \rangle \omega$ lie outside the subset $\bZ$, namely in $\Pi \bzeta$. It is then possible to choose $g$ such that $\langle\!\langle f, g \rangle\!\rangle_\omega \neq 0$.

It is clear from the form of the additional two terms in \eqref{bilinear R zeta b} that the bilinear form on the whole of $R_{\Pi \bzeta}(\fg^\bC)^\Pi \oplus \fg$ is itself also non-degenerate.
\end{proof}
\end{lemma}

\subsection{The isomorphism \texorpdfstring{$\boldsymbol{\pi_\bzeta} : R_{\Pi\bzeta}(\fg^\bC)^\Pi \SimTo \fg \oplus \gzeta$}{pi zeta}} \label{sec: pi zeta iso}

A rational function $f \in R_{\Pi\bzeta}(\fg^\bC)^\Pi$, as in \eqref{uz def}, is uniquely determined by the coefficients $f^{\uz{y, q}} \in \fg$ for $\uz{y, q} \in \brb{\bzeta_{\rm r}}$ and $f^{\uz{y, q}} \in \fg^\bC$ for $\uz{y, q} \in \brb{\bzeta_{\rm c}}$ at each of its real and complex poles, along with the constant term $f^{\rm c} \in \fg$. It is therefore convenient to introduce the vector space in which the coefficients of such rational functions live.
Explicitly, we associate with the zeroes of $\omega$ the real vector space
\begin{equation} \label{gzeta def}
\gzeta \coloneqq \prod_{\uz{y, q} \in \brb{\bzeta_{\rm r}}} \fg \times \prod_{\uz{y, q} \in \brb{\bzeta_{\rm c}}} \fg^\bC
\end{equation}
where $\fg^\bC$ is regarded as a real vector space. Its dimension is
\begin{equation} \label{dim zeroes}
\dim \gzeta = \dim \fg \sum_{y \in \bzeta_{\rm r}} m_y + 2 \dim \fg \sum_{y \in \bzeta_{\rm c}} m_y = \dim \fg \sum_{y \in \Pi \bzeta} m_y.
\end{equation}

We now have an obvious isomorphism
\begin{align} \label{pi zeta def}
\boldsymbol{\pi_\bzeta} : R_{\Pi\bzeta}(\fg^\bC)^\Pi &\overset{\cong}\longrightarrow \fg \oplus \gzeta, \notag\\
f &\longmapsto \Big( f^{\rm c}, \big( f^{\uz{y, q}} \big)_{\uz{y, q} \in \brb{\bzeta}} \Big)
\end{align}
which takes a rational function in $R_{\Pi\bzeta}(\fg^\bC)^\Pi$ and returns its constant term in $\fg$ and the coefficients at each of its poles as an element of $\gzeta$.

We also extend this map to an isomorphism $\boldsymbol{\pi_\bzeta} : R_{\Pi\bzeta}(\fg^\bC)^\Pi \oplus \fg \SimTo \fg \oplus \gzeta \oplus \fg$ by letting it act trivially on the additional copy of $\fg$ introduced in \S\ref{sec: vs Rg}.

We define the symmetric bilinear form, cf. \cite[(4.16)]{Lacroix:2020flf},
\begin{subequations} \label{bilinear g zeta}
\begin{equation} \label{bilinear g zeta a}
\langle\!\langle \cdot, \cdot \rangle\!\rangle_{\fg \oplus \gzeta \oplus \fg} : \big( \fg \oplus \gzeta \oplus \fg \big) \times \big( \fg \oplus \gzeta \oplus \fg \big) \longrightarrow \bR,
\end{equation}
given for any $\ms U = (\ms U^{\uz{y,q}})_{\uz{y,q} \in \brb{\bzeta}}, \ms V = (\ms V^{\uz{y,q}})_{\uz{y,q} \in \brb{\bzeta}} \in \gzeta$ and $\ms x, \ms x', \ms y, \ms y' \in \fg$ by
\begin{align} \label{bilinear g zeta b}
&\langle\!\langle (\ms x, \ms U, \ms x'), (\ms y, \ms V, \ms y') \rangle\!\rangle_{\fg \oplus \gzeta \oplus \fg} \notag\\
&\qquad\qquad \coloneqq \sum_{y \in \bzeta} \sum_{\substack{p, q=0\\ p+q \geq m_y - 1}}^{m_y-1} \frac{2}{|\Pi_y|} \Re \bigg( \alpha_{p,q} \big\langle \ms U^{\uz{y,p}}, \ms V^{\uz{y,q}} \big\rangle \bigg) + \langle \ms x, \ms y' \rangle + \langle \ms x', \ms y \rangle,
\end{align}
\end{subequations}
where $\alpha_{p,q} \coloneqq - \frac{1}{(p+q+1-m_y)!} \partial_{\xi_y}^{p+q+1-m_y} \psi_y(\xi_y)$ for $p,q = 0, \ldots, m_y$ is symmetric under the exchange of $p$ and $q$. Here we wrote $\omega = \psi_y(\xi_y) \xi_y^{m_y} \dr \xi_y$ in the local coordinate $\xi_y$ at $y \in \bzeta$ where $\psi_y(\xi_y)|_y \neq 0$ using the fact that $\omega$ has a zero of order $m_y$ at $y$. We also denote by $\Pi_y \subseteq \Pi$ the stabiliser subgroup of $y$ under the action of $\Pi$ on $\CP$, and $|\Pi_y|$ is its order. Explicitly, we have $|\Pi_y| = 2$ for any real point $y \in \bzeta_{\rm r}$ and $|\Pi_y| = 1$ for any complex point $y \in \bzeta_{\rm c}$.

The following is an immediate generalisation of \cite[Lemma 4.3]{Lacroix:2020flf}.

\begin{lemma} \label{lem: pi zeta ip}
For any $f, g \in R_{\Pi \bzeta}(\fg^\bC)^\Pi$ and $\ms u, \ms v \in \fg$ we have
\begin{equation*}
\langle\!\langle \boldsymbol{\pi_{\bzeta}} (f, \ms u), \boldsymbol{\pi_{\bzeta}} (g, \ms v) \rangle\!\rangle_{\fg \oplus \gzeta \oplus \fg}
= \langle\!\langle (f, \ms u), (g, \ms v) \rangle\!\rangle_\omega.
\end{equation*}
\begin{proof}
The first term in the bilinear form \eqref{bilinear R zeta b} can be rewritten as
\begin{equation} \label{bilinear pairing zeros}
\langle\!\langle f, g \rangle\!\rangle_\omega = \sum_{x \in \bZ} \mathrm{res}_x \langle f, g \rangle \omega = - \sum_{y \in \Pi \bzeta} \mathrm{res}_y \langle f, g \rangle \omega = - \sum_{y \in \bzeta} \frac{2}{|\Pi_y|} \Re\big( \!\mathrm{res}_y \langle f, g \rangle \omega \big)
\end{equation}
where in the second equality we used the residue theorem and the fact that the poles of the meromorphic $1$-form $\langle f, g \rangle\omega$ belong to the set $\Pi \bz \sqcup \Pi \bzeta$.
The right hand side of \eqref{bilinear pairing zeros} can be evaluated more explicitly as follows
\begin{align*}
\langle\!\langle f, g \rangle\!\rangle_\omega &= - \sum_{y \in \bzeta} \sum_{p=0}^{m_y-1} \frac{2}{|\Pi_y|} \Re\Big( \!\mathrm{res}_y \big\langle f^{\uz{y,p}} \xi_y^{-p-1}, \psi_y(\xi_y) \xi_y^{m_y} g \big\rangle \dr \xi_y \Big)\\
&= - \sum_{y \in \bzeta} \sum_{p, q=0}^{m_y-1} \frac{2}{|\Pi_y|} \Re \bigg( \frac{1}{p!} \partial_{\xi_y}^p \big( \psi_y(\xi_y) \xi_y^{m_y-q-1} \big) \big|_y \big\langle f^{\uz{y,p}}, g^{\uz{y,q}} \big\rangle \bigg).
\end{align*}
In the first equality we used the fact that $g \omega$ is regular at $y \in \bzeta$ so that the only contribution to the residue is from the pole term $f^{\uz{y,p}} \xi_y^{-p-1}$ at $y$ in $f$ and we wrote $\omega$ locally in the coordinate $\xi_y$. In the second equality we took the residue and used the fact that the poles of $g$ at $x \neq y$ in the expression $\partial_{\xi_y}^p \big( \psi_y(\xi_y) \xi_y^{m_y} g \big) \big|_y$ with $p = 0, \ldots, m_y-1$ vanish since $\psi_y(\xi_y) \xi_y^{m_y}$ has a zero of order $m_y$ at $y$. Finally, we note that $\frac{1}{p!} \partial_{\xi_y}^p \big( \psi_y(\xi_y) \xi_y^{m_y-q-1} \big) \big|_y = \frac{1}{(p+q+1-m_y)!} \partial_{\xi_y}^{p+q+1-m_y} \psi_y(\xi_y)$ if $p+q \geq m_y - 1$ and is zero otherwise, from which the result now follows.
\end{proof}
\end{lemma}

\subsection{The isomorphism \texorpdfstring{$\jb_\bz : R_{\Pi\bzeta}(\fg^\bC)^\Pi \SimTo \ff^\perp$}{alpha}} \label{sec: j hat iso}

Let us introduce a linear map
\begin{align} \label{j z def}
\jb_{\bz} : R_{\Pi\bzeta}(\fg^\bC)^\Pi &\longrightarrow \fd, \notag\\
f &\longmapsto \bigg( \sum_{p=0}^{n_x - 1} \frac{1}{p!} \big( \partial^p_{\xi_x} f \big) \big|_x \otimes \varepsilon_x^p \bigg)_{x \in \bz},
\end{align}
which takes a rational function with poles at the zeroes of $\omega$ and returns the first $n_x$ terms in its Taylor expansion at each pole $x \in \bz$ of $\omega$ in the local coordinate $\xi_x$, where $\xi_x = z-x$ if $x \in \bz'$ and $\xi_\infty = z^{-1}$ if $\infty \in \bz$. The linear map \eqref{j z def} cannot be an isomorphism on dimensional grounds by \eqref{dim R plus g}. However, we will show below in Proposition \ref{prop: Cauchy-Vandermonde} that it is injective. Before doing so we will show that \eqref{j z def} maps the bilinear form \eqref{bilinear R zeta}, or rather its restriction to $R_{\Pi\bzeta}(\fg^\bC)^\Pi$, to the bilinear form \eqref{form on gz rc} on $\fd$.

\begin{lemma} \label{lem: bilinear forms 1}
For any $f, g \in R_{\Pi\bzeta}(\fg^\bC)^\Pi$, we have $\langle\!\langle \jb_{\bz} f, \jb_{\bz} g \rangle\!\rangle_{\fd} = \langle\!\langle f, g \rangle\!\rangle_\omega$.
\begin{proof}
Let $f, g \in R_{\Pi\bzeta}(\fg^\bC)^\Pi$.
First note that by using the reality conditions at all of the poles of $\omega$ we may rewrite the bilinear form \eqref{bilinear R zeta b} more explicitly as a sum over the independent poles, namely
\begin{equation} \label{bilinear pairing poles}
\sum_{x \in \bZ} \mathrm{res}_x \langle f, g \rangle \omega = \sum_{x \in \bz} \frac{2}{|\Pi_x|} \Re \big( \!\mathrm{res}_x \langle f, g \rangle \omega \big). 
\end{equation}
Recall that $\Pi_x \subseteq \Pi$ denotes the stabiliser subgroup of $x \in \bz$.

Recall the explicit expression \eqref{omega def} for the meromorphic $1$-form, and in particular its expansion \eqref{omega expansion} at each pole $x \in \bz$. We then have
\begin{align*}
\langle\!\langle f, g \rangle\!\rangle_\omega &= \sum_{x \in \bz'} \sum_{p=0}^{n_x-1} \frac{2}{|\Pi_x|} \Re \bigg( \frac{\ell^x_p}{p!} \big( \partial_{\xi_x}^p \langle f, g \rangle \big)\big|_x \bigg)
+ \sum_{p=0}^{n_{\infty}-1} \frac{\ell^{\infty}_p}{p!} \big( \partial^p_{\xi_\infty}\langle f, g \rangle \big)\big|_\infty\\
&= \sum_{x \in \bz} \sum_{p=0}^{n_x-1} \sum_{q=0}^p \frac{2}{|\Pi_x|} \Re \bigg( \ell^x_p \bigg\langle \frac{1}{q!} (\partial_{\xi_x}^q f)|_x, \frac{1}{(p-q)!} (\partial_{\xi_x}^{p-q} g)|_x \bigg\rangle \bigg)\\
&= \sum_{x \in \bz} \sum_{q,r=0}^{n_x-1} \frac{2}{|\Pi_x|} \Re \bigg( \ell^x_{q+r} \bigg\langle \frac{1}{q!} (\partial_{\xi_x}^q f)|_x, \frac{1}{r!} (\partial_{\xi_x}^r g)|_x \bigg\rangle \bigg)
= \langle\!\langle \jb_{\bz} f, \jb_{\bz} g \rangle\!\rangle_{\fd}
\end{align*}
where in the second last step we changed variable from $p$ to $r = p-q$ and used the convention that $\ell^x_p = 0$ for $p \geq n_x$. The last equality is by definition \eqref{form on gz rc} of the bilinear form on $\fd$ and of the map $\jb_{\bz}$ in \eqref{j z def}.
\end{proof}
\end{lemma}

Any $\ms v \in \fg$ defines a constant function on $R_{\Pi\bzeta}(\fg^\bC)^\Pi$. Explicitly, in the notation of \eqref{uz def} we have $f^{\rm c} = \ms v$ and $f^{\uz{y, q}} = 0$ for every $\uz{y,q} \in \brb{\Pi \bzeta}$. By abuse of notation we will denote this rational function also as $\ms v$. Its image under $\jb_\bz$ is the element $\jb_\bz \ms v = \Delta \ms v \in \ff$ of the diagonal subalgebra $\ff = \Delta \fg \subset \fd$. Moreover, any element of $\ff$ can be represented in this way.
Let
\begin{equation} \label{v def}
\fv \coloneqq \jb_{\boldsymbol{z}}\big( R'_{\Pi\bzeta}(\fg^\bC)^\Pi \big).
\end{equation}

\begin{proposition} \label{prop: Cauchy-Vandermonde}
The linear map $\jb_{\bz}$ in \eqref{j z def} is an isomorphism onto its image $\ff^\perp$. In particular, we have the direct sum decomposition $\ff^\perp = \ff \dotplus \fv$.
\begin{proof}
We will first show that $\jb_{\bz}\big( R_{\Pi\bzeta}(\fg^\bC)^\Pi \big) \subset \ff^\perp$.
To see this, let $f \in R_{\Pi\bzeta}(\fg^\bC)^\Pi$ and $\Delta \ms v \in \ff$ be arbitrary. Then by Lemma \ref{lem: bilinear forms 1} we have
\begin{equation*}
\langle\!\langle \jb_{\bz} f, \Delta \ms v \rangle\!\rangle_{\fd} = \langle\!\langle \jb_{\bz} f, \jb_{\bz} \ms v \rangle\!\rangle_{\fd} = \langle\!\langle f, \ms v \rangle\!\rangle_\omega = 0,
\end{equation*}
where the last step follows by the residue theorem since $\langle f, \ms v \rangle \omega$ is a meromorphic $1$-form with poles contained in the subset $\bZ$.

It remains to show that the map $\jb_\bz$ is injective. The result will then follow by virtue of \eqref{dim Rg 2} which can be rewritten as $\dim R_{\Pi\bzeta}(\fg^\bC)^\Pi = \dim \fd - \dim \ff = \dim \ff^\perp$ using \eqref{dim poles}. Equivalently, with the help of the bijection \eqref{pi zeta def} it is enough to show that $\jb_\bz \circ \boldsymbol{\pi_\bzeta}^{-1} : \fg \oplus \gzeta \to \fd$ is injective.
The coefficients of the expansions to order $n_x$ at all the poles $x \in \bZ$ are given by
\begin{equation*}
\frac{1}{p!} ( \partial_z^p f)|_x = \sum_{\uz{y,q} \in \brb{\Pi \bzeta'}} C^{\up{x, p}}_{\quad\;\; \uz{y,q}} f^{\uz{y,q}} + \sum_{q=-1}^{m_\infty - 1} C^{\up{x, p}}_{\quad\;\; \uz{\infty,q}} f^{(\infty, q)}
\end{equation*}
for all $\up{x,p} \in \bsb{\bZ}$, where we have incorporated the constant term $f^{(\infty, -1)} \coloneqq f^{\rm c}$ into the $(-1)^{\rm st}$ term of the second sum and introduced the coefficients
\begin{equation} \label{Cauchy mat}
C^{\up{x, p}}_{\quad\;\; \uz{y,q}} \coloneqq \binom{p+q}{p} \frac{(-1)^p}{(x-y)^{p+q+1}}, \qquad C^{\up{x, p}}_{\quad\;\; (\infty, q)} \coloneqq \binom {q+1} p x^{q+1-p}
\end{equation}
for all $\up{x,p} \in \bsb{\bZ}$ and $\uz{y,q} \in \brb{\Pi \bzeta}$ where $q = 0, \ldots, m_y - 1$ for all $y \in \bzeta \setminus \{ \infty \}$ and $q = -1, \ldots, m_\infty - 1$ for $y = \infty$.
The expressions in \eqref{Cauchy mat} are the components of what is known as a confluent Cauchy-Vandermonde matrix, see for instance \cite[Definition 13]{VAVRIN1997271}.
By combining \eqref{dim zeroes} and \eqref{poles vs zeroes} we find $\dim(\fg \oplus \gzeta) = \dim \fd - \dim \fg$. The matrix specified by the components \eqref{Cauchy mat} is of dimension $\dim \fd \times \dim(\fg \oplus \gzeta)$ so removing $\dim \fg$ columns (by removing the highest order term in the expansion at any of the poles $x \in \bZ$) we obtain a square confluent Cauchy-Vandermonde matrix. The result now follows form the fact that a \emph{square} confluent Cauchy-Vandermonde matrix is invertible \cite[Corollary 19]{VAVRIN1997271}.

The last part follows from applying the injective linear map $\jb_\bz$ to the direct sum decomposition \eqref{rat func split off constant}.
\end{proof}
\end{proposition}

Even though we will not explicitly need it in order to construct the action for the Lax connection of the degenerate $\cE$-model in \S\ref{sec: E-model} below, it is useful to try to extend the injective linear map \eqref{j z def} to an isomorphism $R_{\Pi \bzeta}(\fg^\bC)^\Pi \oplus \fg \overset{\cong}\longrightarrow \fd$. This will be useful in constructing the $\cE$-operator $\cE : \fd \to \fd$ on all of $\fd$ in \S\ref{sec: E-operator} below. Explicitly, we want to construct an isomorphism
\begin{subequations} \label{alpha def}
\begin{equation} \label{alpha def map}
\jhb_\bz \coloneqq \jb_\bz \oplus \rb_\bz : R_{\Pi \bzeta}(\fg^\bC)^\Pi \oplus \fg \overset{\cong}\longrightarrow \fd,
\end{equation}
where $\rb_\bz : \fg \to \fd$ is a linear map whose image is complementary to $\ff^\perp$ in $\fd$. We will now give a general procedure for constructing such a linear map by requiring that its image $\ft \coloneqq \mathrm{im} \rb_\bz$ be isotropic and perpendicular to $\fv$, i.e. such that
\begin{equation} \label{alpha def prop 1}
\fd = \ff^\perp \dotplus \ft, \qquad \ft \subset (\ft \dotplus \fv)^\perp, 
\end{equation}
and with the property that, for any $\ms x, \ms y \in \fg$,
\begin{equation} \label{alpha def prop 2}
\langle\!\langle \rb_\bz \ms x, \Delta \ms y \rangle\!\rangle_{\fd} = \langle \ms x, \ms y \rangle.
\end{equation}
\end{subequations}

For simplicity we will suppose in the following argument that $\omega$ does not have a pole at infinity. The construction could also be adapted to that case.
Recall that $m_\infty$ denotes the order of the zero of $\omega$ at infinity. We let $P : \bC \to \bC$ be a generic polynomial of order $m_\infty + 1$, which therefore contains $m_\infty + 2$ arbitrary coefficients in $\bC$. If $\ms x \in \fg$ then $\ms x \, P$ is a $\fg$-valued polynomial. Notice, however, that $\ms x \, P$ does \emph{not} lie in $R_{\Pi \bzeta}(\fg^\bC)^\Pi$ since rational functions in $R_{\Pi \bzeta}(\fg^\bC)^\Pi$, which are of the form \eqref{uz def}, contain a polynomial of order at most $m_\infty$ (and not $m_\infty + 1$). Nevertheless, by a slight abuse of notation we will write
\begin{equation*}
\jb_\bz (\ms x \, P) = \bigg( \sum_{p=0}^{n_x - 1} \frac{1}{p!} \big( \ms x (\partial^p_{\xi_x} P)|_x \big) \otimes \varepsilon_x^p \bigg)_{x \in \bz} \in \fd
\end{equation*}
for the Taylor expansion of $\ms x \, P$ at each $x \in \bz$ up to order $n_x - 1$, cf. \eqref{j z def}. We then define the linear map
\begin{equation} \label{map rho}
\rb_\bz : \fg \longrightarrow \fd,\qquad
\ms x \longmapsto \jb_\bz (\ms x \, P).
\end{equation}
By construction $\ft = \mathrm{im} \rb_\bz$ is a complement to $\ff^\perp = \mathrm{im} \jb_\bz$ in $\fd$ since $\ms x \, P \not\in R_{\Pi \bzeta}(\fg^\bC)^\Pi$. Next, we will argue how to fix the coefficients of the polynomial $P$ by imposing the second property in \eqref{alpha def prop 1} and the property \eqref{alpha def prop 2}.

We will first show how to ensure that $\jb_{\bz} (\ms x \, P) \in \fv^\perp$. Suppose first that $m_\infty = 0$, i.e. that $\omega$ does not vanish at infinity. Then any $g \in R'_{\Pi \bzeta}(\fg^\bC)^\Pi$, specifically \emph{without} constant term, has a zero of order at least $1$ at infinity so that the $1$-form $\langle \ms x \, P, g \rangle \omega$ is regular at infinity. In particular, all of its poles lie in $\bZ$. By the same computation as in the proof of Lemma \ref{lem: bilinear forms 1} one shows that $\langle\!\langle \jb_{\bz} (\ms x \, P), \jb_{\bz} g \rangle\!\rangle_{\fd} = \langle\!\langle \ms x \, P, g \rangle\!\rangle_\omega = 0$ where the last step is by the residue theorem. Hence $\jb_{\bz} (\ms x \, P) \in \fv^\perp$ where in this case $P$ is an arbitrary polynomial of degree $1$, i.e. with two arbitrary coefficients.

Suppose now that $m_\infty > 0$. Then we can write any $g \in R'_{\Pi \bzeta}(\fg^\bC)^\Pi$ as $g = g' + g_\infty$ where $g_\infty$ is a $\fg^\bC$-valued polynomial of order $m_\infty$ with no constant term, which therefore contains $m_\infty$ arbitrary coefficients, and $g'$ contains no polynomial term. Consider the equation $\langle\!\langle \jb_{\bz} (\ms x \, P), \jb_{\bz} g \rangle\!\rangle_{\fd} = 0$. By the exact same reasoning as above we deduce $\langle\!\langle \jb_{\bz} (\ms x \, P), \jb_{\bz} g' \rangle\!\rangle_{\fd} = 0$ whereas the condition $\langle\!\langle \jb_{\bz} (\ms x \, P), \jb_{\bz} g_\infty \rangle\!\rangle_{\fd} = 0$ imposes a triangular system of $m_\infty$ linear equations on the coefficients of the polynomial $P$. By solving these equations we can then ensure that $\jb_{\bz} (\ms x \, P) \in \fv^\perp$ where $P$ is a polynomial of degree $m_\infty + 1$ but with only two free coefficients.

The two remaining coefficients in the polynomial $P$ can be fixed by requiring that $\ft$ is isotropic, which amounts to imposing the condition $\langle\!\langle \rb_\bz \ms x, \rb_\bz \ms y \rangle\!\rangle_{\fd} = 0$ for any $\ms x, \ms y \in \fg$, and the additional property \eqref{alpha def prop 2}.

Note that the linear map \eqref{map rho} could be naturally incorporated into the linear map \eqref{j z def} by extending the space of rational functions $R_{\Pi \bzeta}(\fg^\bC)^\Pi$ to include the polynomial functions of the form $\ms x \, P$ for $\ms x \in \fg$. In other words, if we define the space $\widehat{R}_{\Pi \bzeta}(\fg^\bC)^\Pi \cong R_{\Pi \bzeta}(\fg^\bC)^\Pi \oplus \fg$ consisting of rational functions of the form \eqref{uz def} but with the pole at infinity of order $m_\infty + 1$ rather than $m_\infty$ then the isomorphism \eqref{alpha def map} can equally be described as an isomorphism
\begin{equation*}
\jhb_\bz : \widehat{R}_{\Pi \bzeta}(\fg^\bC)^\Pi \overset{\cong}\longrightarrow \fd,
\end{equation*}
defined in exactly the same way as \eqref{j z def}, namely by Taylor expanding in the local coordinate $\xi_x$ at each $x \in \bz$ up to the order $n_x - 1$. This discussion relates back to the observation made in Remark \ref{rem: stronger poles in L} and in the paragraph after \eqref{flatness of Lc}, namely that the $\fg$-valued gauge field $\cL$, which after going on-shell was valued in $R_{\Pi \bzeta}(\fg^\bC)^\Pi$, could in fact be taken to live in the larger space $\widehat{R}_{\Pi \bzeta}(\fg^\bC)^\Pi$ just defined.

The key property of the isomorphism \eqref{alpha def map}, generalising that of Lemma \ref{lem: bilinear forms 1} and which we shall use to construct the $\cE$-operator in the next section, is the following.

\begin{lemma} \label{lem: bilinear forms 2}
For any $f, g \in R_{\Pi\bzeta}(\fg^\bC)^\Pi$ and $\ms u, \ms v \in \fg$ we have
\begin{equation*}
\langle\!\langle \jhb_\bz (f, \ms u), \jhb_\bz (g, \ms v) \rangle\!\rangle_{\fd} = \langle\!\langle (f, \ms u), (g, \ms v) \rangle\!\rangle_\omega.
\end{equation*}
\begin{proof}
We have
\begin{align*}
\langle\!\langle \jhb_\bz (f, \ms u), \jhb_\bz (g, \ms v) \rangle\!\rangle_{\fd}
&= \langle\!\langle \jb_\bz f, \jb_\bz g \rangle\!\rangle_{\fd} + \langle\!\langle \rb_\bz \ms u, \jb_\bz g \rangle\!\rangle_{\fd}
+ \langle\!\langle \jb_\bz f, \rb_\bz \ms v \rangle\!\rangle_{\fd}\\
&= \langle\!\langle f, g \rangle\!\rangle_{\omega} + \langle\!\langle \rb_\bz \ms u, \Delta g^{\rm c} \rangle\!\rangle_{\fd}
+ \langle\!\langle \Delta f^{\rm c}, \rb_\bz \ms v \rangle\!\rangle_{\fd}\\
&= \langle\!\langle f, g \rangle\!\rangle_{\omega} + \langle \ms u, g^{\rm c} \rangle
+ \langle f^{\rm c}, \ms v \rangle = \langle\!\langle (f, \ms u), (g, \ms v) \rangle\!\rangle_\omega.
\end{align*}
where in the first step we used the isotropy of $\ft$. In the second line we have used Lemma \ref{lem: bilinear forms 1} for the first term and the fact that $\ft \perp \fv$ for the last two terms. The very last step is by definition \eqref{bilinear R zeta b} of the bilinear form on $R_{\Pi\bzeta}(\fg^\bC)^\Pi \oplus \fg$.
\end{proof}
\end{lemma}

\subsection{The \texorpdfstring{$\cE$}{E}-operator} \label{sec: E-operator}

We now wish to define an operator $\cE : \fd \to \fd$ which is symmetric with respect to the bilinear form \eqref{form on gz rc} on $\fd$. To do so, we will first construct an operator \cite{Lacroix:2020flf}
\begin{equation} \label{Etilde def}
\widetilde{\cE} : \fg \oplus \gzeta \oplus \fg \overset{\cong}\longrightarrow \fg \oplus \gzeta \oplus \fg
\end{equation}
which is symmetric with respect to the bilinear form \eqref{bilinear g zeta} on $\fg \oplus \gzeta \oplus \fg$. This may then be transferred to $\fd$ using the isomorphism $\jhb_\bz$ in \eqref{alpha def map} and the isomorphism $\boldsymbol{\pi_\bzeta} : R_{\Pi \bzeta}(\fg^\bC)^\Pi \oplus \fg \SimTo \fg \oplus \gzeta \oplus \fg$ from \S\ref{sec: pi zeta iso}.

Recall from \S\ref{sec: 4d action and field} that we assumed we were given a partition $\bzeta = \bzeta_+ \sqcup \bzeta_-$ of the set of zeroes of $\omega$ such that $\sum_{y \in \bzeta_+} m_y = \sum_{y \in \bzeta_-} m_y$. The latter condition means, in particular, that the vector space $\gzeta$ introduced in \eqref{gzeta def} splits into a direct sum
\begin{equation} \label{gzeta decomp}
\gzeta = \fg^{\brb{\bzeta_+}} \dotplus \fg^{\brb{\bzeta_-}}
\end{equation}
of two subspaces with $\dim \fg^{\brb{\bzeta_+}} = \dim \fg^{\brb{\bzeta_-}}$. In \S\ref{sec: solving bulk eom} we then introduced signs $\epsilon_y = \pm 1$ for each $y \in \bzeta$ which we used to impose the condition \eqref{E-model condition} on the components of the gauge field. Given this data, we introduce the involution
\begin{align} \label{Etilde def explicit}
\widetilde{\cE} : \fg \oplus \gzeta \oplus \fg &\overset{\cong}\longrightarrow \fg \oplus \gzeta \oplus \fg, \notag\\
\Big( f^{\rm c}, \big( f^{\uz{y,q}} \big)_{\uz{y,q} \in \brb{\bzeta}}, \ms u \Big) &\longmapsto \Big( \ms u, \big( \epsilon_y f^{\uz{y,q}} \big)_{\uz{y,q} \in \brb{\bzeta}}, f^{\rm c} \Big).
\end{align}
In other words, we leave the elements of $\fg^{\brb{\bzeta_+}}$ fixed, we change the sign of those in $\fg^{\brb{\bzeta_-}}$ and we flip the two additional copies of $\fg$.
One easily checks that the involution \eqref{Etilde def explicit} is symmetric with respect to the bilinear form \eqref{bilinear g zeta}.

Following \cite{Lacroix:2020flf}, consider the isomorphism
\begin{equation} \label{C def}
\cC \coloneqq \jb_\bz \boldsymbol{\pi_{\bzeta}}^{-1} : \fg \oplus \gzeta \overset{\cong}\longrightarrow \ff^\perp.
\end{equation}
We use this to transfer \eqref{Etilde def} to an operator on $\fd$. Explicitly, we have the following.

\begin{lemma} \label{lem: E def}
The operator $\cE \coloneqq \cC \widetilde{\cE} \cC^{-1} : \fd \overset{\cong}\longrightarrow \fd$ is an involution and is symmetric with respect to the bilinear form \eqref{form on gz rc} on $\fd$.
\begin{proof}
The involution property is immediate from that of $\widetilde{\cE}$.

By combining Lemmas \ref{lem: pi zeta ip} and \ref{lem: bilinear forms 2} we have $\langle\!\langle \cC \ms U, \cC \ms V \rangle\!\rangle_\fd = \langle\!\langle \ms U, \ms V \rangle\!\rangle_{\fg \oplus \gzeta \oplus \fg}$ for any $\ms U, \ms V \in \fg \oplus \gzeta \oplus \fg$. Therefore, for any $\ms x, \ms y \in \fd$ we deduce
\begin{align*}
\langle\!\langle \ms x, \cE \ms y \rangle\!\rangle_{\fd}
&= \langle\!\langle \ms x, \cC \widetilde{\cE} \cC^{-1} \ms y \rangle\!\rangle_{\fd}
= \langle\!\langle \cC^{-1} \ms x, \widetilde{\cE} \cC^{-1} \ms y \rangle\!\rangle_{\fg \oplus \gzeta \oplus \fg}\\
&= \langle\!\langle \widetilde{\cE} \cC^{-1} \ms x, \cC^{-1} \ms y \rangle\!\rangle_{\fg \oplus \gzeta \oplus \fg}
= \langle\!\langle \cC \widetilde{\cE} \cC^{-1} \ms x, \ms y \rangle\!\rangle_{\fd}
= \langle\!\langle \cE \ms x, \ms y \rangle\!\rangle_{\fd}.
\end{align*}
Hence $\cE$ is also symmetric, as required.
\end{proof}
\end{lemma}

\begin{lemma}
The restriction of the symmetric bilinear form $\langle\!\langle \cdot, \cE \cdot \rangle\!\rangle_{\fd} : \fd \times \fd \to \bR$ to $\ff \subset \fd$ is non-degenerate.
\begin{proof}
Let $\ms v \in \ff$ and suppose that $\langle\!\langle \ms u, \cE \ms v \rangle\!\rangle_{\fd} = 0$ for all $\ms u \in \ff$. Then $\cE \ms v \in \ff^\perp$ which is a contradiction since $\cE \ff = \ft$ is by definition a complement of $\ff^\perp$ in $\fd$.
\end{proof}
\end{lemma}

\subsection{Solving the constraint between \texorpdfstring{$\cL$}{L} and \texorpdfstring{$h$}{h}} \label{sec: E-model}

Having introduced all of the necessary ingredients in the previous sections, we are finally in a position to complete the final step of the construction of $2$d integrable field theories from $4$d Chern-Simons theory.
Recall from \S\ref{sec: 2d action} that in order to write down the final $2$d action as in \eqref{ec:2daction} we need to have a solution $\cL = \cL(h)$ of the constraint \eqref{constraint L h} satisfying the transformation property \eqref{ec:equivariance}. We will now show how to construct solutions which give rise to integrable \emph{degenerate} $\cE$-models.

Let $\fv_\pm$ denote the eigenspaces of $\cE$ restricted to $\fv$ with eigenvalues $\pm 1$. These are the images of the subspaces $\fg^{\brb{\bzeta_\pm}}$ under the isomorphism $\cC$ defined in \eqref{C def}. Equivalently, we can describe $\fv_\pm$ as the image of the spaces $R'_{\Pi\bzeta_\pm}(\fg^\bC)^\Pi$ of $\Pi$-equivariant $\fg^\bC$-valued rational functions with poles in $\Pi\bzeta_\pm$ (see the start of \S\ref{sec: vs Rg}), namely
\begin{equation} \label{v pm def}
\fv_\pm \coloneqq \jb_{\boldsymbol{z}}\big( R'_{\Pi\bzeta_\pm}(\fg^\bC)^\Pi \big).
\end{equation}
We have $\dim \fv_+ = \dim \fv_- = \tfrac 12 \dim \fd - \dim \fg$. Following \cite{Klimcik:2019kkf, Klimcik:2021bqm}, define the projection operators $W^\pm_h : \fd \to \fd$ by, see in particular \cite[(3.23)]{Klimcik:2021bqm},
\begin{equation} \label{projector W degenerate}
\ker W^\pm_h = \mathrm{Ad}_{h^{-1}} \fk, \qquad
\mathrm{im} W^\pm_h = \ff \oplus \fv_\pm.
\end{equation}

The constraint \eqref{constraint L h} explicitly says $B_\pm \coloneqq \mathrm{Ad}_h (\jb_\bz \cL_\pm) - \partial_\pm h h^{-1} \in C^\infty(\Sigma, \fk)$, which can be rewritten as $\jb_\bz \cL_\pm = \mathrm{Ad}_{h^{-1}} B_\pm + h^{-1} \partial_\pm h$. Applying the operator $W^\pm_h$ to both sides, and using the fact that $\mathrm{Ad}_{h^{-1}} B_\pm$ is valued in $\ker W^\pm_h$, we get
\begin{equation} \label{W on jL}
W^\pm_h(\jb_\bz \cL_\pm) = W^\pm_h(h^{-1} \partial_\pm h).
\end{equation}
On the other hand, in terms of light-cone components the condition \eqref{E-model condition} is equivalent to the statement that $\cL^{\uz{y,q}}_\pm = 0$ for any $y \in \bzeta_\mp$ and $q = 0, \ldots, m_y-1$. We can rewrite this in terms of the $\widetilde{\cE}$-operator \eqref{Etilde def explicit} as $\widetilde{\cE} \big( \boldsymbol{\pi_\bzeta} (\cL_\pm -\cL^{\rm c}_\pm) \big) = \pm \boldsymbol{\pi_\bzeta} (\cL_\pm - \cL^{\rm c}_\pm)$. Applying the isomorphism $\cC$ from \eqref{C def} on both sides and using the definition of the $\cE$-operator in Lemma \ref{lem: E def} we then obtain the equivalent condition
\begin{equation} \label{deg E-model condition}
\cE ( \jb_\bz \cL_\pm - \Delta \cL^{\rm c}_\pm) = \pm ( \jb_\bz \cL_\pm - \Delta \cL^{\rm c}_\pm)
\end{equation}
which implies that $\jb_\bz \cL_\pm - \Delta \cL^{\rm c}_\pm \in \fv_\pm$. Since $\Delta \cL^{\rm c}_\pm \in \ff$ it follows that $\jb_\bz \cL_\pm \in \mathrm{im} W^\pm_h$ so that the left hand side of \eqref{W on jL} just becomes $\jb_\bz \cL_\pm$. In other words, we arrive at the following solution
\begin{equation} \label{constraint solution W}
\jb_\bz \cL_\pm(h) = W^\pm_h(h^{-1} \partial_\pm h)
\end{equation}
of the constraint \eqref{constraint L h}.

It remains to show that this solution satisfies the desired transformation property \eqref{ec:equivariance}. For this we need the following lemma.

\begin{lemma} \label{lem: properties of W}
For any $g \in C^\infty(\Sigma, G)$ and $k \in C^\infty(\Sigma, K)$ we have
\begin{equation*}
W^\pm_{kh\Delta(g)^{-1}} = \mathrm{Ad}_{\Delta(g)} \circ W^\pm_h \circ \mathrm{Ad}_{\Delta(g)^{-1}}.
\end{equation*}
\begin{proof}
The projection operator $W^\pm_{k h\Delta(g)^{-1}} : \fd \to \fd$ is defined by
\begin{equation*}
\ker W^\pm_{k h\Delta(g)^{-1}} = \mathrm{Ad}_{\Delta(g)} \mathrm{Ad}_{h^{-1}} \fk, \qquad
\mathrm{im} W^\pm_{k h\Delta(g)^{-1}} = \ff \oplus \fv_\pm
\end{equation*}
where in the first equality we have used the fact that $\mathrm{Ad}_{k^{-1}} \fk = \fk$ since $k$ is valued in $K$.
The statement now follows using the fact that $\fv$ is invariant under the adjoint action of the diagonal subgroup $F = \mathrm{im} \Delta$.
\end{proof}
\end{lemma}

The property \eqref{ec:equivariance} now easily follows using Lemma \ref{lem: properties of W}, namely we have
\begin{align*}
\jb_\bz \cL_\pm(kh\Delta(g)^{-1}) &= W^\pm_{k h \Delta(g)^{-1}}\big( \Delta(g) h^{-1} k^{-1} \partial_\pm (k h \Delta(g)^{-1}) \big)\\
&= \mathrm{Ad}_{\Delta(g)} \circ W^\pm_h \big( \mathrm{Ad}_{h^{-1}} (k^{-1} \partial_\pm k) + h^{-1} \partial_\pm h - \Delta(g)^{-1} \partial_\pm \Delta(g) \big)\\
&= \Delta(g) \big( W^\pm_h ( h^{-1} \partial_\pm h ) \big) \Delta(g)^{-1} - \partial_\pm \Delta(g) \Delta(g)^{-1}\\
&= {}^{\Delta(g)} \big( \jb_\bz \cL_\pm(h) \big),
\end{align*}
where in the third equality we have used the fact that $\mathrm{Ad}_{h^{-1}} (k^{-1} \partial_\pm k)$ is valued in $\mathrm{Ad}_{h^{-1}}\fk = \ker W^\pm_h$ and $\Delta(g)^{-1} \partial_\pm \Delta(g)$ is valued in $\ff$.

Substituting the solution \eqref{constraint solution W} into the $2$d action \eqref{ec:2daction} we get the degenerate $\cE$-model action \cite[(3.22)]{Klimcik:2021bqm}
\begin{align} \label{E-model action}
S_{2d}(h) &= \frac{1}{2}\int \Big(\langle\!\langle h^{-1}\partial_-h,W^+_h (h^{-1}\partial_+h) \rangle\!\rangle \notag\\
&\qquad\qquad\qquad - \langle\!\langle h^{-1}\partial_+h,W^-_h (h^{-1}\partial_-h) \rangle\!\rangle\Big)\dr \sigma^{+}\wedge \dr \sigma^{-}-\frac{1}{2}I^{\mathrm{WZ}}[h] \,.
\end{align}
This model can be defined for more general Lie algebras $\fd$ and $\cE$-operators than the ones considered in this article. In general, the action \eqref{E-model action} does not describe a $2$d integrable field theory. However, in the present case where $\fd$ is the defect Lie algebra \eqref{defect alg} and the $\cE$-operator is as defined in \S\ref{sec: E-operator}, namely when the data originates from $4$d Chern-Simons theory as reviewed in \S\ref{sec: 2d IFT review}, the action \eqref{E-model action} describes an integrable field theory by construction. In particular, since the right hand side of \eqref{constraint solution W} takes values in $\ff \oplus \fv_\pm \subset \ff^\perp$ and $\jb_\bz : R_{\Pi \bzeta}(\fg^\bC)^\Pi \SimTo \ff^\perp$ is an isomorphism by Proposition \ref{prop: Cauchy-Vandermonde}, we can apply its inverse to both sides to obtain the Lax connection. Namely, denoting this inverse by $\boldsymbol{p} : \ff^\perp \SimTo R_{\Pi \bzeta}(\fg^\bC)^\Pi$ we have
\begin{equation} \label{Lax connection E-model}
\cL_\pm(h) = \boldsymbol{p} \big( W^\pm_h(h^{-1} \partial_\pm h) \big),
\end{equation}
which provides a Lax connection for the integrable degenerate $\cE$-model \eqref{E-model action}.

Although it is immediate by construction that \eqref{Lax connection E-model} satisfies the flatness equation \eqref{flatness of Lh}, it is instructive to show this explicitly. We will need the following lemma which is an immediate generalisation of \cite[Lemma 4.6 \& Remark 4.7]{Lacroix:2020flf}. Recall the definition of the real vector spaces $\fv_\pm$ in \eqref{v pm def}.

\begin{lemma} \label{lem: p property}
For any $\ms u_\pm \in \ff \oplus \fv_\pm$ we have $\boldsymbol{p} [\ms u_+, \ms u_-] = [\boldsymbol{p} \ms u_+, \boldsymbol{p} \ms u_-]$.
\begin{proof}
Let $\ms u_\pm \in \ff \oplus \fv_\pm$ which we can write as $\ms u_\pm = \jb_\bz f_\pm$ for some $f_\pm \in R_{\Pi\bzeta_\pm}(\fg^\bC)^\Pi$. Then we have
\begin{align*}
[\boldsymbol{p} \ms u_+, \boldsymbol{p} \ms u_-] = [f_+, f_-] = \boldsymbol{p} \jb_\bz [f_+, f_-] = \boldsymbol{p} [\jb_\bz f_+, \jb_\bz f_-] = \boldsymbol{p} [\ms u_+, \ms u_-].
\end{align*}
To see the second step, first observe that $[f_+, f_-] \in R_{\Pi \bzeta}(\fg^\bC)^\Pi$ by virtue of the fact that the poles of $f_+$, which lie in $\bzeta_+$, are disjoint from those of $f_-$, which lie in $\bzeta_-$. Therefore $\jb_\bz$ has a well defined action on $[f_+, f_-]$ and we can  insert the identity in the form $\mathrm{id} = \boldsymbol{p} \jb_\bz$ out front. In the third step we then used the fact that the linear map $\jb_\bz$, defined by taking the truncated Taylor expansions at the points $x \in \bz$ is in fact a morphism of Lie algebras.
\end{proof}
\end{lemma}

The equations of motion of the degenerate $\cE$-model \eqref{E-model action} take the form of a flatness equation \cite[(2.8)]{Klimcik:2021bqm}
\begin{equation*}
\partial_+ \big( W^-_h(h^{-1} \partial_- h) \big) - \partial_- \big( W^+_h(h^{-1} \partial_+ h) \big) + \big[ W^+_h(h^{-1} \partial_+ h), W^-_h(h^{-1} \partial_- h) \big] = 0
\end{equation*}
in $\fd$. Applying the linear map $\boldsymbol{p} : \ff^\perp \SimTo R_{\Pi \bzeta}(\fg^\bC)^\Pi$ to both sides and using Lemma \ref{lem: p property} on the commutator term we obtain the desired flatness equation \eqref{flatness of Lh} for the Lax connection with light-cone components \eqref{Lax connection E-model}.

\begin{remark}
The action of the non-degenerate $\cE$-model \cite[(2.5)]{Klimcik:2021bqm} can be written in exactly the same form as in \eqref{E-model action} but where the projectors $W^\pm_h$ are now defined by the conditions \cite[(2.6)]{Klimcik:2021bqm}
\begin{equation*}
\ker W^\pm_h = \mathrm{Ad}_{h^{-1}} \fk, \qquad
\mathrm{im} W^\pm_h = \fv_\pm
\end{equation*}
instead of \eqref{projector W degenerate}. Of course, this action for the non-degenerate $\cE$-model is equivalent to the one derived in \cite{Lacroix:2020flf} (see \cite[\S 2.2]{Lacroix:2020flf}). The action in the form \eqref{E-model action} can be derived directly in exactly the same way as above. Explicitly, assuming that $\omega$ has a double pole at infinity, as in \cite{Lacroix:2020flf}, one first fixes the $F$ symmetry by setting the edge mode at infinity equal to the identity. As recalled at the start of this section, this then removes the constant term in both components of the Lax connection. The exact same procedure as above then applies, with the absence of constant terms in the Lax connection reducing \eqref{deg E-model condition} to $\cE ( \jb_\bz \cL_\pm) = \pm ( \jb_\bz \cL_\pm )$ which was the condition used in \cite{Lacroix:2020flf}. In particular, this condition now implies that $\jb_\bz \cL_\pm \in \fv_\pm$.
\end{remark}

\section{Examples} \label{sec: examples}

In \S\ref{sec: review} and \S\ref{sec:reductionto2d} we presented a general construction of integrable degenerate $\cE$-models from $4$d Chern-Simons theory, with the final $2$d action given in \eqref{E-model action}. In practice, starting from a choice of meromorphic $1$-form $\omega$ one should build the associated defect Lie algebra $\fd$, identify its non-degenerate bilinear form $\langle\!\langle \cdot, \cdot \rangle\!\rangle_\fd$ and work out the Lie group structure of the defect Lie group $D$. The real vector space of rational functions $R'_{\Pi\bzeta}(\fg^\bC)^\Pi$, as defined in \S\ref{sec: vs Rg}, may then be used to explicitly construct the subspaces $\fv_\pm$ of $\fd$. By making a choice of isotropic Lie subalgebra $\fk \subset \fd$ one is then able to explicitly construct the projectors $W^\pm_h$ defined by \eqref{projector W degenerate} in terms of which the action \eqref{E-model action} is expressed. In this section we apply this procedure to explicitly construct the pseudo-chiral model and the bi-Yang-Baxter model.

\subsection{Pseudo-chiral model}

The first example we consider is that of a meromorphic $1$-form with a $4^{\rm th}$ order pole at the origin and two simple zeroes at $\pm a$ with $a>0$, namely
\begin{equation*}
\label{ec:omega4pole}
\omega = \frac{a^2-z^2}{z^4} \dr z\,. 
\end{equation*}
The defect Lie algebra \eqref{defect alg} for this choice of meromorphic 1-form is given by
\begin{equation}
\label{ec:dla4pole}
    \fd = \fg \otimes \mathbb{R}[\varepsilon]/(\varepsilon^4) \,.
\end{equation}
We denote the elements of $\fd$ by $\ms u^p \coloneqq \ms u \otimes \varepsilon^p$ with $p=0,1,2,3$. The bilinear form \eqref{form on gz rc} on the defect Lie algebra reads
\begin{equation}
\label{ec:pd4 form}
    \langle\!\langle \ms u^p,\ms v^q\rangle \!\rangle_{\fd} =
    \begin{cases}
    \begin{array}{cc}
        a^2\langle \ms u,\ms v\rangle & \text{if} \quad p+q=3 \\
        -\langle \ms u,\ms v\rangle  & \text{if} \quad p+q=1 \\
        0 & \text{otherwise} \,.
    \end{array}
    \end{cases}
\end{equation}

Next, let us describe the Lie group structure on the defect Lie group $D$ with Lie algebra \eqref{ec:dla4pole}. It is given by the $3^{\rm rd}$ order jet bundle $J^3G$ of the Lie group $G$, which in the right trivialisation is isomorphic to $G \times \fg \times \fg \times \fg$. That is, a general element $h \in D \cong G \times \fg \times \fg \times \fg$ can be expressed as a tuple $h=(g,\ms u, \ms v,\ms w)$, with $g\in G$ and $\ms u, \ms v,\ms w\in \fg$. The group product and inverse on $D$ are then given by\footnote{The multiplication law and the inverse differ from \cite{Vizman,Lacroix:2020flf} by a normalization convention.} \cite{Vizman}
\begin{align*}
    (g,\ms u, \ms v,\ms w)(\Tilde{g},\ms x,\ms y,\ms z)&=\left(g\Tilde{g}, \ms u +\mathrm{Ad}_g \ms x,\ms v + \mathrm{Ad}_g \ms y+\frac{1}{2}[\ms u, \mathrm{Ad}_g \ms x], \right. \\
    &\qquad \left. \ms w+\mathrm{Ad}_g \ms z + \frac{2}{3}[\ms u, \mathrm{Ad}_g \ms y]+\frac{1}{3}[\ms v, \mathrm{Ad}_g \ms x]+\frac{1}{6}[\ms u,[\ms u, \mathrm{Ad}_g \ms x]]\right),\\
     (g,\ms u, \ms v, \ms w)^{-1}&=\left(g^{-1},-\mathrm{Ad}_g^{-1}\ms u,-\mathrm{Ad}_g^{-1} \ms v,-\mathrm{Ad}_g^{-1} \ms w+\frac{1}{3}\mathrm{Ad}_g^{-1}[\ms u, \ms v]\right) \,.
\end{align*}

To specify the kernels of the operators $W_h^{\pm}$, as in \eqref{projector W degenerate}, we need to make a choice of Lagrangian subalegbra $\fk \subset \fd$. A natural choice in the present case is
\begin{equation} \label{k choice pseudo-chiral}
\fk = \fg \otimes \varepsilon^2\mathbb{R}[\varepsilon]/(\varepsilon^4),
\end{equation}
which is easily seen to be Lagrangian. One additional nice feature of this choice of $\fk$ is that it is an ideal in $\fd$ and thus, for any $h\in D$ we have $\mathrm{Ad}_h^{-1}\fk = \fk$. Hence, 
\begin{equation}
    \ker W^{\pm}_h= \fg \otimes \varepsilon^2\mathbb{R}[\varepsilon]/(\varepsilon^4)=\{\ms y^{2}+\ms z^3 \,|\, \ms y,\ms z \in \fg\}\,.
\end{equation}
On the other hand, specifying the image of $W^{\pm}_h$ requires identifying the diagonal subalgebra $\ff = \mathrm{im} \Delta$ and the subspaces $\fv_\pm$ defined in \eqref{v pm def}. Starting with $\ff$, the diagonal embedding for the defect Lie algebra \eqref{ec:dla4pole} is simply $\ms w \mapsto \ms w^0$, so that
\begin{equation}
\label{f 4 order pole}
    \ff = \mathrm{im} \Delta = \{ \ms w^0 \,|\, \ms w \in \fg\}\,.
\end{equation}
To describe $\fv_\pm$ we begin by identifying the spaces of rational functions $R'_{\Pi \bzeta_\pm}(\fg^\bC)^\Pi$ corresponding to the meromorphic 1-form \eqref{ec:omega4pole}. By partitioning the set of zeroes $\Pi \bzeta = \bzeta = \{a, -a\}$ of $\omega$ as $\Pi\bzeta_\pm = \bzeta_\pm = \{\pm a\}$, we find
\begin{equation}
    R'_{\Pi \bzeta_\pm}(\fg^\bC)^\Pi = \left\{\frac{\ms x}{z\mp a} \, \bigg | \, \ms x\in \fg\right\} \,,
\end{equation}
so that expanding such rational functions to $4^{\rm th}$ order at the origin gives
\begin{align}
\label{vplus 4 order pole}
    \fv_+ & = \jb_z\left(R'_{\Pi \bzeta_+}(\fg^\bC)^\Pi\right)=\left\{-\frac{\ms x^0}{a}-\frac{\ms x^1}{a^2}-\frac{\ms x^2}{a^3}-\frac{\ms x^3}{a^4} \,\,\bigg|\,\, \mathsf{x}\in \fg\right\},\\
\label{vminus 4 order pole}    
   \fv_- &=\jb_z\left(R'_{\Pi \bzeta_-}(\fg^\bC)^\Pi\right)= \left\{\frac{\ms x^0}{a}-\frac{\ms x^1}{a^2}+\frac{\ms x^2}{a^3}-\frac{\ms x^3}{a^4} \,\,\bigg|\,\, \mathsf{x}\in \fg\right\}\,. 
\end{align}
Then $\mathrm{im} W^{\pm}_h = \ff \oplus\fv_\pm$. With the explicit expressions for the image and kernel of the projectors, we may proceed with the computation of $W^{\pm}_h (h^{-1}\partial_\pm h)$. 

In order to simplify the discussion, we start by fixing both the $K$-symmetry and the $F$-symmetry. First, we note that the Lie group $K$ with Lie algebra $\fk$ is identified with the subgroup $\{\mathrm{id}\} \times \{0\} \times \fg \times \fg$ of $G \times \fg \times \fg \times \fg$. On the other hand, the Lie group $F$ with Lie algebra $\ff$ is identified with the subgroup $G\times\{0\}\times\{0\}\times\{0\}$. Fixing both of these gauge symmetries, implies that our physical degree of freedom will be described by a representative of the class of $h \in C^\infty(\Sigma, D)$ in the double coset $K \setminus D / F$. By a slight abuse of notation, we will also denote it by $h=(\mathrm{id},\ms u,0,0)$. We then have
\begin{equation}
\label{hdh for 4 pole}
    h^{-1}\dr h = \dr \ms u^1-\frac{1}{2}[\ms u,\dr \ms u]^2+\frac{1}{6}[\ms u,[\ms u,\dr \ms u]]^3 \,.
\end{equation}
In order to find the explicit action of $W^\pm_h$ on $h^{-1}\partial_\pm h$, we decompose the latter with respect to the direct sum decomposition $\fd = \ker W^\pm_h \dotplus \mathrm{im} W^\pm_h$. Focusing first on $h^{-1}\partial_+ h$, we look for $\ms w, \ms x,\ms y,\ms z \in \fg$ such that
\begin{equation}
\label{hdh decomposition}
    h^{-1}\partial_+ h = \left(\ms y^2 + \ms z^3\right) + \left(\ms w^0 -\frac{\ms x^0}{a}-\frac{\ms x^1}{a^2}-\frac{\ms x^2}{a^3}-\frac{\ms x^3}{a^4}\right)\,,
\end{equation}
which will then give
\begin{equation*}
    W^+_h (h^{-1}\partial_+ h) = \ms w^0 -\frac{\ms x^0}{a}-\frac{\ms x^1}{a^2}-\frac{\ms x^2}{a^3}-\frac{\ms x^3}{a^4}.
\end{equation*}
Explicitly decomposing the $\sigma^+$-component of \eqref{hdh for 4 pole} as in \eqref{hdh decomposition} we find
\begin{subequations} \label{Wpm explicit pseudo-chiral}
\begin{equation}
    W^+_h (h^{-1}\partial_+ h) = \partial_+ \ms u^1+ \frac{\partial_+\ms u^2}{a}+\frac{\partial_+ \ms u^3}{a^2}\,.
\end{equation}
By a completely analogous argument we get
\begin{equation}
    W^-_h (h^{-1}\partial_- h) =  \partial_- \ms u^1- \frac{\partial_-\ms u^2}{a}+\frac{\partial_- \ms u^3}{a^2}\,.
\end{equation}
\end{subequations}
Using the expression for the bilinear form \eqref{ec:pd4 form} we then obtain
\begin{equation}
    \langle\!\langle W^\pm_h (h^{-1}\partial_\pm h) , h^{-1}\partial_\mp h\rangle\!\rangle = \pm a\langle \partial_+ \ms u,\partial_- \ms u\rangle
\pm \frac{a^2}{2}\langle \ms u, [\partial_+ \ms u, \partial_- \ms u] \rangle.
\end{equation}

Finally, we compute the Wess-Zumino term \eqref{ec:wesszumino}. Since the Cartan $3$-form is cubic in $\widehat{h}^{-1} \dr\widehat{h}$, with $\widehat{h} \in C^\infty(\Sigma \times I, D)$, and the latter has no term of Takiff degree $0$ by \eqref{hdh for 4 pole}, or its analogue for $\widehat{h} = (\rm{id}, \widehat{\ms u}, 0, 0)$, it follows from the explicit form \eqref{ec:pd4 form} of the bilinear form on $\fd$ that only the term cubic in $\dr \widehat{\ms u}$ can contribute, so that
\begin{equation}
     I^{\mathrm{WZ}}[h]=-\frac{1}{6}\int_{\Sigma\times I}a^2\langle \dr  \widehat{\ms u},[\dr  \widehat{\ms u},\dr  \widehat{\ms u}]\rangle = \frac{a^2}{6}\int_{\Sigma}\langle \ms u,[\dr \ms u,\dr \ms u]\rangle \,.
\end{equation}

The $2$d action \eqref{E-model action} of the integrable degenerate $\cE$-model corresponding to the meromorphic 1-form \eqref{ec:omega4pole}, the choice of Lagrangian subalgebra $\fk \subset \fd$ in \eqref{k choice pseudo-chiral} and the split $\bzeta_\pm = \{\pm a\}$ of the zeroes of $\omega$, is therefore the $\sigma$-model with target space $K \setminus D / F \cong \fg$ and action given by
\begin{equation}
\label{ec:pseudochiralaction}
    S[\ms u]=\int_\Sigma \Big( a\langle \partial_+ \ms u,\partial_- \ms u\rangle+\frac{a^2}{3}\langle \ms u,[\partial_+ \ms u,\partial_- \ms u]\rangle \Big) \dr \sigma^+ \wedge \dr \sigma^-
\end{equation}
for $\ms u \in C^\infty(\Sigma, \fg)$, which is the pseudo-chiral model of Zakharov and Mikhailov \cite{Zakharov1978}. 

\subsubsection{Lax connection}

Having found the action of the $2$d integrable field theory, we now proceed with the computation of its Lax connection. Its light-cone components are given by \eqref{Lax connection E-model} where $\boldsymbol{p} : \ff^\perp \SimTo R_{\Pi \bzeta}(\fg^\bC)^\Pi$ is the inverse of $\jb_z$ and
\begin{equation} \label{f perp decomp}
\ff^\perp = \ff\oplus \fv_+ \oplus \fv_-
\end{equation}
with $\ff$, $\fv_+$ and $\fv_-$ given in \eqref{f 4 order pole}, \eqref{vplus 4 order pole} and \eqref{vminus 4 order pole}, respectively. Hence, the action of $\boldsymbol{p}$ on an element of $\ff^\perp$ decomposed with respect to \eqref{f perp decomp} is simply
\begin{equation}
    \boldsymbol{p} \left(\ms w^0 -\frac{\ms x^0}{a}-\frac{\ms x^1}{a^2}-\frac{\ms x^3}{a^3}-\frac{\ms x^3}{a^4}+\frac{\ms y^0}{a}-\frac{\ms y^1}{a^2}+\frac{\ms y^3}{a^3}-\frac{\ms x^3}{a^4}\right) = \ms w + \frac{\ms x}{z-a}+ \frac{\ms y}{z+a}\,.
\end{equation}
The action of $\boldsymbol{p}$ on $W^\pm_{h}(h^{-1}\partial_\pm h)$ given in \eqref{Wpm explicit pseudo-chiral} can now be computed to give the light-cone components of the Lax connection \eqref{Lax connection E-model}, namely we find
\begin{equation}
    \cL_\pm = \frac{-a^2}{z\mp a}\partial_\pm \ms u  \mp a \partial_\pm \ms u \,.
\end{equation}
The zero curvature equation for this Lax connection is equivalent to 
\begin{equation}
    \partial_+\partial_- \ms u - \frac{a}{2}[\partial_+ \ms u,\partial_- \ms u]=0\,,
\end{equation}
which corresponds to the equation of motion of $\ms u$ for the action \eqref{ec:pseudochiralaction}, thus proving the Lax integrability of the model.

\subsection{Bi-Yang-Baxter \texorpdfstring{$\sigma$}{s}-model} 

The second example we consider is the bi-Yang-Baxter $\sigma$-model \cite{KlimcikBi,Klimcik:2014bta}. Following the conventions used in \cite{Delduc:2015xdm,Delduc:2019whp}, we take the meromorphic 1-form
\begin{equation}
\label{byb model omega}
    \omega=\frac{16 K z}{\zeta^2(z-z_+)(z-z_-)(z-\tilde{z}_+)(z-\tilde{z}_-)} \dr z
\end{equation}
where $K\in \mathbb{R}$. The four simple poles $z_\pm, \tilde{z}_\pm \in \bC$ and the coefficient $\zeta \in \mathbb{R}$ are related to the two real deformation parameters $\eta$ and $\tilde{\eta}$ of the model by
\begin{gather} \label{ec:zpluszminus}
    z_\pm =\frac{-2\rho \pm i \eta}{\zeta}\,, \quad \tilde{z}_\pm=-\frac{2+2\rho \pm i \tilde{\eta}}{\zeta}\,, \quad \rho=-\tfrac{1}{2}\left(1-\frac{\eta^2-\tilde{\eta}^2}{4}\right),\\
    \zeta^2=\left(1+\frac{(\eta+\tilde{\eta})^2}{4}\right)\left(1+\frac{(\eta-\tilde{\eta})^2}{4}\right)\,.
\end{gather}
Note also that $\omega$ has two simple zeroes, at $0$ and $\infty$. The defect Lie algebra \eqref{defect alg} for this choice of meromorphic 1-form is given by
\begin{equation}
\label{byb defect lie algebra}
    \fd = \big(\fg^{\mathbb{C}}\otimes \mathbb{C}[\varepsilon]/(\varepsilon)\big) \times \big(\fg^{\mathbb{C}}\otimes \mathbb{C}[\Tilde{\varepsilon}]/(\Tilde{\varepsilon})\big) \cong \fg^\bC \times \fg^\bC,
\end{equation}
where we recall that each factor is treated as a real vector space.
Therefore elements of $\fd$ are given by tuples $(\ms u,\ms v)\in \fd$ with $\mathsf{u},\mathsf{v}\in \fg^{\mathbb{C}}$.
The bilinear form \eqref{form on gz 1} on the defect Lie algebra reads
\begin{equation}
\label{ec:byb form}
    \langle\!\langle (\mathsf{u},\mathsf{v}),(\tilde{\mathsf{u}},\tilde{\mathsf{v}})\rangle \!\rangle_{\fd} =\tfrac{4K}{\eta}\Im\langle \mathsf{u},\tilde{\mathsf{u}}\rangle +\tfrac{4K}{\tilde{\eta}}\Im\langle \mathsf{v},\tilde{\mathsf{v}}\rangle.
\end{equation}
The defect Lie group $D$ with Lie algebra $\fd$ is simply $G^{\bC}\times G^{\bC}$, a general element of which is a tuple $(h,\Tilde{h})$ with $h,\tilde{h} \in G^{\bC}$.

To specify the kernels of the projection operators defined by \eqref{projector W degenerate} we need to choose a Lagrangian subalgebra $\fk \subset \fd$. Following \cite{Delduc:2019whp}, we take two skew-symmetric solutions $R,\tilde{R}\in \mathrm{End}\,\fg$ to the modified Yang-Baxter equation with $c=\ms i$ in terms of which we define the Lie subalgebra
\begin{equation}
\label{ec:byblagsubalgebra}
    \fk \coloneqq \fg_R\times \fg_{\tilde{R}}=\{((R-\ms i)\mathsf{x},(\tilde{R}-\ms i)\mathsf{y})\,|\, \mathsf{x},\mathsf{y}\in \fg\}
\end{equation}
which is seen to be Lagrangian with respect to the bilinear form \eqref{ec:byb form}. To simplify the discussion we will gauge fix the $K$-symmetry. Let $K= G_R \times G_{\tilde{R}} \subset G^{\bC}\times G^{\bC}$ be the Lie group with Lie algebra $\fk = \fg_R\times \fg_{\tilde{R}}$. Following \cite{Delduc:2019whp}, we assume that the direct sum decomposition $\fg^{\bC} \times \fg^{\bC}= \fk \dotplus (\fg \times \fg)$ lifts to the group level, that is, $G^{\bC}\times G^{\bC} = K (G \times G)$ so that a natural parametrisation of the quotient $K \setminus (G^\bC \times G^\bC)$ is then given by $G\times G$. In this way, our physical degrees of freedom will be described by a representative of the class of $h \in C^\infty(\Sigma, G^\bC \times G^\bC)$ in the coset $K \setminus (G^\bC \times G^\bC)$ which we denote by $(g,\tilde{g})$ with $g,\tilde{g} \in C^\infty(\Sigma, G)$. Hence, from \eqref{projector W degenerate} we have
\begin{equation}
    \mathrm{ker}\, W^\pm_{(g,\Tilde{g})} = \mathrm{Ad}_{(g,\Tilde{g})}^{-1}\fg_R \times \fg_{\Tilde{R}}=\left\{\left((R_g-\ms i)\mathrm{Ad}_g^{-1}\mathsf{x},(\tilde{R}_{\tilde{g}}-\ms i)\mathrm{Ad}_{\tilde{g}}^{-1}\mathsf{y}\right)\,|\, \mathsf{x},\mathsf{y}\in \fg\right\}\,,
\end{equation}
where we have defined $R_g = \mathrm{Ad}_g^{-1} \circ R \circ \mathrm{Ad}_g$, and similarly for $\Tilde{R}_{\Tilde{g}}$.

On the other hand, the images of $W^{\pm}_{(g,\Tilde{g})}$ are given in terms of the subalgebra $\ff = \mathrm{im} \Delta$ and the subspaces $\fv_\pm$ defined in \eqref{v pm def}. The diagonal embedding for the defect Lie algebra \eqref{byb defect lie algebra} is simply $\ms a \mapsto \ms (\ms a,\ms a)$, so that
\begin{equation}
\label{f byb model}
    \ff =\{(\ms a,\ms a) \,|\, \mathsf{a}\in \fg\}\,.
\end{equation}
To determine $\fv_\pm$ we must first identify the space of rational functions $R'_{\Pi \bzeta_\pm}(\fg^\bC)^\Pi$ corresponding to the meromorphic 1-form \eqref{byb model omega}. Fixing the partition of the set of zeroes $\Pi \bzeta = \bzeta = \{0, \infty\}$ of $\omega$ to be $\Pi\bzeta_+ = \bzeta_+ = \{\infty\}$ and $\Pi\bzeta_- = \bzeta_- = \{0\}$ we have
\begin{equation}
    R'_{\Pi \bzeta_+}(\fg^\bC)^\Pi = \left\{\ms b z \,  | \, \ms b\in \fg\right\} \,,\quad R'_{\Pi \bzeta_-}(\fg^\bC)^\Pi = \left\{\frac{\ms b}{z} \,   | \, \ms b\in \fg\right\}.
\end{equation}
Expanding such rational functions at the set of independent poles $\bz = \{ z_+, \tilde{z}_+ \}$ of $\omega$ yields
\begin{align}
\label{vplus byb model}
    \fv_+ & = \jb_z\left(R'_{\Pi \bzeta_+}(\fg^\bC)^\Pi\right)=\left\{\left(\ms b z_+,\ms b\tilde{z}_+\right)\,|\, \ms b\in \fg\right\}, \\
\label{vminus byb model}    
   \fv_- &=\jb_z\left(R'_{\Pi \bzeta_-}(\fg^\bC)^\Pi\right)= \left\{\left(\frac{\ms b}{z_+},\frac{\ms b}{\tilde{z}_+}\right)\,|\, \ms b\in \fg \right\}\,.
\end{align}
We then have $\mathrm{im} W^{\pm}_{(g,\Tilde{g})} = \ff \oplus\fv_\pm$ and we may now proceed with the computation of $W^{\pm}_{(g,\Tilde{g})}(j_\pm,\Tilde{\jmath}_\pm)$ where we defined $j_{\pm}\coloneqq g^{-1}\partial_\pm g$ and $\Tilde{\jmath}_\pm \coloneqq \Tilde{g}^{-1}\partial_\pm \Tilde{g}$. 

In order to find the explicit action of $W^\pm_{(g,\Tilde{g})}$ on $(j_\pm,\Tilde{\jmath}_\pm)$, we decompose the latter with respect to the direct sum decomposition $\fd = \ker W^\pm_{(g,\Tilde{g})} \dotplus \mathrm{im} W^\pm_{(g,\Tilde{g})}$. Explicitly, focusing first on $(j_+, \tilde{\jmath}_+)$, we look for $\ms a, \ms b, \ms x, \ms y \in \fg$ such that 
 \begin{equation}
 \label{ec:uv}
     \left(j_+,\Tilde{\jmath}_+\right)=(\ms a, \ms a)+\left(\ms b z_+,\ms b \Tilde{z}_+\right)+\left((R_g-\ms i)\mathrm{Ad}_g^{-1}\mathsf{x},(\tilde{R}_{\Tilde{g}}-\ms i)\mathrm{Ad}_{\Tilde{g}}^{-1}\mathsf{y}\right).
\end{equation}
To match the notation from \cite{Delduc:2019whp} it is convenient to introduce
\begin{equation}
    J_\pm = \frac{1}{1\pm\frac{\eta}{2} R_g \pm\frac{\Tilde{\eta}}{2}R_{\Tilde{g}}}(j_\pm - \Tilde{\jmath}_\pm)\,,
\end{equation}
in terms of which the solution to \eqref{ec:uv} can be conveniently written as
\begin{gather*}
    \ms a = j_+ + \bigg( \rho - \frac{\eta}{2} R_g \bigg) J_+ = \tilde{\jmath}_+ + \bigg( 1+\rho + \frac{\tilde{\eta}}{2} \tilde{R}_{\tilde{g}} \bigg) J_+, \\
    \ms b = \frac{\zeta}{2} J_+,
    \qquad \mathrm{Ad}_{g}^{-1} \ms x = \frac{\eta}{2} J_+,
    \qquad \mathrm{Ad}_{\Tilde{g}}^{-1} \ms y  = -\frac{\tilde{\eta}}{2} J_+\,.
\end{gather*}
In particular, given that $W^+_{(g,\Tilde{g})}$ is a projector, its action on \eqref{ec:uv} is given by the first two terms on the right hand side, namely we have
\begin{equation}
\label{ec:wpbybm}
    W_{(g,\Tilde{g})}^+ (j_+, \Tilde{\jmath}_+) 
    = \bigg( j_+ - \frac{\eta}{2} (R_g - \mathrm{i}) J_+, \tilde{\jmath}_+ + \frac{\tilde{\eta}}{2} (\tilde{R}_{\tilde{g}} - \mathrm{i}) J_+ \bigg) \,.
\end{equation}
Similarly, to compute the action of $W^-_{(g,\Tilde{g})}$ on $(j_-,\Tilde{\jmath}_-)$ we look again for $\ms a, \ms b, \ms x, \ms y \in \fg$ but this time such that 
 \begin{equation}
 \label{ec:uv2}
     (j_-,\Tilde{\jmath}_-)=(\ms a, \ms a)+\left(\frac{\ms b}{z_+},\frac{\ms b}{\Tilde{z}_+}\right)+\left((R_g-\ms i)\mathrm{Ad}_g^{-1}\mathsf{x},(\tilde{R}_{\Tilde{g}}-\ms i)\mathrm{Ad}_{\Tilde{g}}^{-1}\mathsf{y}\right) \,.
 \end{equation}
Doing so, we find
\begin{equation}
\label{ec:wmbybm}
    W_{(g,\Tilde{g})}^- (j_-,\Tilde{\jmath}_-)
    = \bigg( j_- + \frac{\eta}{2} (R_g - \mathrm{i}) J_-, \tilde{\jmath}_- - \frac{\tilde{\eta}}{2} (\tilde{R}_{\tilde{g}} - \mathrm{i}) J_- \bigg)\,.
\end{equation}
Using the expression for the bilinear form \eqref{ec:byb form} we then obtain
\begin{equation}
    \langle\!\langle  W_{(g,\Tilde{g})}^\pm (j_\pm,\Tilde{\jmath}_\pm ), (j_\mp,\Tilde{\jmath}_\mp)\rangle \!\rangle_{\fd} = \pm 2K \langle j_+ - \tilde{\jmath}_+, J_- \rangle
\end{equation}
where we used the fact that $\langle j_- - \tilde{\jmath}_-, J_+ \rangle = \langle j_+ - \tilde{\jmath}_+, J_- \rangle$ which follows from the skew-symmetry of $R$ and $\tilde R$.

On the other hand, it is immediate to verify that the Wess-Zumino term vanishes identically. We thus find that the $2$d action \eqref{E-model action} is given by
\begin{equation}
    S[g,\Tilde{g}]=K\int_{\Sigma}\left\langle j_+-\Tilde{\jmath}_+,J_-\right\rangle \dr \sigma \wedge\dr \tau,
\end{equation}
matching the action of the bi-Yang-Baxter $\sigma$-model as written in \cite[(2.2)]{Delduc:2015xdm}. 

\subsubsection{Lax connection}

The Lax connection is given by \eqref{Lax connection E-model}, which for this specific example becomes 
\begin{equation}
\cL_\pm\big( (g,\Tilde{g}) \big) = \boldsymbol{p} \big( W^\pm_{(g,\Tilde{g})}(j_\pm,\Tilde{\jmath}_\pm) \big)\,,
\end{equation}
where $\boldsymbol{p} : \ff^\perp \SimTo R_{\Pi \bzeta}(\fg^\bC)^\Pi$ is the inverse of $\jb_\bz$ with
\begin{equation} \label{f perp biYB}
\ff^\perp = \ff\oplus \fv_+ \oplus \fv_-
\end{equation}
with $\ff$, $\fv_+$ and $\fv_-$ given in \eqref{f byb model}, \eqref{vplus byb model} and \eqref{vminus byb model}, respectively, so that the action of $\boldsymbol{p}$ on an element in $\ff^\perp$ decompose with respect to \eqref{f perp biYB} is simply
\begin{equation}
    \boldsymbol{p} \left((\ms a,\ms a)+(\ms b z_+,\ms b \Tilde{z}_+)+\left(\frac{\ms c}{z_+},\frac{\ms c}{\Tilde{z}_+}\right)\right) = \ms a + \ms b z + \frac{\ms c}{z}\,.
\end{equation}
Therefore, decomposing $W^\pm_{(g,\Tilde{g})}(j_\pm,\Tilde{\jmath}_\pm)$ with respect to \eqref{f perp biYB} we find 
\begin{equation}
   \cL_+ = B_++\frac{\zeta}{2}z J_+   \,,\quad \cL_-=B_-+\frac{\zeta}{2}z^{-1}J_-
\end{equation}
where we have defined 
\begin{equation}
    B_{\pm}= j_\pm + \bigg( \rho \mp \frac{\eta}{2}R_g \bigg) J_\pm \,,
\end{equation}
with $\rho$ defined in \eqref{ec:zpluszminus}. The expressions for the components of the Lax connection coincide, up to a conventional sign, with \cite[(2.18)]{Delduc:2015xdm}.

\chapter{Integrable Deformations From Twistor Space}

\label{sec:intdeformations}

This work was done in collaboration with Lewis Cole, Ryan Cullinan, Ben Hoare and Daniel Thompson \cite{Cole:2023umd}.

\section{Introduction}

Developing a systematic understanding of the landscape of integrable 2-dimensional field theories [IFT$_{2}$] has been a longstanding challenge.
In the case of non-linear sigma-models with target space a group manifold $G$, with associated Lie algebra $\mathfrak{g}$, a significant advance was the introduction of integrable deformations of both the principal chiral model [PCM] and the Wess-Zumino-Witten [WZW] model, known respectively as $\eta$-, or Yang-Baxter [YB]
\cite{Klimcik:2008eq}, and $\lambda$-models \cite{Sfetsos:2013wia}.
These discoveries have paved the way to wider classes of integrable deformations with applications to worldsheet string theory and holography (for a recent survey see, e.g.,~\cite{Hoare:2021dix}).

At the classical level, the integrability of these models can be captured by a flat $\mathfrak{g}^\mathbb{C}$-valued Lax connection $\mathrm{D} = \dr + \mathcal{L}$ which depends meromorphically on a spectral parameter $\zeta\in \mathbb{CP}^1$.
The Poisson algebra of the spatial component $\mathcal{L}_\sigma$ can be entirely characterised by a single meromorphic function $\varphi(\zeta) $ known as the {\em twist function}.
The form of this algebra implies that an infinite tower of higher-spin conserved charges in involution can be constructed \cite{Lacroix:2017isl}.

In a remarkable sequence of works \cite{Costello:2013zra,Costello:2017dso,Costello:2018gyb,Costello:2019tri,Delduc:2019whp} it was shown that, by geometrising the spectral parameter plane and considering it as part of space-time, such 2-dimensional integrable models have a 4-dimensional origin as a Chern-Simons type (CS$_4$) theory.

The action of CS$_4$ for a $\mathfrak{g}^\mathbb{C}$-valued gauge field $A$ is defined as
\begin{equation}
\label{eq:4dCS}
S_{\mathrm{CS}_4} = \frac{1}{2\pi \mathrm{i}} \int_{\Sigma \times\mathbb{CP}^1} \omega \wedge \mathrm{Tr} \Big(A \wedge \dr A + \frac{2}{3} A \wedge A \wedge A \Big) ~,
\end{equation}
in which $\omega$ is a meromorphic differential on $\mathbb{CP}^1$.
To recover the integrable 2-dimensional theories above, this meromorphic differential should be specified in terms of the relevant twist function as
\begin{equation}
\omega = \varphi(\zeta) \, \dr \zeta ~.
\end{equation}
To fully define the theory, the action should be complemented with a choice of boundary conditions on the gauge field at the location of the poles of $\omega$.
With suitable boundary conditions in place this 4-dimensional theory localises to a 2-dimensional integrable theory defined on the worldsheet $\Sigma$.
For the case of $\eta$- and $\lambda$-models, the relevant boundary conditions were constructed in \cite{Delduc:2019whp}.
See also \cite{Lacroix:2021iit} for a recent review.

Four-dimensional Chern-Simons theory can in turn be understood \cite{Bittleston:2020hfv} as coming from a reduction of 6-dimensional holomorphic Chern-Simons [hCS$_{6}$] on Euclidean twistor space,
\begin{equation}
\label{eq:6dCS}
S_{\mathrm{hCS}_6} = \frac{1}{2\pi \mathrm{i}} \int_{ \bb{PT}} \Omega \wedge \mathrm{Tr} \Big(\cA \wedge \bar{\pd} \cA + \frac{2}{3} \cA \wedge \cA \wedge \cA \Big) ~.
\end{equation}
An action of the form \eqref{eq:6dCS} was first considered in \cite{Witten:1992fb} as the cubic open string field theory action for the type B topological string.
In the context of type B topological string theory, the target space-time is necessarily Calabi-Yau which ensures it is complemented with a trivial canonical bundle, admitting a globally holomorphic top form $\Omega$.
Twistor space however is not Calabi-Yau and as such does not possess a trivial canonical bundle.
Therefore, to study \eqref{eq:6dCS} on $\bb{PT}$ we instead require that $\Omega$ is a meromorphic $(3,0)$-form on $ \bb{PT}$ which, in this work, is assumed to be nowhere vanishing.
Schematically, this process can be understood as defining a non-compact Calabi-Yau 3-fold by excising the poles of $\Omega$ from $\bb{PT}$, which we can now take to be a consistent target space of our type B topological string \cite{Costello:2023hmi}.
This action has also been studied in \cite{Penna:2020uky} with a focus on dimensionally reduced gravity and supergravity.

The connection between hCS$_{6}$ and CS$_{4}$ can be immediately anticipated since $\bb{PT}$ is diffeomorphic to $\mathbb{E}^4 \times \mathbb{CP}^1$ (see appendix \ref{appendix:twistors} for twistor space conventions).
Identifying the $\mathbb{CP}^1$ with the spectral parameter plane, we specify a reduction ansatz, known as {\em symmetry reduction}, which identifies the worldsheet $\Sigma \hookrightarrow \mathbb{E}^4$.
The details, however, are subtle.
While $\omega$ has both zeroes and poles, $\Omega$ only has poles (the zeroes arise from the data that specifies the symmetry reduction).
Moreover, suitable boundary conditions on $\cA$ are less well understood and currently only explicitly known for a limited class of theories, primarily Dirichlet-type boundary conditions that give, e.g., the principal chiral model with Wess-Zumino term.
The generalisation away from Dirichlet boundary conditions necessary to recover lines of continuous integrable deformations, including the aforementioned $\eta$- and $\lambda$-models, is not known.
Indeed, obstacles to the construction of integrable deformations from 6 dimensions were highlighted in \cite{He:2021xoo}.

Starting in 6 dimensions, an appealing prospect is to swap the order of symmetry reduction and localisation.
Indeed performing the integration over $\mathbb{CP}^1$ (localising to the poles of $\Omega$) results in a 4-dimensional theory.
Here the avatar of integrability is that the equations of motions can be recast as an anti-self-dual Yang-Mills [ASDYM] equation.
The connection between integrable equations and ASDYM has long been known (see \cite{Mason:1991rf}), and ASDYM has been shown to provide a 4-dimensional analogue of 2-dimensional rational CFTs (the WZW model in particular) \cite{Nair:1990aa,Nair:1991ab,Losev:1995cr}.
We then anticipate a return to the same 2-dimensional integrable model by performing symmetry reduction.
This gives rise to a diamond correspondence of theories
\begin{equation*}
\label{diagram:diamonddef}
\begin{tikzpicture}
\node at (0,2) {$\mathbf{hCS_6}$};
\node at (-2,0) {$\mathbf{CS_4}$};
\node at (2,0) {$\mathbf{IFT_4}$};
\node at (0,-2) {$\mathbf{IFT_2}$};
\draw[->,very thick,decorate, decoration={snake, segment length=12pt, amplitude=2pt}] (-0.4,1.6)--(-1.6,0.4);
\draw[->,very thick] (0.4,1.6)--(1.6,0.4);
\draw[->,very thick,decorate, decoration={snake, segment length=12pt, amplitude=2pt}] (1.6,-0.4)--(0.4,-1.6);
\draw[->,very thick] (-1.6,-0.4)--(-0.4,-1.6);
\end{tikzpicture}
\end{equation*}
in which the wavy arrows indicate symmetry reduction and straight arrows indicate localisation (integration over $\mathbb{CP}^1$).
In the context of integrable deformations of IFT$_2$, the right-hand side of this diamond is less well understood.
As such, a key goal of the present work is to develop this side of the diamond for deformed models, and in particular for $\lambda$-type deformations of the WZW and coupled WZW models.

Briefly, the key results of this work are:
\newcommand\circled[1]{\footnotesize\tikz[baseline=(char.base)]{%
\node[shape=circle,fill=black!20,draw,inner sep=2pt] (char) {#1};}}
\begin{enumerate}
\item We establish the consequence of a new class of boundary conditions for hCS$_{6}$.
These reduce to a wider class of boundary conditions in CS$_{4}$ than have previously been considered (relaxing the assumption of an isotropic subalgebra of the defect algebra).

\item Integrating over $\mathbb{CP}^1$ results in a novel multi-parametric IFT$_{4}$ whose equations of motion can be recast in terms of an anti-self-dual Yang-Mills connection.
This new IFT$_4$ exhibits two semi-local symmetries, which can be understood as the residual symmetries preserving the boundary conditions.
For each of these two semi-local symmetries, the Noether currents can be used to construct two inequivalent Lax formulations of the dynamics.

\item Upon symmetry reduction, this IFT$_4$ descends to the 2-field $\lambda$-type IFT$_2$ of \cite{Georgiou:2017jfi} providing a new multi-parametric sigma-model example of the Ward conjecture \cite{Ward:1985}.
Generically the semi-local symmetries of the IFT$_4$ reduce to global symmetries of the IFT$_2$ and the two Lax formulations of the IFT$_4$ give rise to two Lax connections of IFT$_2$.

\item When the symmetry reduction constraints are aligned to these semi-local symmetries, the IFT$_2$ symmetries are enhanced to either affine or fully local (gauge) symmetries.
In the latter case, the IFT$_2$ becomes the standard (1-field) $\lambda$-model.
\end{enumerate}

\section{Holomorphic 6-Dimensional Chern-Simons Theory}

Our primary interest in this work will be the hCS$_6$ diamond containing the $\lambda$-deformed IFT$_2$ originally constructed in \cite{Sfetsos:2013wia}.
By proposing a carefully chosen set of boundary conditions, we will be able to find a diamond of theories that arrives at a multi-parametric class of integrable $\lambda$-deformations between coupled WZW models.

To this end we restrict our study of hCS$_6$, defined by the action
\begin{equation}\label{eq:hCS6}
S_{\mathrm{hCS}_6} = \frac{1}{2 \pi \mathrm{i}} \int_{\bb{PT}} \Omega \wedge \mathrm{Tr} \Big( \cA \wedge \bar{\pd} \cA + \frac{2}{3} \cA \wedge \cA \wedge \cA \Big) ~,
\end{equation}
to the case where the $(3,0)$-form is given by, in the basis of $(1,0)$-forms defined in appendix \ref{appendix:basis},
\begin{equation}\label{eq:Omegaintdef}
\Omega = \frac{1}{2} \, \Phi \, e^0 \wedge e^A \wedge e_A ~, \qquad
\Phi = \frac{K}{
\langle \pi \alpha \rangle
\langle \pi \tilde{\alpha} \rangle
\langle \pi \beta \rangle^2} ~.
\end{equation}
Here, we view $\bb{PT}$ as diffeomorphic to $\mathbb{E}^4 \times \mathbb{CP}^1$ and adopt the standard coordinates $x^{AA'}$ on $\mathbb{E}^4$ and homogeneous coordinates $\pi_{A^\prime}$ on $\mathbb{CP}^1$.
The constant spinors $\alpha$, $\tilde\alpha$ and $\beta$ should be understood as part of the definition of the model.
See appendix \ref{appendix:twistors} for further details of twistor notation and conventions.
The gauge field is similarly written in the basis of $(0,1)$-forms as
\begin{equation}
\cA = \cA_0 \, \bar{e}^0 + \cA_A \, \bar{e}^A ~,
\end{equation}
and the action is invariant under shifts of $\cA$ by any $(1,0)$-form, i.e.\ $\cA \mapsto \cA + \rho$ where $\rho \in \Omega^{(1,0)}(\bb{PT})$.

The first step in studying the 6-dimensional theory is to impose conditions ensuring the vanishing of the `boundary' term that appears in the variation of the action
\begin{equation}
0 = \int_{\bb{PT}} \bar{\pd} \Omega \wedge \mathrm{Tr} \big( \cA \wedge \delta \cA \big) ~.
\end{equation}
Since $\Omega$ is meromorphic, as opposed to holomorphic, this receives contributions from the poles at $\alpha$, $\tilde{\alpha}$, and $\beta$.
We assume that Dirichlet conditions $\cA_A \vert_{\pi = \beta} = 0$ are imposed at the second-order pole.
At the first-order poles, we can then evaluate the integral over $\mathbb{CP}^1$ to obtain
\unskip\footnote{To compute the boundary variation of the action, we have used the identities $e^C \wedge e_C \wedge \bar{e}^A \wedge \bar{e}^B = -2 \, \vol_4 \, \varepsilon^{AB}$ (where $\vol_4 = \dr x^0 \wedge \dr x^1 \wedge \dr x^2 \wedge \dr x^3 $) and
$$
\frac{1}{2 \pi \mathrm{i}} \int_{\mathbb{CP}^1} e^0 \wedge \bar{e}^0 \, \bar{\pd}_0 \bigg( \frac{1}{\langle \pi \alpha \rangle} \bigg) f(\pi) = f(\alpha) ~.
$$}
the condition
\begin{equation}
\frac{1}{\langle \alpha \tilde \alpha \rangle \langle \alpha \beta \rangle^2 } \int_{\mathbb{E}^4} \vol_4 \, \varepsilon^{AB} \mathrm{Tr} \big( \cA_A \delta \cA_B \big) \big\vert_{\pi = \alpha } = \frac{1}{\langle \alpha \tilde \alpha \rangle \langle \tilde \alpha \beta \rangle^2 } \int_{\mathbb{E}^4} \vol_4 \, \varepsilon^{AB} \mathrm{Tr} \big( \cA_A \delta \cA_B \big) \big\vert_{\pi = \tilde \alpha } ~.
\end{equation}
For reasons that will shortly become apparent, let us introduce a unit norm spinor $\mu^A$ about which we can expand any spinor $X^A$ as
\begin{align}
\label{ec:mucontraction}
X^A &= [X\hat \mu] \mu^A - [X \mu ]\hat{\mu}^A ~.
\end{align}
Expanding the gauge field components in terms of the basis $\mu^A$ and $\hat{\mu}^A$, and solving locally pointwise on ${\mathbb{E}^4}$, this condition may be written as
\begin{equation}\label{eq:boundaryvariation}
\frac{1}{\langle\alpha\beta\rangle^2}\mathrm{Tr} \big( [{\cal A} \mu] [\delta{\cal A} \hat\mu] - [{\cal A} \hat\mu] [\delta{\cal A} \mu] \big) \big\vert_{\pi=\alpha}=
\frac{1}{\langle\tilde\alpha\beta\rangle^2}\mathrm{Tr} \big( [{\cal A} \mu] [\delta{\cal A} \hat\mu] - [{\cal A} \hat\mu] [\delta{\cal A} \mu] \big)\vert_{\pi=\tilde\alpha} ~.
\end{equation}
The boundary conditions we are led to consider are
\begin{equation} \label{eq:boundarycondition}
[{\cal A} \mu]\vert_{\pi=\alpha} = \sigma \frac{\langle\alpha\beta\rangle}{\langle\tilde\alpha\beta\rangle} [{\cal A} \mu]\vert_{\pi=\tilde\alpha} ~, \qquad
[{\cal A} \hat \mu]\vert_{\pi=\alpha} = \sigma^{-1} \frac{\langle\alpha\beta\rangle}{\langle\tilde\alpha\beta\rangle}[{\cal A} \hat \mu]\vert_{\pi=\tilde\alpha}~,
\end{equation}
where we have introduced the free parameter $\sigma$, which will play the role of the deformation parameter in the IFT$_{4}$.

Let us note that these boundary conditions are invariant under the following discrete transformations
\begin{equation}\label{z21}
\quad \alpha \leftrightarrow \tilde\alpha ~, \qquad \sigma \mapsto \sigma^{-1} ~,
\end{equation}
\begin{equation}\label{z22}
\quad\mu \mapsto \hat \mu ~, \qquad \sigma \mapsto \sigma^{-1} ~.
\end{equation}
These will descend to transformations that leave the IFT$_4$ invariant.

\subsection{Residual Symmetries and Edge Modes}\label{ssec:gsem}
A general feature of Chern-Simons theory with a boundary is the emergence of propagating edge modes as a consequence of the violation of gauge symmetry by boundary conditions.
A similar effect underpins the emergence of the dynamical field content of the lower dimensional theories that descend from hCS$_{6}$.
Generally, group-valued degrees of freedom, here denoted by $h$ and $\tilde{h}$, would be sourced at the locations of the poles of $\Omega$.
If however, the boundary conditions~\eqref{eq:boundarycondition} admit residual symmetries, then these will result in symmetries of the IFT$_4$ potentially mixing the $h$ and $\tilde{h}$ degrees of freedom.
These may be global symmetries, gauge symmetries, or semi-local symmetries depending on the constraints imposed by the boundary conditions.
It is thus important to understand the nature of any residual symmetry preserved by the boundary conditions \eqref{eq:boundarycondition}.

Gauge transformations act on the hCS$_6$ gauge field as
\begin{equation}
\hat{g} :  \quad \cA \mapsto \hat{g}^{-1} \cA \hat{g} + \hat{g}^{-1} \bar{\pd} \hat{g} ~.
\end{equation}
In the bulk, i.e.\ away from the poles of $\Omega$, these are unconstrained, but at the poles they will only leave the action invariant if they preserve the boundary conditions.
For later convenience, we will denote the values of the gauge transformation parameters at the poles by
\begin{equation}
\hat{g} \vert_{\alpha} = r ~, \quad
\hat{g} \vert_{\tilde{\alpha}} = \tilde{r} ~, \quad
\hat{g} \vert_{\beta} = \ell^{-1} ~.
\end{equation}
Firstly, the transformation acting at $\beta$ must preserve the constraint $\cA_A \vert_\beta = 0$.
Initially, one might suppose that only constant $\ell$ would preserve this boundary condition, but in fact it is sufficient for $\ell$ to be holomorphic with respect to the complex structure defined by $\beta$
\begin{equation}\label{eq:lconstraints}
\beta^{A^\prime} \pd_{A A^\prime} \ell = 0 \quad \Rightarrow \quad
\frac{1}{\langle \alpha \beta \rangle} \alpha^{A^\prime} \pd_{A A^\prime} \ell = \frac{1}{\langle \tilde\alpha \beta \rangle} \tilde\alpha^{A^\prime} \pd_{A A^\prime} \ell ~.
\end{equation}
These differential constraints arise from the fact that the anti-holomorphic vector fields $\bar{\pd}_A = \pi^{A^\prime} \pd_{A A^\prime}$ are valued in $\mathcal{O}(1)$.
In other words, they depend explicitly on the $\mathbb{CP}^1$ coordinate (see appendix \ref{appendix:twistors} for more details).

Secondly, the transformations acting at $\alpha$ and $\tilde{\alpha}$ must preserve the boundary conditions \eqref{eq:boundarycondition}, implying the constraints
\begin{equation}\label{eq:rconstraints}
\begin{aligned}
\tilde{r} & = r \ , \\
\frac{1}{\langle\alpha\beta\rangle}\mu^A \alpha^{A^\prime} \pd_{A A^\prime} r & = \frac{\sigma }{\langle\tilde\alpha\beta\rangle} \mu^A \tilde{\alpha}^{A^\prime} \pd_{A A^\prime} r \ , \\
\frac{1}{\langle\alpha\beta\rangle} \hat{\mu}^A \alpha^{A^\prime} \pd_{A A^\prime} r & = \frac{\sigma^{-1} }{\langle\tilde\alpha\beta\rangle} \hat{\mu}^A \tilde{\alpha}^{A^\prime} \pd_{A A^\prime} r ~.
\end{aligned}
\end{equation}
These residual symmetries are neither constant (i.e.\ global symmetries) nor fully local (i.e.\ gauge symmetries).
Instead, we expect that our IFT$_4$ should exhibit two semi-local symmetries subject to the above differential constraints, akin to the semi-local symmetries of the 4d WZW model first identified in \cite{Nair:1990aa,Nair:1991ab}
\unskip\footnote{Complementary to this perspective, the WZW$_4$ algebra can also be obtained as a global symmetry of five-dimensional K\"ahler Chern-Simons on a manifold with boundary \cite{Banados:1996yj}.}.

\paragraph{Symmetry reduction}
As we progress around the diamond, we will perform `symmetry reduction' (see \S~\ref{sec:4dhcsred} and \S~\ref{sec:4dIFTto2dIFT} for details).
In essence, this will mean we restrict to fields and gauge parameters that are independent of two directions, i.e.\ they obey the further differential constraints (where $\gamma^{A^\prime}$ is some constant spinor)
\begin{equation}
\mu^A \gamma^{A^\prime} \pd_{A A^\prime} \hat{g} = 0 ~, \quad
\hat{\mu}^A \hat{\gamma}^{A^\prime} \pd_{A A^\prime} \hat{g} = 0 ~.
\end{equation}
We can then predict some special points in the lower dimensional theories by considering how these differential constraints interact with those imposed by the boundary conditions.
Generically, these four differential constraints (two from the boundary conditions and two from symmetry reduction) will span a copy of $\mathbb{E}^4$ at each pole, meaning that only constant transformations (i.e.~global symmetries) will survive.
However, if the symmetry reduction is carefully chosen, the two sets of constraints may partially or entirely coincide.
In the case that they entirely coincide, the lower dimensional symmetry parameter will be totally unconstrained, meaning that the IFT$_2$ will possess a gauge symmetry.
Alternatively, if the constraints partially coincide then the lower dimensional theory will have a symmetry with free dependence on half the coordinates, e.g.\ the chiral symmetries of the 2d WZW model.

\section{Localisation of \texorpdfstring{hCS$_{6}$}{hCS6} to \texorpdfstring{IFT$_4$}{IFT4}}\label{S:Localisation of 6d to 4d}

\begin{tcolorbox}
\begin{minipage}[c]{0.4\linewidth}

\begin{equation*}
\begin{tikzpicture}[scale=0.8]
\node at (0,2) {$\mathbf{hCS_6}$};
\node at (-2,0) {$\mathbf{CS_4}$};
\node at (2,0) {$\mathbf{IFT_4}$};
\node at (0,-2) {$\mathbf{IFT_2}$};
\draw[->,very thick,decorate, decoration={snake, segment length=13pt, amplitude=2pt}] (-0.4,1.6)--(-1.6,0.4);
\draw[->,very thick,red] (0.4,1.6)--(1.6,0.4);
\draw[->,very thick,decorate, decoration={snake, segment length=13pt, amplitude=2pt}] (1.6,-0.4)--(0.4,-1.6);
\draw[->,very thick] (-1.6,-0.4)--(-0.4,-1.6);
\end{tikzpicture}
\end{equation*}

\end{minipage}
\begin{minipage}[c]{0.59\linewidth}
Let us now proceed in navigating the top right-hand side of the diamond.
By integrating over $\mathbb{CP}^1$ we will `localise' hCS$_6$ on $\bb{PT}$ to an effective theory defined on $\mathbb{E}^4$.
This resulting theory is `integrable' in the sense that its equations of motion can be encoded in an anti-self-dual connection.
\end{minipage}

\end{tcolorbox}

The localisation analysis is naturally presented in terms of new variables $\cA^\prime$ and $\hat{h}$, which are related to the fundamental field by
\begin{equation}
\label{ec:redundancy}
\cA = \hat{h}^{-1} \cA^\prime \hat{h} + \hat{h}^{-1} \bar{\pd} \hat{h} ~.
\end{equation}
However, there is some redundancy in this new parametrisation.
There are internal gauge transformations (leaving $\cA$ invariant) given by
\begin{equation}\label{eq:internalgauge}
\hat{g} : \quad \cA^\prime \mapsto \hat{g}^{-1} \cA^\prime \hat{g} + \hat{g}^{-1} \bar{\pd} \hat{g} ~, \qquad
\hat{h} \mapsto \hat{g}^{-1} \hat{h} ~.
\end{equation}
These allow us to impose the constraint $\cA^\prime_0 = 0$, i.e.~it has no leg in the $\mathbb{CP}^1$-direction.
This is done so that $\cA'$ may be interpreted as an anti-self-dual Yang-Mills connection on $\bE^4$.

There are still internal gauge transformations that are $\mathbb{CP}^1$-independent, and we can use these to fix the value of $\hat{h}$ at one pole.
We will therefore impose the additional constraint $\hat{h} \vert_{\beta} = \mathrm{id}$ so that we have resolved this internal redundancy.
The values of $\hat{h}$ at the remaining poles
\begin{equation}
\hat{h} \vert_{\alpha} = h ~, \qquad
\hat{h} \vert_{\tilde{\alpha}} = \tilde{h} ~,
\end{equation}
will be dynamical edge modes as a consequence of the violation of gauge symmetry by boundary conditions.
As we will now see, the entire action localises to a theory on $\mathbb{E}^4$ depending only on these edge modes.

The hCS$_6$ action is written in these new variables as
\begin{equation}
\begin{aligned}\label{eq:hcsNewVars}
S_{\mathrm{hCS}_6} & = \frac{1}{2 \pi \mathrm{i}} \int_{\bb{PT}} \Omega \wedge \mathrm{Tr} \big( \cA^\prime \wedge \bar{\pd} \cA^\prime \big)
+ \frac{1}{2 \pi \mathrm{i}} \int_{\bb{PT}} \bar{\pd} \Omega \wedge \mathrm{Tr} \big( \cA^\prime \wedge \bar{\pd} \hat{h} \hat{h}^{-1} \big) \\
& \hspace{3em} - \frac{1}{6 \pi \mathrm{i}} \int_{\bb{PT}} \Omega \wedge \mathrm{Tr} \big( \hat{h}^{-1} \bar{\pd} \hat{h} \wedge \hat{h}^{-1} \bar{\pd} \hat{h} \wedge \hat{h}^{-1} \bar{\pd} \hat{h} \big) ~.
\end{aligned}
\end{equation}
The cubic term in $\cA^\prime$ has dropped out since we have imposed $\cA^\prime_0 = 0$.
Inspecting the terms in our action involving $\hat{h}$, we see that the second term localises to the poles due to the anti-holomorphic derivative acting on $\Omega$.
The third term similarly localises to the poles.
For this, we consider a manifold whose boundary is $\bb{PT}$.
\unskip\footnote{More generally, a manifold whose boundary is a disjoint union of copies of $\bb{PT}$.}
We take the $7$-manifold $\bb{PT} \times [0, 1]$ and extend our field $\hat{h}$ over this interval. We do this by choosing a smooth homotopy to a constant map, such that it's restriction to $\bb{PT}\times \{0\}$ coincides with $\hat h$. Denoting this extension with the same symbol, we see that the third term in our action may be equivalently written as
\begin{equation}
- \frac{1}{6 \pi \mathrm{i}} \int_{\bb{PT} \times [0, 1]} \dr \bigg[ \Omega \wedge \mathrm{Tr} \big( \hat{h}^{-1} \dr \hat{h} \wedge \hat{h}^{-1} \dr \hat{h} \wedge \hat{h}^{-1} \dr \hat{h} \big) \bigg] ~.
\end{equation}
Then, using the closure of the Wess-Zumino $3$-form and the fact that all of the holomorphic legs on $\bb{PT}$ are saturated by $\Omega$, this is equal to
\begin{equation}
S_{\mathrm{WZ}_4} = - \frac{1}{6 \pi \mathrm{i}} \int_{\bb{PT} \times [0, 1]} \bar{\pd} \Omega \wedge \mathrm{Tr} \big( \hat{h}^{-1} \dr \hat{h} \wedge \hat{h}^{-1} \dr \hat{h} \wedge \hat{h}^{-1} \dr \hat{h} \big) ~.
\end{equation}
Therefore, this contribution also localises, meaning that the only information contained in the field $\hat{h} : \bb{PT} \to G$ are its values
\unskip\footnote{For higher order poles in $\Omega$, the $\mathbb{CP}^1$-derivatives of $\hat{h}$ would also contribute to the action.}
at the poles of $\Omega$.
Explicitly, this contribution is given by
\unskip\footnote{In principle there are also contributions from the double pole at $\beta$ both in this term and the second term in the action~\eqref{eq:hcsNewVars}.
Since this is a double pole, these contributions may depend on $\partial_0\hat h\vert_\beta$, which is unconstrained.
However, one can check that they vanish using just the boundary conditions $\mathcal{A}_A\vert_\beta = 0$ and internal gauge-fixing $\hat h\vert_\beta = \mathrm{id}$.
Alternatively, we may use part of the residual external gauge symmetry to fix $\partial_0\hat h\vert_\beta = 0$, which ensures such contributions vanish.}
\begin{equation}
\begin{aligned}\label{eq:WZ4gen}
S_{\mathrm{WZ}_4} & = -\frac{K}{\langle \alpha \tilde{\alpha}\rangle}\int_{\mathbb{E}^4 \times [0, 1]} \hspace{-1em} \vol_4 \wedge \dr \rho \, \varepsilon^{AB} \bigg[ \frac{1}{ \langle \alpha \beta \rangle^2} \, \mathrm{Tr} \big( h^{-1} \pd_\rho h \cdot \alpha^{A^\prime} h^{-1} \pd_{A A^\prime} h \cdot \alpha^{B^\prime} h^{-1} \pd_{B B^\prime} h \big) \\
& \hspace{140pt} - \frac{1}{ \langle \tilde{\alpha} \beta \rangle^2} \, \mathrm{Tr} \big( \tilde{h}^{-1} \pd_\rho \tilde{h} \cdot \tilde{\alpha}^{A^\prime} \tilde{h}^{-1} \pd_{A A^\prime} \tilde{h} \cdot \tilde{\alpha}^{B^\prime} \tilde{h}^{-1} \pd_{B B^\prime} \tilde{h} \big) \bigg]
\\
& = \frac{K}{6\langle \alpha \tilde{\alpha} \rangle} \int_{\mathbb{E}^4 \times [0, 1]} \bigg[ \frac{1}{ \langle \alpha \beta \rangle^2} \mu_{\alpha} \wedge \mathrm{Tr} \big( h^{-1} \dr h \big)^3 -\frac{1}{ \langle \tilde\alpha \beta \rangle^2} \mu_{\tilde\alpha} \wedge \mathrm{Tr} \big( \tilde{h}^{-1} \dr \tilde{h} \big)^3 \bigg] ~,
\end{aligned}
\end{equation}
where
\begin{align}\label{eq:hol20}
\mu_{\alpha} &= \varepsilon_{AB} \alpha_{A^\prime} \alpha_{B^\prime} \, \dr x^{A A^\prime} \wedge \dr x^{B B^\prime} ~, \qquad
\mu_{\tilde\alpha} = \varepsilon_{AB} \tilde\alpha_{A^\prime} \tilde\alpha_{B^\prime} \, \dr x^{A A^\prime} \wedge \dr x^{B B^\prime} ~,
\end{align}
are the $(2,0)$-forms with respect to the complex structure on $\bE^4$ defined by $\alpha_{A'}$ and $\tilde{\alpha}_{A'}$ respectively.
\unskip\footnote{Here we are using the fact that $\bE^4$ is endowed with a hyper-K\"ahler structure such that there is a $\mathbb{CP}^1$ space of complex structures (see appendix \ref{appendix:twistors}).}

Knowing that the latter two terms in the action~\eqref{eq:hcsNewVars} localise to the poles, we are one step closer to deriving the IFT$_4$.
There are two unresolved problems: the first term is still a genuine bulk term; and the second term contains $\cA^\prime$, rather than being written exclusively in terms of the fields $h$ and $\tilde{h}$.
Both of these problems will be resolved by invoking the bulk equations of motion for $\cA^\prime$.
This will completely specify its $\mathbb{CP}^1$-dependence, and, combined with the boundary conditions, we will then be able to solve for $\cA^\prime$ in terms of the edge modes $h$ and $\tilde{h}$.

Varying the first term in the action, which is the only bulk term, we find the equation of motion $\bar{\pd}_0 \cA^\prime_A = 0$, which implies that these components are holomorphic.
Combined with the knowledge that $\cA^\prime_A$ has homogeneous weight $1$, we deduce that the $\mathbb{CP}^1$-dependence is given by $\cA^\prime_A = \pi^{A^\prime} A_{A A^\prime}$ where $A_{A A^\prime}$ is $\mathbb{CP}^1$-independent.

Turning our attention to the boundary conditions, we first consider the double pole where we have imposed $\cA_A \vert_{\beta} = 0$.
Recalling that $\hat{h} \vert_{\beta} = \mathrm{id}$, this simply translates to $\cA^\prime_A \vert_{\beta} = 0$.
This tells us that $\cA^\prime_A = \langle \pi \beta \rangle B_A$ for some $B_A$, hence $A_{AA'} = \beta_{A'} B_{A}$.
Therefore, we have that
\begin{equation}\label{eq:calAtoB}
{\cal A}_A = \langle \pi \beta \rangle \mathrm{Ad}_{\hat{h}}^{-1}B_A + \pi^{A'} \hat{h}^{-1}\partial_{A A'} \hat{h} ~.
\end{equation}

The solution for $B_A$ found by solving the remaining two boundary conditions \eqref{eq:boundarycondition} is written more concisely if we introduce some notation.
We will make extensive use of the operators
\begin{equation}
U_{\pm} = \left( 1- \sigma^{\pm 1} \Lambda \right)^{-1} ~, \qquad \Lambda = \mathrm{Ad}_{\tilde{h}}^{-1} \mathrm{Ad}^{\phantom{-1}}_{h\phantom{\tilde{h}}} ,  
\end{equation}
which enjoy the useful identities
\begin{equation}
U_+^{T} + U_- = \mathrm{id} ~, \qquad U_\pm \Lambda = - \sigma^{\mp 1} U^T_{\mp} ~,
\end{equation}
where transposition is understood to be with respect to the ad-invariant inner product on $\mathfrak{g}$.
In terms of the components of $\hat{h}^{-1}\partial_{A A'} \hat{h}$, defined with useful normalisation factors,
\def\hj{\,\widehat{\jmath}\,}
\def\tj{\,\widetilde{\jmath}\,}
\def\htj{\,\widehat{\phantom{\rule{-1.5pt}{7.5pt}}\smash{\widetilde{\jmath}}}\,}
\begin{equation}\begin{aligned}\label{eq:licurrents}
j =\langle\alpha\beta\rangle^{-1} \mu^A \alpha^{A'} h^{-1 }\partial_{A A'} h ~, \qquad
\hj =\langle\alpha\beta\rangle^{-1} \hat{\mu}^A \alpha^{A'} h^{-1 }\partial_{A A'} h ~, \\
\tj =\langle\tilde{\alpha}\beta\rangle^{-1} \mu^A \tilde{\alpha}^{A'} \tilde{h}^{-1 }\partial_{A A'} \tilde{h} ~,
\qquad \htj =\langle\tilde{\alpha}\beta\rangle^{-1} \hat{\mu}^A \tilde{\alpha}^{A'} \tilde{h}^{-1 }\partial_{A A'} \tilde{h} ~,
\end{aligned}\end{equation}
we find that the solutions to the remaining boundary conditions may be written as
\begin{align}\label{eq:bdef}
\mathrm{Ad}_h^{-1} B_A &= \widehat{b} \mu_A - b \hat{\mu}_A ~, & b&= U_+ (j - \sigma \tj ) ~, &
\widehat{b}&= U_- (\hj - \sigma^{-1}\htj) ~, \\
\mathrm{Ad}_{\tilde{h}}^{-1} B_A &= \widehat{\widetilde{b}} \mu_A - \widetilde{b} \hat{\mu}_A ~, & \widetilde{b} &=U_-^T(\tj - \sigma^{-1} j) ~, &
\widehat{\widetilde{b}}& = U_+^T(\htj - \sigma \hj) ~.
\label{eq:bdef2}
\end{align}
Note that $b =\mathrm{Ad}_h^{-1} [B \mu ]$, $\widehat{b} =\mathrm{Ad}_h^{-1} [B \hat\mu ]$, etc., and that $b$, $\widetilde{b}$, $\widehat{b}$ and $\widehat{\widetilde{b}}$ are related as
\begin{equation}
\label{eq:brelations}
\widetilde{b} - \tj = \sigma^{-1}(b-j) ~, \qquad
\widehat{\widetilde{b}} - \htj = \sigma(\widehat{b} - \hj) ~.
\end{equation}

Recovering the IFT$_4$ is then simple.
The first term in the action \eqref{eq:hcsNewVars} vanishes identically on shell, and we can substitute in our solution for $\cA^\prime$ in terms of $h$ and $\tilde{h}$ to get a 4d theory only depending on these edge modes.
This results in the action
\begin{align} \label{eq:IFT4}
\nonumber S_{\mathrm{IFT}_4} &= \frac{1}{2 \pi \mathrm{i}} \int_{\bb{PT}} \bar{\pd} \Omega \wedge \mathrm{Tr} \big( \cA^\prime \wedge \bar{\pd} \hat{h} \hat{h}^{-1} \big) + S_{\mathrm{WZ}_4} \\
&= \frac{K}{\langle \alpha \tilde{\alpha} \rangle } \int_{\mathbb{E}^4} \vol_4 \, \mathrm{Tr} \big( b ( \hj - \Lambda^T \htj ) - \widehat{b} ( j - \Lambda^T \tj ) \big) + S_{\mathrm{WZ}_4}\\
\nonumber &= \frac{K}{\langle \alpha \tilde{\alpha} \rangle } \int_{\mathbb{E}^4 }\vol_4 \, \mathrm{Tr} \Big( j ( U^{T}_+ - U_{-} ) \hj + \tj ( U^{T}_+ - U_{-} ) \htj - 2 \sigma \tj \, U_{+}^{T} \, \hj + 2 \sigma^{-1} j \, U_{-} \, \htj \Big)+ S_{\mathrm{WZ}_4} ~,
\end{align}
where
\begin{equation}
\begin{aligned}\label{eq:wz4}
S_{\mathrm{WZ}_4} = \frac{K}{\langle \alpha \tilde{\alpha} \rangle} \int_{\mathbb{E}^4 \times [0, 1]} \hspace{-1em} \vol_4 \wedge \dr \rho \, \mathrm{Tr} \big( h^{-1} \pd_\rho h \cdot [j,\hj] - \tilde{h}^{-1} \pd_\rho \tilde{h} \cdot [\tj,\htj] \big) ~.
\end{aligned}
\end{equation}

Observe that the 4d IFT~\eqref{eq:IFT4} with~\eqref{eq:wz4} is mapped into itself under the formal transformation
\begin{equation}\label{eq:formal1}
h \leftrightarrow \tilde{h} ~, \qquad \alpha \leftrightarrow \tilde \alpha~, \qquad \sigma \mapsto \sigma^{-1} ~,
\end{equation}
interchanging the positions of the two poles.
This directly follows from the invariance~\eqref{z21} of the hCS$_6$ boundary conditions.
On the other hand, looking at how the transformation~\eqref{z22} descends to the IFT$_4$, we find
\unskip\footnote{Note that to derive this we use that $\hat{\hat\mu} = - \mu$ following the ``quaternionic conjugation'' defined in appendix \ref{appendix:twistors}.}
\begin{equation}\label{eq:formal2}
j \mapsto \hj~, \qquad \hj \mapsto -j~, \qquad \tj \mapsto \htj~, \qquad \htj \mapsto - \tj ~, \qquad \sigma \mapsto \sigma^{-1} ~.
\end{equation}
It is then straightforward to check that the action~\eqref{eq:IFT4} with~\eqref{eq:wz4} is invariant under this transformation.

Let us emphasise that, to our knowledge, the IFT$_4$ described by the action~\eqref{eq:IFT4} with~\eqref{eq:wz4} has not been considered in the literature before.
In the following subsections we will study some properties of this model starting with its symmetries, and moving onto its equations of motion and their relation to the 4d ASDYM equations.

\subsection{Symmetries of the \texorpdfstring{IFT$_4$}{IFT4}}

Having derived the action functional for the IFT$_4$, we will now examine those symmetries that leave this action invariant.
While they may not be obvious from simply looking at the action, in \S~\ref{ssec:gsem} we leveraged the hCS$_6$ description to predict the symmetries of the IFT$_4$.
These may then be verified by explicit computation.

To this end, we recall that the hCS$_6$ gauge transformations act as
\begin{equation}
\hat{g} : \quad \cA \mapsto \hat{g}^{-1} \cA \hat{g} + \hat{g}^{-1} \bar{\pd} \hat{g} ~,
\end{equation}
and we denoted the value of this transformation parameter at the poles by
\begin{equation}
\hat{g} \vert_{\alpha} = r ~, \quad
\hat{g} \vert_{\tilde{\alpha}} = \tilde{r} ~, \quad
\hat{g} \vert_{\beta} = \ell^{-1} ~.
\end{equation}
Tracing through the derivation above, we find that these result in an induced action on the IFT$_4$ fields,
\begin{equation}\label{eq:ift4symmetries}
(\ell, r, \tilde{r}) : \quad h \mapsto \ell h r \ , \quad \tilde{h} \mapsto \ell \tilde{h} \tilde{r} \ ,
\end{equation}
where $\ell$, $r$ and $\tilde{r}$ obey the constraints~\eqref{eq:lconstraints} and~\eqref{eq:rconstraints} respectively.
One can explicitly verify that the IFT$_4$ is indeed invariant under these transformations.
Key to this is exploiting a Polyakov-Wiegmann identity such the variation of $S_{\mathrm{WZ}_{4}}$ in eq.~\eqref{eq:wz4} produces a total derivative.
This gives a contribution on $\mathbb{E}^4$ that cancels the variation of the remainder of eq.~\eqref{eq:IFT4}.
Useful intermediate results to this end are
\begin{equation}\begin{gathered}
\mathrm{Ad}_h \mapsto \mathrm{Ad}_\ell \mathrm{Ad}_h \mathrm{Ad}_r ~, \quad \mathrm{Ad}_{\tilde{h}} \mapsto \mathrm{Ad}_\ell \mathrm{Ad}_{\tilde{h}} \mathrm{Ad}_r ~, \quad U_\pm \mapsto \mathrm{Ad}_r^{-1} U_\pm \mathrm{Ad}_r ~, \\
b \mapsto \mathrm{Ad}_r^{-1} ( b
+ \langle\alpha\beta \rangle^{-1} \mathrm{Ad}_h^{-1} \mu^A \alpha^{A'} \ell^{-1}\partial_{AA'}\ell ) ~, \quad
\hat{b} \mapsto \mathrm{Ad}_r^{-1} ( \hat{b} + \langle\alpha\beta \rangle^{-1} \mathrm{Ad}_h^{-1} \hat{\mu}^A \alpha^{A'} \ell^{-1}\partial_{AA'}\ell ) ~,
\end{gathered}\end{equation}
in which the constraints~\eqref{eq:lconstraints} and~\eqref{eq:rconstraints} have been used.

We can also derive the Noether currents corresponding to these residual semi-local symmetries directly from hCS$_6$.
The variation of the action under an infinitesimal gauge transformation $\delta \cA = \bar{\pd} \hat{\epsilon} + \comm{\cA}{\hat{\epsilon}}$ is
\begin{equation}
\delta S_{\text{6dCS}} = \frac{1}{2 \pi \mathrm{i}} \int_{\bb{PT}} \bar{\pd} \Omega \wedge \mathrm{Tr} \big( \cA \wedge \bar{\pd} \hat{\epsilon} \big) ~.
\end{equation}
First let us consider a transformation that descends to the $\ell$-symmetry, i.e.~one for which
$$
\hat{\epsilon} \vert_{\alpha} = \hat{\epsilon} \vert_{\tilde{\alpha}} = 0 ~, \quad \hat{\epsilon} \vert_{\beta} = \epsilon^{(\ell)} ~.
$$
The only contribution to the variation localises to $\beta$ and is given by
\begin{equation}
\delta_\ell S_{\text{6dCS}} \propto \int_{\mathbb{E}^4} \vol_4 \, \pd_0 \bigg( \frac{1}{\langle \pi \alpha \rangle \langle \pi \tilde{\alpha} \rangle} \varepsilon^{AB} \mathrm{Tr} \big( \cA_A \bar{\pd}_B \hat{\epsilon} \big) \bigg) \bigg\vert_{\beta} ~.
\end{equation}
Since $\cA_A \vert_\beta \propto \langle \pi \beta \rangle$ (recall that we fix $\hat{h} \vert_\beta = \text{id}$) the only way the integrand will be non-vanishing is for the $\pd_0$ operator to act on $\cA_A $.
Noting that $\pd_0 \langle \pi \beta \rangle \vert_\beta = 1$ we have that $\pd_0 \cA_A \vert_\beta = B_A$, and hence the variation becomes
\begin{equation}
\delta_\ell S_{\text{6dCS}} \propto \int_{\mathbb{E}^4} \vol_4 \, \mathrm{Tr} \Big( B^B \beta^{B^\prime} \pd_{B B^\prime} \epsilon^{(\ell)} \Big) ~.
\end{equation}
If we think of the $\ell$-symmetry as a global transformation, then this would provide the conservation law associated to the Noether current, i.e.
\begin{equation}
\beta^{B^\prime} \partial_{B B'} B^B = 0 ~,
\end{equation}
and indeed we will see later that this conservation law follows from the equations of motion of the IFT$_{4}$.
As the parameter $\epsilon^{(\ell)}$ is allowed to be holomorphic with respect to the complex structure defined by $\beta$, the interpretation is more akin to a Kac-Moody current.

For the case corresponding to the $r$-symmetry we have
$$
\hat{\epsilon} \vert_{\alpha} = \hat{\epsilon} \vert_{\tilde{\alpha}} = \epsilon^{(r)} \, , \quad \hat{\epsilon} \vert_{\beta} = 0 ~.
$$
In this case the variation receives two contributions with an opposite sign
\begin{equation}
\delta_{r} S_{\text{6dCS}} \propto \int_{\mathbb{E}^4} \vol_4 \, \varepsilon^{AB} \, \mathrm{Tr} \bigg(\frac{1}{\langle \alpha \beta\rangle^2} \cA_A \bar{\pd}_B \hat{\epsilon} \big\vert_{\alpha} - \frac{1}{\langle \tilde\alpha \beta\rangle^2}\cA_A \bar{\pd}_B \hat{\epsilon} \big\vert_{\tilde{\alpha}} \bigg) ~.
\end{equation}
Integrating by parts gives
\begin{equation}
\delta_{r} S_{\text{6dCS}} \propto \int_{\mathbb{E}^4} \vol_4 \, \mathrm{Tr} \bigg( \epsilon^{(r)} \, \bigg( \frac{ \alpha^{A^\prime}}{\langle \alpha \beta \rangle^2 } \pd_{A A^\prime} \cA^A \vert_{\alpha} - \frac{ \tilde{\alpha}^{A^\prime}}{\langle \tilde\alpha \beta \rangle^2 } \pd_{A A^\prime} \cA^A \vert_{\tilde{\alpha}} \bigg) \bigg) ~.
\end{equation}
Introducing new currents defined by
\begin{equation}
\langle \alpha \beta \rangle\, C_{A} = \cA_A \vert_{\alpha} ~, \quad \langle \tilde\alpha \beta \rangle\, \widetilde{C}_{A} = \cA_A \vert_{\tilde \alpha} ~,
\end{equation}
the conservation law associated to the $r$-symmetry is given by
\begin{equation}
\frac{1}{\langle \alpha \beta \rangle}\alpha^{A^\prime} \pd_{A A^\prime} C^A - \frac{1}{\langle \tilde\alpha \beta \rangle}\tilde{\alpha}^{A^\prime} \pd_{A A^\prime} \widetilde{C}^A= 0 ~.
\end{equation}
Recalling from eq.~\eqref{eq:calAtoB} that ${\cal A}_A = \langle \pi \beta \rangle \mathrm{Ad}_{\hat{h}}^{-1}B_A + \pi^{A'} \hat{h}^{-1}\partial_{A A'} \hat{h}$, we can relate the $B$ current to the $C$ and $\widetilde{C}$ currents as follows
\begin{align}\label{eq:Cdef}
C_A &= \mathrm{Ad}_{h}^{-1}B_A + \frac{1}{ \langle \alpha \beta \rangle} \alpha^{A'}h^{-1}\partial_{A A'} h = (\widehat{b}- \hj) \mu_A - (b-j) \hat{\mu}_A ~, \\ \label{eq:Ctdef}
\widetilde{C}_A &=\mathrm{Ad}_{\tilde{h}}^{-1}B_A + \frac{1}{ \langle \tilde\alpha \beta \rangle} \tilde\alpha^{A'} \tilde{h}^{-1}\partial_{A A'} \tilde{h} = \sigma (\widehat{b}- \hj) \mu_A - \sigma^{-1} (b-j) \hat{\mu}_A~,
\end{align}
where we have used the identities \eqref{eq:brelations}.
The transformation of these currents under the $(\ell,r)$-symmetries is given by
\begin{equation}
\begin{aligned}
\label{ec:Btransfrl}
(\ell,r): \quad B_{A} & \mapsto \mathrm{Ad}_\ell B_A - \langle \alpha \beta\rangle^{-1} \alpha^{A'} \partial_{A A'} \ell \ell^{-1} ~, \\
(\ell,r): \quad C_{A} & \mapsto \mathrm{Ad}_r^{-1} C_A
+ \langle \alpha \beta\rangle^{-1} \alpha^{A'} r^{-1} \partial_{A A'} r ~, \\
(\ell,r): \quad \widetilde{C}_{A} & \mapsto \mathrm{Ad}_r^{-1} \widetilde{C}_A + \langle \tilde \alpha \beta\rangle^{-1} \tilde\alpha^{A'} r^{-1} \partial_{A A'} r ~.
\end{aligned}
\end{equation}
As a consequence notice that the 4d ($\mathbb{CP}^1$-independent) gauge field introduced above, $A_{AA'} =\beta_{A'} B_A$, is invariant under the right action, whereas the left action acts as a conventional gauge transformation
\begin{equation}
\label{ec:Atransfrl}
(\ell,r): \quad A_{AA' } \mapsto \mathrm{Ad}_\ell A_{AA'} - \partial_{A A'} \ell \ell^{-1} ~,
\end{equation}
albeit semi-local rather than fully local since $\ell$ is constrained as in eq.~\eqref{eq:lconstraints}.
The transformation of these currents and the 4d ASD connection also follows from the hCS$_6$ description.
While the original gauge transformations act on $\cA$, we observe that $r$ and $\tilde{r}$ become right-actions on $\hat{h}$, leaving $\cA^\prime$ and $A_{A A^\prime}$ invariant.
By comparison, after fixing $\hat{h} \vert_\beta = \text{id}$, it is only a combination of the `internal' transformations and the original gauge transformations that preserve this constraint.
In particular, $\ell$ has an induced action on $h$, $\tilde{h}$ and $\cA^\prime$, thus leading to the above transformations of $B_A$ and $A_{A A^\prime}$.

As we will show momentarily, the equations of motion of the theory correspond to anti-self duality of the field strength of the connection $A_{A A^\prime}$, hence it immediately follows that the equations of motion are preserved by the symmetry transformations~\eqref{ec:Atransfrl}.
To close the section we note that the action is concisely given in terms of the currents as
\begin{align} \label{eq:actionIFT4currents}
S_{\mathrm{IFT}_4}
&= \frac{K}{\langle \alpha \tilde{\alpha} \rangle } \int_{\mathbb{E}^4} \vol_4 \, \epsilon^{AB} \mathrm{Tr} \big( \mathrm{Ad}^{-1}_h B_A (C_B - \Lambda^T \widetilde{C}_B ) \big) + S_{\mathrm{WZ}_4} ~.
\end{align}

\subsection{Equations of Motion, 4d ASDYM and Lax Formulation}
The equations of motion of the IFT$_4$ can be obtained in a brute force fashion by performing a variation of the action \eqref{eq:IFT4}.
This calculation requires the variation of the operators $U_\pm$
\begin{equation}
\delta U_\pm(X) = U_\pm(\delta X) + U_\pm\left( [ \tilde{h}^{-1}\delta \tilde{h}, U_{\mp }^{T}(X) ] \right) - U_{\mp }^T\left( [h^{-1}\delta h , U_\pm(X)] \right) ~,
\end{equation}
but is otherwise straightforward.
The outcome is that the equations of motion can be written as
\begin{equation}\begin{aligned}\label{eq:reom}
& -\frac{\mu^A \alpha^{A'}}{\langle \alpha \beta \rangle} \partial_{AA'} \widehat{b} +
\frac{\hat\mu^A \alpha^{A'}}{\langle \alpha \beta \rangle} \partial_{AA'} b + [\hj , b ] - [j , \widehat{b}] - [ \widehat{b}, b] = 0 ~, \\
& -\frac{\mu^A \tilde\alpha^{A'}}{\langle \tilde{\alpha} \beta \rangle} \partial_{AA'} \widehat{\widetilde{b}} +
\frac{\hat\mu^A \tilde\alpha^{A'}}{\langle \tilde{\alpha} \beta \rangle} \partial_{AA'} \widetilde{b} + [\htj , \widetilde{b} ] - [\tj , \widehat{\widetilde{b}}] - [ \widehat{\widetilde{b}}, \widetilde{b}] = 0~,
\end{aligned}\end{equation}
in which we invoke the definitions of $b$, $\widetilde{b}$, $\widehat{b}$ and $\widehat{\widetilde{b}}$ above in eqs.~\eqref{eq:bdef} and \eqref{eq:bdef2}.
These equations can be written in terms of the current $B_A$ as
\begin{equation} \label{eq:SDYM1}
\begin{aligned}
\alpha^{A^\prime} \pd_{A A^\prime} B^A + \frac{1}{2} \langle \alpha \beta \rangle \comm{B_A}{B^A} & = 0 ~, \\
\tilde{\alpha}^{A^\prime} \pd_{A A^\prime} B^A + \frac{1}{2} \langle \tilde{\alpha} \beta \rangle \comm{B_A}{B^A} & = 0 ~.
\end{aligned}
\end{equation}
Taking a weighted sum and difference equations gives
\begin{equation} \label{eq:SDYM2}
\begin{gathered}
\big( \langle \tilde{\alpha} \beta \rangle \alpha^{A^\prime} - \langle \alpha \beta \rangle \tilde{\alpha}^{A^\prime} \big) \pd_{A A^\prime} B^A = -\langle \alpha \tilde\alpha \rangle \beta^{A'} \partial_{AA'}B^A = 0 ~, \\
\big( \langle \tilde{\alpha} \beta \rangle \alpha^{A^\prime} + \langle \alpha \beta \rangle \tilde{\alpha}^{A^\prime} \big) \pd_{A A^\prime} B^A + \langle \tilde{\alpha} \beta \rangle \langle \alpha \beta \rangle \comm{B_A}{B^A} = 0 ~.
\end{gathered}
\end{equation}
the first of which is the anticipated conservation equation for the $\ell$-symmetry.
Making use of the definitions of $C$ and $\widetilde{C}$ in \eqref{eq:Cdef}, the equations of motion are equivalently expressed as
\begin{equation}\begin{split}\label{eq:Ceqm} \alpha^{A'}\partial_{AA'}C^A + \frac{1}{2}\langle\alpha\beta\rangle [C_A,C^A] = 0 ~,
\\ \tilde\alpha^{A'}\partial_{AA'}\widetilde{C}^A + \frac{1}{2}\langle\tilde\alpha\beta\rangle [\widetilde{C}_A,\widetilde{C}^A] = 0 ~.
\end{split}\end{equation}
Noting that $[C_A,C^A] = [\widetilde{C}_A,\widetilde{C}^A]$ we can take the difference of these equations to obtain
\begin{equation}
\label{eq:Cconservation}
\frac{1}{\langle\alpha\beta\rangle} \alpha^{A'}\partial_{AA'}C^A - \frac{1}{\langle\tilde\alpha\beta\rangle} \tilde\alpha^{A'}\partial_{AA'}\widetilde{C}^A = 0 ~,
\end{equation}
which is the anticipated conservation law for the $r$-symmetry.

\paragraph{ASDYM.} We will now justify the claim that this theory is integrable by constructing explicit Lax pair formulations of the dynamics in two different fashions.
First we will show the equations of motion \eqref{eq:SDYM1} can be recast as the anti-self-dual equation for a Yang-Mills connection.
Before demonstrating that this holds for our particular model, let us highlight that this follows from the construction of hCS$_6$ by briefly reviewing the Penrose-Ward correspondence \cite{Ward:1977ta}.
Recalling that we have resolved one of the hCS$_6$ equations of motion $\cF^\prime_{0A} = 0$ to find $\cA^\prime_A = \pi^{A^\prime} A_{A A^\prime}$, it follows that the remaining system of equations should be equivalent to the vanishing of the other components of the field strength $\cF^\prime_{AB} = 0$.
We may express this in terms of the anti-holomorphic covariant derivative $\bar{D}^\prime_A = \bar{\pd}_A + \cA^\prime_A$ as $\comm{\bar{D}^\prime_A}{\bar{D}^\prime_B} = 0$, which may also be written as
\begin{equation}\label{eq:contraction}
\pi^{A^\prime} \pi^{B^\prime} \comm{D_{A A^\prime}}{D_{B B^\prime}} = 0 ~.
\end{equation}
This is equivalent to the vanishing of $\pi^{A^\prime} \pi^{B^\prime} F_{A A^\prime B B^\prime}$ where $F$ is the field strength of the 4d connection $A_{A A^\prime}$.
To make contact with the anti-self-dual Yang-Mills equation, note that an arbitrary tensor that is anti-symmetric in Lorentz indices, e.g.~$F_{\mu \nu}$, can be expanded in spinor indices as
\begin{equation}
F_{A A^\prime B B^\prime} = \varepsilon_{AB} \, \phi_{A^\prime B^\prime} + \varepsilon_{A^\prime B^\prime} \, \tilde{\phi}_{AB} ~.
\end{equation}
Here, $\phi$ and $\tilde{\phi}$ are both symmetric and correspond to the self-dual and anti-self-dual components of the field strength respectively.
Explicitly computing the contraction~\eqref{eq:contraction}, we find that the remaining equation is simply $\phi = 0$ which is indeed the anti-self-dual Yang-Mills equation.
In effect, this argument demonstrates that a holomorphic gauge field on twistor space (which is gauge-trivial in $\mathbb{CP}^1$) is in bijection with a solution to the 4-dimensional anti-self-dual Yang-Mills equation -- this statement is the Penrose-Ward correspondence.

Now, returning to the case at hand, recall that our connection is of the form $A_{A A^\prime} = \beta_{A^\prime} B_A$, so the anti-self-dual Yang-Mills equation specialises to
\begin{equation}\label{eq:sadymsp}
\langle \pi \beta \rangle \big( \pi^{A^\prime} \pd_{A A^\prime} B_B - \pi^{B^\prime} \pd_{B B^\prime} B_A + \langle \pi \beta \rangle \comm{B_A}{B_B} \big) = 0 ~.
\end{equation}
This should hold for any $\pi^{A^\prime} \in \mathbb{CP}^1$ and we can extract the key information by expanding $\pi^{A^\prime}$ in the basis formed by $\alpha^{A^\prime}$ and $\tilde{\alpha}^{A^\prime}$, that is
\begin{equation}
\pi^{A^\prime} = \frac{1}{\langle \alpha \tilde{\alpha} \rangle} \Big( \langle \pi \tilde{\alpha} \rangle \alpha^{A^\prime} - \langle \pi \alpha \rangle \tilde{\alpha}^{A^\prime} \Big) ~.
\end{equation}
Substituting into~\eqref{eq:sadymsp}, we find two independent equations with $\mathbb{CP}^1$-dependent coefficients $\langle \pi \beta \rangle \langle \pi \tilde{\alpha} \rangle$ and $\langle \pi \beta \rangle \langle \pi \alpha \rangle$ respectively.
These are explicitly given by the two equations in eq.~\eqref{eq:SDYM1}.
Therefore, as expected, the equations of motion for this IFT$_4$ are equivalent to the anti-self-dual Yang-Mills equation for $A_{A A^\prime} = \beta_{A^\prime} B_A$.

Let us comment on the relation to Ward's conjecture which postulates that many
\unskip\footnote{The original conjecture \cite{Ward:1985} states that ``many (and perhaps all?)'' integrable models arise in this manner.
However, a notable absentee of this proposal is the Kadomtsev-Petviashvilii (KP) equation, see \cite{Mason17} for a discussion.}
integrable models arise as reductions of the ASDYM equations.
It is clear that that the equations of motion for the $\lambda$-deformations explored in this paper arise as symmetry reductions of the ASDYM equations for the explicit form of the connection given above.
On the other hand, a generic ASDYM connection can be partially gauge-fixed such that the remaining degrees of freedom are completely captured by the so-called Yang's matrix, which is the fundamental field of the WZW$_{4}$ model.
In this case, the equations of motion of the WZW$_{4}$ model, known as Yang's equations, are the remaining ASDYM equations.
It is natural to ask whether a generic ASDYM connection can also be partially gauge-fixed to take the explicit form found in this paper.
This would provide a 1-to-1 correspondence between solutions of the ASDYM equations and solutions to our 4d IFT.

\paragraph{B-Lax.}
The anti-self-dual Yang-Mills equation is also `integrable' in the sense that it admits a Lax formalism.
Using the basis of spinors $\mu^A$ and $\hat{\mu}^A$, we define the Lax pair $L$ and $M$ by
\begin{equation} \label{eq:4dLax}
L^{(B)} =\langle \pi \hat{\gamma} \rangle^{-1} \hat{\mu}^A \pi^{A^\prime} D_{A A^\prime} ~, \qquad
M^{(B)} = \langle \pi \gamma \rangle^{-1} \mu^A \pi^{A^\prime} D_{A A^\prime} ~,
\end{equation}
where the normalisations are for later convenience.
\unskip\footnote{The constant spinors $\gamma$ and $\hat\gamma$ appear in the symmetry reduction and will be introduced in \S~\ref{sec:4dhcsred}.}
It is important to emphasise that here $\pi^{A'}$ is not just an {\em ad hoc} spectral parameter.
It is introduced directly as a result of the hCS$_6$ equations of motion and is the coordinate on $\mathbb{CP}^1 \hookrightarrow \bb{PT}$.
The vanishing of $\comm{L^{(B)} }{M^{(B)} } = 0$ for any $\pi^{A^\prime} \in \mathbb{CP}^1$ is equivalent to the anti-self-dual Yang-Mills equation.

\paragraph{C-Lax.}
Let us now turn to the equations of motion cast in terms of the $C$-currents in eq.~\eqref{eq:Ceqm}.
Evidently, looking at the definition of these currents eq.~\eqref{eq:Cdef}, we see that when $\sigma = 1$ we have $\widetilde{C} = C$ and the equations of motion have the same form as the $B$-current equations \eqref{eq:SDYM1}.
Therefore, when $\sigma=1$, we can package the $C$-equations in terms of a ASDYM connection $A^{(C)} _{A A'} = \beta_{A'} C_A$.
Away from this point, when $\widetilde{C} \neq C$ it is not immediately evident if these equations follow from an ASDYM connection.
Regardless, we can still give these equations a Lax pair presentation as follows.

Letting $\varrho \in \mathbb{C}$ denote a spectral parameter we define
\begin{equation}\begin{aligned}
L^{(C)} = \frac{1}{n_L} \hat{\mu}^A \left( \frac{\alpha^{A'}}{ \langle \alpha \beta \rangle} (1+\varrho) + \frac{\sigma^{-1} \tilde{\alpha}^{A'}}{ \langle \tilde\alpha \beta \rangle} (1-\varrho)\right) \partial_{A A'} + \frac{1}{n_L} \hat{\mu}^A C_A ~, \\
M^{(C)} = \frac{1}{n_M} \mu^A \left( \frac{\alpha^{A'}}{ \langle \alpha \beta \rangle} (1+\varrho) + \frac{\sigma \tilde{\alpha}^{A'}}{ \langle \tilde\alpha \beta \rangle} (1-\varrho)\right) \partial_{A A'} + \frac{1}{n_M} \mu^AC_A ~.
\label{eq:other4dLax}
\end{aligned}
\end{equation}
Noting that $\mu^B \widetilde{C}_B = \sigma^{-1} \mu^B C_B$ and $\hat\mu^B \widetilde{C}_B = \sigma \hat\mu^B C_B$ one immediately sees that the terms inside $[L^{(C)}, M^{(C)} ]$ linear in $\varrho$ yield the conservation law eq.~\eqref{eq:Cconservation}.
The contributions independent of $\varrho$, combined with eq.~\eqref{eq:Cconservation}, give either of eq.~\eqref{eq:Ceqm}.
It will be convenient to fix the overall normalisation of these Lax operators to be
\begin{align}
n_L = \frac{\langle\alpha\hat\gamma\rangle }{\langle \alpha \beta \rangle } (1+\varrho) + \frac{\langle\tilde \alpha\hat\gamma\rangle }{\langle \tilde \alpha \beta \rangle }\sigma^{-1} (1- \varrho) ~, \qquad n_M = \frac{\langle\alpha \gamma\rangle }{\langle \alpha \beta \rangle } (1+\varrho) + \frac{\langle\tilde \alpha\gamma\rangle }{\langle \tilde \alpha \beta \rangle }\sigma (1- \varrho) ~.
\end{align}

Unlike the spectral parameter $\pi_{A^\prime}$ entering the $B$-Lax, there is no clear way to associate the spectral parameter of the $C$-Lax, $\varrho$, with the $\mathbb{CP}^1$ coordinate alone.
Indeed, under a natural assumption, we will see that $\varrho$ has origins from both the $\mathbb{CP}^1$ geometry {\em and} the parameters that enter the boundary conditions.

The existence of a second Lax formulation of the theory, distinct from the ASDYM equations encoded via the $B$-Lax, is an unexpected feature.
We will see shortly that, upon symmetry reduction, this twin Lax formulation is inherited by the IFT$_{2}$.

\subsection{Reality Conditions and Parameters}
\def\zsf{\mathsf{z}}
\def\hzsf{\hat{\mathsf{z}}}
\def\wsf{\mathsf{w}}
\def\hwsf{\hat{\mathsf{w}}}

To understand how the reality of the action of the IFT$_4$~\eqref{eq:IFT4} with~\eqref{eq:wz4}, as well as the dependence on the parameters $K$, $\sigma$, $\alpha_{A'}$, $\tilde\alpha_{A'}$, $\beta_{A'}$, $\mu^A$ and $\hat\mu^A$, let us denote our coordinates
\begin{equation}\begin{aligned}\label{eq:newcoordinates}
\wsf & = \frac{\langle\alpha\beta\rangle}{\langle\alpha\tilde\alpha\rangle[\mu\hat\mu]}\hat\mu_A \tilde \alpha_{A'} x^{A A'} ~,
\qquad &
\hwsf & = -\frac{\langle\alpha\beta\rangle}{\langle\alpha\tilde\alpha\rangle[\mu\hat\mu]} \mu_A \tilde \alpha_{A'} x^{A A'} ~,
\\
\zsf & = -\frac{\langle\tilde\alpha\beta\rangle}{\langle\alpha\tilde\alpha\rangle[\mu\hat\mu]} \hat\mu_A \alpha_{A'} x^{A A'} ~, \qquad &
\hzsf & = \frac{\langle\tilde\alpha\beta\rangle}{\langle\alpha\tilde\alpha\rangle[\mu\hat\mu]} \mu_A \alpha_{A'} x^{A A'} ~,
\end{aligned}\end{equation}
such that
\begin{equation}
j = h^{-1}\partial_{\wsf} h ~, \qquad
\hj = h^{-1} \partial_{\hwsf} h ~, \qquad
\tj = \tilde{h}^{-1} \partial_{\zsf} \tilde{h} ~, \qquad
\htj = \tilde{h}^{-1} \partial_{\hzsf} \tilde{h} ~.
\end{equation}
In this subsection we let $\mu^A$ and $\hat\mu^A$ be an unconstrained basis of spinors, i.e.,~not related by Euclidean conjugation or of fixed norm.
This means the action~\eqref{eq:IFT4} with~\eqref{eq:wz4} comes with an extra factor of $[\mu\hat\mu]^{-1}$.
Writing the volume element $\vol_4 = \frac1{12} \varepsilon_{AB}\varepsilon_{CD}\varepsilon_{A'C'}\varepsilon_{B'D'} dx^{AA'} \wedge dx^{BB'} \wedge dx^{CC'}\wedge dx^{DD'}$ in terms of the coordinates $\{\wsf,\hwsf,\zsf,\hzsf\}$ we find
\begin{equation}
\vol_4
= \frac{\langle \alpha \tilde\alpha\rangle^2[\mu\hat\mu]^2}{\langle \alpha \beta\rangle^2\langle \tilde\alpha \beta\rangle^2} \dr \wsf \wedge \dr \hwsf \wedge \dr \zsf \wedge \dr \hzsf
= \frac{\langle \alpha \tilde\alpha\rangle^2[\mu\hat\mu]^2}{\langle \alpha \beta\rangle^2\langle \tilde\alpha \beta\rangle^2} \vol_4' ~.
\end{equation}
Substituting into the action~\eqref{eq:IFT4} with \eqref{eq:wz4}, we see that the IFT$_4$ now only depends explicitly on two parameters
\begin{equation}\label{eq:kprime}
K' =\frac{\langle \alpha \tilde\alpha\rangle[\mu\hat\mu]}{\langle\alpha\beta\rangle^2 \langle\tilde\alpha\beta\rangle^2} K \quad \textrm{and} \quad \sigma ~.
\end{equation}
Moreover, the action is invariant under the following $GL(1;\mathbb{C})$ space-time symmetry
\begin{equation}\label{eq:spacetimesym}
\zsf \to e^{\vartheta} \zsf ~, \qquad
\wsf \to e^{\vartheta} \wsf ~, \qquad
\hzsf \to e^{-\vartheta} \hzsf ~, \qquad
\hwsf \to e^{-\vartheta} \hwsf ~,
\end{equation}
where $\vartheta\in\mathbb{C}$.

To find a real action we should impose reality conditions on the coordinates $\{\wsf,\hwsf,\zsf,\hzsf\}$, the fields $h$ and $\tilde{h}$, and the parameters $K'$ and $\sigma$.
We start by observing four sets of admissible reality conditions simply found by inspection of the 4d IFT.
Note that, implicitly, we will not assume Euclidean reality conditions for $x^{AA'}$.
Starting from Euclidean reality conditions we complexify and take different split signature real slices.
We will then turn to the hCS$_6$ origin of the different reality conditions.

Introducing $\Theta$, the lift of an antilinear involutive automorphism $\theta$ of the Lie algebra $\mathfrak{g}$ to the group $G$, the four sets of reality conditions are as follows:
\begin{enumerate}
\item In the first case, the coordinates are all real, $\bar \wsf = \wsf$, $\bar{\hwsf} = \hwsf$, $\bar \zsf = \zsf$, $\bar{\hzsf} = \hzsf$; $K'$ and $\sigma$ are real; and the group-valued fields are elements of the real form, $\Theta(h) = h$ and $\Theta(\tilde h) = \tilde h$.
Under conjugation we have $U_\pm \to U_\pm$.
\item In the second case, the coordinates conjugate as $\bar \wsf = \hwsf$, $\bar \zsf = \hzsf$; $K'$ is imaginary and $\sigma$ is a phase factor; and the group-valued fields are elements of the real form, $\Theta(h) = h$ and $\Theta(\tilde h) = \tilde h$.
Under conjugation we have $U_\pm \to U_\mp$.
\item In the third case, the coordinates conjugate as $\bar \wsf = \hzsf$, $\bar{\zsf} = \hwsf$; $K'$ and $\sigma$ are real; and the group-valued fields are related by conjugation $\Theta(h) = \tilde h$.
Under conjugation we have $U_\pm \to U_\pm^T$.
\item In the final case, the coordinates conjugate as $\bar \wsf = \zsf$, $\bar{\hwsf} = \hzsf$; $K'$ is imaginary and $\sigma$ is a phase factor; and the group-valued fields are related by conjugation $\Theta(h) = \tilde h$.
Under conjugation we have $U_\pm \to U_\mp^T$.
\end{enumerate}
The action~\eqref{eq:IFT4} with \eqref{eq:wz4} is real for each of these sets of reality conditions.
To determine the signature for each set of reality conditions, we note that
\unskip\footnote{Conjugating in Euclidean signature we find the reality conditions
$$\bar \wsf = \frac{\langle\hat\alpha\hat\beta\rangle}{\langle\hat\alpha \hat{\tilde\alpha}\rangle}\left(\frac{\langle \alpha\hat{\tilde\alpha}\rangle}{\langle \alpha\beta\rangle} \hwsf
+ \frac{\langle \tilde\alpha\hat{\tilde\alpha}\rangle}{\langle \tilde\alpha\beta\rangle} \hzsf\right) ~, \qquad
\bar \zsf = \frac{\langle\hat{\tilde\alpha}\hat\beta\rangle}{\langle \hat{\alpha}\hat{\tilde\alpha}\rangle}\left(\frac{\langle \hat{\alpha}\tilde\alpha\rangle}{\langle \tilde\alpha\beta\rangle} \hzsf - \frac{\langle \alpha\hat{\alpha}\rangle}{\langle \alpha\beta\rangle} \hwsf\right)~.$$
As an example, let us take $\hat\alpha = \tilde\alpha$, in which case the reality conditions simplify to $\bar \wsf =  \frac{\langle\hat\alpha\hat\beta\rangle}{\langle \hat\alpha\beta\rangle}\hzsf$ and $\bar \zsf =  \frac{\langle\alpha\hat\beta\rangle}{\langle \alpha\beta\rangle} \hwsf$.
Substituting into the metric we find $\frac{2\langle\alpha\hat\alpha\rangle[\mu\hat\mu]}{\langle\alpha\beta\rangle\langle\hat\alpha\hat\beta\rangle}\left(\dr\wsf \dr\bar\wsf + \psi\bar\psi \dr\zsf \dr\bar\zsf\right)$ where $\psi = \frac{\langle\alpha\beta\rangle}{\langle\alpha\hat\beta\rangle}$.
Since the prefactor is real and positive, this indeed has Euclidean signature.
Note that these reality conditions are distinct from case 3 above, and they do not imply reality of the 4d IFT.}
\begin{equation}
\varepsilon_{AB}\varepsilon_{A'B'}\dr x^{AA'}\dr x^{BB'} = \frac{2\langle\alpha\tilde\alpha\rangle[\mu\hat\mu]}{\langle\alpha\beta\rangle\langle\tilde\alpha\beta\rangle} (\dr\wsf \dr\hzsf - \dr\zsf \dr\hwsf) ~,
\end{equation}
It is then straightforward to see that the four sets of reality conditions above all correspond to split signature.
Note that for the metric to be real, we require the prefactor to be real in cases 1 and 4 and imaginary in cases 2 and 3.
We will see that this is indeed the case when we comment on the hCS$_6$ origin.

In cases 1 and 4 the reality conditions are preserved by an $SO(1,1)$ space-time symmetry~\eqref{eq:spacetimesym} with $\vartheta \in \mathbb{R}$.
On the other hand, in cases 2 and 3, the reality conditions are preserved by an $SO(2)$ space-time symmetry with $|\vartheta| = 1$.
In \S~\ref{sec:4dIFTto2dIFT}, we will be interested in symmetry reducing while preserving the space-time symmetry, recovering an action on $\mathbb{R}^2$ or $\mathbb{R}^{1,1}$ that is invariant under the Euclidean or Poincar\'e groups respectively.
We have freedom in how we do this since the action is not invariant under $SO(2)$ rotations acting on $(\zsf, \wsf)$ and $(\hzsf, \hwsf)$.
Therefore, we can choose symmetry reduce along different directions in each of these planes, in principle introducing an additional two parameters.
We should note that in the Euclidean case, since the two planes are related by conjugation, we will break the reality properties of the action unless the two symmetry reduction directions are also related by conjugation, reducing the number of parameters by one for a real action.
This is not an issue in the Lorentzian case since the coordinates are real, hence we expect to find a four-parameter real Lorentz-invariant IFT$_2$.
We will indeed see that this is the case in \S~\ref{sec:4dIFTto2dIFT}.

\paragraph{Origin of reality conditions from hCS$_6$.}

Let us now briefly describe the origin of the different sets of reality conditions from 6 dimensions.
It is shown in~\cite{Bittleston:2020hfv} that for the hCS$_6$ action to be real we require that
\begin{equation}
\overline{C(\Phi)} = C(\Phi) ~,
\end{equation}
where $\Phi$ is defined in eq.~\eqref{eq:Omegaintdef} and $C$ is a conjugation that acts on the coordinates $(x,\pi)$, not on the fixed spinors $\alpha$, $\tilde\alpha$ and $\beta$.
\unskip\footnote{Conjugation in Euclidean signature can be defined as $C(\mu_A) = \hat\mu_A = \epsilon_A{}^B \bar \mu_B$, $C(\gamma_A') = \hat \gamma_{A'} = \varepsilon_{A'}{}^{B'} \bar \gamma_{B'}$ and $C(x^{AA'}) = (\epsilon^T)^A{}_B \bar x^{BB'} \varepsilon_{B'}{}^{A'}$ with $\varepsilon_{1}{}^{2} = -1$, while in split signature, we take $C(\mu_A) = \bar \mu_A$, $C(\gamma_{A'}) = \bar \gamma_{A'}$ and $C(x^{AA'}) = \bar x^{AA'}$.
We will restrict our attention to Euclidean and split signatures since there are no ASD connections in Lorentzian signature~\cite{Bittleston:2020hfv}.}
In Euclidean signature this constraint becomes
\begin{equation}
\frac{\bar K}{\langle \pi\hat{\alpha} \rangle \langle \pi\hat{\tilde\alpha} \rangle \langle \pi\hat\beta \rangle^2 }
=
\frac{K}{\langle \pi\alpha \rangle \langle \pi\tilde\alpha \rangle \langle \pi \beta\rangle^2 } ~.
\end{equation}
We immediately see that this has no solutions since the double pole at $\beta$ is mapped to $\hat\beta$ and $\hat\beta = \beta$ has no solutions.
On the other hand, in split signature we have
\begin{equation}
\frac{\bar K}{\langle \pi\bar{\alpha} \rangle \langle \pi\bar{\tilde\alpha} \rangle \langle \pi\bar\beta \rangle^2 }
=
\frac{K}{\langle \pi\alpha \rangle \langle \pi\tilde\alpha \rangle \langle \pi \beta\rangle^2 } ~.
\end{equation}
This can be solved by taking $K$ and $\beta$ to be real, and $\alpha$ and $\tilde \alpha$ to either both be real or to form a complex conjugate pair.

We also need to ask that the boundary conditions~\eqref{eq:boundarycondition} are compatible with the reality conditions.
Imposing $C^* (\mathcal{A}_A) = \theta(\mathcal{A}_A)$, we can either take $\mu$ and $\hat\mu$ to either both be real or to form a complex conjugate pair.
The two choices of reality conditions for $(\alpha,\tilde\alpha)$ and the two for $(\mu,\hat\mu)$ give a total of four sets of reality conditions, which we anticipate will recover those in the list presented above.
With the same ordering, we have the following:
\begin{enumerate}
\item In the first case, we take real $(\alpha,\tilde \alpha)$ and real $(\mu,\hat\mu)$.
Analysing the boundary conditions we find that $\mathcal{A}_A$ is valued in the real form at the poles, implying that $h$ and $\tilde{h}$ are as well, and that $\sigma$ is real.
Since both $\langle \alpha\tilde\alpha \rangle$ and $[\mu\hat\mu]$ are real, real $K$ implies that $K'$ is real using eq.~\eqref{eq:kprime}.
\item In the second case, we take real $(\alpha,\tilde \alpha)$ and complex conjugate $(\mu,\hat\mu)$.
Analysing the boundary conditions we find that $\mathcal{A}_A$ is valued in the real form at the poles, implying that $h$ and $\tilde{h}$ are as well, and that $\sigma$ is a phase factor.
Since $\langle \alpha\tilde\alpha \rangle$ is real and $[\mu\hat\mu]$ is imaginary, real $K$ implies that $K'$ is imaginary using eq.~\eqref{eq:kprime}.
\item In the third case, we take complex conjugate $(\alpha,\tilde \alpha)$ and complex conjugate $(\mu,\hat\mu)$.
Analysing the boundary conditions we find that $\mathcal{A}_A$ at $\alpha$ is the conjugate of $\mathcal{A}_A$ at $\tilde\alpha$, implying that $h$ and $\tilde{h}$ are also conjugate, and that $\sigma$ is real.
Since both $\langle \alpha\tilde\alpha \rangle$ and $[\mu\hat\mu]$ are imaginary, real $K$ implies that $K'$ is real using eq.~\eqref{eq:kprime}.
\item In the final case, we take complex conjugate $(\alpha,\tilde \alpha)$ and real $(\mu,\hat\mu)$.
Analysing the boundary conditions we find that $\mathcal{A}_A$ at $\alpha$ is the conjugate of $\mathcal{A}$ at $\tilde\alpha$, implying that $h$ and $\tilde{h}$ are also conjugate, and that $\sigma$ is a phase factor.
Since $\langle \alpha\tilde\alpha \rangle$ is imaginary and $[\mu\hat\mu]$ is real, real $K$ implies that $K'$ is imaginary using eq.~\eqref{eq:kprime}.
\end{enumerate}
Finally, one can also check that in split signature, the different reality conditions for $(\alpha,\tilde\alpha)$ and $(\mu,\hat\mu)$ imply the different reality conditions for the coordinates $\{\wsf,\hwsf,\zsf,\hzsf\}$ given above.

As implied above, see also~\cite{Bittleston:2020hfv}, a real action in split signature in 4 dimensions is useful for symmetry reducing and constructing real 2d IFTs since both Euclidean and Lorentzian signature in 2 dimensions can be accessed. However, the lack of a real action in Euclidean signature raises questions about the quantisation of the IFT$_4$ itself. We will briefly return to the issue of quantisation in \S~\ref{sec:outlookintdef}.

\subsection{Equivalent Forms of the Action and its Limits}
In this section we describe alternative, but equivalent ways of writing the action of the 4d IFT~\eqref{eq:IFT4} with~\eqref{eq:wz4}, and consider two interesting limits of the theory.
These constructions are motivated by analogous ones that are important in the context of the 2d $\lambda$-deformed WZW model.

First, let us note that the IFT$_4$~\eqref{eq:IFT4} with~\eqref{eq:wz4} can be written in the following two equivalent forms
\begin{equation}\begin{split}
S_{\mathrm{IFT}_4} & = K' \int_{\mathbb{E}^4 }\vol'_4 \, \mathrm{Tr} \big( ( j - \sigma \tj) ( U_{+}^T - U_- ) ( \hj + \sigma^{-1} \htj) - \sigma \tj \hj + \sigma^{-1} j \, \htj \big)+ S_{\mathrm{WZ}_4}
\\
& \! = K' \int_{\mathbb{E}^4 }\vol'_4 \, \mathrm{Tr} \big( (\mathrm{Ad}_h j - \mathrm{Ad}_{\tilde h} \tj) ( \widetilde U_{+}^T - \widetilde U_- ) (\mathrm{Ad}_h \hj - \mathrm{Ad}_{\tilde{h}} \htj) + \mathrm{Ad}_{\tilde h}\tj \, \mathrm{Ad}_h \hj - \mathrm{Ad}_h j \,\mathrm{Ad}_{\tilde h} \htj \big)+ S_{\mathrm{WZ}_4} ~,
\end{split}\end{equation}
where
\begin{equation}\begin{gathered}\label{eq:tildeulam}
U_\pm = \big(1 - \sigma^{\pm1} \Lambda\big)^{-1} ~, \qquad \Lambda = \mathrm{Ad}_{\tilde{h}}^{-1} \mathrm{Ad}_h ~,
\\
\widetilde U_\pm = \big(1 - \sigma^{\pm1} \widetilde \Lambda\big)^{-1} ~, \qquad \widetilde \Lambda = \mathrm{Ad}_h \mathrm{Ad}_{\tilde{h}}^{-1} ~.
\end{gathered}\end{equation}
Written in this way, it is straightforward to see that the symmetries of the 4d IFT are given by transformations of the form~\eqref{eq:ift4symmetries} with
\begin{equation}\begin{split}
(\partial_{\wsf} - \sigma \partial_{\zsf}) r = (\partial_{\hwsf} - \sigma^{-1} \partial_{\hzsf}) r = 0 ~, &
\qquad
(\partial_{\wsf} - \partial_{\zsf}) \ell = (\partial_{\hwsf} - \partial_{\hzsf})\ell = 0 ~,
\end{split}\end{equation}
which, as expected, coincide with~\eqref{eq:lconstraints} and~\eqref{eq:rconstraints} upon using the definitions~\eqref{eq:newcoordinates}.

We can also introduce auxiliary fields $B^A$, $C^A$ and $\widetilde{C}^A$ to rewrite the action as
\begin{equation}\begin{split}\label{eq:gwzwform}
S_{\mathrm{IFT}_4} = K'
\int_{\mathrm{E}_4} \vol_4' \,\mathrm{Tr}\big( & j \hj - 2j \mathrm{Ad}_h^{-1}[B\hat\mu] +2 \hj [C\mu] - 2 [C\mu] \mathrm{Ad}_h^{-1}[B\hat\mu]
\\ & \vphantom{\int} + \tj \htj - 2\htj \mathrm{Ad}_{\tilde h}^{-1} [B\mu]+ 2\tj [\widetilde{C}\hat\mu] -2 [\widetilde{C}\hat\mu]\mathrm{Ad}_{\tilde h}^{-1} [B\mu]
\\ & + 2 [B\mu] [B\hat\mu] + 2\sigma^{-1} [C\mu] [\widetilde{C}\hat\mu] \big) + S_{\mathrm{WZ}_4} ~.
\end{split}
\end{equation}
Here we take the auxiliary fields $B^A$, $C^A$ and $\widetilde{C}^A$ to be undetermined.
Varying the action and solving their equations of motion, we find that on-shell, they are given by the expressions introduced above in eqs.~\eqref{eq:bdef} and~\eqref{eq:Ctdef}.
Moreover, substituting their on-shell values back into~\eqref{eq:gwzwform} we recover the 4d IFT.
Using the symmetry~\eqref{z21}, we note that the action can also be written in a similar equivalent form, in which tilded and untilded quantities are swapped, $\sigma \to \sigma^{-1}$, $K' \to - K'$ and $\widetilde{B} = B$.
This can also be seen by making the off-shell replacements $[B\mu] \to [B\mu]$, $[B\hat\mu] \to \mathrm{Ad}_{h} ([C\hat\mu] + \hj)$, $[C\mu]\to \sigma [\widetilde C\mu]$ and $[\widetilde C\hat \mu] \to \mathrm{Ad}_{\tilde h}^{-1}[B\hat\mu] - \htj$, all of which are compatible with the on-shell values of the auxiliary fields.

\medskip

The first limit we consider is $\sigma \to 0$, in which the action becomes
\begin{equation} \label{eq:IFT4s0}
S_{\mathrm{IFT}_4} \vert_{\sigma\rightarrow 0}= \mathring{S}_{\mathrm{IFT}_4} = K'\int_{\mathbb{E}^4 }\vol'_4 \, \mathrm{Tr} \big( j \hj + \tj \htj - 2 \mathrm{Ad}_h j \, \mathrm{Ad}_{\tilde{h}} \htj\big) + S_{\mathrm{WZ}_4} ~.
\end{equation}
This has the form of a current-current coupling between two building blocks that could be described as `holomorphic WZW$_4$' of the form
\begin{equation}\label{eq:hWZW4}
S_{\mathrm{hWZW}4}[h ,\alpha ] = \int_{\mathbb{E}^4} \vol_4' \, \mathrm{Tr} \big( j \hj \big) - \int_{\mathbb{E}^4\times [0,1]} \vol_4'\wedge d\rho\, \mathrm{Tr} \big( h^{-1} \partial_\rho h [j, \hj]\big) ~.
\end{equation}
This somewhat unusual theory has derivatives only in the holomorphic two-plane singled out by the complex structure on $\bE^4$ defined by $\alpha$ (i.e.~only $\partial_{\wsf}$  and $\partial_{\hwsf}$ enter), although the field depends on all coordinates of $\mathbb{E}^4$.
This structure is quite different (both in the kinetic term and Wess-Zumino term) from the conventional WZW$_{4}$~\cite{Donaldson:1985zz} for which the action
\unskip\footnote{This is also the 4d IFT that was found in \cite{Bittleston:2020hfv,Penna:2020uky} from hCS$_6$ with two double poles at $\pi = \alpha$ and $\pi = \beta$, with Dirichlet boundary conditions.} is
\begin{equation}\begin{split}
S_{\mathrm{WZW}_4}[h, \alpha, \beta ] & = \int_{\mathbb{E}^4} \mathrm{Tr}\big(h^{-1} dh \wedge \star h^{-1} dh \big) + \frac{1}{6} \int_{\mathbb{E}^4 \times [0,1]} \hspace{-1em}\hspace{-1em} \varpi_{\alpha,\beta}\wedge \mathrm{Tr}\big((h^{-1} dh )^3\big) ~,
\\
\varpi_{\alpha,\beta} & = \epsilon_{AB}\alpha_{A'}\beta_{B'}\dr x^{AA'} \wedge \dr x^{BB'} ~.
\end{split}\end{equation}
The K\"ahler point of the theory is achieved when $\beta =\hat{\alpha}$ such that $\varpi_{\alpha,\beta}$ is the K\"ahler form associated to the complex structure defined by $\alpha$.
In fact, the WZ term of our holomorphic WZW$_4$ is of this general form with $\beta = \alpha$ such that $ \varpi_{\alpha,\beta}$ defines a $(2,0)$-form.
However even in the $\beta = \alpha$ case, the kinetic term does not match that of the holomorphic WZW$_4$ action.

Returning to the holomorphic WZW$_4$, we can establish that the theory is invariant under a rather large set of symmetries.
Since only $\wsf$ and $\hwsf$ derivatives enter, it is immediate that the transformation $h \mapsto l( \hzsf, \zsf) h r(\hzsf ,\zsf)$ leaves the action eq.~\eqref{eq:hWZW4} invariant.
These are further enhanced, as in a WZW$_2$, to
\begin{equation}
(\ell , r): \quad h \mapsto \ell(\zsf,\hzsf, \wsf ) \hspace{0.1em} h \hspace{0.1em} r(\zsf,\hzsf, \hwsf) ~.
\end{equation}
From this perspective holomorphic WZW$_4$ can be considered the embedding of a WZW$_2$ in 4 dimensions.
Similarly, we have a symmetry for the holomorphic WZW$_4$ for $\tilde{h}$
\begin{equation}
(\tilde{\ell} , \tilde{r}):  \quad  \tilde{h} \mapsto \tilde{\ell}( \hzsf, \hwsf,\wsf ) \hspace{0.1em} \tilde{h} \hspace{0.1em} \tilde{r}(\zsf, \hwsf,\wsf) ~.
\end{equation}
The interaction term, $\mathrm{Ad}_h j \, \mathrm{Ad}_{\tilde{h}} \htj$, in the action~\eqref{eq:IFT4s0} preserves the right actions, but breaks the enhanced independent $\ell, \tilde{\ell}$ left actions.
Instead a new `diagonal' left action emerges
\begin{equation}
(\ell , r, \tilde{r}): \quad h \mapsto \ell(\zsf + \wsf, \hzsf + \hwsf ) \hspace{0.1em} h \hspace{0.1em} r(\zsf,\hzsf, \hwsf) ~, \quad
\tilde h \mapsto \ell(\zsf + \wsf, \hzsf + \hwsf ) \hspace{0.1em} \tilde h \hspace{0.1em} \tilde r(\zsf, \wsf,\hwsf) ~.
\end{equation}
It is important to emphasise that here the right actions on $h$ and $\tilde{h}$ are independent ($r$ and $\tilde{r}$ are not the same).
This stems from the enlargement of the residual symmetries of the 6-dimensional boundary conditions.
The constraints of eq.~\eqref{eq:rconstraints} are relaxed such that gauge parameters at different poles are unrelated but are chiral.

In this limit the currents associated to the left and right action become
\begin{equation}\begin{aligned}
B_A &\vert_{\sigma\rightarrow 0} = \mathring{B}_A = \mathrm{Ad}_{\tilde{h}} \htj \mu_A - \mathrm{Ad}_h j \hat{\mu}_A ~, \\
C_A &\vert_{\sigma\rightarrow 0} =\mathring{C}_A = - \big(\hj - \Lambda^{-1} \htj \big) \mu_A ~, \\
\widetilde{C}_A &\vert_{\sigma\rightarrow 0} =\mathring{\widetilde{C}}_A= \big( \tj - \Lambda j \big) \hat\mu_A ~.
\end{aligned}\end{equation}
The conservation laws become
\begin{equation}\begin{gathered}
\partial_{\wsf} (\hj - \Lambda^{-1} \htj) = 0 ~, \qquad
\partial_{\hzsf} (\tj - \Lambda j) = 0 ~,
\\
\partial_{\hwsf}(\mathrm{Ad}_h j) - \partial_{\wsf}(\mathrm{Ad}_{\tilde h}\htj) + \partial_{\zsf} (\mathrm{Ad}_{\tilde{h}}\htj) - \partial_{\hzsf} (\mathrm{Ad}_h j) = 0 ~.
\end{gathered}\end{equation}

To compute the $\mathcal{O}(\sigma)$ correction to the action~\eqref{eq:IFT4s0} we first note that
\begin{equation}
B_A = \mathring{B}_A + \sigma \left( \mathrm{Ad}_h \mathring{\widetilde{C}}_A + \mathrm{Ad}_{\tilde{h}}\mathring{C}_A \right) + \mathcal{O}(\sigma^2) ~,
\end{equation}
and that the combination $C_A - \Lambda^T \widetilde{C}_A = \mathring{C}_A - \Lambda^T \mathring{\widetilde{C}}_A $ is independent of $\sigma$.
Then from the expression of the IFT${}_4$ action in terms of currents \eqref{eq:actionIFT4currents}, we see that the leading correction to $\mathring{S}_{\mathrm{IFT}_4}$ is given by
\begin{align} \label{eq:perturbation}
2 \sigma K' \int_{\mathbb{E}^4} \vol'_4 \, \epsilon^{AB} \mathrm{Tr} \big( \mathring{ \widetilde{C}_A} \mathring{C}_B \big)
&=
-2\sigma K'\int_{\mathbb{E}^4} \vol_4 \, \mathrm{Tr}\big( (\tj - \Lambda j) (\hj - \Lambda^{-1} \htj) \big) ~,
\end{align}
i.e.~the perturbing operator is the product of two currents associated to the right-acting symmetries.

The second limit we consider is $\sigma \to 1$.
Recall that in this limit, we have that $\widetilde{C} = C$ from eqs.~\eqref{eq:Cdef} and~\eqref{eq:Ctdef}, and a symmetry emerges interchanging $B$ and $C$, as well as $h$ and $\tilde{h}^{-1}$.
This is also evident if we set $\sigma =1$ in~\eqref{eq:gwzwform}.
An alternative way to take $\sigma \to 1$ is to first set $h = \exp(\frac{\nu}{K'})$ and $\tilde{h} = \exp(\frac{\tilde\nu}{K'})$, along with $\sigma = e^{\frac{1}{K'}}$ and take $K' \to \infty$.
In this limit the 4d IFT becomes
\begin{equation}\begin{split}\label{eq:natd}
S_{\mathrm{IFT}_4}\vert_{K'\rightarrow \infty} = - \int_{\mathbb{E}^4 }\vol'_4 \, \mathrm{Tr} \big( (\partial_{\wsf} \nu - \partial_{\zsf}\tilde \nu) \frac{1}{ 1 - \operatorname{ad}_\nu + \operatorname{ad}_{\tilde\nu} } (\partial_{\hwsf} \nu - \partial_{\hzsf}\tilde \nu) \big) ~,
\end{split}\end{equation}
which is reminiscent of a 4d version of the non-abelian T-dual of the principal chiral model, albeit with two fields instead of one.
If instead we take the limit in the action with auxiliary fields~\eqref{eq:gwzwform}, also setting $[C\mu] = [B\mu] + \mathcal{O}(K'^{-1})$ and $[\widetilde{C}\hat\mu] = [B\hat\mu] + \mathcal{O}(K'^{-1})$, we find
\begin{equation}\begin{split}\label{eq:natdlim}
S_{\mathrm{IFT}_4}\vert_{K'\rightarrow \infty} = \int_{\mathrm{E}_4} \vol_4' \,\mathrm{Tr}\Big( & 2\nu\big(\partial_{\wsf} [B\hat\mu] - \partial_{\hat{\wsf}} [B\mu] + [[B\hat\mu],[B\mu]]\big)
\\ & \vphantom{\int} +2 \tilde\nu\big( \partial_{\hat{\zsf}} [B\mu]-\partial_{\zsf} [B\hat\mu] + [[B\mu],[B\hat\mu]]\big)-2[B\mu] [B\hat\mu] \Big) ~,
\end{split}
\end{equation}
after integrating by parts.
Integrating out the auxiliary field $B^A$, we recover the action~\eqref{eq:natd}.
It would be interesting to instead integrate out the fields $\nu$ and $\tilde{\nu}$ to give a 4d analogue of 2d non-abelian T-duality.
However, note that, unlike in 2 dimensions, $\nu$ and $\tilde{\nu}$ do not enforce the flatness of a 4d connection, hence there is no straightforward way to parametrise the general solution to their equations.

\section{Symmetry Reduction of \texorpdfstring{hCS$_6$}{hCS6} to \texorpdfstring{CS$_4$}{CS4}}
\label{sec:4dhcsred}

\begin{tcolorbox}
\begin{minipage}[c]{0.4\linewidth}

\begin{equation*}
\begin{tikzpicture}[scale=0.8]
\node at (0,2) {$\mathbf{hCS_6}$};
\node at (-2,0) {$\mathbf{CS_4}$};
\node at (2,0) {$\mathbf{IFT_4}$};
\node at (0,-2) {$\mathbf{IFT_2}$};
\draw[->,very thick, red, decorate, decoration={snake, segment length=13pt, amplitude=2pt}] (-0.4,1.6)--(-1.6,0.4);
\draw[->,very thick] (0.4,1.6)--(1.6,0.4);
\draw[->,very thick,decorate, decoration={snake, segment length=13pt, amplitude=2pt}] (1.6,-0.4)--(0.4,-1.6);
\draw[->,very thick] (-1.6,-0.4)--(-0.4,-1.6);
\end{tikzpicture}
\end{equation*}

\end{minipage}
\begin{minipage}[c]{0.59\linewidth}
Returning to hCS$_{6}$, we now descend via the top left-hand side of the diamond by performing a symmetry reduction of the action. Doing so, we find the resulting theory is CS$_4$.
\end{minipage}

\end{tcolorbox}

The idea of symmetry reduction is to take a truncation of a $d$-dimensional theory specified by a $d$-form Lagrangian ${\cal L}^d$ depending on a set of fields $\{\Phi\}$ to obtain a lower dimensional theory.
We assume here that we are reducing along two directions singled out by vector fields $V_i$, $i=1,2$.
The reduction procedure imposes that all fields are invariant,
$L_{V_i} \Phi = 0$, with dynamics now specified by the $d-2$-form Lagrangian
${\cal L}^{d-2}= V_1 \lrcorner V_2 \lrcorner{\cal L}^{d}$.
While similar in spirit to a dimensional reduction, there is no requirement that $V_i$ span a compact space, hence there is no scale separation in this truncation.

In order to perform the symmetry reduction, we will introduce a unit norm spinor $\gamma_{A'}$ about which we can expand any spinor $X_{A'}$ as
\begin{equation}
X_{A'} = \langle X \hat \gamma \rangle \gamma_{A'}- \langle X \gamma \rangle \hat{\gamma}_{A' } ~.
\end{equation}
The spinor $\gamma_{A^\prime}$ defines another complex structure on $\mathbb{E}^4$ which coincides with the complex structure on $\mathbb{E}^4 \subset \bb{PT}$ at the point $\pi_{A^\prime} = \gamma_{A^\prime}$.
It coincides with the opposite complex structure -- swapping holomorphic and anti-holomorphic -- at the antipodal point $\pi_{A^\prime} = \hat{\gamma}_{A^\prime}$. Using the spinor $\mu^A$, we can define a basis of one-forms adapted to this complex structure,
\begin{equation}\label{eq:ComplexCoords}
\begin{aligned}
\dr z &= \mu_A \gamma_{A'} \dr x^{A A' } ~, \qquad & \dr\bar{z} &= \hat{\mu}_A \hat{\gamma}_{A'} \dr x^{A A' } ~, \\
\dr w &= \hat{\mu}_A \gamma_{A'} \dr x^{A A' } ~, \qquad & \dr \bar{w} &= - \mu_A \hat{\gamma}_{A'} \dr x^{A A' } ~.
\end{aligned}
\end{equation}
We will perform symmetry reduction along the vector fields dual to $\dr z$ and $\dr \bar{z}$,
\begin{equation}
\pd_z = \hat{\mu}^A \hat{\gamma}^{A^\prime} \pd_{A A^\prime} ~, \quad
\pd_{\bar{z}} = \mu^A \gamma^{A^\prime} \pd_{A A^\prime} ~.
\end{equation}
The symmetry reduction along the $ \partial_z$ and $ \partial_{\bar{z}}$ directions takes us from a theory on $\bb{PT}$ to a theory on $\Sigma \times \mathbb{CP}^1$ in which $w,\bar{w}$ are coordinates on the worldsheet $\Sigma$.

To perform this reduction, it is expedient \cite{Bittleston:2022cmz} to make use of the invariance of the action \eqref{eq:hCS6}  under the addition of any $(1,0)$-form to ${\cal A} \mapsto \hat {\cal A} = {\cal A} + \rho_0 e^0 + \rho_A e^A$. By choosing $\rho_A$ appropriately, we can ensure that $\hat{\cal A}$ has no $\dr z$ or $\dr \bar{z}$ legs and is given by
\begin{equation}\label{eq:Ashifte}
\hat {\cal A} = \hat{ {\cal A}}_w \dr w +\hat{ {\cal A}}_{\bar{w}} \dr \bar{w} + {\cal A}_0 \bar{e}^0 ~,
\end{equation}
where these components are related to those of $\cA$ by
\begin{equation}\label{eq:AwAwb}
\hat{{\cal A}}_w = - \frac{[{\cal A} \mu]}{\langle \pi \gamma \rangle } ~, \qquad \hat{{\cal A}}_{\bar{w}} = - \frac{[{\cal A} \hat{\mu}]}{\langle \pi \hat{\gamma} \rangle } ~.
\end{equation}
An important feature to note is that $\hat {\cal A}$ necessarily has singularities at $\gamma$ and $\hat{\gamma}$.
While at the 6-dimensional level this is a mere gauge-choice artefact, it plays a crucial role in the construction of the CS$_4$ theory.

In these variables, the boundary  variation and boundary condition of hCS$_6$ are restated as
\begin{equation}\label{eq:bv2}
r_+ \mathrm{Tr} \big( \hat{{\cal A}}_w \delta \hat{ {\cal A}}_{\bar{w}} - \hat{{\cal A}}_{\bar{w}} \delta \hat{{\cal A}}_w \big) \big\vert_{\pi=\alpha}= - r_- \mathrm{Tr} \big(\hat{ {\cal A}}_w \delta \hat{ {\cal A}}_{\bar{w}} - \hat{ {\cal A}}_{\bar{w}} \delta \hat{{\cal A}}_w \big) \big\vert_{\pi=\tilde\alpha} ~,
\end{equation}
\begin{equation}
\label{eq:6dBC}
\hat{ {\cal A}}_w \vert_{\pi = \alpha} = t s \hat{ {\cal A}}_w \vert_{\pi = \tilde{\alpha}} ~, \qquad \hat{ {\cal A}}_{\bar{w}} \vert_{\pi = \alpha} = t^{-1} s \hat{ {\cal A}}_{\bar{w}} \vert_{\pi = \tilde{\alpha}} ~,
\end{equation}
where we have introduced the combinations
\begin{equation}\begin{aligned}\label{eq:rpmst}
r_+ & = K \frac{\langle \alpha \gamma \rangle \langle \alpha \hat{\gamma} \rangle }{\langle \alpha \tilde\alpha \rangle \langle \alpha \beta \rangle^2 } ~, & \qquad r_- & = - K \frac{\langle \tilde\alpha \gamma \rangle \langle \tilde\alpha \hat{\gamma} \rangle }{\langle \alpha \tilde\alpha \rangle \langle \tilde\alpha \beta \rangle^2 } ~,
\\ s & = \sqrt{-\frac{r_-}{r_+}} = \frac{\langle\alpha\beta\rangle}{\langle\tilde\alpha\beta\rangle}\sqrt{\frac{\langle \tilde\alpha \gamma \rangle \langle \tilde\alpha \hat{\gamma} \rangle}{\langle \alpha \gamma \rangle \langle \alpha \hat{\gamma} \rangle }} ~,
& \qquad
t & = \sigma s \frac{\langle\tilde\alpha\beta\rangle\langle \alpha \hat{\gamma} \rangle }{\langle\alpha\beta\rangle\langle \tilde\alpha \hat{\gamma} \rangle} ~.
\end{aligned}\end{equation}
Upon symmetry reduction to CS$_4$, $r_\pm$ will correspond to the residues of the 1-form $\omega$.

Since the shifted gauge field $\hat{\mathcal{A}}$ manifestly has no $\dr z$ or $\dr \bar{z}$ legs, and we impose $L_z\hat{\mathcal{A}}=L_{\bar{z}} \hat{\mathcal{A}} =0 $, the contraction by $\pd_z$ and $\pd_{\bar{z}}$ only hits $\Omega$. It then follows that the symmetry reduction yields
\begin{equation}
\label{ec:4dCSaction1}
S_{\mathrm{CS}_4} = \frac{1}{2 \pi \mathrm{i}} \int_{\Sigma \times \mathbb{CP}^1} \omega \wedge \mathrm{Tr} \bigg( \hat {\cal A} \wedge \dr \hat {\cal A} + \frac{2}{3} \hat {\cal A} \wedge \hat {\cal A} \wedge \hat {\cal A} \bigg) ~,
\end{equation}
in which the meromorphic 1-form on $\mathbb{CP}^1$ is given by
\begin{equation}\label{eq:omega4}
\omega = \partial_{\bar z} \, \lrcorner \partial_{z} \, \lrcorner \, \Omega = \Phi \, \varepsilon_{AB} \left( \partial_{\bar z} \, \lrcorner \, e^A \right) \left( \partial_{z} \, \lrcorner \, e^B \right) \, e^0 =  - K \, \frac{
\langle \pi \gamma \rangle
\langle \pi \hat{\gamma} \rangle}{
\langle \pi \alpha \rangle
\langle \pi \tilde{\alpha} \rangle
\langle \pi \beta \rangle^2} \, e^0 ~.
\end{equation}
To compare with the literature, we will also translate to inhomogeneous coordinates on $\mathbb{CP}^1$.
The $\mathbb{CP}^1$ coordinate itself will be given by $\zeta = \pi_{2^\prime} / \pi_{1^\prime}$ on the patch $\pi_{1^\prime} \neq 0$.
We also specify representatives for the various other spinors in our theory.
Without loss of generality we can choose
\begin{equation}~\label{eq:inhom1}
\alpha_{A'} = (1, \alpha_+) ~, \qquad \tilde{\alpha}_{A'} = (1, \alpha_-) ~, \qquad \beta_{A'}= (0,1) ~,
\end{equation}
such that
\begin{equation}
\langle \tilde{\alpha} \beta \rangle = \langle \alpha \beta \rangle = 1 ~, \qquad \langle \tilde\alpha \alpha \rangle = \alpha_+ - \alpha_- = \Delta \alpha ~.
\end{equation}
We also denote the inhomogeneous coordinates for $\gamma_{A'}$ and $\hat{\gamma}_{A'}$ by
\begin{equation}~\label{eq:inhom2}
\gamma_+ = \frac{\gamma_{2^\prime}}{ \gamma_{1^\prime}} ~, \qquad \gamma_- = \frac{\hat{\gamma}_{2^\prime}}{\hat{\gamma}_{1^\prime}} = -\frac{\overline{\gamma_{1^\prime}} }{ \overline{\gamma_{2^\prime}} } ~, \qquad \gamma_{1^\prime} \overline{\gamma_{2^\prime}} = \frac{1}{\gamma_+ - \gamma_-} = \frac{1}{\Delta \gamma} ~.
\end{equation}
Then, the meromorphic $1$-form $\omega$ is written in inhomogeneous coordinates as
\begin{equation}
\omega = \frac{K}{\Delta \gamma } \frac{ (\zeta- \gamma_+) (\zeta- \gamma_-)}{ (\zeta -\alpha_+)(\zeta -\alpha_-) } \dr \zeta = \varphi(\zeta) \, \dr \zeta ~.
\end{equation}

To complete the specification of the theory we simply note that the 6d boundary conditions immediately descend to
\begin{equation}
\label{eq:4dBC}
\hat{{\cal A}}_w \mid_{\pi = \alpha} = t s \hat{{\cal A}}_w \mid_{\pi = \tilde{\alpha}} ~, \qquad \hat{{\cal A}}_{\bar{w}} \mid_{\pi = \alpha} = t^{-1} s \hat{{\cal A}}_{\bar{w}} \mid_{\pi = \tilde{\alpha}} ~.
\end{equation}

Before we discuss the residual symmetries of the CS$_4$ models, let us make two related observations.
First, fixing the shift symmetry to ensure $\hat{\cal A}$ is horizontal with respect to the symmetry reduction introduces poles into our gauge field $\hat{\cal A}$ at $\gamma$ and $\hat{\gamma}$.
Thus, despite starting with potentially smooth field configurations in 6 dimensions we are forced to consider singular ones in 4 dimensions.
We can understand the origin of these singularities by recalling the holomorphic coordinates on $\bE^4$ with respect to the complex structure on $\bb{PT} = \mathbb{CP}^3 \backslash \mathbb{CP}^1$.
Described in detail in appendix~\ref{appendix:twistors}, $\bb{PT}$ is only diffeomorphic to $\bE^4 \times \mathbb{CP}^1$, and the complex structure is more involved in these coordinates.
The holomorphic coordinates on $\bE^4$ with respect to this complex structure are given by $v^A = \pi_{A^\prime} x^{A A^\prime}$, which align with our coordinates $\{ z, w \}$ at $\pi \sim \gamma$ and $\{ \bar{z}, \bar{w} \}$ at $\pi \sim \hat{\gamma}$.
It is precisely at these points that we are forced to introduce poles by the symmetry reduction procedure.

Second, in line with the singular behaviour in the gauge field, we have also introduced zeroes in $ \omega $ at $\pi \sim \gamma$ and $\pi \sim \hat{\gamma}$ whereas $\Omega$ in 6 dimensions was nowhere vanishing.
Of course, given the pole structure of $\Omega$, the introduction of two zeroes was inevitable given the Riemann-Roch theorem.

\subsection{Residual Symmetries and the Defect Algebra}

Let us take a moment to consider the residual symmetries of these CS$_4$ models.
Here the residual symmetry preserving the boundary condition~\eqref{eq:4dBC} is generically constrained to only include constant modes,
\begin{equation}
\label{eq:6dgaugetransfred}
r = \tilde{r} ~, \qquad
(1-ts) \partial_w r = 0 ~, \qquad
(1-ts^{-1}) \partial_{\bar w} r = 0 ~.
\end{equation}
At the special `diagonal' point in parameter space where $t=s=1$, notice these differential equations are identically solved and we find a local gauge symmetry.
This enhancement of residual gauge freedom matches with previous considerations in the context of CS$_{4}$, where diagonal boundary conditions of the form $A\vert_\alpha = A\vert_{\tilde{\alpha}}$ are known to give rise to the $\lambda$-deformed WZW as an IFT$_{2}$, a theory that depends on a single field $h$.
The residual gauge symmetries are those satisfying $\hat{g}\vert_{\alpha}=\hat{g}\vert_{\tilde{\alpha}}$ and can be used to reduce the number of fields appearing in the resulting IFT$_2$ to one (see \S~5.4 in \cite{Delduc:2019whp}).

Another interesting point occurs when we take $t = s$ or $t=s^{-1}$ in which case we retain a chiral residual symmetry.
When $t=0$ the boundary conditions admit an enlarged residual symmetry as there is no requirement that $r= \hat{g}\vert_\alpha$ and $\tilde{r}= \hat{g}\vert_{\tilde{\alpha}} $ match.
Instead they must be chiral and of opposite chiralities i.e.~$\partial_w r = \partial_{{\bar w}} \tilde{r} = 0$.
As mentioned earlier, for more generic values of $t$ and $s$ the residual symmetries will be constrained, preventing them from being used to eliminate any degrees of freedom.
While these boundary conditions have not been yet considered for $t,s\neq 1$ and $t \neq 0$, we will see that they give rise to the multi-parametric class of $\lambda$-deformations between coupled WZW models introduced in \cite{Georgiou:2017jfi}.

To make further contact with the literature, it is helpful to rephrase the boundary conditions \eqref{eq:4dBC} in terms of a defect algebra, which in the case at hand is simply the Lie algebra $\mathfrak{d} = \mathfrak{g}\oplus\mathfrak{g}$ equipped with an ad-invariant pairing
\begin{equation}\label{eq:innerint}
\langle\!\langle \mathbb{X}, \mathbb{Y} \rangle\!\rangle = r_+ \mathrm{Tr} (x_1\, y_1) + r_- \mathrm{Tr} (x_2 \, y_2)
~, \qquad \mathbb{X} = (x_1,x_2) ~, \mathbb{Y} = (y_1,y_2) ~.
\end{equation}
We map our boundary conditions into this algebra by defining $\mathbb{A}_w = (\hat{\cal A}_w \vert_{\pi = \alpha} , \hat{\cal A}_w \vert_{\pi = \tilde{\alpha}} ) $ and $\mathbb{A}_{\bar w} = (\hat{\cal A}_{\bar w} \vert_{\pi = \alpha} , \hat{\cal A}_{\bar w} \vert_{\pi = \tilde{\alpha}} ) $ such that the requirement that the boundary variation vanishes locally can be recast as
\begin{equation}
0 =\langle\!\langle\mathbb{A}_w, \delta \mathbb{A}_{\bar{w}} \rangle\!\rangle - \langle\!\langle \mathbb{A}_{\bar{w}}
, \delta \mathbb{A}_w \rangle\!\rangle ~.
\end{equation}
The boundary conditions~\eqref{eq:4dBC} read
\begin{equation}\begin{aligned}\label{eq:4dBCdoubled}
\mathbb{A}_{w} \in \mathfrak{l}_t &=\textrm{span} \{ \left( t s x, x \right) \mid x \in \mathfrak{g} \} ~, \\
\mathbb{A}_{\bar{w}} \in \mathfrak{l}_{t^{-1}} &=\textrm{span} \{ \left( t^{-1} s x, x \right) \mid x \in \mathfrak{g} \} ~.
\end{aligned}\end{equation}
Since $\langle\!\langle \mathfrak{l}_t , \mathfrak{l}_{t^{-1} }\rangle\!\rangle= 0 $ the boundary conditions are suitable, however it should be noted that $\mathfrak{l}_t$ is itself neither a subalgebra nor an isotropic subspace of $\mathfrak{d}$.
This is more general than boundary conditions previously considered
\unskip\footnote{Of course in the limit $t,s \rightarrow 1$ $\mathfrak{l}_t$ revert to defining the diagonal isotropic subalgebra.
In the special case where $t\rightarrow 0,\infty$ we recover chiral Dirichlet boundary conditions considered in \cite{Costello:2019tri,Ashwinkumar:2023zbu}.} in the context of 4-dimensional Chern Simons theory.
In particular, we might expect that generalising \cite{Benini:2020skc,Lacroix:2020flf,Liniado:2023uoo} to boundary conditions defined by subspaces that are neither a subalgebra nor an isotropic subspace of $\mathfrak{d}$ will lead to novel families of 2-dimensional integrable field theories.

It is worth highlighting that these boundary conditions still define maximal isotropic subspaces, but now inside the space of algebra-valued $1$-forms, rather than just the defect algebra.
Consider the space of $\fg$-valued $1$-forms on $\Sigma \times \mathbb{CP}^1$, equipped with the symplectic structure
\unskip\footnote{As defined, this is not quite a symplectic structure since it is degenerate -- for example, it vanishes on the subspace of $1$-forms which only have legs along $\mathbb{CP}^1$.
A more careful treatment would involve restricting the symplectic form to a subspace where it is non-degenerate, but we will neglect this for the purpose of our brief discussion.}
\begin{equation}
\mathcal{W}(X, Y) = \int_{\Sigma \times \mathbb{CP}^1} \dr \omega \wedge \mathrm{Tr} \big( X \wedge Y \big) \ , \quad X, Y \in \Omega^1(\Sigma \times \mathbb{CP}^1) \otimes \fg ~.
\end{equation}
The boundary conditions above define maximal isotropic subspaces with respect to this symplectic structure, that is half-dimensional subspaces $\mathcal{Y} \subset \Omega^1(\Sigma \times \mathbb{CP}^1) \otimes \fg$ such that $\mathcal{W}(X, Y) = 0$ for all $X, Y \in \mathcal{Y}$.
Indeed, this is required for them to be `good' boundary conditions.
The isotropic subspaces of the defect algebra described earlier are then special cases of these subspaces.

\section{Symmetry Reduction of \texorpdfstring{IFT$_4$}{IFT4} to \texorpdfstring{IFT$_2$}{IFT2} }
\label{sec:4dIFTto2dIFT}

\begin{tcolorbox}
\begin{minipage}[c]{0.4\linewidth}

\begin{equation*}
\begin{tikzpicture}[scale=0.8]
\node at (0,2) {$\mathbf{hCS_6}$};
\node at (-2,0) {$\mathbf{CS_4}$};
\node at (2,0) {$\mathbf{IFT_4}$};
\node at (0,-2) {$\mathbf{IFT_2}$};
\draw[->,very thick,decorate, decoration={snake, segment length=13pt, amplitude=2pt}] (-0.4,1.6)--(-1.6,0.4);
\draw[->,very thick] (0.4,1.6)--(1.6,0.4);
\draw[->,very thick, red, decorate, decoration={snake, segment length=13pt, amplitude=2pt}] (1.6,-0.4)--(0.4,-1.6);
\draw[->,very thick] (-1.6,-0.4)--(-0.4,-1.6);
\end{tikzpicture}
\end{equation*}

\end{minipage}
\begin{minipage}[c]{0.59\linewidth}
In this section we will apply the same symmetry reduction previously applied to hCS$_6$ to the IFT$_4$.
In doing so we derive the IFT$_2$ corresponding to the CS$_4$ model described above.

\end{minipage}

\end{tcolorbox}

Recalling that the reduction requires that the fields $h$ and $\tilde{h}$ depend only on $w, \bar{w}$ and not on $z, \bar{z}$, we can simply set $\partial_z = \partial_{\bar{z}} = 0$ in the action eq.~\eqref{eq:IFT4}.
To compare with the literature, when discussing 2-dimensional theories we will define $\partial_+ \equiv \partial_w $ and $\partial_- \equiv \partial_{\bar{w}}$ (implicitly rotating to 2d Minkowski space where the action is rendered real for real parameters) and denote
\begin{equation}
J_\pm = h^{-1} \partial_\pm h ~, \qquad \widetilde{J}_\pm =\tilde{h}^{-1} \partial_\pm \tilde{h} ~.
\end{equation}
To evaluate the symmetry reduction, denoted by $\leadsto$, of the IFT$_{4}$ action we first note that
\begin{equation}
j \leadsto \frac{\langle \alpha \gamma \rangle }{\langle \alpha \beta \rangle } J_+ ~, \quad \hj \leadsto \frac{\langle \alpha \hat\gamma \rangle }{\langle \alpha \beta \rangle } J_-\, , \quad \tj \leadsto \frac{\langle \tilde\alpha \gamma \rangle }{\langle \tilde \alpha \beta \rangle } \widetilde{J}_+\, , \quad \htj \leadsto \frac{\langle \tilde\alpha \hat \gamma \rangle }{\langle \tilde \alpha \beta \rangle } \widetilde{J}_- ~.
\end{equation}
The resulting 2-dimensional action is given by
\begin{equation}\label{eq:IFT2}
\begin{aligned}
S_{\mathrm{IFT}_4} \leadsto S_{\mathrm{IFT}_2} \ & = \int_\Sigma \vol_2 \,\mathrm{Tr}\big( r_+ \, J_+ ( U^{T}_+ - U_{-} ) J_{-} - r_- \, \widetilde{J}_+ ( U^{T}_+ - U_{-} ) \widetilde{J}_{-} + r_+ \, {\cal L}_{\mathrm{WZ}}(h) + r_- \, {\cal L}_{\mathrm{WZ}}(\tilde h) \\
& \hspace{60pt}- 2 \, t \sqrt{- r_+ \cdot r_-} \, \widetilde{J}_+ U_+^{T}J_{-} + 2 \, t^{-1} \sqrt{- r_+ \cdot r_-} \, J_+ U_{-} \widetilde{J}_{-}\big) ~,
\end{aligned}
\end{equation}
where $\vol_2 = \dr\bar{w}\wedge\dr w = \dr\sigma^-\wedge\dr\sigma^+$.
This theory, depending on two $G$-valued fields, $h$ and $\tilde{h}$, and four independent parameters, $r_\pm$, $t$ and $\sigma$, exactly matches a theory introduced in \cite{Georgiou:2017jfi} as a multi-field generalisation of the $\lambda$-deformed WZW model \cite{Sfetsos:2013wia}.
To make a precise match with \cite{Georgiou:2017jfi} we relate their fields $(g_1,g_2)$ to our fields $(h, \tilde{h}^{-1})$.
The model in \cite{Georgiou:2017jfi} is defined by two WZW levels $k_{1,2}$ and by two deformation matrices which we take to be proportional to the identity with constants of proportionality $\lambda_{1,2}$.
The mapping of parameters is then
\begin{equation}~\label{eq:GSparam}
\lambda_1 = \sigma t^{-1} ~, \qquad \lambda_2= t ~, \qquad k_1 = r_+ ~, \qquad k_2 =- r_- ~, \qquad \lambda_0 = \sqrt{k_1/ k_2} = \sqrt{-r_-/r_+} = s^{-1} ~.
\end{equation}
In 2d Minkowski space, the Lagrangian~\eqref{eq:IFT2} is real if the parameters $r_\pm$, $s$ $t$ and $\sigma$ are all real.
Assuming $K$ and $\sigma$ are real, this is the case if $r_+$ and $r_-$ have the opposite sign and the parameters $\alpha_\pm$ and $\gamma_\pm$ lie on the same line in $\mathbb{C}$, which we can take to be the real line without loss of generality.
This follows since $r_\pm$, $s$ and $t$ are all expressed as ratios of differences of $\alpha_\pm$ and $\gamma_\pm$, hence are invariant under translations and scalings.
\unskip\footnote{Note that this is the subgroup of $SL(2,\mathbb{C})$ transformations that preserves the choice $\beta_{A'} = (0,1)$.}

\subsection{Limits}

The four-parameter model has a number of interesting limits, many of which are discussed in \cite{Georgiou:2017jfi}.
Here, we briefly summarise some key ones.
First, let us take $t \to 0$.
In order to have a well-defined limit we keep $\sigma t^{-1} = \lambda$ finite, implying $\sigma \to 0$ as well.
\unskip\footnote{An analogous limit is to take $\sigma \to 0$ and keep $t$ finite.}
The resulting 2d Lagrangian is given by
\begin{equation}\label{eq:IFT2cc}
S_{\mathrm{IFT}_2}\vert_{t,\sigma\to 0} = \int_\Sigma \vol_2 \,\mathrm{Tr}\big( r_+ \, J_+ J_{-} - r_- \, \widetilde{J}_+ \widetilde{J}_{-} - 2 \, \lambda s r_+ J_+ \Lambda^{-1} \widetilde{J}_{-} + r_+ \, {\cal L}_{\mathrm{WZ}}(h) + r_- \, {\cal L}_{\mathrm{WZ}}(\tilde h) \big) ~.
\end{equation}
This current-current deformation preserves half the chiral symmetry of the $G_{r_+} \times G_{-r_-}$ WZW model, which corresponds to the UV fixed point $\lambda = 0$.
Indeed, this model can be found by taking chiral Dirichlet boundary conditions in 4d CS~\cite{Costello:2019tri,Ashwinkumar:2023zbu}, corresponding to the special case $t=0$ in the boundary conditions we find from symmetry reduction~\eqref{eq:4dBCdoubled}.
Assuming $-r_- > r_+$, in the IR we have that $\lambda = s^{-1}$.
At this point the Lagrangian can be written as
\begin{equation}\label{eq:IFT2cc2}
\begin{aligned}
S_{\mathrm{IFT}_2}\vert_{t,\sigma\to 0, t\sigma^{-1} = s} & = \int_\Sigma \vol_2 \,\mathrm{Tr}\big( r_+ (\mathrm{Ad}_h J_+ - \mathrm{Ad}_{\tilde{h}} \tilde{J}_+) (\mathrm{Ad}_h J_{-} - \mathrm{Ad}_{\tilde{h}}\tilde{J}_-) - (r_- + r_+) \, \widetilde{J}_+ \widetilde{J}_{-} \\
& \hspace{60pt} + r_+ \, {\cal L}_{\mathrm{WZ}}(\tilde{h}^{-1} h) + (r_-+r_+) \, {\cal L}_{\mathrm{WZ}}(\tilde h) \big) ~.
\end{aligned}
\end{equation}
Redefining $h \to \tilde{h} h$, we find the $G_{r_+} \times G_{-r_- - r_+}$ WZW model.
In the case of equal levels $r_- = -r_+$ this reduces to the $G_{r_+}$ WZW model.

The equal-level, $r_- = -r_+$, version of~\eqref{eq:IFT2cc}, whose classical integrability was first shown in~\cite{Bardakci:1996gs}, is canonically equivalent~\cite{Georgiou:2017oly} and related by a path integral transformation~\cite{Hoare:2019mcc} to the $\lambda$-deformed WZW model.
Indeed, from the point of view of 4d CS, these two models have the same twist function.
To recover~\eqref{eq:IFT2cc} with equal levels, we take chiral Dirichlet boundary conditions, $t=0$, $s=1$ in~\eqref{eq:4dBCdoubled}, while to recover the $\lambda$-deformed WZW model we take diagonal boundary conditions $t=s=1$.

It follows that if we take $t = s =1$ in eq.~\eqref{eq:IFT2}, we expect to recover the $\lambda$-deformed WZW model.
Indeed, setting $r_- = - r_+$ and $t=1$, the Lagrangian~\eqref{eq:IFT2} becomes
\begin{equation}\label{eq:lambdamodel}
S_{\mathrm{IFT}_2}\vert_{t=s=1} = \int_\Sigma \vol_2 \,\mathrm{Tr}\big( r_+ \, (J_+ - \widetilde{J}_+) ( U_{+ }^T - U_{-}) (J_- - \widetilde{J}_-) + r_+ \, {\cal L}_{\mathrm{WZ}}(h \tilde{h}^{-1}) \big) ~.
\end{equation}
As explained in subsection~\ref{ssec:gsem}, at this point in parameter space the symmetry reduction directions are aligned such that the constrained symmetry transformations~\eqref{eq:rconstraints} become a gauge symmetry of the IFT$_2$.
This allows us to fix $\tilde h =1$, recovering the standard form of the $\lambda$-deformed WZW model~\cite{Sfetsos:2013wia} with $\sigma$ playing the role of $\lambda$.
Further taking $\sigma \to 0$, we recover the $G_{r_+}$ WZW model.

Another point in parameter space where we expect a gauge symmetry to emerge is when the symmetry reduction preserves the left-acting symmetry.
This corresponds to setting $t=\sigma$ and $s=1$, i.e.~$r_- = - r_+$.
Doing so we find
\begin{equation}\label{eq:lambdamodel2}
S_{\mathrm{IFT}_2}\vert_{t=\sigma,s=1} = \int_\Sigma \vol_2 \,\mathrm{Tr}\big( r_+ ( \mathrm{Ad}_h J_+ - \mathrm{Ad}_{\tilde h} \widetilde{J}_+) ( \widetilde U_{+ }^T - \widetilde U_{-} ) (\mathrm{Ad}_h J_{-} - \mathrm{Ad}_{\tilde h} \widetilde{J}_{-}) + r_+ \, {\cal L}_{\mathrm{WZ}}(\tilde{h}^{-1} h) \big) ~,
\end{equation}
where we recall that $\widetilde U_\pm$ are defined in eq.~\eqref{eq:tildeulam}.
This Lagrangian is invariant under a left-acting gauge symmetry as expected, which can be used to fix $\tilde h = 1$.
We again recover the standard form of the $\lambda$-deformed WZW model with $\sigma$ playing the role of $\lambda$.
The CS$_4$ description of this limit is analysed in appendix \ref{appendix:alt4dCSlambda}.

Before we move onto the integrability of the 2d IFT and its origin from the 4d IFT, let us briefly note the symmetry reduction implications of the formal transformations~\eqref{eq:formal1} and~\eqref{eq:formal2}, which in turn descended from the discrete invariances of the hCS$_6$ boundary conditions~\eqref{z21} and~\eqref{z22}.
The first~\eqref{eq:formal1} implies that the 2d IFT is invariant under
\begin{equation}
r_+ \leftrightarrow r_- ~, \qquad \sigma \to \sigma^{-1}~,\qquad t \to t^{-1} ~,\qquad h \leftrightarrow \tilde h ~.
\end{equation}
recovering the `duality' transformation of \cite{Georgiou:2017jfi}.
Since the second involves interchanging $w$ and $z$, it tells us the parameters are transformed if we symmetry reduce requiring that the fields $h$ and $\tilde{h}$ only depend on $z,\bar{z}$, instead of $w,\bar{w}$.
We find that $\sigma \to \sigma^{-1}$ and $t \to t \sigma^{-2}$.

\subsection{Integrability and Lax Formulation}
\def\P{B}
\def\Q{C}
The analysis of \cite{Georgiou:2017jfi} shows that the equations of motion of \eqref{eq:IFT2} are best cast in terms of auxiliary fields
\unskip\footnote{To avoid conflict of notation $\P_\pm,\Q_\pm$ here correspond to $A_\pm, B_\pm$ of \cite{Georgiou:2017jfi}.}
$\P_\pm,\Q_\pm$ which are related to the fundamental fields by
\begin{equation}\begin{aligned}\label{eq:auxiliaries}
J_- = \mathrm{Ad}_h^{-1} \P_- - \lambda_0^{-1}\lambda_2^{-1}\Q_- ~, \qquad \tilde{J}_- = \lambda_0 \lambda_1^{-1}\mathrm{Ad}_{\tilde{h}}^{-1}\P_- - \Q_- ~, \\
J_+ = \lambda_0^{-1} \lambda_1^{-1}\mathrm{Ad}_h^{-1} \P_+ - \Q_+ ~, \qquad
\tilde{J}_+= \mathrm{Ad}_{\tilde{h}}^{-1}\P_+ - \lambda_0 \lambda_2^{-1}\Q_+ ~.
\end{aligned}
\end{equation}
The equations of motion for $h$ and $\tilde{h}$, together with the Bianchi identities obeyed by their associated Maurer-Cartan forms, can be repackaged into the flatness of two Lax connections with components
\begin{equation}\label{eq:twoLax}
{\cal L}^{1}_\pm = \frac{2 \zeta_{\mathrm{GS}}}{\zeta_{\mathrm{GS}} \mp 1} \frac{1 - \lambda_0^{\mp 1} \lambda_1}{1- \lambda_1^2} \P_\pm ~, \qquad
{\cal L}^{2}_\pm = \frac{2 \zeta_{\mathrm{GS}}}{\zeta_{\mathrm{GS}} \mp 1} \frac{1 - \lambda_0^{\pm 1} \lambda_2}{1- \lambda_2^2} \Q_\pm ~.
\end{equation}
Here $\zeta_{\mathrm{GS}}$ is the spectral parameter used in \cite{Georgiou:2017jfi}.
Taken together, the flatness of this pair of Lax connections implies both the Bianchi identities and the equations of motions.
However, if one is prepared to enforce the definition~\eqref{eq:auxiliaries} of auxiliary fields in terms of fundamental fields (such that the Bianchi equations are automatically satisfied) then either Lax will generically (i.e.~away from special points in parameter space such as $\lambda_i=1$) imply the equations of motion
\unskip\footnote{It is less evident in contrast if {\em all} the non-local conserved charges of the theory can be obtained from a single Lax.}
of the theory.

We can relate this discussion to the construction above by symmetry reducing the 4d Lax operators \eqref{eq:4dLax} and \eqref{eq:other4dLax}.
First we note that the currents corresponding to the $(\ell, r)$-symmetries reduce to simple combinations of the auxiliary fields introduced in eq.~\eqref{eq:auxiliaries}
\begin{equation} \begin{aligned}
B_A &\leadsto \frac{\langle \alpha \hat \gamma \rangle}{\langle \alpha \beta\rangle} \P_- \mu_A - \frac{\langle \tilde \alpha \gamma \rangle}{\langle \tilde\alpha \beta\rangle} \P_+
\hat{\mu}_A ~, \\
C_A &\leadsto \frac{\langle \tilde \alpha \hat \gamma \rangle}{\langle \tilde \alpha \beta\rangle} \sigma^{-1} \Q_- \mu_A- \frac{\langle \alpha \gamma \rangle}{\langle \alpha \beta\rangle} \Q_+
\hat{\mu}_A ~.
\end{aligned}
\end{equation}
Notice that all explicit appearances of the operators $U_\pm$ have dropped out such that these currents reduce exactly to the 2-dimensional auxiliary gauge fields.

Using the complex coordinates adapted for symmetry reduction defined in eq.~\eqref{eq:ComplexCoords}, and introducing a specialised inhomogeneous coordinate on $\mathbb{CP}^1$ given by $\varsigma = \langle \pi \hat{\gamma} \rangle / \langle \pi \gamma \rangle$, the 4d $B$-Lax pair \eqref{eq:4dLax} may be written as
\begin{equation}
L^{(B)} = D_{\bar{w}} - \varsigma^{-1} D_{z} \ , \quad
M^{(B)} = D_{w} + \varsigma D_{\bar{z}} \ .
\end{equation}
We can symmetry reduce the 4d Lax pairs, $L^{(B/C)},M^{(B/C)}$ of eqs.~\eqref{eq:4dLax} and \eqref{eq:other4dLax} to obtain
\begin{equation}
\begin{aligned}
L^{(B)} & \leadsto \partial_{-} + ( \langle \beta \gamma \rangle - \varsigma^{-1} \langle \beta \hat\gamma \rangle ) \frac{\langle \alpha \hat\gamma \rangle}{\langle \alpha \beta \rangle} \P_- ~,
& M^{(B)} & \leadsto \partial_+ + ( \varsigma \langle \beta \gamma \rangle - \langle \beta \hat\gamma \rangle ) \frac{\langle \tilde{\alpha} \gamma \rangle}{\langle \tilde{\alpha} \beta \rangle }\P_+ ~, \\
L^{(C)} & \leadsto \partial_{-} - \frac{1}{(1-\varrho) + \lambda_0 \lambda_2 ( 1+\varrho) }\Q_- \, ,
& M^{(C)} & \leadsto \partial_+ - \frac{1}{(1+ \varrho) + \lambda_0^{-1} \lambda_2 ( 1- \varrho) }\Q_+ ~.
\end{aligned}
\end{equation}
Now using the inhomogeneous coordinates introduced in eqs.~\eqref{eq:inhom1} and~\eqref{eq:inhom2}, and the relations between parameters~\eqref{eq:GSparam}, the 4d Lax operators immediately descend upon symmetry reduction to the 2d Lax connections~\eqref{eq:twoLax}, provided the 4d and 2d spectral parameters are related as
\begin{equation}
\begin{aligned}
& L^{(B)} \leadsto \partial_{-} + {\cal L}^1_- \, , \quad M^{(B)} \leadsto \partial_{+} + {\cal L}^1_+ \, , \quad \zeta_{\mathrm{GS}} = \frac{\bar{\gamma}_2 + \gamma_1 \varsigma}{-\bar{\gamma}_2 + \gamma_1 \varsigma} ~, \\
& L^{(C)} \leadsto \partial_{-} + {\cal L}^2_- \, , \quad M^{(C)} \leadsto \partial_{+} + {\cal L}^2_+ \, , \quad \zeta_{\mathrm{GS}} = \frac{1-\lambda_2^2}{( \lambda_0 - \lambda_0^{-1})\lambda_2 - (1- \lambda_0 \lambda_2)(1-\lambda_0^{-1}\lambda_2)\varrho } ~.
\end{aligned}
\end{equation}
The relation between $\zeta_{\mathrm{GS}}$ and $\varsigma$ can be recast in   the standard $\mathbb{CP}^1$ homogeneous coordinate $\pi \sim ( 1 , \zeta)$ as
\begin{equation}\label{eq:zgsident}
\zeta_{\mathrm{GS}} = \frac{\gamma_+ - \gamma_-}{2\zeta -(\gamma_+ + \gamma_-) } ~,
\end{equation}
such that if we choose to fix $\gamma_\pm = \pm 1$ then $ \zeta_{\mathrm{GS}} = \zeta^{-1} $.
If we make the assumption that the $\zeta_{\mathrm{GS}}$ entering in the two different Lax formulations have the same origin then we can map between between $\varrho$ and the $\mathbb{CP}^1$ homogeneous coordinate
\begin{equation}
\varrho = -\frac{1+\zeta}{2} \frac{1+ts}{1-ts} + \frac{1-\zeta}{2} \frac{1+ts^{-1}}{1-ts^{-1}} ~.
\end{equation}
Therefore, under this assumption, we see that $\varrho$ depends on the parameter $t$, which is part of the specification of boundary conditions and not just geometric data of $\mathbb{CP}^1$.
Indeed $\varrho$ becomes constant when $t\rightarrow1$, hence there is no spectral parameter dependence left.
In contrast, when $t \rightarrow 0$, we have $\varrho \rightarrow - \zeta $.

\section{Localisation of \texorpdfstring{CS$_{4}$}{CS4} to \texorpdfstring{IFT$_2$}{IFT2} }\label{sec:4to2}

\begin{tcolorbox}
\begin{minipage}[c]{0.4\linewidth}

\begin{equation*}
\begin{tikzpicture}[scale=0.8]
\node at (0,2) {$\mathbf{hCS_6}$};
\node at (-2,0) {$\mathbf{CS_4}$};
\node at (2,0) {$\mathbf{IFT_4}$};
\node at (0,-2) {$\mathbf{IFT_2}$};
\draw[->,very thick,decorate, decoration={snake, segment length=13pt, amplitude=2pt}] (-0.4,1.6)--(-1.6,0.4);
\draw[->,very thick] (0.4,1.6)--(1.6,0.4);
\draw[->,very thick,decorate, decoration={snake, segment length=13pt, amplitude=2pt}] (1.6,-0.4)--(0.4,-1.6);
\draw[->,red,very thick] (-1.6,-0.4)--(-0.4,-1.6);
\end{tikzpicture}
\end{equation*}

\end{minipage}
\begin{minipage}[c]{0.59\linewidth}

Finally, we localise the CS$_4$ theory obtained by symmetry reduction of hCS$_6$ in \S~\ref{sec:4dhcsred}.
This will result in a 2-dimensional theory on $\Sigma$, which matches the IFT$_2$ derived from symmetry reduction of the IFT$_4$ in \S~\ref{sec:4dIFTto2dIFT}.

\end{minipage}

\end{tcolorbox}
In the following discussion we will make use of the $\mc{E}$-model formulation of CS$_{4}$ \cite{Benini:2020skc, Lacroix:2020flf, Liniado:2023uoo}.
In this approach we accomplish localisation via algebraic means, constructing from the data of our CS$_{4}$ a \textit{defect algebra} and \textit{projectors}.
The choice of boundary conditions in CS$_{4}$ corresponds to a choice of two mutually orthogonal subspaces of our defect algebra, from which we can then write down the action and Lax connection for the corresponding 2d IFT.
To obtain the IFT$_{2}$ we could also follow an analogous route to that taken \S~\ref{S:Localisation of 6d to 4d}, namely integrating out the $\mathbb{CP}^1$ directions directly.
Details of this approach are presented in appendix \ref{S:4d CS to 2d IFT}.

The gauge field $\hat{\cA}$ is related to the 2-dimensional Lax connection by a change of variables that takes the form of a gauge transformation, which importantly does not preserve the boundary conditions,
\begin{equation}
\label{ec:4dAparametrisation}
\hat{\mc{A}} = \hat{h}^{-1} \dr \hat{h} + \hat{h}^{-1} \mathcal{L} \hat{h} ~.
\end{equation}
As before, we fix the redundancy in this parametrisation  by demanding that $\mathcal{L} $ has no legs in the $\mathbb{CP}^1$ direction, though may of course depend on it functionally, and that $\hat{h}\vert_\beta = \text{id}$.
The bulk equations of motion, $\omega \wedge F[\hat{\cal A}] = 0$,  ensures that $\mathcal{L}$ is flat and meromorphic in $\zeta$ with analytic structure mirroring $\omega$.
The key idea is that the field $\hat{h}$ evaluated at the poles serve as edge modes that become the degrees of freedom of the IFT$_{2}$, and the boundary conditions will determine the form of the Lax connection $\mathcal{L}$ in terms of these fields.
However, the complete construction requires a more careful treatment, especially when $\Omega$ has higher order poles \cite{Benini:2020skc}.

Let us start by recalling the boundary conditions eq.~\eqref{eq:4dBCdoubled} phrased in terms of $\mathbb{A}_w$ and $\mathbb{A}_{\bar{w}}$ valued in the defect algebra $\mathfrak{d} =\mathfrak{g} + \mathfrak{g}$.
These are that $\mathbb{A}_w \in \mathfrak{l}_t$ and $\mathbb{A}_{\bar{w}} \in \mathfrak{l}_{t^{-1}}$, where these subspaces are mutually orthogonal $\langle\!\langle \mathfrak{l}_t , \mathfrak{l}_{t^{-1} }\rangle\!\rangle= 0$.
Given these boundary conditions for the gauge field at the simple poles, we introduce a group valued field $\mathbbm{h} = (h\vert_\alpha,h\vert_{\tilde\alpha}) \in \mathbb{D} =\exp \mathfrak{d}$ and an algebra element $\mathbb{L} = (\mathcal{L}\vert_\alpha,\mathcal{L}\vert_{\tilde\alpha}) \in \mathfrak{d} $ such that
\begin{equation}
\begin{aligned}
\mathbb{A}_w &= \mathbbm{h}^{-1} \partial_w \mathbbm{h} + \mathbbm{h}^{-1} \mathbb{L}_w \mathbbm{h} \in \mathfrak{l}_t \\
\mathbb{A}_{\bar{w}}&= \mathbbm{h}^{-1} \partial_{\bar{w}} \mathbbm{h} + \mathbbm{h}^{-1} \mathbb{L}_{\bar{w}} \mathbbm{h} \in \mathfrak{l}_{t^{-1}}
\end{aligned}
\end{equation}
$\mathbb{L}$ can be understood as the 2d Lax connection lifted to the double by evaluating it at the poles of the spectral parameter.
From this, and the known singularity structure of $\hat{{\cal A}}$, we will recover the full Lax connection.

It is important to emphasise that most previous treatments have assumed that $\mathbb{A}_w$ and $\mathbb{A}_{\bar{w}}$ lie in the same subspace, and moreover that this space is a maximal isotropic subalgebra $\mathfrak{l} \subset \mathfrak{d}$.
The only exception that we know of are the chiral Dirichlet boundary conditions of \cite{Costello:2019tri,Ashwinkumar:2023zbu}, which are a special case of our boundary conditions.
Taking $\mathfrak{l}$ to be an isotropic subalgebra of $\mathfrak{d}$ ensures that the resulting IFT$_{2}$ has a residual gauge symmetry given by the left action of $\exp{(\mathfrak{l})}$ on $\mathbbm{h}$.
This can be fixed by setting $\mathbbm{h}\in \mathbb{D}/ \exp(\mathfrak{l})$.
For example, if we take $r_+ = - r_-$, then $\mathfrak{l}=\mathfrak{g}_{\text{diag}}$, the diagonally embedded $\mathfrak{g}$ in $\mathfrak{d}= \mathfrak{g} +\mathfrak{g}$, is a suitable isotropic subalgebra and denoting
$\mathbbm{h} = ( h, \tilde{h})$ the residual symmetry can be fixed by setting $\tilde{h} = \text{id}$.
Here, however, we do not have any such residual gauge symmetry in general and the 2-dimensional theory will depend on the entire field content in $\mathbbm{h}$.

As above, we switch notation in 2 dimensions to $\partial_w = \partial_+$ and $\partial_{\bar{w}} = \partial_-$.
The IFT$_{2}$ action is \cite{Liniado:2023uoo, Klimcik:2021bqm}\footnote{The action below is related to the action in \cite{Liniado:2023uoo} with the redefinition $\mathbbm{h}\to \mathbbm{h}^{-1}$. This is due to the convention on gauge transformations. Indeed, there they consider $A^h = h A h^{-1}-\mathrm{d}hh^{-1}$ in contrast to our choice $A^h=h^{-1}Ah + h^{-1}\mathrm{d}h$.}\footnote{Note that we have chosen $\frac{1}{2\pi i}$ as an overall coefficient in equation \eqref{ec:4dCSaction1}, in contrast to $\frac{i}{4\pi}$ considered in \cite{Liniado:2023uoo}.}
\begin{equation}\begin{aligned}\label{eq: 2d E-model action}
S_{\mathrm{2d}}(\mathbbm{h}) & =\int \Big(\langle \!\langle \partial_-\mathbbm{h}\mathbbm{h}^{-1},W_\mathbbm{h}^{+}(\partial_+\mathbbm{h}\mathbbm{h}^{-1})\rangle \!\rangle\\
& \qquad\qquad -\langle \!\langle \partial_+\mathbbm{h}\mathbbm{h}^{-1},W_\mathbbm{h}^{-}(\partial_-\mathbbm{h}\mathbbm{h}^{-1})\rangle \!\rangle\Big) \mathrm{d}\sigma^-\wedge \mathrm{d}\sigma^{+}+S_{\mathrm{WZ}}[\mathbbm{h}] ~.
\end{aligned}\end{equation}
The projectors $W_\mathbbm{h}^{\pm}: \mathfrak{d} \rightarrow \mathfrak{d}$ are defined via their kernel and image
\begin{equation}
\operatorname{Ker} W_\mathbbm{h}^{\pm}= \mathrm{Ad}_{\mathbbm{h}}\mathfrak{l}_{t^{\pm 1}} ~,
\qquad
\textrm{Im} W_\mathbbm{h}^{\pm}=  \bigg\{  \left( \frac{\mathsf{y}}{\alpha_+ - \gamma_\pm } , \frac{\mathsf{y}}{\alpha_- - \gamma_\pm } \right)  \, \bigg\vert \ \mathsf{y} \in \mathfrak{g} \, \bigg\}~,
\end{equation}
where we recall that $\zeta =\gamma_\pm$ are the zeroes of $\omega$.
Care is required to correlate the zeroes of omega with the pole structure of ${\cal A}$, which has been determined by our symmetry reduction data.
In the case at hand for instance, the zeroes at $\pi = \gamma$ and $\pi = \hat\gamma$ are associated to poles in ${\cal A}_w$ and in ${\cal A}_{\bar{w}}$ respectively.

In order to unpack the IFT$_{2}$ action, let us outline the calculation of
$W_\mathbbm{h}^{+}(\partial_+\mathbbm{h}\mathbbm{h}^{-1})$.
Defining the useful combinations
\unskip\footnote{In terms of these parameters, the relations~\eqref{eq:rpmst} become
\begin{equation*}
r_+ = K \frac{ u_+ v_+ }{\Delta \gamma \Delta \alpha } ~,
\qquad
r_- = -K \frac{ u_- v_- }{\Delta \gamma \Delta \alpha} ~,
\qquad
s = \sqrt{\frac{u_- v_- }{u_+ v_+}} ~,
\qquad
t = \sigma s \frac{u_+}{u_-} ~.
\end{equation*}
}
\begin{equation}\label{eq:upmvpm}
v_\pm = \alpha_\pm - \gamma_+ ~, \qquad u_\pm = \alpha_\pm - \gamma_- ~,
\end{equation}
we can parameterise the kernel and image of $W_\mathbbm{h}^{+}$ as
\begin{equation}
\textrm{Ker} W_\mathbbm{h}^{+}= \bigg\{ \left( \frac{\mathrm{Ad}_h \mathsf{x}}{v_+ } , \frac{\sigma^{-1} \mathrm{Ad}_h \Lambda^{-1} \mathsf{x}}{v_- } \right)\, \bigg\vert  \ \mathsf{x}  \in \mathfrak{g}\,    \bigg\}  \, , \qquad
\textrm{Im} W_\mathbbm{h}^{+}= \bigg\{ \left( \frac{\mathsf{y}}{v_+ } , \frac{\mathsf{y}}{v_- } \right)\, \bigg\vert  \ \mathsf{y} \in \mathfrak{g} \bigg\} ~.
\end{equation}
We decompose $\partial_+\mathbbm{h}\mathbbm{h}^{-1}$ into the kernel and image by solving
\begin{equation}
\left(\mathrm{Ad}_h J_+,\mathrm{Ad}_h \Lambda^{-1}\widetilde{J}_+ \right) =
\left( \frac{\mathrm{Ad}_h \mathsf{x}}{v_+ } , \frac{\sigma^{-1} \mathrm{Ad}_h \Lambda^{-1} \mathsf{x}}{v_- } \right) +
\left( \frac{\mathsf{y}}{v_+ } , \frac{\mathsf{y}}{v_- } \right) ~.
\end{equation}
This yields
\begin{equation}
\mathsf{y} = \mathrm{Ad}_h U_+ \left(J_+ v_+ -\sigma \widetilde{J}_+ v_- \right) =  v_- \P_+ ~,
\end{equation}
in which we see the reappearance of the auxiliary combinations encountered earlier in eq.~\eqref{eq:auxiliaries}.
It follows that
\begin{equation}
W_\mathbbm{h}^{+}(\partial_+\mathbbm{h}\mathbbm{h}^{-1}) = \left(\frac{v_-}{v_+}\P_+ , \P_+ \right) ~,
\end{equation}
from which we can evaluate (trace implicit)
\begin{equation}\label{eq:WpLL}
\begin{aligned}
& \langle \!\langle \partial_-\mathbbm{h}\mathbbm{h}^{-1},W_\mathbbm{h}^{+}(\partial_+\mathbbm{h}\mathbbm{h}^{-1})\rangle \!\rangle = \frac{v_-}{v_+ }r_+ J_- \mathrm{Ad}_h^{-1}\P_+ + r_- \widetilde{J}_- \Lambda \mathrm{Ad}_h^{-1}\P_- \\
& \qquad = r_+J_+ U_+^T J_- + r_- \widetilde{J}_+ U_- \widetilde{J}_- - t \sqrt{-r_- r_+} \widetilde{J}_+ U_+^T J_- + t^{-1} \sqrt{-r_- r_+} J_+ U_- \widetilde{J}_-
~.
\end{aligned}
\end{equation}
In a similar fashion we find that
\begin{equation}
W_\mathbbm{h}^{-}(\partial_-\mathbbm{h}\mathbbm{h}^{-1}) = \left(\P_- , \frac{u_+}{u_-} \P_- \right) ~,
\end{equation}
and
\begin{equation}\label{eq:WmLL}
\begin{aligned}
& \langle \!\langle \partial_+\mathbbm{h}\mathbbm{h}^{-1},W_\mathbbm{h}^{-}(\partial_-\mathbbm{h}\mathbbm{h}^{-1})\rangle \!\rangle\\
&\qquad = r_+ J_+ U_- J_- + r_- \widetilde{J}_- U_+^T \widetilde{J}_++ t \sqrt{-r_- r_+} \widetilde{J}_+ U_+^T J_- - t^{-1} \sqrt{-r_- r_+} J_+ U_-\widetilde{J}_- ~.
\end{aligned}
\end{equation}
Taking the difference of eq.~\eqref{eq:WpLL} and eq.~\eqref{eq:WmLL} we find that the Lagrangian of the 2-dimensional action eq.~\eqref{eq: 2d E-model action} exactly matches the IFT$_{2}$ obtained previously in eq.~\eqref{eq:IFT2} by descent on the other side of the diamond.
This explicitly verifies our diamond of theories.

Let us note that this IFT$_2$ has also been constructed from CS$_4$ in a two-step process in \cite{Bassi:2019aaf}.
First, a more general 2-field model based on a twist function with additional poles and zeroes, and the familiar isotropic subalgebra boundary conditions, is constructed.
Second, a special decoupling limit is taken, where a subset of these poles and zeroes collide.
It remains to understand how to recover our boundary conditions~\eqref{eq:4dBC} from those considered in \cite{Bassi:2019aaf}.

To complete the circle of ideas we can also directly obtain a Lax formulation from CS$_4$.
This is essentially achieved by undoing the map into the defect algebra as follows.
Given an element $\mathbb{X} =(x,y) \in \mathfrak{d}$, we determine $a,b \in \mathfrak{g}$ such that
\begin{equation}
(x,y) = \left( \frac{a}{u_+}+ \frac{b}{v_+} , \frac{a}{u_+}+ \frac{b}{v_-} \right) ~.
\end{equation}
We introduce a map $\wp$ into the space of $\mathfrak{g}$-valued meromorphic functions
\begin{align}
\wp: \mathbb{X}\mapsto \frac{a}{\zeta - \gamma_-} + \frac{b}{\zeta - \gamma_+ } ~,
\end{align}
in terms of which the components of the Lax connection are given by
\begin{equation}
{\cal L}_\pm = \wp W_\mathbbm{h}^{\pm}(\partial_\pm\mathbbm{h}\mathbbm{h}^{-1}) = \frac{\alpha_\mp-\gamma_\pm}{\zeta - \gamma_\pm}\P_\pm ~.
\end{equation}
In this way we recover the Lax $\mathcal{L}^1$ that we obtained from symmetry reduction of the 4d ASDYM Lax pair with the identification~\eqref{eq:zgsident}.  There does not appear an algebraic derivation in this spirit of the other Lax  $\mathcal{L}^2$.
This in contrast to the derivation of this model from CS$_4$ in \cite{Bassi:2019aaf} where the extra data associated to the additional poles means that both Lax in~\eqref{eq:twoLax} can be directly constructed.
More generally, this highlights an interesting question that we leave for the future about the integrability and the counting of conserved charges, beyond the existence of a Lax connection, when we consider boundary conditions not based on isotropic subalgebras.

\subsection{RG Flow}
Let us recall the RG equations given in \cite{Georgiou:2017jfi}
\begin{equation}\label{eq:RGlambda}
\dot{\lambda}_i= - \frac{c_{G}}{2 \sqrt{ k_1 k_2} } \frac{ \lambda_i^2 ( \lambda_i - \lambda_0 ) ( \lambda_i - \lambda_0^{-1} ) }{ (1- \lambda_i^2)^2 } ~, \quad i=1,2~,
\end{equation}
where dot indicates the derivative with respect to RG `time' $\frac{d}{d\log \mu}$ and $c_{G}$ is the dual Coxeter number. The levels $k_1$ and $k_2$ and $\lambda_0 = \sqrt{k_1/k_2}$ are RG invariants.
In this section we will interpret this flow in terms of the data that is more natural from the perspective of 4d CS, namely the poles and zeroes of the differential
\begin{equation}
\omega = \frac{K}{\Delta \gamma } \frac{ (\zeta- \gamma_+) (\zeta- \gamma_-)}{ (\zeta -\alpha_+)(\zeta -\alpha_-) } \dr\zeta = \varphi(\zeta) \, \dr \zeta ~, \label{eq:omega}
\end{equation}
and the boundary conditions of the theory.

Using the map between parameters given in eq.~\eqref{eq:GSparam} we can infer from eq.~\eqref{eq:RGlambda} a flow on the parameters $\{ t, \alpha_\pm , \gamma_\pm, K\}$.
Let us first consider the parameter $t=\lambda_2$.
As discussed in \cite{Georgiou:2017jfi}, there is a flow from $t=0$ in the UV to $t = \lambda_0$ in the IR (assuming that $\lambda_0<1$).
Explicitly the flow equation
\begin{equation}
\dot{t} = \frac{c_{G}}{ 2 k_2 \lambda_0 } \frac{t^2}{(1-t^2)^2} (t- \lambda_0)(t- \lambda_0^{-1}) ~,
\end{equation}
has the solution
\begin{equation}
f(\lambda_0, t) + f(\lambda_0^{-1}, t) +t + t^{-1} =\frac{c_{G}}{ 2\sqrt{k_1 k_2} } \log \mu / \mu_{t_0} ~, \quad f(x,t)= x \log \left( \frac{t^{-1}-x}{t- x} \right) ~.
\end{equation}
The interesting observation is that the boundary conditions
\begin{equation}
\label{ec:tzeroBC}
\begin{aligned}
\mathbb{A}_{w} \in \mathfrak{l}_t &=\textrm{span} \{ \left( t \lambda_0^{-1} x, x \right) \mid x \in \mathfrak{g} \} ~, \\
\mathbb{A}_{\bar{w}} \in \mathfrak{l}_{t^{-1}} &=\textrm{span} \{ \left( \lambda_0^{-1} x, t x \right) \mid x \in \mathfrak{g} \} ~,
\end{aligned}\end{equation}
display algebraic enhancements at the fixed points.
In the UV, $t=0$ limit, these boundary conditions become chiral, $\mathbb{A}_w \in \mathfrak{g}_R \subset\mathfrak{d}$ and $\mathbb{A}_{\bar{w}} \in \mathfrak{g}_L \subset\mathfrak{d}$.
While $\mathfrak{g}_{L,R}$ are now subalgebras, neither are isotropic with respect to the inner product \eqref{eq:innerint}.
In non-doubled notation the UV limit becomes
\begin{equation} \label{eq:UVBC}
\hat{{\cal A}}_w \vert_\alpha = 0 ~, \qquad \hat{{\cal A}}_{\bar{w}} \vert_{\tilde{\alpha}} = 0 ~.
\end{equation}
On the other hand in the IR limit, $t= \lambda_0$, we see that $\mathbb{A}_{w} \in \mathfrak{g}_{\textrm{diag}} \subset\mathfrak{d}$, again a subalgebra, but only an isotropic one for $k_1 = k_2$, i.e.~$r_+ = -r_-$.
In non-doubled notation the IR limit becomes
\unskip\footnote{The seemingly more democratic boundary condition of $t=1$,
$$
\sqrt{k_1} \hat{{\cal A}}_{ w} \vert_\alpha = \sqrt{k_2} \hat{{\cal A}}_{ w} \vert_{\tilde{\alpha}} ~, \quad \sqrt{k_1} \hat{{\cal A}}_{\bar w} \vert_\alpha = \sqrt{k_2} \hat{{\cal A}}_{\bar{w}} \vert_{\tilde{\alpha}} ~
$$
which does define an isotropic space of $\mathfrak{d}$ (not a subalgebra however) is {\em not} attained along this flow.
}
\begin{equation} \label{eq:IRBC}
\hat{{\cal A}}_w \vert_\alpha = \hat{{\cal A}}_w \vert_{\tilde{\alpha}} ~, \qquad k_1 \hat{{\cal A}}_{\bar w} \vert_\alpha = k_2 \hat{{\cal A}}_{\bar{w}} \vert_{\tilde{\alpha}} ~.
\end{equation}
While in general, there are no residual gauge transformations preserving the boundary conditions, in the UV and IR limits we notice chiral boundary symmetries emerging.
For example, in the IR these are those satisfying $g^{-1} \pd_{\bar{w}} g = 0$, which corresponds to $t = s^{-1}$ in eq.~\eqref{eq:6dgaugetransfred}.

Let us now turn to the action of RG on the differential $\omega$.
An immediate observation is that the RG invariant WZW levels are given by monodromies about simple poles
\unskip\footnote{The monodromy about the double pole at infinity is trivially RG invariant since the sum of all the residues vanishes.}
\begin{equation}
\pm k_{1,2}= r_\pm = \frac{1}{2\pi i} \oint_{\alpha_\pm} \omega = \textrm{res}_{\zeta = \alpha_\pm } \varphi(\zeta) ~,
\end{equation}
exactly in line with the conjecture of Costello (reported and supported by Derryberry \cite{Derryberry:2021rne}).
While there are more parameters in $\omega$ than there are RG equations, we can form the ratios of poles and zeroes
\begin{equation}
q_\pm = \frac{\alpha_\pm - \gamma_+ }{\alpha_\pm -\gamma_-} = \frac{v_\pm}{u_\pm} ~,
\end{equation}
in terms of which the RG system of \cite{Georgiou:2017jfi} translates to
\begin{equation}
\dot{q}_\pm = -\frac{c_{G}}{2 K } \frac{(1+q_\mp)}{(-1 + q_\mp)} q_\pm ~, \qquad \dot{K}= -\frac{c_{G}}{ 2 } \frac{q_-+q_+}{(1-q_-)(1-q_+)} ~.
\end{equation}
The RG invariants are given by
\begin{equation}\label{eq:rginvariants}
k_1 k_2 = \frac{ K^2 q_- q_+}{(q_+ -q_-)^2} ~, \qquad \frac{k_1 }{k_2} = \lambda_0^{2} = \frac{q_+}{(1-q_+)^2}\frac{(1-q_-)^2}{q_-} ~,
\end{equation}
which allows us to retain either of $q_\pm$ as independent variables.
We can directly solve these equations
\begin{equation}
\begin{aligned}
\sqrt{k_1 k_2} \frac{q_+ - q_- }{\sqrt{q_+ q_-}} + k_1 \log q_+ - k_2 \log q_- =\frac{c_{G}}{ 2 } \log \mu / \mu_{q_0} ~,
\end{aligned}
\end{equation}
and a remarkable feature, also conjectured by Costello, is that this quantity is precisely the contour integral between zeroes
\begin{equation}
\frac{\dr }{\dr \log \mu} \int_{\gamma_-}^{\gamma_+}\omega = \frac{c_{G} }{2} ~.
\end{equation}

\begin{figure}
\centering
\includegraphics[width=5cm]{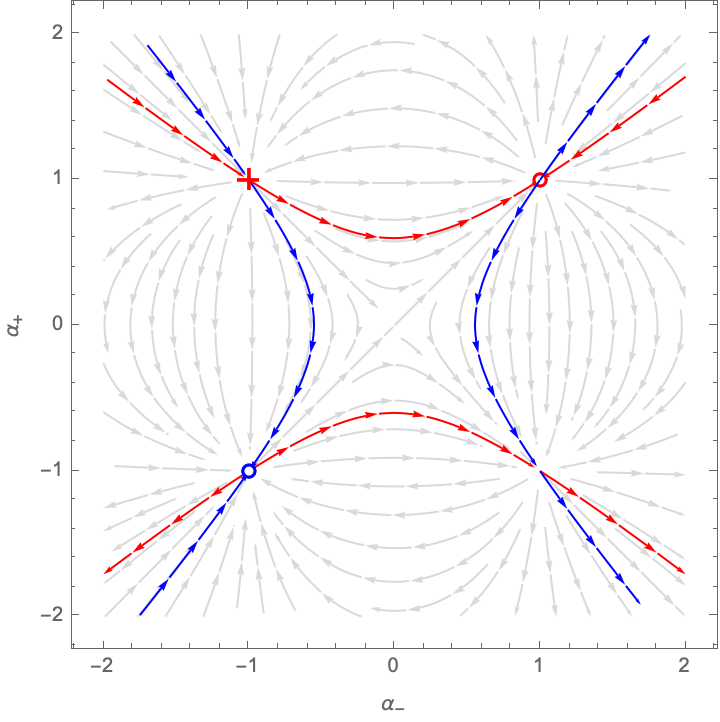}
\includegraphics[width=5cm]{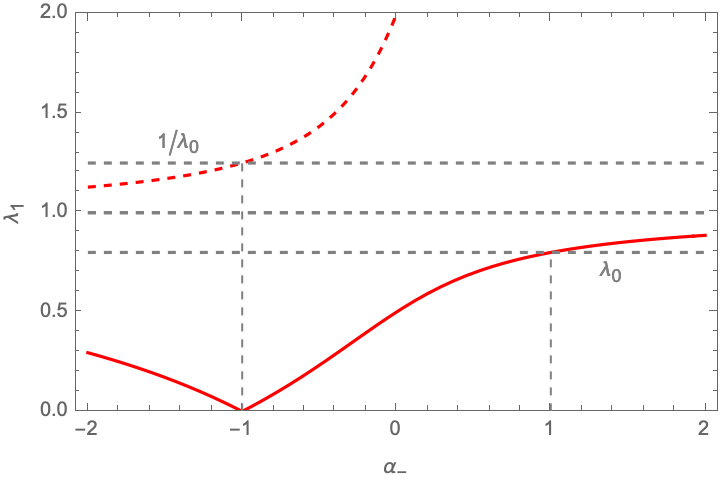}
\caption{\footnotesize{Left: RG flow across the $\alpha_+$, $\alpha_-$ plane with arrows directed to the IR.
The highlighted parabola are the solutions that lie on the locus of the RG invariant quantity $\lambda_0^2 =k_1 /k_2$, plotted here for $\lambda_0= 0.8$ (red) and $\lambda_0 =1.2$ (blue).
Right: The value of $\lambda_1$ plotted along the red loci of the left panel (upper branch solid and lower branch dotted).
In both cases $\lambda_1 \rightarrow 1$ asymptotically as $\alpha_- \rightarrow\pm \infty$.
Of note is the flow displayed by the upper red branch between the UV fixed point $(\alpha_-, \alpha_+) = (-1,1)$ with $ \lambda_1 =0$ and the IR fixed point $(\alpha_-, \alpha_+) = (1,1)$ with $ \lambda_1 =\lambda_0$.}}
\label{fig:enter-label}
\end{figure}

To best understand the action of the RG flow on the locations of the poles directly, we replace $K$ with the RG invariant $k_2$ (or $k_1$), and fix the zeroes to be located at $\gamma_\pm=\pm 1$.
This yields the RG invariant relation
\begin{equation}
1-\alpha_+^2 - \lambda_0^2 (1-\alpha_-^2) = 0 ~,
\end{equation}
and a flow equation
\begin{equation}
\label{eq:alphaminusdot}
\dot\alpha_- = \frac{c_{G} } {8 k_2} \frac{ \alpha_+ (1-\alpha_-^2)^2}{\alpha_- - \alpha_+} ~,
\end{equation}
the solution of which is
\begin{equation}
\frac{\alpha_+ - \alpha_-}{1- \alpha_+^2} + \frac{1}{2}\log \frac{\alpha_+ +1 }{\alpha_+ - 1} -\frac{1}{2 \lambda_0^2}\log \frac{\alpha_- +1 }{\alpha_- - 1 } = \frac{c_{G}}{4 k_1 } \log \mu/\mu_{\alpha_0} ~.
\end{equation}
As illustrated in fig.~\ref{fig:enter-label}, this system displays a finite RG trajectory linking fixed points.
In the UV limit the poles accumulate to different zeroes, and in the IR the poles accumulate to the same zero.
Let us consider the upper red trajectory of fig.~\ref{fig:enter-label} in which we choose $\lambda_0<1$ and pick the positive branch of the solution $\alpha_+ = + \sqrt{1- \lambda_0^2 (1-\alpha_-^2)}$.
With this choice we see that there are finite fixed points
\unskip\footnote{There are also fixed points to the RG flow at $\alpha_+=0$ with $\alpha_-^2 = 1 - \frac{k_2}{k_1}$ however by assumption $k_2>k_1$, and so these do not correspond to real values of $\alpha_-$ and consequently $\lambda_1$ is imaginary.
We do not consider such complex limits here.}
such that the right hand side of eq.~\eqref{eq:alphaminusdot} vanishes at
\begin{equation}
\textrm{UV}: \quad (\alpha_- , \alpha_+) = (-1,1) ~, \quad \lambda_1 = 0 ~, \qquad \textrm{IR}: \quad (\alpha_- , \alpha_+) = (1,1) \, , \quad \lambda_1 = \lambda_0 ~, \qquad
\end{equation}
in which we recall the map
\begin{equation}
\lambda_1 = \left( \frac{(1+\alpha_-)(-1+\alpha_+)}{(-1+\alpha_-)(1+\alpha_+)} \right)^{\frac{1}{2}} ~.
\end{equation}

One of the appealing features of the IFT$_{2}$~\eqref{eq:IFT2} is that it provides a classical Lagrangian interpolation that includes its own UV and IR limits \cite{Georgiou:2017jfi}.
That is to say these CFTs can be obtained directly from the Lagrangian eq.~\eqref{eq:IFT2} by tuning the parameters of the theory to their values at the end points of the RG flow.
Given the interpretation of these RG flows as describing poles colliding with zeroes it is natural to expect that a similar interpolation can be obtained directly in 4d by taking limits of the differential $\omega$ in eq.~\eqref{eq:omega}.

Here we will explore how this works for the IFT$_2$~\eqref{eq:IFT2} in the IR.
The limit we will consider is to collide the poles at $\alpha_\pm$ with the zero at $\gamma_+$, following the upper red trajectory in fig.~\ref{fig:enter-label}.
This corresponds to taking $q_\pm \to 0$.
In order to be consistent with the RG invariants ~\eqref{eq:rginvariants}, we take this limit as
\begin{equation}
q_+ = k_1 \epsilon + \mathcal{O}(\epsilon^2) ~, \qquad q_- = k_2 \epsilon + \mathcal{O}(\epsilon^2) ~, \qquad K = k_1 - k_2 + \mathcal{O}(\epsilon) ~, \qquad \epsilon \to 0 ~.
\end{equation}
Taking this limit in~\eqref{eq:omega}, and redefining the spectral parameter such that the remaining pole and zero are fixed to $1$ and $-1$ respectively, yields
\begin{equation}
\label{ec:IRomega1}
\omega \rightarrow  \frac{k_1- k_2}{2} \frac{ \zeta +1}{\zeta -1 }d\zeta ~.
\end{equation}

Let us consider the implication of this limit from the CS$_4$ perspective.
Given that the pole structure of $\omega$ is modified in this limit, so will the double $\mathfrak{d}$, and thus we should be careful in our interpretation of the boundary conditions.
If we take $\omega$ to be given by \eqref{ec:IRomega1}
and consider the boundary conditions \eqref{ec:tzeroBC} with $t=0$, the condition $\mathbb{A}_{w} \in \mathfrak{l}_{t}$ becomes $\hat{\cA}_{w}\vert_{\alpha}=0$.
From eq.~\eqref{eq:AwAwb} we know that $\hat{\cA}_{w}$ has a pole at $\gamma_+$.
In other words, we can write
\begin{equation}
\hat{\cA}_{w}= \frac{(\zeta - \alpha_+)}{(\zeta - \gamma_+)}\Xi(\zeta) ~,
\end{equation}
with $\Xi(\zeta)$ regular as $\alpha_+ \to \gamma_+$.
Hence, in the IR there is no boundary condition for $\hat{\cA}_{w}$ at $\zeta=1$.
On the other hand, the boundary condition $\mathbb{A}_{\bar w} \in \mathfrak{l}_{t^{-1}}$ for $t=0$ is $\hat{\cA}_{\bar w}\vert_{\bar \alpha}=0$ which in the limit $\alpha_+,\gamma_+ \to 1$ becomes a chiral boundary condition for the $\bar w$ component
\begin{equation}
\hat{\cA}_{\bar w}\vert_{\zeta=1} = 0 ~.
\end{equation}
For this choice of boundary condition one can localise the CS$_4$ action following the procedure described in either \S~\ref{sec:4dhcsred} or appendix \ref{S:4d CS to 2d IFT}, and the resulting two dimensional IFT is the WZW model at level $k_2-k_1$.

In contrast, from the 2-dimensional perspective it is known that the full result at this IR fixed point is actually a product WZW model on $G_{k_2}\times G_{k_1-k_2}$ \cite{Georgiou:2017jfi}.
This indicates that there is some delicacy in taking the IR limit directly as a Lagrangian interpolation in 4d even when it is possible in 2d.
One reason for this is there is also the freedom to perform redefinitions of the spectral parameter, which can, in general, produce non-equivalent limits of $\omega$.
Such limits are known as decoupling limits~\cite{Delduc:2019bcl,Bassi:2019aaf} in the literature, and have been investigated for the UV fixed point of the bi-Yang-Baxter model in \cite{Kotousov:2022azm}.

\section{Discussion and Outlook}\label{sec:outlookintdef}

In this work we have constructed a diamond of integrable models related by localisation and symmetry reduction.
Starting from holomorphic Chern-Simons theory with the meromorphic (3,0)-form~\eqref{eq:Omegaintdef}, we have found a new choice of admissible boundary conditions, which leads to a well-defined 6-dimensional theory.
This generalises the analysis carried out in \cite{Bittleston:2020hfv,Penna:2020uky} to a new class of boundary conditions not of Dirichlet type.

By first viewing twistor space, $\bb{PT}$, as a $\bb{CP}^{1}$ bundle over $\bb{E}^{4}$, we solved the equations of motion along the $\bb{CP}^{1}$ fibres.
In doing so we fully specified the dependence of the integrand on the $\bb{CP}^{1}$ fibre and thus could perform a fibrewise integration along those directions.
Consequently our 6-dimensional theory then localised to the poles of $\Omega$, leading to a new 4-dimensional theory on $\mathbb{E}^4$ given by the action~\eqref{eq:IFT4}.
Indeed, this 4-dimensional theory is ‘integrable’ in the sense that its equations of motion can be encoded in an anti-self-dual connection, as expected from the Penrose-Ward correspondence.
Moreover, this new IFT$_4$ exhibits two semi-local symmetries, which can be understood as the residual symmetries preserving the boundary conditions of hCS$_6$.
For each of these semi-local symmetries, the Noether currents can be used to construct inequivalent Lax formulations of the dynamics.

On the other hand, symmetry reducing hCS$_6$ along two directions of the $\mathbb{E}^4$ in $\bb{PT}\cong \mathbb{CP}^1\times \mathbb{E}^4$, leads to an effective CS$_4$ theory on $\mathbb{CP}^1\times \mathbb{E}^2$.
Under this procedure, the meromorphic $(3,0)$-form reduces to the meromorphic $(1,0)$-form used in \cite{Delduc:2019whp} to construct the $\lambda$-model, whereas the 6-dimensional boundary conditions reduce to a class of boundary conditions in CS$_4$ that have not been previously considered.
Specifically, they relax the assumption of an isotropic subalgebra of the defect algebra.
By performing the standard localisation procedure of CS$_4$ we obtain the 2-field $\lambda$-type IFT$_2$ introduced in \cite{Georgiou:2017jfi}.

Notably, this same multi-parametric class of integrable $\lambda$-deformations between coupled WZW models can be obtained by symmetry reduction (along the same directions) of the novel IFT$_4$ mentioned above.
Furthermore, the semi-local symmetries of the IFT$_4$ reduce to global symmetries of the IFT$_2$ and the two Lax formulations of the IFT$_4$ give rise to two Lax connections for the IFT$_2$.
When the directions of the symmetry reduction are aligned to these semi-local symmetries, the IFT$_2$
symmetries are enhanced to either affine or fully local (gauge) symmetries.
In the latter case, the
IFT$_2$ becomes the standard (1-field) $\lambda$-model.

\medskip

This work opens up a range of interesting further directions.
There are a selection of direct generalisations that can be made to incorporate the wide variety of integrable deformations known in the literature.
Perhaps the most interesting outcome of this would be the construction of swathes of new four-dimensional integrable field theories.
Our work focused on the case where $\Omega$ was nowhere vanishing; it would be interesting to explore the relaxation of this condition together with its possible boundary conditions, and how the ASDYM equations are modified.
Moreover, one might hope that the study of boundary conditions in hCS$_{6}$ could lead to a full classification of the landscape of integrable sigma-models in 2d, and perhaps result in theories not yet encountered in the literature.

From the perspective of the IFT$_2$, there is a close relationship between the notions of Poisson-Lie symmetry, duality and integrability \cite{Vicedo:2015pna,Hoare:2015gda,Sfetsos:2015nya,Klimcik:2015gba}.
This poses an interesting question as to the implications of such dualities for both the IFT$_4$ and hCS$_6$.
For the model considered here, we might seek to understand the semi-local symmetries of IFT$_{4}$ in the context of the $q$-deformed symmetries expected to underpin the IFT$_{2}$.

In this work, the integrable models we have studied can be viewed as descending from the open string sector of a type B topological string.
An interesting direction for future work is to consider the closed string sector \cite{Bershadsky:1993cx} and its possible integrable descendants.
A tantalising prospect is to understand the closed string counterparts of the integrable deformations we have considered in the context of the non-linear graviton construction for self-dual space-times \cite{Penrose:1976js, Hitchin:1979rts,Ward:1980am}.

By coupling the open and closed string sectors \cite{Costello:2015xsa, Costello:2018zrm, Costello:2019jsy, Costello:2021bah} one can find an anomaly free quantization to all loop orders in perturbation theory of the coupled hCS$_{6}$-BCOV action.
This mechanism has already proven a powerful tool in the context of the top-down approach to celestial holography \cite{Costello:2023hmi,Bittleston:2022nfr}.
This could provide an angle of attack to address the important questions of when the IFT$_4$ can be quantised, if the IFT$_4$ is {\em quantum} integrable, and if there is a higher dimensional origin of IFT$_{2}$ as quantum field theories.

\section{Appendix}

\subsection{Appendix A: Conventions}

\subsubsection{Twistor Space and Homogeneous Coordinates}
\label{appendix:twistors}

We begin by reviewing the construction of twistor space originally introduced by Penrose \cite{Penrose:1967wn,Penrose:1968me,Penrose1969SolutionsOT}.
For further references see \cite{Adamo:2017qyl, Sharma:2022arl}.
The projective spin bundle $ \mathbb{CP}^1\hookrightarrow \mathbb{PS}_\bC \twoheadrightarrow \bC^4 $ can be trivialised with coordinates $x^{A A^\prime} \in \bC^4$ and $\pi_{A^\prime} \in \mathbb{CP}^1$ where this is defined relative to the equivalence relation $\pi_{A^\prime} \sim r \pi_{A^\prime}$ for $r \in \bC \backslash \{ 0 \}$ and in which indices run over $A \in \{ 1, 2 \}$ and $A^\prime \in \{ 1^\prime, 2^\prime \}$.

Twistor space $\mathbb{PT}_{\mathbb{C}}$ is defined to be the image of $\mathbb{PS}_{\mathbb{C}}$ under the projection
\begin{equation}
\begin{aligned}
& p : \mathbb{PS}_{\mathbb{C}} \to \mathbb{CP}^3 ~, \quad \mathbb{PT}_{\mathbb{C}} \coloneqq p (\mathbb{PS}_{\mathbb{C}}) \subset \mathbb{CP}^3 ~, \\
& p : (x^{A A^\prime}, \pi_{A^\prime}) \mapsto (\omega^{A}, \pi_{A^\prime}) = (x^{A A^\prime} \pi_{A^\prime}, \pi_{A^\prime}) ~.
\end{aligned}
\end{equation}
Here, we are using homogeneous coordinates $Z^\alpha = (\omega^{A}, \pi_{A^\prime}) \in \mathbb{CP}^3$ defined up to the equivalence relation $Z^\alpha \sim r Z^\alpha$ for $r \in \bC \backslash \{ 0 \}$.
Notice that the image of $\mathbb{PS}_\bC$ covers all of $\mathbb{CP}^3$ save for a $\mathbb{CP}^1$ worth of points defined by $\pi_{A^\prime} = (0, 0)$.
Hence
\begin{equation}
\mathbb{PT}_\bC = \mathbb{CP}^3 \, \backslash \, \mathbb{CP}^1 = \big\{ Z^\alpha = (\omega^{A}, \pi_{A^\prime}) \ \vert \ Z^\alpha \sim r Z^\alpha \ , \ \pi_{A^\prime} \neq (0, 0) \big\} ~.
\end{equation}

In this work we shall be predominantly interested in the restriction to Euclidean twistor space $\mathbb{PT}_{\mathbb{E}}$.
This follows similarly to above but now starting with the Euclidean projective spin bundle obtained by taking a real slice of the base $\bC^4$.
On this slice, we can be more explicit about the base manifold coordinates $x^{A A^\prime} \in \bE^4$, which can be written in terms of four real variables $x^0, x^2, x^2, x^3 \in \bR$ as
\begin{equation}
\bE^4 ~: \quad x^{A A^\prime} =
\frac{1}{\sqrt{2}}
\begin{pmatrix}
x^0 + \mathrm{i} x^1 & \mathrm{i} x^3 - x^2 \\
\mathrm{i} x^3+ x^2 & x^0 - \mathrm{i} x^1
\end{pmatrix} ~.
\end{equation}
Our choice of orientation is such that $\vol_4 = \dr x^0 \wedge \dr x^1 \wedge \dr x^2 \wedge \dr x^3 $.
For this special case of Euclidean signature,
\unskip\footnote{For brevity, we suppress the subscript $\mathbb{E}$ in the main document where we will always work in Euclidean signature, unless specified otherwise.} the image of the projection map, hence the twistor space is unchanged,
\begin{equation}
\mathbb{PT}_\bE = \mathbb{CP}^3 \, \backslash \, \mathbb{CP}^1 = \big\{ Z^\alpha = (\omega^{A}, \pi_{A^\prime}) \ \vert \ Z^\alpha \sim r Z^\alpha \ , \ \pi_{A^\prime} \neq (0, 0) \big\} ~.
\end{equation}
Furthermore, for Euclidean space the projection map is invertible, hence $\mathbb{PT}_\bE$ and $\mathbb{PS}_\bE$ are diffeomorphic as real manifolds.
Importantly, the complex structure on $\mathbb{PT}_\bE$ inherited from being a subset of $\mathbb{CP}^3$ is not trivially expressed in terms of the coordinates on $\mathbb{PS}_\bE$.
Nonetheless, we are free to use two choices of coordinates on $\mathbb{PT}_\bE$, viewing it either as a subset of $\mathbb{CP}^3$, or as diffeomorphic to $\mathbb{PS}_\bE = \bE^4 \times \mathbb{CP}^1$.

We will now give this diffeomorphism concretely by introducing additional notation for the $\mathbb{CP}^3$ coordinates on $\mathbb{PT}_\bE$.
We define the quaternionic conjugation operation (denoted by $\hat{\bcdot}$) to be
\begin{equation}
\begin{aligned}
\omega^{A} = (\omega^1, \omega^2) & \mapsto \hat{\omega}^{A} = (-\overline{\omega^2}, \overline{\omega^1}) ~, \\
\pi_{A^\prime} = (\pi_{1^\prime}, \pi_{2^\prime}) & \mapsto \hat{\pi}_{A^\prime} = (-\overline{\pi_{2^\prime}}, \overline{\pi_{1^\prime}}) ~.
\end{aligned}
\end{equation}
Notice that this operation squares to $-1$ so must be applied four times to return to the original value, hence the name.

We also define inner products given by
\begin{equation}
\begin{aligned}
& \Vert \omega \Vert^2 = [\omega \hat{\omega}] = \omega^A \hat{\omega}_A = \varepsilon_{AB} \omega^A \hat{\omega}^B = \varepsilon^{AB} \omega_B \hat{\omega}_A ~, \\
& \Vert \pi \Vert^2 = \langle \pi \hat{\pi} \rangle = \pi^{A^\prime} \hat{\pi}_{A^\prime} = \varepsilon_{A^\prime B^\prime} \pi^{A^\prime} \hat{\pi}^{B^\prime} = \varepsilon^{A^\prime B^\prime} \pi_{B^\prime} \hat{\pi}_{A^\prime} ~.
\end{aligned}
\end{equation}
These may be explicitly written in terms of the components as
\begin{equation}
\begin{aligned}
& \Vert \omega \Vert^2 = \omega^1 \overline{\omega^1} + \omega^2 \overline{\omega^2} = \omega_1 \overline{\omega_1} + \omega_2 \overline{\omega_2} ~, \\
& \Vert \pi \Vert^2 = \pi_{1^\prime} \overline{\pi_{1^\prime}} + \pi_{2^\prime} \overline{\pi_{2^\prime}} = \pi^{1^\prime} \overline{\pi^{1^\prime}} + \pi^{2^\prime} \overline{\pi^{2^\prime}} ~.
\end{aligned}
\end{equation}
Here, we are using the raising and lowering conventions
\begin{equation}
\begin{aligned}
\omega^A & = \varepsilon^{AB} \omega_B ~, &
\omega_A & = \varepsilon_{AB} \omega^B ~, &
\varepsilon^{AC} \varepsilon_{CB} & = \delta^A_B ~, \\
\pi^{A^\prime} & = \varepsilon^{A^\prime B^\prime} \pi_{B^\prime} ~, &
\pi_{A^\prime} & = \varepsilon_{A^\prime B^\prime} \pi^{B^\prime} ~, &
\varepsilon^{A^\prime C^\prime} \varepsilon_{C^\prime B^\prime} & = \delta^{A^\prime}_{B^\prime} ~.
\end{aligned}
\end{equation}
Explicitly, the values of the anti-symmetric tensors are chosen to be
\begin{equation}
\varepsilon_{12} = +1 ~, \quad
\varepsilon_{1^\prime 2^\prime} = +1 ~, \quad
\varepsilon^{12} = -1 ~, \quad
\varepsilon^{1^\prime 2^\prime} = -1 ~.
\end{equation}
Using these conventions, the action of the conjugation on the inverted indices is
\begin{equation}
\hat{\omega}_A = (-\overline{\omega_2}, \overline{\omega_1}) ~, \quad
\hat{\pi}^{A^\prime} = (-\overline{\pi^{2^\prime}}, \overline{\pi^{1^\prime}}) ~.
\end{equation}

Now, with this notation in place, we can explicitly construct the diffeomorphism from $\mathbb{PT}_\bE$ to $\mathbb{PS}_\bE$.
The inverse of the projection map is given by
\begin{equation}
\begin{aligned}
& p^{-1} : \mathbb{PT}_\bE \to \mathbb{PS}_\bE ~, \\
& p^{-1} : (\omega^{A}, \pi_{A^\prime}) \mapsto (x^{A A^\prime}, \pi_{A^\prime}) = \bigg( \frac{\hat{\omega}^A \pi^{A^\prime} - \omega^A \hat{\pi}^{A^\prime}}{\Vert \pi \Vert^2} , \pi_{A^\prime} \bigg) ~.
\end{aligned}
\end{equation}
When doing calculations on Euclidean twistor space, we have two natural choices of homogeneous coordinates.
We can view $\mathbb{PT}_\bE$ as a subspace of $\mathbb{CP}^3$ and work with the coordinates $Z^\alpha = (\omega^{A}, \pi_{A^\prime})$ defined up to the equivalence relation $Z^\alpha \sim r Z^\alpha$ for $r \in \bC \backslash \{ 0 \}$.
Alternatively, we can view $\mathbb{PT}_\bE$ as diffeomorphic to $\mathbb{PS}_\bE = \bE^4 \times \mathbb{CP}^1$ and use the coordinates $(x^{A A^\prime}, \pi_{A^\prime})$ defined up to the equivalence relation $\pi_{A^\prime} \sim r \pi_{A^\prime}$ for $r \in \bC \backslash \{ 0 \}$.

\subsubsection{Basis of Forms and Vector Fields}
\label{appendix:basis}
Thinking of $\mathbb{PT}_\bE$ as diffeomorphic to $\mathbb{PS}_\bE = \bE^4 \times \mathbb{CP}^1$ and using the coordinates $(x^{A A^\prime}, \pi_{A^\prime})$, we can define a basis of $1$-forms by
\begin{equation}
\begin{aligned}
& e^0 = \langle \pi \dr \pi \rangle ~, &&
& e^A = \pi_{A^\prime} \dr x^{A A^\prime} ~, \\
& \bar{e}^0 = \frac{\langle \hat{\pi} \dr \hat{\pi} \rangle}{\langle \pi \hat{\pi} \rangle^2} ~, &&
& \bar{e}^A = \frac{\hat{\pi}_{A^\prime} \dr x^{A A^\prime}}{\langle \pi \hat{\pi} \rangle} ~.
\end{aligned}
\end{equation}
This basis of $1$-forms is split into the $(1,0)$-forms and $(0,1)$-forms with respect to the complex structure inherited from $\mathbb{CP}^3$.
They should also be understood as being valued in line bundles, since although they are normalised to have weight zero under the rescaling $\hat{\pi}_{A^\prime} \sim \bar{r} \, \hat{\pi}_{A^\prime}$, they carry weight under $\pi_{A^\prime} \sim r \pi_{A^\prime}$.

The dual basis of vector fields is given by
\begin{equation}
\label{eq:twistorderivativesint}
\begin{aligned}
& \pd_0 = \frac{\hat{\pi}_{A^\prime}}{\langle \pi \hat{\pi} \rangle} \frac{\pd}{\pd \pi_{A^\prime}} ~, &&
& \pd_A = - \frac{\hat{\pi}^{A^\prime} \pd_{A A^\prime}}{\langle \pi \hat{\pi} \rangle} ~, \\
& \bar{\pd}_0 = - \langle \pi \hat{\pi} \rangle \, \pi_{A^\prime} \frac{\pd}{\pd \hat{\pi}_{A^\prime}} ~, &&
& \bar{\pd}_A = \pi^{A^\prime} \pd_{A A^\prime} ~.
\end{aligned}
\end{equation}
This basis of $1$-forms, and their duals, enjoy the structure equations,
\begin{equation}
\begin{aligned}
\bar{\pd} e^A = e^0 \wedge \bar{e}^A ~, \quad
\pd \bar{e}^A = e^A \wedge \bar{e}^0 ~,\\ 
\comm{\bar{\pd}_0}{\pd_A} = \bar{\pd}_A ~, \quad
\comm{\bar{\pd}_A}{\pd_0} = \pd_A ~.
\end{aligned}
\end{equation}

\subsubsection{Hyper-K{\"a}hler Structure}
Recall that $\bE^4$ can be given a hyper-K\"ahler structure consisting of a triplet $\mathcal{J}_i$, $i=1,2,3$ of complex structures obeying $\mathcal{J}_i^2 = - \textrm{id}$ and $\mathcal{J}_1\mathcal{J}_2\mathcal{J}_3 =\textrm{id} $.
For each $\mathcal{J}_i $ the corresponding K\"ahler forms $\varpi_i$ are self-dual with respect to the metric $ds^2 = \sum_{\mu=0}^3 dx^\mu \otimes dx^\mu$ and may be given explicitly as
\begin{equation}
\varpi_i= \frac{1}{2} \epsilon_{ijk} dx^j\wedge dx^k + dx^0 \wedge dx^i \, .
\end{equation}
For completeness, in the coordinate basis $\{ dx^\mu \}$, $\mu =0 \dots 3$, i.e.~$\mathcal{J}_i dx^\nu = (\mathcal{J}_i)_\mu{}^\nu dx^\mu$, the components of the complex structures are given as 
\begin{equation*}
    (\mathcal{J}_1)_\mu{}^\nu = \left(
\begin{array}{cccc}
 0 & 1 & 0 & 0 \\
 -1 & 0 & 0 & 0 \\
 0 & 0 & 0 & 1 \\
 0 & 0 & -1 & 0 \\
\end{array}
\right) ~, \quad 
  (\mathcal{J}_2)_\mu{}^\nu = \left(
\begin{array}{cccc}
 0 & 0 & 1 & 0 \\
 0 & 0 & 0 & -1 \\
 -1 & 0 & 0 & 0 \\
 0 & 1 & 0 & 0 \\
\end{array}
\right) ~, \quad  
 (\mathcal{J}_3)_\mu{}^\nu = \left(
\begin{array}{cccc}
 0 & 0 & 0 & 1 \\
 0 & 0 & 1 & 0 \\
 0 & -1 & 0 & 0 \\
 -1 & 0 & 0 & 0 \\
\end{array}
\right) ~.
\end{equation*}

Taking combinations of these $\mathcal{J}_i$, we find a space of complex structures parameterised by $\pi_{A'} \sim (1 ,\zeta)   \in \mathbb{CP}^1$, 
\begin{equation}
\mathcal{J}_\pi = \frac{\zeta + \bar{\zeta}}{ 1+ \zeta \bar{\zeta}} \mathcal{J}_3 - \frac{\mathrm{i}(\bar\zeta-\zeta ) }{ 1+ \zeta \bar{\zeta}} \mathcal{J}_2 + \frac{1 - \zeta \bar{\zeta} }{ 1+ \zeta \bar{\zeta}} \mathcal{J}_1\, , \qquad \mathcal{J}_\pi^2 = - \textrm{id}\, .
\end{equation}
We can also define this $\bb{CP}^1$-dependent complex structure in spinor notation with $\pi_{A'} \sim (1 ,\zeta)$ as 
\begin{equation}
\mathcal{J}_\pi = - \frac{\mathrm{i}}{\Vert \pi \Vert^2} (\pi^{A^\prime} \hat{\pi}_{B^\prime} + \hat{\pi}^{A^\prime} \pi_{B^\prime}) \, \pd_{A A^\prime} \otimes \dr x^{A B^\prime} = \mathrm{i} \, \pd_A \otimes e^A - \mathrm{i} \, \bar{\pd}_A \otimes \bar{e}^A \ .
\end{equation}
Thinking of $\mathbb{PT}_\bE$ as the bundle of complex structures over $\bE^4$, the coordinates $v^A = \pi_{A^\prime} x^{A A^\prime}$ are holomorphic coordinates with respect to the complex structure $\mathcal{J}_\pi$. We may denote these coordinates by $v^A = (z_\pi, w_\pi)$. 
The K\"ahler form corresponding to ${\cal J}_\pi$ can equally be expressed as
\begin{equation}
\begin{aligned}
\varpi_\pi & = \frac{1}{2} ({\cal J}_\pi)_{A A^\prime B B^\prime} \, \dr x^{A A^\prime} \wedge \dr x^{B B^\prime} = - \frac{\mathrm{i}}{\Vert \pi \Vert^2} \varepsilon_{AB} \, \pi_{A'} \hat{\pi}_{B'} \, \dr x^{A A'} \wedge \dr x^{B B'} \\
& = -\frac{\mathrm{i}}{1+\zeta \bar{\zeta}}\left(\dr z_\pi \wedge \dr \bar{z}_\pi + \dr w_\pi \wedge \dr \bar{w}_\pi\right) \, .
\end{aligned}
\end{equation}
Holomorphic self-dual $(2,0)$- and $(0,2)$-forms are given by
\begin{equation}
\begin{aligned}
\mu_\pi = \varepsilon_{AB} \, \pi_{A'}\pi_{B'} \, \dr x^{A A'} \wedge \dr x^{B B'} = 2 \, \dr z_\pi \wedge \dr w_\pi \, , \\
\bar{\mu}_\pi = \varepsilon_{AB} \, \hat{\pi}_{A'} \hat{\pi}_{B'} \, \dr x^{A A'} \wedge \dr x^{B B'} = 2 \, \dr \bar{z}_\pi \wedge \dr \bar{w}_\pi \, .
\end{aligned}
\end{equation}

\subsection{Appendix B: Alternative Localisation of \texorpdfstring{CS$_4$}{CS4} to \texorpdfstring{IFT$_2$}{IFT2}}\label{S:4d CS to 2d IFT}
In \S \ref{sec:4to2} the defect algebra approach to 4d CS (for a complete discussion see \cite{Lacroix:2020flf, Liniado:2023uoo}) was utilised in the passage to the 2d IFT.
In this appendix we will perform the localisation of 4d CS to the 2d IFT via a more standard route, mirroring the localisation approach from 6d hCS to the 4d IFT, showing that we also land on \eqref{eq:IFT2}. \\

We recall the action \eqref{ec:4dCSaction}
\begin{equation}
\label{ec:4dCS}
S_{\mathrm{CS}_4} = \frac{1}{2 \pi \mathrm{i}} \int_{\Sigma \times \mathbb{CP}^1} \omega \wedge \mathrm{Tr} \bigg( \hat {\cal A} \wedge \dr \hat {\cal A} + \frac{2}{3} \hat {\cal A} \wedge \hat {\cal A} \wedge \hat {\cal A} \bigg) ~,
\end{equation}
with the  meromorphic one form of eq. \eqref{eq:omega4}  and boundary conditions as per eq. \eqref{eq:6dBC}. 

We work with the parametrisation of $\hat{\cA}$ of eq.\eqref{ec:4dAparametrisation}, recalled here for convenience   
\begin{equation}
 \hat{\cA}_{\bar{\zeta}} = \hat{h}^{-1}\partial_{\bar{\zeta}} \hat{h}\, , \quad \hat{\cA}_{I}  = \hat{h}^{-1} \mc{L}_{I} \hat{h} + \hat{h}^{-1}(\partial_{I} \hat{h}) \, , \quad I = w , \bar{w}    ~ .  
\end{equation} 
Viewing $\hat{\cA} ={\cal L}^{\hat{h}} $ as the formal gauge transform of ${\cal L}$ by $\hat{h}$, we use the following identity satisfied by the Chern-Simons density

\begin{equation} \text{CS}(\hat{\cA}) = \text{CS}(\mc{L}^{\hat{h}}) = \text{CS}(\mc{L}) - \dr \, \mathrm{Tr} \big( \hat{J} \wedge \hat{h}^{-1} \mc{L} \, \hat{h}  \big) - \frac{1}{6} \mathrm{Tr} \big( \hat{J} \wedge [ \hat{J} , \hat{J}\,  ] \big) ~ , 
\end{equation}
in which $\hat{J} = \hat{h}^{-1} \dr \hat{h} 
 $.    Noting that on shell  $\text{CS}(\mc{L})= \mc{L} \wedge \dr \mc{L} $ and $\omega \wedge (\partial_{\zeta} \mc{L}) \wedge \mc{L} = 0$  we then   arrive at the following action
\begin{equation}\label{4dCSii}
S_{\mathrm{CS}_4} = - \frac{1}{2 \pi \mathrm{i}} \int \dr \omega \wedge \mathrm{Tr} \big( \hat{J} \wedge \hat{h}^{-1} \mc{L} \, \hat{h}  \big) - \frac{1}{12 \pi \mathrm{i}} \int \omega \wedge \big( \hat{J} \wedge [ \hat{J} , \hat{J} \,  ]  \big) ~ .
\end{equation}
 In this form we see how our action will localise at the poles of $\omega$ giving a 2d theory 
\begin{equation}\label{4d localised action}
{S} = {r}_{+} \int_{\Sigma} \mathrm{Tr} \big( \hat{J} \wedge \hat{h}^{-1}\mc{L} \; \hat{h} \big) {|}_{\alpha} + r_{-} \int_{\Sigma}\mathrm{Tr} \big( \hat{J} \wedge \hat{h}^{-1} \mc{L}\; \hat{h} \big) {|}_{\tilde{\alpha}} +\text{WZ terms} \; ,
\end{equation}
where we recall
$$r_+ = K \frac{\langle \alpha \gamma \rangle \langle {\alpha \hat{\gamma}} \rangle}{\langle \alpha \tilde{\alpha} \rangle \langle \alpha \beta \rangle^{2}} \; \; \;\;\; \text{and} \; \; \; \; \; r_{-}= -K \frac{\langle \tilde{\alpha} \gamma \rangle \langle \tilde{\alpha} \hat{\gamma} \rangle}{\langle \alpha \tilde{\alpha} \rangle \langle \tilde{\alpha} \beta \rangle^{2}} \; .$$
 
To complete the construction  we need to specify the meromorphic structure of $\mc{L}$ that ensures the theory is well defined given the  form of $\omega$ and is compatible with the boundary conditions. This requires that
\begin{equation} {\mc{L}}_{w} = \frac{\langle \pi \beta \rangle}{\langle \pi \gamma \rangle}{\mc{M}}_{w} + {\mc{N}}_{w}\; \; , \qquad \mc{L}_{\bar{w}} = \frac{\langle \pi \beta \rangle }{\langle \pi \hat{\gamma} \rangle } {\mc{M}}_{\bar{w}} + {\mc{N}}_{\bar{w}} ~ , 
\end{equation}
where $\mc{M}_{I}$, $\mc{N}_{I} \in {C}^{\infty}(\Sigma,\tf{g})$.
The boundary conditions in this parametrisation read,
\begin{equation}
\begin{aligned}
   &\hat{h}{\mid}_{\beta} = \text{id} \,,  \quad  {\cal L}\vert_\beta = 0 \;  , \; \;  \; \;    \\
   &\mathrm{Ad}_h^{-1} {\cal L}_w \vert_\alpha  + J_w  ~=  ts \left(  \mathrm{Ad}_{\tilde{h}} ^{-1} {\cal L}_w  \vert_{\tilde{\alpha}} + \tilde{J}_w\right)   ,\\  
    &\mathrm{Ad}_h^{-1} {\cal L}_{\bar{w}} \vert_\alpha  + J_{\bar{w}}  ~=  t^{-1} s \left(  \mathrm{Ad}_{\tilde{h}} ^{-1} {\cal L}_{\bar{w}}  \vert_{\tilde{\alpha}} + \tilde{J}_{\bar{w}}\right)   ,
    \end{aligned}
\end{equation}
in which we use the definitions,
\begin{equation}
\hat{h}{\mid}_{\alpha}= h \; , \qquad \hat{h}{\mid}_{\tilde\alpha} = \tilde{h} \; , \qquad   \hat{J}{\mid}_{\alpha}= J \; , \qquad \hat{J}{\mid}_{\tilde{\alpha}}= \tilde{J}  ~ . 
\end{equation}
Solving these conditions uniquely determines  the Lax connection
\begin{equation}\begin{aligned}\label{Lax solvedi}
{\mc{M}}_{w}&= \frac{\la \alpha \gamma \ra}{ \la \alpha \beta \ra}  \mathrm{Ad}_{h} \left[ 1 - \sigma \, \mathrm{Ad}^{-1}_{\tilde{h}} \mathrm{Ad}_{h} \right]^{-1} \left( ts \tilde{J}_{w} -{J}_{w} \right) \; , \qquad &
{\mc{N}}_{w}&=0 \; , \\
{\mc{M}}_{\bar{w}}&= \frac{\la \alpha \hat{\gamma} \ra}{\la \alpha \beta \ra} \mathrm{Ad}_{h} \left[ 1 - \sigma^{-1}\mathrm{Ad}^{-1}_{\tilde{h}} \mathrm{Ad}_{h} \right]^{-1} \left( t^{-1}s \tilde{{J}}_{\bar{w}} -{J}_{\bar{w}} \right) \; , \qquad &
{\mc{N}}_{\bar{w}}&=0 \; .
\end{aligned}\end{equation}
where we have introduced the parameter $\sigma = t \sqrt{ \frac{\la \alpha \gamma \ra \la \tilde{\alpha} \hat{\gamma} \ra}{\la \tilde{\alpha} \gamma \ra \la {\alpha} \hat{\gamma} \ra}}$. It will also be useful to state the alternative forms
\begin{equation}\begin{aligned}\label{Lax solvedii}
\mc{M}_w=& \frac{\la \tilde{\alpha} {\gamma} \ra}{\la \tilde{\alpha} \beta \ra}\mathrm{Ad}_{\tilde{h}} \left[ 1 - \sigma^{-1}  \mathrm{Ad}^{-1}_{h} \mathrm{Ad}_{\tilde{h}} \right]^{-1} \left({t}^{-1}{s}^{-1} J_{w} - \tilde{J}_{w}\right) \; , \\
\mc{M}_{\bar{w}}=& \frac{\la \tilde{\alpha} \hat{{\gamma}} \ra}{\la \tilde{\alpha} \beta \ra} \mathrm{Ad}_{\tilde{h}} \left[ 1 - \sigma \mathrm{Ad}^{-1}_{h} \mathrm{Ad}_{\tilde{h}} \right]^{-1} \left({t}{s}^{-1} J_{\bar{w}} - \tilde{J}_{\bar{w}}\right) \; .
\end{aligned}\end{equation}
Inserting \eqref{Lax solvedi} and \eqref{Lax solvedii} into \eqref{4d localised action} we obtain
\begin{equation}\begin{aligned}
S =& \; { r_{+}} \int_{\Sigma} \text{vol}_{2} \, \mathrm{Tr} \Big( {J}_{w}  \left[ 1 - \sigma^{-1} \mathrm{Ad}^{-1}_{\tilde{h}} \mathrm{Ad}_{h} \right]^{-1} \left( t^{-1}s \tilde{J}_{\bar{w}} - {J}_{\bar{w}} \right) \Big) \\
& - { r_{+}} \int_{\Sigma} \text{vol}_{2} \, \mathrm{Tr} \Big( {J}_{\bar{w}}  \left[ 1-\sigma \mathrm{Ad}^{-1}_{\tilde{h}}\mathrm{Ad}_{h} \right]^{-1} \left( ts \tilde{J}_{w} - {J}_{w} \right) \Big) \\
& - { r_{-}} \int_{\Sigma} \text{vol}_{2} \, \mathrm{Tr} \Big( \tilde{{J}}_{w}  \left[ 1- \sigma \mathrm{Ad}^{-1}_{h} \mathrm{Ad}_{\tilde{h}} \right]^{-1} \left( ts^{-1} {{J}}_{\bar{w}} -{\tilde{J}}_{\bar{w}} \right) \Big) \\
& + { r_{-} }\int_{\Sigma} \text{vol}_{2} \, \mathrm{Tr} \Big(\tilde{{J}}_{\bar{w}}  \left[ 1-\sigma^{-1} \mathrm{Ad}_{h}^{-1} \mathrm{Ad}_{\tilde{h}} \right]^{-1} \left( t^{-1}s^{-1} {J}_{w} -{\tilde{J}}_{w} \right) \Big) \\
& + \text{WZ terms} \; ,
\end{aligned}\end{equation}
where $\vol_2 = d{\bar w} \wedge d w$. Expanding out this action, collecting together terms, and Wick rotating to Minkowski space, we arrive at the action \eqref{eq:IFT2}.

\subsection{Appendix B: Alternative \texorpdfstring{CS$_4$}{CS4} Setup for the \texorpdfstring{$\lambda$}{lambda}-Model} \label{appendix:alt4dCSlambda}

In this section, we will consider an alternative symmetry reduction of our hCS$_6$ setup which also recovers the $\lambda$-deformed IFT$_2$.
In order to recover a 1-field IFT$_2$, we need one of the semi-local residual symmetries of the IFT$_4$ to become a gauge symmetry under symmetry reduction.
Let us denote the symmetry reduction vector fields by $V_1$ and $V_2$.
Taking the example of the residual left-action parameterised by $\ell$, this must obey the constraint $\beta^{A^\prime} \pd_{A A^\prime} \ell = 0$.
In order for this to become a gauge symmetry of the IFT$_2$, the symmetry reduction constraints $L_{V_1} \ell$ and $L_{V_2} \ell = 0$ must coincide with the pre-existing constraints on $\ell$.
This means that we must choose to symmetry reduce along the vector fields
\begin{equation}
V_1 = \mu^A \beta^{A^\prime} \pd_{A A^\prime} ~, \qquad
V_2 = \hat{\mu}^A \beta^{A^\prime} \pd_{A A^\prime} ~.
\end{equation}

Following the recipe described elsewhere in this paper, we deduce that the CS$_4$ $1$-form is given by
\unskip\footnote{Since $\beta^{A^\prime}$ appears in both of our symmetry reduction vector fields, the two zeroes from symmetry reduction have cancelled the double pole.
Similarly, the boundary condition $\cA_A \vert_{\beta} = 0$ can be interpreted as a simple zero in each component of the gauge field.
These simple zeroes cancel the simple poles introduced in symmetry reduction, leaving a gauge field with no singularities.}
\begin{equation}
\omega = K \frac{1}{(\zeta - \alpha_+) (\zeta - \alpha_-)} \, \dr \zeta ~.
\end{equation}
In the 4d CS description, we can already see that we have eliminated one degree of freedom relative to other symmetry reductions.
The symmetry reduction zeroes have eliminated the double pole at $\beta$, effectively removing one field from the IFT$_2$.

Furthermore, if we denote the surviving coordinates on $\Sigma$ by $y^1 = \hat{\mu}^A \hat{\beta}^{A^\prime} \pd_{A A^\prime}$ and $y^2 = -\mu^A \hat{\beta}^{A^\prime} \pd_{A A^\prime}$, the boundary conditions reduce to
\begin{equation}
\hat{\mathcal{A}}_1 \vert_\alpha = \sigma \hat{\mathcal{A}}_1 \vert_{\tilde{\alpha}} ~, \qquad
\hat{\mathcal{A}}_2 \vert_\alpha = \sigma^{-1} \hat{\mathcal{A}}_2 \vert_{\tilde{\alpha}} ~.
\end{equation}
Since the localisation from CS$_4$ to the IFT$_2$ has been described in detail elsewhere, we will be brief in this section.
In the parametrisation
\begin{equation}
\hat{\mathcal{A}} = \hat{h}^{-1} \cL \hat{h} + \hat{h}^{-1} \dr \hat{h} ~,
\end{equation}
we fix the constraints $\cL_{\bar{\zeta}} = 0$ and denote the values of $\hat{h}$ at the poles by $\hat{h} \vert_\alpha = h$ and $\hat{h} \vert_{\tilde{\alpha}} = \text{id}$.
We can then use the bulk equations of motion and the boundary conditions to solve for $\cL_1$ and $\cL_2$ in terms of $h$.
We find the solutions
\begin{equation} \label{eq:alt_lax}
\cL_1 = (\sigma - \mathrm{Ad}_h^{-1})^{-1} h^{-1} \pd_1 h ~, \qquad
\cL_2 = (\sigma^{-1} - \mathrm{Ad}_h^{-1})^{-1} h^{-1} \pd_2 h ~.
\end{equation}

Finally, the action localises to 2d and is given, up to an overall factor of $K/(\alpha_+ - \alpha_-)$, by
\begin{equation}
- \int_\Sigma \dr y^1 \wedge \dr y^2 \, \mathrm{Tr} \bigg( h^{-1} \pd_1 h \cdot \frac{1 + \sigma \mathrm{Ad}_h^{-1}}{1 - \sigma \mathrm{Ad}_h^{-1}} h^{-1} \pd_2 h \bigg) - \frac{1}{6} \int_{\Sigma \times [0, 1]} \mathrm{Tr} \big( h^{-1} \dr h \wedge h^{-1} \dr h \wedge h^{-1} \dr h \big) ~.
\end{equation}
This can be recognised as the $\lambda$-deformed IFT$_2$.

\chapter{Integrable Coset Models from Twistor Space}

\label{sec:gaugingthediamond}

This work was done in collaboration with Lewis Cole, Ryan Cullinan, Ben Hoare and Daniel Thompson \cite{Cole:2024ess}.

Quantum field theories (QFTs) in two dimensions have both direct applications in condensed matter systems and as the worldsheet theories of strings, and can provide a tractable sandpit for the study of quantum field theory more generally. Special examples are provided by conformal field theories (CFTs) and integrable field theories (IFTs), for which powerful infinite-dimensional symmetries enable us to exactly determine certain key properties and observables.  

One longstanding goal has been to provide a constructive origin of these integrable systems from some putative parent theory, perhaps in higher dimensions. For instance, Ward suggested \cite{Ward:1985gz} that all integrable equations may arise as reductions of the 4d anti-self-dual Yang-Mills (ASDYM) equation.
Given a choice of complex structure on $\bR^{4}$, the ASDYM equation are
\begin{align}
\label{eq:F20}
F^{2,0} &= 0 = F_{z w} \ , \\
\label{eq:F02}
F^{0,2} &=  0 = F_{\bar{z}\bar{w}} \ , \\
\label{eq:F11}
\varpi \wedge F^{1,1} &= 0 = F_{z\bar{z}} + F_{w\bar{w}} \ ,
\end{align}
where $\varpi$ is the K{\"a}hler form.  There are (at the very least) two senses in which ASDYM can be viewed as an integrable theory in its own right.  First is that the ASDYM equations can be exactly solved by the ADHM construction \cite{ADHM}.  Second is that these equations admit a zero curvature formulation in terms of a Lax pair of differential operators:
$$L= \nabla_z -\zeta \nabla_{\bar{w}} \, , \quad M= \nabla_w  + \zeta \nabla_{\bar{z}}  \, ,  \qquad [L,M] = 0 \quad \forall \, \zeta \iff F = -\star F\, .  $$  
Accordingly, in this work, we will denote four-dimensional QFTs whose equations of motion can be recast as the anti-self duality of some connection as IFT$_{4}$.

A prominent example in this class of theories is the 4d Wess-Zumino-Witten model (WZW$_{4}$) \cite{Donaldson:1985zz, Nair:1990aa, Nair:1991ab, Losev:1995cr}, which arises as a partial gauge fixing of the ASDYM equation. Up to a gauge transformation, one may parameterise a generic connection that solves equations \eqref{eq:F20} and \eqref{eq:F02} as $A = - \bar{\pd} g g^{-1}$, where the group-valued field $g$ becomes the fundamental field of WZW$_{4}$. The remaining ASDYM equation \eqref{eq:F11} becomes $\varpi \wedge \pd ( \bar{\pd} g g^{-1} ) = 0$, which is the equation of motion of WZW$_{4}$, also known as Yang's equation. The more well-known WZW$_{2}$ also arises as a reduction of WZW$_{4}$, and Yang's equation reduces to the familiar holomorphic conservation law characterising this CFT$_{2}$.
Another example is found by solving equations \eqref{eq:F20} and \eqref{eq:F11}, leaving equation \eqref{eq:F02} as the dynamical equation of motion. In this case, the IFT$_{4}$ is known as the LMP model \cite{Leznov:1986mx, Parkes:1992rz}, which gives the pseudo-dual of the principal chiral model after reduction.

Alternatively, motivated by the similarity between Reidemeister moves in knot theory and the Yang-Baxter equation that underpins integrability, Witten suggested \cite{Witten:1989wf} that integrable models might have a description in terms of Chern-Simons theory. The realisation of this idea came some years later, with Costello's understanding \cite{Costello:2013zra, Costello:2013sla} that the gauge theory description should combine the topological nature of Chern-Simons theory with the holomorphic nature of the spectral parameter characterising IFTs. The theory proposed in \cite{Costello:2013zra, Costello:2013sla} was extended and developed in a sequence of papers \cite{Costello:2017dso, Costello:2018gyb, Costello:2019tri} describing a Chern-Simons theory, which we denote by CS$_{4}$, defined over a four-manifold $\Sigma \times C$ with the action
\begin{equation}
S_{\text{CS}_{4}}[A] = \frac{1}{2 \pi \mathrm{i}} \int_{ \Sigma\times C } \omega \wedge \mathrm{Tr} \left( A \wedge \dr A + \frac{2}{3} A \wedge A \wedge A \right) \, .
\end{equation}
Here, $\omega$ is a meromorphic differential on the complex curve $C$, which we will take to be $C = \mathbb{CP}^1$.
Specifying boundary conditions at the poles of $\omega$, the dynamics can be `localised' to take place on $\Sigma$, which is identified with the space-time of the IFT$_{2}$, and the curve $C$ is associated to spectral parameter of the Lax connection of the integrable model (see \cite{Lacroix:2021iit} for a pedagogical introduction).

\begin{figure}
\centering
\begin{tikzpicture}
\node at (0,2) {$\mathbf{hCS_6}$};
\node at (-2,0) {$\mathbf{CS_4}$};
\node at (2,0) {$\mathbf{IFT_4}$};
\node at (0,-2) {$\mathbf{IFT_2}$ / $\mathbf{CFT_2}$ };
\draw[->,very thick,decorate, decoration={snake, segment length=12pt, amplitude=2pt}] (-0.4,1.6)--(-1.6,0.4);
\draw[->,very thick] (0.4,1.6)--(1.6,0.4);
\draw[->,very thick,decorate, decoration={snake, segment length=12pt, amplitude=2pt}] (1.6,-0.4)--(0.4,-1.6);
\draw[->,very thick] (-1.6,-0.4)--(-0.4,-1.6);
\end{tikzpicture}
\caption{The diamond correspondence of integrable avatars, in which wavy arrows indicate a descent by reduction and straight arrows involve localisation i.e. integration over $\mathbb{CP}^1$.}
\label{diagram:diamond}
\end{figure}
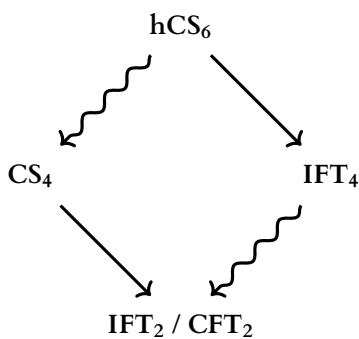

An elegant origin of both the CS$_{4}$ and the ASDYM descriptions was provided in the work of Bittleston and Skinner \cite{Bittleston:2020hfv} in terms of a six-dimensional holomorphic Chern-Simons theory (hCS$_{6}$), first proposed in \cite{Kcostello, Costello:2021bah}. The theory is defined over (the Euclidean slice of) Penrose's twistor space \cite{Penrose:1967wn} with the action functional
\begin{equation}
S_{\text{hCS}_{6}}[\cA] = \frac{1}{2 \pi \mathrm{i}} \int_{\mathbb{PT}} \Omega \wedge \mathrm{Tr} \left( \cA \wedge \bar{\partial} \cA + \frac{2}{3} \cA \wedge \cA \wedge \cA \right) \, ,
\end{equation}
in which $\Omega$ is some meromorphic $(3,0)$ form. This action is supplemented by a choice of boundary conditions at the poles of $\Omega$. The various lower-dimensional descriptions follow from exploiting the fibration structure $\mathbb{CP}^1 \hookrightarrow \mathbb{PT} \twoheadrightarrow \mathbb{R}^4$. Reducing along two directions within $\mathbb{R}^4$, hCS$_6$ descends to CS$_{4}$. Alternatively, one may instead first choose to localise over $\mathbb{CP}^1$, and this leads to IFT$_{4}$ of the ASDYM description. Indeed, the integrability properties of ASDYM are fundamentally tied to this twistorial origin and  evidence suggests that at a quantum level this twistor space is the natural arena to consider \cite{Costello:2021bah, Bittleston:2022nfr}.  
 Applying the reduction along $\bR^{4}$ to this IFT$_{4}$ produces an IFT$_{2}$ which may also be recovered by localising the CS$_{4}$ description. In this way, we have a diamond correspondence of theories illustrated in Figure \ref{diagram:diamond}.
Other recent work on hCS$_{6}$ includes \cite{Penna:2020uky, Costello:2021bah, Cole:2023umd}.

Given an IFT$_{2}$ or CFT$_{2}$ it is sometimes possible to obtain another I/CFT$_{2}$ via gauging. Perhaps the most famous example is the GKO $G/H$ coset CFTs \cite{Goddard:1984vk}, which can be given a Lagrangian description by taking a WZW$_{2}$ CFT on $G$ and gauging a (vectorially acting) $H$ subgroup \cite{Gawedzki:1988hq, Karabali:1988au, Bardakci:1990lbc, Witten:1991mm}. This motivates the core question of this work:
\begin{equation*}
\text{\em How can the diamond correspondence be gauged?}
\end{equation*}
Resolving this question dramatically expands the scope of theories that can be given a higher-dimensional avatar. A significant clue is given by the rather remarkable Polyakov-Wiegmann (PW) identity, which shows that the $G/H$ gauged WZW model is actually equivalent to the difference of a $G$ WZW model and an $H$ WZW model. This points towards a general resolution that integrable gauged models might be obtained as differences of ungauged models. This is less obvious than it might first seem; it was noted in \cite{Losev:1995cr} that for a PW identity to apply for WZW$_{4}$ one requires that the gauging is performed by connections with field strength restricted to be type $(1,1)$. The six-dimensional origin of such a constraint is rather intriguing and will be elucidated in this paper. In the context of CS$_{4}$, Stedman recently proposed \cite{Stedman:2021wrw} considering the difference of CS$_{4}$ to give rise to gaugings of IFT$_{2}$. We will recover this construction as a reduction of hCS$_{6}$ theory in the present work, as well as uncovering some additional novelties in the CS$_{4}$ description.

At the top of the diamond, we will consider a theory of two-connections $\mathcal{A} \in \Omega^{1}(\mathbb{PT}) \otimes \fg$ and $\mathcal{B} \in \Omega^{1}(\mathbb{PT}) \otimes \fh$ for a subalgebra $\fh \subset \fg$. The action of this theory is
\begin{equation}
S_{\mathrm{ghCS_6}} = S_{\mathrm{hCS}_{6}}[\mathcal{A}] - S_{\mathrm{hCS}_{6}}[\mathcal{B}] + S_{\mathrm{int}}[\mathcal{A},\mathcal{B} ] \, ,
\end{equation}
in which the term $S_{\mathrm{int}}$ couples the two gauge fields. We will develop this story by means of two explicit examples: choosing $\Omega$ to have two double poles, we will study the diamond relevant to the gauged WZW theory; and with $\Omega$ containing a single fourth-order pole we will study the gauged LMP model. This seemingly simple setup gives rise to a rich story whose results we will now summarise:
\begin{enumerate}
\item Our investigations indicate that general gaugings of the WZW$_{4}$ model break integrability in four dimensions. Integrability is preserved if the gauge field $B$ is constrained to satisfy two of the three anti-self-dual Yang-Mills equations, namely $F^{2,0}[B] = 0$ and $F^{0,2}[B] = 0$.
\item The two gauge fields $\cA$ and $\cB$ of ghCS$_{6}$ source various degrees of freedom in the gauged WZW$_{4}$. In particular, as well as the fundamental field $g$ and the 4d gauge field $B$, auxiliary degrees enter as Lagrange multipliers for $F^{2,0}[B] = 0$ and $F^{0,2}[B] = 0$.
\item Reducing by two dimensions, we recover a variety of IFT$_{2}$ including the special case of gauged WZW$_{2}$. In general, we find a coupled model between a gauged IFT$_{2}$ and a Hitchin system \cite{Hitchin1987} involving the gauge field $B$ and a pair of adjoint scalar fields. These scalars may source a potential for the gauged WZW$_{2}$ in which case we recover the complex sine-Gordon model and more broadly the homogeneous sine-Gordon models \cite{Fernandez-Pousa:1996aoa}. At the special point associated to the 2d PCM, Lagrange multipliers ensure that the gauge field is flat and hence trivial --- this is essential as the gauged PCM is not generically integrable.
\item We also use this formalism to perform an integrable gauging of the LMP model. Just as in the gauging of WZW$_{4}$, the field strength of the gauge field must be constrained to obey two of the anti-self-dual Yang-Mills equations, this time $F^{2,0}[B] = 0$ and $\varpi \wedge F^{1,1}[B] = 0$. It is noteworthy that the two equations which are enforced by Lagrange multipliers agree with the two equations that are identically solved in the ungauged case. This is true for both the WZW$_{4}$ and the LMP model. In addition, we show that the gauged LMP model obeys a PW-like identity such that it may be expressed the difference of an LMP model on $\fg$ and $\fh$.
\end{enumerate}
Let us outline the structure of this paper.
We begin in section \S\ref{sec:ungaugedWZWdiamond} with a review of the diamond correspondence of theories for the ungauged WZW model.
In section \S\ref{sec:gaugedWZWdiamond}, we introduce the gauging of this diamond concentrating in particular on the right hand side. We recover the gauged IFT$_{4}$ and demonstrate that its equations of motion may be rewritten as ASDYM.
The wide array of IFT$_{2}$ are explored in section \S\ref{sec:newIFT} where we also show that they are integrable and provide the associated Lax connection.
Following the gauging of the WZW$_{4}$, section \S\ref{sec:6dCSto4dCS} fleshes out the left hand side of the diamond, connecting to four-dimensional Chern-Simons by first reducing, and then to IFT$_{2}$ by localisation.
Section \S\ref{Gauged LMP action section} describes the diamond in the context of the gauged LMP theory.
We conclude with a brief outlook in section \S \ref{sec:outlookgauged}. Although the subject matter necessarily entails a degree of technical complexity we have endeavoured to keep the main presentation streamlined and complement this with a number of technical appendices.

\section{The ungauged WZW diamond}
\label{sec:ungaugedWZWdiamond}

In this section, we briefly describe the diamond correspondence of theories in which the two-dimensional theory is the WZW$_{2}$ CFT. This is a summary of some analysis first presented in \cite{Bittleston:2020hfv} which will serve to fix conventions and recapitulate key steps relevant to later sections.

\subsection{\texorpdfstring{hCS$_6$}{hCS6} with double poles}

We begin at the top of the diamond with 6d holomorphic Chern-Simons theory (hCS$_6$) whose fundamental field is an algebra-valued connection $\cA \in \Omega^{0,1}(\mathbb{PT}) \otimes \fg$. The six-dimensional action is given by
\begin{equation}\label{eq:hCS6gauged}
S_{\mathrm{hCS}_6}[\cA ] = \frac{1}{2 \pi \mathrm{i}} \int_{\mathbb{PT}} \Omega \wedge \mathrm{Tr} \Big( \cA \wedge \bar{\pd} \cA + \frac{2}{3} \cA \wedge \cA \wedge \cA \Big) \, ,
\end{equation}
in which we have introduced a meromorphic $(3,0)$-form $\Omega$. As a real manifold, there is an isomorphism $\mathbb{PT} \cong \bR^{4} \times \CP$ and we will introduce coordinates $x^{a \dot{a}} \in \bR^{4}$ and $\pi_{a} \in \CP$. In these coordinates, the meromorphic $(3,0)$-form (which has two double poles at $\alpha_{a} , \beta_{a} \in \CP$) is given by\footnote{Spinor contractions are defined to be $\langle \alpha \beta \rangle = \epsilon^{a b} \alpha_a \beta_b$, see appendix~\ref{sec:spinorconventions} for further details of spinor conventions.}
\begin{equation}\label{eq:Omegagauged}
\Omega = \frac{1}{2} \Phi(\pi) \, \epsilon_{\dot{a}\dot{b}} \, \pi_{a} \dr x^{a \dot{a}} \wedge \pi_b \dr x^{b \dot{b}} \wedge \langle \pi \dr \pi \rangle ~, \qquad
\Phi = \frac{\langle \alpha \beta \rangle^2 }{\langle \pi \alpha \rangle^2 \langle \pi \beta \rangle^2} ~.
\end{equation}
The poles of $\Omega$ in $\CP$ play the role of boundaries in hCS$_{6}$ because total derivatives pick up a contribution from $\bar{\pd} \Omega$ which is a distribution with support at these poles. To ensure a well defined variational principal, we impose boundary conditions on the gauge field at these poles given by
\begin{equation} \label{eq:bc1}
\cA \vert_{\pi = \alpha} = 0 \, , \qquad
\cA \vert_{\pi = \beta} = 0 \, .
\end{equation}
Turning to the symmetries of this model, the theory is invariant under gauge transformations acting as
\begin{equation} \label{eq:gt1}
\hat{\gamma}: \quad {\cal A} \mapsto ({\cal A} )^{\hat{\gamma}} = \hat{\gamma }^{-1} \mathcal{A} \hat{\gamma } + \hat{\gamma }^{-1} \bar{\pd} \hat{\gamma } \, ,
\end{equation}
so long as they preserve the boundary conditions. This amounts to restrictions on the allowed transformations at the poles of $\Omega$ which are given by
\begin{equation}
\pi^a \pd_{a \dot{a}} \hat{\gamma} \vert_{\pi = \alpha} = 0 \, , \quad
\pi^a \pd_{a \dot{a}} \hat{\gamma} \vert_{\pi = \beta} = 0 \, .
\end{equation}

\subsection{Localisation of \texorpdfstring{hCS$_6$}{hCS6} with double poles to \texorpdfstring{WZW$_{4}$}{WZW4}}

Surprisingly, all of the physical degrees of freedom in hCS$_{6}$ can be captured by a four-dimensional integrable field theory (IFT$_{4}$). This field theory is derived by localising the hCS$_{6}$ action, integrating out the $\CP$ and landing on a theory on $\bR^{4}$. For the choice of meromorphic $(3,0)$-form $\Omega$ and boundary conditions given above, this 4d theory is WZW$_{4}$. This localisation is possible because of the substantial gauge symmetry in Chern-Simons theories. Indeed, the dynamical fields arise precisely where this gauge symmetry is broken, namely at the poles of $\Omega$. Fields capturing these degrees of freedom are known as `edge modes' which enter via the field redefinition
\begin{equation}
\mathcal{A} = ( \mathcal{A}' )^{\hat{g}} = \hat{g}^{-1} \mathcal{A}' \hat{g} + \hat{g}^{-1} \bar{\pd} \hat{g} \, .
\end{equation}
Expressing the action $S_{\mathrm{hCS}_6}[\cA]$ in terms of the fields $\cA'$ and $\hat g$ one obtains
\begin{equation}
\begin{aligned} \label{eq:hcsA'g'}
S_{\mathrm{hCS}_6}[\cA] & = S_{\mathrm{hCS}_6}[\cA^{\prime}]
+ \frac{1}{2 \pi \mathrm{i}} \int_{\mathbb{PT}} \bar{\pd} \Omega \wedge \mathrm{Tr} \big( \cA^\prime \wedge \bar{\pd} \hat{g} \hat{g}^{-1} \big) \\
& \hspace{4em} - \frac{1}{6 \pi \mathrm{i}} \int_{\mathbb{PT}\times[0,1]} \bar \partial\Omega \wedge \mathrm{Tr} \big(\hat{g}^{-1} \dr \hat{g} \wedge \hat{g}^{-1} \dr \hat{g} \wedge \hat{g}^{-1} \dr \hat{g} \big)~,
\end{aligned}
\end{equation}
where, with a slight abuse of notation, we are also denoting by $\hat g$ a smooth homotopy to a constant map in the last term (such abuse will be perpetuated later without further comment).
Notably, the edge mode $\hat{g}$ only appears in this action against the 4-form $\bar{\pd} \Omega$ which is a distribution with support at the poles of $\Omega$. This means that the action only depends on $\hat{g}$ through its value (and $\CP$-derivative) at the poles of $\Omega$ which we will denote by
\begin{equation}
\hat{g} \vert_{\pi = \alpha} = g \ , \qquad
\hat{g}^{-1} \pd_{0} \hat{g} \vert_{\pi = \alpha} = u \ , \qquad
\hat{g} \vert_{\pi = \beta} = \tilde{g} \ , \qquad
\hat{g}^{-1} \pd_{0} \hat{g} \vert_{\pi = \beta} = \tilde{u} \ .
\end{equation}
Let us consider the symmetries of the theory in this new parameterisation. The gauge transformation \eqref{eq:gt1} acts trivially on $\cA^{\prime}$ whilst $\hat{g}$ transforms with a right-action as
\begin{equation}\label{eq:gt2}
\hat{\gamma}: \quad \mathcal{A}' \mapsto \mathcal{A}' \, , \qquad
\hat{g} \mapsto \hat{g} \hat{\gamma} \, .
\end{equation}
In addition, the new parameterisation has introduced a redundancy (which we dub an internal gauge symmetry) acting as
\begin{equation} \label{eq:gt3}
\check \gamma: \quad \cA' \mapsto \check \gamma^{-1}\cA' \check \gamma + \check \gamma^{-1}\bar \partial \check \gamma \, , \qquad
\hat g \mapsto \check \gamma^{-1}\hat g \, .
\end{equation}
We can exploit these symmetries to impose gauge fixing conditions on the fields $\cA^{\prime}$ and $\hat{g}$. Let us fix $\cA'$ such that it has no $\mathbb{CP}^1$-leg, and fix the value of $\hat{g}$ at $\pi = \beta$ to the identity.\footnote{At this point, we may further fix the $\mathbb{CP}^1$-derivative of $\hat{g}$ at both $\pi = \alpha$ and $\pi = \beta$ to zero. However, such terms drop out of the action in this ungauged case anyway without specifying this fixing.} The surviving edge mode at the other pole $g = \hat{g} \vert_{\pi = \alpha}$ will become the fundamental field of the WZW$_{4}$.

Returning to the action \eqref{eq:hcsA'g'}, the first term is a genuine six-dimensional bulk term which we eliminate by going on-shell.
The bulk equation of motion imposes holomorphicity of $\cA'$, which may be solved in terms of a series of $\CP$-independent components $A'_{a \dot{a}}$ as
\begin{equation}
\cA' = \pi^a A'_{a \dot{a}} \bar{e}^{\dot{a}} \ , \qquad
\bar{e}^{\dot{a}} = \frac{\hat{\pi}_a \dr x^{a \dot{a}}}{\langle \pi \hat{\pi} \rangle } \, .
\end{equation}
In this expression, $\bar{e}^{\dot{a}}$ is a basis $(0,1)$-form on twistor space introduced in appendix~\ref{sec:TwistorSpace}.
This completely specifies the $\CP$-dependence of $\cA^{\prime}$, and the boundary conditions eq.~\eqref{eq:bc1} may be solved to determine $A'_{a \dot{a}}$ in terms of $g$,
\begin{equation} \label{eq:A'}
A'_{a \dot{a}} = - \frac{\beta_a \alpha^b }{\langle \alpha \beta \rangle }\pd_{b \dot{a}} g g^{-1} \, .
\end{equation}
From these components, we can construct a 4d connection $A' = A'_{a \dot{a}} \dr x^{a \dot{a}}$, and this parameterisation of $A'$ in terms of $g$ is known in the literature as Yang's parameterisation ($g$ being called Yang's matrix).
This solution for $\cA^{\prime}$ may now be substituted into the action and the integral over $\CP$ can be computed explicitly. The second and third term of \eqref{eq:hcsA'g'} localise to a four-dimensional action, and the detailed manipulations are presented in appendix~\ref{appendix on localisation formulae}. We land on the WZW$_{4}$ theory defined by
\begin{equation} \label{eq:WZW4}
\begin{aligned}
S_{\text{WZW}_{4}} & = \frac{1}{2} \int_{\bR^{4}} \, \mathrm{Tr} \big( g^{-1} \dr g \wedge \star g^{-1} \dr g \big)
+ \int_{\bR^{4} \times [0,1]} \omega_{\alpha , \beta} \wedge \mathcal{L}_{\text{WZ}}[g ] \ .
\end{aligned}
\end{equation}
In the second term, we have introduced a 2-form defined by
\begin{equation} \omega_{\alpha,\beta} =\frac{1}{\langle \alpha \beta \rangle } \alpha_a \beta_b \, \epsilon_{\dot a \dot b} \, \dr x^{a \dot a} \wedge \dr x^{b\dot b} \ ,
\end{equation}
and the WZ 3-form
\begin{equation}
\mathcal{L}_{\text{WZ}}[g] = \frac{1}{3} \mathrm{Tr} \big( \tilde{g}^{-1} \dr \tilde{g} \wedge \tilde{g}^{-1} \dr \tilde{g} \wedge \tilde{g}^{-1} \dr \tilde{g} \big) ,
\end{equation}
defined, as is usual, using a suitable extension $\tilde{g}$ of $g$. The equation of motions of this theory are given by
\begin{equation} \label{eq:WZW4EOM}
0 = \dr \left( \star - \omega_{\alpha , \beta} \wedge\right) \dr g g^{-1} \, \quad \Leftrightarrow\, \quad \epsilon^{\dot{a}\dot{b}} \beta^a \pd_{a\dot{a}}\left( \alpha^b \pd_{b \dot{b}} g g^{-1} \right) = 0 \, .
\end{equation}

The six-dimensional gauge transformations (constrained by boundary conditions) descend to semi-local symmetries of this action ($\gamma_L = \hat \gamma \vert_{\beta}$ and $ \gamma_R= \hat \gamma \vert_{\alpha}$) which act as
\begin{equation}
g \rightarrow \gamma_L^{-1} \cdot g\cdot \gamma_R \, , \qquad
\alpha^a \pd_{a \dot{a}} \gamma_R = 0 \, , \qquad
\beta^a \pd_{a \dot{a}} \gamma_L = 0 \ .
\end{equation}
Of particular interest is the case where $\beta = \hat{\alpha}$ (i.e.\ the poles of $\Omega$ are antipodal on $\mathbb{CP}^1$) in which case $\omega_{\alpha,\hat{\alpha}} = \varpi$ is proportional to the K{\"a}hler form on $\bR^4$. Here, we are referring to the K{\"a}hler form with respect to the complex structure $\mathcal{J}_\alpha$ which is defined\footnote{Recall that $\bR^4$ is a hyper-K{\"a}hler manifold which has a $\mathbb{CP}^1$'s worth of complex structures, see appendix~\ref{sec:spinorconventions}, eq.~\eqref{eq:complexStruct} .} by the point $\alpha \in \mathbb{CP}^1$. In this case, the semi-local symmetries can be interpreted as a holomorphic left-action and anti-holomorphic right-action (akin to the two-dimensional WZW current algebra).

\subsection{Interpretation as ADSYM} \label{Yang's Matrix Section}

A 4d Yang-Mills connection $A'$ with curvature $F[A'] = \dr A' + A' \wedge A'$ is said to be anti-self dual if it obeys $F = - {\star} F$. After converting to bi-spinor notation, the anti-self-dual Yang-Mills (ASDYM) equations can be expressed as
\begin{equation}\label{eq:ASDYM}
\pi^a \pi^b F_{a\dot{a} b \dot{b}} = 0 \, , \qquad \forall \, \pi_{a} \in \mathbb{CP}^1 \, .
\end{equation}
This contains three independent equations which can be extracted by introducing some basis spinors $\alpha_{a}$ and $\beta_{a}$ satisfying $\langle \alpha \beta \rangle \neq 0$. The three independent equations are then expressed in terms of contractions with these basis spinors as
\begin{align}
\label{eq:Faa} \alpha^a \alpha^b F_{a\dot{a} b \dot{b}} & = 0 \, ,\\
\label{eq:Fbb} \beta^a \beta^b F_{a\dot{a} b \dot{b}} & = 0 \, , \\
\label{eq:Fab} (\alpha^a \beta^b + \beta^a \alpha^b) F_{a\dot{a} b \dot{b}} & = 0 \, .
\end{align}
The six-dimensional origin of WZW$_4$ (and indeed all such constructed IFT$_{4}$) ensures that the connection $A'$ introduced in the previous section satisfies the ASDYM equation when evaluated on solutions to the WZW$_{4}$ equation of motion. This follows from the six-dimensional equation $\Omega \wedge \mathcal{F}[\mathcal{A}^{\prime}] = 0$ which encodes both the holomorphicity of $\cA^{\prime}$ and eq. \eqref{eq:ASDYM}. To see this explicitly for WZW$_4$ where the connection $A'$ in given by eq.~\eqref{eq:A'}, we note that the $\beta$-contracted eq.~\eqref{eq:Fbb} holds because $\langle \beta \beta \rangle = 0$, and the $\alpha$-contracted eq.~\eqref{eq:Faa} holds due to the Maurer-Cartan identity. The remaining eq.~\eqref{eq:Fab} yields
the equation of motion of WZW$_4$ \eqref{eq:WZW4EOM}.

\subsection{Reduction of \texorpdfstring{WZW$_{4}$}{WZW4} to \texorpdfstring{WZW$_{2}$}{WZW2}}\label{sec:wzw4wzw2}

Next, we will apply a two-dimensional reduction to WZW$_{4}$ specified by two vector fields $V_{i}$ on $\bR^{4}$ with $i = 1, 2$. The idea of reduction is to restrict to field configurations which are invariant under the flow of these vector fields. The two-dimensional dynamics of the reduced theory will be specified by the Lagrangian $\mathcal{L}_{\text{IFT}_2} = (V_1 \wedge V_2) \vee \mathcal{L}_{\text{IFT}_4}$ where $\mathcal{L}_{\text{IFT}_4}$ is the Lagrangian density of the parent theory and we denote the contraction of a vector field $V$ with a differential form $X$ by $V \vee X$.

Let us introduce a pair of unit norm spinors $\gamma_a$ and $\kappa_{\dot a}$ and define the basis of $1$-forms on $\bR^4$
\begin{equation}\label{eq:ComplexCoords}
\dr z = \gamma_a \kappa_{\dot a} \dr x^{a \dot a } ~, \qquad
\dr\bar{z} = \hat{\gamma}_a \hat{\kappa}_{\dot a} \dr x^{a \dot a } ~, \qquad
\dr w= \gamma_a \hat \kappa_{\dot a} \dr x^{a \dot a } ~, \qquad
\dr \bar{w} = - \hat \gamma_a \kappa_{\dot a} \dr x^{a\dot a } ~.
\end{equation}
These are adapted to the complex structure $\mathcal{J}_{\gamma}$ defined by $\gamma_a \in \CP$. We choose to reduce along the vector fields dual to $\dr z$ and $\dr \bar z$ by demanding that $\pd_z g = \pd_{\bar{z} }g = 0$.\footnote{In this case for reality we have $\mathcal{L}_{\text{IFT}_2} = \mathrm{i} (\partial_z \wedge \partial_{\bar{z}}) \vee \mathcal{L}_{\text{IFT}_4}$. } Then, contracting the WZW$_{4}$ Lagrangian with these vector fields results in the two-dimensional action of a principal chiral model (PCM) plus Wess-Zumino (WZ) term:
\begin{equation} \label{eq:PCM+WZ}
S_{\text{PCM}+ \mathscr k \text{WZ}_2}[g] = \frac{1}{2 } \int_\Sigma \mathrm{Tr} \big( g^{-1}\dr g \wedge \star g^{-1}\dr g\big) ~ + ~ \frac{ \mathrm{i} \mathscr k }{3} \int_{\Sigma\times[0,1]} \mathrm{Tr}(\hat g^{-1}\dr \hat g\wedge \hat g^{-1}\dr \hat g\wedge \hat g^{-1}\dr \hat g) \, .
\end{equation}
In this action, the relative coefficient between the WZ-term and the PCM term is given by
\def\ab{\langle \alpha \beta \rangle}
\def\bg{\langle \beta \gamma \rangle}
\def\bgh{\langle \beta \hat \gamma \rangle}
\def\ag{\langle \alpha \gamma \rangle}
\def\ga{\langle \gamma \alpha \rangle}
\def\gb{\langle \gamma \beta\rangle}
\def\agh{\langle \alpha \hat \gamma \rangle}
\def\rma{\mathrm{a}}
\def\rmb{\mathrm{b}}
\def\aa{\mathsf{a}}
\def\bb{\mathsf{b}}
\begin{equation}
\mathscr k = \frac{ \alpha +\beta}{\alpha - \beta} \, , \qquad
\alpha = \frac{\ga}{\agh} \, , \qquad \,
\beta = \frac{\gb}{\bgh} \, .
\end{equation}

Varying the basis spinor $\gamma_{a}$ in these expressions changes the choice of reduction vector fields and interpolates between a family of two-dimensional theories.
The WZW$_{2}$ CFT limit is obtained when $\mathscr k \rightarrow 1$ with $ \alpha \beta $ held fixed. This can be achieved by starting at the K{\"a}hler point in 4d, with $\beta_{a} = \hat{\alpha}_{a}$, and choosing the reduction to be aligned with the complex structure, i.e.\ setting $\gamma_{a} = \alpha_{a}$. An alternative reduction which turns off the WZ term is achieved by setting $\beta = -\alpha $.

For general choices of reduction, the four-dimensional semi-local symmetries descend to a global $G_L\times G_R$ symmetry; this is because, for example, the conditions $\alpha^a \pd_{a \dot{a}} \gamma_R = 0$ and $\pd_z \gamma_R = \pd_{\bar{z}} \gamma_R = 0$ generically contain four independent constraints leaving only constant solutions. However, when the reduction is taken to the CFT point, this system of four constraints is not linearly independent, and chiral symmetries emerge satisfying $\pd_w \gamma_R = 0$ (and vice versa for $\gamma_L$).

\paragraph{Lax connection.}
A virtue of this approach is that a $\mathfrak{g}^\mathbb{C}$-valued Lax connection for the dynamics of the resultant IFT$_{2}$ may be derived from the 4d connection $A'$:
\begin{equation}\label{eq:LaxA}
\begin{aligned}
\mathscr{L}_{\bar{w}} &= \frac{1}{\langle \pi \hat{\gamma } \rangle } \hat{\kappa}^{\dot{a}} \pi^a (\pd_{a \dot{a}} + A'_{a \dot{a}} )= \pd_{\bar{w} } + \frac{\ (\beta-\zeta) }{ ( \alpha-\beta ) }\pd_{\bar{w} } g g^{-1} \, , \\
\mathscr{L}_{w} &= \frac{1}{\langle \pi \gamma \rangle } \kappa^{\dot{a}} \pi^a (\pd_{a \dot{a}} + A'_{a \dot{a}} )= \pd_w + \frac{\alpha (\beta-\zeta) }{ \zeta ( \alpha-\beta ) }\pd_w gg^{-1} \, ,
\end{aligned}
\end{equation}
where the spectral parameter is given as $\zeta = \frac{\langle \gamma\pi \rangle}{\langle \pi \hat{\gamma} \rangle}$. Flatness of this connection for all values of $\zeta$ invokes the field equation of the PCM + WZ theory
\begin{equation}
\alpha \pd_{\bar{w}} ( \pd_w g g^{-1} ) - \beta \pd_w ( \pd_{\bar{w} } g g^{-1} ) = 0 \, \quad \Leftrightarrow \quad \dr (\star- \mathrm{i} \mathscr k ) \dr g g^{-1 } = 0 \, .
\end{equation}
Notice that in the CFT limit $\mathscr k \rightarrow 1$ with $\beta \rightarrow \infty$, $\alpha \rightarrow 0$ the Lax connection becomes chiral and spectral parameter independent.

\subsection{Reduction of \texorpdfstring{hCS$_{6}$}{hCS6} to \texorpdfstring{CS$_{4}$}{CS4}}

Instead of first integrating over $\mathbb{CP}^1$ and then reducing to two dimensions, one could instead directly apply the reduction to hCS$_{6}$. This produces CS$_{4}$ with action
\begin{equation}
S_{\text{CS}_{4}}[A] = \frac{1}{2 \pi \mathrm{i}} \int_{\Sigma \times \mathbb{CP}^1} \omega \wedge \mathrm{Tr} \left( A \wedge \dr A + \frac{2}{3} A \wedge A \wedge A \right) \, .
\end{equation}
Here $\Sigma$ is the $\bR^2 \subset \bR^4$ with coordinates by $w, \bar{w}$, and the meromorphic 1-form $\omega$ is given by
\begin{equation}
\omega =\mathrm{i} (\pd_z \wedge \pd_{\bar{z}}) \vee \Omega \, .
\end{equation}
A crucial feature here is that this contraction introduces zeroes in $\omega$ to complement its poles, as required by the Riemann-Roch theorem. For the case at hand, $\omega$ is given explicitly by
\begin{equation}
\omega =\mathrm{i} \frac{\langle \alpha \beta \rangle^2 \langle\pi \gamma \rangle\langle\pi \hat \gamma \rangle
}{
\langle \pi \alpha \rangle^2
\langle \pi \beta \rangle^2} \langle \pi \dr \pi \rangle \, ,
\end{equation}
and the zeros are introduced at the points $\pi_{a} = \gamma_{a}, \hat{\gamma}_{a}$.
The details of the reduction show that, whilst our six-dimensional gauge field was regular, the connection $A$ entering in CS$_{4}$ develops poles at the zeros of $\omega$. In particular, the component $A_w$ will have a simple pole at $\pi_{a} = \gamma_{a}$ and $A_{\bar{w}}$ will have a simple pole at $\pi_{a} = \hat{\gamma}_{a}$. The four-dimensional Chern-Simons connection is subject to the same boundary conditions as its parent, namely it vanishes at the points $\alpha$ and $\beta$ in $\mathbb{CP}^1$. Subsequent localisation of CS$_{4}$ then gives the same PCM+WZ theory derived by reducing WZW$_{4}$.

\section{The gauged WZW diamond} \label{sec:gaugedWZWdiamond}

We now come to the main results of this paper. In this section, we will construct a diamond correspondence of theories which realises the gauged WZW$_{2}$ model, i.e.\ the $G/H$ coset CFT.

\subsection{Gauged WZW Models}
First let us review the gauging of the WZW model and the crucial Polyakov-Wiegmann identity.
Letting $G$ be a Lie group and $g \in C^{\infty}(\Sigma,G)$ a smooth $G$-valued field, the WZW$_2$ action is\footnote{To minimise factors of imaginary units we momentarily adopt Lorentzian signature.
Schematically, we have $S_{\text{Lorentz}} = \mathrm{i} S_{\text{Euclid}}|_{\star\to \mathrm{i}\star}$.}
\begin{equation}
S_{\text{WZW}_2}[g] = \frac{1}{2} \int_\Sigma \mathrm{Tr}_{\fg} \big( g^{-1}\dr g \wedge \star g^{-1}\dr g\big) ~ + ~ \frac{1}{3} \int_{\Sigma\times[0,1]} \mathrm{Tr}_{\fg} (\hat g^{-1} \dr \hat g \wedge \hat g^{-1} \dr \hat g \wedge \hat g^{-1} \dr \hat g)\, .
\end{equation}
Gauging a vectorial $H$-action of the principal chiral model term is straightforward. We introduce an $\fh$-valued connection $B \in \Omega^{1}(\Sigma) \otimes \fh$ transforming as
\begin{equation}
\ell \in C^{\infty}(\Sigma,H) : \quad B \mapsto \ell^{-1} B \ell + \ell^{-1} \dr \ell \, , \quad g \mapsto \ell^{-1} g \ell \, ,
\end{equation}
with field strength $F[B]= \dr B + B \wedge B$.
The principal chiral term is then gauged by
replacing the exterior derivatives with covariant derivatives $\dr g \to \Dr g = \dr g + [B, g]$.
Less trivially, the gauge completion of the WZ 3-form is \cite{Witten:1991mm, Figueroa-OFarrill:1994uwr,Figueroa-OFarrill:1994vwl,Figueroa-OFarrill:2005vws} 
\begin{equation} \label{eq:gWZWlag}
\mathcal{L}_{\text{gWZ}}[g, B ]= \mathcal{L}_{\text{WZ}}[g ] + \dr \, \mathrm{Tr}_\fg ( g^{-1} \dr g \wedge B + \dr g g^{-1} \wedge B + g^{-1} B g \wedge B ) \, .
\end{equation}
Adding these two pieces together gives the gauged WZW$_2$ action,
\begin{equation} \label{eq:gaugedwzw2}
S_{\text{gWZW}_2}[g,B] = S_{\text{WZW}_2}[g] + \int_\Sigma \mathrm{Tr}_\fg ( g^{-1}\dr g \wedge (1- \star) B + \dr g g^{-1}\wedge (1+\star)B + B \wedge \star B + g^{-1} B g \wedge (1 - \star)B )\, .
\end{equation}
Notice that chiral couplings between currents and gauge fields emerge from combinations of the PCM and WZ contributions.
A remarkable feature,
\begin{equation}
\mathcal{L}_{\text{WZ}}[g_1 g_2] = \mathcal{L}_{\text{WZ}}[g_1 ] + \mathcal{L}_{\text{WZ}}[ g_2] + \dr \, \mathrm{Tr}_\fg \left(\dr g_2 g_2^{-1} \wedge g_1^{-1}\dr g_1 \right) \, ,
\end{equation}
ensures that \eqref{eq:gaugedwzw2} can instead be cast as a difference of two WZW$_2$ models. To see this we choose a parameterisation of the gauge field $B$ in terms of two smooth $H$-valued fields
\begin{equation}
\label{eq:Btoab}
B = \frac{1+\star}{2} a^{-1} \dr a
+ \frac{1-\star}{2}b^{-1}\dr b \, , \quad
a,b\in C^{\infty}(\Sigma,H) \, .
\end{equation}
In two dimensions, this is not a restriction on the field content of the gauge field, but simply a way of parameterising the two independent components of $B$. With such a parameterisation, if we then further define $\tilde{g} = a g b^{-1} \in C^{\infty}(\Sigma,G)$ and $\tilde{h} = a b^{-1}\in C^{\infty}(\Sigma,H)$ the gauged model \eqref{eq:gaugedwzw2} can be written as the difference of two WZW$_2$ models:
\begin{equation}\label{eq:PW2d}
S_{\text{gWZW}_2}[g,B ] = S_{\text{WZW}_2}[\tilde{g}] - S_{\text{WZW}_2}[\tilde{h}] \,.
\end{equation}
This is known as the Polyakov-Wiegmann (PW) identity \cite{Polyakov:1983tt}.

\subsection{Gauging of the \texorpdfstring{WZW$_4$}{WZW4} model} \label{sec:gaugingWZW4}

Let us now consider the four-dimensional WZW$_4$ model, given by eq.~\eqref{eq:WZW4}.
The gauging procedure follows in the exact same manner, producing an analogous gauged WZW$_4$ action,
\begin{equation} \label{eq:gWZW4}
S_{\text{gWZW}_4}^{(\alpha,\beta)}[g,B] = \frac{1}{2} \int_{\mathbb{R}^4} \mathrm{Tr} (g^{-1} \nabla g \wedge \star g^{-1 }\nabla g )
+ \int_{\mathbb{R}^4 \times [0,1] } \omega_{\alpha \beta } \wedge \mathcal{L}_{\text{gWZ}}[g, B] \,.
\end{equation}
Here, we denote the covariant derivative by $\nabla g = \dr g + [B, g]$. A critical difference between two and four dimensions is the applicability of the PW identity, as was pointed out by \cite{Losev:1995cr}. In two dimensions, this mapping relies on the relation \eqref{eq:Btoab}. To extend it to four dimensions, we consider the operator on $1$-forms
\begin{equation}
{\cal J}_{\alpha,\beta} (\sigma ) = - \mathrm{i} \star (\omega_{\alpha,\beta } \wedge \sigma) \, .
\end{equation}
One may check that $ {\cal J}_{\alpha,\beta}^2 = -\mathrm{id}$, so that we can introduce useful projectors
\begin{equation}\label{eq:projectors}
P = \frac{1}{2}\left(\mathrm{id} - \mathrm{i}{\cal J} \right) \, \qquad \bar{P} = \frac{1}{2}\left(\mathrm{id} + \mathrm{i}{\cal J}\right) \ ,
\end{equation}
which furnish a range of identities detailed in appendix~\ref{sec:appendixprojectors}. With these in mind, we can write a four-dimensional analogue to \eqref{eq:Btoab}, 
\begin{equation}\label{eq:B4toab}
B = P \left( a^{-1} \dr a \right) + \bar{P} \left( b^{-1} \dr b \right) \, \quad a, b \in C^{\infty}(\mathbb{R}^4,H)\,.
\end{equation}
With this parameterisation of the gauge field, it is indeed possible to use the composite fields $\tilde g=a g b^{-1}\in C^{\infty}(\mathbb{R}^4,G)$ and $\tilde h=ab^{-1}\in C^{\infty}(\mathbb{R}^4,H)$ to express the gauged WZW$_4$ action in a fashion akin to eq.~\eqref{eq:PW2d} as
\begin{equation}
S_{\text{gWZW}_4}^{(\alpha,\beta)}[g,B] = S_{\text{WZW}_4}^{(\alpha,\beta)}[\tilde{g}] - S_{\text{WZW}_4}^{(\alpha,\beta)}[\tilde{h}] \, .
\end{equation}
However, unlike in two dimensions, this parameterisation of the gauge field eq.~\eqref{eq:B4toab} is not generic. It implies a restriction on the connection, namely that its curvature satisfies
\begin{equation} \label{eq:Bconstraint}
\alpha^a\alpha^b F_{a\dot a b\dot b}[B]=0\,,\quad \beta^{a}\beta^b F_{a \dot{a} b\dot{b}}[B] = 0\, .
\end{equation}
This can be thought of as analogue to imposing that $F$ be strictly a $(1,1)$-form (which indeed this becomes when $\beta =\hat{\alpha}$ and the WZW$_{4}$ is taken at the K{\"a}hler point). It is noteworthy that these constraints on the background gauge field agree with two of the three ASDYM equations; the same two equations that were identically satisfied by the Yang parameterisation of the connection $A'$. In the forthcoming analysis, we will see how this arises from the hCS$_{6}$ construction.

\subsection{A six-dimensional origin} \label{sec:6dorigin}

We now turn to the six-dimensional holomorphic Chern-Simons theory on twistor space that will descend to the above gauged WZW models in two and four dimensions.
Given the factorisation of gWZW$_2$ to the difference of WZW$_2$ models, a natural candidate here is to consider simply the difference of hCS$_6$ theories to generalize the six-dimensional action introduced in \cite{Kcostello, Bittleston:2020hfv, Penna:2020uky, Costello:2021bah}. Indeed, a similar idea was proposed by \cite{Stedman:2021wrw} in the construction of 2d coset models from the difference of CS$_4$ theories. However, how this should work in six dimensions is less clear as the factorisation of gWZW$_4$ requires the curvature of the gauge field to be constrained.

The fundamental fields of our theory are two connections $\cA \in \Omega^{0,1}(\mathbb{PT}) \otimes \fg$ and $\cB \in \Omega^{0,1}(\mathbb{PT}) \otimes \fh$, which appear in the six-dimensional action
\begin{equation} \label{eq:new6daction}
S_{\mathrm{ghCS_6}}[\cA , \cB] = S_{\mathrm{hCS_6}}[\cA] - S_{\mathrm{hCS_6}}[\cB]
- \frac{1}{2\pi \mathrm{i}} \int_{\mathbb{PT}} \bar{\partial} \Omega \wedge \mathrm{Tr} \big( \cA \wedge \cB \big) \, ,
\end{equation}
where the functional $S_{\mathrm{hCS_6}}$ is defined in eq.~\eqref{eq:hCS6gauged}. As well as the bulk hCS$_{6}$ functionals, we have also included a coupling term between the two connections which contributes on the support of $\bar{\pd} \Omega$, i.e.\ at the poles of $\Omega$. We will shortly provide a motivation for this boundary term related to the boundary conditions we will impose on the theory.

This definition is slightly imprecise; strictly speaking, the inner product denoted by `$\mathrm{Tr}$' should be defined separately for each algebra, i.e.~$\mathrm{Tr}_\fg$ and $\mathrm{Tr}_\fh$. In the coupling term, where $\mathcal{B}$ enters inside $\mathrm{Tr}_\fg$, we should first act on $\cB$ with some Lie algebra homomorphism from $\fh$ to $\fg$, and in principle this homomorphism could be chosen differently at each pole of $\Omega$. We discuss more general gaugings, beyond the vectorial gauging hereby considered, in appendix~\ref{Sec:General gaugings}.

To complete the specification of the theory, we must supply boundary conditions which ensure the vanishing of the boundary term in the variation of \eqref{eq:new6daction},
\begin{equation}
\delta S_{\mathrm{ghCS_6}} \big\vert_{\text{bdry}} = \frac{1}{2 \pi \mathrm{i}} \int_{\mathbb{PT}} \bar{\partial} \Omega \wedge \mathrm{Tr} \big( (\delta \cA + \delta \cB) \wedge (\cA - \cB) \big) \, .
\end{equation}
Since $\bar{\partial} \Omega$ only has support at the poles of $\Omega$, the integral over $\CP$ may be computed explicitly in this term. As well as contributions proportional to delta-functions on $\CP$, this will also include $\CP$-derivatives of delta-functions since the poles in $\Omega$ are second order. Using the localisation formula in the appendix~\ref{appendix on localisation formulae}, we find
\begin{equation}
\begin{aligned}
\delta S_{\mathrm{ghCS_6}} \big\vert_{\text{bdry}} = - \int_{\bR^{4}} \bigg[
& \frac{\alpha_{a} \beta_{b} \Sigma^{ab}}{\langle \alpha \beta \rangle} \wedge \mathrm{Tr} \big( (\delta \cA + \delta \cB) \wedge (\cA - \cB) \big) \\
& \qquad + \frac{1}{2} \alpha_{a} \alpha_{b} \Sigma^{ab} \wedge \pd_{0} \mathrm{Tr} \big( (\delta \cA + \delta \cB) \wedge (\cA - \cB) \big)
\bigg] + \alpha \leftrightarrow \beta \, .
\end{aligned}
\end{equation}
In this expression, we introduce a basis for the self-dual 2-forms defined by $\Sigma^{ab} = \varepsilon_{\dot{a} \dot{b}} \dr x^{a \dot{a}} \wedge \dr x^{b \dot{b}}$.
Let us also introduce an orthogonal decomposition of $\fg$ such that
\begin{equation}
\fg = \fh \oplus \fk \ , \qquad
\mathrm{Tr} \big( X \cdot Y \big) = \mathrm{Tr} \big( X^{\fh} \cdot Y^{\fh} \big) + \mathrm{Tr} \big( X^{\fk} \cdot Y^{\fk} \big) \ .
\end{equation}
To attain the vanishing of the boundary variation, we consider the boundary conditions
\begin{equation} \label{eq:ABboundarycond}
\cA^{\fk} \big\vert_{\alpha , \beta} = 0 \ , \qquad
\cA^{\fh} \big\vert_{\alpha , \beta} = \cB \big\vert_{\alpha , \beta} \ , \qquad
\pd_{0} \cA^{\fh} \big\vert_{\alpha , \beta} = \pd_{0} \cB \big\vert_{\alpha , \beta} \ .
\end{equation}
This completes our definition of the gauged hCS$_{6}$ theory.

One might choose to think of the boundary term in the variation as being a potential for a `symplectic' form\footnote{Precedent in the literature dictates that we denote the symplectic form as $\Omega$; we trust that context serves to disambiguate from the meromorphic differential $\Omega$.} \footnote{This is slightly loose as the 2-form is degenerate, so properly speaking we should restrict to symplectic leaves.}
\begin{equation}
\Theta = \delta S_{\mathrm{ghCS_6}} \big\vert_{\text{bdry}} \ , \qquad
\Omega = \delta \Theta = - \frac{1}{2 \pi \mathrm{i}} \int_{\mathbb{PT}} \bar{\partial} \Omega \wedge \Big( \mathrm{Tr}_\fg \big( \delta \cA \wedge \delta \cA \big) - \mathrm{Tr}_\fh \big( \delta \cB \wedge \delta \cB \big) \Big) \ ,
\end{equation}
such that our boundary conditions define a Lagrangian (i.e.\ maximal isotropic) subspace.
We should like to really interpret this as a symplectic form on an appropriate space of fields defined over $\mathbb{R}^{4}$.
Evaluating the integral over $\CP$ and writing $\Omega = \Omega_{\cA} - \Omega_{\cB}$, this symplectic form is given by
\begin{equation}
\Omega_{\cA} = \int_{\bR^{4}} \bigg[ \frac{\alpha_{a} \beta_{b} \Sigma^{ab}}{\langle \alpha \beta \rangle} \wedge \mathrm{Tr}_{\fg} \big( \delta \cA \wedge \delta \cA \big) \big\vert_{\alpha} + \frac{1}{2} \alpha_{a} \alpha_{b} \Sigma^{ab} \wedge \pd_{0} \mathrm{Tr}_{\fg} \big( \delta \cA \wedge \delta \cA \big) \big\vert_{\alpha} \bigg] + \alpha \leftrightarrow \beta \ ,
\end{equation}
with an analogous expression for $\Omega_{\cB}$.
Because our boundary conditions are identical at each pole, we concentrate now only on the contributions associated to the pole at $\alpha$. The symplectic form is not sensitive to the entire field configuration $\cA \in \Omega^{1} (\mathbb{PT}) \otimes \fg$, but rather to the evaluation of $\cA$ at the poles and its first derivatives,
\begin{equation}
\vec{\cA} = \big( \cA \vert_{\alpha} , \pd_{0} \cA \vert_{\alpha} \big) \ .
\end{equation}
This data may be interpreted as defining a 1-form (more precisely a $(0,1)$-form with respect to the complex structure defined by $\alpha$) on $\bR^{4}$ valued in the algebra\footnote{The dimension of $\vec{\fg}$ is $2 \dim(G)$, so it must be isomorphic to $\bR^{\dim(G)} \oplus \bR^{\dim(G)}$ as a vector space. The Lie algebra structure may be derived by considering consecutive infinitesimal gauge transformations. In the CS$_{4}$ literature, these structures have been studied under the name `defect Lie algebra' \cite{Benini:2020skc, Lacroix:2020flf}.} $\vec{\fg} = \fg \ltimes \bR^{\dim(G)}$. With this in mind, it is more accurate to say that the contribution from the pole at $\pi_{a} = \alpha_{a}$ in $\Omega$ is a symplectic form on the space of configurations
\begin{equation}
\big( \vec{\cA} , \vec{\cB} \big) \in \Omega^{0,1} (\bR^{4}) \otimes \big( \vec{\fg} \oplus \vec{\fh} \big) \ .
\end{equation}
This symplectic form may be succinctly written by introducing an inner product on the algebra $\vec{\fg} \oplus \vec{\fh}$, and our boundary conditions describe an isotropic subspace with respect to this inner product.\footnote{ This need not be the case, as our boundary conditions could generically intertwine constraints on the algebra and spacetime components, meaning they could not be captured by a subspace of the algebra alone. They would always, however, define an isotropic subspace of $\Omega^{0,1} (\bR^{4}) \otimes \big( \vec{\fg} \oplus \vec{\fh} \big)$ by definition. Examples of this more general type of boundary condition can be found in \cite{Cole:2023umd}.}

To be explicit we associate $\mathbb{R}^{\dim{G}}$ with the dual $\fg^\ast$ and denote the natural pairing of the algebra and its dual with $\llparenthesis\bullet , \bullet \rrparenthesis$. We let $\vec{X} = (x, \tilde{x}) \in \vec{\fg} $ such that bracket on $\vec{\fg}$ is defined by
\begin{equation}
[\vec{X}, \vec{Y}]_{\vec{\fg}} = ( [x,y] , \mathrm{ad}_x^\ast \tilde{y}-\mathrm{ad}^\ast_y \tilde{x} )\, ,
\end{equation}
where the co-adjoint action is $\llparenthesis x, \mathrm{ad}^\ast_y \tilde{x}\rrparenthesis = \llparenthesis[x,y], \tilde{x}\rrparenthesis $. We equip $\vec{\fg}$ with the inner product
\begin{equation}\label{eq:inner}
\langle \vec{X}, \vec{Y}\rangle_{\vec{\fg}} = \frac{\langle \beta \hat{\alpha} \rangle}{\langle \alpha \beta \rangle \langle \alpha \hat{\alpha} \rangle} \mathrm{Tr}_\fg(x \cdot y)+ \frac{1}{2} \Big( \llparenthesis x, \tilde{y}\rrparenthesis + \llparenthesis y,\tilde{x}\rrparenthesis \Big)
\, ,
\end{equation}
such that the relevant contribution to the symplectic form is given by
\begin{equation}
\Omega_{\cA} = \int_{\bR^{4}} \mu_\alpha \wedge \biprod{\delta \vec{\cA}}{\delta \vec{\cA}}_{\vec{\fg}} \ .
\end{equation}
where $\mu^\alpha= \alpha_{a} \alpha_{b} \Sigma^{ab} $ is the $(2,0)$-form defined by the complex structure associated to $\alpha \in \CP$.

In a similar fashion we will let $\vec{U} = (u ,\tilde{u})$ and $\vec{V} = (v, \tilde{v})$ be elements of $\vec{\fh}$ which is equipped with a bracket and pairing via the same recipe. We consider the commuting direct sum $\vec{\fg} \oplus \vec{\fh}$ equipped with pairing and bracket
\begin{equation}
\biprodb{(\vec{X}, \vec{U})}{(\vec{Y}, \vec{V})} = \langle \vec{X}, \vec{Y}\rangle_{\vec{\fg}} - \langle \vec{U}, \vec{V}\rangle_{\vec{\fh}} \, , \qquad \commb{(\vec{X}, \vec{U})}{(\vec{Y}, \vec{V})} = \big( \comm{\vec{X}}{\vec{Y}}_{\vec{\fg}} \, , \comm{\vec{U}}{\vec{V} }_{\vec{\fh}} \big) \,,
\end{equation}
such that the total symplectic form coming from the pole at $\alpha$ is just
\begin{equation}
\Omega = \int_{\bR^{4}} \mu_\alpha \wedge \biprodb{(\delta \vec{\cA} , \delta \vec{\cB})}{(\delta \vec{\cA} , \delta \vec{\cB})} \ .
\end{equation}
Then, our boundary conditions can be expressed as $\big( \vec{\cA} , \vec{\cB} \big) \in \Omega^{0,1} (\bR^{4}) \otimes L$ where we introduce a subspace
\begin{equation}
L = \big\{ (\vec{X}, \vec{U}) \in \vec{\fg} \oplus \vec{\fh} \mid x = u \, , \ P_{\fh}^\ast \tilde{x} = \tilde{u} \big\} \ ,
\end{equation}
in which $P_{\fh}^\ast$ is the dual to the projector $P_\fh$ into the subgroup i.e.\ $\llparenthesis x, P_{\fh}^\ast \tilde{x}\rrparenthesis= \llparenthesis P_{\fh} x , \tilde{x}\rrparenthesis$. As $L$ is defined by $\dim \fg+ \dim \fh$ constraints, it is half-dimensional and it is also isotropic with respect to $\biprodb{\bcdot}{\bcdot}$, hence defining a Lagrangian subspace. Moreover, assuming that $G/H$ is reductive, $L$ is a subalgebra\footnote{If $\fg = \fh + \fk$ is not assumed to be reductive then the stabiliser of $L$ consists of elements of the form 
\begin{equation*}
\textrm{stab}_L = \big\{ (\vec{X}, \vec{U}) \in \vec{\fg} \oplus \vec{\fh} \mid x= u \, , \ P_{\fh}^\ast \tilde{x} = \tilde{u} \, , \ [u, \fk]= 0 \, , \ ([\fh,\fk ] , \tilde{u})=0 \big\} \, .
\end{equation*}
\vspace{-1.75em}
}. Pre-empting the following section, this analysis indicates that there will be a residual $\vec{\fh}$ gauge symmetry associated to the pole at $\alpha$, and similarly at $\beta$.

We can make one further observation\footnote{We thank A. Arvanitakis for this suggestion.} of the role of the boundary contribution from a symplectic perspective that is best illustrated by a finite-dimensional analogy. Recall that the cotangent bundle ${\cal M} = T^*X$ is a symplectic manifold; if we let $\{x^i\}$ be local coordinates on $X$ and $\{ \xi_i\}$ the components of a 1-form $\xi = \xi_i dx^i \in T_x^*X$, then $p=(x^i, \xi_i)$ provide local coordinates for ${\cal M}$ in terms of which the canonical symplectic form is $\Omega= \dr \xi_i \wedge \dr x^i$. The tautological potential (which admits a coordinate free definition in terms of the projection $\pi: T^*X \rightarrow X$) for this is given by $\Theta = \xi_i \dr x^i$.   The zero section, i.e. points $p=(x^i, \xi_i= 0)$ of $T^*X$ is a Lagrangian and notice that $\Theta$ vanishes trivially here.   Now Weinstein's tubular neighbourhood theorem ensures that in the vicinity of a Lagrangian $L$,  any symplectic manifold ${\cal M}$ locally looks like $T^*L$ with $L$ given by the zero section.    In the case at hand, our boundary conditions are of the schematic form $\xi = \cA - \cB  = 0 $, and the effect of including the specific boundary contribution to the Lagrangian ensures that the resultant symplectic potential is the tautological one.

To close this section, let us comment that at the special point for which $\alpha_{a} = \hat{\beta}_{a}$, one of the terms in the inner product eq.~\eqref{eq:inner} vanishes. This allows for a larger class of admissible boundary conditions, even for the ungauged model, including the examples
\begin{equation}\label{natd}
\cA \big\vert_{\hat{\alpha}} = 0 \ , \qquad
\pd_{0} \cA \big\vert_{\alpha} = 0 \ \qquad \textrm{or} \qquad
\pd_{0} \cA \big\vert_{\hat{\alpha}} = 0 \ , \qquad
\pd_{0} \cA \big\vert_{\alpha} = 0 \, .
\end{equation}
We leave these for future development.

\subsection{Localisation to \texorpdfstring{gWZW$_{4}$}{gWZW4}}
\label{sec:localtogwzw4}

The localisation procedure follows in a similar fashion to the ungauged model. However, given that there are now two gauge fields $\cA$ and $\cB$, some care is required to account for degrees of freedom and residual symmetries.

We introduce a new pair of connections $\cA'\in \Omega^{0,1}(\mathbb{PT}) \otimes \fg$ and $\cB'\in \Omega^{0,1}(\mathbb{PT}) \otimes \fh$, along with group valued fields $\hat g \in C^{\infty}(\mathbb{PT}, G)$ and $\hat h \in C^{\infty}(\mathbb{PT}, H)$ related to the original gauge fields by
\begin{equation}
\begin{aligned} \label{eq:redundancyredux}
\cA = \hat{g}^{-1} \cA^\prime \hat{g} + \hat{g}^{-1} \bar{\pd} \hat{g} \equiv \cA^\prime{}^{\hat{g}} ~,\\
\cB = \hat{h}^{-1} \cB^\prime \hat{h} + \hat{h}^{-1} \bar{\pd} \hat{h} \equiv \cB^\prime{}^{\hat{h}} ~.
\end{aligned}
\end{equation}
The redundancy in this parameterisation is given by the action of $\check \gamma \in C^{\infty}(\mathbb{PT},G)$ and $\check \eta \in C^{\infty}(\mathbb{PT},H)$:
\begin{align}
\label{eq:intgamma}
& \cA' \mapsto \check \gamma^{-1} \cA' \check \gamma + \check \gamma^{-1}\bar \partial \check \gamma \,, \qquad
\hat g \mapsto \check \gamma^{-1} \hat g \,,\\
\label{eq:inteta}
& \cB' \mapsto \check \eta^{-1} \cB' \check \eta + \check \eta^{-1}\bar \partial \check \eta \,, \qquad \hspace{0.3em}
\hat h \mapsto \check \eta^{-1} \hat h \,,
\end{align}
which leave $\cA$ and $\cB$ invariant. As before, this is partially used to fix away the $\mathbb{CP}^1$ legs
\begin{equation}
\label{eq:noCP1constraint}
\cA_0^\prime = \cB_0^\prime = 0 \,.
\end{equation}
The localisation procedure will produce a four-dimensional boundary theory with fields given by the evaluations of $\hat g, \hat h$ and their $\mathbb{CP}^1$-derivatives at the poles $\alpha$ and $\beta$ of $\Omega$. Since the $\mathbb{CP}^1$-derivatives will have an important role, we give them names, 
\begin{equation}
\label{ec:uandv}
\hat u = \hat g^{-1} \partial_0 \hat g \,,\quad \hat v = \hat h^{-1} \partial_0 \hat h \,.
\end{equation}
After fixing \eqref{eq:noCP1constraint}, we note that there is still some remaining symmetry given by internal gauge transformations \eqref{eq:intgamma} and \eqref{eq:inteta} which are $\mathbb{CP}^1$-independent. We use this residual symmetry to fix the values
\begin{equation}
\label{eq:ghatinf}
\hat g|_{\beta}=\mathrm{id} \,, \quad \hat h|_{\beta}=\mathrm{id}\,.
\end{equation}
On the other hand, the action \eqref{eq:new6daction} is invariant under gauge transformations acting on $\cA$ and $\cB$ which preserve the boundary conditions \eqref{eq:ABboundarycond}. These are given by smooth maps $\hat \gamma \in C^{\infty}(\mathbb{PT},G)$ and $\hat \eta \in C^{\infty}(\mathbb{PT},H)$ satisfying\footnote{This requires that $G/H$ is reductive meaning $[\fh, \fk] \subset \fk$.}
\begin{equation}
\hat{\gamma }\vert_{\alpha, \beta} = \hat{\eta} \vert_{\alpha, \beta} \,, \quad \partial_0 \hat{\gamma }\vert_{\alpha, \beta} = \partial_0 \hat{\eta} \vert_{\alpha, \beta} \,.
\end{equation}
The induced action of these gauge transformations on the new field content is
\begin{align}
& \quad \cA' \mapsto \cA' \, , \qquad
\hat g \mapsto \hat g \hat \gamma \, , \qquad
\hat u \mapsto \hat \gamma^{-1} \hat u \hat \gamma + \hat \gamma^{-1}\partial_0 \hat \gamma \\
&\quad \cB' \mapsto \cB' \, , \qquad \hspace{0.25em}
\hat h \mapsto \hat h \hat \eta\, , \qquad
\hat v \mapsto \hat \eta^{-1} \hat v\hat \eta + \hat \eta^{-1}\partial_0 \hat \eta\,.
\end{align}
We want to use this symmetry to further fix degrees of freedom. Note that whereas the right action on the fields $\hat g$ and $\hat h$ at $\alpha$ is entirely unconstrained, we would like the action at $\beta$ to preserve the gauge fixing condition \eqref{eq:ghatinf}. This is achieved by performing both an internal and external gauge transformation simultaneously, and requiring $\hat \gamma|_{\beta}=\check \gamma$ and $\hat \eta|_{\beta}=\check \eta$. This results in an induced left action on the fields $\hat g$ and $\hat h$ at $\alpha$. In summary, introducing some notation for simplicity, we have our boundary degrees of freedom
\begin{align}
& \hat g|_{\alpha} \coloneqq g \,, \qquad \hat g|_{\beta} = \mathrm{id}\,,\qquad \hat u|_{\alpha}\coloneqq u \,,\qquad \hat u|_{\beta}\coloneqq \tilde u \\
&
\label{eq:gaugefixh1}\hat h|_{\alpha} \coloneqq h \,, \qquad \hat h|_{\beta} = \mathrm{id}\,,\qquad \hat v|_{\alpha}\coloneqq v \,,\qquad \hat v|_{\beta}\coloneqq \tilde v \,,
\end{align}
and boundary gauge transformations
\begin{align}
& \hat \gamma \vert_{\alpha} = \hat \eta \vert_{\alpha} \coloneqq r \, , \qquad \quad \hspace{0.2em}
\hat{\gamma}^{-1} \partial_0 \hat \gamma \vert_{\alpha} = \hat{\eta}^{-1} \partial_0 \hat \eta \vert_{\alpha} \coloneqq \epsilon \ , \\
& \hat \gamma \vert_{\beta} = \hat \eta \vert_{\beta} \coloneqq \ell^{-1} \,,\qquad
\hat{\gamma}^{-1} \partial_0 \hat \gamma|_{\beta} = \hat{\eta}^{-1} \partial_0 \hat \eta|_{\beta} \coloneqq \tilde \epsilon \ ,
\end{align}
which act on the boundary fields as
\begin{align}
& g \mapsto \ell g r \,,\quad u \mapsto r^{-1}ur+\epsilon \,, \quad \tilde u \mapsto \ell \tilde u \ell^{-1}+\tilde \epsilon \\
&h \mapsto \ell h r \,,\quad v \mapsto r^{-1}vr+\epsilon \,, \quad \tilde v \mapsto \ell \tilde v \ell^{-1}+\tilde \epsilon\,.
\end{align}
with $\ell,r \in C^{\infty}(\mathbb{R}^4,H)$ and $\epsilon, \tilde \epsilon \in C^{\infty}(\mathbb{R}^4, \fh)$. Based on our expectation of a gauge theory containing a $G$-valued field and a vectorial $H$-gauge symmetry, we use the above symmetries to fix
\begin{equation} \label{eq:gaugefixh2}
h = \text{id} \, , \qquad v = \tilde{v} = 0\,.
\end{equation}
We are thus left with a residual symmetry $r = \ell^{-1}$ acting as
\begin{eqnarray}
g \mapsto \ell g \ell^{-1} \, , \quad u \mapsto \ell u \ell^{-1} \, , \quad \tilde{u} \mapsto \ell \tilde{u} \ell^{-1} \, , \quad B \mapsto \ell B \ell^{-1} - \dr \ell \ell^{-1} \, ,
\end{eqnarray}
which will become the gauge symmetry of our 4d theory.

We now proceed with the localisation of the six-dimensional action.
As with the ungauged model, the first step is to write the action in terms of $\cA', \cB'$ and $\hat{g}, \hat{h}$. Given that the localisation formula \eqref{eq:localisationformula} introduces at most one $\partial_0$ derivative, all dependence on $\hat h$ will drop due to our gauge fixing choices \eqref{eq:gaugefixh1} and \eqref{eq:gaugefixh2}. Hence, there will be no contribution from $S_{\mathrm{hCS}_6}[\cB]$ to the four-dimensional action.
As per eq.~\eqref{eq:hcsA'g'} we find that the bulk equations (i.e. contributions to the variation of the action that are not localised to the poles of $\Omega$) enforce $\bar \partial_0 \cA'_{\dot a}=\bar \partial_0 \cB'_{\dot a}= 0$. This implies that the components $\cA'_{\dot a},\cB'_{\dot a}$ are holomorphic, which (combined with the fact that they have homogeneous weight 1) allows us to deduce that
\begin{equation}
\cA'_{\dot a} = \pi^a A'_{a\dot a} \, , \quad \cB'_{\dot a} = \pi^a B'_{a\dot a} \, ,
\end{equation}
in which $A'_{a\dot a},B'_{a\dot a}$ are $\mathbb{CP}^1$-independent. Imposing this bulk equation, and the gauge fixings described above, the remaining contributions in \eqref{eq:new6daction} are given by
\begin{equation} \label{eq:almostlocalised}
\begin{aligned}
& S_{\text{ghCS}_6}[\cA , \cB ] = \frac{1}{2 \pi \mathrm{i}} \int_{\mathbb{PT}} \bar{\pd} \Omega \wedge \mathrm{Tr} \big( \cA^\prime \wedge \bar{\pd} \hat{g} \hat{g}^{-1} - (\hat g^{-1} \cA^{\prime} \hat{g} + \hat{g}^{-1} \bar \partial \hat{g})\wedge \cB^{\prime} \big) \\
& \hspace{9em} - \frac{1}{6 \pi \mathrm{i}} \int_{\mathbb{PT} \times[0,1]} \bar \partial\Omega \wedge \mathrm{Tr} \big(\hat{g}^{-1} \dr \hat{g} \wedge \hat{g}^{-1} \dr \hat{g} \wedge \hat{g}^{-1} \dr \hat{g} \big) \, .
\end{aligned}
\end{equation}

In the ungauged model, the next step was to solve the boundary conditions for $\cA'$ in terms of $\hat{g}$. Here, the boundary conditions do not fully determine $A'_{a\dot a}, B'_{a\dot a}$ and instead relate them as\footnote{The boundary conditions on the $\mathbb{CP}^1$ derivatives of the gauge fields impose
\begin{equation}
\frac{\alpha^{a}}{\langle \alpha \beta \rangle} \big( \nabla_{a \dot{a}} g g^{-1} \big)^{\fh} = - \beta^{a} \nabla_{a \dot{a}} \tilde{u}^{\fh} \ , \qquad
\frac{\beta^{a}}{\langle \alpha \beta \rangle} \big( g^{-1} \nabla_{a \dot{a}} g \big)^{\fh} = - \alpha^{a} \nabla_{a \dot{a}} u^{\fh} \ , 
\end{equation}
however we will not invoke these since they will follow as equations of motion of the 4d theory due to the addition of the boundary term in the gauged hCS$_6$ action~\eqref{eq:new6daction}.
For more details see appendix~\ref{Sec:General gaugings}.}
\begin{equation}
\label{eq:bcsol1}
A'_{a \dot{a}} = B'_{a\dot a} + \Theta_{a\dot a}\coloneqq B'_{a\dot{a}} - \frac{1}{\langle \alpha \beta \rangle} \beta_a \alpha^b \nabla_{b \dot{a}} g g^{-1}\,,
\end{equation}
where the covariant derivative is given by $\nabla_{a\dot a} g g^{-1}=\partial_{a\dot a}g g^{-1} + B'_{a\dot a} - \mathrm{Ad}_g B'_{a\dot a}$.
Equation \eqref{eq:bcsol1} allows us to express \eqref{eq:almostlocalised} entirely in terms of $\cB'$, $\Theta = \pi^a \Theta_{a \dot a} \bar e^{\dot a}$ and $\hat g$. Many of the terms combine to produce a gauged Wess-Zumino Lagrangian contribution (eq.~\eqref{eq:gWZWlag}) with the result
\begin{equation}
S_{\mathrm{ghCS}_6}[\cA , \cB ] = \frac{1}{2 \pi \mathrm{i}} \int_{\mathbb{PT}} \bar{\pd} \Omega \wedge \mathrm{Tr} \big(\Theta \wedge \left( \nabla \hat{g} \hat{g}^{-1} - \cB^\prime \right)\big) - \frac{1}{2 \pi \mathrm{i}} \int_{\mathbb{PT} \times[0,1]} \bar \partial\Omega \wedge \mathcal{L}_{\mathrm{gWZ}}[\hat g,\cB'] \,.
\end{equation}
Given that both $B_{a\dot{a}} $ and $\Theta_{a\dot{a}}$ are $\mathbb{CP}^1$-independent, we have that
\begin{eqnarray}
\int_{\mathbb{PT}} \bar{\pd} \Omega \wedge \mathrm{Tr} (\Theta \wedge \cB^\prime) = 0 \, ,
\end{eqnarray}
with cancelling contributions from the two end points of the integral. Hence we are left with a manifestly covariant result \begin{equation}
S_{\text{ghCS}_6}[\cA , \cB] = \frac{1}{2 \pi \mathrm{i}} \int_{\mathbb{PT}} \bar{\pd} \Omega \wedge \mathrm{Tr} \big( \Theta \wedge \left( \nabla \hat{g} \hat{g}^{-1} \right)
- {\cal L}_{\mathrm{gWZ}}[\hat{g},\cB^\prime ] \big) \, .
\end{equation}
Application of the localisation formula in the appendix~\eqref{ec:localisationformula} yields the four-dimensional action
\begin{equation}\label{4d action II}
\begin{split}
S_{\text{IFT}_{4}} =& \frac{1}{2} \int_{\bR^{4}} \mathrm{Tr} \big(
\nabla g g^{-1} \wedge {\star} \nabla g g^{-1} \big) + \int_{\bR^{4} \times [0,1]} \omega_{\alpha , \beta} \wedge \cL_{\text{gWZ}} [g, B'] \\ &\qquad - \int_{\bR^{4}}
\mu_\alpha \wedge \mathrm{Tr} ( u \cdot F[B'] ) + \mu_\beta \wedge \mathrm{Tr} ( \tilde{u} \cdot F[B'] )
\ .
\end{split}
\end{equation}
At this point only the $\fh$-components of $u$ and $\tilde u$ contribute to the action, and so henceforth, to ease notation and without loss of generality, we set their $\fk$-components to zero. 

Something rather elegant has occurred; we have found the localisation of the six-dimensional theory returns not only the gauging of the WZW$_{4}$ model, but also residual edge modes serving as Lagrange multipliers that constrain the field strength to obey exactly those conditions of eq.~\eqref{eq:Bconstraint} which ensure the theory can be written as the difference of WZW$_{4}$ models.
The constraints $F^{2,0}=0$ and $F^{0,2}=0$ have also been imposed by Lagrange multipliers in the context of 5d K{\"a}hler Chern-Simons theory \cite{Nair:1990aa,Nair:1991ab}. This theory bears a similar relationship to WZW$_{4}$ as 3d Chern-Simons theory bears to WZW$_{2}$.   This poses a natural question:   what is the direct relationship between this 5d K{\"a}hler Chern-Simons theory  and 6d holomorphic Chern-Simons theory?  We suspect the mechanism here is rather similar to that which relates  CS$_{4}$ and CS$_{3}$  \cite{Yamazaki:2019prm}; we comment further on this in the outlook.

\subsection{Equations of motion and ASDYM} \label{sec:EOMandASDYM}

Making use of the projectors previously introduced in eq.~\eqref{eq:projectors}, the equations of motion then read
\begin{equation}
\label{ec:4deom}
\begin{aligned}
\delta B': &\quad 0 = \bar{P} \nabla g g^{-1}\vert_{\fh} - P g^{-1} \nabla g \vert_{\fh} + \star \left( \mu_\alpha \wedge \nabla u+ \mu_\beta \wedge \nabla \tilde{u} \right) \ , \\
\delta g: &\quad 0 = \nabla \star \nabla g g^{-1} - \omega_{\alpha , \beta}\wedge \nabla (\nabla g g^{-1}) + 2 \omega_{\alpha , \beta}\wedge F[B'] \ , \\
\delta u: &\quad 0 = \mu_\alpha \wedge F[B'] \ , \\
\delta \tilde{u}: &\quad 0 = \mu_\beta \wedge F[B'] \ . \\
\end{aligned}
\end{equation}
We can exploit the projectors to extract from the $B'$ equation of motion the two independent contributions:
\begin{equation}
\delta B': \quad \begin{aligned}
0 = \bar{P} \left( \nabla g g^{-1} \vert_{\fh} + \star( \mu_\beta \wedge \nabla \tilde{u} ) \right) \\
0 = P \left( g^{-1}\nabla g \vert_{\fh} - \star( \mu_\alpha\wedge \nabla u ) \right) \label{eq:Beqm}
\end{aligned} \ . 
\end{equation}
In fact, these are exactly the conditions that arise from the $\mathbb{CP}^1$ derivative components of the boundary condition $\pd_0 \cA^\fh \vert_{\alpha, \beta} = \pd_0 \cB \vert_{\alpha, \beta}$.

Making use of the identity
\begin{equation}
\nabla( \omega_{\alpha \beta }\wedge \star( \mu_\beta \wedge \nabla \tilde{u} ) ) = \nabla( \mu_\beta \wedge \nabla \tilde{u} ) = \mu_\beta \wedge F[B'] \cdot \tilde{u} \ , 
\end{equation}
we obtain an on-shell integrability condition for the first of eq.~\eqref{eq:Beqm}, namely that
\begin{equation}
\nabla ( \omega_{\alpha \beta }\wedge \bar{P} (\nabla g g^{-1} \vert_{\fh} ) ) = 0 \, .
\end{equation}
Hence, using the projection of the $\delta g$ equation of motion into $\fh$, we have that $\omega_{\alpha,\beta} \wedge F[B']= 0$ follows on-shell.

Let us turn back to the ASDYM equations which we can recast as
\begin{equation}
\mu_\alpha \wedge F = 0 \, , \quad \mu_\beta \wedge F = 0\, , \quad \omega_{\alpha, \beta} \wedge F = 0 \, .
\end{equation}
In differential form notation, the solution of the boundary condition eq.~\eqref{eq:bcsol1} can be written as
\begin{equation}
A' = B' - \bar{P}( \nabla g g^{-1}) \, .
\end{equation}
By virtue of the identities obeyed by the projectors, eq. \eqref{eq:projectorid}, and the covariant Maurer-Cartan identity obeyed by $R^\nabla = \nabla g g^{-1} $,
\begin{equation}
\nabla R^\nabla - R^\nabla \wedge R^\nabla = (1-\mathrm{Ad}_g) F[B'] \, ,
\end{equation}
one can readily establish
\begin{eqnarray}
\mu_\beta \wedge F[A'] &=& \mu_\beta \wedge F[B'] \, , \\
\mu_\alpha \wedge F[A'] &=&\mu_\alpha \wedge \mathrm{Ad}_g F[B'] \, , \\
2\omega_{\alpha \beta } \wedge F[A'] &=& 2 \omega_{\alpha \beta } \wedge F[B'] + 2 \omega_{\alpha \beta } \wedge \nabla \bar{P} (R^\nabla ) \\
&=& 2 \omega_{\alpha \beta } \wedge F[B'] - \nabla (\star \nabla g g^{-1}) + \omega_{\alpha \beta } \wedge \nabla (\nabla g g^{-1}) \, .
\end{eqnarray}
Hence we conclude that the $\delta g, \delta u, \delta \tilde{u}$ equations of motion are equivalent to the ASDYM equations for the connection $A'$. Demanding that the $B'$ connection is also ASD requires in addition that $\omega_{\alpha,\beta}\wedge F[B']= 0$, and as shown above this is indeed a consequence of the $B'$ equations of motion.

\subsection{Constraining then reducing}

We now proceed to the bottom of the diamond by reduction of the IFT$_4$. In this section, we shall first implement the constraints imposed by the Lagrange multipliers $u,\tilde{u}$ in the 4d theory and then reduce.
While not the most general reduction, this will allow us to directly recover the gauged WZW coset CFT.
In section \ref{sec:newIFT}, we will investigate more general reductions, in particular what happens if we reduce without first imposing constraints.

Imposing the reduction ansatz that $\pd_z = \pd_{\bar{z}} =0$ in the complex coordinates of eq.~\eqref{eq:ComplexCoords}, we have that the solution to the constraints $B= P(a^{-1} \dr a) + \bar{P}(b^{-1} \dr b) $ becomes
\begin{equation}\label{eq:Bincomplex}
\begin{aligned}
B' = B'_{a \dot{a}} \dr x^{a \dot{a}}&=\frac{1}{\alpha - \beta } \left(\alpha
b^{-1}\partial_{w} b - \beta a^{-1} \partial_{w} a \right) \, \dr w - \frac{1}{\alpha - \beta } \left(\beta b^{-1} \partial_{\bar{w}} b - \alpha a^{-1} \partial_{\bar{w}} a \right) \, \dr \bar{w} \\
& \qquad + \frac{1}{\alpha - \beta } \left( b^{-1}\pd_{\bar{w}} b - a^{-1}\pd_{\bar{w}}a \right) \dr z + \frac{\alpha \beta }{\alpha -\beta } \left( b^{-1}\pd_{w} b - a^{-1}\pd_{w} a \right) \dr \bar{z} \ .
\end{aligned}
\end{equation}
For simplicity, let us first consider the K\"ahler point and align the reduction to the complex structure (implemented simply by taking $\alpha \rightarrow 0$ and $\beta \rightarrow \infty$). In this scenario, the reduction ansatz enforces that $B'_z = 0$ and $B'_{\bar{z}} = 0$ with the remaining components of $B'$ parameterising a generic two-dimensional gauge field. Effectively, we can simply ignore the constraints altogether but impose $B'_z =0 $ and $B'_{\bar{z}} =0 $ as part of the specification of a reduction ansatz. This could be interpreted as demanding $D_z = D_{\bar{z}} = 0 $ acting on fields. In this case it is immediate that the 4d gauged WZW reduces to a 2d gauged WZW.

Away from the K\"ahler point and aligned reduction, i.e. not fixing $\alpha$ and $\beta$, one must keep account of contributions coming from $B'_z$ and $B'_{\bar{z}}$. We can still view $B'_w $ and $B'_{\bar{w}}$ components of eq.~\eqref{eq:Bincomplex} as a parametrisation of a generic 2d gauge field, but there is no way in which we can view the $B'_z$ and $B'_{\bar{z}}$ as a local combination of the $B'_w$ and $B'_{\bar{w}}$. We forced to work with the variables $a$ and $b$ rather than a 2d gauge field. Fortunately, however, the reduction can still be performed immediately if we use the composite fields $\tilde g=a g b^{-1}$ and $\tilde h=ab^{-1}$.
These composite variables are invariant under the $\fh$-gauge symmetry, but a new would-be-affine symmetry emerges under $a \rightarrow \ell a$, $b \rightarrow b r^{-1}$ with $\alpha^b \partial_{b \dot{b}} r = \beta^{b}\partial_{b \dot{b}} \ell = 0$. These leave $B', g, h$ invariant but act as $\tilde{g} \rightarrow \ell\tilde{g} r $ and $\tilde{h} \rightarrow \ell \tilde{h} r $. At the K\"ahler point and aligned reduction, these symmetries descend to affine symmetries, but in general descend only to global transformations. Recall that the 4d gWZW becomes
\begin{equation}
S_{\text{gWZW}_4}^{(\alpha,\beta)}[g,B'] = S_{\text{WZW}_4}^{(\alpha,\beta)}[\tilde{g}] - S_{\text{WZW}_4}^{(\alpha,\beta)}[\tilde{h}] \, .
\end{equation}
It is then immediate that this reduces to the difference of PCM+WZ theories of eq.~\eqref{eq:PCM+WZ} with WZ coefficient $\mathscr{k}$:
\begin{equation}
S_{ \textrm{IFT}_{2}}[ \tilde{g} ,\tilde{h}] = S_{\text{PCM}+ \mathscr k \text{WZ}_2}[\tilde{g}] - S_{\text{PCM}+ \mathscr k \text{WZ}_2}[\tilde{h} ]\,.
\end{equation}
Away from the CFT point, $\mathscr{k} = 1$, this cannot be recast in terms of a deformation of the gauged WZW expressed as a local function of $B',g$.

\paragraph{Lax formulation.}
To obtain the Lax of the resultant IFT$_{2}$ we first note that the four-dimensional gauge fields, upon solving the constraints on $B'$, are gauge equivalent to
$$
A'_{a\dot{a}} = -\frac{1}{\ab} \beta_a \alpha^b \pd_{b\dot{a}} \tilde{g} \tilde{g} ^{-1} \, , \quad B'_{a\dot{a}} = -\frac{1}{\ab} \beta_a \alpha^b \pd_{b\dot{a}} \tilde{h} \tilde{h} ^{-1}\,.
$$
Thus, we may simply follow the construction of the Lax from the ungauged model of eq.~\eqref{eq:LaxA}, with the connection $A'$ producing a Lax for the $S_{\text{PCM}+ \mathscr k \text{WZ}_2}[\tilde{g}]$ and the $B'$ producing one for $S_{\text{PCM}+ \mathscr k \text{WZ}_2}[\tilde{h}]$.

\section{More General \texorpdfstring{IFT$_{2}$}{IFT2} from \texorpdfstring{IFT$_{4}$}{IFT4}: Reducing then Constraining}
\label{sec:newIFT}

In the previous section, we reduced from the gauged WZW$_{4}$ model to an IFT$_2$, but prior to reduction we enforced the constraints imposed by the Lagrange multiplier fields.
These constraints determine implicit relations between the components of the gauge field as per eq.~\eqref{eq:Bincomplex}.
In the simplest case, where we work at the K\"ahler point and align the reduction directions with the complex structure, the constraints enforce $B'_z = B'_{\bar{z}}= 0$.
However, if we do not impose the constraints in 4d, the standard reduction ansatz would only require that $B'_z$ and $B'_{\bar{z}}$ are functionally independent of $z$ and $\bar{z}$, a weaker condition. 

\def\PPu{\Phi}
\def\PPb{\bar\Phi}
In this section, we shall explore the consequences of reducing without first constraining.
Denoting the reduction with $\rightsquigarrow$ we anticipate that the lower-dimensional description will include additional fields as\footnote{Note, we are dropping the prime on the 2d gauge field $B$.}
\begin{equation}
\label{ec:2dfieldvar}
\begin{aligned}
&B'_w(w,\bar w,z,\bar z) \rightsquigarrow B_w(w,\bar w)\,,\quad B'_{\bar w}(w,\bar w,z,\bar z) \rightsquigarrow B_{\bar w}(w,\bar w)\,,\\
&B'_z(w,\bar w,z,\bar z) \rightsquigarrow \PPb(w,\bar w) \,,\quad B'_{\bar z}(w,\bar w,z,\bar z) \rightsquigarrow \PPu (w,\bar w)\,,
\end{aligned}\end{equation}
where $\PPu$ and $\PPb$ will be adjoint scalars in the lower-dimensional theory (sometimes called Higgs fields in the literature).
These will enter explicitly in the lower-dimensional theory through the reduction of covariant derivatives
\begin{equation}
\nabla_{z}g {g}^{-1} \rightsquigarrow \PPb - g \PPb {g}^{-1} \,, \qquad \nabla_{\bar{z}}g {g}^{-1} \rightsquigarrow \PPu - g \PPu {g}^{-1}\,.
\end{equation}
On-shell the 4d gauge field $B'$ is ASD and couples to matter in the gWZW$_{4}$ model.
It is well-known that the reduction of an ASDYM connection leads to the Hitchin system, and we shall see this feature in the lower-dimensional dynamics below.

The two-dimensional Lagrangian that arises from reducing eq.~\eqref{4d action II} without first constraining is
\unskip\footnote{2d Lagrangians are defined as $S_{\text{IFT}_2} = 2 \mathrm{i} \int_{\mathbb{R}^2} \dr w \wedge \dr \bar{w} \, L_{\text{IFT}_2}$.
We denote
\begin{equation*}L_{\text{gWZ}} = L_{\text{WZ}}(g) + \mathrm{Tr}\big(( g^{-1}\partial_w g + \partial_w g g^{-1})B_{\bar w} - (g^{-1}\partial_{\bar w} g + \partial_{\bar w} g g^{-1})B_{w} + B_w\mathrm{Ad}_g B_{\bar w} - B_w \mathrm{Ad}_g^{-1} B_{\bar w}\big),
\end{equation*}
where
$
\int \dr w \wedge \dr \bar{w} \, L_{\text{WZ}}(g) = \int_{\mathbb{R}^2\times[0,1]} \mathcal{L}_{\text{WZ}}(\hat g) = \frac{1}{3}\int_{\mathbb{R}^2\times[0,1]}\mathrm{Tr}\big(\hat g^{1}\dr \hat g \wedge \hat g^{1}\dr \hat g \wedge \hat g^{1}\dr \hat g\big)
$.}
\def\aa{\mathsf{a}}
\def\bb{\mathsf{b}}
\begin{equation}
\begin{split}\label{eq:2daction}
L_{\text{IFT}_2}
&= \frac{1}{2}\mathrm{Tr}\big( g^{-1}D_w g g^{-1}D_{\bar w}g\big)+\frac{1}{2}\frac{\alpha + \beta }{\alpha - \beta } L_{\text{gWZ}}
+\mathrm{Tr}\big( \PPu\PPb + \frac{\alpha }{\alpha - \beta } \PPu\mathrm{Ad}_g\PPb -\frac{\beta }{\alpha - \beta } \PPu\mathrm{Ad}_g^{-1}\PPb\big) \\
& \qquad + \frac{1}{\alpha - \beta }\mathrm{Tr}\big(\PPu\,(g^{-1}D_{\bar w}g + D_{\bar w}gg^{-1}) + \alpha \beta \, \PPb\,(g^{-1}D_{w}g +D_{w}gg^{-1})\big) \\
& \qquad + \mathrm{Tr}\big(\tilde u ( F_{\bar w w} - \beta ^{-1} D_{\bar w}\PPu - \beta D_{w}\PPb -[\PPb,\PPu])\big) + \mathrm{Tr}\big( u ( F_{\bar w w} -\alpha ^{-1}  D_{\bar w}\PPu -\alpha  D_{w}\PPb  -[\PPb,\PPu])\big) \,,
\end{split}
\end{equation}
where we denote the 2d covariant derivative as $ D = \dr + \operatorname{ad}_B$ and note that we have rescaled $\tilde u \to \frac{\tilde u}{\langle \beta\gamma\rangle\langle\beta\hat\gamma\rangle}$ and $u \to \frac{u}{\langle \alpha\gamma\rangle\langle\alpha\hat\gamma\rangle}$.
The fields of the IFT$_2$ are $g\in G$ and $B_{w,\bar{w}},\PPu,\PPb,u,\tilde{u} \in \mathfrak{h}$.
\def\kk{\mathscr k}
In addition to the overall coupling, the IFT$_2$ eq.~\eqref{eq:2daction} only depends on a single parameter.
This can be seen by introducing
\unskip\footnote{Here, we have implicitly assumed that $\alpha\beta \geq 0$, which implies that $|\kk| \geq 1$.
The other regime of interest, $\alpha \beta \leq 0$ and $|\kk|\leq1$ is related by an analytic continuation $\kk' \to - \mathrm{i} \kk'$.}
\begin{equation}
\kk = \frac{\alpha +\beta }{\alpha - \beta }\,, \qquad
\kk' = - \frac{2\sqrt{\alpha \beta}}{\alpha - \beta}\,, \qquad
\kk^2 - \kk'{}^2 = 1\,,
\end{equation}
rescaling $\PPu \to \sqrt{\alpha \beta }\PPu$ and $\PPb \to \frac{1}{\sqrt{\alpha \beta }} \PPb$, and defining
$X^- = \kk'{}^{-1}(u + \tilde u)$ and $\tilde X^+ = \kk'{}^{-1}(u - \tilde u)$.
The Lagrangian eq.~\eqref{eq:2daction} can be rewritten as
\def\Op{\mathcal{O}}
\begin{equation}\label{eq:actk}
\begin{split}
L_{\text{IFT}_2}
&= \frac{1}{2}\mathrm{Tr}\big( g^{-1}D_w gg^{-1}D_{\bar w}g\big)+\frac{\kk}{2} L_{\text{gWZ}} + \mathrm{Tr}\big(\PPu \Op \PPb + \PPu \,V_{\bar w} + \PPb \, V_{w}\big)
\\ & \qquad
+ \mathrm{Tr}\big( X^- ( \kk' ( F_{\bar w w} - [\PPb,\PPu] ) + \kk (D_w\PPb + D_{\bar w} \PPu))\big)
+ \mathrm{Tr}\big( \tilde X^+ (D_{w}\PPb - D_{\bar w}\PPu)\big)\,,
\end{split}
\end{equation}
where
\begin{equation}\begin{gathered}
\Op = 1 - \frac{\kk+1}{2} \mathrm{Ad}_g + \frac{\kk-1}{2} \mathrm{Ad}_g^{-1} \,,
\qquad
V_{w,\bar{w}} = -\frac{\kk'}{2} (g^{-1}D_{w,\bar{w}} g + D_{w,\bar{w}} gg^{-1}) \,.
\end{gathered}\end{equation}
Note that the CFT points $\kk = 1$ or $\kk = - 1$ correspond to taking $\gamma_{a} \to \hat{\alpha}_{a}$ or $\gamma_{a} \to \alpha_{a}$, i.e.\ when the zeroes of the twist function coincide with the poles.

By construction, as the reduction of gWZW$_{4}$, the equations of motion of this theory are equivalent to the zero curvature of Lax connections, whose components are given by the $\dr w$ and $\dr\bar{w}$ legs of the 4d gauge fields.
Explicitly, these Lax connections are given by
\begin{align}
&\begin{aligned}\label{eq:laxka}
\mathcal{L}^{(A)}_w & = \partial_w + B_w -\frac{\kk+1}{2} K_w - \frac{1}{\zeta} \big(\PPu + \frac{\kk'}{2}K_w\big) \,,
\\
\mathcal{L}^{(A)}_{\bar w} & = \partial_{\bar w} + B_{\bar w} + \frac{\kk-1}{2} K_{\bar w} + \zeta \big(\PPb + \frac{\kk'}{2}K_{\bar w}\big) \,,
\end{aligned}
\\
&\begin{aligned}\label{eq:laxkb}
\mathcal{L}^{(B)}_w & = \partial_w + B_w - \frac{1}{\zeta} \PPu \,,
\qquad \qquad
\mathcal{L}^{(B)}_{\bar w} = \partial_{\bar w} + B_{\bar w} + \zeta \PPb \,,
\end{aligned}\end{align}
where we have also redefined the spectral parameter $\zeta \to \sqrt{\alpha\beta}\zeta$ compared to section \ref{sec:wzw4wzw2}
and we have introduced the currents
\begin{equation}
K_w = D_w g g^{-1} + \frac{\kk-1}{\kk'}(1 - \mathrm{Ad}_g )\PPu \,,
\qquad
K_{\bar w} = D_{\bar w} g g^{-1} - \frac{\kk+1}{\kk'}(1 - \mathrm{Ad}_g) \PPb \,.
\end{equation}

\subsection{Lax formulation}
Before analysing the Lagrangian eq.~\eqref{eq:actk} in more detail, let us show explicitly that its equations of motion are indeed equivalent to the zero-curvature condition for the Lax connections eq.~\eqref{eq:laxka} and eq.~\eqref{eq:laxkb}.
The equations of motion that follow from the Lagrangian eq.~\eqref{eq:actk} varying $\tilde{X}^+$, $X^-$ and $g$ are
\begin{equation}\begin{aligned}\label{eq:xpxmg}
& \delta \tilde X^+: & & \mathcal{E}_+ \equiv D_{w}\PPb - D_{\bar w}\PPu = 0 \,,
\\
& \delta X^-: & & \mathcal{E}_- \equiv\kk' \big(F_{\bar w w} - [\PPb,\PPu]\big) + \kk \big(D_w\PPb + D_{\bar w} \PPu) = 0\,,
\\
&\delta gg^{-1}: & & \mathcal{E}_g \equiv \frac{\kk-1}{2}\Big(D_w K_{\bar w} + \frac{\kk+1}{\kk'}[\PPb,K_w]\Big)
- \frac{\kk+1}{2}\Big(D_{\bar w} K_{w} - \frac{\kk-1}{\kk'}[\PPu,K_{\bar w}]\Big)
\\ &&& \hspace{200pt} + \frac{\kk }{\kk'}\mathcal{E}_- - \frac{1}{\kk'} \big(D_w\PPb + D_{\bar w} \PPu\big) = 0 \,.
\end{aligned}
\end{equation}
We also have the Bianchi identity following from the zero-curvature of the Maurer-Cartan form $\dr g g^{-1}$
\begin{equation}\begin{split}\label{eq:bianchi}
&\mathcal{Z} \equiv D_w K_{\bar w} + \frac{\kk+1}{\kk'} [\PPb,K_w] - D_{\bar w} K_w + \frac{\kk-1}{\kk'} [\PPu,K_{\bar w}] + [K_{\bar w},K_w] +
\frac{1}{\kk'}(1-\mathrm{Ad}_g)(\mathcal{E}_- + \mathcal{E}_+) = 0\,.
\end{split}\end{equation}

The zero curvature of the A-Lax eq.~\eqref{eq:laxka} gives rise to three equations that are linear combinations of the equations of motion eq.~\eqref{eq:xpxmg} and the Bianchi identity eq.~\eqref{eq:bianchi}:
\begin{equation}\begin{aligned}
0 & = \frac{\kk-1}{2} \mathcal{Z}' - \mathcal{E}_g + \frac{\kk}{\kk'}\mathcal{E}_- - \frac{1}{\kk'}\mathcal{E}_+ \,,\\
0 & = \kk'^2 \mathcal{Z}' - 2 \kk \mathcal{E}_g + 2\kk'\mathcal{E}_- \,,
\\
0 & = \frac{\kk+1}{2} \mathcal{Z}' - \mathcal{E}_g + \frac{\kk}{\kk'}\mathcal{E}_- + \frac{1}{\kk'}\mathcal{E}_+\,,
\end{aligned}\end{equation}
where we have defined $\mathcal{Z}' \equiv \mathcal{Z} - \frac{1}{\kk'}(1-\mathrm{Ad}_g) (\mathcal{E}_- + \mathcal{E}_+ )$.
On the other hand, the zero curvature of the B-Lax~\eqref{eq:laxkb} defines the Hitchin system:
\begin{equation}\begin{aligned}
\label{eq:HitchinSystem}
0 &= D_{\bar{w }} \PPu \,, \qquad
0 = F_{\bar{w} w} - [\PPb, \PPu]\,, \qquad
0 = D_w \PPb \, ,
\end{aligned}\end{equation}
which can be rewritten as the three equations $\mathcal{E}_\pm = 0$ and $\mathcal{E}_0 \equiv D_{w}\PPb + D_{\bar w}\PPu = 0$.
Therefore, the two Lax connections give rise to five independent equations, which are linear combinations of the equations of motion~\eqref{eq:xpxmg}, the Bianchi identity~\eqref{eq:bianchi}, and the additional equation $\mathcal{E}_0 = 0 $.

To recover this final equation from the equations of motion, let us consider the variational equations for $B_w$, $B_{\bar w}$, $\PPb$ and $\PPu$
\begin{equation}\begin{aligned}
& \delta B_w: & & \mathcal{E}_B \equiv \kk' D_{\bar w} X^- - [\PPb,\tilde X^+ + \kk X^-] + \frac{\kk-1}{2}P_{\mathfrak{h}}K_{\bar w} + \frac{\kk+1}{2}P_{\mathfrak{h}}\mathrm{Ad}_g^{-1} K_{\bar w} - \frac{\kk+1}{\kk'}P_{\mathfrak{h}}(1-\mathrm{Ad}_g^{-1})\PPb = 0 \,,
\\
& \delta B_{\bar w}: & & \mathcal{E}_{\bar B} \equiv \kk' D_{ w} X^- - [\PPu,\tilde X^+ - \kk X^-] + \frac{\kk+1}{2}P_{\mathfrak{h}}K_{w} + \frac{\kk-1}{2}P_{\mathfrak{h}}\mathrm{Ad}_g^{-1} K_{w} - \frac{\kk-1}{\kk'}P_{\mathfrak{h}}(1-\mathrm{Ad}_g^{-1})\PPu = 0 \,,
\\
&\delta \PPu: & & \mathcal{E}_{\PPu} \equiv D_{\bar w} (\tilde X^+ - \kk X^-) + \kk' [\PPb,X^-] - \frac{\kk'}{2} P_{\mathfrak{h}}(1+\mathrm{Ad}_g^{-1})K_{\bar w} + P_{\mathfrak{h}}(1-\mathrm{Ad}_g^{-1})\PPb =0 \,,
\\
&\delta \PPb: & & \mathcal{E}_{\PPb} \equiv D_w (\tilde X^+ + \kk X^-) + \kk' [\PPu,X^-] + \frac{\kk'}{2}P_{\mathfrak{h}}(1+\mathrm{Ad}_g^{-1})K_w - P_{\mathfrak{h}}(1-\mathrm{Ad}_g^{-1})\PPu =0 \,.
\end{aligned}\end{equation}
These can be understood as a first-order system of equations for $\tilde{X}^+$ and $X^-$.
Consistency of the system implies that they should satisfy the integrability conditions $[D_{\bar w},D_w] \tilde{X}^+ = [F_{\bar w w},\tilde{X}^+]$ and $[D_{\bar w},D_w] X^- = [F_{\bar w w},X^-]$.
We find that
\begin{equation}\begin{split}
&\kk' [D_{\bar w},D_w] X^- - \kk' [F_{\bar w w},X^-] = [X^+ ,\mathcal{E}_+]
+[X^-,\mathcal{E}_-] + P_{\mathfrak{h}} (1-\mathrm{Ad}_g^{-1}) \mathcal{E}_g + \kk
P_{\mathfrak{h}} \mathrm{Ad}_g^{-1} \mathcal{Z} \,,
\end{split}\end{equation}
hence, using the Bianchi identity~\eqref{eq:bianchi}, this vanishes on the equations of motion for $\tilde X^+$, $X^-$ and $g$~\eqref{eq:xpxmg}.
On the other hand, we have
\begin{equation}
\kk' [D_{\bar w},D_w] \tilde X^+ - \kk' [F_{\bar w w},\tilde X^+] =
[X^+,\mathcal{E}_-] + [X^-,\mathcal{E}_+]
+ \frac{2\kk}{\kk'}\mathcal{E}_-
- \frac{2}{\kk'}\mathcal{E}_0
-P_{\mathfrak{h}} (1+\mathrm{Ad}_g^{-1})\mathcal{E}_g + \kk P_{\mathfrak{h}}\mathrm{Ad}_g^{-1}\mathcal{Z} \, .
\end{equation}
Here we see that in addition to the Bianchi identity~\eqref{eq:bianchi} and equations of motion~\eqref{eq:xpxmg}, we also require $\mathcal{E}_0 = 0$, recovering the final equation from the Lax system.

\subsection{Relation to known models}

As we will shortly see, if we take $H$ to be abelian, the Lagrangian~\eqref{eq:actk} can be related to known models, including the homogeneous sine-Gordon models and the PCM plus WZ term.
However, for non-abelian $H$~\eqref{eq:actk} has not been considered before, and defines a new integrable field theory in two dimensions.
Moreover, by integrating out $\PPu$, $\PPb$ and the gauge field $B_{w,\bar{w}}$, it leads to an integrable sigma model for the fields $g$, $\tilde X^+$ and $X^-$.
We leave the study of these models for future work.

To recover a sigma model from the Lagrangian \eqref{eq:actk} for abelian $H$, in addition to integrating out $B_w$ and $B_{\bar w}$, we have two options.
The first is to integrate out $\PPu$ and $\PPb$.
The second is to solve the constraint imposed by the Lagrange multiplier $\tilde X^+$.
For abelian $H$ the Lagrangian~\eqref{eq:actk} simplifies to
\begin{equation}\label{eq:actkab}
\begin{split}
L_{\text{IFT}_2}^{\text{ab}}
&= \frac{1}{2}\mathrm{Tr}\big( g^{-1}D_w g g^{-1}D_{\bar w}g\big)+\frac{\kk}{2}L_{\text{gWZ}} + \mathrm{Tr}\big(\PPu\Op \PPb + \PPu\,V_{\bar w} + \PPb \, V_{w}\big)
\\ & \qquad + \mathrm{Tr}\big(( X^- ( \kk' F_{\bar w w} + \kk (\pd_w\PPb + \pd_{\bar w} \PPu ))\big)
+ \mathrm{Tr}\big( \tilde X^+ ( \partial_{w}\PPb - \partial_{\bar w}\PPu)\big) \,.
\end{split}
\end{equation}
This takes the form of the first-order action in the Buscher procedure, and it follows that the two sigma models will be T-dual to each other with dual fields $X^+$ and $\tilde X^+$.
Explicitly the Lagrangians, before integrating out $B_w$ and $B_{\bar w}$, are
\begin{equation}\label{eq:actkintphi}
\begin{split}
L_{\text{IFT}_2}^{\tilde{X}}
&= \frac{1}{2}\mathrm{Tr}\big( g^{-1}D_w g g^{-1}D_{\bar w}g\big)+\frac{\kk}{2} L_{\text{gWZ}}
+ \kk' \mathrm{Tr}\big( X^- F_{\bar w w} \big)
\\ & \qquad + \mathrm{Tr}\big( (\pd_w \tilde X^+ - V_w + \kk \pd_w X^- ) \Op^{-1} (\pd_{\bar w} \tilde X^+ + V_{\bar w} - \kk \pd_{\bar w} X^-) \big) \,,
\end{split}
\end{equation}
and
\begin{equation}\label{eq:actk2}
\begin{split}
L_{\text{IFT}_2}^{X}
&= \frac{1}{2}\mathrm{Tr}\big( g^{-1}D_w gg^{-1}D_{\bar w}g\big) +\frac{\kk}{2} L_{\text{gWZ}} + \kk' \mathrm{Tr}\big( X^- F_{\bar w w} \big)
\\ & \qquad + \frac14\mathrm{Tr}\big(\partial_w X^+\Op \partial_{\bar w} X^+ + 2\partial_w X^+ ( V_{\bar w} - \kk \partial_{\bar w} X^-) + 2\partial_{\bar w} X^+ ( V_{w} - \kk \partial_w X^-)\big)\,,
\end{split}
\end{equation}
where in the second we have locally solved the constraint imposed by the Lagrange multiplier $\tilde{X}^+$ by setting
\begin{equation}\label{eq:locsol}
\PPu = \frac12 \partial_w X^+ \,, \qquad
\PPb = \frac12\partial_{\bar w} X^+ \,,\qquad X^+ \in \mathfrak{h} \,.
\end{equation}

As mentioned above, the first approach can also be straightforwardly applied for non-abelian $H$.
Generalising the second approach is more subtle.
The constraint imposed by the Lagrange multiplier $\tilde{X}^+$ in the Lagrangian~\eqref{eq:actk} implies that
\begin{equation}\label{eq:covflat}
D_{w}\PPb - D_{\bar w}\PPu = 0.
\end{equation}
Typically the full solution to this equation would be expressed in terms of path-ordered exponentials of $B_w$ and $B_{\bar w}$.
To avoid non-local expressions, we can restrict $\PPu$ and $\PPb$ to be valued in the centre of $\mathfrak{h}$, denoted $\mathcal{Z}(\mathfrak{h})$.
Note that this is not a restriction if $H$ is abelian.
With this restriction, the Lagrangian~\eqref{eq:actk} again simplifies to~\eqref{eq:actkab}, and the constraint~\eqref{eq:covflat} then becomes $\partial_{ w}\PPb - \partial_{\bar w}\PPu = 0$, which we can again locally solve by~\eqref{eq:locsol} now with $X^+ \in \mathcal{Z}(\mathfrak{h})$, similarly leading to the Lagrangian~\eqref{eq:actk2}.

\medskip

\paragraph{Relation to PCM plus WZ term.}

Taking $H$ to be abelian, we can relate the Lagrangian~\eqref{eq:actkab} to that of the PCM + WZ term for $G \times H$ through a combination of T-dualities and field redefinitions.
We start by parametrising
\begin{equation}
g = e^{\frac12\tau} g e^{\frac12\tau}\,, \qquad \tau \in \mathfrak{h}\,,
\end{equation}and setting $\partial_{\bar w,w}\tau \to 2C_{w,\bar{w}}$.
We also integrate by parts and set $\partial_{w}X^- \to 2\Psi$ and $\partial_{\bar w}X^- \to 2\bar \Psi$.
To maintain equivalence with the Lagrangian that we started with, we add $\mathrm{Tr}\big(\tilde{\tau} (\partial_w C_{\bar w} - \partial_{\bar w} C_w)\big) + \mathrm{Tr}\big(\tilde{X}^- (\partial_w \bar\Psi - \partial_{\bar w} \Psi)\big)$, i.e., the Lagrange multipliers $\tilde \tau$ and $\tilde X^-$ locally impose $C_{w,\bar w} = \frac12\partial_{\bar w,w}\tau$, $\Psi = \frac12\partial_w X^-$ and $\bar\Psi = \frac12\partial_{\bar w}X^-$.
We can then redefine the fields as
\unskip\footnote{To arrive at this field redefinition, we first look for the shifts of $B_{w,\bar w}$, $C_{w,\bar w}$, $\Psi$ and $\bar \Psi$ that decouple $\Phi$ and $\bar \Phi$ from all other fields apart from $\tilde{X}^+$.
Since both $C_{w}$ and $C_{\bar w}$ transform in the same way, as do $\Psi$ and $\bar \Psi$, we can then easily compute the transformation of $\tilde\tau$, $\tilde X^-$ and $\tilde X^+$ by demanding that the triplet of terms $\mathrm{Tr}\big( \tilde{\tau}F_{w\bar{w}}(C) + \tilde{X}^-F_{w\bar{w}}(\Psi) + \tilde{X}^+F_{w\bar{w}}(\Phi)\big)$ is invariant up to a simple rescaling, i.e., it becomes $\mathrm{Tr}\big( \tilde{\tau}F_{w\bar{w}}(C) + \kk' \tilde{X}^-F_{w\bar{w}}(\Psi) + \frac{1}{\kk'}\tilde{X}^+F_{w\bar{w}}(\Phi)\big)$.}
\begin{equation}\begin{aligned}\label{tildetrans}
B_w & \to B_w - \frac{\kk}{\kk'}\PPu \,, \qquad &
C_w & \to C_w - \frac{1}{\kk'}\PPu \,, \qquad &
\Psi & \to \Psi + \frac{\kk}{\kk'^2} \PPu \,,
\\ B_{\bar w} & \to B_{\bar w} + \frac{\kk}{\kk'}\PPb \,, \qquad &
C_{\bar w} & \to C_{\bar w} - \frac{1}{\kk'}\PPb\,, \qquad &
\bar\Psi & \to \bar\Psi + \frac{\kk}{\kk'^2} \PPb \,,
\\
\tilde{X}^+ & \to \frac{1}{\kk'}\tilde{X}^+ - \frac{\kk}{\kk'} \tilde{X}^- + \frac{1}{\kk'}\tilde{\tau}\,, \qquad &
\tilde{X}^- & \to \kk' \tilde{X}^-\,, \qquad &
\tilde{\tau} & \to \tilde{\tau}\,.
\end{aligned}\end{equation}
Doing so, we arrive at the following Lagrangian
\begin{equation}\begin{split}
L_{\text{IFT}_2}^{\text{ab}} & = \frac{1}{2}\mathrm{Tr}\big( g^{-1}\pd_w g g^{-1}\pd_{\bar w} g \big)
+ \frac{\kk}{2} L_{\text{WZ}}(g)
\\ & \qquad + \frac{1-\kk}{2}\mathrm{Tr}\big(g^{-1}\pd_w g (C_{\bar w} - B_{\bar w})
+ \pd_{\bar w} g g^{-1} ( C_{w} + B_{w})
+ (C_{w} + B_{w}) \mathrm{Ad}_g(C_{\bar w} - B_{\bar w}) \big)
\\ & \qquad + \frac{1+\kk}{2}\mathrm{Tr}\big(g^{-1}\pd_{\bar w} g( C_{w} - B_{w})
+ \pd_{w} g g^{-1}(C_{\bar w} + B_{\bar w})
+ (C_{w} - B_{w}) \mathrm{Ad}_g^{-1} (C_{\bar w} + B_{\bar w}) \big)
\\
& \qquad + \mathrm{Tr}\big( B_w B_{\bar w}
+ C_w C_{\bar w}
+ \kk C_w B_{\bar w}
- \kk B_w C_{\bar w} \big)
\\ & \qquad + \mathrm{Tr}\big( \tilde{\tau} (\partial_w C_{\bar w} - \partial_{\bar w} C_w)\big)
+ \kk' \mathrm{Tr}\big( \tilde{X}^- (\partial_w \bar \Psi - \partial_{\bar w} \Psi)\big)
+ 2\kk' \mathrm{Tr}\big( \Psi B_{\bar w} - B_w,\bar\Psi\big)
\\ & \qquad + \frac{1}{\kk'} \mathrm{Tr}\big( \tilde{X}^+ (\partial_w \bar \Phi - \partial_{\bar w} \Phi)\big)
- \frac{2}{\kk'^2}\mathrm{Tr}\big(\Phi\bar\Phi\big) \,.
\end{split}\end{equation}
The final steps are to integrate out $\tilde \tau$, $\Psi$ and $\bar \Psi$, and $\Phi$ and $\bar \Phi$, leading us to set
\begin{equation}
C_{w,\bar w} = \frac12\partial_{w,\bar{w}}\tau, \qquad B_{w, \bar w} = - \frac12\partial_{w,\bar{w}}\tilde X^-, \qquad \Phi = - \frac{\kk'}{2}\pd_w\tilde{X}^+, \quad \bar\Phi = \frac{\kk'}{2}\pd_{\bar w}\tilde{X}^+.
\end{equation}
Redefining $g \to e^{-\frac12(\tau + \tilde{X}^-)} g e^{-\frac12(\tau - \tilde{X}^-)}$, we find the difference of the PCM plus WZ term Lagrangians for $G$ and $H$
\begin{equation}\begin{split}\label{eq:pcmpluswz}
L_{\text{PCM} + \kk \text{WZ}_2} & = \frac{1}{2}\mathrm{Tr}\big(g^{-1}\pd_w g g^{-1}\pd_{\bar w} g \big)
+ \frac{\kk}{2} L_{\text{WZ}}(g)
- \frac12 \mathrm{Tr}\big(\partial_w \tilde{X}^+ \partial_{\bar w} \tilde{X}^+ \big),
\end{split}\end{equation}
where we recall that for abelian $H$ the WZ term vanishes.

To summarise, starting from the sigma model~\eqref{eq:actk2} we T-dualise in $\tau$, $X^+$ and $X^-$, we then perform a GL$(3)$ transformation on the dual coordinates, and finally T-dualise back in $\tau$ to recover~\eqref{eq:pcmpluswz}, the difference of the PCM plus WZ term Lagrangians for $G$ and $H$.
This relation may have been anticipated since this is the model we would expect to find starting from the ghCS$_6$ action~\eqref{eq:new6daction} and instead imposing the boundary conditions $\mathcal{A}|_{\alpha,\beta} = \mathcal{B}|_{\alpha,\beta} = 0$.

\medskip

\paragraph{\texorpdfstring{$\kk \to 1$}{k → 1} limit.}
As we have seen, the $\kk \to 1$ limit is special since if we first constrain and then reduce we recover the gauged WZW coset CFT.
By first reducing and then constraining, we can recover massive integrable perturbations of these theories.
We consider the setup where $\PPu$ and $\PPb$ are restricted to lie in $\mathcal{Z}(\mathfrak{h})$ and solve the constraint imposed by the Lagrange multiplier $\tilde{X}^+$ by~\eqref{eq:locsol}.
Taking $\kk\to 1$ the Lagrangian~\eqref{eq:actk2} simplifies further to
\begin{equation}\label{eq:ppwave}
\begin{split}
L_{\text{IFT}_2}
&= \frac{1}{2}\mathrm{Tr}\big( g^{-1}D_w gg^{-1}D_{\bar w}g\big)+\frac{1}{2} L_{\text{gWZ}}
\\ & \qquad + \frac14\mathrm{Tr}\big(\partial_w X^+(1-\mathrm{Ad}_g) \partial_{\bar w} X^+ - 2\partial_w X^+ \partial_{\bar w} X^- - 2\partial_{\bar w} X^+ \partial_w X^-\big)\,,
\end{split}
\end{equation}
This is reminiscent of a sigma model for a pp-wave background, with the kinetic terms for the transverse fields described by the gauged WZW model for the coset $G/H$, except that the would-be light-cone coordinates $X^+$ and $X^-$ have $\dim \mathcal{Z}(\mathfrak{h})$ components.
Nevertheless, we still have the key property that the equation of motion for $X^-$ is $\partial_w\partial_{\bar w}X^+ = 0$, whose general solution is $X^+ = Y(w) + \bar Y(\bar w)$. Substituting into the Lagrangian~\eqref{eq:ppwave} we find
\begin{equation}\label{eq:gsg}
\begin{split}
L_{\text{IFT}_2}
&= \frac{1}{2}\mathrm{Tr}\big( g^{-1}D_w g g^{-1}D_{\bar w}g\big)+\frac{1}{2} L_{\text{gWZ}} + \frac14 \mathrm{Tr}\big( Y'\bar Y' - Y'\mathrm{Ad}_g\bar Y'\big) \,.
\end{split}
\end{equation}
In the special case that $Y = w \Lambda$ and $\bar Y = \bar{w}\bar\Lambda$ this is the gauged WZW model for the coset $G/H$ deformed by a massive potential $V = \mathrm{Tr}(\Lambda \mathrm{Ad}_g \bar \Lambda) - \mathrm{Tr}(\Lambda\bar\Lambda)$ as studied in \cite{Park:1994bx}.
Taking the limit $\mathscr{k}\rightarrow1$ directly at the level of the Lax connection given by eq.~\eqref{eq:laxka}, keeping track of the definitions of the currents $K_w,K_{\bar{w}}$ which depend on $\kk$, we find
\begin{equation}
\mathcal{L}_w \rightarrow \partial_w + B_w - D_w g g^{-1} + \frac{1}{2\zeta} \Lambda \, ,\qquad
\mathcal{L}_{\bar{w}} = \partial_{\bar{w}} + B_{\bar w} - \frac{\zeta}{2} \mathrm{Ad}_g \bar\Lambda \,,
\end{equation}
recovering the Lax given in \cite{Park:1994bx,Fernandez-Pousa:1996aoa}.

When $G$ is compact and $H = U(1)^{\mathrm{rk}_G}$, $\Lambda$ and $\bar\Lambda$ can be chosen such that these models have a positive-definite kinetic term and a mass gap.
These are known as the homogeneous sine-Gordon models \cite{Fernandez-Pousa:1996aoa}.
For $G=SU(2)$ and $H=U(1)$ the homogeneous sine-Gordon model becomes the complex sine-Gordon model after integrating out the gauge fields $B_w$ and $B_{\bar w}$. Note that if $\mathcal{Z}(\mathfrak{h})$ is one-dimensional and $Y(w)$ and $\bar Y(\bar w)$ are both non-constant then we can always use the classical conformal symmetry to reach $Y = w \Lambda$ and $Y' = \bar{w}\Lambda$, hence recovering a constant potential.
For higher-dimensional $\mathcal{Z}(\mathfrak{h})$, this is not the case.

\subsection{Example: \texorpdfstring{$SL(2)/U(1)_V$}{SL(2)/U(1)V}}

To illustrate the features of this construction, let us consider the example of $SL(2)/U(1)_V$ for which the 2d gauged WZW describes the trumpet CFT. To be explicit we use $\mathfrak{sl}(2)$ generators
\begin{equation}
T_1= \left(
\begin{array}{cc}
1 & 0 \\
0 & -1 \\
\end{array}
\right) \, , \quad T_2 = \left(
\begin{array}{cc}
0 & 1 \\
1 & 0 \\
\end{array}
\right) \, , \quad T_3 = \left(
\begin{array}{cc}
0 & 1 \\
-1 & 0 \\
\end{array}
\right)\, ,
\end{equation}
and parametrise the group element as
\begin{equation}
g= \left(
\begin{array}{cc}
\cos (\theta ) \sinh (\rho )+\cosh (\rho ) \cos (\tau ) &
\sin (\theta ) \sinh (\rho )+\cosh (\rho ) \sin (\tau )
\\
\sin (\theta ) \sinh (\rho )-\cosh (\rho ) \sin (\tau ) &
\cosh (\rho ) \cos (\tau )-\cos (\theta ) \sinh (\rho )
\\
\end{array}
\right) \, .
\end{equation}
We choose the $U(1)$ vector action generated by $T_3$ such that
\begin{equation}
\delta g = \epsilon [g, T_3] \quad \Rightarrow \quad \delta \rho=\delta \tau = 0 \, , \quad \delta \theta = \epsilon,
\end{equation}
hence we gauge fix by setting $\theta = 0 $. The analysis here is simplified by the observation that there is no WZ term since there are no 3-forms on the two-dimensional target space.

\paragraph{The CFT point.}
For orientation, we first work at the CFT point corresponding to $\kk=1$. Recall from the discussion in \S\ref{sec:gaugedWZWdiamond}, that first constraining in 4d and then reducing, enforces $\PPb = \PPu = 0 $ and the Lagrange multiplier sector vanishes.
This gives the conventional gauged WZW model described by a target space geometry
\begin{equation}
\dr s^2 = \dr \rho^2 + \coth^2\rho \, \dr \tau^2 \, .
\end{equation}
Let us now consider the IFT$_{2}$ that results from taking the same reduction that would lead to the CFT, but now in our reduction ansatz set $\PPu = \frac{m}{2} T_3 $ and $\PPb = -\frac{m}{2} T_3 $.
The Lagrangian that follows is
\begin{equation}\label{eq:csg}
L_{\text{CsG}} = \pd_w \rho \pd_{\bar{w}} \rho + \coth^2 \rho \, \pd_w \tau \pd_{\bar{w} } \tau - m^2 \sinh^2 \rho \, .
\end{equation}
This theory is well known as the complex sinh-Gordon model, a special case of the integrable massive deformations of $G/H$ gauged WZW models known as the homogeneous sine-Gordon models \cite{Park:1994bx,Fernandez-Pousa:1996aoa}.

\paragraph{Unconstrained reduction: integrating out \texorpdfstring{$\PPu$, $\PPb$ and $B_{w,\bar{w}}$}{Phi, Phi-bar, and B}.}
We now turn to the more general story, away from the CFT point, by considering the reduction without first imposing constraints. Taking the IFT$_{2}$~\eqref{eq:actkab} and integrating out $\PPu$, $\PPb$ and the gauge field $B_{w,\bar{w}}$ while retaining $X^-$ and $\tilde{X}^+$, results in the sigma model with target space metric and B-field
\begin{equation}\begin{aligned}\label{eq:sl2backorig}
\dr s^2 &= \dr \rho^2 + \coth^2\rho \, \dr \tau^2 + \textrm{csch}^2\rho \left( \dr \tilde{X}^+{}^2 - \dr X^-{}^2 \right) \, \\
B_2 &= \mathcal{V} \wedge \dr \tilde{X}^+ \, ,\qquad \mathcal{V} = \kk \textrm{csch}^2\rho \, \dr X^- + \kk' \coth^2\rho \, \dr \tau \,.
\end{aligned}\end{equation}

\paragraph{Unconstrained reduction: the dual.}
On the other hand, if we solve the constraint imposed by the Lagrange multiplier $\tilde{X}^+$ setting $\PPu = \frac12\pd_{w} X^+$ and $\PPb = \frac12\pd_{\bar{w}} X^+$, we find the sigma model with target space geometry
\begin{equation}\begin{aligned}\label{sl2back}
\dr s^2 &= \dr \rho^2 + \coth^2\rho \, \dr \tau^2 - \textrm{csch}^2\rho \, \dr X^-{}^2 + \sinh^2\rho \, (\dr X^++\mathcal{V})^2  \, ,  \\
B_2 &= 0\, .
\end{aligned}
\end{equation}
This can of course be recognised as the T-dual of~\eqref{eq:sl2backorig} along $\tilde{X}^+$.
In the limit $\kk \to 1$~\eqref{sl2back} becomes the pp-wave background
\begin{equation}\begin{aligned}
\dr s^2 &= \dr \rho^2 + \coth^2\rho \, \dr \tau^2 + \sinh^2\rho \, \dr X^+{}^2 +2\dr X^+ \dr X^-  \, ,
\\
B_2 &= 0\, ,
\end{aligned}\end{equation}
and if we light-cone gauge fix, $X^+ = m (w-\bar w)$, in the associated sigma model we recover the complex sinh-Gordon Lagrangian~\eqref{eq:csg} as expected.

\paragraph{Relation to PCM plus WZ term.}

Finally we demonstrate a relation between the models above and the PCM plus WZ term.
Let us start with the metric and B-field for the PCM plus WZ term for $G = \text{GL}(2)$
\begin{equation}\begin{aligned}
\dr s^2 & = \dr \tilde{X}^+{}^2 + \dr \rho^2 - \cosh^2\rho \dr \tau^2 + \sinh^2 \rho \dr \tilde X^-{}^2 \, ,
\\
B & = \kk \cosh^2\rho \, \dr \tau \wedge \dr \tilde X^- \, .
\end{aligned}\end{equation}
Note that $\dr B = \kk \sinh 2\rho \, \dr\rho\wedge\dr \tau \wedge \dr \tilde X^-$, which is proportional to the volume for $SL(2)$.
We first T-dualise $\tau \to \tilde \tau$, and then perform the following field redefinition
\begin{equation}
\tilde X^+ \to \kk' \tilde X^+ + \frac{\kk}{\kk'}\tilde X^- - \tilde \tau \ , \qquad
\tilde X^- \to \frac{1}{\kk'} \tilde X^- \ .
\end{equation}
It is straightforward to check that this is the inverse transformation to~\eqref{tildetrans}.
Finally, T-dualising back, $\tilde X^+ \to X^+$, $\tilde X^- \to X^-$ and $\tilde \tau \to \tau$, we precisely recover the background~\eqref{sl2back}, demonstrating that it can be understood as a generalised TsT transformation of the PCM plus WZ term.

\subsection{The LMP limit}
The PCM plus WZ term admits a limit in which it becomes the 2d analogue of the LMP model, otherwise known as the pseudodual of the PCM \cite{Zakharov:1973pp}, see, e.g.~\cite{Hoare:2018jim}.
It is possible to generalise this limit to the gauged model~\eqref{eq:actk} by setting
$g = \exp(\varepsilon U)$, $\kk = \varepsilon^{-1} \ell$, $\tilde X^+ \to \varepsilon^2 \tilde X^+$, $X^- \to \varepsilon^3 X^- - \varepsilon P_{\mathfrak{h}} U$ rescaling the Lagrangian by $\varepsilon^{-2}$, and taking $\varepsilon \to 0$.
Implementing this limit in~\eqref{eq:actk} we find
\begin{equation}\label{eq:actklmp}
\begin{split}
L_{\text{IFT}_2}^{\text{LMP}}
&= \frac{1}{2}\mathrm{Tr}\big( D_w U D_{\bar w} U + [\PPu,U][\PPb,U]\big) - \frac{\ell}{6} \mathrm{Tr}\big((D_w U + [\PPu,U][U,(D_{\bar w}U - [\PPb,U]]\big)
\\ & \qquad
+ \ell\mathrm{Tr}\big(X^- ( F_{\bar w w} - [\PPb,\PPu] + D_w\PPb + D_{\bar w} \PPu)\big)
+ \mathrm{Tr}\big((\tilde X^+ ( D_{w}\PPb - D_{\bar w}\PPu)\big)
\\ & \qquad + \frac{1}{2\ell} \mathrm{Tr}\big(U ( F_{\bar w w} - [\PPb,\PPu] -D_w\PPb - D_{\bar w} \PPu)\big) \,.
\end{split}
\end{equation}
Similarly we can take the limit in the Lax connections~\eqref{eq:laxka} and~\eqref{eq:laxkb}.
The B-Lax~\eqref{eq:laxkb} is unchanged, while the A-Lax~\eqref{eq:laxka} becomes
\begin{equation}
\begin{aligned}\label{eq:laxkalmp}
\mathcal{L}^{(A)}_w & = \partial_w + B_w -\frac{\ell}{2} K^{\text{LMP}}_w- \frac{1}{\zeta} \big(\PPu + \frac{\ell}{2}K^{\text{LMP}}_w\big) \,,
\\
\mathcal{L}^{(A)}_{\bar w} & = \partial_{\bar w} + B_{\bar w} + \frac{\ell}{2} K^{\text{LMP}}_{\bar w} + \zeta \big(\PPb + \frac{\ell}{2}K^{\text{LMP}}_{\bar w}\big) \,,
\end{aligned}
\end{equation}
where
\begin{equation}
K^{\text{LMP}}_w = D_w U + [\PPu,U] \,,
\qquad
K^{\text{LMP}}_{\bar w} = D_{\bar w} U - [\PPb,U] \,.
\end{equation}
As we will see in \S\ref{Gauged LMP action section} this model can also be found directly from 6d hCS and 4d CS by considering a twist function with a single fourth-order pole.

As in the gauged WZW case, we can again find an integrable sigma model from~\eqref{eq:actklmp} by integrating out $\PPu$, $\PPb$ and the gauge field $B_{w,\bar w}$.
For abelian $H$ we can also construct the dual model by solving the constraint imposed by the Lagrange multiplier $\tilde{X}^+$ and integrating out $B_w$ and $B_{\bar w}$.
For $SL(2)/U(1)_V$ the resulting backgrounds can be found by taking the LMP limit
\begin{equation}\begin{aligned}
& \rho \to \varepsilon \rho - \frac16 \varepsilon^3 \rho\tau^2 \,, \qquad
\tau \to \varepsilon \tau -\frac13 \varepsilon^3 \rho^2 \tau \,, \qquad
(\dr s^2, B_2) \to \varepsilon^{-2} (\dr s^2, B_2) \,, \qquad \kk \to \varepsilon^{-1}\ell \,,
\\ & X^- \to \varepsilon^3 X^- - \varepsilon \tau \,, \qquad
\tilde{X}^+ \to \varepsilon^2 \tilde{X}^+ \,, \qquad
X^+ \to X^+ \,, \qquad \varepsilon \to 0 \,,
\end{aligned}\end{equation}
in eqs.~\eqref{eq:sl2backorig} and~\eqref{sl2back}.
This limit breaks the manifest global symmetry given by shifts of the coordinate $\tau$.
This is in agreement with the fact that the Lagrangian~\eqref{eq:actklmp} is not invariant under $U \to U + H_0$ ($H_0 \in \mathfrak{h}$), while its gauged WZW counterpart~\eqref{eq:actk} is invariant under $g \to h_0 g h_0$ ($h_0 \in H$) for abelian $H$.

Curiously, we can actually take a simplified LMP limit
\begin{equation}\begin{aligned}
& \rho \to \varepsilon \rho \,, \qquad
\tau \to \varepsilon \tau \,, \qquad
(\dr s^2, B_2) \to \varepsilon^{-2} (\dr s^2, B_2) \,, \qquad \kk \to \varepsilon^{-1}\ell \,,
\\ & X^- \to \varepsilon^3 X^- - \varepsilon \tau \,, \qquad
\tilde{X}^+ \to \varepsilon^2 \tilde{X}^+ \,, \qquad
X^+ \to X^+ \,, \qquad \varepsilon \to 0 \,,
\end{aligned}\end{equation}
in the backgrounds~\eqref{eq:sl2backorig} and~\eqref{sl2back} that preserves this global symmetry. Taking this limit in eq.~\eqref{eq:sl2backorig} we find
\begin{equation}\begin{aligned}\label{eq:lmporig}
\dr s^2 & = \dr \rho^2 + \dr \tau^2 + \frac{1}{\rho^{2}}\dr \tilde X^+{}^2 + \frac{2}{\rho^{2}} \dr X^- \dr \tau \,,
\\ B_2 &= \mathcal{V} \wedge \dr\tilde{X}^+ \,, \qquad
\mathcal{V} = \frac{\ell}{\rho^2}\dr X^- + \left(\ell - \frac{1}{2\ell\rho^2} \right)\dr \tau \,,
\end{aligned}\end{equation}
while the limit of eq.~\eqref{sl2back} is
\begin{equation}\begin{aligned}
\dr s^2 & = \dr \rho^2 + \dr \tau^2 + \rho^{2}(\dr \tilde X^+ + \mathcal{V} )^2 + \frac{2}{\rho^{2}} \dr X^- \dr \tau \,,
\\
B_2 &= 0\, .
\end{aligned}\end{equation}
As for the gauged WZW case these two backgrounds above can also be constructed as a generalised TsT transformation of the background for the LMP model on GL$(2)$
\begin{equation}\begin{aligned}
\dr s^2 & = \dr \tilde X^+{}^2 + \dr \rho^2 - \dr \tau^2 +\rho^{2}\dr \tilde X^-{}^2\,,
\\
B_2 &= \ell \rho^2 d\tau \wedge \dr \tilde X^-\, .
\end{aligned}\end{equation}
Explicitly, if we first T-dualise $\tau\to\tilde\tau$, then perform the following field redefinition
\begin{equation}
\tilde{X}^+ \to \ell \tilde{X}^+ + \frac{1}{2\ell^2}\tilde{X}^- - \tilde{\tau} \,, \qquad \tilde{X}^- \to \frac{1}{\ell} \tilde{X}^- \,,\qquad\tilde\tau \to \tilde \tau - \frac{1}{2\ell^2}\tilde X^- \,,
\end{equation}
and finally T-dualise back,
\unskip\footnote{Note that here the order of T-dualities matters. In particular, we cannot first T-dualise $\tilde\tau$ after the coordinate redefinition since it turns out to be a null coordinate.}
$\tilde X^+ \to X^+$, $\tilde X^- \to X^-$ and $\tilde \tau \to \tau$, we recover the background~\eqref{eq:lmporig}.

\section{Reduction to \texorpdfstring{gCS$_4$}{gCS4} and localisation}
\label{sec:6dCSto4dCS}

Having discussed the right hand side of the diamond, we briefly describe the left hand side that follows from first reducing to obtain a gauged 4d Chern-Simons theory on $\mathbb{R}^2\times \mathbb{CP}^1$ and then integrating over $\mathbb{CP}^1$ to localise to a two-dimensional field theory on $\mathbb{R}^2$. We show that the resulting IFT$_2$ matches \eqref{eq:2daction}.

We recall the six-dimensional coupled action
\begin{equation}
\label{ec:recall6daction}
S_{\text{ghCS}_6}[\cA , \cB] = S_{\text{hCS}_6}[\cA] - S_{\text{hCS}_6}[\cB] - \frac{1}{2\pi \mathrm{i} }\int_{\mathbb{PT} } \bar{\partial} \Omega \wedge \mathrm{Tr}\left( \cA \wedge \cB \right) \, .
\end{equation}
We note that the three terms in the action are invariant under the transformations $\cA\mapsto \hat \cA = \cA + \rho^{\cA}_{\dot a} e^{\dot a}+\rho^{\cA}_{0}e^{0}$ and $\cB\mapsto \hat \cB=\cB + \rho^{\cB}_{\dot a} e^{\dot a}+\rho^{\cB}_{0}e^{0}$, given that both $\Omega$ and $\bar\partial \Omega$ are top forms in the holomorphic directions. By choosing $\rho^{\cA}$ and $\rho^{\cB}$ appropriately, we can ensure that neither $\hat \cA$ nor $\hat \cB$ have $\dr z$ or $\dr \bar z$ legs, so
\begin{align}
\hat \cA &= \hat \cA_{w} \dr w + \hat \cA_{\bar w} \dr \bar w+\cA_0 \bar e^{0} \quad \text{with}\quad \hat \cA_{w}=-\frac{[\cA \kappa]}{\langle \pi \gamma\rangle} \,, \,\hat \cA_{\bar w}=-\frac{[\cA \hat\kappa]}{\langle \pi \hat\gamma\rangle} \ , \\
\hat \cB &= \hat \cB_{w} \dr w + \hat \cB_{\bar w} \dr \bar w+\cB_0 \bar e^{0} \quad \text{with}\quad \hat \cB_{w}=-\frac{[\cB \kappa]}{\langle \pi \gamma\rangle} \,, \,\hat \cB_{\bar w}=-\frac{[\cB \hat\kappa]}{\langle \pi \hat\gamma\rangle} \,.
\end{align}
To perform the reduction we follow the procedure outlined in \S \ref{sec:wzw4wzw2}. Namely, we contract the six-dimensional Lagrangian of \eqref{ec:recall6daction} with the vector fields $\partial_z$ and $\partial_{\bar z}$, and restrict to gauge connections which are invariant under the flow of these vector fields. Thus, since the shifted gauge fields $\hat \cA$ and $\hat \cB$ manifestly have no $\dr z$ or $\dr \bar z$ legs, and we are restricting to field configurations satisfying $L_{\partial_z}\hat \cA=L_{\partial_z}\hat \cB=L_{\partial_{\bar z}}\hat \cA=L_{\partial_{\bar z}}\hat \cB=0$, the contraction by $\partial_z$ and $\partial_{\bar z}$ only hits $\Omega$ in the first two terms, and $\bar \partial \Omega$ in the third. In particular, we find
\begin{equation}
(\partial_z \wedge \partial_{\bar z})\vee \Omega = \frac{\langle \alpha \beta\rangle^2}{2}\frac{\langle \pi \gamma\rangle \langle \pi \hat \gamma\rangle}{\langle \pi \alpha\rangle^2\langle \pi \beta\rangle^2}e^0\,, \qquad (\partial_z \wedge \partial_{\bar z})\vee \bar \partial \Omega=- \frac{\langle \alpha \beta\rangle^2}{2}\bar\partial_0\bigg(\frac{\langle \pi \gamma\rangle \langle \pi \hat \gamma\rangle}{\langle \pi \alpha\rangle^2\langle \pi \beta\rangle^2}\bigg)e^0 \wedge \bar e^{0}\,.
\end{equation}
Hence, the six-dimensional action reduces to a four-dimensional coupled Chern-Simons action
\begin{equation}
\label{ec:4dgCS}
S_{\text{gCS}_4}[\hat A,\hat B]=\int_{X}\omega \wedge \mathrm{CS}[\hat A]-\int_{X}\omega \wedge \mathrm{CS}[\hat B]-\frac{1}{2\pi i}\int_{X}\bar \partial \omega \wedge \langle \hat A, \hat B\rangle\,,
\end{equation}
where $X = \mathbb{CP}^1 \times \bR^{2}$,
\begin{equation}
\omega=\frac{\langle \alpha \beta\rangle^2}{2}\frac{\langle \pi \gamma\rangle \langle \pi \hat \gamma\rangle}{\langle \pi \alpha\rangle^2\langle \pi \beta\rangle^2}e^0 \ ,
\end{equation}
and $\hat A$ and $\hat B$ are the restrictions of $\hat \cA$ and $\hat \cB$ to $X$. Similarly, the boundary conditions \eqref{eq:ABboundarycond} descend to analogous boundary conditions on $\hat A$ and $\hat B$. The action \eqref{ec:4dgCS} has been considered before in \cite{Stedman:2021wrw}, albeit not with the choice of $\omega$ hereby discussed.

With the gauged 4d Chern-Simons action at hand, we may now localize. The procedure is entirely analogous to the one depicted in \S \ref{sec:localtogwzw4} so we shall omit some of the details. We begin by reparametrising our four-dimensional gauge fields $\hat A$ and $\hat B$ in terms of a new pair of connections $\hat A', \hat B'$ and smooth functions $\hat g \in C^{\infty}(X,G)$ and $\hat h \in C^{\infty}(X,H)$. We use the redundancy in the reparametrisation to fix $\hat A'_0 = \hat B'_0=0$. The boundary degrees of freedom of the resulting IFT$_2$ will be \emph{a priori} be given by the evaluation of $\hat g$, $\hat h$, $\hat u $ and $\hat v$ at $\alpha$ and $\beta$. However, as in the 6d setting, we have some residual symmetry we can use to fix $\hat g|_{\beta}=\mathrm{id}$, $\hat h|_{\alpha,\beta}=\mathrm{id}$, and similarly, $\hat v|_{\alpha,\beta}=0$. We are thus left with
\begin{equation}
\hat g|_{\alpha}\coloneqq g \,,\quad \hat u|_{\alpha}\coloneqq u\,, \quad \hat u|_{\beta}=\tilde u\,.
\end{equation}
In terms of these variables, the bulk equations of motion of gCS$_4$ theory imply
\begin{equation}
\label{ec:4dCSbulkeom}
\bar{\pd}_{0} \hat A^{\prime}_{i} = 0 \ , \qquad
\bar{\pd}_{0} \hat B^{\prime}_{i} = 0 \ ,
\end{equation}
away from the zeroes of $\omega$, namely $\gamma$ and $\hat \gamma$.
The on-shell gCS$_4$ action can be thus written as
\begin{equation} \label{eq:almostlocalised4d}
\begin{aligned}
& S_{\text{gCS}_4}[\hat A' , \hat B' ] = \frac{1}{2 \pi \mathrm{i}} \int_{X} \bar{\pd} \omega \wedge \mathrm{Tr} \big( \hat A^\prime \wedge \bar{\pd} \hat{g} \hat{g}^{-1} -(\hat g^{-1} \hat A^{\prime} \hat{g} + \hat{g}^{-1} \bar \partial \hat{g})\wedge \hat B^{\prime} \big) \\
& \hspace{9em} - \frac{1}{6 \pi \mathrm{i}} \int_{X\times[0,1]} \bar \partial\omega \wedge \mathrm{Tr} \big(\hat{g}^{-1} \dr \hat{g} \wedge \hat{g}^{-1} \dr \hat{g} \wedge \hat{g}^{-1} \dr \hat{g} \big) \, .
\end{aligned}
\end{equation}
To obtain the IFT$_2$ we begin by looking at the bulk equations of motion \eqref{ec:4dCSbulkeom}. Liouville's theorem shows that the only bounded, holomorphic functions on $\CP$ are constant functions. We are after something a little more general than this, however, as we do not require the components of our gauge field to be bounded at the zeroes of $\omega$.
Indeed, we allow the $w$-component to have a pole at $\pi \sim \gamma$ and the $\bar{w}$-component to have a pole at $\pi \sim \hat{\gamma}$.
With this analytic structure in mind, we can parameterise the solution of the bulk equation for $\hat B^{\prime}$ by
\begin{equation}
\label{ec:expforB}
\hat B^{\prime}_{w} = B_{w} + \frac{\langle \pi \hat{\gamma} \rangle}{\langle \pi \gamma \rangle} \, \Phi \ , \qquad
\hat B^{\prime}_{\bar{w}} = B_{\bar{w}} - \frac{\langle \pi \gamma \rangle}{\langle \pi \hat{\gamma} \rangle} \, \bar \Phi \,,
\end{equation}
where we have conveniently used the field variables introduced in \eqref{ec:2dfieldvar} to ease comparison with \eqref{eq:2daction} after localisation to the IFT$_2$. In particular, under $\pi$-independent gauge transformations $B_{w}, B_{\bar w}$ have the transformation of 2d gauge fields, whilst $\Phi$ and $\bar \Phi$ transform as adjoint scalars.

Note that in the singular piece of these solutions, we have chosen to align the zero of each with the pole of the other. Notice that this choice is also completely general, since moving the zeros in the singular pieces amounts to field redefinitions between $B_w$ and $\Phi$, respectively, $B_{\bar w}$ and $\bar \Phi$.
This is convenient since the flatness condition on $\hat B^{\prime}$ reproduces Hitchin's equations,
\begin{equation}
F_{w \bar{w}}[\hat B^{\prime}] = F_{w \bar{w}}[B] - [ \Phi , \bar \Phi] - \frac{\langle \pi \gamma \rangle}{\langle \pi \hat{\gamma} \rangle} D_{w} \bar \Phi - \frac{\langle \pi \hat{\gamma} \rangle}{\langle \pi \gamma \rangle} D_{w} \Phi \ .
\end{equation}
On the other hand, for the $\hat A^{\prime}$ gauge field a convenient choice of parameterisation when solving the bulk equation of motion \eqref{ec:4dCSbulkeom} is
\begin{equation}
\label{ec:expforA}
\hat A^{\prime}_{i} = \frac{\langle \pi \alpha \rangle}{\langle \pi \gamma \rangle} \frac{\langle \beta \gamma \rangle}{\langle \beta \alpha \rangle} \, U_{i} + \frac{\langle \pi \beta \rangle}{\langle \pi \gamma \rangle} \frac{\langle \alpha \gamma \rangle}{\langle \alpha \beta \rangle} \, V_{i} \,,\quad i=w,\bar w\,.
\end{equation}
This parametrisation, in which we have chosen the coefficients such that one term vanishes at $\pi \sim \alpha$ while the other vanishes at $\pi \sim \beta$, is adapted to the boundary conditions which can be solved for $U_{i}$ and $V_{i}$ to yield
\begin{equation}
\label{ec:Aprimefinal}
\begin{aligned}
\hat A^{\prime}_{w} & = \hat B^{\prime}_{w}- \frac{\langle \pi \beta \rangle}{\langle \pi \gamma \rangle} \frac{\langle \alpha \gamma \rangle}{\langle \alpha \beta \rangle} \bigg( D_{w} g g^{-1} + \frac{\langle \alpha \hat{\gamma} \rangle}{\langle \alpha \gamma \rangle}(1- \mathrm{Ad}_{g} )\Phi \bigg)   \, \\
\hat A^{\prime}_{\bar{w}} & = \hat B^{\prime}_{\bar{w}} - \frac{\langle \pi \beta \rangle}{\langle \pi \hat{\gamma} \rangle} \frac{\langle \alpha \hat{\gamma} \rangle}{\langle \alpha \beta \rangle} \bigg(  D_{\bar{w}} g g^{-1} - \frac{\langle \alpha \gamma \rangle}{\langle \alpha \hat{\gamma} \rangle}(1- \mathrm{Ad}_{g} ) \bar\Phi \bigg)  \ .
\end{aligned}
\end{equation}
Replacing \eqref{ec:expforB} and \eqref{ec:Aprimefinal} in \eqref{eq:almostlocalised4d} and integrating\footnote{To do so, we use the localisation formula in homogeneous coordinates
\begin{equation}
\label{ec:4d2dlocalformula}
\frac{1}{2 \pi \mathrm{i}} \int_{X} \bar{\pd} \omega \wedge Q = - \frac{1}{2}\int_{\mathbb{R}^2}
\bigg[
\frac{
\langle \alpha \gamma\rangle\langle \beta \hat \gamma \rangle + \langle \alpha \hat \gamma\rangle \langle \beta \gamma\rangle
}{
\langle \alpha \beta \rangle
} Q \vert_{\alpha}
+ \langle \alpha \gamma\rangle\langle \alpha \hat \gamma\rangle (\pd_{0} Q) \vert_{\alpha} \bigg]
\quad + \quad \alpha \leftrightarrow \beta \ .
\end{equation}
for any $Q \in \Omega^2(X)$.} along $\mathbb{CP}^1$
we recover the IFT$_2$ given in \eqref{eq:2daction}.

\section{Gauged LMP action} \label{Gauged LMP action section}

In the previous sections, we analysed a ghCS$_6$ setup where the meromorphic $(3,0)$-form $\Omega$ had two double poles, showing that such a theory leads to a gauged WZW$_4$ upon localisation to $\mathbb{R}^{4}$. To highlight some of the universal features of this procedure, we will now focus on another example in which the meromorphic $(3,0)$-form will have a single fourth order pole. Such a configuration of the ungauged hCS$_{6}$ was shown in \cite{Bittleston:2022cmz} to lead to the LMP action for ASDYM \cite{Leznov:1986mx}, \cite{Parkes:1992rz}.

\subsection{LMP action from \texorpdfstring{hCS$_6$}{hCS6}}
Let us start by reviewing the ungauged localisation of hCS$_{6}$ with a fourth order pole. We start with the action and $(3,0)$-form defined by
\begin{equation}\label{hCS Action}
S_{\mathrm{hCS_6}}[\cA] = \frac{1}{2 \pi \mathrm{i}} \int_{\mathbb{PT}} \Omega \wedge \text{CS}(\cA) \; , \qquad
\Omega = k \, \frac{{e}^{0} \wedge {e}^{\ensuremath\dot{a}} \wedge {e}_{\ensuremath\dot{a}}}{{\la \pi \alpha \ra }^{4}} \; .
\end{equation}
As is usual in hCS$_{6}$, we must impose boundary conditions on the gauge field $\cA$ to ensure the vanishing of the boundary variation
\begin{equation}
\delta S_{\mathrm{hCS_6}} \big\vert_{\text{bdry}} = \frac{1}{2 \pi \mathrm{i}} \int_{\mathbb{PT}} \bar{\partial} \Omega \wedge \text{tr}(\delta \mc{A} \wedge \mc{A} ) \; .
\end{equation}
Evaluating the above integral is achieved by making use of the localisation formula (see appendix~\ref{appendix on localisation formulae})
\begin{equation} \label{4th order pole locln}
\frac{1}{2 \pi \mathrm{i}} \int_{\mathbb{PT}} \bar{\pd} \Omega \wedge Q = \frac{k}{6}
\int_{\bR^{4}} \alpha_{a} \alpha_{b} \Sigma^{ab}\wedge \pd_{0}^{3} Q \big\vert_{\alpha} \ .
\end{equation}
Then, one finds that the boundary variation vanishes if we impose the boundary conditions
\begin{equation}
{\mc{A}}{|}_{\pi = \alpha} = 0 \; \; \; \; \text{and} \; \; \; \; {\partial}_{0} {\mc{A}} {|}_{\pi = \alpha} = 0 \; .
\end{equation}

\paragraph{Admissible gauge transformations.}
We now check which residual gauge symmetries survive with the preceding choice of boundary conditions.
We proceed in a familiar fashion, introducing a new parameterisation of our gauge field $\mc{A}$ as
\begin{equation}\label{parameterisation of cA}
\mc{A} = \hat{g}^{-1} \mc{A}' \hat{g} + \hat{g}^{-1} \bar{\partial} \hat{g} \; , \qquad
{\mc{A}}'_{0}=0 \ .
\end{equation}
This parameterisation has both external and internal gauge symmetries which act as
\begin{equation}
\begin{aligned}
\text{External} \quad \hat{\gamma} :& \quad \mc{A} \mapsto \mc{A}^{\hat{\gamma}}  \, , \quad \mc{A}^\prime \mapsto \mc{A}^\prime \, , \quad    \hat{g} \mapsto \hat{g} \hat{\gamma} \ , \\
\text{Internal} \quad \check{\gamma} :& \quad \mc{A} \mapsto \mc{A}\, , \quad \mc{A}^\prime  \mapsto \mc{A}^\prime{}^{\check{\gamma}} \, , \quad    \hat{g} \mapsto \check{\gamma}^{-1} \hat{g} \ .
\end{aligned}
\end{equation}
The internal gauge transformations must satisfy $\bar{\partial}_{0}\ensuremath\check{\gamma} = 0$ to preserve the condition ${\mc{A}}'_{0}=0$. These transformations leave the value of $\mc{A}$ invariant and as such they are fully compatible with the boundary conditions. We will use the internal gauge symmetry to fix $\hat{g}{|}_{\pi = \alpha} = \mathrm{id}$. The story for the external gauge symmetries is slightly different, under external gauge transformations $\mc{A} \mapsto \mc{A}^{\hat{\gamma}}$ and so the value of $\mc{A}$ at the poles is not generically invariant. As such we must proceed with caution: we require our boundary conditions to be invariant under external gauge transformations, imposing constraints on the admissible symmetries at $\pi = \alpha$. This limits the amount of symmetry available for gauge fixing. The gauge transformation of the first boundary condition reads
\begin{equation}
{\mc{A}}^{\hat{\gamma}}{|}_{\pi = \alpha } = \left( \hat{\gamma}^{-1} \mc{A} \hat{\gamma} + \hat{\gamma}^{-1} \bar{\partial} \hat{\gamma} \right){|}_{\pi = \alpha } = 0 \quad \Longrightarrow \quad
{\gamma}^{-1} \alpha^{a} \partial_{a \ensuremath\dot{a}} \gamma = 0 \; ,
\end{equation}
where we have defined
$$\hat{\gamma}{|}_{\pi = \alpha} = \gamma \; .$$
Here, we have derived the fact that at $\pi = \alpha$ we restrict our gauge transformations such that they are holomorphic on $\mathbb{R}^{4}$ with respect to the complex structure given by the point $\pi = \alpha$. Another way of stating this is that our admissible external gauge symmetries on $\mathbb{PT}$ localise to semi-local symmetries in the effective theory on $\mathbb{R}^{4}$. However this restriction is derived from only one half of the boundary conditions. Introducing the notation
$$\hat{\mathbf{\Gamma}}:=\hat{\gamma}^{-1} \partial_{0} \hat{\gamma} \; ,$$
the gauge transformation of the second boundary condition reads
\begin{equation}
\begin{aligned}
\partial_{0} \mc{A}^{\hat{\gamma}} {|}_{\pi = \alpha} =& \; \partial_{0}\left( \hat{\gamma}^{-1} \mc{A} \hat{\gamma} + \hat{\gamma}^{-1} \bar{\partial} \hat{\gamma} \right){|}_{\pi = \alpha }= 0 \; , \\
=& \; \left( \left[ \hat{\gamma}^{-1} \mc{A} \, \hat{\gamma}, \, \hat{\mathbf{\Gamma}} \right] + \hat{\gamma}^{-1} \partial_{0} {\mc{A}} \hat{\gamma} + \bar{\partial} \hat{\mathbf{\Gamma}} + \left[ \hat{\gamma}^{-1} \bar{\partial} \hat{\gamma} , \hat{\mathbf{\Gamma}} \right] +\hat{\gamma}^{-1} \partial_{\dot{a}} \hat{\gamma} \; \bar{e}^{\dot{a}} \right){|}_{\pi = \alpha} \ . \\
\end{aligned}
\end{equation}
Imposing the original boundary conditions we arrive at the constraint equation
\begin{equation}
{\alpha}^{a} {\partial}_{a \dot{a}} {\mathbf{\Gamma}} + {\gamma}^{-1}\hat{\alpha}^{a} \partial_{a \dot{a}} \gamma = 0 \; ,
\end{equation}
where we have used $\langle \alpha \hat{\alpha} \rangle = 1$ and defined
$$\hat{\mathbf{\Gamma}}{|}_{\pi= \alpha} = \mathbf{\Gamma} \; .$$
One solution is that the external gauge transformations are global symmetries of the localised effective theory $\dr_{\mathbb{R}^{4}} \gamma = 0$, and $\mathbf{\Gamma}$ is holomorphic on $\mathbb{R}^{4}$ with respect to choice of complex structure given by the point $\alpha \in \mathbb{CP}^{1}$.

Tentatively, our localised theory should have 4 degrees of freedom, known as ‘edge modes',
\begin{equation}
\underline{\mathbf{u}}:= ( g , \mathbf{u}^{1} , \mathbf{u}^{2}, \mathbf{u}^{3}) \; .
\end{equation}
where
\begin{equation}
g = \hat{g} {|}_{\pi = \alpha} \; , \; \; \; \mathbf{u}^{1} : = {\hat{g}}^{-1} \pd_{0}\hat{g} {|}_{\pi = \alpha} \; , \; \; \; \mathbf{u}^{2} : = \hat{g}^{-1} \pd^{2}_{0} \hat{g} {|}_{\pi = \alpha}\; , \; \; \; \mathbf{u}^{3} : = \hat{g}^{-1} \pd^{3}_{0} \hat{g} {|}_{\pi = \alpha}\; .
\end{equation}
However, some of these fields are spurious and can be gauged fixed away using the admissible gauge symmetries. We have already used the internal gauge symmetry to fix $g= \mathrm{id}.$. Furthermore, the second and third $\pd_{0}$-derivatives of the external gauge transformations are unconstrained by the boundary conditions, so they can be used to gauge fix $\mathbf{u}^{2} = \mathbf{u}^{3} = 0$. This leaves us with one dynamical degree of freedom in the localised theory on $\mathbb{R}^{4}$, namely $\mathbf{u}^{1}: \mathbb{R}^{4} \rightarrow \tf{g}$ which we will now denote by $\mathbf{u}$ for brevity. In conclusion, after gauge fixing we have
\begin{equation}\label{Gauge fixing u}
\underline{\mathbf{u}}= ( \mathrm{id} , \mathbf{u} , 0,0 ) \; .
\end{equation}

\paragraph{Solving the boundary conditions.}
Using the boundary conditions, we will solve for $\mc{A} '$ in the parametrisation \eqref{parameterisation of cA} in terms of the edge modes.
The first boundary condition tells us
\begin{equation}
\mc{A}'{|}_{\pi=\alpha} = 0 \quad \Rightarrow \quad
\alpha^{a}{A}_{a \ensuremath\dot{a}} = 0 \quad \Rightarrow \quad
A_{a \ensuremath\dot{a}} = \alpha_{a} C_{\dot{a}} \ .
\end{equation}
Then, the second boundary condition equation is written as
\begin{equation}
\partial_{0} \mc{A}' {|}_{\pi=\alpha} + \bar{\partial} \mathbf{u} = 0 \; ,
\end{equation}
which allows us to conclude that
\begin{equation}
{C}_{\ensuremath\dot{\alpha}} = \alpha^{a} \partial_{\alpha \ensuremath\dot{\alpha}} \mathbf{u} \; .
\end{equation}
We now have all the ingredients to localise the hCS$_{6}$ action to $\mathbb{R}^{4}$.

\paragraph{Localisation to \texorpdfstring{$\mathbb{R}^{4}$}{R4}.}
We can write the action \eqref{hCS Action} in the new variables as
\begin{equation}
S = \frac{1}{2 \pi \mathrm{i}} \int_{\mathbb{PT}} \bar{\partial} \Omega \wedge \mathrm{Tr} ( \mc{A}' \wedge \bar{\partial} \hat{g} \hat{g}^{-1} )
- \frac{1}{6 \pi \mathrm{i}} \int_{\mathbb{PT} \times [0,1]} \bar{\pd} \Omega \wedge \mathrm{Tr} \left( (\hat{g}^{-1} \dr \hat{g})^{3} \right) \; ,
\end{equation}
where in the second term we have extended $\mathbb{PT}$ to the 7-manifold $\mathbb{PT} \times [0,1]$, whose boundary is a disjoint union of two copies of $\mathbb{PT}$. We have also extended our fields via a smooth homotopy $\hat{g} \rightarrow \hat{g}(t)$ so that $\hat{g}(0) = \mathrm{id}$ and $\hat{g}(1) = \hat{g}$. Applying the localisation formula \eqref{4th order pole locln} and the choice of gauge fixing \eqref{Gauge fixing u}, we arrive at the spacetime action
\begin{equation}\label{LMP Action}
S_{\text{LMP}}[\mathbf{u}] = \frac{k}{3} \int_{\mathbb{R}^{4}} \frac{1}{2} \mathrm{Tr} ( \dr \mathbf{u} \; \wedge \star \dr \mathbf{u} ) + \frac{1}{3} \alpha_{a} \alpha_{b} \Sigma^{ab} \wedge \mathrm{Tr}( \, \mathbf{u} \left[ \dr \mathbf{u} , \dr \mathbf{u} \right] \, ) \; .
\end{equation}
We identify the action \eqref{LMP Action} as the LMP model for ASDYM. Upon performing reduction to $\mathbb{R}^{2}$ the above action becomes the pseudo-dual of the PCM \cite{Liniado:2023uoo}.

\subsection{Gauged LMP action from \texorpdfstring{ghCS$_{6}$}{ghCS6}}
In the previous subsection, we derived the LMP action from hCS$_6$. Next, we shall consider the same fourth order pole structure for gauged hCS$_{6}$. The starting point is to calculate the boundary variation and make a choice of isotropic subspace such that it vanishes.

\paragraph{Boundary conditions.}
Starting from the action
\begin{equation} \label{4th pole ghCS6}
S_{\mathrm{ghCS_6}}[\cA , \cB] = S_{\mathrm{hCS_6}}[\cA] - S_{\mathrm{hCS_6}}[\cB]
- \frac{1}{2\pi \mathrm{i}} \int_{\mathbb{PT}} \bar{\partial} \Omega \wedge \mathrm{Tr} \big( \cA \wedge \cB \big) \, ,
\end{equation}
the boundary variation is given by
\begin{equation}
\delta S_{\mathrm{ghCS_6}} \big\vert_{\text{bdry}} = \frac{1}{2 \pi \mathrm{i}} \int_{\mathbb{PT} } \bar{\partial} \Omega \wedge \mathrm{Tr} \big( \delta \cA \wedge (\cA - \cB) - \delta \cB \wedge (\cB - \cA) \big) \; .
\end{equation}
Following in a parallel fashion to the hCS$_{6}$ case, one finds a suitable choice of boundary conditions is given by
\begin{equation}
\cA {|}_{\pi = \alpha} = \cB{|}_{\pi = \alpha} \; , \quad \partial_{0}\cA {|}_{\pi = \alpha} = \partial_{0}\cB{|}_{\pi = \alpha} \; , \quad \partial^{2}_{0}\cA^{\tf{h}} {|}_{\pi = \alpha} = \partial^{2}_{0}\cB{|}_{\pi = \alpha}
\; , \quad \partial^{3}_{0}\cA^{\tf{h}} {|}_{\pi = \alpha} = \partial^{3}_{0}\cB{|}_{\pi = \alpha} \;.
\end{equation}

\paragraph{Gauge fixing.}
Gauge fixing will once again prove dividends in completing the localisation calculation, as such, we will consider the set of admissible gauge transformations respecting our boundary conditions. Performing a gauge transformation on the first boundary condition, one arrives at
\begin{equation}
\left( \hat{\gamma}^{-1}\cA \hat{\gamma} + \hat{\gamma}^{-1} \bar{\pd} \hat{\gamma} \right){|}_{\pi = \alpha} = \left( \hat{\eta}^{-1}\cB \hat{\eta} + \hat{\eta}^{-1} \bar{\pd} \hat{\eta} \right){|}_{\pi = \alpha} \ ,
\end{equation}
from which one concludes that the admissible gauge transformations must obey
$\hat{\gamma}{|}_{\alpha} = \hat{\eta}{|}_{\alpha} $. Running through systematically, the second boundary condition requires
\begin{equation}
\begin{aligned}
\left( \left[ \hat{\gamma}^{-1} \cA \, \hat{\gamma} + \hat{\gamma}^{-1} \bar{\pd} \hat{\gamma}, \, \hat{\mathbf{\Gamma}} \right] + \right. & \left. \hat{\gamma}^{-1} \pd_{0} {\mc{A}} \hat{\gamma} + \bar{\partial} \hat{\mathbf{\Gamma}} +\hat{\gamma}^{-1} \partial_{\dot{a}} \hat{\gamma} \; \bar{e}^{\dot{a}} \right){|}_{\pi = \alpha} \\
&= \left( \left[ \hat{\eta}^{-1} \cB \, \hat{\eta} + \hat{\eta}^{-1} \bar{\pd} \hat{\eta}, \, \hat{\mathbf{N}} \right] + \hat{\eta}^{-1} \pd_{0} {\mc{B}} \hat{\eta} + \bar{\partial} \hat{\mathbf{N}} +\hat{\eta}^{-1} \partial_{\dot{a}} \hat{\eta} \; \bar{e}^{\dot{a}} \right){|}_{\pi = \alpha} \; ,
\end{aligned}
\end{equation}
where we have denoted $\hat{\mathbf{\Gamma}} = \hat{\gamma}^{-1} \pd_{0} \hat{\gamma}$ and $\hat{\mathbf{N}} = \hat{\eta}^{-1} \pd_{0} \hat{\eta}$. Making use of the original boundary condition and the constraint $\hat{\gamma}{|}_{\alpha} = \hat{\eta}{|}_{\alpha} $, we conclude that admissible gauge transformations must also obey $\hat{\mathbf{\Gamma}}{|}_{\pi = \alpha} = \hat{\mathbf{N}}{|}_{\pi = \alpha}$.
In a similar fashion, from the third boundary condition
we conclude that $\hat{\mathbf{\Gamma}}^{(2)}_{\tf{h}}{|}_{\alpha} = \hat{\mathbf{N}}^{(2)}{|}_{\alpha}$ where $\hat{\mathbf{\Gamma}}^{(2)} := \hat{\gamma}^{-1} \pd_{0}^{2} \hat{\gamma}$ and $\hat{\mathbf{N}}^{(2)} := \hat{\eta}^{-1} \pd_{0}^{2} \hat{\eta}$.
Finally, from the fourth boundary condition
we find $\hat{\mathbf{\Gamma}}^{(3)}_{\tf{h}}{|}_{\alpha} = \hat{\mathbf{N}}^{(3)}{|}_{\alpha}$ where $\hat{\mathbf{\Gamma}}^{(3)}:= \hat{\gamma}^{-1} \pd_{0}^{3} \hat{\gamma}$ and $\hat{\mathbf{N}}^{(3)} := \hat{\eta}^{-1} \pd_{0}^{3} \hat{\eta}$.

Armed with the admissible gauge symmetries of our theory, we set ourselves the task of gauge fixing our degrees of freedom. Naively, we would think there are 8 degrees of freedom in our theory,
\begin{equation}
\begin{aligned}
\underline{\mathbf{u}} &:= \left( g , \mathbf{u} , \mathbf{u}^{2}, \mathbf{u}^{3} \right) \; , \\
\underline{\mathbf{v}} &:= \left( h , \mathbf{v}^{1} , \mathbf{v}^{2}, \mathbf{v}^{3} \right) \ . \\
\end{aligned}
\end{equation}
First, one considers the internal gauge symmetries of $\cA$ and $\cB$ which one can use to set both $g$ and $h$ to the identity. Next, one should note that the $H$-valued external gauge transformations of $\cB$ parameterised by $\hat{\eta}$ are unconstrained at the point $\pi = \alpha$. As such, one can gauge fix $\mathbf{v}^{i} = 0$ for $i = 1,2,3$. Now, as our external gauge transformations of $\cA$ parameterised by $\hat{\gamma}$ are constrained to coincide with $\hat{\eta}$ at $\pi = \alpha$, and we have used these symmetries in our choice of gauge fixing, we find that we are unable to gauge fix $\mathbf{u}^{i}$. As such, each of these degrees of freedoms will appear as fields in our effective theory on $\mathbb{R}^{4}$. In summary, after gauge fixing one has,
\begin{equation}\label{Gauge fixed u and v}
\begin{aligned}
\underline{\mathbf{u}} &= \left( \mathrm{id} , \mathbf{u} , \mathbf{u}^{2}, \mathbf{u}^{3} \right) \; , \\
\underline{\mathbf{v}} &= \left( \mathrm{id} , 0 , 0, 0 \right) \ . \\
\end{aligned}
\end{equation}

\paragraph{Solving the boundary conditions.}
The first boundary condition reads
\begin{equation}
\left( \hat{g}^{-1} \cA ' \hat{g} + \hat{g}^{-1} \bar{\pd} \hat{g} \right){|}_{\pi = \alpha} = ( \hat{h}^{-1} \cB ' \hat{h} + \hat{h}^{-1} \bar{\pd} \hat{h} ){|}_{\pi = \alpha} \ .
\end{equation}
Given our choice of gauge fixing \eqref{Gauge fixed u and v} and the explicit solutions e.g.\ $\cA^{\prime}_{\dot{a}} = \pi^{a} A_{a \dot{a}}$, this implies
\begin{equation}\label{1st BC}
\cA ' {|}_{\pi = \alpha}= \cB'{|}_{\pi = \alpha} \quad \Rightarrow \quad
\alpha^{a}{A}_{a \ensuremath\dot{a}} = \alpha^{a} {B}_{a \ensuremath\dot{a}} \quad \Rightarrow \quad
{A}_{a \ensuremath\dot{a}} = {B}_{a \ensuremath\dot{a}} -{\alpha}_{a}{Q}_{\ensuremath\dot{a}} \;.
\end{equation}
We can then use the second boundary condition to solve for ${Q}_{\ensuremath\dot{a}}$,
\begin{equation}
\partial_{0} \cA {|}_{\pi = \alpha} = \partial_{0} \cB {|}_{\pi = \alpha} \; , \qquad \implies \qquad
{Q}_{\ensuremath\dot{a}} = -{\alpha}^{a} \left( \left[ {B}_{a \ensuremath\dot{a}} , \mathbf{u} \right] + \pd_{a \dot{a}} \mathbf{u} \right) = - \alpha^{a} \nabla_{a \ensuremath\dot{a}} \mathbf{u} \; .
\end{equation}
These two boundary conditions are sufficient to solve for $A_{a \dot{a}}$ in terms of the other degrees of freedom,
\begin{equation}
{A}_{a \ensuremath\dot{a}} = {B}_{a \ensuremath\dot{a}} + {\alpha}_{a} \alpha^{b} \nabla_{b \ensuremath\dot{a}} \mathbf{u} \; .
\end{equation}

\paragraph{Localisation to \texorpdfstring{$\mathbb{R}^{4}$}{R4}.}
Writing the action \eqref{4th pole ghCS6} in terms of the new field variables,
one can see the only terms that will contribute to the effective action given our choice of gauge \eqref{Gauge fixed u and v} will be
\begin{equation}
\begin{aligned}
S_{\mathrm{ghCS_6}} = \frac{1}{2 \pi i} \int_{\mathbb{PT}} \bar{\pd} \Omega \wedge \mathrm{Tr}( \cA' \wedge \bar{\pd} \hat{g} \hat{g}^{-1} - \hat{g}^{-1} \cA' \hat{g} \wedge \cB' - {\hat{g}^{-1} \bar{\pd} \hat{g} \wedge \cB'} ) - \frac{1}{6\pi i}\int_{\mathbb{PT} \times [0,1]} \bar{\pd}\Omega \wedge \mathrm{Tr}\left((\hat{g}^{-1} \dr \hat{g} )^{3} \right) \; .
\end{aligned}
\end{equation}
The localisation calculation of the gauged model is slightly more involved than the ungauged case due to the additional degrees of freedom appearing. However, as seen in the calculations in previous sections we expect $\mathbf{u}^{2}$ and $\mathbf{u}^{3}$ to appear only as Lagrange multipliers, in particular imposing self-duality type constraints for our gauge field $B$. With this tenet in mind, one can show that the 4d theory is given by
\begin{equation}\label{Gauged LMP action}
\begin{aligned}
S_{\text{gLMP}}[\mathbf{u},B] = k \int_{\mathbb{R}^{4}} \vol_{4} \frac{1}{2} \mathrm{Tr} \, (\nabla^{a \ensuremath\dot{a}} \mathbf{u} \nabla_{a \ensuremath\dot{a}} \mathbf{u}&) + \frac{1}{3} \epsilon^{\ensuremath\dot{a} \ensuremath\dot{b}} \, \mathrm{Tr} \, ( \mathbf{u}  \left[ \alpha^{a} \nabla_{a \ensuremath\dot{a}} \mathbf{u} , \alpha^{b} \nabla_{b \ensuremath\dot{b}} \mathbf{u}  \right] ) +  \mathbf{u} \; \epsilon^{\ensuremath\dot{a} \ensuremath\dot{b}} \hat{\alpha}^{a} \hat{\alpha}^{b} {F}_{a \ensuremath\dot{a} b \ensuremath\dot{b}} (B) \\
& + \frac{1}{2}\mathbf{u}^{2} \epsilon^{\ensuremath\dot{a} \ensuremath\dot{b}} \left( {\alpha}^{a} \hat{\alpha}^{b} + \hat{\alpha}^{a} \alpha^{b} \right) {F}_{a \ensuremath\dot{a} b \ensuremath\dot{b}}(B) + \tilde{\mathbf{u}}^{3} \epsilon^{\ensuremath\dot{a} \ensuremath\dot{b}} \alpha^{a} \alpha^{b} {F}_{a \ensuremath\dot{a} b \ensuremath\dot{b}}(B) \; ,
\end{aligned}
\end{equation}
where we have performed a field redefinition $\mathbf{u}^{3} \rightarrow \tilde{\mathbf{u}}^{3}:= \frac{1}{6}( \mathbf{u}^{3} + 2\left[ \mathbf{u} , \mathbf{u}^{2} \right])$. Upon reducing along a particular $\mathbb{R}^{2}$ subgroup, and appropriately performing redefinitions of our fields and parameters, one finds that the gauged LMP action matches the two-dimensional action eq.~\eqref{eq:actklmp}.

\paragraph{Implementing the Lagrange multipliers.}
In section \ref{Yang's Matrix Section} we reviewed how solutions to the ASDYM can be formulated in terms of a Yang's matrix after a partial gauge fixing of the ASD connection. In this section we will look to integrate out our Lagrange multiplier fields present in the action eq.~\eqref{Gauged LMP action} by solving the self duality constraints they impose in a similar vein.
Indeed, one may understand the LMP equations of motion as the remaining ASDYM equation after these two constraints have been solved. This is analogous to the statement that the WZW$_{4}$ equation of motion is the remaining ASDYM equation for Yang's matrix.

The equation of motion found by varying $\tilde{\mathbf{u}}^{3}$ is an integrability condition along the 2-plane defined by $\alpha^{a}$, and it maybe be solved by
\begin{equation}
\epsilon^{\ensuremath\dot{a} \ensuremath\dot{b}} \alpha^{a} \alpha^{b} {F}_{a \ensuremath\dot{a} b \ensuremath\dot{b}}(B) = 0 \qquad \implies \qquad
\alpha^{a}{B}_{a \ensuremath\dot{a}} = {h}^{-1} \alpha^{a} \partial_{a \ensuremath\dot{a}}{h} \; ,
\end{equation}
where $h \in C^{\infty}(\mathbb{R}^{4}) \otimes H$. It is helpful to parameterise the remaining degrees of freedom in $B_{a \dot{a}}$ in terms of a new field $C_{\dot{a}}$, defined by the relation
\begin{equation}
B_{a \dot{a}} = h^{-1} \pd_{a \dot{a}} h - \alpha_{a} \, h^{-1} C_{\dot{a}} h \ .
\end{equation}
Then, the $\mathbf{u}^{2}$ equation of motion becomes
\begin{equation}
\epsilon^{\ensuremath\dot{a} \ensuremath\dot{b}} \left( {\alpha}^{a} \hat{\alpha}^{b} + \hat{\alpha}^{a} \alpha^{b} \right) {F}_{a \ensuremath\dot{a} b \ensuremath\dot{b}}(B) = 0 \qquad \Longleftrightarrow \qquad
\epsilon^{\ensuremath\dot{a} \ensuremath\dot{b}} \alpha^{a} \pd_{a \dot{a}} C_{\dot{b}} = 0 \ .
\end{equation}
This may be solved explicitly by ${C}_{\ensuremath\dot{a}} = \alpha^{a} \partial_{a \ensuremath\dot{a}} f$ for $f \in C^{\infty}(\mathbb{R}^{4}) \otimes \tf{h}$, such that the gauge field $B$ is given by
\begin{equation}\label{on-shell gauge field B}
{B}_{a\ensuremath\dot{a}} = {h}^{-1} \partial_{a \ensuremath\dot{a}} h + {h}^{-1} X_{a \ensuremath\dot{a}} h \; , \; \text{where} \; \; {X}_{a \ensuremath\dot{a}} = - {\alpha}_{a} {\alpha}^{b} \partial_{{b} \ensuremath\dot{a}} f \; .
\end{equation}
Reinserting this expression into the action eq.~\eqref{Gauged LMP action}, the resulting theory may be written as a difference of two LMP actions. This can be done by performing a field redefinition $h \mathbf{u} {h}^{-1} = v - f$, for $v \in C^{\infty}(\mathbb{R}^{4}) \otimes \tf{g}$, such that one arrives at the action
\begin{equation}
\begin{aligned}
S_{\text{gLMP}}[\mathbf{u},B] = \mathscr{k} \int_{\mathbb{R}^{4}} \frac{1}{2} \mathrm{Tr} ( \dr v \wedge \star \dr v ) + & \frac{1}{3} \alpha_{a} \alpha_{b} \Sigma^{ab} \wedge \mathrm{Tr}(v \left[ \dr v , \dr v \right] ) \\
& - \mathscr{k} \int_{\mathbb{R}^{4}} \frac{1}{2} \mathrm{Tr} ( \dr f \wedge \star \dr f ) + \frac{1}{3} \alpha_{a} \alpha_{b} \Sigma^{ab} \wedge \mathrm{Tr}(f \left[ \dr f , \dr f \right] ) \; .
\end{aligned}
\end{equation}
This demonstrates the conclusion
\begin{equation}
S_{\text{gLMP}}[\mathbf{u},B] = S_{\text{LMP}}[v] - S_{\text{LMP}}[f] \; .
\end{equation}

\section{Outlook}\label{sec:outlookgauged}

The construction presented in this work has led us to new integrable field theories in both four and two dimensions.
We conclude by highlighting a number of interesting future directions prompted by these results.

Motivated by the observation that the gauged WZW model on the coset $G/H$ in two dimensions can be written as the difference of WZW models for the groups $G$ and $H$, we took the difference of two hCS$_6$ theories as our starting point.
The boundary conditions~\eqref{eq:ABboundarycond} led us to add a boundary term resulting in the action~\eqref{eq:new6daction}.
It is worth highlighting that the boundary variation vanishes on the boundary conditions~\eqref{eq:ABboundarycond} whether or not the boundary term is included, and the contribution of the boundary term to the IFT$_4$ vanishes if we invoke all the boundary conditions.
However, while the algebraic boundary conditions, $\mathcal{A}^{\mathfrak{k}}|_{\alpha,\beta}=0$ and $\mathcal{A}^{\mathfrak{h}}|_{\alpha,\beta}=\mathcal{B}_{\alpha,\beta}$ can be straightforwardly solved, this is not the case for the differential one $\partial_0\mathcal{A}^{\mathfrak{h}}|_{\alpha,\beta}=\partial_0\mathcal{B}_{\alpha,\beta}$.
Therefore, we relaxed this condition meaning that the contribution of the boundary term no longer vanishes.
Remarkably, for the specific boundary term added in~\eqref{eq:new6daction}, the constraints implied by the differential boundary condition now follow as on-shell equations of motion, leading to fully consistent IFT$_4$ and IFT$_2$.

There are compelling reasons to follow this strategy, including that the symplectic potential becomes tautological upon including the boundary term.
However, it remains to understand why the differential boundary condition can be consistently dropped for this particular choice of boundary term, and a systematic interpretation of this is an open question.
To address this, it would be appropriate to pursue a more formal study, complementing a homotopic analysis (along the lines done for CS$_4$ in~\cite{Benini:2020skc}) with a symplectic/Hamiltonian study of the 6d holomorphic Chern-Simons theory (similar to \cite{Vicedo:2019dej}  in the context of CS$_4$).  

A second arena for formal development is the connection between 6d holomorphic Chern-Simons and five-dimensional K\"ahler Chern-Simons (KCS$_{5}$) theory \cite{Nair:1990aa,Nair:1991ab}.  This should mirror the relationship between CS$_{4}$ and CS$_{3}$ theories described by Yamazaki \cite{Yamazaki:2019prm}.  To make this suggestion precise in the present context one may consider a Kaluza-Klein expansion around the $U(1)$ rotation in the $\mathbb{CP}^1$ that leaves fixed the location of the double poles, retaining the transverse coordinate as part of the bulk five-manifold of KCS$_{5}$. The details of this are left for future study.

It would also be interesting to explore the new integrable IFT$_4$ and IFT$_2$ that we have constructed.
$G/H$ coset CFTs in two dimensions have a rich spectrum of paraferminonic operators \cite{Bardakci:1990lbc,BARDAKCI1991439}. It would be very interesting to establish the lift or analogue of these objects in the context of the IFT$_4$.
The natural framework for this is likely to involve the study of co-dimension one defects and associated higher-form symmetries.

For abelian $H$ we find IFT$_2$ that, in the $\kk \to 1$ limit, are related to massive integrable perturbations of the $G/H$ gauged WZW models known as homogeneous sine-Gordon models~\cite{Park:1994bx,Fernandez-Pousa:1996aoa}.
These include the sine-Gordon and complex sine-Gordon models as special cases, two of the most well-understood IFT$_2$.
There is nothing in our construction that prohibits non-abelian $H$ and it would be interesting to study the resulting models in more detail.
The homogeneous sine-Gordon models before gauging are closely related to the non-abelian Toda equations~\cite{Leznov:1982ew,Hollowood:1994vx}, for which an alternative derivation from CS$_4$ involving both order and disorder defects was presented in~\cite{Fukushima:2021ako}.
It would be instructive to understand the relationship between the two approaches.
 
An important class of IFT$_2$ are the symmetric space sigma models.  These can be constructed either by restricting fields to parameterise $G/H$ directly or by gauging a left action of H on the PCM.
These theories have been realised in CS$_{4}$ through branch cut defects \cite{Costello:2019tri} and recently in hCS$_{6}$ \cite{Cole:2024noh}.
One might explore the realisation the gauging construction of such models within the current framework, and generalise to $\mathbb{Z}_4$ graded semi-symmetric spaces (relevant for applications of CS$_4$ to string worldsheet theories \cite{Costello:2020lpi,PhysRevD.109.106015}).      

When $G/H$ is a symmetric space, an alternative class of massive integrable perturbations of the $G/H$ gauged WZW model are known as the symmetric space sine-Gordon models \cite{Bakas:1995bm,Fernandez-Pousa:1996aoa}.
In the landscape of IFT$_2$ these are related to the $\lambda\to0$ limit~\cite{Hollowood:2014rla,Hoare:2015gda} of the $\lambda$-deformation of the symmetric space sigma model~\cite{Sfetsos:2013wia}.
Note that $\kk \to 1$ and $\lambda \to 0$ both correspond to conformal limits and it would be instructive to explore the relation between the two constructions.
More generally, it would be interesting to generalise the construction in this work to deformed models, in particular splitting one or both double poles in the meromorphic (3,0)-form $\Omega$ into simple poles, or dual models, for example considering the alternative boundary conditions~\eqref{natd}.

Finally, recently novel approaches to constructing IFT$_3$ using higher Chern-Simons theory in 5d has been explored in \cite{Schenkel:2024dcd,Chen:2024axr}. 
Given that there is an overlap between the models that can be obtained from these constructions and from hCS$_6$, or more precisely its reduction to five dimensions, CS$_5$ on the mini-twistor correspondence space $\mathbb{PN}$ ~\cite{Bittleston:2020hfv}, it would be exciting to understand the link between the two, and investigate the existence of categorical generalisations of hCS$_6$.

\section{Appendix}

\subsection{Appendix A: Spinor and differential form conventions }
\label{sec:spinorconventions}

We work on $\mathbb{R}^4$ and define coordinates in bispinor notation as
\begin{equation}
x^{a \dot{a}} = \frac{1}{\sqrt{2}} \begin{pmatrix}
x_0 +\mathrm{i} x_1 & x_2 +\mathrm{i} x_3 \\ -x_2 + \mathrm{i} x_3 & x_0 - \mathrm{i} x_1
\end{pmatrix} \ .
\end{equation}
We fix orientation such that $\star 1 = \vol_4= \dr x_0 \wedge \dr x_1 \wedge \dr x_2 \wedge\dr x_3$. For 1-forms $\sigma = \sigma_{a \dot{a}} \dr x^{a \dot{a}} $ and $\tau = \tau_{a \dot{a}} \dr x^{a \dot{a}} $ we have
\begin{equation}
\star^2 \sigma = - \sigma \ , \quad \sigma \wedge \star \tau = - \star \sigma \wedge \tau = \vol_4 \epsilon^{ab} \epsilon^{\dot{a} \dot{b}} \sigma_{a \dot{a}}\tau_{b \dot{b}} \ , \quad \dr \star \sigma = \vol_4 \epsilon^{ab} \epsilon^{\dot{a} \dot{b}} \pd_{a \dot{a}}\sigma_{b \dot{b}} \ .
\end{equation}
Contraction of spinors is given by
\begin{equation}
\langle \alpha \beta \rangle = \alpha_1 \beta_2 - \alpha_2 \beta_1 = \alpha^a \beta_a \ ,
\end{equation}
and spinor indices are raised as
\begin{equation}
\alpha^a = \epsilon^{ab }\alpha_b \ , \quad \epsilon^{12} = - \epsilon^{21} =-1 \ .
\end{equation}
We define $\epsilon_{12} = + 1$ such that $\epsilon^{a b} \epsilon_{b c} = \delta^a_{c}$. The (quaternionic) conjugation of a spinor $\alpha_a = (\alpha_1, \alpha_2)$ is defined to be $\hat{\alpha}_a = (- \bar{\alpha}_2 , \bar{\alpha}_1) $. Identical definitions hold for the anti-chiral spinors with dotted indices and contraction denoted with square brackets though these do not play a role in this work.

A basis for self-dual 2-forms is given by
\begin{equation}
\Sigma^{ab} = \epsilon_{\dot{a} \dot{b}} \dr x^{a \dot{a}} \wedge \dr x^{b\dot{b}} \ ,
\end{equation}
from which, given any two spinors, we can define self-dual forms
\begin{equation}
\Sigma_{\alpha,\beta} = \alpha_a \beta_b \Sigma^{ab} = \alpha_a \beta_b \epsilon_{\dot{a} \dot{b}} \dr x^{a \dot{a}} \wedge \dr x^{b\dot{b}} \ , \qquad \star \Sigma_{\alpha,\beta} = \Sigma_{\alpha,\beta} \ .
\end{equation}
As they will play key roles, we denote
\begin{equation}
\omega_{\alpha, \beta} = \frac{1}{\langle \alpha \beta \rangle } \Sigma_{\alpha,\beta} \ , \quad \mu_\alpha = \Sigma_{\alpha,\alpha }\ , \quad \mu_\beta = \Sigma_{\beta,\beta} \ .
\end{equation}

$\mathbb{R}^4$ is equipped with a hyper-K\"ahler structure and has a $\mathbb{CP}^1$s worth of complex structures. We can compactly express the complex structure corresponding to a spinor $\gamma_a$ as
\begin{equation}\label{eq:complexStruct}
{\mathcal J}_\gamma = - i (\gamma^a \pd_{a\dot{a}} ) \otimes (\hat{\gamma}_b \dr x^{b \dot{a}} )- i (\hat{\gamma}^a \pd_{a\dot{a}} ) \otimes (\gamma_b \dr x^{b \dot{a}} ) \ ,
\end{equation}
for which adapted complex coordinates are given by
\begin{equation}
\begin{aligned}
\dr z = \gamma_a \kappa_{\dot a} \dr x^{a \dot a } ~, \qquad &
\dr\bar{z} = \hat{\gamma}_a \hat{\kappa}_{\dot a} \dr x^{a \dot a } ~, \qquad
\dr w = \gamma_a \hat \kappa_{\dot a} \dr x^{a \dot a } ~, \qquad
\dr \bar{w} = - \hat \gamma_a \kappa_{\dot a} \dr x^{a\dot a } ~.
\end{aligned}
\end{equation}
With these coordinates we have that
\begin{align}
\mu_\alpha & = -2 \langle \alpha \gamma\rangle^2 \dr \bar{w} \wedge \dr \bar{z}- 2 \langle \alpha \gamma\rangle \langle \alpha \hat{\gamma}\rangle ( \dr z \wedge \dr \bar{z} +\dr w \wedge \dr \bar{w} )-2 \langle \alpha \hat{\gamma}\rangle^2 \dr w \wedge \dr z \ , \\
\omega_{\alpha, \beta} & = -2 \frac{\langle\alpha \hat\gamma \rangle \langle\beta \hat{ \gamma} \rangle }{\langle\alpha \beta \rangle } \dr w \wedge \dr z -2 \frac{\langle\alpha \gamma \rangle \langle\beta \gamma \rangle }{\langle\alpha \beta \rangle } \dr \bar{w} \wedge \dr \bar{z} - \frac{ \langle\alpha \gamma \rangle \langle\beta \hat{ \gamma} \rangle +\langle\alpha \hat\gamma \rangle \langle\beta \gamma \rangle }{\langle\alpha \beta \rangle} ( \dr z \wedge \dr \bar{z} +\dr w \wedge \dr \bar{w} ) \ .
\end{align}
Notice that, if we align the spinor $\alpha$ to $\gamma$ and $\beta$ to $\hat{\gamma}$, then $\omega_{\gamma, \hat{\gamma}}$ is (proportional to) the corresponding K\"ahler form $\varpi$ of type $(1,1)$ and $\mu_\gamma$ is a holomorphic $(2,0)$-form and $\mu_{\hat{\gamma}}$ is $(0,2)$-form.

\subsection{Appendix B: Twistor space} \label{sec:TwistorSpace}
We work on the Euclidean slice of Penrose's twistor space, $\mathbb{PT}_{\mathbb{E}}$. Starting from the twistor space of complexified Minkowski space,
\begin{equation}
\mathbb{PT} = \mathbb{CP}_3 \setminus \mathbb{CP}^{1} = \{ Z^\alpha = (\omega^{\dot{a}} , \pi_a ) \} \vert \pi_a \neq 0 \ , Z^\alpha \sim r Z^\alpha ~~ r\in \mathbb{C}^\times \} \ ,
\end{equation}
we obtain $\mathbb{PT}_{\mathbb{E}}$ by making a choice of reality conditions, in particular, by selecting the slice of $\mathbb{PT}$ invariant under the anti-holomorphic (quartic) involution acting on the holomorphic coordinates as ${Z}^{\alpha} \mapsto \hat{Z}^{\alpha} = (\hat{\omega}^{\dot{a}} , \hat{\pi}_{a})$. This choice of reality conditions induces a double fibration and we find that Euclidean twistor space can be viewed as the holomorphic vector bundle $\mathbb{PT}_{\mathbb{E}} \cong \mc{O}(1) \oplus \mc{O}(1) \rightarrow \mathbb{CP}^{1}$, where the holomorphic coordinates along the fibre direction are given by the incidence relations $\omega^{\dot{a}} = {x}^{a \dot{a}} {\pi}_{a}$. With this we choose a basis of $(1,0)$-forms and $(0,1)$-forms
\begin{equation}
\begin{aligned}
& e^0 = \langle \pi \dr \pi \rangle ~, &&
& e^{\dot{a}} = \pi_{a} \dr x^{a \dot{a} } ~, \\
& \bar{e}^0 = \frac{\langle \hat{\pi} \dr \hat{\pi} \rangle}{\langle \pi \hat{\pi} \rangle^2} ~, &&
& \bar{e}^{\dot{a}} = \frac{\hat{\pi}_{a} \dr x^{\dot{a} a }}{\langle \pi \hat{\pi} \rangle} ~,
\end{aligned}
\end{equation}
and their dual vector fields
\begin{equation}
\label{eq:twistorderivatives}
\begin{aligned}
& \pd_0 = \frac{\hat{\pi}_{a}}{\langle \pi \hat{\pi} \rangle} \frac{\pd}{\pd \pi_{a}} ~, &&
& \pd_{\dot{a}} = - \frac{\hat{\pi}^{a} \pd_{a \dot{a} }}{\langle \pi \hat{\pi} \rangle} ~, \\
& \bar{\pd}_0 = - \langle \pi \hat{\pi} \rangle \, \pi_{a} \frac{\pd}{\pd \hat{\pi}_{a}} ~, &&
& \bar{\pd}_{\dot{a}} = \pi^{a} \pd_{a \dot{a} } ~.
\end{aligned}
\end{equation}
It is important to note that this basis of $1$-forms, and their duals, satisfy the structure equations,
\begin{equation}
\begin{aligned}
\bar{\pd} e^{\dot{a}} = e^0 \wedge \bar{e}^{\dot{a}} ~, \quad
\pd \bar{e}^{\dot{a}} = e^{\dot{a}} \wedge \bar{e}^0 ~,\\
\comm{\bar{\pd}_0}{\pd_{\dot{a}}} = \bar{\pd}_{\dot{a}} ~, \quad
\comm{\bar{\pd}_{\dot{a}}}{\pd_0} = \pd_{\dot{a}} ~.
\end{aligned}
\end{equation}

\subsubsection{Homogeneous and inhomogeneous coordinates}

Homogeneous coordinates on $\CP$ will be denoted by $\pi_{a} = (\pi_{1} , \pi_{2})$, which are defined up to the equivalence relation $\pi_{a} \sim s \, \pi_{a}$ for any non-zero $s \in \bC^{\ast}$.
These have the advantage of being globally defined on $\CP$ but can lead to technical complications in certain calculations.
It can also be useful to work with inhomogeneous coordinates on two patches covering $\CP \cong S^{2}$.
Introducing an arbitrary spinor $\gamma_{a}$ that satisfies $\langle \gamma \hat{\gamma} \rangle = 1$, the two patches covering $\CP$ will be defined as
\begin{equation}
U_{1} = \{ \pi_{a} \mid \langle \pi \hat{\gamma} \rangle \neq 0 \} \ , \qquad
U_{2} = \{ \pi_{a} \mid \langle \pi \gamma \rangle \neq 0 \} \ .
\end{equation}
Inhomogeneous coordinates may be defined on each patch by
\begin{equation}
\zeta = \frac{
\langle \gamma \pi \rangle
}{
\langle \pi \hat{\gamma} \rangle
} \ , \qquad
\xi = \frac{
\langle \pi \hat{\gamma} \rangle
}{
\langle \gamma \pi \rangle
} \ , \qquad
\xi = \zeta^{-1} \ .
\end{equation}
In this section, we restrict our attention to $U_{1}$ and the inhomogeneous coordinate $\zeta$, knowing that an analogous discussion holds for the other patch.
The complex conjugate of the inhomogeneous coordinate $\zeta$ is
\begin{equation}
\bar{\zeta} = - \frac{
\langle \hat{\pi} \hat{\gamma} \rangle
}{
\langle \gamma \hat{\pi} \rangle
} \ .
\end{equation}

Forms and vector fields on $\CP$ written in these coordinates are related to one another by
\begin{equation}
\begin{aligned}
\dr \zeta & = \frac{
e^{0}
}{
\langle \pi \hat{\gamma} \rangle^{2}
} \ , \quad & \quad
\dr \bar{\zeta} & = \frac{
\langle \pi \hat{\pi} \rangle^{2}
}{
\langle \gamma \hat{\pi} \rangle^{2}
} \bar{e}^{0} \ , \\
\pd_{\zeta} & = \langle \pi \hat{\gamma} \rangle^{2} \pd_{0} \ , \quad & \quad
\pd_{\bar{\zeta}} & = \frac{
\langle \gamma \hat{\pi} \rangle^{2}
}{
\langle \pi \hat{\pi} \rangle^{2}
} \bar{\pd}_{0} \ .
\end{aligned}
\end{equation}
It is also helpful to define a weight zero basis of $(1,0)$-forms on $\bR^{4} \subset \PTE$ by
\begin{equation}
\theta^{\dot{a}} = \frac{
e^{\dot{a}}
}{
\langle \pi \hat{\gamma} \rangle
}
= \dr x^{a \dot{a}} \, \gamma_{a} + \zeta \, \dr x^{a \dot{a}} \, \hat{\gamma}_{a} \ .
\end{equation}
Likewise, the weight zero basis of $(0,1)$-forms on $\mathbb{R}^{4} \subset \mathbb{PT}$ are defined by
\begin{equation}
\bar{\theta}^{\dot{a}} = \la \pi \hat{\gamma} \ra \, \bar{e}^{\dot{a}} = \frac{1}{1+\zeta \bar{\zeta}}\left( dx^{a \dot{a}} \, \hat{\gamma}_{a} - \bar{\zeta} \, d x^{a \dot{a}} \, \gamma_{a} \right) \; .
\end{equation}
Given a point on $\CP$ defined by $\alpha_{a}$ in homogeneous coordinates, we denote the corresponding point in the inhomogeneous coordinate $\zeta$ by
\begin{equation}
\alpha = \frac{
\langle \gamma \alpha \rangle
}{
\langle \alpha \hat{\gamma} \rangle
}
= \zeta \vert_{\pi_{a} \sim \alpha_{a}} \ .
\end{equation}
We also have the relation
\begin{equation} \label{eq:hom_inhom_identity}
\frac{
\langle \pi \alpha \rangle
}{
\langle \pi \hat{\gamma} \rangle
\langle \hat{\gamma} \alpha \rangle
}
= (\zeta - \alpha) \ .
\end{equation}

\subsection{Appendix C: Projector technology}
\label{sec:appendixprojectors}
We consider the operator on $1$-forms on $\mathbb{R}^4$ given by
\begin{equation}
{\cal J}_{\alpha,\beta} (\sigma ) = - \mathrm{i} \star (\omega_{\alpha,\beta } \wedge \sigma) \ , \quad {\cal J}_{\alpha,\beta}^2 = -\mathrm{id} \ ,
\end{equation}
which allows us to define projectors
\begin{equation}
P = \frac{1}{2}\left(\mathrm{id} - \mathrm{i}{\cal J} \right) \, \qquad \bar{P} = \frac{1}{2}\left(\mathrm{id} + \mathrm{i}{\cal J}\right) \ .
\end{equation}
For this to define a complex structure on the real Euclidean slice of $\mathbb{R}^{4} \subset \mathbb{C}^{4}$ we require that  $\cal J$ maps Euclidean-real $1$-forms to Euclidean-real $1$-forms.
While not true for general $\alpha$ and $\beta$, this is the case if we take $\alpha = \gamma$ and $\beta =\hat{\gamma}$.
Then ${\cal J}_{\gamma,\hat{\gamma}}$ is the complex structure ${\cal J}_\gamma$, see~eq.~\eqref{eq:complexStruct}. The projectors $P$ and $\bar{P}$ project onto the $(1,0)$ and $(0,1)$ components thus realising the Dolbeault complex.

These projectors satisfy a range of useful identities:
\begin{align}
\label{eq:projectorid}
&\bar{P} ( \star (\mu_\alpha\wedge \sigma) ) = 0 \ , \qquad P( \star (\mu_\beta \wedge \sigma) ) = 0 \ ,\hspace{-25pt} &&\mu_\beta \wedge \bar{P}(\sigma) = 0 \ , \qquad \mu_\alpha \wedge P(\sigma) = 0 \ ,
\\ \label{eq:projectoridII}
&\omega_{\alpha, \beta }\wedge \bar{P}(\sigma) = -\star \bar{P}(\sigma) \ , && \omega_{\alpha, \beta } \wedge P(\sigma) = \star P(\sigma) \ ,
\\
\label{eq:projectoridIII}
&\omega_{\alpha, \beta }\wedge \bar{P}(\sigma) \wedge \tau = \omega_{\alpha, \beta } \wedge \sigma \wedge P(\tau) \ , && \omega_{\alpha, \beta }\wedge \bar{P}(\sigma) \wedge \bar{P}(\tau) = 0 \ .
\end{align}
To move between form and component notation is useful to observe that
\begin{equation}\label{eq:projecttocomp}
P(\sigma)_{a \dot{a}} = - \frac{1}{\la \alpha \beta \ra} \alpha_a \beta^b \sigma_{b \dot{a}} \ , \qquad \bar{P} (\sigma)_{a \dot{a}} = \frac{1}{\la \alpha \beta \ra} \beta_a \alpha^b \sigma_{b \dot{a}} \ .
\end{equation}
Further relations, useful for analysing the $\mathbb{CP}^{1}$-derivative boundary conditions, are
\begin{align}\label{eq:CP1compid}
& 2 \alpha^{a}\sigma_{a\dot{a}} e^{\dot{a}}\vert_{\alpha} =   \star ( \mu_\alpha \wedge \sigma) \ ,  && \hspace{-80pt} \beta^{a}\tau_{a\dot{a}} e^{\dot{a}}\vert_{\alpha} = -\langle \alpha \beta \rangle P(\tau) \ , \\
& 2\beta^{a}\sigma_{a\dot{a}} e^{\dot{a}}\vert_{\beta} =   \star ( \mu_\beta\wedge \sigma) \ , && \hspace{-80pt} \alpha^{a}\tau_{a\dot{a}} e^{\dot{a}}\vert_{\beta} = \langle \alpha \beta \rangle \bar{P}(\tau) \ .
\end{align}

As an application of this projector technology let us consider the (ungauged) WZW$_4$ model, for which the equations of motion can be cast in terms of the right-invariant Maurer-Cartan form $R= \dr g g^{-1}$, which obeys $\dr R = R\wedge R $, as
\begin{equation}
\dr \star \bar{P}(R) = \frac{1}{2} \dr\; (\star - \omega_{\alpha,\beta} \wedge)\; \dr g g^{-1} = 0 \ .
\end{equation}
We now consider a Yang-Mills connection $A = - \bar{P}(X) $. The equations for this to be anti-self dual are
\begin{equation}
\mu_\beta \wedge F[A] = 0 \ , \qquad \mu_\alpha \wedge F[A]= 0 \ , \qquad \omega_{\alpha,\beta} \wedge F[A] = 0\ .
\end{equation}
The first of these vanishes identically by virtue of the fact that $\mu_\beta \wedge A = 0 $. Since $ \mu_\alpha \wedge A = - \mu_\alpha \wedge X$, the second yields a Bianchi identity
\begin{equation}
\mu_\alpha \wedge F[A]= - \mu_\alpha \wedge \left( \dr X - X \wedge X\right)\ ,
\end{equation}
hence is solved by $X=R$. The final equation returns the equations of motion as
\begin{equation}
\omega_{\alpha,\beta} \wedge F[A] = -\dr( \omega_{\alpha,\beta}\wedge \bar{P}(R) ) +\omega_{\alpha,\beta}\wedge \bar{P}(R) \wedge \bar{P}(R) = \dr \star \bar{P}(R) \ .
\end{equation}

At the K{\"a}hler point $\beta = \hat{\alpha} = \hat{\gamma}$, we can simply write the ASDYM equations as
\begin{equation}
F^{2,0} = 0 \ , \qquad
F^{0,2} = 0 \ , \qquad
\varpi \wedge F^{1,1} = 0 \ .
\end{equation}
In this case, the connection given by $A = - \bar{\pd} g g^{-1}$ is of type $(0,1)$, hence $F^{2,0}=0$ automatically, $F^{0,2}=0$ is zero by the Bianchi identity and the equations of motion of WZW$_{4}$ are
\begin{equation}
\varpi \wedge \pd ( \bar{\pd} g g^{-1} ) = 0 \ .
\end{equation}

\subsection{Appendix D: Derivation of localisation formulae}
\label{appendix on localisation formulae}

In this work we are required to evaluate integrals of the form
\begin{equation}\label{General form of integral}
I = \frac{1}{2 \pi \mathrm{i}} \int_{\PTE} \bar{\pd} \Omega \wedge Q \ , \qquad
Q \in \Omega^{0,2} (\PTE) \ .
\end{equation}
In this appendix, we will derive general formulae for these integrals for the cases in which $\Omega$ has either two double poles or a single fourth-order pole. To compute these integrals efficiently we will work in inhomogeneous coordinates and make use of the identities
\begin{equation} \label{eq:inhom_pole_der}
\pd_{\bar{\zeta}} \bigg( \frac{1}{\zeta - \alpha} \bigg) = - 2 \pi \mathrm{i} \, \delta^{2}(\zeta - \alpha) \ , \qquad
\int_{\CP} \dr \zeta \wedge \dr \bar{\zeta} \, \delta^{2}(\zeta - \alpha) \, f(\zeta) = f(\alpha) \ .
\end{equation}

\subsubsection{Two double poles}
We consider the $(3,0)$-form given by
\begin{equation}
\Omega = \frac{1}{2}
\frac{
\langle \alpha \beta \rangle^{2}
}{
\langle \pi \alpha \rangle^{2}
\langle \pi \beta \rangle^{2}
}
e^0 \wedge e^{\dot{a}} \wedge e_{\dot{a}} = \frac{1}{2}
\frac{
(\alpha - \beta)^{2}
}{
(\zeta - \alpha)^{2}
(\zeta - \beta)^{2}
}
\dr \zeta \wedge \theta^{\dot{a}} \wedge \theta_{\dot{a}} \ .
\end{equation}
Substituting this into the integral gives
\begin{equation}
I = - \frac{1}{2} \frac{1}{2 \pi \mathrm{i}}
\int_{\PTE} \dr \zeta \wedge \dr \bar{\zeta} \, \pd_{\bar{\zeta}}
\bigg( \frac{
(\alpha - \beta)^{2}
}{
(\zeta - \alpha)^{2}
(\zeta - \beta)^{2}
} \bigg)
\wedge \theta^{\dot{a}} \wedge \theta_{\dot{a}} \wedge Q \ .
\end{equation}
Then, using the identity \eqref{eq:inhom_pole_der} gives
\begin{equation}
I = - \frac{(\alpha - \beta)^{2}}{2}
\int_{\PTE} \dr \zeta \wedge \dr \bar{\zeta} \,
\bigg[ \frac{
\pd_{\zeta} \delta(\zeta - \alpha)
}{
(\zeta - \beta)^{2}
}
+ \frac{
\pd_{\zeta} \delta(\zeta - \beta)
}{
(\zeta - \alpha)^{2}
} \bigg]
\wedge \theta^{\dot{a}} \wedge \theta_{\dot{a}} \wedge Q \ .
\end{equation}
Since the integral is symmetric under $\alpha \leftrightarrow \beta$ we will only compute the first term explicitly.
Integrating by parts and evaluating the integral over $\CP$ gives
\begin{equation} \label{eq:checkpoint}
I = \frac{(\alpha - \beta)^{2}}{2}
\int_{\bR^{4}} \pd_{\zeta}
\bigg( \frac{
\theta^{\dot{a}} \wedge \theta_{\dot{a}} \wedge Q
}{
(\zeta - \beta)^{2}
} \bigg) \bigg\vert_{\alpha}
\quad + \quad \alpha \leftrightarrow \beta \ .
\end{equation}
We first distribute the $\pd_{\zeta}$ derivative, leaving the 2-form $Q$ completely general, resulting in
\begin{equation}
\begin{aligned}
I = \frac{(\alpha - \beta)^{2}}{2}
\int_{\bR^{4}}
\bigg[
& \frac{-2}{(\zeta - \beta)^{3}} \theta^{\dot{a}} \wedge \theta_{\dot{a}} \wedge Q
+ \frac{2}{(\zeta - \beta)^{2}} \hat{\gamma}_{a} \dr x^{a \dot{a}} \wedge \theta_{\dot{a}} \wedge Q \\
& \qquad + \frac{
\theta^{\dot{a}} \wedge \theta_{\dot{a}}
}{
(\zeta - \beta)^{2}
} \wedge \pd_{\zeta} Q
\bigg] \bigg\vert_{\alpha}
\quad + \quad \alpha \leftrightarrow \beta \ .
\end{aligned}
\end{equation}
The overall factor of $(\alpha - \beta)^{2}$ outside the integral cancels with the denominators in the integrand.
We now make use of \eqref{eq:hom_inhom_identity} to return to spinor notation and introduce self-dual 2-forms defined by $\Sigma^{ab} = \varepsilon_{\dot{a} \dot{b}} \dr x^{a \dot{a}} \wedge \dr x^{b \dot{b}}$ to write
\begin{equation}
\begin{aligned}
I = \frac{1}{2}
\int_{\bR^{4}}
\bigg[
& \frac{
-2 \langle \hat{\gamma} \beta \rangle
}{
\langle \alpha \beta \rangle
\langle \alpha \hat{\gamma} \rangle
} \alpha_{a} \alpha_{b} \Sigma^{ab} \wedge Q \vert_{\alpha}
+ \frac{2}{
\langle \alpha \hat{\gamma} \rangle
}
\hat{\gamma}_{a} \alpha_{b} \Sigma^{ab} \wedge Q \vert_{\alpha} \\
& \qquad + \alpha_{a} \alpha_{b} \Sigma^{ab} \wedge \frac{
\pd_{\zeta} Q
}{
\langle \pi \hat{\gamma} \rangle^{2}
} \bigg\vert_{\alpha} \bigg]
\quad + \quad \alpha \leftrightarrow \beta \ .
\end{aligned}
\end{equation}
Expanding $\alpha_{a}$ in the basis formed by $\hat{\gamma}_{a}$ and $\beta_{a}$, we see that one component of the first term cancels the second term, and only a term proportional to $\alpha_{a} \beta_{b} \Sigma^{ab}$ survives.
In the third term of the integral, we recognise the combination $\pd_{0}$ acting on $Q$ and make this replacement.
In conclusion, we have the general formula
\begin{equation} \label{eq:localisationformula}
\frac{1}{2 \pi \mathrm{i}} \int_{\PTE} \bar{\pd} \Omega \wedge Q = \int_{\bR^{4}}
\bigg[
\frac{
\alpha_{a} \beta_{b} \Sigma^{ab}
}{
\langle \alpha \beta \rangle
} \wedge Q \vert_{\alpha}
+ \frac{1}{2} \alpha_{a} \alpha_{b} \Sigma^{ab} \wedge (\pd_{0} Q) \vert_{\alpha} \bigg]
\quad + \quad \alpha \leftrightarrow \beta \ ,
\end{equation}
or in differential form notation
\begin{equation}
\frac{1}{2 \pi \mathrm{i}} \int_{\PTE} \bar{\pd} \Omega \wedge Q = \int_{\bR^{4}}
\bigg[
\omega_{\alpha,\beta} \wedge Q \vert_{\alpha}
+ \frac{1 }{2} \mu_\alpha \wedge (\pd_{0} Q) \vert_{\alpha} \bigg]
\quad + \quad \alpha \leftrightarrow \beta \ .
\end{equation}
It is also helpful to specialise to 2-forms of the form $Q = \pi^{a} \pi^{b} Q_{a \dot{a} b \dot{b}} \bar{e}^{\dot{a}} \wedge \bar{e}^{\dot{b}}$, which we will often encounter.
In this case, we may make use of the identity
\begin{equation} \label{eq:es_to_vol4}
e^{\dot{c}} \wedge e_{\dot{c}} \wedge \bar{e}^{\dot{a}} \wedge \bar{e}^{\dot{b}} = - 2 \, \vol_{4} \, \varepsilon^{\dot{a} \dot{b}} \ .
\end{equation}
and its generalisation valid for any spinors $\alpha_{a}$ and $\beta_{a}$
\begin{equation}
\alpha_{a} \beta_{b} \Sigma^{ab} \wedge \bar{e}^{\dot{a}} \wedge \bar{e}^{\dot{b}} = - 2 \, \vol_{4} \, \frac{\langle \alpha \hat{\pi} \rangle \langle \beta \hat{\pi} \rangle}{\langle \pi \hat{\pi} \rangle^{2}} \, \varepsilon^{\dot{a} \dot{b}} \ .
\end{equation}
Using these identities on the above formula for $Q = \pi^{a} \pi^{b} Q_{a \dot{a} b \dot{b}} \bar{e}^{\dot{a}} \wedge \bar{e}^{\dot{b}}$ gives
\begin{equation}
\label{ec:localisationformula}
\frac{1}{2 \pi \mathrm{i}} \int_{\PTE} \bar{\pd} \Omega \wedge Q = - \int_{\bR^{4}} \vol_{4} \,
\bigg[
\frac{
\varepsilon^{\dot{a} \dot{b}} (\alpha^{a} \beta^{b} + \beta^{a} \alpha^{b})
}{
\langle \alpha \beta \rangle
} Q_{a \dot{a} b \dot{b}} \vert_{\alpha}
+ \varepsilon^{\dot{a} \dot{b}} \alpha^{a} \alpha^{b}
(\pd_{0} Q_{a \dot{a} b \dot{b}}) \vert_{\alpha} \bigg]
\quad + \quad \alpha \leftrightarrow \beta \ .
\end{equation}
Finally, we specialise to the case when $Q_{a \dot{a} b \dot{b} } = X_{a \dot{a}} Y_{b\dot{b}} $, for which the answer can again be recast in differential form notation as
\begin{equation}
\label{ec:localisationformula2}
\frac{1}{2 \pi \mathrm{i}} \int_{\PTE} \bar{\pd} \Omega \wedge Q = \int_{\bR^{4}}
\bigg[
\omega_{\alpha,\beta} \wedge X\wedge Y \vert_{\alpha}
+ \frac{1 }{2} \mu_\alpha \pd_{0} \wedge (X\wedge Y ) \vert_{\alpha} \bigg]
\quad + \quad \alpha \leftrightarrow \beta \ .
\end{equation}

To apply these formulae we also need the following $\mathbb{CP}^{1}$-derivatives:
\begin{align}
\pd_0 ( \dr \hat{g} \hat{g}^{-1} ) &= \hat{g} \dr \hat{u} \hat{g}^{-1} \ , \\
\pd_0 ( \hat{g}^{-1}\dr \hat{g} ) &= \dr \hat{u} + [ \hat{g}^{-1}\dr \hat{g}, \hat{u} ] \ , \\
\pd_0 ( A ) &= \pd_0 (B) = 0 \ , \\
\pd_0 (\hat{g}^{-1} A \hat{g} ) &= [\hat{g}^{-1} A \hat{g} ,\hat{u} ] \ , \\
\pd_0 \frac{1}{3} \mathrm{Tr} (\hat{g}^{-1} \dr\hat g)^3 &= \dr ~\mathrm{Tr}( \hat{u} (\hat{g}^{-1} \dr\hat g)^2) \ ,
\end{align}
where we have defined $\hat{u} = \hat{g}^{-1}\pd_0 \hat{g}$.

\subsubsection{Fourth-order pole}
In section \ref{Gauged LMP action section}, we consider a different $(3,0)$-form given by
\begin{equation}\label{4th order}
\Omega = k \frac{{e}^{0} \wedge {e}^{\dot{a}}\wedge {e}_{\dot{a}}}{{\langle \pi \alpha \rangle}^{4}} = \frac{k'}{{\la \hat{\gamma} \alpha \ra}^{4}} \frac{d \zeta \wedge \theta^{\dot{a}} \wedge \theta_{\dot{a}}}{(\zeta - \alpha)^{4}} \ .
\end{equation}
Substituting this into the general integral expression above gives
\begin{equation}
I = - \frac{k}{{\la \hat{\gamma} \alpha \ra}^{4}} \frac{1}{2 \pi \mathrm{i}}
\int_{\PTE} \dr \zeta \wedge \dr \bar{\zeta} \, \pd_{\bar{\zeta}}
\bigg( \frac{1}{(\zeta - \alpha)^{4}} \bigg)
\wedge \theta^{\dot{a}} \wedge \theta_{\dot{a}} \wedge Q \ .
\end{equation}
Then, using the identity \eqref{eq:inhom_pole_der}, we find
\begin{equation}
I = - \frac{k}{6 {\la \hat{\gamma} \alpha \ra}^{4}}
\int_{\PTE} \dr \zeta \wedge \dr \bar{\zeta} \,
\Big( \pd_{\zeta}^{3} \delta (\zeta - \alpha) \Big)
\wedge \theta^{\dot{a}} \wedge \theta_{\dot{a}} \wedge Q \ .
\end{equation}
Integrating by parts and evaluating the integral over $\CP$ gives
\begin{equation}
I = \frac{k}{6 {\la \hat{\gamma} \alpha \ra}^{4}}
\int_{\bR^{4}} \pd_{\zeta}^{3} \Big( \theta^{\dot{a}} \wedge \theta_{\dot{a}} \wedge Q \Big) \Big\vert_{\alpha} \ .
\end{equation}
In order to distribute the $\pd_\zeta$ derivatives, it is helpful to use the identities
\begin{equation}
\theta^{a} \big\vert_{\alpha} = \frac{\alpha_{a} \dr x^{a \dot{a}}}{\langle \hat{\gamma} \alpha \rangle} \ , \qquad
\pd_{\zeta} \theta^{a} \big\vert_{\alpha} = \hat{\gamma}_{a} \dr x^{a \dot{a}} \ , \qquad
\pd_{\zeta}^{2} \theta^{a} \big\vert_{\alpha} = 0 \ .
\end{equation}
Distributing the three $\pd_{\zeta}$ derivatives gives
\begin{equation}
I = \frac{k}{6 {\la \hat{\gamma} \alpha \ra}^{4}}
\int_{\bR^{4}} \bigg[
\frac{\alpha_{a} \alpha_{b} \Sigma^{ab}}{\langle \alpha \hat{\gamma} \rangle^{2}} \wedge \pd_{\zeta}^{3} Q \big\vert_{\alpha}
+ 6 \, \frac{\alpha_{a} \hat{\gamma}_{b} \Sigma^{ab}}{\langle \alpha \hat{\gamma} \rangle} \wedge \pd_{\zeta}^{2} Q \big\vert_{\alpha}
+ 6 \, \hat{\gamma}_{a} \hat{\gamma}_{b} \Sigma^{ab} \wedge \pd_{\zeta} Q \big\vert_{\alpha}
\bigg] \ .
\end{equation}
Converting this expression back into homogeneous coordinates (and using the fact that $Q$ is a $(0,2)$-form on twistor space hence $\hat{\alpha}_{a} \dr x^{a \dot{a}} \wedge Q \vert_{\alpha} = 0$) this integral becomes
\begin{equation}
I = \frac{k}{6}
\int_{\bR^{4}} \alpha_{a} \alpha_{b} \Sigma^{ab}\wedge \pd_{0}^{3} Q \big\vert_{\alpha} \ .
\end{equation}

\subsection{Appendix E: Localisation derivation with general gaugings}\label{Sec:General gaugings}

In this appendix we describe in more detail the derivation of the gauged WZW$_4$ model from the gauged hCS$_6$ theory and the application of the localisation formulae in appendix \ref{General form of integral}. We will do this in a more general manner, allowing the gauging of an $H$ subgroup that acts as
\begin{equation}\label{eq:asymaction}
g\mapsto \rho_\beta(\ell) g \rho_\alpha(\ell^{-1}) \ , \quad B \mapsto \ell B \ell^{-1} - \dr \ell \ell^{-1} \ , \quad \ell \in H \subset{G} \ ,
\end{equation}
where $\rho_i: H \rightarrow G$ are group homomorphisms (algebra homomorphisms will be denoted by the same symbol). The covariant derivative is then given by \begin{equation}
\nabla g g^{-1} =\dr g g^{-1} + B_\beta - g B_\alpha g^{-1} \mapsto \rho_\beta(\ell) ( \nabla g g^{-1})\rho_\beta(\ell^{-1}) \ ,
\end{equation}
in which we ease the notation by setting $B_i =\rho_i(B)$.

The starting point is the six-dimensional theory
\begin{equation}\label{actds}
S_{\mathrm{ghCS_6}}[\cA , \cB] = S_{\mathrm{hCS_6}}[\cA] - S_{\mathrm{hCS_6}}[\cB] + S_{\mathrm{bdry}}[\cA , \cB] \ ,
\end{equation}
where we take the boundary interaction term to be
\begin{equation}\label{bdry}
S_{\mathrm{bdry}}[\cA , \cB]= -\frac{q}{2\pi \mathrm{i} }\int_{\mathbb{PT} } \bar{\partial} \Omega \wedge \mathrm{Tr}_\mathfrak{g}\left( \cA \wedge \rho(\cB) \right) \ .
\end{equation}
Here we have introduced a parameter $q$, which will ultimately be set to one, to keep track of the contributions from this boundary term.
To specify this term we include an algebra homomorphism $\rho$ that only needs to be defined piecewise on the components of the support of $\bar{\pd}\Omega$.
We could choose to dispense with the higher-dimensional covariance and simply add different boundary terms specified only at the location of the poles, but it is convenient to formally consider $\rho$ to be a defined as a piecewise map that takes values $\rho \vert_{\pi = \alpha , \beta } = \rho_{\alpha, \beta } $.

To define a six-dimensional theory requires imposing conditions that ensure the vanishing of the boundary variation
\begin{equation}
\int_{\mathbb{PT} } \bar{\pd} \Omega \wedge \left( \mathrm{Tr}_{\mathfrak{g}}\left(\delta\cA \wedge (\cA - q \rho (\cB) )+q \rho(\delta \cB) \wedge \cA \right) - \mathrm{Tr}_{\mathfrak{h}}(\delta \cB \wedge \cB) \right) \ .
\end{equation}
We are required to cancel a term involving the inner product on the algebra $\mathfrak{h}$ with one on $\mathfrak{g}$, which can be achieved by demanding
\begin{equation}
\mathrm{Tr}_\mathfrak{g} ( \rho(x) \rho(y) )\vert_{\alpha,\beta} = \mathrm{Tr}_\mathfrak{h} ( x y) \quad \forall x,y \in \mathfrak{h} \ .
\end{equation}
Note that as a consequence this implies
\begin{equation}\label{eq:anomv2}
\mathrm{Tr}_\mathfrak{g} ( \rho_\alpha (x) \rho_\alpha (y) ) = \mathrm{Tr}_\mathfrak{h} ( x y) = \mathrm{Tr}_\mathfrak{g} ( \rho_\beta (x) \rho_\beta (y) ) \ ,
\end{equation}
which is the familiar anomaly-free condition required to construct a gauge-invariant extension to the WZW model with the gauge symmetry~\eqref{eq:asymaction}. With this condition satisfied, the boundary term produced by variation is given by
\begin{equation}
\int_{\mathbb{PT} } \bar{\pd} \Omega \wedge \left( \mathrm{Tr}_{\mathfrak{g}}\left(\delta\cA \wedge (\cA - q \rho (\cB) )+q \rho(\delta \cB) \wedge (\cA - q^{-1} \rho(\cB))\right) \right) \ ,
\end{equation}
and is set to zero by the conditions
\begin{equation} \label{eq:ABboundarycond2}
\cA^{\fk} \big\vert_{\alpha , \beta} = 0 \ , \qquad
\cA^{\fh} \big\vert_{\alpha , \beta} =\rho(\cB) \big\vert_{\alpha , \beta} \ , \qquad
\pd_{0} \cA^{\fh} \big\vert_{\alpha , \beta} = \rho( \pd_{0}\cB) \big\vert_{\alpha , \beta} \ .
\end{equation}
If we impose all of these conditions from the outset, the contribution from the explicit boundary term $S_{\mathrm{bdry}}[\cA , \cB]$ would vanish.
However, from a four-dimensional perspective the $\mathbb{CP}^{1}$-derivative boundary conditions lead to constraints relating derivatives of the fundamental fields to the 4-dimensional gauge field $B$ that comes from $\mathcal{B}$.
While these can be formally solved for $B$, our aim is to construct a gauged IFT$_4$ with a gauge field.
Therefore, we only impose the conditions $\cA^{\fk} \big\vert_{\alpha , \beta} = 0$ and $\cA^{\fh} \big\vert_{\alpha , \beta} =\rho(\cB) \big\vert_{\alpha , \beta}$, which can be solved for the 4-dimensional gauge field $A$ that comes from $\mathcal{A}$ and substituted into the Lagrangian without concern. Doing this, we find that $S_{\mathrm{bdry}}[\cA , \cB]$ does contribute, and when $q=1$ in particular, it provides a gauge invariant completion of the action. Importantly, the $\mathbb{CP}^{1}$-derivative boundary conditions that we have not imposed have not been forgotten, instead when $q=1$ they are recovered as on-shell equations in this four-dimensional theory. This provides an alternative view of the procedure; when $q=1$ the explicit boundary term~\eqref{bdry} is serving to implement the constraints arising from $\pd_{0} \cA^{\fh} \big\vert_{\alpha , \beta} = \rho( \pd_{0}\cB) \big\vert_{\alpha , \beta}$ at the Lagrangian level.
We can see this explicitly by observing that if we just impose $\cA^{\fk} \big\vert_{\alpha , \beta} = 0$ and $\cA^{\fh} \big\vert_{\alpha , \beta} =\rho(\cB) \big\vert_{\alpha , \beta}$ then
\begin{equation}\begin{split}
    & \Big(\delta\cA \wedge (\cA - q \rho (\cB) )+q \rho(\delta \cB) \wedge \cA - \rho(\delta \cB) \wedge \rho(\cB)\Big)|_{\alpha,\beta} = 0 \ ,
    \\
    & \partial_0 \Big(\delta\cA \wedge (\cA - q \rho (\cB) )+q \rho(\delta \cB) \wedge \cA - \rho(\delta \cB) \wedge \rho(\cB)\Big)|_{\alpha,\beta}
    \\ & \qquad =  (1-q)\delta(\partial_0\cA - \rho( \partial_0\cB) )\wedge \rho(\cB) |_{\alpha,\beta}  + (1+q)\rho(\delta\cB) \wedge (\partial_0 \cA - \rho(\partial_0 \cB))|_{\alpha,\beta} \ .
\end{split}\end{equation}
Therefore, for $q=1$ we see that the boundary equations of motion for $\mathcal{B}$ are precisely the $\mathbb{CP}^{1}$-derivative boundary conditions $\pd_{0} \cA^{\fh} \big\vert_{\alpha , \beta} = \rho( \pd_{0}\cB) \big\vert_{\alpha , \beta}$.

The localisation proceeds as follows. First, we change parametrisation $\cA = \cA'{}^{\hat{g}}$ and $\cB =\cB'{}^{\hat{h} } $ fixing some of the redundancy by demanding that $\cA'$ and $\cB'$ have no $\mathbb{CP}^{1}$ legs. Second, we fix some of the residual symmetry preserved by the boundary conditions to set $\hat{g} \vert_{\beta} = \hat{h} \vert_{\alpha,\beta} = \mathrm{id} $ and $ \pd_0 \hat{h } \vert_{\alpha,\beta} = 0 $. The remaining fields are $\hat{g}\vert_{\alpha} =g $, $\hat{g}^{-1} \pd_0 \hat{g}\vert_{\alpha} = u $, $\hat{g}^{-1} \pd_0 \hat{g}\vert_{\beta} = \tilde{u} $ and the four-dimensional gauge fields $A$ and $B$ that arise from $\cA'$ and $\cB'$ once their holomorphicity is imposed.

We may now directly apply the localisation formulae~\eqref{ec:localisationformula2} to show that the hCS$_6$ terms localise, before imposing boundary conditions, to give
\begin{equation}
\begin{split}
S_{\mathrm{hCS}_6}[\mathcal{A}] \simeq & \int_{\mathbb{R}^4} \omega_{\alpha,\beta} \wedge \mathrm{Tr}_\fg ( A^g \wedge g^{-1} \dr g ) - \omega_{\alpha,\beta}\wedge \mathcal{L}_{\textrm{WZ}}[g] \\
& \qquad + \frac{1 }{2} \mu_\alpha\wedge \mathrm{Tr}_\fg ( A^g \wedge \dr u) + \frac{1 }{2} \mu_\beta \wedge \mathrm{Tr}_\fg ( A \wedge \dr \tilde{u}) \ ,
\end{split}
\end{equation}
while $ S_{\mathrm{hCS}_6}[\cB]$ yields zero in this gauge. Let us first consider the terms involving $\omega_{\alpha,\beta}$. Since the gauge completion of the WZ term is
\begin{equation}
\label{eq:generalgwzw1}
\mathcal{L}_{\textrm{gWZ}}[g,B] = \mathcal{L}_{\textrm{WZ}}[g] + \mathrm{Tr}_\mathfrak{g} \left( g^{-1} \dr g \wedge
B_\alpha + \dr g g^{-1} \wedge B_\beta + g^{-1} B_\beta g B_\alpha \right) \ ,
\end{equation}
we may express them (trace implicit) as
\begin{equation}\begin{split}
& \omega_{\alpha,\beta} \wedge \left( A^g \wedge g^{-1} \dr g - \mathcal{L}_{\textrm{WZ}}[g] \right) \\
= \ \, & \omega_{\alpha,\beta} \wedge \left( A^g \wedge g^{-1} \dr g - \mathcal{L}_{\textrm{gWZ}}[g,B] + g^{-1} \dr g \wedge B_\alpha + \dr g g^{-1} B_\beta +g^{-1} B_\beta g B_\alpha \right) \\
= \ \, & \omega_{\alpha,\beta} \wedge \left( (A^g- B_\alpha) \wedge g^{-1} \nabla g - \mathcal{L}_{\textrm{gWZ}}[g,B] + A^g
\wedge B_\alpha - A \wedge B_\beta \right) \ .
\end{split}\end{equation}
In differential form notation, the algebraic boundary conditions of eq.~\eqref{eq:ABboundarycond2} become
\begin{align}\label{eq:BCsol}
A = B_\beta - \bar{P}( \nabla g g^{-1} ) \ , \qquad A^g = P(g^{-1}\nabla g) + B_\alpha \ .
\end{align}
It follows that
\begin{equation}\begin{split}
& \omega_{\alpha,\beta} \wedge \left( A^g \wedge g^{-1} \dr g -\mathcal{L}_{\textrm{WZ}}[g] \right) \\
=\ \, & \omega_{\alpha,\beta} \wedge \left( P( g^{-1} \nabla g) \wedge g^{-1} \nabla g - \mathcal{L}_{\textrm{gWZ}}[g,B] + A^g
\wedge B_\alpha - A \wedge B_\beta \right) \\
=\ \, & -\frac{1}{2} g^{-1} \nabla g \wedge\star( g^{-1} \nabla g) -\omega_{\alpha,\beta} \wedge \left( \mathcal{L}_{\textrm{gWZ}}[g,B] - A^g
\wedge B_\alpha + A \wedge B_\beta \right)
\ .
\end{split}\end{equation}
Here, in the last line, we made use of the identity $
\omega \wedge P(\sigma)\wedge \sigma = - \frac{1}{2}\sigma\wedge \star \sigma $ for a 1-form $\sigma$. To treat the terms involving $\mu_{\alpha}$ and $\mu_\beta$ we combine the algebraic boundary conditions~\eqref{eq:BCsol} with the properties $\mu_\alpha \wedge P(X) = \mu_\beta \wedge \bar{P}(X) = 0 $ such that $\mu_a\wedge A^g= \mu_\alpha B_\alpha$ and $\mu_\beta \wedge A= \mu_\beta B_\beta$. In summary, we find
\begin{equation}
\begin{split}
S_{\mathrm{hCS}_6}[\mathcal{A}] &\simeq \int_{\mathbb{R}^4} -\frac{1}{2} \mathrm{Tr}_\fg\left(g^{-1}\nabla g \wedge \star g^{-1}\nabla g \right) - \omega_{\alpha,\beta}\wedge\left(\mathcal{L}_{\textrm{gWZ}}[g,B] + \mathrm{Tr}_\fg( A \wedge B_\beta - A^g B_\alpha ) \right) \\
& \qquad \quad + \frac{1}{2} \mu_\alpha\wedge \mathrm{Tr}( B_\alpha \wedge \dr u) + \frac{1 }{2} \mu_\beta \wedge \mathrm{Tr}( B_\beta \wedge \dr \tilde{u}) \ .
\end{split}
\end{equation}

The localisation of the explicit boundary term yields, after using $\mu_a\wedge A^g= \mu_\alpha B_\alpha$,
\begin{equation}
\begin{split}
S_{\mathrm{bdry}}[\cA , \cB] \simeq & - q \int_{\mathbb{R}^4}\omega_{\alpha,\beta} \wedge \mathrm{Tr}_\fg ( A^g B_\alpha - A B_\beta )\\
& \qquad \qquad + \frac{1}{2} \mu_\alpha\wedge \mathrm{Tr}_\fg ( (\dr u + [B_\alpha , u]) B_\alpha ) + \frac{1 }{2} \mu_\beta\wedge \mathrm{Tr}_\fg ( (\dr \tilde{u} + [B_\beta , \tilde{u}]) B_\beta ) \ .
\end{split}
\end{equation}
The significance of the boundary term now becomes clear.
It serves to ensure manifest gauge invariance when we do not impose the $\mathbb{CP}^{1}$-derivative boundary conditions.
When $q=1$ the terms $\omega_{\alpha,\beta} \wedge \mathrm{Tr}( A^g B_\alpha - A B_\beta )$ directly cancel. The contributions of the entire localised action that are wedged against $\mu_\alpha$ sum to
\begin{equation}
\mu_\alpha \wedge \mathrm{Tr}_\fg \left( (1- q) ~\dr u\wedge B_\alpha + 2 q ~ u F[B]_\alpha -2q ~\dr ( B_\alpha u) \right) \ .
\end{equation}
We see that for $q=1$ we find a gauge-invariant field strength together with a total derivative term that we discard.
The terms wedged against $\mu_\beta$ give a similar contribution.
Hence the fully localised action becomes
\begin{equation}
\begin{split}
S \simeq & \int_{\mathbb{R}^4}
-\frac{1}{2} \mathrm{Tr}_\fg\left(g^{-1}\nabla g \wedge \star g^{-1}\nabla g \right) - \omega_{\alpha,\beta}\wedge \mathcal{L}_{\textrm{gWZ}}[g,B] \\
& \qquad + \mu_\alpha\wedge \mathrm{Tr}_\fg( u F[B]_\alpha ) + \mu_\beta \wedge \mathrm{Tr}_\fg( \tilde{u} F[B]_\beta) \ .
\end{split}
\end{equation}
Noting that the components of $u$ and $\tilde{u}$ in the complement of $\fh$ decouple, we can view $u$ and $\tilde{u}$ as $\fh$-valued and write
\begin{equation}
\begin{split}
S \simeq & \int_{\mathbb{R}^4}
- \frac{1}{2} \mathrm{Tr}_\fg\left(g^{-1}\nabla g \wedge \star g^{-1}\nabla g \right) - \omega_{\alpha,\beta}\wedge \mathcal{L}_{\textrm{gWZ}}[g,B] \\
& \qquad + \mu_\alpha\wedge \mathrm{Tr}_\fh( u F[B] ) + \mu_\beta \wedge \mathrm{Tr}_\fh( \tilde{u} F[B]) \ .
\end{split}
\end{equation}

\chapter{Holomorphic 2-Chern-Simons Theory}
\label{chap:5d2CS}

This work was done in collaboration with Hank Chen \cite{Chen:2024axr}.

\section{Introduction} 

Integrable models represent a unique domain of exploration where complex systems exhibit a remarkable level of order. This is typically expressed through an infinite number of symmetries, which in turn, allow for the construction of an infinite number of independently conserved charges. The existence of such a large number of conserved quantities imposes stringent constraints on the system, which may render it soluble to some extent; it is precisely this solvable character what makes integrable systems so special. 

However, the strength of integrable theories is also their greatest weakness: identifying the infinite number of conserved charges is at the same time, the most complicated challenge. Significant progress was made in this direction with the introduction of the Lax formalism \cite{Lax:1968fm} for two-dimensional models. Indeed, this framework provides a systematic way to identify and construct the conserved charges. Specifically, in the case of two dimensional field theories, one looks for a $\mathfrak{g}^{\mathbb{C}}$-valued \emph{Lax connection} which is on-shell flat and depends meromorphically on a $\mathbb{C}P^1$-valued parameter known as the \emph{spectral parameter}. If such an object is found, then its holonomy along Cauchy slices can be used as a generating functional for the conserved charges, by expanding it in powers of the spectral parameter $z\in \mathbb{C}P^1$.

Similarly, in the context of four-dimensional systems, integrability has also been expressed in terms of the \emph{on-shell} flatness of some connection; the prototypical example being the anti-self-dual Yang-Mills (ASDYM) equations. In fact, it has long been established that two-dimensional integrable models arise as symmetry reductions of the ASDYM equations. Of course, anti-self-duality is not quite a flatness condition on the curvature, but instead, the requirement $F=-\star F$. However, the Penrose-Ward correspondence \cite{Ward:1977ta} identifies solutions to the ASDYM equations on four-dimensional space-time, with holomorphic vector bundles over twistor space. In particular, for a vector bundle to be holomorphic, the corresponding connection must be flat.

Despite the inherent elegance of these geometric characterizations of integrability, both suffer from the same issue: although they provide a systematic procedure for constructing the conserved charges once the on-shell flat connection is found, they offer no instruction whatsoever on how to find the connection in the first place.  

In 2013, Costello conceives a beautifully unifying framework to address this problem, originally, for the case of two-dimensional models\footnote{To be precise, the first series of papers dealt with discrete integrable systems in two-dimensions, whereas only the last article of the series discusses two-dimensional field theories. In particular, the latter is the one we will be interested in.} \cite{Costello:2013zra, Costello:2013sla}, which was further refined in a series of seminal papers written in collaboration with Witten and Yamazaki \cite{Costello:2017dso, Costello:2018gyb, Costello:2019tri}. Loosely speaking, one starts with three-dimensional Chern-Simons (CS$_3$) theory, whose fundamental field $A$ is a gauge connection which is \emph{on-shell} flat: at the very least, a tempting candidate to feature in this construction. There are nonetheless, two immediate concerns. First, we are looking for a Lax connection which is defined over a $2$-fold, whereas the gauge field $A$ lives in a three dimensional manifold. This is not \emph{really} a problem, as we can always gauge fix one of the components of the connection to zero. The real issue, is that we want the fields of the two-dimensional theory to be sections of a $G$-bundle over spacetime, not the connection itself! 

To resolve these issues, Costello complexifies and compactifies one of the real directions of the $3$-fold where CS$_3$ is defined, to obtain a $4$-fold $X=\mathbb{C}P^1 \times \Sigma$. Furthermore, he introduces a disorder operator $\omega$ which is the key object that defines $4$d Chern-Simons theory (CS$_4$), whose action is given by
\begin{equation}
    S_{\mathrm{CS}_4}=\int_X \omega \wedge \langle A, \mathrm{d}A+\tfrac{2}{3}A\wedge A\rangle \,.
\end{equation}
The disorder operator $\omega$ is a meromorphic $(1,0)$-form with poles (and eventually zeros) on $\mathbb{C}P^1$. These singularities, which are punctures on the Riemmann sphere, act as a boundary of the form $\{\text{poles of }\omega\}\times \Sigma$ for the $4$-fold $X$. Conceptually, by introducing a boundary on the theory we are partially breaking gauge symmetry. Now gauge transformations identify field configurations which are physically indistinguishable; said differently, they kill \emph{would-be} degrees of freedom. Thus, breaking the gauge symmetry of the theory will \emph{resurrect} \cite{Tong:2016kpv} these would-be degrees of freedom exactly where the symmetry was broken, namely, \emph{on the boundary} $\partial X = \{\text{poles of }\omega\}\times \Sigma$. These boundary degrees of freedom which we will refer to as \emph{edge-modes} are the fields of the two-dimensional theory. Crucially, the bulk of the theory is unaltered, so the flatness of the gauge field $A$, which will then become the Lax connection of the boundary theory, will be implied by construction, and with this, the integrability of the theory. 

Four-dimensional Chern-Simons theory has been an incredibly successful program since its conception. Most of the previously known two-dimensional integrable field theories (IFTs) have been constructed from this perspective, together with new models as well \cite{Delduc:2019whp, Costello:2020lpi,Fukushima:2020kta,Fukushima:2020tqv,Fukushima:2021ako, Ashwinkumar:2020gxt, Caudrelier:2020xtn, Berkovits:2024reg}. See also \cite{Lacroix:2021iit} for a review of its original formulation. Moreover, alternative approaches have been constructed in order to include more general choices of the meromorphic $(1,0)$-form $\omega$ using techniques from homotopical methods \cite{Benini_2019,Lacroix:2021iit, Liniado:2023uoo}. In addition, the framework has also been extended to the case in which $\mathbb{C}P^1$ is replaced with higher genus Riemmann surfaces \cite{Lacroix:2023qlz, Lacroix:2024wrd}. 

In an interesting turn of events, inspired by an idea suggested in a seminar by Costello, Bittleston and Skinner show that CS$_4$ can be described from an even more general perspective, starting with six-dimensional holomorphic Chern-Simons theory (hCS$_6$) on twistor space \cite{Bittleston:2020hfv}. Indeed, holomorphic Chern-Simons theory had first been considered in \cite{Witten:1992fb} to describe the open string sector of the type B topological string. In this context, the target space must be Calabi-Yau, which ensures the existence of a globally defined holomorphic $(3,0)$-form. Twistor space however, is not Calabi-Yau and it therefore does not admit a trivial canonical bundle. Costello's suggestion to overcome this problem was to consider a meromorphic $(3,0)$-form rather than a holomorphic one; in other words, introducing disorder defects. 

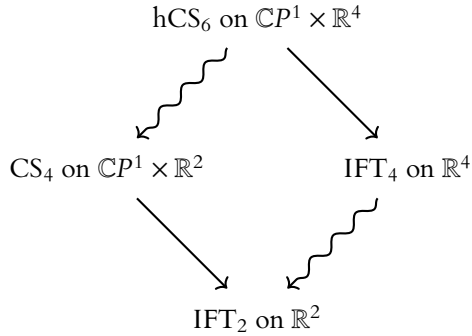
\begin{figure}[ht]
\label{fig:diamond}
\centering
\begin{tikzpicture}
\node at (0,2) {hCS$_6$ on $\mathbb{C}P^1\times \mathbb{R}^4$};
\node at (-2,0) {CS$_4$ on $\mathbb{C}P^1\times \mathbb{R}^2$};
\node at (2,0) {IFT$_4$ on $\mathbb{R}^4$};
\node at (0,-2) {IFT$_2$ on $\mathbb{R}^2$};
\draw[->,thick,decorate, decoration={snake, segment length=12pt, amplitude=2pt}] (-0.4,1.6)--(-1.6,0.4);
\draw[->,thick] (0.4,1.6)--(1.6,0.4);
\draw[->,thick,decorate, decoration={snake, segment length=12pt, amplitude=2pt}] (1.6,-0.4)--(0.4,-1.6);
\draw[->, thick] (-1.6,-0.4)--(-0.4,-1.6);
\end{tikzpicture}
\caption{Diamond of correspondence of theories}
\end{figure}

This led to a remarkable generalization of CS$_4$ theory because not only it provided a systematic way to construct four-dimensional integrable field theories (in the sense of ASDYM) but it also managed to include the symmetry reduction of ASDYM to 2d integrable models in a unique coherent scheme. More precisely, Bittleston and Skinner introduce a diamond of correspondence of theories as the one shown in figure \ref{fig:diamond}.

The straight lines represent integration along $\mathbb{C}P^1$. That is, one can morally do the same procedure to go from hCS$_6$ on (Euclidean) twistor space $\mathbb{PT}\cong \mathbb{C}P^1\times \mathbb{R}^4$ to an IFT on $\mathbb{R}^4$ than the one introduced in \cite{Costello:2019tri} to go from CS$_4$ on $\mathbb{C}P^1\times \mathbb{R}^2$ to an IFT on $\mathbb{R}^2$. The wiggly lines, in turn, represent symmetry reduction, which schematically consists of quotienting out a copy of $\mathbb{R}^2$ from the copy of $\mathbb{R}^4$. This was formally implemented in \cite{Bittleston:2020hfv, Penna:2020uky} for various choices of meromorphic $(3,0)$-forms, and it was shown that following both sides of the diamond leads to the same 2d IFT. This work was further generalized in \cite{Cole:2023umd, Cole:2024ess, Cole:2024noh}.

It is clear that some of the fundamental features of integrable structures appearing in field theories can be understood from a gauge theoretic perspective. So far, the strategy to study integrable models in higher dimensions from this perspective, has been to identify the structural similarities between hCS$_6$ on $\mathbb{PT}$ and CS$_4$ on $\mathbb{C}P^1\times \mathbb{R}^2$ and exploit them.

\medskip

In this article we propose a different alternative, based on the "categorical ladder = dimensional ladder" proposal \cite{Crane:1994ty, Baez:1995xq}. It states that higher-dimensional physics can be described by higher categorical structures, and that one can "climb" the dimensions by categorification. In brief, a category can be thought of as a collection of objects together with morphisms on them, such that certain "coherence conditions" are satisfied. Category theory itself can thus be understood as the study of \textit{structure} and the relations between them; for a comprehensive overview of category theory, one can consult \cite{maclane:71}. \textit{Categorification} is then, abstractly, a way to impart relations between structures in a coherent manner. Applying this idea to categories themselves gives rise to the notion of higher categories, which consist of objects, relations between these objects, relations between these relations, and so on. Indeed, as one climbs this categorical ladder, the structures that appear are suited, in each step, for describing higher dimensional data.

As expected, the theory of higher categories can become extremely complicated and abstract very rapidly (see eg. \cite{Lurie:2009}). Nevertheless, the idea of applying such intricate higher categorical tools to study higher-dimensional physics has been very popular in recent years \cite{Gaiotto:2014kfa,Kapustin:2013uxa,Wen:2019,Zhu:2019,Kong:2020wmn,Dubinkin:2020kxo,chen:2022,Delcamp:2023kew,Chen2z:2023}, and has led to many fruitful classification results. Examples include, but are not limited to, the study of higher-dimensional gapped topological phases in condensed matter theory \cite{Johnson-Freyd:2020,KongTianZhou:2020}, as well as the study of anomalies in quantum field theory (QFT) \cite{Freed:2014,Cordova:2018cvg,Benini_2019}. The striking power of functorial topological QFTs (TQFTs) since Atiyah-Segal \cite{Atiyah:1988}, in particular, to produce novel topological invariants \cite{lurie2008classification,Tillmann_2004,Freed_2021,Reutter:2020bav,Douglas:2018} from abstract categorical data cannot be overstated.

Higher category theory has also been used to study gauge theories in higher dimensions, leading to the notion of \emph{higher gauge theories}. These comprise a very rich and intricate system of gauge principles governed by categorical structures \cite{Baez:2002jn,Baez:2004in,Wockel2008Principal2A,Kim:2019owc}, which give rise to observables that are sensitive to the topology and geometry of surfaces, in analogy to the Wilson lines in the usual 3d Chern-Simons setup \cite{Reshetikhin:1991tc,Turaev:1992hq,Witten:1988hc,Guadagnini2008DeligneBeilinsonCA}. The existence of such theories, invites the following question:

\medskip

\begin{quote}
\centering
{\em What is the interplay between higher-gauge theories and higher-dimensional integrable systems?} 
\end{quote}

Our paper is dedicated to answering this question in a deep manner. Specifically, we will consider higher Chern-Simons theory based on a Lie 2-group/Lie group crossed module and its corresponding Lie 2-algebra \cite{Song_2023, Zucchini:2021bnn}. Schematically speaking, every object appearing in regular gauge theory is replaced by its corresponding higher \emph{counterpart}: the Lie group $G$ is replaced by a $2$-group $\mathsf{H}\rightarrow G$, the Lie algebra $\mathfrak{g}$ by a Lie 2-algebra $\fh\rightarrow \fg$, the connection $A$ by a $2$-connection $(A,B)$, and so on and so forth.  Crucially, introducing these higher categorical objects \emph{requires} increasing the dimension of the manifold where the theory is defined. Indeed, 2-Chern-Simons theory is defined over a four dimensional manifold $X$ and it is constructed so that its equations of motion correspond to the (higher) flatness of the 2-connection $(A,B)$ (see \S \ref{sec:higherchernsimons} for details). Our idea, is thus to explore in what sense this categorified flatness condition is related to integrability. 

We thus proceed \emph{a lá} Costello \cite{Costello:2019tri}. We write $X=\mathbb{R}\times Y$ for a $3$-fold $Y$, we complexify and compactify the copy of $\mathbb{R}$, and introduce a disorder operator $\omega$. This procedure defines a five-dimensional theory, which can be localised to a three-dimensional boundary theory on $Y$, with equations of motion equivalent to the higher flatness of the 2-connection. In this article we will focus on a specific choice of meromorphic $(1,0)$-form given by $\omega=z^{-1}\dr z$, inspired by the resulting 2d theory corresponding to this choice of $\omega$ in the context of CS$_4$: the 2d Wess-Zumino-Witten (WZW) model.

\subsubsection{Summary of results}
We explicitly derive the localized 3d theory on $Y$, whose fields are given by a smooth function $g \in C^{\infty}(Y)\otimes G$ and a $1$-form $\Theta \in \Omega^1(Y)\times \fh$ parameterizing the {broken} gauge symmetries at the poles of $\omega$. We prove in \textbf{Proposition \ref{extended2flat}} that the equations of motion of the theory  give rise to a flat $2$-connection $(L,H)$. \textbf{Theorem \ref{descending}} states that the remaining, unbroken symmetries of the 5d bulk 2-Chern-Simons theory on $X=Y\times\mathbb{C}P^1$ descend to residual global symmetries of the 3d theory on $Y$, in complete analogy with the CS$_4$ -- IFT$_2$ story \cite{Costello:2019tri}. We also show in \S \ref{thf} that we recover \textit{exactly} the Chern-Simons/matter coupling studied in the context of 3-brane/5-brane couplings in the A-model \cite{Aganagic:2017tvx}.

By making use of higher groupoids, we prove in \textbf{Theorem \ref{relbdy}} that the higher holonomies $(V,W)$ arising in our 3d theory are invariants of \textit{homotopies relative boundary}, and that they are consistent with the Eckmann-Hilton argument \cite{Gaiotto:2014kfa,Dubinkin:2020kxo}. This can be understood as an improved version of the "surface independence" notion discovered in \cite{Alvarez:1997ma}. Conserved quantities can then be obtained from $(V,W)$, labelled by \textit{categorical characters} \cite{Ganter:2006,Ganter:2014,davydov2017lagrangian,Sean:private,Huang:2024} of the Lie 2-group $\mathbb{G}$ and homotopy classes of surfaces relative boundary in $Y$. If certain technical conditions are met, the above results are fortified in \textbf{Theorem \ref{bordinv}} to show that the higher holonomies in fact define bordism invariants. These can be understood as a "global" version of the statement of conservation for the currents $J =(L,H)$; a more detailed explanation (with images) is given in \S \ref{holonomies}.

Finally, by analyzing the conserved Noether charges associated to the (infinitesimal) symmetries of the 3d action, we prove in \textbf{Theorem \ref{gradedbrackets}} that the currents $(L,H)$ form an infinite dimensinoal Lie 2-algebra. Moreover, by leveraging a transverse holomorphic foliation (THF) on $Y$, \textbf{Theorem \ref{affine2alg}} further illustrates that they form a {\it Lie 2-algebra extension} \cite{Angulo:2018}; we call these current algebras the \textbf{affine Lie 2-algebras of planar currents}. The existence of such an infinite dimensional symmetry algebra of our three-dimensional model is the origin of its integrable structure. 

It is worth emphasizing that both the existence of a spectral-parameter independent on-shell flat $2$-connection, and every result we have proven in \S \ref{dgaffinecurrents}, are direct higher homotopy analogues of the properties of the 2d WZW model: the anti/chiral anti/holomorphic currents satisfy a centrally extended current algebra, which is the underlying affine Lie algebra of the 2d WZW model \cite{KNIZHNIK198483}.

\subsubsection{Organization of the paper}

The paper is organized as follows. In section \S \ref{sec:higherchernsimons} we introduce the concepts of a Lie group crossed-module and it's corresponding Lie 2-algebra. This will allow us to define a homotopy Maurer-Cartan theory, of which 2-Chern-Simons theory is a particular example. We proceed with the introduction of the 2-Chern-Simons action and describe some of its properties. In \S \ref{sec:h2csth} we define \emph{holomorphic} 2-Chern-Simons theory by introducing the disorder operator $\omega$. We present in full detail the procedure of localisation to a three-dimensional boundary theory. In \S \ref{3dift} we study the family of three-dimensional actions obtained in \S \ref{sec:h2csth} and in particular, we show that the equations of motion are equivalent to the higher flatness of the $2$-connection $(L,H)$. In \S \ref{residuals} we analyze the symmetries of these theories in depth. In \S \ref{sec:holonomies} we discuss the integrable properties associated to the flatness of the $2$-connection $(L,H)$. Specifically , we show that their $2$-holonomies are conserved and homotopy invariant. Finally, in section \S \ref{dgaffinecurrents} we describe the current algebra associated to the symmetries of the $3$d theory studied in \S \ref{residuals}. This is a graded affine Lie $2$-algebra with a central extension, which we interpret as the higher version of the affine Lie algebra that appears in 2d WZW. We conclude the article with an outline of future interesting directions to explore.

\section{Higher Chern-Simons Theory}
\label{sec:higherchernsimons}

We will define higher Chern-Simons theory from the perspective of \emph{homotopy Maurer-Cartan} theories. As we will see, these theories are naturally described using higher structures, yet they all share a common ingredient: their equations of motion are flatness of some (eventually higher) connection. In this section we introduce the necessary mathematical background and definitions, together with the 2-Chern-Simons action and some of its properties.

The notion of a higher principal bundle with connection has been developed recently in the literature \cite{Baez:2004in,Martins:2006hx,chen:2022,Porst2008Strict2A}. The kinematical data attached to such higher-gauge theories is captured by the structure of \textit{homotopy $L_\infty$-algebras}, which are graded vector spaces equipped with higher $n$-nary Lie brackets \cite{Kim:2019owc,Baez:2004} satisfying Koszul identities \cite{Bai_2013}. In this paper, we will focus our attention on $L_2$-algebras/strict Lie 2-algebras and their corresponding Lie 2-groups, which we now define. 

\begin{definition}\label{2grpdef}
A \textbf{(Lie) group crossed-module} $\mathbb{G}=(\mathsf{H}\xrightarrow{\bar\mu_1}G,\rhd)$ consists of two (Lie) groups $\mathsf{H},G$, a (Lie) group homomorphism $\bar\mu_1:\mathsf{H}\rightarrow G$ and an (smooth) action $\rhd:G\rightarrow \operatorname{Aut}\mathsf{H}$ such that the following conditions
\begin{equation}
    \bar\mu_1(x\rhd y) = x\bar\mu_1(y)x^{-1}\,,\qquad (\bar\mu_1(y))\rhd y'=yy'y^{-1}\label{pfeif2}
\end{equation}
are satisfied for each $x\in G$ and $y,y'\in \mathsf{H}$.
\end{definition}
The infinitesimal approximation of a Lie 2-group gives rise to a Lie 2-algebra, which is a strict 2-term $L_\infty$-algebra \cite{Baez:2003fs}. More precisely, we have

\begin{definition}\label{lie2alg}
Let $\mathbb{K}$ denote a field of characteristic zero (such as $\mathbb{R}$ or $\mathbb{C}$). An $\boldsymbol{L_2}$-\textbf{algebra} $\mathfrak{G}=\fh\rightarrow \fg$ over $\mathbb{K}$ consist of two Lie algebras $\big(\fh,[-,-]_\fh\big)$ and $\big(\fg,[-,-]\big)$ over $\mathbb{K}$ and the tuple of maps,
\begin{equation*}
    \mu_1: \fh\rightarrow\fg,\qquad \mu_2: \fg\wedge\fh \rightarrow \fh\, ,\footnotemark
\end{equation*}
\footnotetext{Here $\wedge$ denotes the skew-symmetric tensor product and $\odot$ denotes the symmetric tensor product.} subject to the following conditions for each $\sfx, \sfx',\sfx'' \in\fg$ and $\sfy,\sfy'\in\fh$:
\begin{enumerate}
    \item The $\fg$-equivariance of $\mu_1$ and the Peiffer identity,
    \begin{equation*}
        \mu_1(\mu_2(\sfx,\sfy))=[\sfx,\mu_1(\sfy)]\,,\qquad \mu_2(\mu_1(\sfy),\sfy') =  [\sfy,\sfy']_\fh =- \mu_2(\mu_1(\sfy'),\sfy)\,.\label{pfeif1}
\end{equation*}
\noindent Note $\operatorname{ker}\mu_1\subset \fh$ is an Abelian ideal due to the Peiffer identity.

\item Graded Jacobi identities,
\begin{align*}
\label{ec:gradjacobid}
        0&=[\sfx,[\sfx',\sfx'']]+[\sfx',[\sfx'',\sfx]]+[\sfx'',[\sfx,\sfx']], \\
        0&= \mu_2(\sfx,\mu_2(\sfx',\sfy)) - \mu_2(\sfx',\mu_2(\sfx,\sfy)) - \mu_2([\sfx,\sfx'],\sfy)\,.
    \end{align*}
\end{enumerate}
Moreover, we call $\fG$ \textbf{balanced} \cite{Zucchini:2021bnn} if it is equipped with a graded symmetric non-degenerate bilinear form $\langle -,-\rangle: \fg\odot\fh \rightarrow \mathbb{K}$ which is invariant
\begin{equation}
    \langle \sfx,\mu_2(\sfx',\sfy)\rangle = \langle [\sfx,\sfx'],\sfy\rangle\,,\qquad \langle \mu_1(\sfy),\sfy'\rangle = \langle \mu_1(\sfy'),\sfy\rangle\label{inv}
\end{equation}
for each $\sfx,\sfx'\in\fg$ and $\sfy,\sfy'\in \fh$.
\end{definition}
\noindent The unary and binary brackets $\mu_1,\mu_2$ are integrated respectively to a Lie group map $\bar\mu_1:\mathsf{H}\rightarrow G$ and an action $G\rightarrow\operatorname{Der}\fh$ of $G$ on the degree-(-1) Lie algebra $\fh$. As an abuse of notation, we shall denote this action also by $\rhd$.\footnote{This action $\rhd:G\rightarrow\operatorname{Der}\fh$ is part of the 2-adjoint action of $\mathbb{G}$ on its own Lie 2-algebra $\fG=\operatorname{Lie}\mathbb{G}$ \cite{Bai_2013,Angulo:2018,chen:2022}.}

Based on the structure of a balanced Lie 2-algebra $\big(\fG,\mu_1,\mu_2,\langle -,-\rangle\big)$, we construct 2-Chern-Simons theory as a homotopy Maurer-Cartan theory from the Batalin–Vilkovisky (BV) formalism \cite{Jurco:2018sby}. We will also describe the relation with the derived superfield formalism considered in \cite{Zucchini:2021bnn}. 

\subsection{Homotopy Maurer-Cartan Theory}\label{hMC}

Three-dimensional Chern-Simons theory can be described as the simplest example of what is known as a \emph{homotopy Maurer-Cartan theory}. These were first introduced in \cite{Zwiebach:1992ie} in the context of string field theory and are structured upon $L_\infty$-algebras. Indeed, CS$_3$ is a homotopy Maurer-Cartan theory for an $L_1$-algebra, that is, a plain Lie algebra. We will therefore define 2-Chern-Simons theory by considering the simplest graded structure within the $L_\infty$-algebra, namely, an $L_2$-algebra. 

We thus begin by turning the Lie $2$-algebra $\fG=\fh\rightarrow\fg$ into a differential graded (dg) algebra by tensoring $\fG$ with the de Rham complex $\Omega^\ast(X)$ over a space $X$. This gives rise to a Lie 2-algebra with the graded components
\begin{equation*}
    \cL_n = (\Omega^\bullet(X)\otimes\fG)_n = \bigoplus_{i+j=n} \Omega^i(X)\otimes \fg_j,\qquad \fg_{-1}=\fh,~\fg_0=\fg,
\end{equation*}
together with the differential $\ell_1 = \mathrm{d}-\mu_1$ and $\ell_n = \mu_n\otimes\wedge^n$ for all $n\geq 2$.

\begin{definition}

An element $\mathcal{A}\in\cL_1$ is a \textbf{Maurer-Cartan element} if its curvature vanishes, 
\begin{equation}
\label{ec:curvature}
   \mathscr{F}(\cA) = \sum_{n=1}^2 \frac{1}{n!}\ell_n(\mathcal{A},\dots,\mathcal{A})=0 \,.
\end{equation}
We denote the space of Maurer-Cartan elements by $\mathsf{MC}\subset\Omega^\bullet(X)\otimes \fG$.

\end{definition}

Let us see what this means by writing out these objects explicitly. An element $\cA \in \cL_1$, by definition, is of the form $\mathcal{A}=(A,B)$ where $A\in \Omega^1(X)\otimes\fg$ and $B \in \Omega^2(X)\otimes\fh$. The curvature \eqref{ec:curvature} is thus
\begin{align*}
   \mathscr{F}(\mathcal{A})=  \ell_1(\mathcal{A}) + \frac{1}{2}\ell_2(\mathcal{A},\mathcal{A})&= \dr A - \mu_1(B) + \frac{1}{2}[A, A]  + \dr B +\mu_2(A, B)\,.
\end{align*}
Organizing this quantity by degree, we see that we obtain two equations
\begin{equation}
\label{ec:2flatnessss}
    \mathcal{F}=\dr A + \tfrac{1}{2}[A,A] - \mu_1(B) =0\,,\quad \mathcal{G}=\dr B + \mu_2(A,B) =0\,,
\end{equation}
where $\cF$ and $\cG$ are known as the \textbf{fake-curvature} and \textbf{2-curvature} respectively. Equation \eqref{ec:2flatnessss} then corresponds to the \textbf{fake-/2-flatness conditions} \cite{Kim:2019owc,Baez:2004in,Martins:2006hx,Chen:2022hct}.

\medskip

\begin{remark}
    In the case of CS$_3$, a degree $1$ element $\mathcal{A} \in \cL_1$ is simply a $\fg$-valued $1$-form $A$. In particular, it will be a Maurer-Cartan element if it's curvature $F(A) = \mathrm{d}A+\tfrac{1}{2}[A,A]$ vanishes. 
\end{remark}

\medskip

The goal is to define an action whose variational principle is associated to the zero-curvature condition; in other words, the minimal locus of the action consists only of Maurer-Cartan elements. This brings us to the following definition

\begin{definition}

We define the \textbf{2-Maurer-Cartan action} as the action functional

\begin{equation}
\label{ec:2mcaction}
    S_\text{2MC}[\mathcal{A}]\coloneqq \sum_{m=1}^2\frac{1}{(m+1)!}\int_X \langle \mathcal{A},\ell_m(\mathcal{A},\dots,\mathcal{A})\rangle \,.
\end{equation}
In particular, for an arbitrary variation $\delta \cA$ we find that $\delta S_{\mathrm{2MC}}[\cA]=0$ if and only if $\mathscr{F}(\mathcal{A}) =0$. 
    
\end{definition}

\begin{remark}
    Note that $\cA=(A,B) \in \cL_1$  consists of a $\mathfrak{g}$-valued $1$-form $A$ and an $\mathfrak{h}$-valued $2$-form $B$. Similarly $\mathscr{F}=(\mathcal{F},\mathcal{G}) \in \cL_2$ consists of a $\fg$-valued $2$-form $\cF$ and an $\fh$-valued $3$-form $\cG$. Since the pairing on the balanced Lie algebra $\mathfrak{G}$ pairs elements of $\fg$ with those of $\fh$, we conclude that the integrand in \eqref{ec:2mcaction} is a $4$-form, and thus $X$ must be a $4$-manifold. This explains why dg Lie algebras (dgla's), at the very least, are {\it required} to define a 4d analogue of Chern-Simons theory.
\end{remark}
Going back to \eqref{ec:2mcaction}, we make use of the invariance property \eqref{inv} to bring the 2-Maurer-Cartan action into the form 
\begin{align}
    S_\text{2CS}[\mathcal{A}] = \int_X \langle \dr A + \tfrac{1}{2}[A, A]- \mu_1(B), B\rangle \,,\label{2chernsimons}
\end{align}
which is what we shall refer to in the rest of the paper as \textbf{2-Chern-Simons theory} (2CS) \cite{Zucchini:2021bnn}. This theory is also called the "4d BF-BB theory" in some literature \cite{Baez:1995ph,Geiller:2022fyg,Girelli:2021zmt}. By construction, the equations of motion of \eqref{2chernsimons} will be precisely $\mathscr{F}(\mathcal{A})=0$. Explicitly, it is simple to check that 
\begin{align*}
    \delta_B S_\text{2CS}[A,B]=0 \,: &\quad \mathcal{F}=\dr A + \tfrac{1}{2}[A,A] - \mu_1(B)=0\,,\\
    \delta_A S_\text{2CS}[A,B]=0\,:&\quad \mathcal{G}=\dr B + \mu_2(A, B) = 0\,.
\end{align*}

\subsubsection{Gauge Symmetries of 2CS Theory}
\label{sec:gsof2CS}

As in any gauge theory, we will be interested in gauge symmetries of the action. These are given by a higher analogue of usual gauge transformations and are characterized by a smooth map $h \in C^{\infty}(X,G)$ and a $1$-form $\Gamma \in \Omega^{1}(X)\otimes \mathfrak{h}$ which act on the gauge fields as \cite{Baez:2004in}
\begin{equation}
\begin{aligned}
\label{ec:2gaugetr1}
    &A \mapsto A^{(h,\Gamma)}=\operatorname{Ad}_h^{-1}A + h^{-1}\mathrm{d}h + \mu_1(\Gamma) \\
    &B \mapsto B^{(h,\Gamma)}= h^{-1}\rhd B + \mathrm{d}\Gamma + \mu_2(A^{(h,\Gamma)}, \Gamma) - \tfrac{1}{2}[\Gamma,\Gamma] \,.    
\end{aligned}
\end{equation}
We call the simultaneous transformations of the gauge fields  a  $\boldsymbol{2}$-\textbf{gauge transformation}. Note that the above expressions correspond to finite gauge transformations of the gauge fields, which in this article will play a crucial role. In the literature people often consider only infinitesimal gauge transformations, which can be obtained by a formal expansion $h \sim 1+\alpha$ and neglecting terms quadratic in $\alpha$ and $\Gamma$, see for instance \textbf{Prop. 2.9} of \cite{Baez:2004in}. With a simple computation, it can be shown that under these gauge transformations, the fake-curvature and the $2$-curvature transform covariantly, 
\begin{align*}
    \mathcal{F}\big(A^{(h,\Gamma)},B^{(h,\Gamma)}\big) = \operatorname{Ad}_h^{-1}\mathcal{F}(A,B)\,,\qquad \mathcal{G}\big(A^{(h,\Gamma)},B^{(h,\Gamma)}\big) = h^{-1}\rhd \mathcal{G}(A,B) + \mu_2(\mathcal{F}(A,B), \Gamma) \,,
\end{align*}
whence the transformations \eqref{ec:2gaugetr1} leave the subspace $\mathsf{MC}\subset\Omega^\bullet(X)\otimes\fG$ of Maurer-Cartan elements invariant.

\begin{remark}\label{homtrans}
Notice \eqref{ec:2gaugetr1} includes a class of transformations that translates the 1-form field. As such, the value of $A$ is only defined modulo the image of $\mu_1$ up to 2-gauge redundancy. However, to properly treat this seemingly large gauge freedom in the Batalin–Vilkovisky (BV) formalism, one would have to make use of {\it homotopy transfer}. Given the BV(-BRST) $L_\infty$-algebra $\cL$ encoding all of the fields in the theory, it was argued in \cite{Arvanitakis:2020rrk} that integrating out degrees of freedom is equivalent to projecting onto its cohomology $H(\cL)$. If $H(\cL)\neq0$ is not trivial, then the homotopy transfer theorem states that higher homotopy brackets --- in particular a trilinear tertiary "homotopy map" $\tilde\ell_3$ \cite{Chen:2012gz} --- will be induced on $H(\cL)$ through this projection \cite{stasheff2018,Arvanitakis:2020rrk}.

We shall explain the mathematics behind homotopy transfer in \S \ref{sec:homtransf}. But the upshot is that integrating out/gauging this shift symmetry $A\rightarrow A+\mu_1\Gamma$ away will in general lead to a non-trivial deformation of the theory, and the 2-Chern-Simons theory will end up differing perturbatively from the 4d BF theory. Though if $\mu_1$ \textit{were} invertible, then $A$ can indeed be completely gauged away. In this case, the cohomology of the BV-BRST complex $\cL$ would be trivial, hence performing a homotopy transfer to project away the field $A$ leads to a trivial theory, at least perturbatively. In this case, the 2CS theory becomes the so-called 4d BF-BB theory, which is conjectured to coincide with the Crane-Yetter-Broda TQFT \cite{Baez:1995ph} that hosts only non-perturbative degrees-of-freedom. It is sensitive only to the ($G$-)bordism class of the underlying 4-manifold. 
\end{remark}

We are interested in the variation of 2CS theory under a gauge transformation, which we describe in the following proposition.

\begin{proposition}
\label{prop:gvar}
Under the 2-gauge transformations \eqref{ec:2gaugetr1} the 2CS action transforms as
\begin{equation}
\label{ec:actionundergt}
\begin{aligned}
S_{\mathrm{2CS}}\big[A^{(h,\Gamma)},B^{(h,\Gamma)}\big] &= S_{\mathrm{2CS}}[A,B] \\
&\hspace{.7cm}+\int_X\mathrm{d}\big[\langle \mathrm{Ad}_h^{-1}F(A) ,\Gamma\rangle + \tfrac{1}{2} \langle\mathrm{Ad}_h^{-1}A+h^{-1}\mathrm{d}h,[\Gamma,\Gamma]\rangle + \tfrac{1}{2}L_{\mathrm{CS}}(\Gamma)\big]    
\end{aligned}
\end{equation}
where
\begin{equation}
\label{ec:basicCS}
  L_{\mathrm{CS}}(\Gamma) = \langle\mu_1(\Gamma),\dr\Gamma+\tfrac{1}{3}[\Gamma,\Gamma]\rangle\,. 
\end{equation}
  
\end{proposition}
\begin{proof}
Let us begin by simplifying the expressions that appear in the Lagrangian $L_{\mathrm{2CS}}\big[A^{(h,\Gamma)},B^{(h,\Gamma)}\big]$ conveniently. First, we note that
\begin{equation}
    F\big(A^{(h,\Gamma)}\big) = F(A^h)+F(\Gamma)+[A^h,\mu_1(\Gamma)]
\end{equation}
where $A^h = h^{-1}Ah+h^{-1}\dr h$. On the other hand, we can write
\begin{equation}
    B^{(h,\Gamma)}=h^{-1}\rhd B + F(\Gamma)+\mu_2(A^h,\Gamma)\,.
\end{equation}
Thus, we find that
\begin{equation}
    F\big(A^{(h,\Gamma)}\big)-\tfrac{1}{2}\mu_1\big(B^{(h,\Gamma)}\big) = \mathrm{Ad}_h^{-1}\big[F(A)-\tfrac{1}{2}\mu_1(B)\big] + \tfrac{1}{2}\big(\mu_1(F(\Gamma))+[A^h,\mu_1\Gamma]\big)\,,
\end{equation}
where we have used the covariance $F(A^h)=\mathrm{Ad}_h^{-1}F(A)$ of the usual curvature. Hence we have
\begin{equation}
\begin{aligned}    L_{\mathrm{2CS}}\big[A^{(h,\Gamma)},B^{(h,\Gamma)}\big]= L_{\mathrm{2CS}}[A,B] &+ \langle \mathrm{Ad}_h^{-1}\big[F(A)-\tfrac{1}{2}\mu_1(B)\big], F(\Gamma)+\mu_2(A^h,\Gamma)\rangle \\
& \hspace{1cm}+\tfrac{1}{2}\langle \mathrm{Ad}_h^{-1}\mu_1(B)+ F(\Gamma)+\mu_2(A^h,\Gamma),F(\Gamma)+\mu_2(A^h,\Gamma)\rangle \,.
\end{aligned}
\end{equation}
The terms containing $B$ cancel each other. Expanding the last term and doing some algebraic manipulations we get
\begin{equation}
\label{ec:formula1}
\begin{aligned}
    \tfrac{1}{2}\langle F(\Gamma)+\mu_2(A^h,\Gamma),F(\Gamma) &+\mu_2(A^h,\Gamma)\rangle = \\
    &\tfrac{1}{2}\dr L_{\mathrm{CS}}(\Gamma)+\tfrac{1}{2}\langle [A^h,\mu_1(\Gamma)],\mu_2(A^h,\Gamma)\rangle + \langle \dr \Gamma,\mu_2(A^h,\Gamma)\rangle\,,
\end{aligned}
\end{equation}
where we have used the Jacobi identity to eliminate terms cubic in $\Gamma$ and $L_{\mathrm{CS}}(\Gamma)$ is given in \eqref{ec:basicCS}. The remaining term is after some manipulations,
\begin{equation}
\label{ec:formula2}
\begin{aligned}
    \langle \mathrm{Ad}_h^{-1}F(A),F(\Gamma)+\mu_2(A,\Gamma)\rangle 
    &= \dr\langle A^h,\tfrac{1}{2}[\Gamma,\Gamma]\rangle +  \dr\langle F(A^{h}),\Gamma \rangle -\langle \dr \Gamma,\mu_2(A^h,\Gamma)\rangle   \\
    & \hspace{5.5cm} +\tfrac{1}{4}\langle [A^h,A^h],[\Gamma,\Gamma]\rangle
\end{aligned}
\end{equation}
where we have used the Bianchi identity $\dr F(A^h)+[A^h,F(A^h)]=0$. Putting together \eqref{ec:formula1} and \eqref{ec:formula2} we see that the non-exact terms cancel using the graded Jacobi identity, and we arrive to the desired result.

\end{proof}

    Note that the gauge variation is a total boundary term. In other words, violation to gauge non-invariance of 2-Chern-Simons theory is completely hologrpahic, in contrast with CS$_3$ whose gauge variation contains the well-known Wess-Zumino-Witten term, which must be defined in the $3$d bulk.

\subsubsection{Secondary gauge transformations}
\label{secondarygaugetransf}
A special feature of 2-gauge theory in general is that there are redundancies in the \textit{2-gauge transformations} $(h,\Gamma)$ \eqref{ec:2gaugetr1} itself \cite{chen:2022,Baez:2004,Soncini:2014ara,Kapustin:2013uxa}. This can be attributed to the fact that the gauge symmetries in \eqref{2chernsimons} form in actuality a 2-group, whose objects are the gauge parameters $(h,\Gamma)$.\footnote{This is an example of a categorical {\it gauge} symmetry, which is distinct from the categorical global symmetries that have been discussed recently in various places, eg. \cite{Gaiotto:2014kfa,Delcamp:2023kew}.} We call these \textbf{secondary gauge transformations}, which can be formalized in a categorical context as in \cite{Baez:2004,Baez:2004in,Soncini:2014ara}.

More explicitly, these secondary gauge transformations are in essence deformations/"morphisms" $(h,\Gamma)\rightarrow (h',\Gamma')$ such that $(h',\Gamma')$ remains a 2-gauge transformation as in \eqref{ec:2gaugetr1}, 
\begin{equation}\label{ec:2ndgt}
    (A^{(h,\Gamma)},B^{(h,\Gamma)}) \rightarrow (A^{(h',\Gamma')},B^{(h',\Gamma')}) 
\end{equation}
In full generality, these morphisms are labelled by a pair of fields (the so-called "ghosts-of-ghosts") $\varphi\in C^\infty(X)\otimes\mathsf{H}, E\in\Omega^1(X)\otimes\fh.$ Essentially, the 0-form $\varphi$ induces a translation $h\rightarrow \bar\mu_1(\varphi)h$ along the base while the 1-form $E$ induces a translation $\Gamma\rightarrow\Gamma+ E$ along the faces. We leave the interested reader to find the detailed differential constraints these fields must satisfy in the relevant literature \cite{Soncini:2014ara,Baez:2004in}. 

As we are in the context of a {\it strict} 2-gauge theory, it is safe to not make crucial use of these data. For {\it weak/semistrict} 2-gauge theories, on the other hand, it has been noted that they in fact play a key role \cite{Baez:2004,Soncini:2014ara}, even in the case where the 2-group is finite \cite{Kapustin:2013uxa}. Though this requires one to always be on-shell of the fake-flatness condition $F - \mu_1B=0$ \cite{Kim:2019owc}, which is undesirable in our considerations.    

\subsection{The derived superfield formulation}
We pause here momentarily to comment that the above higher-gauge theory computations can be reformulated using the general operational framework developed in \cite{Zucchini:2019pbv,Zucchini:2019rpp,Zucchini:2021bnn}. This formalism makes use of \emph{derived} structures, and provides a direct link with ordinary gauge theory. We will content ourselves with presenting the key expressions, the details can be found in \cite{Zucchini:2021bnn}.

Given the Lie $2$-group $\mathbb{G}$ we define the derived Lie $2$-group $\mathsf{D}\mathbb{G}$ as the set of maps $\mathbb{R}[1]\rightarrow \mathbb{G}$ labelled by the semidirect product $G \ltimes \fh$,
\begin{equation}
    \alpha \mapsto (x,e^{\alpha\cdot \sfy})\,, \qquad \alpha\in \mathbb{R}\,,\qquad x\in G,~\sfy\in\fh \,.
\end{equation}
In particular, $\mathsf{D}\mathbb{G}$ is a graded Lie group; it shall make appearance again in \S \ref{dgaffinecurrents}. Similarly, one can also define the corresponding derived Lie 2-algebra $\mathsf{D}\fG$ as the space of maps $\mathbb{R}[1]\rightarrow \fg\ltimes \fh$.
 
We may now describe our higher connection and higher gauge transformations as elements valued in $\mathsf{D}\mathfrak G$ and $\mathsf{D}\mathbb{G}$ respectively. Explicitly, for the $2$-connection we have a degree-$1$ polyform
\begin{equation}
    \cA (\alpha)=(A,\alpha \cdot B) \in (\Omega^\bullet(X)\otimes \mathsf{D}\mathfrak G)_1 \,,
\end{equation}
whereas for the $2$-gauge transformation, a degree-$0$ polyform
\begin{equation}
    U(\alpha)=(h, e^{\alpha\cdot\Gamma}) \in (\Omega^\bullet(X)\otimes \mathsf{D}\mathbb{G})_0 \,.
\end{equation}
Note $\Gamma\in\Omega^1(X)\otimes\fh$, and $e^{\alpha\cdot\Gamma} = 1 + \alpha\cdot\Gamma + \frac{1}{2}[\alpha\cdot\Gamma,\alpha\cdot\Gamma]$ terminates due to the Jacobi identity \cite{Baez:2004in}. The tuples $(A,B),(h,\Gamma)$ that we have used previously can therefore be understood respectively as the labels for these derived polyforms $\mathcal{A},U$. 

With these expressions at hand, we may write a general $2$-gauge transformation \eqref{ec:2gaugetr1} in the compact form
\begin{equation}
\label{ec:superfieldgauge}
    \mathcal{A} \to \mathcal{A}^U =~ _2\operatorname{Ad}_U^{-1}\mathcal{A} + U^{-1}\tilde{\mathrm{d}} U\,
\end{equation}
as a function of the derived parameter $\alpha$, where the expressions for the $2$-adjoint action $_2\operatorname{Ad}$  and the Maurer-Cartan form $U^{-1}\tilde{\mathrm{d}}U$ are given, respectively, in eqs. (3.2.8) and (3.2.14) of \cite{Zucchini:2021bnn}, and $\tilde{\mathrm{d}}= \mathrm{d}-\mu_1\tfrac{\partial}{\partial \alpha}$. Finally, we may induce a pairing $(-,-)$ on $\mathsf{D}\mathfrak G$ from the pairing $\langle -,- \rangle$ on $\mathfrak G$ (see eq. (3.2.15) in \cite{Zucchini:2021bnn}) to write the four-dimensional $2$CS action \eqref{2chernsimons} as
\begin{equation}
    S_{2\mathrm{CS}}[\cA] = \int_X (\cA,\tilde{\mathrm{d}}\cA + \tfrac{1}{3}[\cA,\cA]) \,.
\end{equation}
As such, the derived superfield formalism allows us to rewrite expressions in higher gauge theory as analogues of those in ordinary gauge theory.

\medskip

In addition to the compactness of this derived operational framework, it also introduces an \textit{internal} grading attached to the graded field algebra: in contrast to ordinary gauge theory, derived gauge theory has two gradings \cite{Zucchini:2019pbv, Zucchini:2019rpp}.
This additional grading endows both connections and gauge transformations with \emph{ghostlike} partners, in a way similar to the AKSZ formulation of BV theory \cite{Alexandrov:1995kv,Ikeda:2012pv,Calaque:2021sgp}. The derived formalism therefore gives a useful and unifying description of higher gauge theory, also known as "non-Abelian bundle gerbes with connection" \cite{Baez:2004in,Nikolaus2011FOUREV,Schreiber:2013pra}.

\section{Holomorphic 2-Chern-Simons Theory}
\label{sec:h2csth}

Having introduced 2-Chern-Simons theory together with its equations of motion and its behaviour under gauge transformations, we are in conditions to define its holomorphic variant. Following \cite{Costello:2019tri}, we take our $4$-manifold $X=\mathbb{R} \times Y$ and we complexify and compactify the copy of $\mathbb{R}$ to $\mathbb{C}P^1$. Taking coordinates $z,\bar z$ on $\mathbb{C}P^1$ and $x^i$ with $i=1,2,3$ on $Y$, we define \textbf{holomorphic 2-Chern-Simons theory} (h2CS) by the action functional 
\begin{equation}
\label{ec:h2cs}
    S_\text{h2CS}[A,B] = \frac{1}{2 \pi i} \int_X \omega \wedge \langle F(A) - \tfrac{1}{2}\mu_1(B), B\rangle
\end{equation}
where $\omega \in \Omega^{(1,0)}(X)$ is a meromorphic $(1,0)$-form on $\mathbb{C}P^1$. In this article we will make a particular choice of $\omega$ given by 
\begin{equation}
    \omega = \frac{\mathrm{d}z}{z} \,.
\end{equation}
Note that $\omega$ is nowhere vanishing and has simple poles at $z=0$ and $z=\infty$. For future use, we write down the gauge fields in components as 
\begin{align}
    & A= A_{\bar z} \mathrm{d}\bar z + A_i \mathrm{d}x^i \\
    & B= B_{i\bar z}\,\mathrm{d}x^i \wedge \mathrm{d}\bar z + B_{ij}\,\mathrm{d}x^i \wedge \mathrm{d}x^j \,,
\end{align}
where we have ignored the $z$-components of the gauge fields provided $\omega \in \Omega^{(1,0)}(X)$.  

Without any additional constraints, the action \eqref{ec:h2cs} is not well defined: although the singularities in $\omega$ do not break the local integrability of the action (see \textbf{Lemma 2.1} in \cite{Benini:2020skc}), boundary terms arise under an arbitrary field variation. Indeed, we have
\begin{equation}
\label{ec:fullvar}
    \delta S_{\mathrm{h2CS}} = \frac{1}{2\pi i}\int_X  \omega\wedge \langle \delta \mathcal{F},B \rangle - \omega\wedge \langle \delta A, \mathcal{G} \rangle + \mathrm{d}\omega \wedge \langle \delta A,B \rangle.
\end{equation}
The first two terms are bulk terms which give rise to the equations of motion. The third term, however, is a distribution localized at the poles of $\omega$. Indeed, we can integrate along $\mathbb{C}P^1$, to find (see \textbf{Lemma 2.2} in \cite{Delduc:2019whp})
\begin{equation}
\label{ec:bdaryvar}
    \frac{1}{2\pi i}\int_{\mathbb{C}P^1\times Y} \mathrm{d}\omega \wedge \langle \delta A,B\rangle = \int_Y \iota_Y^*\langle \delta A,B\rangle|_{z=0}-\iota_Y^*\langle \delta A,B\rangle_{z=\infty}\,,
\end{equation}
where $\iota_Y: Y\hookrightarrow X$ is an embedding of the 3-manifold $Y$ into $X=\mathbb{C}P^1 \times Y$.
Thus, to have a well defined action principle, namely $\delta S_{\text{h2CS}}=0$, we must impose boundary conditions on the gauge fields $A$ and $B$ at $z=0,\infty$. Throughout the bulk of the article we will make different choices of boundary conditions depending on the additional structure we introduce on $Y$. However, the main example to which we will consistently come back to simplify discussions is given by the choice
\begin{equation}
\begin{aligned}
    \label{ec:bconditions1}
    & A_1|_{z=0}=A_2|_{z=0}=B_{12}|_{z=0}=0 \,,\\
    & A_3|_{z=\infty}=B_{13}|_{z=\infty}=B_{23}|_{z=\infty}=0 \,,
\end{aligned} 
\end{equation}
and we restrict to variations satisfying the same boundary conditions. It is immediate to verify that this choice of boundary conditions makes \eqref{ec:bdaryvar} vanish.

\begin{remark}
    The quantity \eqref{ec:bdaryvar} defines the {\it symplectic form} $\varpi_\text{blk}$ on the on-shell fields (i.e. Maurer-Cartan elements, c.f. \S \ref{hMC}) $\mathsf{MC}_\text{blk}\subset \iota_Y^* \Omega(X)\otimes \fG$ of the bulk $5$d theory. In particular, a choice of boundary conditions such that $\delta S_{\mathrm{h2CS}}=0$ is then equivalent to a choice of a Lagrangian subspace $\mathcal{L}\subset \mathsf{MC}_\text{blk}$ such that
\begin{equation*}
    \varpi_\text{blk}\vert_\mathcal{L} =0.
\end{equation*}
Indeed, fields $(\iota_Y^*A,\iota_Y^*B)\in\mathcal{L}$ by definition will satisfy $\delta S_\text{2CS}[A,B] = 0$. The variational quantities such as $\delta A,\delta B$, are naturally interpreted as covectors on {\it field space}. This is made precise in the covariant phase space formalism \cite{Julia:2002df,Geiller:2022fyg}.
\end{remark}


In the above language, the boundary conditions \eqref{ec:bconditions1} correspond to the choice of Lagrangian subspace
\begin{equation*}
    \mathcal{L} = \operatorname{Span}\{A_3,B_{23},B_{13}\}\vert_{z=0} \cup \operatorname{Span}\{A_1,A_2,B_{12}\}\vert_{z=\infty}\,,
\end{equation*}
which is simply given by the components of the fields that survive at the respective poles $z=0,z=\infty$. We shall see in \S \ref{3dift} that this example is in fact part of a covariant family of such "chiral" Lagrangian subspaces $\mathcal{L}_\ell\subset\mathsf{MC}_\text{blk}$, which are parameterized by a unit vector $\ell\in S^2$ called the \textbf{chirality}. Moreover, in \S \ref{covbcs} we shall see that, if additional structure such as a \emph{transverse holormophic foliation} is imposed on $Y$, the different choices of the chirality vector $\ell$ can lead to very different three dimensional actions. 

\medskip

Going back to our h2CS action, the bulk equations of motion are readily obtained from \eqref{ec:fullvar}. Indeed, they imply the Maurer-Cartan condition --- namely the fake- and 2-flatness, see \S \ref{hMC} --- of the 2-connection $\mathcal{A}=(A,B)$. In components, these are 
\begin{align}
    \label{ec:eom1}
   & \partial_{\Bar{z}}A_i - \partial_{i}A_{\bar z}+[A_{\bar z},A_i]-\mu_1(B_{\Bar{z}i})=0 \\
   \label{ec:eom2}
    & \partial_j A_i - \partial_i A_j +[A_j,A_i]-\mu_1(B_{ij})=0  \\
    \label{ec:eom3}
    & \partial_{i}B_{j\Bar{z}}+ \partial_{\bar z}B_{ij}+\mu_2(A_i,B_{j \Bar{z}})+\mu_2(A_{\bar z},B_{ij}) =0 \\
    \label{ec:eom4}
    & \partial_{i}B_{jk}+\mu_2(A_i,B_{jk}) =0
\end{align}
where $i,j,k\in \{1,2,3\}$ and where we have explicitly separated the $\bar z$ component equations for later convenience. Preempting what will happen after localisation to a three-dimensional boundary theory, equations \eqref{ec:eom1} and \eqref{ec:eom3} will imply that $A$ and $B$ are holomorphic on $\mathbb{C}P^1$ and therefore constant along $\mathbb{C}P^1$. On the other hand, equations \eqref{ec:eom2} and \eqref{ec:eom4} will result on fake flatness and 2-flatness of the three-dimensional localised fields.

\subsection{Gauge Symmetries of \texorpdfstring{$\mathrm{h2CS}$}{V}}
\label{sec:gaugesym5d}

As discussed in \S \ref{sec:gsof2CS}, 2-Chern-Simons theory is invariant (modulo boundary terms) under an arbitrary 2-gauge transformation. In the holomorphic case, the introduction of the meromorphic $1$-form $\omega$ breaks this gauge symmetry, in the sense that only a subset of gauge transformations leave \eqref{ec:h2cs} invariant. In particular, it is precisely the breaking of this gauge symmetry which gives rise to the edge-modes on the boundary, which will be the fields of our three-dimensional theory. Moreover, the transformations which remain symmetries of the 5d action will descend to symmetries of the 3d boundary theory.

This is entirely analogous to what happens in the case of 4d Chern-Simons theory, as well as in holomorphic Chern-Simons theory on Twistor space. Notably, in these cases, the gauge transformations which remain symmetries of the action after the introduction of $\omega$ are those which preserve the boundary conditions imposed on the gauge fields. We will show that this is not true any more in h2CS theory. 

Indeed, in proposition \ref{prop:gvar} we have shown that under the 2-gauge transformation \eqref{ec:2gaugetr1} the variation is a total boundary term 
\begin{equation}
\label{ec:omega}
    \Omega= \langle \mathrm{Ad}_h^{-1}F(A) ,\Gamma\rangle + \tfrac{1}{2} \langle\mathrm{Ad}_h^{-1}A+h^{-1}\mathrm{d}h,[\Gamma,\Gamma]\rangle + \tfrac{1}{2}L_{\mathrm{CS}}(\Gamma)\,.
\end{equation}
With the introduction of the meromorphic $1$-form we have that under a finite gauge transformation
\begin{equation}
\begin{split}
        S_{\text{h2CS}}\big[A^{(h,\Gamma)},B^{(h,\Gamma)}\big]
        &=S_{\text{h2CS}}[A,B]+\frac{1}{2\pi i}\int_X \omega \wedge \mathrm{d}\Omega[A,B,h,\Gamma] \\
        &= S_{\text{h2CS}}[A,B] + \int_{Y}\iota_Y^*\Omega[A,B,h,\Gamma]|_{z=0}-\iota_Y^*\Omega[A,B,h,\Gamma]|_{z=\infty}\,,
\end{split}
\end{equation}
where in the second line we have integrated by parts to get the distribution $\mathrm{d}\omega$ and then we have integrated along $\mathbb{C}P^1$ to localize at the poles. Let us consider the boundary conditions \eqref{ec:bconditions1} and restrict to gauge transformations which preserve these boundary conditions. In other words, we constraint our gauge parameters $(h,\Gamma)$ by requiring that $A^{(h,\Gamma)}$ and $B^{(h,\Gamma)}$ satisfy the same boundary conditions than $A$ and $B$. From \eqref{ec:2gaugetr1} we find the set of constraints
\begin{align}
   z=0: \quad & h^{-1}\partial_1h + \mu_1(\Gamma_1) = h^{-1}\partial_2 h + \mu_1(\Gamma_2) = \partial_1 \Gamma_2 - \partial_2 \Gamma_1 -[\Gamma_1,\Gamma_2] =0  \label{ec:gtbczero} \\    z=\infty:\quad & h^{-1}\partial_3h  + \mu_1(\Gamma_3)=\partial_i \Gamma_3 - \partial_3 \Gamma_i + \mu_2(A^{(h,\Gamma)}_i,\Gamma_3) - [\Gamma_i,\Gamma_3] =0 \,,\quad i=1,2\,. \label{ec:gtbcinfty}
\end{align}
Writing down $\iota_Y^*\Omega$ explicitly, using the boundary conditions \eqref{ec:bconditions1} and the constraints \eqref{ec:gtbczero} and \eqref{ec:gtbcinfty} we find
\begin{align}
     \int_{Y}\iota_Y^*\Omega[A,B,h,\Gamma]|_{z=0}&-\iota_Y^*\Omega[A,B,h,\Gamma]|_{z=\infty} =\int_Y \mathrm{vol}_3 \big[ \langle \mu_1 (\partial_3 \Gamma_1),\Gamma_2 \rangle |_{z=0} \nonumber\\
     &\qquad -\big(\langle A_2,h\rhd[\Gamma_1,\Gamma_3]\rangle-\langle [A_1,A_2],h\rhd \Gamma_3\rangle -\langle \mu(\Gamma_1),\partial_3 \Gamma_2\rangle\big)|_{z=\infty}\big]\,.\label{ec:deltaSgamma1}
\end{align}
In other words, we see that preserving the boundary conditions is not enough to attain the symmetry. This is in fact a surprising result, because of equation \eqref{ec:bdaryvar}. Indeed there we have imposed the same boundary conditions on the gauge fields and its variations and this led to the vanishing of $\delta S$ on-shell. In particular, if we take the arbitrary variations $\delta A,\delta B$ to be infinitesimal gauge transformations then this would imply that \eqref{ec:deltaSgamma1} should vanish on-shell, which it does if we neglect terms quadratic in $\Gamma$.\footnote{The attentive reader might worry about the term $\langle [A_1,A_2],h\rhd \Gamma_3\rangle$ not being quadratic in $\Gamma$. However, the claim is that $\delta S =0$ \emph{on-shell}. In particular, using the equations of motion and the boundary conditions, one can show that this term indeed vanishes for infinitesimal gauge transformations.} The failure of the vanishing of the boundary variation in going from infinitesimal to finite gauge transformations can be attributed to the presence of the 3d Chern-Simons $L_{\mathrm{CS}}(\Gamma)$ term appearing in \eqref{ec:omega}, which we know is sensitive to the global structure of the $3$-manifold $Y$.

This implies that if we want our transformations to be symmetries of h2CS theory, we must impose further constraints on the gauge parameters $(h,\Gamma)$. In order to do this in a consistent manner, we introduce some structure. Recall that in \S \ref{sec:h2csth} we considered Lagrangian subspaces to define boundary conditions. In particular, the condition that the gauge transformations should preserve the boundary conditions can be expressed as follows. Given a choice of Lagrangian subspace $\mathcal{L} \subset\mathsf{MC}_\text{blk}$ of the on-shell Maurer-Cartan fields $(A,B)$, we require 
\begin{equation}\label{def2alg}
    \big(\iota_Y^*A^{(h,\Gamma)},\iota_Y^*B^{(h,\Gamma)}\big)\in \mathcal{L}\,,\qquad \forall~ (\iota_Y^*A,\iota_Y^*B)\in\mathcal{L}\,.
\end{equation}
This imposes differential and algebraic constraints on the derived 2-gauge parameters $(h,\Gamma)\in \mathsf{D}\cL_0 =(\Omega^\bullet(X)\otimes \mathsf{D}\mathbb{G})_0$, and thus it defines a {\it defect subcomplex} $\mathsf{D}\cL_\text{def}\subset \mathsf{D}\cL_0$. This is in fact a derived Lie 2-subgroup of $\mathsf{D}\cL_0$, since the composition law in the derived 2-group $\mathsf{D}\cL_0$ satisfies\footnote{This follows from the closure of the 2-gauge algebra \cite{Martins:2010ry,Mikovic:2016xmo,Baez:2004in,chen:2022}. For strict Lie 2-algebras $\fG=\operatorname{Lie}\fG$, this is true in general. However, if $\fG$ were weak, then it is known that the 2-gauge algebra closes only on-shell of the fake-flatness condition \cite{Kim:2019owc}.}
\begin{equation*}
    \big(\iota_Y^*A^{(h,\Gamma)\cdot (h',\Gamma')},\iota_Y^*B^{(h,\Gamma)\cdot (h',\Gamma')}\big) = \bigg(\iota_Y^*\big(A^{(h,\Gamma)}\big)^{ (h',\Gamma')},\iota_Y^*\big(B^{(h,\Gamma)}\big)^{(h',\Gamma')}\bigg).
\end{equation*}
Therefore, if $(h,\Gamma),(h',\Gamma')\in\mathsf{D}\cL_\text{def}$ lie in the defect subcomplex, then \eqref{def2alg} state that so does $(h,\Gamma)\cdot (h',\Gamma')$. In the example considered above, we have that $\mathsf{D}\cL_{\mathrm{def}}$ is given by the set of 2-gauge transformations which satisfy the boundary constraints \eqref{ec:gtbczero} and \eqref{ec:gtbcinfty}. 

As these constraints are localized at the poles, are independent, and only involve $Y$-dependencies, they are constraints on 2-gauge transformations of the \textbf{archipelago type} \cite{Delduc:2019whp}. Formally, they are finite 2-gauge transformations that are non-trivial only at a neighborhood around each puncture $z=0,z=\infty$. The precise definition is the following.
\begin{definition}
    We say a 2-gauge parameter $(h,\Gamma)\in\mathsf{D}\cL_0$ is of \textbf{archipelago type} if and only if there exists open discs $D_z\subset\mathbb{C}P^1$ of radius $R_z$ containing each puncture $z=0,\infty$ such that:
    \begin{enumerate}
        \item $(h,\Gamma)=(1,0)$ is trivial away from $D_0\coprod D_\infty$,
        \item the restriction $(h,\Gamma)\vert_{Y\times D_z}$ depends on $Y$ and the radial coordinate $0<r<R_z$ in $D_z$, and
        \item there exists an open disc $V_z\subset D_z$ such that $(h,\Gamma)\vert_{Y\times V_z}$ depends only on $Y$.
    \end{enumerate}
\end{definition}
It suffices to consider 2-gauge transformations of archipelago type in $\mathsf{D}\cL_\text{def}$; the restriction map $-\vert_{Y\times V_0} \times -\vert_{Y\times V_\infty}$ to the two open discs gives a surjection 
\begin{equation*}
    \mathsf{D}\cL_\text{def} \rightarrow \mathsf{D}\cL_\text{def}^0\times \mathsf{D}\cL_\text{def}^\infty \subset (\Omega^\bullet(Y)\otimes\mathsf{D}\mathbb{G})_0^{2\times}
\end{equation*}
onto those $(h,\Gamma)\in\mathsf{D}\cL_\text{def}^0$, depending only on $Y$, satisfying \eqref{ec:gtbczero} and those $(h',\Gamma')\in\mathsf{D}\cL_\text{def}^\infty$ satisfying \eqref{ec:gtbcinfty}. Conversely, the elements in $\mathsf{D}\cL_\text{def}^0\times \mathsf{D}\cL_\text{def}^\infty$ can always be patched along $\mathbb{C}P^1$ into a 2-gauge transformation of archipelago type, by using bump functions along the radial coordinate on $D_z\supset V_z$.

Let us use this structure to impose the additional constraints on the gauge parameters so they become symmetries of h2CS theory. We therefore introduce a subcomplex $\mathsf{D}\cL_\text{sym}\subset\mathsf{D}\cL_\text{def}$ consisting of those derived gauge parameters which are symmetries of the action. In other words, we define $\mathsf{D}\cL_\text{sym}$ such that
\begin{equation}
    (h,\Gamma)\in\mathsf{D}\cL_\text{sym} \quad \text{if and only if}\quad  \Omega[A,B,h,\Gamma]\vert_{z=0} - \Omega[A,B,h,\Gamma]\vert_{z=\infty}=0 \,.
\end{equation}
The additional constraints we impose on $(h,\Gamma)$ will be dictated by the requirement that $\mathsf{D}\cL_\text{sym}\subset \mathsf{D}\cL_\text{def}$ forms a derived Lie 2-subgroup. More precisely, we have
\begin{proposition}
The derived Lie 2-group of defect symmetries $\mathsf{D}\cL_\text{sym}$ associated to h2CS and the Lagrangian subspace defined by \eqref{ec:bconditions1} is  
\begin{equation}
    \mathsf{D}\cL_\text{sym} \cong \mathsf{D}\cL_\text{sym}^0\times \mathsf{D}\cL_\text{sym}^\infty\,,\qquad \begin{cases} \mathsf{D}\cL_\text{sym}^0 = \{(h,\Gamma)\in\mathsf{D}\cL_\text{def}^0\mid \Gamma_1,\Gamma_2 \in \Omega^1(Y)\otimes \ft\} \\ \mathsf{D}\cL_\text{sym}^\infty = \{(h',\Gamma')\in \mathsf{D}\cL_\text{def}^\infty\mid  \Gamma_3'=0\}\end{cases}\label{defectgaugeconstraints}
\end{equation}
where $\ft\subset\fh$ is the maximal Abelian subalgebra of $\fh$.  
\end{proposition}
\begin{proof}
     Since these constraints concern only the 1-form gauge parameters $\Gamma$, it is easy to see that $\mathsf{D}\cL_\text{sym}\subset\mathsf{D}\cL_\text{def}$ is a derived 2-subgroup and the constraints are closed under 2-group composition law. To see that \eqref{defectgaugeconstraints} indeed eliminates \eqref{ec:deltaSgamma1}, we note that $\Gamma'_3=0$ eliminates the first two terms evaluated at $z=\infty$. Moreover, for the last term at $z=\infty$, we first notice that if $\Gamma_3'=0$ then \eqref{ec:gtbcinfty} implies
\begin{equation*}
    \partial_{i} \Gamma'_3 - \partial_3 \Gamma'_{i}=- \partial_3 \Gamma'_{i}=0\,,\quad i=1,2\,,
\end{equation*}
hence the last term automatically drops. Finally, for the term evaluated at $z=0$ we have
\begin{equation}
   2\langle \mu_1(\partial_3\Gamma_1),\Gamma_2\rangle=\langle \mu_1(\partial_3\Gamma_1),\Gamma_2\rangle-\langle \mu_1(\partial_3\Gamma_2),\Gamma_1\rangle= \langle h^{-1}\partial_3h, [\Gamma_1,\Gamma_2]\rangle \label{defectcomputation}
\end{equation}
where we have repeatedly integrated by parts and used the boundary constraints \eqref{ec:gtbczero}. In particular, this term drops iff $\Gamma_1,\Gamma_2$ are Abelian.
\end{proof}

We will prove in \S \ref{residuals} that $\mathsf{D}\cL_\text{sym}$ indeed descends to global symmetries of the 3d field theory upon localization. In fact, we will also demonstrate in \S \ref{dgaffinecurrents} that these global symmetries exhibit conserved Noether charges that satisfy a differential graded analogue of the affine Virasoro Lie algebra that lives in the 2d Wess-Zumino-Witten model \cite{KNIZHNIK198483}.

\subsection{Field Reparametrization}
\label{sec:fieldreparam}

We now begin the localization procedure for 5d h2CS theory \eqref{ec:h2cs}, to obtain a 3-dimensional boundary theory. This will proceed, in spirit, entirely analogous to that in the context of CS$_4$ theory \cite{Costello:2019tri} and hCS$_6$ theory on twistor space \cite{Bittleston:2020hfv}. However, this has not been done explicitly before in the context of 2-Chern-Simons theory, and we will therefore describe it in detail.

Towards this, we first introduce new field variables $A' \in \Omega^1(X)\otimes \mathfrak{g}$, $\hat g \in C^{\infty}(X,G)$, $B'\in \Omega^{2}(X)\otimes \mathfrak{h}$ and $\hat \Theta \in \Omega^1(X)\otimes \mathfrak{h}$ that reparameterize our original fields $(A,B)$ in \eqref{ec:h2cs}, with
\begin{align}
\label{ec:reparA}
    &A= \hat g^{-1}A'\hat g  +\hat g^{-1}\mathrm{d} \hat g + \mu_1(\hat \Theta)\,,\\
\label{ec:reparB}
    &B = \hat g^{-1}\rhd B' + \mathrm{d}\hat \Theta + \mu_2(A,\hat \Theta) -\tfrac{1}{2}[\hat \Theta, \hat \Theta] \,.
\end{align}
This expression implies that, if $(\hat g,\hat \Theta) \in \mathsf{D}\cL_{\mathrm{sym}}$ then $S_{\text{2CS}}[A,B]=S_{\text{2CS}}[A',B']$, in which case such reparametrizations are just a gauge redundancy. Crucially, however, we will intentionally take $(\hat g,\hat \Theta) \notin \mathsf{D}\cL_{\mathrm{sym}}$. 

On the other hand, the reparametrisations \eqref{ec:reparA} and \eqref{ec:reparB} are in fact, redundant. More precisely, we can perform transformations on $A',B',\hat g$ and $\hat \Theta$ which leave $A$ and $B$ invariant. These are characterized by the following proposition
\begin{proposition}
\label{prop:internalsym}
    Given $u \in C^{\infty}(X,G)$ and $\Lambda \in \Omega^1(X) \otimes \mathfrak{h}$, the transformations
    \begin{align}
    \label{ec:actiononu}
    & \hat g\mapsto u^{-1}\hat g,\hspace{3.05cm} A'\mapsto u^{-1}A'u + u^{-1}\mathrm{d}u+ \mu_1(\Lambda)\\
    \label{ec:actionontheta}
    &\hat \Theta\mapsto \hat \Theta -(\hat g^{-1}u)\rhd \Lambda \qquad \qquad B'\mapsto u^{-1}\rhd B' +d\Lambda + \mu_2(A'^{(u,\Lambda)},\Lambda)-\tfrac{1}{2}[\Lambda,\Lambda].
\end{align}
leave the gauge fields $A$ and $B$ invariant. 
\end{proposition}

\begin{proof}
    The proof follows by explicit computation. Indeed, from equation \eqref{ec:reparA} we find
    \begin{align*}
    \hat g^{-1}A' \hat g &\mapsto  \hat g^{-1}A'\hat g + \mathrm{Ad}_{\hat g}^{-1} \mathrm{d}uu^{-1} + \mathrm{Ad}_{\hat{g}^{-1}u} \mu_1(\Lambda)\\
     \hat g^{-1}\mathrm{d}\hat g &\mapsto  -\mathrm{Ad}_{\hat g}^{-1} \mathrm{d}uu^{-1} + \hat g^{-1}\mathrm{d}\hat g \\
     \mu_1(\Theta) &\mapsto \mu_1(\Theta) - \mathrm{Ad}_{\hat{g}^{-1}u} \mu_1(\Lambda)\,,
\end{align*}
from where invariance of $A$ follows. Similarly, from equation \eqref{ec:reparB} we find
\begin{align}
\label{ec:Binv1}
    \hat g^{-1}\rhd B' &\mapsto \hat{g}^{-1}\rhd \big[ B' + u\rhd (\mathrm{d}\Lambda + \tfrac{1}{2}[\Lambda,\Lambda]) +\rhd\mu_2(A',u\rhd \Lambda)\big] + \mu_2(\mathrm{d}uu^{-1},u\rhd \Lambda) \\
    \label{ec:Binv2}
     \mathrm{d}\hat \Theta &\mapsto  \mathrm{d}\hat \Theta + \mu_2(\hat g^{-1}\mathrm{d}\hat g,\hat g u^{-1}\rhd \Lambda)- \mu_2(\mathrm{d}uu^{-1},u\rhd \Lambda)-\hat g^{-1}u\rhd \mathrm{d}\Lambda\\
     \label{ec:Binv3}
    \mu_2(A,\hat \Theta) &\mapsto \mu_2(A,\hat \Theta)-\mu_2(A,\hat g^{-1}u\rhd \Lambda)\\
    \label{ec:Binv4}
     -\tfrac{1}{2}[\hat\Theta,\hat\Theta] &\mapsto -\tfrac{1}{2}[\hat\Theta,\hat\Theta] + \mu_2(\mu_1(\hat \Theta),\hat g^{-1}u\rhd \Lambda)-\hat g^{-1}u \rhd \tfrac{1}{2}[\Lambda,\Lambda]\,.
\end{align}
Adding all of these together we are left after some cancellations with 
\begin{equation}
    B \mapsto B - \mu_2(A,\hat g^{-1}u\rhd \Lambda) + \mu_2(\hat g^{-1}A'\hat g  +\hat g^{-1}\mathrm{d} \hat g + \mu_1(\hat \Theta),\hat g^{-1}u\rhd \Lambda) =B \,,
\end{equation}
where the last equality follows from the definition of $A$ in \eqref{ec:reparA}. 
\end{proof}

\begin{definition}
    We call the transformations $(u,\Lambda)$ of proposition \ref{prop:internalsym} \textbf{ internal}  2-gauge transformations, and the invariance of $A$ and $B$ under these transformations the \textbf{ reparametrization symmetry}. 
\end{definition}

The idea is to use this reparametrization symmetry to fix the $\bar z$ components of $A'$ and $B'$ to zero. This will allow us to interpret the (gauge fixed) gauge fields as a flat 2-connection on $Y$ after localisation. Thus, we take an internal 2-gauge transformation $(u_1,\Lambda_1)$ such that 
\begin{equation}
    L' = A'^{(u_1,\Lambda_1)}\,,\quad H'=B'^{(u_1,\Lambda_1)}
\end{equation}
satisfy $L'_{\bar z}=0$ and $H'_{i\bar z}=0$. This amounts to a set of constraints on $u_1$ and $\Lambda_1$ given by 
 
\begin{align}
    & L'_{\bar z}= u_1^{-1}A'_{\bar z}u_1 + u_1^{-1}\partial_{\bar z}u_1 + \mu_1(\Lambda_{1,\bar z})=0 \label{1gauconst1} \\
    &H'_{i\bar z}= u_1^{-1}\rhd B'_{i\bar z} + \partial_{i}\Lambda_{1,\bar z}-\partial_{\bar z}\Lambda_{1,i} + \mu_2(L'_i,\Lambda_{1,\bar z})-\mu_2(L'_{\bar z},\Lambda_{1,i})- [\Lambda_{1,i},\Lambda_{1,\bar z}]=0.\label{2gauconst1}
\end{align}

\begin{lemma}
There exist an internal 2-gauge transformation satisfying \eqref{1gauconst1} and \eqref{2gauconst1}.  
\end{lemma}

\begin{proof}
    Consider first the constraint \eqref{1gauconst1}, which can be written as
\begin{equation}
    \partial_{\bar z}u_1 u_1^{-1} = A'_{\bar z} + \mu_1(u_1\rhd \Lambda_{1,\bar z})\,.
\end{equation}
Due to the graded nature of the problem, the most natural constraint is the following:
\begin{equation}
    u_1\rhd \Lambda_{1,\bar z} \in \operatorname{ker}\mu_1\,,\qquad \partial_{\bar z}u_1 u_1^{-1} = A'_{\bar z}\,.
\end{equation}
Since $\operatorname{ker}\mu_1$ is an ideal, this implies $\Lambda_{1,\bar z}\in\operatorname{ker}\mu_1$. The quantity $u_1$ is then only subject to the differential equation $\partial_{\bar z}u_1u_1^{-1}=A'_{\bar z}$, which fixes the $\bar z$-dependence of $u_1$. The above solution also implies
\begin{equation*}
    [\Lambda_{1,i},\Lambda_{1,\bar z}] = -\tfrac{1}{2}\mu_2(\mu_1(\Lambda_{1,\bar z}),\Lambda_{1,i})=0 \,,
\end{equation*}
by the Peiffer identity, hence the quadratic term in \eqref{2gauconst1} drops. Assuming further that each component of $\Lambda_1$ are independent, this secondary differential constraint can be written as a decoupled set of PDEs 
\begin{align}
    &\partial_i \Lambda_{1,\bar z} + \mu_2(L_i',\Lambda_{1,\bar z}) = -u_1^{-1}\rhd B'_{i\bar z}\,,\\
    &\partial_{\bar z} \Lambda_{1,i} + \mu_2(L_{\bar z}',\Lambda_{1,i}) = -u_1^{-1}\rhd B'_{i \bar z}\,.
\end{align}
If we introduce the covariant derivatives on $X=\mathbb{C}P^1\times Y$
\begin{equation}
    Y: \nabla^{L'}_i = \partial_i +\mu_2(L'_i,-) \,,\qquad \mathbb{C}P^1: \bar\partial^{L'} = \partial_{\bar z} + \mu_2(L'_{\bar z},-)\,,
\end{equation}
then these equations take the form of inhomogeneous covariant curvature equations
\begin{equation}
    \nabla^{L'}_i \Lambda_{1,\bar z} = -u^{-1}_1\rhd B'_{i\bar z}\,,\qquad \bar\partial^{L'}\Lambda_{1,i} = -u_1^{-1}\rhd B'_{i \bar z}\,.
\end{equation}
By Fredholm alternative, \textbf{provided the differential operators $\nabla^{L'},\bar\partial^{L'}$ have trivial kernel}, then we can find {\it unique} solutions $(\Lambda_{1,\bar z},\Lambda_{1,i})$ to these inhomogeneous PDEs for each value of the forcing $(u_1\rhd B'_{i\bar z},u_1^{-1}\rhd B'_{i \bar z})$.
\end{proof}

Having used part of the reparametrization symmetry to fix the $\bar z$ components of the fields to zero, we look for any remaining internal gauge transformation which preserves this condition. In other words, we look for $(u_2,\Lambda_2)$ such that 
\begin{equation}
    L \coloneqq L'^{(u_2,\Lambda_2)}\,,\quad H \coloneqq H'^{(u_2,\Lambda_2)} \,
\end{equation}
satisfy $L_{\bar z}=0$ and $H_{i\bar z}=0$. This amounts to
\begin{align}
    & L_{\bar z}=  u_2^{-1}\partial_{\bar z}u_2 + \mu_1(\Lambda_{2,\bar z}) =0 \\
    &H_{i\bar z} = \partial_{i}\Lambda_{2,\bar z}-\partial_{\bar z}\Lambda_{2,i} + \mu_2(L_i,\Lambda_{2,\bar z})-\mu_2(L_{\bar z},\Lambda_{2,i})- [\Lambda_{2,i},\Lambda_{2,\bar z}]=0\,,
\end{align}
which can be achieved by taking $(u_2, \Lambda_2)$ with 
\begin{equation}
    \partial_{\bar z}u_2=0\,,\quad \Lambda_{2,\bar z}=0 \,,\quad \partial_{\bar z}\Lambda_{2,i} =0\,.
\end{equation}
Note that these conditions on $(u_2,\Lambda_2)$ give us complete freedom on $u_2|_Y$ and $\Lambda_{2,i}|_Y$, since they only constrain the $\mathbb{C}P^1$-dependence of the latter. 

Hence, we can use this internal gauge transformation to fix the values of $\hat g$ and $\hat \Theta$ at one of the two punctures, either $z=0$ or $z=\infty$. Taking $(u_2,\Lambda_2)$ such that $u_2|_Y=\hat g|_{z=\infty}$ and $\Lambda_{2,i}|_{Y}=\hat{\Theta}_i|_{z=\infty}$, our fields $\hat g$ and $\hat \Theta$ at the poles become
\begin{equation}
\label{ec:gaugefixing1}
    \hat g|_{z=0}\coloneqq g\,,\quad \hat g|_{z=\infty}=1\,, \quad \hat\Theta|_{z=0}\coloneqq \Theta\,,\quad \hat \Theta_i|_{z=\infty}=0\,.
\end{equation}
To summarize, after fixing the reparametrisation symmetry we have arrived to the expressions
\begin{align}
\label{ec:AintermsofL}
    & A= \hat g^{-1} L \hat g + \hat g^{-1}\dr \hat g + \mu_1(\hat \Theta) \\
    & B = \hat g^{-1}\rhd H + \mathrm{d}\hat \Theta +\mu_2(A,\hat \Theta)-\tfrac{1}{2}[\hat \Theta,\hat\Theta]
\end{align}
where the $\bar z$ components of $L$ and $H$ vanish, and the values of $\hat g$ and $\hat \Theta$ at the poles of $\omega$ are given by \eqref{ec:gaugefixing1}.

\section{Three-Dimensional Field Theories}\label{3dift}

We are ready to construct the main player of this paper: a family of three-dimensional field theories associated to different choices of boundary conditions. To begin with, we write our 5d action \eqref{ec:h2cs} in terms of the reparametrisation fields $L,H,\hat g$ and $\hat \Theta$. Given that the expression of $A$ and $B$ in terms of the latter is formally the same than that of a gauge transformation, we can use \eqref{ec:actionundergt} to write
\begin{equation}
\label{ec:reparaction}
    S_{\mathrm{h2CS}}[A,B] = \frac{1}{2\pi i}\int_X \langle F(L)-\tfrac{1}{2}\mu_1(H),H\rangle + \frac{1}{2\pi i}\int_X\omega \wedge  \mathrm{d}\Omega\big(L,H,\hat g,\hat \Theta\big)
\end{equation}
where we recall the expression for $\Omega$ once again
\begin{equation}
\label{ec:omega12345}
    \Omega\big(L,H,\hat g,\hat \Theta\big) = \langle \hat g^{-1}F(L)\hat g,\hat \Theta\rangle + \langle \hat g^{-1}L \hat g+\tfrac{1}{2}\hat g^{-1}\mathrm{d}\hat g,[\hat\Theta, \hat\Theta]\rangle + \tfrac{1}{2} L_{\mathrm{CS}}\big(\hat\Theta\big)\,.
\end{equation}
Let us note that since $H_{i\bar z}=0$ then $\langle \mu_1(H),H\rangle=0$. On the other hand, the second term can be integrated by parts to find
\begin{equation}
\label{ec:reparaction2}
    S_{\mathrm{2CS}}\big[L,H,\hat g,\hat \Theta\big] = \frac{1}{2\pi i}\int_X \omega \wedge \langle F(L),H\rangle + \frac{1}{2\pi i}\int_X \mathrm{d}\omega \wedge \Omega\big(L,H,\hat g,\hat \Theta\big)\,.
\end{equation}
While the second term in the above expression is effectively three-dimensional due to the presence of $\mathrm{d}\omega$, the first term is a genuine five-dimensional bulk term which we must eliminate to obtain a three-dimensional theory. We do this by going partially on-shell. Indeed, under an arbitrary variation $\delta H$ we find the bulk equation of motion $\partial_{\bar z}L=0$. The remaining term, can be integrated along $\mathbb{C}P^1$ to find (see \textbf{Lemma 2.2} in \cite{Delduc:2019whp})
\begin{equation}
\label{ec:3dift1}
\begin{aligned}
    S_{3d}[L,H,\hat g,\hat \Theta] &= 
    \int_{Y}\iota_Y^*\Omega\big(L,H,\hat g,\hat\Theta\big)|_{z=0}-\iota_Y^*\Omega\big(L, H,\hat g,\hat \Theta\big)|_{z=\infty}\\
    & =\int_Y  \big[\langle g^{-1}F(L) g,\Theta\rangle + \tfrac{1}{2} \langle g^{-1}L g+ g^{-1}\mathrm{d}g,[\Theta,\Theta]\rangle + \tfrac{1}{2} L_{\mathrm{CS}}(\Theta)\rangle\big]_{z=0}
\end{aligned}
\end{equation}
where we have used the expressions \eqref{ec:gaugefixing1} for the evaluations of $\hat g$ and $\hat \Theta$ at $z=0$. Note that term at infinity is identically vanishing due to the gauge fixing conditions \eqref{ec:gaugefixing1}. The action \eqref{ec:3dift1} is three-dimensional but it is still written in terms of $L$, whereas we want our 3d action to be written in terms of $g$ and $\Theta$ only. To do so, we will consider a family of boundary conditions which includes \eqref{ec:bconditions1}, in order to express $L$ in terms of $g$ and $\Theta$ and obtain explicit forms for three-dimensional actions.


\subsection{Covariant family of boundary conditions}\label{covbcs}
We begin with a particularly convenient family of boundary conditions/Lagrangian subspaces $\mathcal{L}\subset\mathsf{MC}_\text{blk}$, which can be understood as a generalization of the example \eqref{ec:bconditions1}. First, recall that the 2-Chern-Simons action is symmetric under global Poincar{\'e} transformations. For a product manifold $X = Y\times \mathbb{C}P^1$, the Poincar{\'e} group admits as subgroups the isometry group $\operatorname{Iso}(Y)$ of $Y$. It can be seen that this isometry group $\operatorname{Iso}(Y)$ is inherited as a global symmetry of the symplectic form $\varpi_\text{blk}$ \eqref{ec:bdaryvar}.

Here, we are going to introduce a family of boundary conditions that breaks only the linear part of the isometries $\operatorname{Iso}(Y)$ (namely, we neglect the translations). For concreteness and simplicity, let us for now think of $Y$ as a real Riemannian 3-manifold, then the linear isometries are given by $O(3)\subset \operatorname{Iso}(Y)$. The symmetry breaking patterns are described by maximal subgroups of $O(3)$. It is known that there is at most one unique maximal Lie (ie. infinite) subgroup $O(2)$ of $O(3)$, from which we obtain the following {\it symmetry breaking pattern}
\begin{equation}
    S^2\hookrightarrow O(3)\rightarrow O(2).
\end{equation}
The fibre $S^2\cong O(3)/O(2)$ is a 2-sphere, which paramterizes a collection $\{\mathcal{L}_\ell\mid \ell\in S^2\}$ of Lagrangian subspaces of $\mathsf{MC}_\text{blk}$. We call this data $\ell\in S^2$ the \textbf{chirality} vector.

\medskip

We now work to explicitly write down the boundary conditions associated to $\mathcal{L}_\ell$. To do so, we first define the 3-dimensional vectors $\vec{A} = (A_1,A_2,A_3)$ and $\vec{B} = (B_{23},B_{13},B_{12})$, which are built locally out of the components of our fields $(A,B)$ along $Y$. For a given triple $\vec{\ell},\vec{n},\vec{m}$ of unit vectors on $Y$, we consider the family of boundary conditions
\begin{align}
\label{ec:covbc1}
    & \vec n\cdot \vec{A}|_{z=0}= \vec m\cdot \vec{A}|_{z=0}= \vec \ell\cdot \vec{B}|_{z=0}=0 \\
    \label{ec:covbc2}
    & \vec \ell\cdot \vec{A}|_{z=\infty}= \vec m\cdot \vec{B}|_{z=\infty}=\vec n\cdot \vec{B}|_{z=\infty}=0 \,,
\end{align}
which is given by the following Lagrangian subspace
\begin{equation}
\label{ec:lagrangiansubspaces}
    \mathcal{L}_\ell = \operatorname{Span}\{\vec \ell\cdot \vec{A},\vec{B}_\perp\}\vert_{z=0} \cup\operatorname{Span}\{\vec{A}_\perp,\vec \ell\cdot \vec{B}\}\vert_{z=\infty},
\end{equation}
where we have used a shorthand to denote $\vec{A}_\perp = (\vec n\cdot \vec{A},\vec m\cdot \vec{A})$.

\begin{remark}
It can be seen by direct computation, that this choice of boundary conditions implies $\delta S_{\text{2CS}}=0$ if and only if the triple $(\vec\ell,\vec n,\vec m)$ defines a global orthonormal frame on $Y$ specified up to orientation by, say, $\vec\ell$. The simple example considered in \eqref{ec:bconditions1} can be recovered with $\vec\ell=(0,0,1)$, and all other Lagrangian subspaces $\mathcal{L}_\ell$ work in the same way. The characterization of the defect symmetry 2-group $\mathsf{D}\cL_\text{def}$ \eqref{defectgaugeconstraints}, thus holds for any member $\mathcal{L}_\ell$ of this $S^2$-family of Lagrangian subspaces up to a global rotation of the framing on $Y$.
\end{remark}

\subsection{The 3d Actions}\label{3dlocalization}
Having chosen our boundary conditions defined through the Lagrangian subspace \eqref{ec:lagrangiansubspaces} we proceed with the evaluation of the three dimensional action \eqref{ec:3dift1}. To do so, we must use our boundary conditions \eqref{ec:covbc1} and \eqref{ec:covbc2} to solve for $L$ in terms of $g$ and $\Theta$. Recall that $A$ and $L$ are related by equation \eqref{ec:AintermsofL}, namely
\begin{equation}
    A = \hat g^{-1}L\hat g + \hat g^{-1}\mathrm{d}\hat g + \mu_1(\hat \Theta)\,.
\end{equation}
Hence, the boundary conditions imply
\begin{align}
    & 0=\vec n \cdot \vec A|_{z=0} = g^{-1} \big(\vec n \cdot \vec L|_{z=0}\big)g+ g^{-1}\big(\vec n \cdot \nabla\big) g + \mu_1\big(\vec n \cdot \vec \Theta\big) \label{chiralbc1}\\
    &0=\vec m \cdot \vec A|_{z=0} = g^{-1} \big(\vec m \cdot \vec L|_{z=0}\big)g+ g^{-1}\big(\vec m \cdot \nabla\big) g + \mu_1\big(\vec m \cdot \vec \Theta\big) \label{chiralbc2}\\
    &0 = \vec \ell \cdot \vec A|_{z=\infty} =\vec \ell \cdot \vec L|_{z=\infty}\label{chiralbc3}
\end{align}
where we have used the fact that $\hat g|_{z=\infty}=1$ and $\hat \Theta|_{z=\infty}=0$. The last constraint \eqref{chiralbc3} is of particular importance, since it will in fact eliminate the component $\operatorname{proj}_\ell L = (\vec \ell \cdot \vec L)\vec\ell=0$ along $\ell$ entirely. This is because $L$ is constrained to be holomorphic on $\mathbb{C} P^1$ according to the bulk equation of motion $\partial_{\bar z}L_i =0$ and hence constant. The boundary condition $\vec \ell \cdot \vec L|_{z=\infty}=0$ then fixes this constant to be zero.


Let us compute each of the terms in $\Omega$ separately, making use of the boundary conditions \eqref{chiralbc1}, \eqref{chiralbc2}, \eqref{chiralbc3}. It will be convenient to introduce the notation 
\begin{equation}
    \vec{a}_\ell = (\vec\ell\cdot \vec a)\vec\ell = \operatorname{proj}_\ell(a)\nonumber
\end{equation}
for the projection of vectors $\vec{a}$ (or vector-valued operator such as $\nabla$) along $\ell$, and for any Lie algebra-valued vectors $\vec{\alpha},\vec{\beta}$ we define
\begin{align*}
    \langle \vec\alpha\cdot\vec\beta\rangle &= \langle \alpha_1,\beta_1\rangle + \langle \alpha_2,\beta_2\rangle+\langle \alpha_3,\beta_3\rangle\\
    [\vec{\alpha}\times \vec{\beta}] &= ([\alpha_2,\beta_3] - [\alpha_2,\beta_3],[\alpha_1,\beta_3] - [\beta_1,\alpha_3],[\alpha_1,\beta_2] - [\beta_1,\alpha_2]).\footnotemark
\end{align*}
\footnotetext{Since we know $(\vec n,\vec m,\vec \ell)$ forms an orthonormal frame on $Y$, the triple product $(-\times -)\cdot -$ satisfies
\begin{equation*}
    (\vec a\times \vec b)\cdot \vec c_\ell = (\vec a_{n}\times \vec b_{m} - \vec a_{m}\times \vec b_{n})\cdot \vec c.
\end{equation*}} Let $\vec k=\nabla gg^{-1}$ denote the {\it right} Maurer-Cartan form. We can then compute
\begin{equation}
\begin{split}
    \iota_Y^* \langle g^{-1}F(L)g,\Theta\rangle 
    &=-\langle \mu_1(\nabla\times\vec\Theta)\cdot\vec\Theta_\ell\rangle  +\langle  \mu_1([\vec\Theta\times\vec\Theta])\cdot \Theta\rangle\\
    &\qquad -\langle \mathrm{Ad}_g^{-1}\nabla_\ell \times \vec k\cdot \vec\Theta\rangle-2\langle \operatorname{Ad}_g^{-1}\vec k_\ell\cdot [\vec\Theta\times \vec\Theta]\rangle \,.
\end{split}
\end{equation}
Second, we have
\begin{equation}
    \iota_Y^*\langle g^{-1}Lg+g^{-1}\mathrm{d}g,\Theta \wedge \Theta\rangle=-2\langle \mu_1(\vec\Theta)\cdot [\vec\Theta \times\vec\Theta]\rangle+\langle \operatorname{Ad}_g^{-1}\vec k_\ell\cdot [\vec\Theta\times\vec\Theta]\rangle\,,
\end{equation}
while the Chern-Simons term $L_{\mathrm{CS}}(\Theta)$ is simply
\begin{align}
    L_{\mathrm{CS}}(\Theta) &= 2\langle \mu_1(\vec\Theta)\cdot\nabla\times\vec\Theta\rangle+2\langle\mu_1(\vec\Theta),[\vec\Theta\times \vec\Theta]\rangle\,.
\end{align}
In combination, we find\footnote{Notice the term $-\langle \mu_1(\nabla\times\vec\Theta)\cdot\Theta_\ell\rangle$ in $\iota_Y^*\langle g^{-1}F(L)g,\Theta\rangle $ combines to change the sign of an identical term in $L_{\mathrm{CS}}(\Theta)$. This allows us to perform an integration by parts to eliminate the appearance of $\Theta_\ell$ altogether in the resulting Lagrangian \eqref{3dintegrand}.} from \eqref{ec:omega12345}
\begin{align}
\iota_Y^*\Omega &= -\langle \mathrm{Ad}_g^{-1}\nabla_\ell\times\vec k\cdot \vec\Theta\rangle+ \langle\mu_1(\vec\Theta)\cdot \nabla_\ell \times \vec\Theta\rangle-\langle\operatorname{Ad}_g^{-1}\vec k_\ell \cdot [\vec\Theta\times\vec\Theta] \rangle\,.\label{3dintegrand}
\end{align}
Finally, we note that we can perform one more simplification by using the invariance of the pairing,
\begin{equation}
    \langle \mu_1(g\rhd\vec\Theta)\cdot \nabla_\ell \times (g\rhd \vec\Theta)\rangle= \langle\mu_1(\vec\Theta)\cdot\nabla_\ell \times\vec\Theta\rangle - \langle \vec k_\ell\cdot g\rhd [\vec\Theta\times\vec\Theta]\rangle,
\end{equation}
from which we construct a family of 3d theories given by
\begin{equation}\label{covariant3daction}
    S_{3d}[g,\Theta] = -\int_Y \langle \nabla_{\ell}\times \vec{k}\cdot \vec{\Theta}^g\rangle - \langle 
    \mu_1(\vec\Theta^g)\cdot \nabla_\ell\times \vec\Theta^g\rangle \,,
\end{equation}
parameterized by ${\ell}\in S^2$, where $\vec\Theta^g = g\rhd\vec\Theta$.

\medskip

We shall show in the following that this theory is {\it completely topological}, in the sense that its equations of motion describe configurations of 2-connections $(L,H)$ that are flat in each real (topological) direction.


\subsubsection{Equations of Motion}\label{3deoms}
The equations of motion of the theory are obtained by varying the fields $g,\Theta$. Indeed, under a variation $\delta\Theta$, we find
\begin{equation}\label{1flat?}
   \delta \Theta: \quad \nabla_\ell \times \vec{k} + \mu_1(\nabla_\ell\times (g\rhd\vec\Theta))=0\,.
\end{equation}
To vary $g$ we define $\delta g$ as the infinitesimal translation of $g$ under right-multiplication $g+\delta_g g = g\cdot(1+\epsilon)$, where $\epsilon\in\mathfrak{g}$ is an infinitesimal Lie algebra element. Using the following formulas
\begin{gather}
    \delta_g (\nabla gg^{-1}) = \operatorname{Ad}_g \nabla \epsilon\,,\qquad \delta_g (g\rhd \vec{\Theta}) = g\rhd \mu_2(\epsilon,\vec{\Theta})\,, \nonumber\\
\delta_g (g^{-1}\nabla g) =  \mathrm{d}\epsilon + [g^{-1}\nabla g,\epsilon]\,, \nonumber
\end{gather}
we compute the variation of the first term neglecting exact boundary terms
\begin{equation}
    \delta_g\langle \nabla_\ell\times \vec{k}\cdot g\rhd\vec{\Theta}\rangle =\langle \epsilon,g^{-1}\rhd \nabla_\ell\cdot (g\rhd (\nabla\times \vec{\Theta}))\rangle \label{variation1}
\end{equation}
where we have used the vector identities
\begin{gather*}
  \nabla\cdot\nabla\times\vec A=0\,,\quad \nabla\cdot(\vec A\times\vec B) = \vec{A}\cdot(\nabla\times\vec{B}) - \vec{B}\cdot(\nabla\times\vec A)\,, \quad
  \nabla\times(f\vec{A})= (\nabla f)\times\vec{A} + f\nabla\times\vec{A}\,.
\end{gather*}
The variation of the second term is
\begin{equation}
    \delta_g \langle \vec{k}_\ell\cdot g\rhd [\vec{\Theta}\times \vec{\Theta}]\rangle = -\langle \epsilon, g^{-1}\rhd (\nabla_\ell \cdot g\rhd [\vec{\Theta}\times\vec{\Theta}])\rangle\,,
\end{equation}
and hence combining with \eqref{variation1} we find the second equation of motion
\begin{equation}\label{2flat}
    \delta g: \quad \nabla_\ell\cdot g\rhd (\nabla\times \vec{\Theta} - [\Theta\times\Theta]) =0.
\end{equation}

\paragraph{Fake- and 2-flatness in 3-dimensions.}
We now work to show that the equations of motion \eqref{1flat?} and \eqref{2flat} of \eqref{covariant3daction} implies the flatness of the 2-connection. Without loss of generality (WLOG), it suffices to prove this statement for one choice of $\ell$, such as $\vec{\ell}=(0,0,1)$ as in \eqref{ec:bconditions1}. This is because $S^2$ is a homogeneous space under $O(3)$ and hence any two choices of the Lagrangian subspace $\mathcal{L}_\ell$ are related by an $O(3)$-action on the 3d theory \eqref{covariant3daction}. For this choice of $\ell$, the equations of motion are given by
\begin{align}
\label{ec:123eomth}
    &\delta \Theta: \quad \partial_3k_i  = -\mu_1(\partial_3(g\rhd\Theta_i))\,,\quad i=1,2\\
    &\delta g: \quad \partial_3\left[g\rhd(\partial_1\Theta_2-\partial_2\Theta_1 - [\Theta_1,\Theta_2])\right]=0\,.\label{2flatx3}
\end{align}

\begin{proposition}\label{extended2flat}
    The equations of motion for \eqref{covariant3daction} are equivalent to fake- and 2-flatness for $(L,H)$ on $Y$.
\end{proposition}
\begin{proof}
    The fake-flatness and 2-flatness conditions for $(\tilde L,\tilde H) \coloneq (\iota_Y^* L,\iota_Y^* H)\vert_{z=0}$ are given by
    \begin{equation}
    \label{ec:formflatness}
        F(\tilde L)-\mu_1(\tilde H)=0\,,\quad \dr \tilde H + \mu_2(\tilde L,\tilde H)=0\,.
    \end{equation}
    The bulk equations of motion \eqref{ec:eom1}, \eqref{ec:eom3} and the boundary conditions \eqref{ec:bconditions1} at $z=\infty$ make $L_3=0, H_{13}=0, H_{23}=0$ on $X=\mathbb{C}P^1\times Y$, so that fake flatness becomes in components
        \begin{equation}
        \partial_1\tilde L_2 - \partial_2\tilde L_1 + [\tilde L_1,\tilde L_2] = \mu_1(\tilde H_{12})\,,\quad \partial_3 \tilde L_1 =0 \,,\quad \partial_3 \tilde L_2 = 0\,, 
    \end{equation}
    and 2-flatness is simply
    \begin{equation}
        \partial_3 \tilde H_{12}=0\,.
    \end{equation}
    On the other hand, the boundary conditions at $z=0$ allow us to compute
    \begin{align}
        &\tilde L_i = -k_i - \mu_1(g\rhd \Theta_i)\,,\quad i=1,2\\
        &\tilde H_{12} = -g\rhd(\partial_1\Theta_2 - \partial_2\Theta_1 - [\Theta_1,\Theta_2])\,,\label{currents}
    \end{align}
    In particular, we find that
    \begin{equation*}
        \partial_3 \tilde H_{12} = -\partial_3 g\rhd(\partial_1\Theta_2-\partial_2\Theta_1 - [\Theta_1,\Theta_2]) 
    \end{equation*}
    vanishes due to \eqref{2flatx3}, and this is precisely the 2-flatness condition. On the other hand, we begin by evaluating 
    \begin{align*}
        \partial_1 \tilde L_2 - \partial_2\tilde L_1 + [\tilde L_1,\tilde L_2] &= -\partial_1k_2 + \partial_2k_1 - [k_1,k_2] + [\mu_1(g\rhd\Theta_1),\mu_1(g\rhd\Theta_2)]\\
        &\qquad -~ \partial_1\mu_1(g\rhd\Theta_2) + \partial_2\mu_1(g\rhd\Theta_1) + [k_1,\mu_1(g\rhd\Theta_2)] - [\mu_1(g\rhd\Theta_1),k_2]\\
        &= -\mu_1\mu_2(k_1,g\rhd\Theta_2) +[k_1,\mu_1(g\rhd\Theta_2)] + \mu_1\mu_2(k_2,g\rhd\Theta_1) - [k_2,\mu_1(g\rhd\Theta_1)] \\
        &\qquad - \mu_1(g\rhd (\partial_1\Theta_2 - \partial_2\Theta_1 - [\Theta_1,\Theta_2])) \\ 
        &= \mu_1(\tilde H_{3}),
    \end{align*}
    where we have used the Maurer-Cartan equations for $k_1,k_2$ as well as the equivariance and Peiffer identities. On the other hand, we have
    \begin{equation*}
        -\partial_3\tilde L_1 = \partial_3(k_1 + \mu_1(g\rhd \Theta_1)),\qquad -\partial_3\tilde L_2 = \partial_3(k_2 + \mu_1(g\rhd \Theta_2)),
    \end{equation*}
    which both vanish due to the equations of motion \eqref{ec:123eomth}. These are precisely the fake-flatness conditions.
\end{proof}
\noindent Note the proof involves local computations, and hence the above statement still holds regardless if $\ell$ aligns with the foliation, given we swap the indices of the (co)vectors involved consistently.

\subsection{Transverse Holomorphic Foliation}\label{thf}
In this section, we are going to relax the assumption that $Y$ is a real Riemannian 3-manifold. We do this by equipping $Y$ with a {\it transverse holomorhic foliation} (THF) $\Phi$, which we take to be along the $x_3$-direction. This means that we have a decomposition of the tangent bundle $TY$ such that each transverse leaf $\operatorname{ker}\mathrm{d}x_3$ has equipped an almost complex structure \cite{Scrdua2017OnTH}. Each local patch $U\subset Y$ can then be given coordinates $(x_3,w,\bar w)$, where $x_3 \in \mathbb{R}$ and $(w,\bar w)\in \mathbb{C}$, such that transitions $(x_3,w,\bar w) \mapsto (x_3',w',\bar w')$ look like \cite{Aganagic:2017tvx}
\begin{equation*}
    x_3'(x_3,w,\bar w),\qquad w'(w),\qquad \bar w'(\bar w)\,.
\end{equation*}
In the following, we will assume the leaves of $\Phi$ never intersect, and that they define an integrable distribution such that $\Phi$ foliates out a complex 2-submanifold $M\subset Y$. 

There are then two directionalities present in our theory: the chirality vector $\ell$ and the direction of the foliation $\Phi$. These directions can either be aligned or misaligned, which lead to distinct theories. We shall study these two cases in more detail in the following.

\subsubsection{Foliation aligned with chirality} 
First, let us suppose the chirality $\ell = \mathrm{d}x_3$ is aligned. The complex structure on the transverse leaves of $\Phi$ allows us to impose boundary conditions which is a mix of topological and holomorphic directions, such as
    \begin{align*}
    & A_w|_{z=0}= A_{\bar w}|_{z=0}= B_{w\bar w}|_{z=0}=0 \\
    & A_3|_{z=\infty}=  B_{3w}|_{z=\infty}= B_{3\bar w}|_{z=\infty}=0\,.
\end{align*}
These boundary conditions can be understood as a partially holomorphic version of \eqref{ec:bconditions1} and indeed lead to the following 3d \textbf{topological-holomorphic} field theory,
\begin{align}\label{holo3daction}
    S_{\mathrm{h}3d} &= \int_Y \operatorname{vol}_Y\left[-\langle \partial_3 k_{\bar w},\Theta^g_{w}\rangle +\langle \partial_3 k_w,\Theta^g_{\bar w}\rangle - \langle \mu_1(\Theta_w^g),\partial_3\Theta_{\bar w}^g\rangle\right].
\end{align}
Writing $w = x_1+ix_2\in M$ in terms of the real coordinates, the above action can be expressed in terms of the lightcone coordinates $x_\pm = x_1 \pm x_2$ on $M$,
\begin{align*}
    S_{\mathrm{h}3d} &= 2\int_Y \operatorname{vol}_Y\left[-\langle \partial_3 k_+,\Theta^g_-\rangle +\langle \partial_3 k_-,\Theta^g_+\rangle - \langle \mu_1(\Theta_+^g),\partial_3\Theta^g_-\rangle\right].
\end{align*}
Notice this action can be reproduced from the action \eqref{covariant3daction} with $\ell=\mathrm{d}x_3$ from a $SO(2)$-rotation
\begin{equation*}
    \hat x_1 \mapsto \frac{1}{\sqrt{2}}(\hat x_1-\hat x_2),\qquad \hat x_2\mapsto \frac{1}{\sqrt{2}}(\hat x_1 + \hat x_2),
\end{equation*}
which can be interpreted as a boost into the lightcone coordinates on $M$. Therefore, the action \eqref{holo3daction} can also be intuitively obtained by "endowing" \eqref{covariant3daction} with a complex structure. This is not a purely cosmetic choice, however; the existence of a complex structure on $M$ plays a significant role in determining the boundary conditions satisfied by the symmetries of our theory (see \S \ref{residuals}).

\subsubsection{Foliation misaligned with chirality}\label{misaligned}
Let us now consider the case where the direction $\mathrm{d}x_3$ of the transverse holomorphic foliation $\Phi$ does {\it not} align with the chirality $d\vec\ell$. We may pick, for instance, the chirality $\vec\ell$ along the holomorphic direction $w$, which leads to the boundary conditions 
\begin{align*}
    & A_3|_{z=0}= A_{\bar w}|_{z=0}= B_{3\bar w}|_{z=0}=0 \\
    & A_{w}|_{z=\infty}=  B_{w\bar w}|_{z=\infty}= B_{3 w}|_{z=\infty}=0 \,.
\end{align*}
The three-dimensional action is then 
\begin{align*}
    S_{\mathrm{h}3d} &= \int_Y \operatorname{vol}_Y\left[-\langle \partial_w k_3,\Theta_{\bar w}^g\rangle +\langle \partial_w k_{\bar w},\Theta_3^g\rangle - \langle\mu_1(\Theta_{\bar w}^g),\partial_w\Theta^g_3\rangle \right].
\end{align*}
Consider the case $\mu_1=0$, whence the final term drops. Using the identities 
\begin{equation}\label{MCidentity}
    \partial_i(\partial_j gg^{-1}) = \operatorname{Ad}_g(\partial_j(g^{-1}\partial_ig)),\qquad \partial_i(g^{-1}\partial_j g) = \operatorname{Ad}_g^{-1}(\partial_j(\partial_i gg^{-1})), 
\end{equation}
we can bring the above action to the form
\begin{equation}\label{CSmatter}
    S_{\mathrm{h}3d} =-\int_Y dw\wedge d\bar w \wedge \mathrm{d}x_3\left[\langle\Theta_{\bar w}, \partial_3m_w\rangle - \langle \Theta_3,\partial_{\bar w}m_w\rangle\right],
\end{equation}
where $m = g^{-1}\mathrm{d}g$ is the {\it left} Maurer-Cartan form. This is in fact nothing but the Chern-Simons/matter theory studied in \cite{Aganagic:2017tvx}, where we take $R= \fg = \operatorname{Lie}G$ as the adjoint $G$-representation and $R^c=\fh$ its conjugate.\footnote{With respect to the degree-1 paring $\langle-,-\rangle$ on the balanced Lie 2-algebra $\fG$.} In other words, we are able to recover the interaction between Chern-Simons theory and matter from our localization procedure on 5d 2-Chern-Simons theory.

It was explained in \cite{Aganagic:2017tvx} that, in the context of the topological A-model, when the flux on a 5-brane vanishes on an embedded 3-brane, then the Chern-Simons theory living on this 3-brane acquires this matter coupling term due to those strings that extend in the ambient 5-brane directions. The fact that we can recover such matter coupling terms from our prescription suggests that the holomorphic 5d 2-Chern-Simons theory may play a part in how the ambient strings couple to the 3-brane in the topological A-model.

\begin{remark}
    We emphasize here that the distinction of aligned vs. misaligned matters only when $Y$ is equipped with a THF $\Phi$. Indeed, if no such data is imposed, then the chirality can always be made to align with $\Phi$ with an $SO(3)$-action on the family $\{\mathcal{L}_\ell\}_{\ell\in S^2}$ of Lagrangians. However, this cannot be done with a THF: indeed, if $\ell$ is tangent to $M$, then one cannot rotate it out of the leaves without destroying the complex structure on $M$. We will see in \S \ref{integrability} that this makes the two types (aligned/misaligned) of theories dynamically distinct. 
\end{remark}

\section{Symmetries of the 3d Theory}\label{residuals}
Let us now study the symmetries of our three dimensional theories. As discussed in \S \ref{sec:gaugesym5d}, the 2-gauge symmetries of h2CS together with the Lagrangian subspace $\mathcal{L}_\ell$ (i.e. the defect) form the derived 2-group of defect symmetries $\mathsf{D}\cL_\text{sym}(\ell)$ characterized in \eqref{defectgaugeconstraints}. The main goal of this section is to examine the {\it global} symmetries $\mathsf{D}\cL_{3d}$ of the localized 3d theory \eqref{covariant3daction}, and we shall prove that it in fact sits inside of $\mathsf{D}\cL_\text{sym}$.

For simplicity, we shall focus on the analysis of the symmetries of the 3d theory in the case where the chirality $\ell$ aligns with the $\mathrm{d}x_3$-direction. If $Y$ has equipped a THF, then the misaligned case can be analogously dealt with, but the resulting symmetries will have different properties.

We begin by first writing down the 3d action \eqref{covariant3daction} with the chirality $\mathrm{d}x_3$,
\begin{equation}
    S_{3d}[g,\Theta] =  \int_Y \langle\partial_3 k_2,\Theta_1^g\rangle -\langle \partial_3 k_1,\Theta^g_2\rangle +\langle \mu_1(\Theta^g_1),\partial_3(\Theta^g_2)\rangle \,,\label{3daction}
\end{equation}
where $\Theta^g = g\rhd \Theta$. As we have already noted in \S \ref{3dlocalization}, the field $\operatorname{proj}_\ell\Theta = \Theta_3$ is completely absent from the theory, so that we will restrict to gauge transformation parameters $\Gamma$ with no $\dr \ell$-leg. We denote the space of 1-forms with no $\mathrm{d}\ell$-leg by $\Omega_\perp^1(Y)\subset\Omega^1(Y)$.

Let us first begin by characterizing the symmetries of the 3d theory \eqref{3daction}. We define the following derived 2-group algebra $(\mathsf{D}\cL_\perp)_\bullet = C^\infty(Y)\otimes G \times \Omega^\bullet_\perp(Y)\otimes \fh$, and consider the subspace $\mathsf{D}\cL_{3d}^R \subset (\mathsf{D}\cL_\perp)_0$ consisting of elements $(\tilde h,\tilde \Gamma)$ satisfying
   \begin{equation}
    \tilde h^{-1}\partial_1 \tilde h + \mu_1(\tilde \Gamma_1) =\tilde h^{-1}\partial_2 \tilde h + \mu_1(\tilde \Gamma_2) = 0 \quad \text{and} \quad 
    \partial_1 \tilde \Gamma_2 - \partial_2\tilde \Gamma_1 - [\tilde \Gamma_1,\tilde \Gamma_2] =0\,.\label{ec:exgtatzero}
\end{equation}
On the other hand, we let $\mathsf{D}\cL_{3d}^L\subset(\mathsf{D}\cL_\perp)_0$ denote the subspace consisting of elements $(h,\Gamma)$ satisfying
\begin{equation}\label{ec:extgtatinfty}
    h^{-1}\partial_3 h = 0 \quad \text{and}\quad \partial_3\Gamma_1=\partial_3\Gamma_2=0\,.
\end{equation}

\begin{lemma}
Let $\mathsf{D}\mathring{\cL}_{3d}^R\subset \mathsf{D}\cL_{3d}^R$ be the subspace for which the components $\tilde \Gamma\in\Omega^1_\perp(Y)\otimes\ft$ are valued in the maximal Abelian subalgebra $\ft\subset\fh$. The 3d theory \eqref{3daction} is invariant under the following transformations:
    \begin{align}
\label{ec:rightaction}
     g \mapsto  g \tilde h, &\qquad \Theta \mapsto \tilde h^{-1}\rhd \Theta+\tilde \Gamma,\qquad (\tilde h,\tilde \Gamma) \in \mathsf{D}\mathring{\cL}_{3d}^R \,\\
     \label{ec:leftaction}
    g\mapsto h^{-1}g, &\qquad \Theta \mapsto \Theta-(h^{-1}g)^{-1}\rhd \Gamma,\qquad (h,\Gamma)\in\mathsf{D}\cL_{3d}^L\,.
\end{align}
Moreover, these transformations commute.
\end{lemma}
\noindent In other words, \eqref{3daction} is invariant under the action of the derived 2-group $\mathsf{D}\cL_{3d} = \mathsf{D}\mathring{\cL}_{3d}^R\times \mathsf{D}\cL_{3d}^L$.
\begin{proof}
    A simple computation using \eqref{ec:extgtatinfty} shows that the action \eqref{3daction} is invariant under a left-action $(h,\Gamma)\in\mathsf{D}\cL_{3d}^L$. For the right-action $(\tilde h,\tilde\Gamma)\in\mathsf{D}\mathring{\cL}^R_{3d}$  \eqref{ec:rightaction}, one can show using \eqref{ec:exgtatzero} that the action transforms as
    \begin{equation*}
        S_{3d}[g,\Theta]\rightarrow S_{3d}[g,\Theta] + \int_Y \langle \mu_1 \partial_3 \tilde\Gamma_1,\tilde\Gamma_2 \rangle.
    \end{equation*}
    The same computation as \eqref{defectcomputation} then removes the remaining term. 

    Now consider a combination of right-acting $(\tilde h,\tilde \Gamma)$ and left-acting $(h,\Gamma)$ symmetry transformations. Of course, their actions clearly commute on $g$, thus we focus on $\Theta$. We have 
\begin{align*}
    \Theta&\mapsto \tilde h^{-1}\rhd \Theta + \tilde \Gamma \mapsto  \tilde h^{-1}\rhd (\Theta - (h^{-1}g)^{-1}\rhd \Gamma)) + \tilde \Gamma \\
    &=  \tilde h^{-1}\rhd \Theta - (h^{-1}g \tilde h)^{-1}\rhd \Gamma + \tilde \Gamma. 
\end{align*}
On the other hand, we have
\begin{equation*}
    \Theta\mapsto \Theta - (h^{-1}g)^{-1}\rhd \Gamma \mapsto  \tilde h^{-1}\rhd \Theta + \tilde \Gamma - (h^{-1}g\tilde h)^{-1}\rhd \Gamma,
\end{equation*}
where we noted that $g\mapsto g\tilde h$ under a right-action. These quantities indeed coincide.
\end{proof}

In section \S \ref{dgaffinecurrents} we will study the conserved Noether currents corresponding to these symmetries, and analyze their homotopy Lie algebra structure in detail. In particular, we will prove that these currents in fact form a centrally extended affine Lie 2-algebra, in complete analogy with the chiral currents in the 2d WZW model \cite{KNIZHNIK198483}.

\subsection{Origin of the Symmetries}

Having characterized the symmetries of the 3d action, we are now in the position to examine how these are actually inherited from the symmetries of the five-dimensional theory discussed in \ref{sec:gaugesym5d}. In other words, we will show that the defect symmetries $\mathsf{D}\cL_\text{sym}$ defined in \eqref{defectgaugeconstraints} descend to the 3d symmetries $\mathsf{D}\cL_{3d}$. 

Recall that in \S \ref{sec:gaugesym5d} we studied the transformation properties of the 5d gauge fields $(A,B)$, whereas in \S \ref{sec:fieldreparam} we studied those of the reparametrisation fields $(A',B';\hat g, \hat \Theta)$. We will call the former {\textbf external} 2-gauge transformations to distinguish them from the latter, which we have dubbed {\textbf internal}. The action of the external 2-gauge transformations on the reparametrization fields is characterized by the following
\begin{proposition}
\label{prop:externalsym}
    The action of an external 2-gauge transformation $(\hat h,\hat \Gamma)$\footnote{We have slightly modified the notation from $(h,\Gamma)$ to $(\hat h,\hat \Gamma)$ for convenience.} given by \eqref{ec:2gaugetr1} induces an action on the reparametrisation fields $\hat g$ and $\hat \Theta$ given by 
    \begin{equation}
     \label{ec:extgt123}
     \hat g \mapsto \hat g \hat h \,,\quad \hat \Theta \mapsto h^{-1}\rhd \hat \Theta +\hat \Gamma \,,
\end{equation}
    and leaves the fields $A'$ and $B'$ invariant. 
\end{proposition}

\begin{proof}
    The proof follows by simply replacing \eqref{ec:extgt123} in \eqref{ec:reparA} and \eqref{ec:reparB} while keeping $A'$ and $B'$ unchanged. 
\end{proof}

We summarize the action of both external and internal gauge transformations on the fields in table \ref{symmtab}. The "mixed" column is there to simply emphasize that, as the fields $(\hat g,\hat \Theta)$ transforms under both internal and external transformations (propositions \ref{prop:internalsym} and \ref{prop:externalsym}), one must in general perform both {\it simultaneously} in order to keep them fixed.

\begin{table}[h]
    \centering
    \begin{tabular}{c|c|c|c}
         & External & Internal & Mixed \\
         \hline
      $(A,B)$   & $\checkmark$ & $\times$  & $\checkmark$ \\ 
        \hline
        $(A',B')$ & $\times$ & $\checkmark$ & $\checkmark$ \\ 
        \hline
        $(\hat g,\hat \Theta)$ & $\checkmark$ & $\checkmark$ & $\times$
    \end{tabular}
    \caption{A table summarizing how different types of symmetries act on the reparameterized fields $(A,B) = (A',B';\hat g,\hat\Theta)$. A check mark $\checkmark$ indicates that the fields under consideration transforms non-trivially, while a cross $\times$ indicates that the fields do not transform. } 
    \label{symmtab}
\end{table}

In \S \ref{sec:fieldreparam} we have shown that for each field configuration $(A,B)$ in h2CS, one can use the internal symmetries to find a \textbf{Lax presentation} $(A,B) = (L,H;\hat g,\hat \Theta)$ such that $(L,H)$ has no $\bar z$-components, and that $(\hat g,\hat \Theta)\vert_{z=\infty}=(1,0)$. We then used this Lax presentation to construct the $S_{3d}$ in \S \ref{3dlocalization}. The goal now is to show that this Lax presentation is stable under an external 2-gauge transformation \eqref{ec:2gaugetr1}. For simplicity, we shall work with the Lagrangian subspace $\mathcal{L}_{\mathrm{d}x_3}$ \eqref{ec:bconditions1}, but our computations can be generalized.

Recall that in \S \ref{3dlocalization}, using the equations of motion and the boundary conditions, we have argued that $L_3,H_{31},H_{32}=0$ are trivial globally. Hence, we look for an external gauge transformation $(h,\Gamma)$ which preserves these conditions, together with $(\hat g,\hat \Theta)\vert_{z=\infty}=(1,0)$. Following table \ref{symmtab} there are two different cases: 

\begin{enumerate}
    \item {\textbf Case 1:} As the external symmetries will transform $(\hat g,\hat\Theta)$, in order to preserve the constraint at $z=\infty$ we may simply ask  $(\hat h,\hat \Gamma)\vert_{z=\infty}=(1,0)$ to be trivial there. This is equivalent to simply doing a 2-gauge transformation in a neighborhood of $z=0$ which will not affect the fields at $z=\infty$.
    
    \item {\textbf Case 2:} Now suppose we take $(\hat h,\hat \Gamma)$ to be generic. In order to preserve $(\hat g,\hat \Theta)\vert_{z=\infty}=(1,0)$, we must then simultaneously perform an internal symmetry $(u,\Lambda)$ to compensate. However, since $(u,\Lambda)$ transforms the reparameterization fields, they must be holomorphic in order to preserve the conditions $L_{\bar z}=0$ and $H_{i\bar z}=0$. As such, they, too, are constants on $\mathbb{C}P^1$ with values equal to
    \begin{equation*}
        u=\hat h\vert_{z=\infty},\qquad \Lambda = \hat\Gamma\vert_{z=\infty}\,.
    \end{equation*}
    Moreover, if we want these transformations to preserve the global condition $L_3,H_{31},H_{32}=0$ of the Lax presentation $(L,H)$ we need
    \begin{align*}
        0&=u^{-1}\partial_3 u + \mu_1\Lambda_3 \\
        0&=\partial_{i}\Lambda_3 - \partial_3\Lambda_{i} + \mu_2(A_{i},\Lambda_3) - \mu_1(A_3,\Lambda_{i}) - [\Lambda_{i},\Lambda_3] \,, \quad i=1,2\,.
    \end{align*}
    
    Consider the second equation. The only object that depends on $\mathbb{C} P^1$ is the field $A$. As such, due to the boundary conditions $A_3\vert_{z=\infty}=0$ and $A_{1,2}\vert_{z=0}=0$, we must in fact have
    \begin{equation*}
        \partial_{i}\Lambda_3 - \partial_3\Lambda_{i} - [\Lambda_{i},\Lambda_3] =- F_{i3}(-\Lambda) = 0 \,,\quad i=1,2
    \end{equation*}    
    everywhere, where $F(\Lambda) = \dr\Lambda + \tfrac{1}{2}[\Lambda
    ,\Lambda]$. On the other hand, the boundary conditions satisfied by $(\hat h,\hat \Gamma)\vert_{z=\infty}$ read
    \begin{equation}
    0=\partial_{i}\hat\Gamma_3 - \partial_3\hat\Gamma_{i} + \mu_2(A_{i},\hat\Gamma_3) - [\hat\Gamma_{i},\hat\Gamma_3]\vert_{z=\infty} = -F_{i3}(-\hat\Gamma) + \mu_2(A_{i},\hat\Gamma_3)\vert_{z=\infty} \,,\quad i=1,2.
    \end{equation}
    But if $(\hat h,\hat\Gamma)\vert_{z=\infty} = (u,\Lambda)$ in order to preserve the condition $(\hat g,\hat \Theta)\vert_{z=\infty} = (1,0)$, then the flatness condition $F(-\Lambda)_{i3}=0$ satisfied by $(u,\Lambda)$ implies
\begin{equation}\label{crowningachievement}
        \mu_2(A_{i},\hat\Gamma_3)\vert_{z=\infty} =0 \,,\quad i=1,2.
    \end{equation}
\end{enumerate}
Clearly, this condition \eqref{crowningachievement} is supplanted by the constraint $\hat\Gamma_3\vert_{z=\infty}=0$ in the defect symmetries $\mathsf{D}\cL_\text{sym}$ \eqref{defectgaugeconstraints}. It is therefore a {\it 2-subgroup} of the external 2-gauge transformations that preserve the Lax presentation!

\subsection{3d Symmetries from 5d Symmetries}

Having analyzed the action of both the external and internal 2-gauge transformations on the Lax presentation, we are ready to describe how these symmetries combine to give the symmetries of the three-dimensional action.

Given that we have obtained $S_{3d}$ by localizing h2CS at $z=0$, one may expect that only the component $\mathsf{D}\cL_\text{def}^0$ of the derived defect 2-group $\mathsf{D}\cL_\text{def} = \mathsf{D}\cL_\text{def}^0 \times \mathsf{D}\cL_\text{def}^\infty$ contributes to the symmetries of $S_{3d}$. Indeed, \eqref{ec:exgtatzero} coincides exactly with the boundary conditions \eqref{ec:gtbczero} that defines $\mathsf{D}\cL_\text{def}^0.$ However, we shall see that the issue is more subtle than that.

Let $\big(\hat h,\hat \Gamma\big)$ be a 2-gauge transformation localized around $z=0$, such that $\big(\hat h,\hat \Gamma\big) = (1,0)$ in a small open neighborhood around $z=\infty$ (cf. "Case 1" analyzed above). Provided such a finite 2-gauge transformation satisfies the boundary condition \eqref{ec:gtbczero}, we can take a limit $\big(\hat h,\hat\Gamma\big)\xrightarrow[z\rightarrow 0]{} \big(\tilde h,\tilde \Gamma\big)$ to arrive at a symmetry transformation which is precisely \eqref{ec:rightaction}. The characterization \eqref{defectgaugeconstraints} then tells us that $\tilde \Gamma_1,\tilde \Gamma_2$ are valued in the maximal Abelian subaglebra $\ft\subset\fh$ in order to preserve the bulk h2CS action. 

\medskip

Now where does the left-acting symmetries \eqref{ec:leftaction} come from? To answer this question, we turn our attention to those 2-gauge transformations $(\hat h',\hat \Gamma')$ that are localized around $z=\infty$ and satisfying the boundary conditions \eqref{ec:gtbcinfty}. As we have already mentioned in the "Case 2" analysis above, the subtlety here is that such transformations must preserve the values of the bulk fields $(\hat g,\hat \Theta)$ at $z=\infty$. If $(\hat h',\hat \Gamma')=(1,0)$ is trivial in a small neighborhood around $z=0$, this forces us to simultaneously perform an {\it internal} gauge transformation in order to compensate.

We thus perform an external and internal transformation simultaneously 
\begin{align}
    \hat g &\mapsto u^{-1}\hat g \hat h'\\
    \hat \Theta &\mapsto  \hat h'^{-1}\rhd \hat\Theta - (u^{-1}\hat g\hat h')^{-1}\rhd\Lambda + \hat \Gamma'\,. 
\end{align}
At the puncture $z=\infty$, to preserve  $\hat g|_{z=\infty}=1$ we must fix $u=\hat h'\vert_{z=\infty}$. This choice of $u$ is $\mathbb{C}P^1$-independent and compatible with preserving the condition $L_{\bar z}=0$. Now, if we evaluate at the puncture $z=0$, recalling $\hat h'|_{z=0}=1$,  we find
\begin{equation}
    \hat g \mapsto u^{-1} \hat g|_{z=0} = \hat h'^{-1}|_{z=\infty} g
\end{equation}
where we have used the fact that $u=\hat h'|_{z=\infty}$. Similarly, at the puncture $z=\infty$ we have for $\hat \Theta$
\begin{align}
    \hat \Theta\vert_{z=\infty}&\to \hat h'^{-1}|_{z=\infty}\rhd \hat \Theta\vert_{z=\infty} - (u^{-1}\hat g\vert_{z=\infty}\hat h'\vert_{z=\infty})^{-1}\rhd\Lambda  +\hat \Gamma'\vert_{z=\infty} \nonumber\\
    & = - \Lambda + \hat \Gamma'\vert_{z=\infty}. 
\end{align}
Thus, we conclude that we must take $\Lambda =\hat \Gamma'|_{z=\infty}$ which, again, is a $\mathbb{C}P^1$-independent choice, compatible with preserving the condition $H_{i\bar z}=0$. Now, if we evaluate at $z=0$ we find, recalling again $(\hat h',\hat \Gamma')\vert_{z=0} = (1,0)$ by hypothesis,
\begin{align}
    \hat \Theta\vert_{z=0}&\to \hat h'^{-1}|_{z=0}\rhd \hat \Theta\vert_{z=0} - (u^{-1}|_{z=0}\hat g\vert_{z=0}\hat h'\vert_{z=0})^{-1}\rhd\Lambda + \hat \Gamma'\vert_{z=0} \nonumber\\
    & =\Theta - (u^{-1} g)^{-1}\rhd\Lambda = \Theta - (\hat h'\vert_{z=\infty}^{-1}g)^{-1}\rhd \hat \Gamma'\vert_{z=\infty},\nonumber
\end{align}
which indeed coincides with the left-action \eqref{ec:leftaction} such that $(\hat h',\hat\Gamma')\xrightarrow[z\rightarrow\infty]{} (h,\Gamma)$. To summarize, we have the following.

\begin{theorem}\label{descending}
    Let $-_\perp: \Omega^1(Y)\rightarrow\Omega^1_\perp(Y)$ denote the map that projects out the $\mathrm{d}x_3$-components of the 1-forms on $Y$. The 5d defect symmetries $\mathsf{D}\cL_\text{sym}$ descends to symmetries $\mathsf{D}\cL_{3d}$ of the 3d theory \eqref{3daction} through the diagram
\[\mathsf{D}\cL_\text{sym} \cong\begin{tikzcd}
	{\mathsf{D}\cL_\text{sym}^0} & {\mathsf{D}(\cL_\text{sym}^0)_\perp} & {\mathsf{D}\mathring{\cL}^R_{3d}} \\
	{\mathsf{D}\cL_\text{sym}^\infty} & {\mathsf{D}(\cL_\text{sym}^\infty)_\perp} & {\mathsf{D}{\cL}^L_{3d}}
	\arrow["{-_\perp}", from=1-1, to=1-2]
	\arrow["\times"{marking, allow upside down}, draw=none, from=1-1, to=2-1]
	\arrow["{z\rightarrow 0}", from=1-2, to=1-3]
	\arrow["\times"{description}, draw=none, from=1-2, to=2-2]
	\arrow["\times"{description}, draw=none, from=1-3, to=2-3]
	\arrow["{-_\perp}", from=2-1, to=2-2]
	\arrow["z\rightarrow\infty", from=2-2, to=2-3]
\end{tikzcd} = \mathsf{D}\cL_{3d} \]
\end{theorem}

We emphasize once again that our analysis above holds for any member $\mathcal{L}_\ell$ in the $S^2$-family of Lagrangian subspaces, hence there is a map
\begin{equation*}
    \mathsf{D}\cL_\text{sym}(\ell)\rightarrow \mathsf{D}\cL_{3d}(\ell)
\end{equation*}
that associates a 5d defect symmetry to a 3d global symmetry for each $\ell\in S^2$.

\paragraph{Misaligned chirality.} To conclude this section, we mention briefly some points of interest when the THF $\mathrm{d}x_3$ of $Y$ does {\it not} align with the chirality. WLOG suppose we take, say, $\ell = w$. In this case, the boundary conditions at $z=0$ would read $$\tilde h^{-1}\partial_{\bar w}\tilde h+\mu_1(\tilde\Gamma_{\bar w})\vert_{z=0}=0,\qquad \tilde h^{-1}\partial_3 \tilde h+\mu_1(\tilde \Gamma_3)\vert_{z=0}=0,\qquad \partial_{\bar w} \tilde \Gamma_3 - \partial_3\tilde \Gamma_{\bar w} - [\tilde\Gamma_{\bar w},\tilde \Gamma_3]\vert_{z=0}=0,$$ while those at $z=\infty$ would read $$h^{-1}\partial_wh+\mu_1\Gamma_w\vert_{z=\infty}=0,\qquad \partial_w\Gamma_{\bar w}\vert_{z=\infty} = \partial_w\Gamma_3\vert_{z=\infty} = 0.$$ 

Here, we see that if $\mu_1=0$, then the boundary conditions at $z=0$ forces $\iota_M^*h\vert_{z=0}$ to merely be {\it holomorphic}, rather than being constant. Similarly, the boundary conditions at $z=\infty$ states that $\iota_M^*\Gamma_w\vert_{z=\infty}$ is \textit{antiholomorphic}. The charges associated to these symmetries therefore acquire holomorphicity properties in the misaligned case. This echoes the statement of Hartog's theorem \cite{Alfonsi_2023}, which forces the negative modes/anti-holomorphic charges in higher-dimensional chiral currents to be split from the positive/holomorphic ones.

\section{Conservation and bordism invariance of the 2-holonomies} \label{sec:holonomies}

In \S \ref{3dift} we constructed a three-dimensional action, and we proved in \textbf{Proposition} \ref{extended2flat} that it's equations of motion are equivalent to the fake- and 2-flatness of $(L,H)$. The aim of this section is to show that this allows for the construction of conserved quantities. 

As the discussion will become quite technical, we begin with a brief informal reminder of how this works in the two-dimensional setting. Consider a theory defined on a cylinder $\Sigma = S^1\times \mathbb{R}$, and assume that its equations of motion can be recast in terms of a flatness equation for some connection $L=L_t \dr t + L_x \dr x$. Conserved quantities can be constructed as follows.

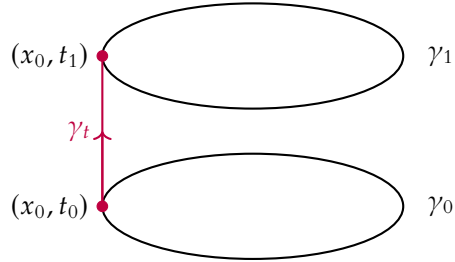
\begin{figure}[h]
    \centering
    \begin{tikzpicture}
        \draw[thick] (0, 1) ellipse (2cm and 0.7cm);

        \draw[thick] (0, -1) ellipse (2cm and 0.7cm);

        \draw[thick, purple] (-2, -1) -- (-2, 1);
        \draw[thick, purple, ->] (-2, -1) -- (-2, 0);  

        \node at (2.5, 1) {$\gamma_1$};
        \node at (2.5, -1) {$\gamma_0$};
        \node[purple] at (-2.3, 0) {$\gamma_t$};

        \node at (-2.7, 1) {$(x_0,t_1)$};
        \node at (-2.7, -1) {$(x_0,t_0)$};

        \filldraw[purple] (-2, 1) circle (2pt);
        \filldraw[purple] (-2, -1) circle (2pt);
    \end{tikzpicture}
    \caption{Schematically, we may understand the path concatenation as follows. We start at $(x_0,t_1)$ we go back to $(x_0,t_0)$ with $\gamma_t^{-1}$, then we go around $(x_0,t_0)$ with $\gamma_0$ and then we go back to $(x_0,t_1)$ with $\gamma_t$. This defines a loop based at $(x_0,t_1)$.}
    \label{fig:pathconc}
\end{figure}

Given a loop $\gamma$ on $\Sigma$, the 1-holonomy of the connection $L$, which is the solution to the parallel transport along $\gamma$, is given by the path-ordered exponential 
\begin{equation}
\label{ec:1holonomy}
    W_\gamma = P \exp \left(-\int_\gamma L\right) \,.
\end{equation}
The flatness of $L$ implies that the 1-holonomy is independent of the path, as long as the endpoints are kept fixed. Formally, we say that it depends solely on the homotopy class relative base point. Intuitively, this can be understood, in the Abelian case, through Stokes theorem. Indeed, if $\gamma_0$ and $\gamma_1$ are two paths with the same endpoints and $\Gamma$ is the surface defined by the smooth homotopy from $\gamma_0$ to $\gamma_1$ we have
\begin{equation}
    \int_{\gamma_0} L - \int_{\gamma_1}L = \int_\Gamma \mathrm{d}L = 0 \quad \Longrightarrow \quad W_{\gamma_0} = W_{\gamma_1}\,.
\end{equation}
We can use this invariance to construct conserved quantities. Indeed, let us consider at a time $t_0$ a loop $\gamma_0$ starting and ending at $x_0$, and at a later time $t_1$ a loop $\gamma_1$ starting and ending at $x_0$ as well. Note that since we are at two different time slices, $\gamma_0$ and $\gamma_1$ are not homotopical relative base point and thus $W_{\gamma_0}\neq W_{\gamma_1}$. However, we can  consider a path $\gamma_{t}$ connecting $(x_0,t_0)$ with $(x_0,t_1)$, such that the path concatenation $\gamma_t\ast\gamma_0\ast \gamma_t^{-1}$ is a loop based at $(x_0,t_1)$, see fig. \ref{fig:pathconc}.  

Given that $\gamma_t\ast\gamma_0\ast \gamma_t^{-1}$ and $\gamma_1$ are homotopical relative base point, flatness of the connection implies $W_{\gamma_t\ast\gamma_{0}\ast \gamma_t^{-1}} = W_{\gamma_1}$. Moreover, 1-holonomies are multiplicative under path concatenation, namely, 
\begin{equation}
W_{\gamma_t\ast\gamma_{0}\ast \gamma_t^{-1}}=  W_{\gamma_t}W_{\gamma_0}W_{\gamma_t}^{-1}= W_{\gamma_1}\,.
\end{equation}
Hence, for each invariant character $\chi_k$ of the group, we find that invariance of $\chi_k$ implies
\begin{equation}
    \chi_k (W_{\gamma_0}) = \chi_k (W_{\gamma_1})\,.
\end{equation}
Thus the quantity $M_k = \chi_k (W_{\gamma_0})$ is conserved.

We now wish to generalize this story to the three-dimensional setting. The mathematics involved in this procedure are subtle, but the essence is the same than in the $2$d case. We will construct $2$-holonomies, which are a higher analogue of \eqref{ec:1holonomy}, defined as a surface ordered exponential of an operator which depends on both $L$ and $H$. We will then show that fake and 2-flatness imply that the $2$-holonomy depends solely on the homotopy class relative boundary of the surface, and we will then use this fact to construct conserved quantities.

Thus, we start with our action $S_{3d}$ given in \eqref{covariant3daction}, whose equations of motion \eqref{1flat?} and \eqref{2flat} imply the fake and $2$-flatness of the $2$-connection $(L,H)$, as shown in \textbf{Proposition} \ref{extended2flat}. We will construct the aforementioned $2$-holonomies as parallel transport operators on the \textit{loop space fibration} $\Omega Y \rightarrow Y$ on $Y$, following \cite{Alvarez:1997ma}. However, our treatment differs from theirs in that our 2-dimensional surface holonomies can be non-Abelian, and are sensitive to the boundary data. 

Associated to any flat $\mathbb{G}$-connection are such 2-holonomies that we construct, and we will prove their conservation and homotopy invariance in a general context. Since we know, from \textbf{Proposition \ref{extended2flat}}, that the currents $(L,H)$ we have obtained specifically from $S_{3d}$ are flat $\mathbb{G}$-connections, these are sufficient to point towards the integrability of $S_{3d}$. However, under certain circumstances, we can in fact prove a much stronger "invariance" property that these specific currents satisfy.

In the following, we shall first review some of the key homotopical aspects of the theory. Our surface holonomies/2-holonomies are constructed as a parallel transport operator defined on the space $\Omega Y$ of loops $S^1\rightarrow Y$ based at a point $y_0\in Y$ \cite{Alvarez:1997ma}. These will then be put together to prove bordism invariance in the aligned case.

\subsection{Surface holonomies from connections on loop spaces}\label{alvarez}
Given a 2-connection $(L,H)$, we can define $\mathbb{G}$-valued integrable charges $(V,W)$ satisfying the following parallel transport equations 
\begin{align}
    0 &= \frac{\dr W}{\dr t} + L_i \frac{\dr\gamma^i}{\dr t}W \implies  W = P\exp\left(-\int_\gamma L\right) \label{paratrans1}\\ 
    0&= \frac{\dr V}{\dr\tau} + \left(\int_0^1 \mathrm{d}t \,\mathcal{A}_i(t)  \frac{\dr\gamma'^i(t)}{\dr\tau}\right)V\implies V = P\exp \left(-\int_{\gamma'}\cA\right)\label{paratrans2},
\end{align}
where $\gamma: [0,1]\rightarrow Y$ is a path on $Y$ based at $y_0\in Y$ and $\gamma':[0,1]\rightarrow PY$ is a \textit{path on path space} based at the constant path $\gamma_0\in PY$ at $y_0$\footnote{This means that $\gamma_0(t)=y_0$ for all $y\in[0,1]$.}. These $(y_0,\gamma_0)$ shall be the initial data for our parallel transport equations \eqref{paratrans1}, \eqref{paratrans2}. 

Geometrically, the first equation describes the usual parallel transport on $Y$, while the second describes a parallel transport over a path $\gamma'$ in the {\it path space} $PY$ of $Y$. The quantities $\cA_i(t)$ are to be understood as the components of a connection on $PY$, defined locally (at a point $\gamma \in PY$) by the formula
\begin{equation}\label{surfacetransport}
   \cA = \int_0^1 \mathrm{d}t\, \mathcal{A}_i(t)\, \delta \gamma^i(t)= \int_0^1\mathrm{d}t\, W_\gamma^{-1}\rhd H_{ij}(t)\,\dot \gamma^j \,\delta \gamma^i(t)\,.
\end{equation}
Its holonomy 
\begin{equation}
    P\exp\left(-\int_{\gamma'}\cA\right) = S\exp\left(-\int_\Sigma (W^{-1}\rhd H)\right)
\end{equation}
serves as the definition of the surface-ordered exponential \cite{Yekuteli:2015}. The above parallel transport equations \eqref{paratrans1}, \eqref{paratrans2} are inspired from \cite{Alvarez:1997ma}, but notably, our approach does not enforce the 1-flatness of $L$. As we shall see, this is necessary for our 2-holonomies to keep track of its boundaries.

\subsubsection{Surface holonomies}

We now describe the construction of surface holonomies presented in \cite{Alvarez:1997ma}. Let $P\rightarrow Y$ be a principal $G$-bundle. We can induce a $G$-bundle $\mathcal{P}=\pi^*P\rightarrow PX$ on the path space simply by pulling-back $P$ along the path space fibration $\pi: PY\rightarrow Y$, sending a path $\gamma$ to its endpoint $\gamma(1)$. In \cite{Alvarez:1997ma}, it was shown that a $G$-connection $\mathcal{A}\in \Omega^1(PY)\otimes\fg$ of $\mathcal{P}$ can be constructed from a $\fg$-valued 2-form $B$ on $Y$ transforming under the adjoint representation of $G$. 

This construction is given as follows. Let $\gamma\in PY$ be a path and let $W_\gamma$ denote the parallel transport operator defined by a $G$-connection $A\in\Omega^1(Y)\otimes\fg$ on $Y$ from $y_0$ to a point on the path $\gamma$. The Lie algebra valued 1-form
\begin{equation}
    \mathcal{A} =\int_0^1 \dr t\,\, (\operatorname{Ad}_{W_{\gamma(t)}}^{-1}B_{ij}(\gamma(t))) \dot{\gamma}^i\delta\gamma^j(t)  
\label{pathconn}
\end{equation}
has the correct transition laws, and hence serves as a local $G$-connection on $PY$. Here, $\delta \gamma^j(t)$ forms a basis for the cotangent space $T^*_\gamma PY$ at $\gamma\in PY$ and can be understood as the lift of the usual basis $\dr x^i$ on $T^*Y$ along the given path $\gamma$. The curvature of this $G$-connection $\mathcal{A}$ is thus computed as $\mathcal{F} = \delta\mathcal{A} + \frac{1}{2}[\mathcal{A},\mathcal{A}]$, which in terms of the 2-form $B$ reads  \cite{Alvarez:1997ma}
\begin{align}
    \mathcal{F} &=\int_0^1 \dr t\, \operatorname{Ad}_{W_{\gamma(t)}}^{-1}(\dr_AB)_{ijk} \dot{\gamma}^i\delta\gamma^j\wedge \delta\gamma^k \nonumber \\
    &\qquad -~\int_0^1\dr t\int_0^1\dr\tau  \big[\operatorname{Ad}_{W_{\gamma(t)}}^{-1}F_{ik}(\gamma(t)), \operatorname{Ad}_{W_{\gamma(\tau)}}^{-1} B_{jm}(\gamma(\tau))\big] \frac{\dr \gamma^i}{\dr t} \frac{\dr\gamma^j}{\dr\tau}\delta\gamma^k\wedge\delta\gamma^m \nonumber\\
    &\qquad +~\int_0^1\dr t\int_0^1\dr\tau  \big[\operatorname{Ad}_{W_{\gamma(t)}}^{-1}B_{ik}(\gamma(t)), \operatorname{Ad}_{W_{\gamma(\tau)}}^{-1}B_{jm}(\gamma(\tau))\big] \frac{\dr\gamma^i}{\dr t} \frac{\dr\gamma^j}{\dr\tau}\delta\gamma^k\wedge\delta\gamma^m. \label{pathcurv}
\end{align}
Due to this form of the curvature, we see that the $G$-connection $\mathcal{A}$ is in general not flat. 

We can then pull $\mathcal{A}$ back along the inclusion $\Omega Y\hookrightarrow PY$ to obtain a $G$-connection on the space $\Omega Y$ of loops based at $y_0$. Closed 2-manifolds (ie. those without boundary) smoothly submersed in $Y$ can be equivalently understood as a loop $\tilde \gamma$ in $\Omega Y$ based at the constant path $\gamma_0:t\mapsto y_0$. Let this loop $\tilde\gamma $ be obtained from lifting a loop $\gamma'$ on $Y$, then the argument for the flatness $F=0$ of the $G$-connection $A$ on $Y$ runs as follows: the values of $\mathcal{A}$ at the beginning and end of the loop are related by an adjoint action,
\begin{equation}\label{looptransport}
    \mathcal{A}[\tilde\gamma(1)] = \operatorname{Ad}_{V_{\gamma'}}^{-1}\mathcal{A}[\tilde\gamma(0)]\,,\qquad V_{\gamma'} =P\exp\left(-\int_{\gamma'}\mathcal{A}\right)\,.
\end{equation}
But given $\tilde\gamma$ is a loop, $\tilde\gamma(1) = \tilde\gamma(0)$ hence $\mathcal{A}$ is independent of the parameterization of the surface only when $W_{\gamma'} = 1$, namely $A$ is a flat $G$-connection.

This flatness condition on $A$ is very restrictive, hence we seek a generalization of this formalism of \cite{Alvarez:1997ma} that relaxes it. We shall do this in the following through the theory of \textit{higher groupoids}, and use it to prove several structural theorems about our improved 2-holonomies.

\subsubsection{2-holonomies as a map of 2-groupoids}\label{holonomies}
We now describe the construction of the higher holonomies \eqref{paratrans1}, \eqref{paratrans2} from the perspective of 2-groupoids. We point the reader to the literature \cite{Whitehead:1941,Brown,Baez:2004,Martins:2007,Kapustin:2013uxa,Ang2018,Bullivant:2017qrv,Wagemann+2021,Bochniak_2021} on 2-groups and their relevance to geometry and homotopy theory.

Let $\mathfrak{G}=\mathfrak{h}\rightarrow\mathfrak{g}$ denote a Lie 2-algebra and let $\mathbb{G}=\mathsf{H}\rightarrow G$ denote its corresponding Lie 2-group. Suppose $(L,H)$ is a 2-$\mathbb{G}$-connection. Solutions $(V,W)$ to the parallel transport equations \eqref{paratrans1}, \eqref{paratrans1} define a map $\mathsf{2Hol}$ of 2-groupoids \cite{Kim:2019owc}:
\[\begin{tikzcd}
P^2Y\rightrightarrows PY\rightrightarrows Y  \arrow[d]\\
\mathsf{H}\rtimes G\rightrightarrows G\rightrightarrows \ast
\end{tikzcd}.\]
This amounts to the following. Fix $y_0\in Y$ and we let this denote the initial condition $\gamma(0),\gamma'(0)=y_0$. Let $\Omega Y$ be the space of loops on $Y$ based at $y_0$, and $P\Omega Y$ denote the space of maps $\Sigma:[0,1]\times S^1\rightarrow Y$, equipped with source $\Sigma\mapsto \Sigma(0,-)$ and target $\Sigma\to \Sigma(1,-)$ maps, such that $\Sigma(-,0) = \gamma_0$ is the group unit (ie. the constant loop at $y_0\in Y$). We take $\mathsf{2Hol}(\ast) = y_0$, and consider the pointed 2-subgroupoid $P^2_{y_0} Y$ of the double path groupoid $P^2Y\rightrightarrows PY\rightrightarrows Y$ given by $$P\Omega Y\rightrightarrows \Omega Y\rightrightarrows \{y_0\}.$$ The statement is then that $\mathsf{2Hol}_{y_0} = (V,W): P^2_{y_0}Y \rightarrow \mathbb{G}$ is a 2-group homomorphism.

\begin{figure}[ht]
    \centering
    \includegraphics[width=0.75\columnwidth]{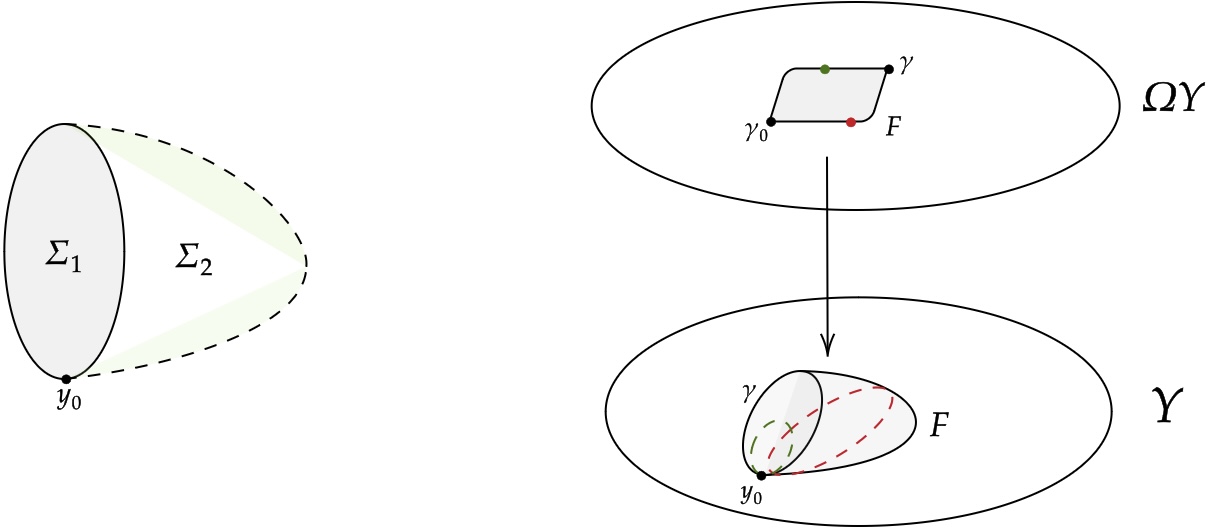}
    \caption{A homotopy relative boundary $F$ between surfaces $\Sigma_1,\Sigma_2\subset Y$ is a 3-manifold with boundary $\Sigma_1\cup_\gamma \overline{\Sigma}_2$. This boundary is the result of $\Sigma_2$ being "capped off" by $\Sigma_1$ at the loop $\gamma=\partial\Sigma_1=\partial\Sigma_2$. It can also be understood as a path homotopy on loop space $\Omega Y$. Recall $\gamma_0$ is the constant loop at $y_0$.}
    \label{bdyhomotopy}
\end{figure}

\begin{definition}
    Let $\Sigma_1,\Sigma_2:[0,1]^2\rightarrow Y$ denote two (smooth) surfaces in $Y$ with the same boundary $\partial\Sigma_1 = \partial\Sigma_2 =\gamma: S^1\rightarrow Y$. A \textbf{homotopy relative boundary} is a (smooth) map $F:[0,1]\times [0,1]^2\rightarrow Y$ such that
    \begin{equation*}
        F(0,-) = \Sigma_1,\qquad F(1,-) = \Sigma_2,\qquad \partial F(t,-) \cong \gamma,
    \end{equation*}
    for all $t\in[0,1]$. See fig. \ref{bdyhomotopy}. We denote homotopy classes of surfaces relative boundary by $[[0,1]^2,Y]_\text{/bdy}$.
\end{definition}

\begin{theorem}\label{relbdy}
    The fake and 2-flatness conditions
    \begin{equation*}
        \dr L + \frac{1}{2}[L, L] - \mu_1(H) = 0,\qquad \dr H + \mu_2(L,H) = 0
    \end{equation*}
    imply the 2-holonomy $V$ descends to a map $[[0,1]^2,Y]_\text{/bdy}\rightarrow \mathsf{H}$ on homotopy classes relative boundary.
\end{theorem}
\begin{proof}
    We consider a submersed 2-manifold $\Sigma\hookrightarrow Y$ with boundary as a 2-mormphsim $\Sigma\in P\Omega Y$ in the 2-group $P^2_{y_0}Y$. WLOG, we can take $\Sigma$ to have source $\gamma_0$ the constant loop on $y_0$ and target $\gamma=\partial\Sigma$ its boundary loop. We shall consider $\Sigma$ equivalently as a "path of loops" $\gamma'(\tau) = \Sigma(\tau,-) \in \Omega Y$ for each $\tau\in [0,1]$.

    The goal now is to prove that fake- and 2-flatness conditions implies that $V_{\Sigma_1} = V_{\Sigma_2}$ whenever $\Sigma_1\simeq\Sigma_2$. For this, we leverage the construction of the $\mathsf{H}$-connection $\mathcal{A}$ on $PY$. This is defined analogous to \eqref{pathconn}, but instead leveraging the group action $\rhd: G\rightarrow\operatorname{Aut}\mathsf{H}$ such that
    \begin{equation*}
        \mathcal{A} =\int_0^1\mathrm{d}t\, \big(W_{\gamma(t)}^{-1}\rhd H_{ij}(\gamma(t))\big)\,\dot \gamma^i\,\delta \gamma^j\,,\qquad W_\gamma = P\exp\left(-\int_\gamma L\right)\,.
    \end{equation*}
    The holonomy is then constructed as usual $V_{\Sigma_1}= P\exp\left(-\int_{\gamma'_1}\mathcal{A}\right)$ where $\gamma'_1:[0,1]\rightarrow \Omega Y$ is the path on loop space corresponding to $\Sigma_1$. A homotopy $F:\Sigma_1\Rightarrow\Sigma_2$ relative boundary can then be understood as a homotopy $[0,1]^2\rightarrow\Omega Y$ between the paths $\gamma_1',\gamma_2'$ such that $\partial F(s,-) \cong \gamma=\partial\Sigma_1=\partial\Sigma_2$ for each $s\in [0,1]$; see fig. \ref{bdyhomotopy}. 
    
    This homotopy $F$ defines a contractible surface $\Gamma\subset\Omega Y$, or equivalently a closed 3-submanifold $S \subset Y$ whose boundary is the gluing $\overline{\Sigma}_2\cup_\gamma \Sigma_1= \overline{\Sigma}_2\#\Sigma_1$ of the two surfaces $\Sigma_1,\Sigma_2$ along $\gamma$, where $\overline{\Sigma}$ denotes the orientation reversal of a surface $\Sigma$. By usual computations, we have (cf. \S 2 of \cite{Alvarez:1997ma})
    \begin{equation}
    \label{Voflgue}
        V_{\overline{\Sigma}_2\cup_\gamma\Sigma_1} = P\exp\left(-\int_{\gamma_2'^{-1}\ast\gamma'_1}\mathcal{A}\right) = P\exp\left(-\int_\Gamma W^{-1}\rhd \mathcal{F}\right),
    \end{equation}
    where $\mathcal{F}=\delta\mathcal{A}+ \frac{1}{2}[\mathcal{A},\mathcal{A}]$ is the curvature, which reads analogously to \eqref{pathcurv} but now instead has the form
    \begin{align}\label{fakecurve}
        \mathcal{F}  &= \int_0^1 \dr t\, \operatorname{Ad}_{W_{\gamma(t)}}^{-1}(\dr_LH(t))_{ijk}\, \frac{\dr\gamma}{\dr t}^i\delta\gamma^j(t)\wedge \delta\gamma^k(t) \\
        & -~\int_0^1\dr t\int_{0}^1 \dr t' \, \left[\operatorname{Ad}_{W_\gamma(t)}^{-1}(F(L(t))_{ik} - \mu_1(H_{ik}(t)))\right] \rhd (W_{\gamma(t')}^{-1}\rhd H_{jl}(t'))\frac{\dr \gamma^i}{\dr t}\frac{\dr \gamma^j}{\dr t'}\delta \gamma^k(t)\wedge\delta\gamma^{l}(t')
    \end{align}
    where we have used the Peiffer identity. This quantity vanishes precisely when $(L,H)$ satisfies fake- and 2-flatness conditions, whence $V_{\overline{\Sigma}_1\cup_\gamma\Sigma_2} = 1$. By definition, $V_{\overline{\Sigma}_1\#\Sigma_2} = V^{-1}_{\Sigma_1} V_{\Sigma_2}=1$ and hence $V_{\Sigma_1}=V_{\Sigma_2}$ for homotopically equivalent surfaces $\Sigma_1\simeq\Sigma_2$ relative boundary, as desired.

\end{proof}

Note homotopies relative boundary between surfaces without boundary reduces to homotopies in the usual sense.

\begin{corollary}
    The 2-holonomy $V$ is consistent with the Eckmann-Hilton argument.
\end{corollary}
\begin{proof}
    The Eckmann-Hilton argument states that charge operators attached to closed submanifolds of codimension larger than one must have commutative fusion rules \cite{Gaiotto:2014kfa}. This comes from the fact that the concatenation of maps $\Sigma_1,\Sigma_2:S^2\rightarrow Y$, given by connected summation $\Sigma_1\#\Sigma_2$, is commutative up to homotopy: there exists a closed contractible 3-submanifold $S\subset Y$ whose boundary is $\partial S = \overline{\Sigma_2\# \Sigma_1} \cup \Sigma_1\#\Sigma_2$. 
    
    The 2-holonomies $V$ are precisely such 2-codimensional operators charged under $\mathsf{H}$, hence we must prove that
    \begin{equation*}
        V_{\Sigma_1}V_{\Sigma_2} = V_{\Sigma_2}V_{\Sigma_1}
    \end{equation*}
    for closed surfaces $\Sigma_1,\Sigma_2\in P\Omega Y$. Toward this, we consider closed submersed 2-submanifolds as loops $\gamma':S^1\rightarrow \Omega Y$ on loop space, where $\gamma'(0) = \gamma'(1)=\gamma_0$ is the constant loop at $y_0$. They are naturally encoded as automorphisms of $\gamma_0$ in the 2-group $P^2_{y_0} Y$, where the constant loop $\gamma_0\in \Omega Y$ is the identity under path concatenation --- in other words, closed 2-submanifolds are precisely the kernel of the boundary map $\partial:P\Omega Y\rightarrow\Omega Y$. Now $\mathsf{2Hol}$ being a map of 2-groupoids means in particular that
    \begin{equation*}
        W\circ\partial = \bar\mu_1\circ V,
    \end{equation*}
    whence on closed 2-submanifolds $\Sigma \in \operatorname{ker}\partial$ the surface holonomym is Abelian,
    \begin{equation*}
        1=W_{\gamma_0} = \bar\mu_1(V_\Sigma)\implies V_\Sigma\in\operatorname{ker}\bar\mu_1\subset\mathsf{H},
    \end{equation*}
     as desired.
    
\end{proof}

We recognize that, in the approach where $\mathcal{A}$ was obtained from pulling-back the $G$-connection on $Y$, \S 3.2.2 of \cite{Alvarez:1997ma} has derived the condition $[B,B]=0$ as part of local integrability. This restricts one to the case where the structure group is a semi-direct product $R[1]\rtimes G$ with an Abelian $G$-module $R$ in degree-(-1), which corresponds precisely to the case $\mu_1=0$ in our setup.

\subsubsection{Whiskering of 2-holonomies}\label{sec:whisker}
\begin{figure}[ht]
    \centering
    \includegraphics[width=0.75\columnwidth]{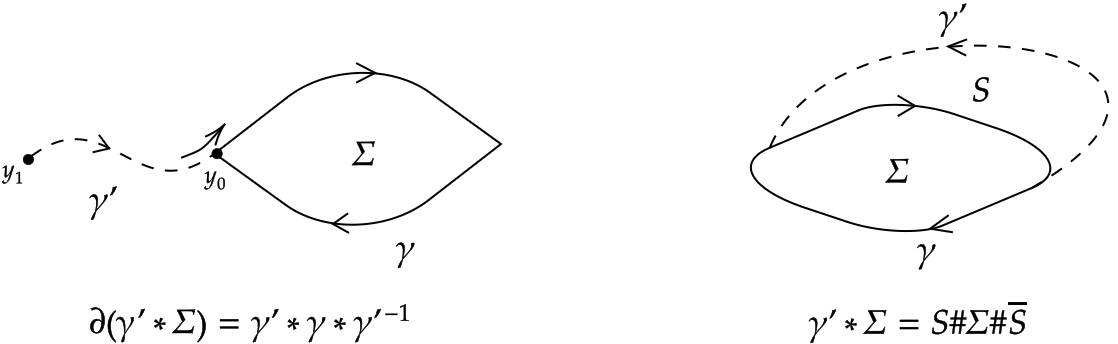}
    \caption{The pictorial representation of the operation of whiskering by $\gamma'$ (dashed). If $\gamma'$ is a path, then it serves to move the base point $y_0$ of the surface $\Sigma$. If $\gamma'$ is a loop bounding $S$, then it serves to change the shape of $\Sigma$. Recall whiskering involves a "walking back".}
    \label{whiskerdiag}
\end{figure}

Now let us examine the operation of \textbf{whiskering} \cite{Baez:2004}. This is an operation in a 2-groupoid by which 1-morphisms are composed with 2-morphisms in order to change the boundary of the 2-morphisms. Geometrically in the double path 2-groupoid $P^2Y$, this can be understood as the gluing of a surface $\Sigma\in P^2 Y$ with a path $\gamma\in PY$, which we consider as an "infinitely thin surface" $1_\gamma\in P^2Y$. The boundary of the composed surface is then $$\partial(\gamma\ast\Sigma) = \gamma\ast\partial\Sigma\ast\gamma^{-1},$$ where $\ast$ is the concatenation of paths in $PY$. This serves to change the base point of $\Sigma$ from $y_0$ to $y_1$; see fig. \ref{whiskerdiag}.

In the 2-group $\mathbb{G}$, this operation is defined by a group action $\rhd: G\rightarrow\operatorname{Aut}\mathsf{H}$ satisfying the equivariance condition $\bar\mu_1(x\rhd y) = x\bar\mu_1(y)x^{-1}$ for each $x\in G,y\in\mathsf{H}$. The fact that the holonomies $(V,W)$ define a 2-groupoid map $\mathsf{2Hol}: P^2Y\rightarrow B\mathbb{G}$ means that whiskering is preserved: for each path $\gamma\in P Y$ based at the initial condition $y_0$, we have
\begin{equation}\label{whisker}
    W_{\gamma\ast \Sigma} = V_\gamma \rhd W_\Sigma,\qquad \forall ~\Sigma\in P\Omega Y.
\end{equation}
This is the statement that changing the base point of $\Sigma$ from $y_0$ to the endpoint $y_1$ of $\gamma$ amounts to a parallel transport of the holonomies $V$. This structure is called a \textbf{balloon} in \cite{Yekuteli:2015}.

For (contractible) loops $\gamma\in\Omega Y$ with a non-empty intersection $\gamma\cap \partial\Sigma\neq \emptyset$, we can use the surface $S$ it encloses to deduce $$\gamma\ast \Sigma = S\#\Sigma\# \overline{S},$$ where we have glued copies of $S$ onto $\Sigma$ along the intersection $\gamma\cap \partial\Sigma$; see fig. \ref{whiskerdiag}. Even though $\gamma$ does not change the base point of $\Sigma$, it changes the shape of the boundary $\partial \Sigma$. Since the holonomy $V$ is only invariant under homotopies \textit{relative boundary}, this induces a non-trivial action
\begin{equation*}
    V_{\gamma\ast \Sigma} = W_{\partial S} \rhd V_\Sigma = \bar\mu_1(V_S)\rhd V_\Sigma = \operatorname{Ad}_{V_S}V_\Sigma,
\end{equation*}
where we have used the Peiffer identity. Further, if $S$ is a {\it closed} 2-submanifold (whose boundary $\partial S = \gamma_0$ just the base point $y_0$ given by the constant loop), then $V_S\in\operatorname{ker}\bar\mu_1$ whence $V_{\gamma\ast\Sigma} = V_\Sigma$. 

This means that, in our setup, the holonomy $V_\Sigma$ indeed {does} depend on the way by which we scan the surface $\Sigma$, in so far as $\Sigma$ has boundary. However, this is consistent with the geometry: a homotopy of $\Sigma$ is {\it not} in general a homotopy relative boundary, and the non-commutativity of the surface holonomies $V_\Sigma$ comes precisely from its boundary.

\subsection{Proof of conservation and bordism invariance of \texorpdfstring{$V$}{V}}\label{integrability}
With the above theory in hand, we are then ready to examine the conservation that arises from fake- and 2-flatness. In \S \ref{holonomies}, we proved that the surface holonomy \eqref{paratrans2} defined a group homomorphism
\begin{equation*}
     [[0,1]^2,Y]_\text{/bdy} \rightarrow \mathsf{H},
\end{equation*}
where $[[0,1]^2,Y]_\text{/bdy}$ denotes the double path 2-groupoid $P^2Y\rightrightarrows PY\rightrightarrows Y$ modulo homotopies relative boundary. This is intimately related to the conservation of surface charges based on $V$. To see this, we follow a geometric argument analogous to that given in \cite{Alvarez:1997ma}. 

Let us introduce a transverse (real or holomorphic) foliation $\Phi$, which determines locally a direction that denotes the "time" $\mathrm{d}u$ along the leaves. The goal is to compare the values of the surface charge $V_{\Sigma_0},V_{\Sigma_1}$ localized on surfaces, with diffeomorphic boundaries $\partial\Sigma_0\cong \partial\Sigma_1$, at two time-slices: $\Sigma_0$ at $u=0$ with base point $y_0 = (m_0,0)\in Y$ and $\Sigma_1$ at $u=1$ with base point $y_1 = (m_1,0)\in Y$. Take $\gamma_u:[0,1]\rightarrow Y$ to be a path along the $u$-direction, connecting the two base points $y_0,y_1$. 

Note here crucially that the fact that our surface holonomy defines invariants of homotopy relative boundary does {\it not} imply $V_{\Sigma_0} = V_{\Sigma_1}$, as $\Sigma_0,\Sigma_1$ have different base points, separated in the $u$-direction. As we have described in \S \ref{sec:whisker}, the way to transport the base point of $\Sigma_0$ to $u=1$ is by whiskering along the path $\gamma_u$,
\begin{equation*}
    \tilde V_{\Sigma_0} = W_{\gamma_u}\rhd V_{\Sigma_0}.
\end{equation*}
Now that both $\tilde V_{\Sigma_0}$ and $V_{\Sigma_1}$ share the same base point, homotopy invariance relative boundary then finally allows us to state
\begin{equation*}
    V_{\Sigma_1} = W_{\gamma_u}\rhd V_{\Sigma_0}.
\end{equation*}

This means that our surface holonomies separated in the $u$-direction differ by a whiskering along $u$. Conserved quantities can therefore be obtained from a certain "trace/character" $\mathcal{X}$ of $V_\Sigma$ that is invariant under whiskering,\footnote{Here we make a technical note that "invariance" is usually only up to higher homotopy when categories are involved, hence $\mathcal{X}(W_{\gamma_u}\rhd V_{\Sigma_0})$ is generally merely "isomorphic" to $ \mathcal{X}(V_{\Sigma_0})$. As far as the authors know, there are at least two such notions of homotopy invariant characters on 2-groups: one is the \textit{graded} character described in \cite{Chen:2023integrable} and another is the \textit{categorical} character \cite{Ganter:2006,Bartlett:2009PhD,Sean:private,Huang:2024}.} 
\begin{equation*}
    \mathcal{X}(V_{\Sigma_1}) = \mathcal{X}(W_{\gamma_u}\rhd V_{\Sigma_0}) = \mathcal{X}(V_{\Sigma_0}).
\end{equation*}
These time-independent \textbf{higher monodromy matrices} 
\begin{equation}
    \mathcal{M}_k([\Sigma]) = \mathcal{X}_k(\mathsf{2Hol}([\Sigma]))\,\label{conservedcharges}
\end{equation}
come labelled by invariant characters $\mathcal{X}_k$ of $\mathbb{G}$, as well as {\it free} homotopy classes (ie. orbits of based homotopy classes under base point-changing whiskering operations) $[\Sigma]$ of surfaces relative boundary in $Y$.

\begin{remark}
    Note that in a similar manner to the 2d WZW construction from CS$_4$ \cite{Costello:2019tri}, the 2-Lax connection we find is independent of the spectral parameter $z\in\mathbb{C}$. This prevents us from series expanding the monodromy matrices \eqref{conservedcharges} in powers series in $z$ in order to construct, in this way, an infinite number of independently conserved quantities. Notably, in the case of the 2d WZW model, a spectral parameter dependent Lax connection can be obtained from CS$_4$ theory by performing a suitable limiting procedure \cite{Costello:2019tri}.\footnote{JL thanks Benoit Vicedo for pointing this out.} Starting with the meromorphic $1$-form $\omega=z^{-2}(z-z_0)(z-z_1)\mathrm{d}z$ for fixed $z_0,z_1 \in \mathbb{CP}^1$, together with Dirichlet boundary conditions on the gauge field at the poles, one obtains the Principal Chiral Model with WZ term, with coefficients proportional to $z_0-z_1$ and $z_0+z_1$ respectively. In the $z_0\to 0$ limit, the 2d action becomes the WZW CFT, but remarkably, the Lax connection obtained in this limit is now spectral parameter dependent. We expect that a $\mathbb{C}P^1$-dependent 2-Lax connection can be constructed using a similar limiting procedure, but we leave this for a future work.
\end{remark}

\medskip

In any case, we will now prove in the following that the currents $(L,H)$ in the 3d theory \eqref{covariant3daction}, which are flat $\mathbb{G}$-connections by \textbf{Proposition \ref{extended2flat}}, can in fact be made to satisfy a stronger form of "conservation" and topological invariance, called bordism invariance.

\begin{figure}[h]
    \centering
    \includegraphics[width=0.7\columnwidth]{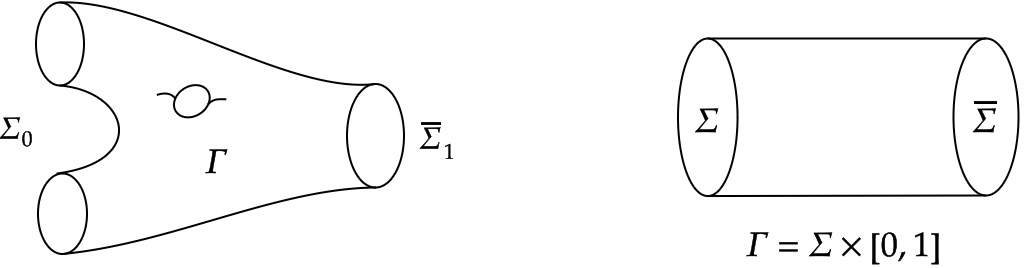}
    \caption{One can in principle define bordisms $\Gamma :\Sigma_0\rightarrow\overline{\Sigma}_1$ with non-trivial topology, or in between surfaces with different number of components, as seen in the left. The right side exhibits the trivial bordism $\Gamma=\Sigma\times [0,1]$ over a surface $\Sigma$.}
    \label{bord}
\end{figure}

\paragraph{Bordism invariance.} By bordism, we mean precisely the following.
\begin{definition}\label{bordbdry}
    A surface $\Sigma\subset Y$ is \textbf{framed} if it is equipped with a trivialization of its normal bundle $N\Sigma \cong TY/T\Sigma$. Let $\Sigma,\Sigma'$ denote two smooth framed 2-manifolds with boundary, a \textbf{ (smooth) framed open bordism} $\Gamma: \Sigma\rightarrow\Sigma'$ is a smooth 3-manifold $\Gamma$ with boundary (framed diffeomorphic to) $\Sigma\coprod \overline{\Sigma}'$, where $\overline{\Sigma}'$ is the orientation reversal of $\Sigma'$. 
\end{definition}
\noindent This is a stronger statement than the conservation of the holonomies. Indeed, bordism invariance is a global statement: conservation can be thought of as being invariant under "very small" bordisms, which is always trivial $\Gamma \cong \Sigma\times [0,1]$. See fig. \ref{bord}.

\begin{remark}
    Conversely, one can always "shrink the cylinder" in a trivial (open) bordism $\Gamma:\Sigma\rightarrow\Sigma'$ to give a homotopy equivalence $\Sigma\simeq\Sigma'$ (relative boundary). One can then ask if these are the only bordisms that do such a thing; for instance, are \textit{$h$-cobordisms} $\Gamma:\Sigma\rightarrow\Sigma'$, where the inclusions $\Sigma,\Sigma'\hookrightarrow\Gamma$ induce homotopy equivalences, homotopic to the trivial bordism, and hence also give rise to homotopy equivalences? This is the so-called \textbf{smooth $h$-cobordism theorem} \cite{Freedman:1990,Smale:1962,Milnor+1965}, and it turned out to be an extremely difficult problem. At dimension three, it is equivalent to the Poincar{\'e} conjecture.
\end{remark}

\begin{figure}[ht]
    \centering
    \includegraphics[width=0.7\columnwidth]{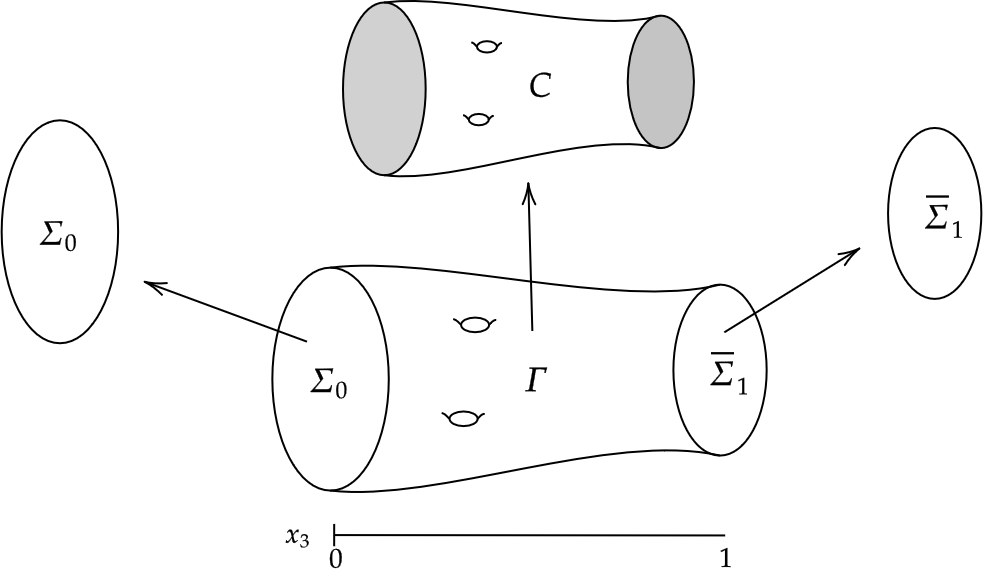}
    \caption{A diagram of a bordism $\Gamma:\Sigma_0\rightarrow \Sigma_1$ and its boundary components $\overline{\Sigma}_1,\Sigma_0,C$.}
    \label{fig:bord}
\end{figure}

Let us first setup the geometry. Let $\Gamma: \Sigma\rightarrow\Sigma'$ denote a bordism between the two surfaces $\Sigma_0,\Sigma_1$, based respectively at $(m_0,0),(m_1,1)\in Y$. The statement of bordism invariance is then given by the condition $V_{\partial\Gamma}=1$. The boundary $\partial\Gamma$ has three pieces: the "bottom cap" $\Sigma_0=\Sigma \times \{0\}$, the "top cap" $\Sigma_1=\Sigma'\times\{1\}$ and the "cylinder tube" $C\cong \partial\Sigma \times\gamma$, hence we must have
\begin{equation}\label{stationarity}
    V_{\overline{\Sigma}_1} V_{C}V_{\Sigma_0} = V^{-1}_{\Sigma_1} V_C V_{\Sigma_0} =(W_\gamma^{-1}\rhd V_{{\Sigma}'}^{-1})V_C V_\Sigma=1,
\end{equation}
where $\overline{\Sigma}$ denotes the orientation reversal of $\Sigma$, and we have used a whiskering along a timelike curve $\gamma$ connecting the base points $(m_0,0),(m'_0,1)$ of the surfaces $\Sigma_0,\Sigma_1$, as in the argument for the conservation of the higher monodromy matrices \eqref{conservedcharges} above. See fig. \ref{fig:bord}.

Let $\Omega^\text{fr;o}_{2}(Y)_\text{/bdy}$ denote the set (Abelian group under disjoint union) of equivalence classes of surfaces in $Y$ with the diffeomorphic boundary, subject to the relation $\Sigma\sim\Sigma'$ iff there exists a framed open bordism $\Gamma$ between them as in \textbf{Definition \ref{bordbdry}}. We now prove the following.
\begin{theorem}\label{bordinv}
    If the chirality $\mathrm{d}\ell$ aligns with the direction $\mathrm{d}u$ of the foliation $\Phi$, then the 2-holonomy $V$ arising from the 3d theory \eqref{covariant3daction} descends to a map $\Omega^\text{fr;o}_2(Y)_\text{/bdy}\rightarrow\mathsf{H}$ on framed open bordism classes.
\end{theorem}
\begin{proof}
    Given the covariance of \eqref{covariant3daction}, we can WLOG choose the chirality $\mathrm{d}\ell = \mathrm{d}x_3$ along the third coordinate. By hypothesis, this identifies $\mathrm{d}x_3$ with the time direction $\mathrm{d}u$ and $\partial_\ell = \partial_u$. 

    Now recall from the proof of \textbf{Proposition \ref{extended2flat}} that the Lax 2-connection $(L,H)$ we obtain from $S_{3d}$ has $L_3=0, H_1=0, H_2=0$ when $\mathrm{d}\ell=\mathrm{d}x_3$. Since we have identified the third coordinate with $u\in\mathbb{R}$, this gives
    \begin{equation*}
        L_u = 0,\qquad \iota_{\partial_u}H=0.
    \end{equation*}
    The fact that $L_u=0$ means that the $W_\gamma=1$ is trivial along timelike curves $\gamma$, and the fact that $\iota_{\partial_u}H=0$ means that $V_C=1$ is trivial on surfaces $C$ whose normal $\hat n$ is spacelike --- such as the cyinder tube piece $C$ of the surface $\partial\Gamma$. This reduces the left-hand side of \eqref{stationarity} down to
    \begin{equation*}
        V_{\Sigma'}^{-1}V_\Sigma=V_{\overline{\Sigma}'}V_\Sigma = V_{\overline{\Sigma}'\# \Sigma},
    \end{equation*}
    where we have used the fact that $V$ respects the gluing of surfaces (cf. \S \ref{holonomies}). However, now that $\Sigma'$ and $\Sigma$ share the same time-slice $u=0$, the surface $\overline{\Sigma}'\#\Sigma$ bounds a 3-submanifold $S$ which is a homotopy relative boundary between $\Sigma'$ and $\Sigma$. We then have 
    $$V_{\overline{\Sigma}'\#\Sigma} = V_{\partial S}=1$$
    by \textbf{Theorem \ref{relbdy}}. This proves \eqref{stationarity}.
\end{proof}

When the two directionalities misalign, however, such as in the case of the Chern-Simons/matter theory \eqref{CSmatter}, then the 2-holonomies are merely invariant under boundary-preserving deformations, but are not framed open bordism invariants. From the QFT literature \cite{Segal1985,Witten:2019,Freed:2014}, bordism invariants define non-perturbative anomalies; a similar sentiment is echoed in \cite{Baez:1995ph} (or {\it Remark \ref{homtrans}}) for the 4d Crane-Yetter-Broda TQFT. An observation {\it Remark \ref{alignedtrvi}} that we shall make in the next section, where we study the homotopy Lie algebra structure of the currents, may provide an explanation for this phenomenon.

\section{Topological-holomorphic 2+1d current algebra}\label{dgaffinecurrents}
Let us now investigate in more detail the algebra of currents in the boundary 3d action $S_{3d}$ given in \eqref{covariant3daction}. Recall that in 4d Chern-Simons theory with the disorder operator $\omega=z^{-1}\dr z$ and chiral boundary conditions, one obtains the two-dimensional Wess-Zumino-Witten CFT \cite{Costello:2019tri}.  In particular, the fields  of the theory form the affine Kac-Moody current algebra. Here, we seek to characterize the 3-dimensional homotopical analogue of these currents. 

\medskip

We begin by recalling the fields and currents in our boundary theory. In general, we have the fields $k = \mathrm{d} gg^{-1}\in\Omega^1(Y)\otimes \fg,\Theta\in\Omega^1(Y)\otimes \fh$ and the following currents
\begin{equation*}
    L = -k -\mu_1(\Theta^g),\qquad H = g\rhd (\dr\Theta - \Theta\wedge \Theta).
\end{equation*}
However, once we fix the Lagrangian $\mathcal{L}_\ell$ with the chirality vector $\vec\ell$, these currents are split up into two parts that are localized at the puncture $z=0$, with their dynamics governed by the boundary theory \eqref{covariant3daction}. To write down the currents, we first introduce a (not necessarily intergrable) subbundle $T^*_\ell Y\subset T^*Y$ of covectors on $Y$ tangent to $\mathrm{d}\ell$, defined by the projection map $\operatorname{proj}_\ell:T^*Y\rightarrow T^*Y_\ell$ that maps $a\mapsto \operatorname{proj}_\ell a = (\vec a\cdot \vec \ell)\vec \ell \cdot \mathrm{d}\ell$.  Since $Y$ is 3d, there is a linear isomorphism between 1-forms and 2-forms on $Y$, and we will use this fact to also define the projection map $\operatorname{proj}_\ell:\Omega^2(Y) \rightarrow \Omega^2_\ell(Y)$ on 2-forms. We can therefore express our currents as
\begin{equation}\label{currentsdefinition}
    L_\perp = -k_\perp - \mu_1(g\rhd \Theta_\perp)\,,\qquad   H_\ell = g\rhd (\operatorname{tor}\Theta_\perp - \frac{1}{2}[\Theta_\perp \times\Theta_\perp])\,,
\end{equation}
where $a_\perp = a - \operatorname{proj}_\ell a$ is the component of the 1- (or 2-)form $a$ normal to $\mathrm{d}\ell$, and $\operatorname{tor}a = \partial_\perp\times a$ denotes the scalar vorticity field of a 2d vector. The equations of motion they satisfy are given by
\begin{equation}\label{eom}
    \partial_{\ell} L_\perp =0\,,\qquad \partial_{\ell} H_\ell  = 0\,,
\end{equation}
as verified in \S \ref{3deoms}. We call the tuple $J = (L_\perp,H_\ell)$ a \textbf{2-current}. Since $\Theta_\ell$ does not at all appear in \eqref{covariant3daction}, we see that $\Theta = \Theta_\perp$, and hence we shall omit the subscript of "$\perp$" on $\Theta$ in the following.

\subsection{The dual 2-currents}
Now it is natural to expect the existence of a {\it dual} 2-current $\tilde J$ which satisfies an analogous set of fake- and 2-flatness equations of motion, and which can be constructed from $J$. This is in parallel with the case of the 2d Wess-Zumino-Witten theory, in which the right- and left-moving chiral currents satisfy dual equations of motion \cite{KNIZHNIK198483}. We will show that this is also the case for our theory \eqref{covariant3daction}.

We being by constructing the currents $\tilde J = (\tilde L_\ell,\tilde H_\perp)$, which are given by
\begin{equation}\label{currentsdefinition1}
    \tilde L_\ell = m_\ell,\qquad \tilde H_\perp = \partial_\ell \star_2\Theta + \mu_2(m_\ell,\star_2\Theta), 
\end{equation}
where $m = g^{-1}\mathrm{d} g$ is the {\it left} Maurer-Cartan form and $\star_2 \Theta$ is the 2d Hodge star operator on $\Theta\in\Omega^1_\ell(Y)$. Let us now prove the following.
\begin{proposition}
    If $J = (L_\perp,H_\ell)$ satisfies \eqref{eom}, then the dual currents $\tilde J = (\tilde L_\ell,\tilde H_\perp)$ satisfy
    \begin{equation}\label{eom1}
       \partial_\perp \tilde L_\ell = -\mu_1(\tilde H_\perp)\,,\qquad \partial_\perp\cdot \tilde H_\perp=0\,.
    \end{equation}
\end{proposition}
\noindent Notice these equations \eqref{eom1} are nothing but the fake and 2-flatness conditions for $\tilde J$.
\begin{proof}
    This is a direct computation.
    \begin{align*}
        \partial_\perp \tilde L_\ell &= \partial_\perp (g^{-1}\partial_\ell g) = \operatorname{Ad}_g^{-1}\partial_\ell (\partial_\perp gg^{-1}) \\ 
        &= -\operatorname{Ad}_g^{-1}\mu_1(\partial_\ell\Theta^g)) = -\mu_1(\mu_2(g^{-1}\partial_\ell g,\Theta) + \partial_\ell\Theta) = -\mu_1\tilde H_\perp,
    \end{align*}
    where we have used the equation $\partial_\ell L_\perp =0$ in \eqref{eom} in the second line. Next, we have
    \begin{align*}
        \partial_\perp\cdot \tilde H_\perp &= \partial_\perp \cdot (\partial_\ell \star_2 \Theta + \mu_2(g^{-1}\partial_\ell,\star_2\Theta)) \\
        &=\partial_\ell (\partial_\perp\cdot\star_2 \Theta) + \mu_2(\partial_\perp(g^{-1}\partial_\ell g)\cdot \star_2\Theta) + \mu_2(g^{-1}\partial_\ell g,\partial_\perp\cdot \star_2\Theta) \\
        &= \partial_\ell (\operatorname{tor} \Theta) - [H_\perp\cdot \star_2\Theta] + \mu_2(g^{-1}\partial_\ell g,\operatorname{tor}\Theta) \\
        &= g^{-1}\rhd (\partial_\ell g\rhd (\operatorname{tor}\Theta)) - [\partial_\ell \star_2\Theta + \mu_2(g^{-1}\partial_\ell,\star_2\Theta)\cdot \star_2\Theta]\,,
    \end{align*}
    where in the third line we have used the fact that $a_\perp\cdot\star_2 b_\perp = a_\perp\times b_\perp$ and the first equation of \eqref{eom1}, and in the fourth line we have used the definition of $\tilde H_\perp$. Consider now just the second term: recall $\star_2^2=-1$ on 1-forms in 2d, 
    \begin{align*}
        [\partial_\ell \star_2\Theta\cdot \star_2\Theta] &= \tfrac{1}{2}\partial_\ell [\star_2\Theta\cdot \star_2\Theta ] = \frac{1}{2}\partial_\ell [\Theta\times \Theta]\\
        [\mu_2(g^{-1}\partial_\ell g \,,\star_2\Theta)\cdot \star_2\Theta] &= \tfrac{1}{2}\mu_2(g^{-1}\partial_\ell g,[\Theta\times \Theta])\,,
    \end{align*}
    where the first line follows from the product rule and the second line follows from the Jacobi identity. The sum of these terms then gives
    \begin{equation*}
        \frac{1}{2}g^{-1} \rhd\partial_\ell (g\rhd [\Theta\times \Theta])\,,
    \end{equation*}
    which combines with $ g^{-1}\rhd (\partial_\ell g\rhd (\operatorname{tor}\Theta))$ to give precisely
    \begin{equation*}
         g^{-1}\rhd (\partial_\ell g\rhd (\operatorname{tor}\Theta - \frac{1}{2}[\Theta\times \Theta])) = g^{-1}\rhd \partial_\ell H_\ell = 0
    \end{equation*}
    by \eqref{eom}.
\end{proof}

We call these sets equations of motion \eqref{eom}, \eqref{eom1} the {\it 2-Lax equations} for the 2-currents $J,\tilde J$. The goal now is to study the differential graded algebraic structures of these topological-holomorphic currents. 

\subsection{Fields as derived 2-group elements in \texorpdfstring{$\mathsf{D}\cL^\bullet$}{D}}\label{fieldcurrents}
Notice the 2-currents $J,\tilde J$, as far as their components are concerned, are just particular "splicing"/splitting of the full 1- and 2-forms $(L,H)$. Therefore, in order to characterize them, it suffices to characterize $(L,H)$ and then perform the slicing. We first introduce the covariant derivative $D = \dr - \mu_2(k,-)$ such that
\begin{equation*}
    g\rhd \dr \Theta = D(g\rhd\Theta)\,.
\end{equation*}
In this way, depending on how the components of the currents are sliced up for $J,\tilde J$, the 2-Lax equations \eqref{eom}, \eqref{eom1} can be written entirely in terms of the fields $(g,\Theta^g)$. This ubiquitous appearance of $\Theta^g = g\rhd\Theta$ instead of the field $\Theta$ itself, both in the actions and the currents, is not a coincidence.

This can be explained in the context of the derived superfield formulation \cite{Zucchini:2021bnn,Jurco:2018sby}. Recall the derived 2-group $\mathsf{D}\mathbb{G}$ as defined in \S \ref{sec:gsof2CS}; our fields $(g,\Theta^g)$ are degree-0 elements of the following graded algebra:
\begin{equation*}
    \mathsf{D}\mathscr{L}^\bullet=\Omega^\bullet(Y) \otimes \mathsf{D}\mathbb{G},\qquad \mathsf{D}\mathscr{L}^0 \cong (C^\infty(Y)\otimes G)\oplus (\Omega^1(Y)\otimes \fh).
\end{equation*}
Due to the 2-group structure inherited from $\mathbb{G}$, this perspective allows us to use the whiskering operation (see \S \ref{sec:whisker}) to write
\[(g,g\rhd\Theta)= 
\begin{tikzcd}
	\ast &&& \ast
	\arrow[""{name=0, anchor=center, inner sep=0}, "g", curve={height=-30pt}, from=1-1, to=1-4]
	\arrow[""{name=1, anchor=center, inner sep=0}, "{\bar\mu_1\left(\exp\int g\rhd\Theta\right)g}"', curve={height=30pt}, from=1-1, to=1-4]
	\arrow["{\exp\int_\gamma g\rhd \Theta}"{description}, shorten <=8pt, shorten >=8pt, Rightarrow, from=0, to=1]
\end{tikzcd}\]
\[
 = \begin{tikzcd}
	\ast && \ast &&& \ast
	\arrow[""{name=0, anchor=center, inner sep=0}, "1", curve={height=-30pt}, from=1-3, to=1-6]
	\arrow[""{name=1, anchor=center, inner sep=0}, "{\bar\mu_1\left(\exp\int \Theta\right)}"', curve={height=30pt}, from=1-3, to=1-6]
	\arrow["g", from=1-1, to=1-3]
	\arrow["{\exp\int_\gamma \Theta}"{description}, shorten <=8pt, shorten >=8pt, Rightarrow, from=0, to=1]
\end{tikzcd} = g\rhd (1,\Theta).
\]
This explains why $\Theta^g=g\rhd\Theta$ keeps showing up in our theory \eqref{covariant3daction} --- it was encoding whiskerings by $g$! The subspace $\Omega^1\otimes\fh\subset \mathsf{D}\mathscr{L}^0$, within which the field $\Theta$ itself lives, then naturally acquires the interpretation of face decorations in the derived 2-group.

\subsubsection{The differential in \texorpdfstring{$\mathsf{D}\mathscr{L}^\bullet$}{D} }
This graded algebra $\mathsf{D}\mathscr{L}^\bullet$ is equipped with the differential $\hat \dr:\mathsf{D}\mathscr{L}^\bullet\rightarrow \mathsf{D}\mathscr{L}^{\bullet+1}$ that satisfies the usual graded Leibniz rule, but also a compatibility with whiskering $\rhd$. Since $\fG$ only has two terms, so does each degree of $\mathsf{D}\mathscr{L}^\bullet$. On elements $(g,g\rhd\Theta)$ of degree-0, we can explicitly write 
\begin{equation}
  \hat\dr(g,\Theta^g) = (\overrightarrow{\Delta} g - \mu_1\Theta^g,\dr\Theta^g - \Theta^g\wedge\Theta^g),\label{differential}  
\end{equation}
where $\overrightarrow{\Delta}g= -\dr gg^{-1}$ is the right Maurer-Cartan form. Note this produces the covariant derivative associated to a connection $\Theta^g$ on a $\mathsf{H}$-bundle over $X$ (recall $\mathsf{H}$ is a $G$-module).

For $g,h\in C^\infty(Y)\otimes G,$ and $ \Theta\in\Omega^1(Y)\otimes\fh$, the compatibility with a whiskering by $g$ can then be written as
\begin{equation*}
    g\rhd \hat{\dr} (h,h\rhd \Theta) = \hat{D} (gh,(gh)\rhd \Theta)\,,\qquad \hat D =  (\dr_1 - [\partial gg^{-1},-]-\mu_1,\dr_2 - \mu_2(\partial gg^{-1},-)),
\end{equation*}
The nilpotency of the differential, $\hat \dr^2=0$, is equivalent to the fake- and 2-flatness of $(L,H)$. This then allows us to identify the (off-shell) currents $$(L,H) = g\rhd \hat{\dr}(1,\Theta) = \hat D(g,\Theta^g) \in \mathsf{D}\mathscr{L}^1$$ as a degree-1 element in this derived complex. 

\subsubsection{Characterizing the currents}
To characterize the on-shell 2-currents, we slice up the dg algebra $\mathsf{D}\mathscr{L}^\bullet$ into two according to the projection $\operatorname{proj}_\ell$ and its complement. This induces a fibration 
\begin{equation*}
    \mathsf{D}\mathscr{L}_\perp^\bullet \cong (\mathsf{D}\mathscr{L}^\bullet/\mathsf{D}\mathscr{L}_\ell^\bullet)\hookrightarrow \mathsf{D}\mathscr{L}^\bullet \rightarrow\mathsf{D}\mathscr{L}^\bullet_\ell\rightarrow 0,
\end{equation*}
where $\mathsf{D}\mathscr{L}_\ell^\bullet = \Omega_\ell^\bullet(Y)\otimes \mathsf{D}\mathbb{G}$. This splits the differential $\hat \dr = \hat \dr_\perp + \hat \dr_\ell$, which are individually nilpotent $\hat \dr_\perp^2,\hat \dr_\ell^2 =0$ on the fields $(g,\Theta^g)\in \mathsf{D}\mathscr{L}^0$, as a consequence of the 2-Lax equations \eqref{eom} and \eqref{eom1}. We can then characterize the current algebras as the following subspaces:
\begin{align}
    \fD &= \{J \in g\rhd \hat{\dr}_\perp(\Omega^1\otimes\fh)\mid \hat \dr_\ell J =0\} \label{homotopycurrents}\\
    \tilde\fD &= \{\tilde J\in g\rhd \hat{\dr}_\ell(\Omega^1\otimes\fh)\mid \hat \dr_\perp \tilde J  =0\}.\label{homotopycurrents1}
\end{align}
Note $\mu_1$ appears in both $\hat \dr_\ell,\hat\dr_\perp $ due to the definition \eqref{differential}. More explicit characterizations are possible given the data of a THF $\Phi$ on $Y$; we will use $\Phi$ more explicitly later.

\subsection{Graded Lie algebra structure of the currents}
In the following, we will extract the graded Lie algebra structure of the currents. We shall do this from the "dual" perspective of the conserved Noether charges associated to the semilocal symmetries of the theory studied in \S \ref{residuals}, and analyzing their transformation properties.

Taking inspiration from covariant field space approach \cite{Geiller:2020edh}, we first smear the currents with gauge fields $\alpha \in C^\infty(Y)\otimes \fg$ and $\Gamma\in\Omega^1(Y)\otimes \fh$, which gives us the local Noether charges (also called "charge aspects"),
\begin{equation*}
    Q_{(\alpha,\Gamma)} = \langle (\alpha,\Gamma),J\rangle = \langle \alpha, H_\ell\rangle + \langle \Gamma,L_\perp\rangle \,.
\end{equation*}
For the currents $J$ of the theory \eqref{covariant3daction}, the above quantities are 2-forms that live in the image of the projection $\operatorname{proj}_\ell$ --- that is, they have no legs along $\mathrm{d}\ell$. On-shell of the equations of motion \eqref{eom}, a simple computation gives
\begin{equation*}
    \dr Q_{(\alpha,\Gamma)} =\langle \partial_\ell\alpha,H_\ell\rangle + \langle \partial_\ell\Gamma,L_\perp\rangle
\end{equation*}
which vanishes provided we impose the boundary conditions  
\begin{equation}\label{bc1}
    \partial_\ell\alpha=0\,,\qquad \partial_\ell\Gamma =0\,.
\end{equation}
In other words, the currents $Q_{(\alpha,\Gamma)}$ generate the infinitesimal {\it left}-acting symmetries $\cL^L_{3d}(\ell)$ of the theory \eqref{covariant3daction}. On the other hand, for the 2-current $\tilde J$, the local charge aspects 
\begin{equation*}
    \tilde Q_{(\alpha,\Gamma)} = \langle(\alpha,\Gamma),\tilde J\rangle = \langle \alpha,\tilde  H_\perp\rangle + \langle \Gamma,\tilde L_\ell\rangle 
\end{equation*}
are 2-forms living in the kernel of $\operatorname{proj}_\ell$ --- meaning they have only legs including $\mathrm{d}\ell$. On-shell of \eqref{eom1}, we find
\begin{align*}
    \dr\tilde Q &= \langle \partial_\perp \alpha, \tilde H_{\perp}\rangle + \langle \partial_\perp \cdot\star_2\Gamma,\tilde L_\ell\rangle - \langle \Gamma,\partial_\perp \tilde L_\ell\rangle  \\ 
    &= \langle \partial_\perp \alpha + \mu_1\Gamma,\tilde H_\perp\rangle + \langle \operatorname{tor}\Gamma, \tilde L_\ell\rangle\,,
\end{align*}
which vanishes provided we impose the boundary conditions\footnote{Recall we have without loss of generality (cf. \S \ref{residuals}) assumed that $\Gamma$ has no component along the chiral direction $\mathrm{d}\ell$, and hence $\Gamma_\perp=\Gamma$.}
\begin{equation}\label{bc2}
    \partial_\perp \alpha + \mu_1\Gamma=0\,,\qquad \operatorname{tor} \Gamma=0\,.
\end{equation}
In other words, the currents $\tilde Q_{(\alpha,\Gamma)}$ generate the (infinitesimal) {\it right}-acting symmetries $\mathring{\cL}_{3d}^R(\ell)$ of the theory \eqref{covariant3daction}.

We now prove the following.
\begin{theorem}\label{gradedbrackets}
    The above conserved Noether charge aspects $Q,\tilde Q$ satisfy the following graded algebra structure:
    \begin{align}
            [Q_{(\alpha,\Gamma)},Q_{(\alpha',\Gamma')}] &= -Q_{([\alpha',\alpha],\mu_2(\alpha',\Gamma)-\mu_2(\alpha,\Gamma'))} \nonumber\\
    &\qquad -\langle \Gamma,\partial \alpha' + \mu_1\Gamma'\rangle + \langle \alpha,\operatorname{tor}\Gamma'\rangle \label{bracket} \\
        [\tilde Q_{(\alpha',\Gamma')},\tilde Q_{(\alpha,\Gamma)}] &= -\tilde Q_{([\alpha,\alpha'],\mu_2(\alpha,\Gamma')-\mu_2(\alpha',\Gamma))}  \nonumber\\
    &\qquad - \langle \Gamma',\partial_\ell\alpha\rangle +\langle \alpha',\partial_\ell \Gamma\rangle,.\label{bracket1}
    \end{align}
    Moreover, $[Q,\tilde Q]=0.$
\end{theorem}
\begin{proof}
    Following \cite{KNIZHNIK198483}, we proceed with the computation of the (graded) Lie algebra bracket between the charges using the general formula \cite{Julia:2002df}
\begin{equation}
    [Q_{(\alpha,\Gamma)}, Q_{(\alpha',\Gamma')}] = -\delta_{(\alpha',\Gamma')}Q_{(\alpha,\Gamma)}\,,
\end{equation}
where $\delta_{(\alpha,\Gamma)}$ denotes a gauge variation. Let us consider first the gauge variations of $Q\in\hat\fD$, which is the Noether charges constructed from the 2-current $J = (L_\perp,H_\ell)$. Recall these charges generate the infinitesimal left variation,
\begin{equation}
\label{ec:infleftaction}
    \delta_{\alpha'} g = -{\alpha'}\cdot g,\qquad \delta_{(\alpha',\Gamma')} \Theta = -g^{-1}\rhd \Gamma',
\end{equation}
we find that
\begin{align*}
    \delta_{(\alpha',\Gamma')} L_\perp &= - [\alpha', L_\perp]+\partial_\perp \alpha' + \mu_1 \Gamma',\\
   \delta_{(\alpha',\Gamma')}  H_\ell &= - \mu_2(\alpha', H_\ell) - \mu_2( L_\perp, \Gamma') - \operatorname{tor}\Gamma',
\end{align*}
in which $(\alpha',\Gamma')$ parameterize the symmetries satisfying the boundary conditions \eqref{bc1}. This then leads to precisely \eqref{bracket}.

Now analogously, we compute the bracket on the dual Noether charges $\tilde Q\in \hat{\tilde \fD}$ constructed from $\tilde J = (\tilde L_\ell,\tilde H_\perp)$. Recalling these generate the infinitesimal right-variation
\begin{equation}
\label{ec:infrightaction}
    \tilde\delta_{\alpha} g = g\cdot\alpha,\qquad \tilde\delta_{(\alpha,\Gamma)} \Theta = -\mu_2(\alpha,\Theta) + \Gamma,
\end{equation}
we can then compute
\begin{align*}
    \tilde\delta_\alpha \tilde L_\ell &= -[\alpha,\tilde L_\ell] + \partial_\ell \alpha, \\ 
    \tilde\delta_{(\alpha,\Gamma)} \tilde H_\perp &= -\mu_2(\alpha,\tilde H_\perp) -\mu_2(\tilde L_\ell,\Gamma) + \partial_\ell\Gamma\,,
\end{align*}
where $(\alpha,\Gamma)$ satisfy the boundary conditions \eqref{bc2}. These then give rise to the brackets \eqref{bracket1}. Finally, to show that $[Q,\tilde Q]=0$ we compute
\begin{equation}
\begin{aligned}
    \tilde \delta_{\alpha'} L_\perp &=-\mathrm{Ad}_g(\partial_\perp \alpha'+\mu_1(\Gamma')) \\
    \tilde \delta_{(\alpha',\Gamma')}H_\ell &= g\rhd \mathrm{tor} \, \Gamma'
\end{aligned}
\end{equation}
where in the variation of $H_\ell$ we have used the graded Jacobi identity and the first boundary condition in \eqref{bc2}. In particular, we note that both of the above vanish due to the boundary conditions \eqref{bc2}. An entirely analogous computation shows that $\delta_\alpha \tilde L_\ell =0$ and $\delta_{(\alpha,\Gamma)}\tilde H_\perp=0$. 
\end{proof}

Notice these two brackets \eqref{bracket}, \eqref{bracket1} take identical forms. In particular, the central terms in both of these expressions can be written collectively as
\begin{equation}
    \langle (\alpha,\Gamma),\hat \dr(\alpha',\Gamma')\rangle\,, \label{central}
\end{equation}
depending on the boundary conditions \eqref{bc1}, \eqref{bc2} that are satisfied by $(\alpha,\Gamma)$ (recall $\Gamma_\ell=0$). Note $\hat \dr$ is the differential on $\fD,\tilde\fD$ defined in \eqref{differential}, which includes the map $\mu_1$. 

These structures run in complete parallel to the story in the 2d WZW model \cite{KNIZHNIK198483}, where the dual charges satisfy identical algebras and they do not have mutual brackets. Moreover, we will show in the following that \eqref{central} can in fact be identified as a Lie 2-algebra 2-cocycle characterizing a central extension sequence.

We emphasize here that $\mathrm{d}\ell$ is the direction of the chirality, so our computations so far are a priori independent of the direction of a THF $\Phi$ of $Y$. However, we will use the data of $\Phi$ in the following to construct a differential for the charge algebra $\hat\fD$ and $\hat{\tilde\fD}$.

\subsection{Central extensions in the affine Lie 2-algebra}
In this section, we prove that the central term \eqref{central} in fact defines a Lie 2-algebra 2-cocycle. Given an arbitrary Lie 2-algebra $\fG$ and an Abelian $\fG$-module $\mathfrak{V}$ --- namely a 2-term complex $\mathfrak{V} = W\xrightarrow{\phi}V$  of vector spaces \cite{Baez:2003fs} with an action $\rho: \fG\rightarrow \mathfrak{gl}(\mathfrak{V})$ by $\fG$ --- the (twisted) cohomology classes $H^2_\rho(\fG,\mathfrak{V})$ of such 2-cocycles have been shown \cite{Angulo:2018} to classify central extension sequences of Lie 2-algebras of the form
\begin{equation*}
    \mathfrak{V}\rightarrow \hat\fG\rightarrow \fG
\end{equation*}
over the ground field $\mathbb{K}$ of characteristic 0. 

For our purposes, it suffices to consider $\mathfrak{V}$ as a trivial $\fG$-module, where $\rho=0$. We first recall the data of a representative 2-cocycle $(s_1,s_2,s_3)$ of an extension class.
\begin{definition}
    Suppose $\mathfrak{V}$ is a trivial $\fG$-module, where $\fG = \fh\xrightarrow{\mu_1}\fg$ is a Lie 2-algebra over the ground field $\mathbb{K}$. A \textbf{Lie 2-algebra 2-cocycle} $s=(s_1,s_2,s_3): \fG\otimes\fG\rightarrow\mathfrak{V}$ with coefficients in $\mathfrak{V}=W\xrightarrow{\phi}V$ is the data of a triple
\begin{equation*}
    s_1: \fg\otimes\fg\rightarrow V,\qquad s_2: \fg\otimes\fh\rightarrow W,\qquad s_3: \fh\rightarrow V,
\end{equation*}
such that
\begin{align}
    0 &= ~\circlearrowright ~ s_1([\sfx_1,\sfx_2],\sfx_3),\label{jac2} \\
    0 &= s_2(\mu_1(\sfy_1),\sfy_2) + s_2(\mu_1(\sfy_2),\sfy_1) \label{cond1}\\
    0&=~ \circlearrowright~ s_2(\mu_1([\sfy_1,\sfy_2]),\sfy_3),\label{jac3} \\
    0&= s_1(\sfx,\mu_1(\sfy)) - \phi s_2(\sfx,\sfy) + s_3(\sfx\rhd \sfy),\label{cond2} \\
    0&= s_2([\sfx_1,\sfx_2],\sfy) - s_2(\sfx_1,\sfx_2\rhd \sfy) + s_2(\sfx_2,\sfx_1\rhd \sfy),\label{jac}\\
    0&= s_2(\sfx,[\sfy_1,\sfy_2]) - s_2(\mu_1(\sfx\rhd \sfy_1),\sfy_2) + s_2(\mu_1(\sfx\rhd \sfy_2),\sfy_1)\label{jac1}
\end{align}
for all $\sfx,\sfx_1,\sfx_2,\sfx_3\in \fg$ and $\sfy,\sfy_1,\sfy_2,\sfy_3\in\fh$, where $\circlearrowright$ denotes a sum over cyclic permutations.
\end{definition}
\noindent The full list of conditions for when $\mathfrak{V}$ is not a trivial $\fG$-module can be found in \cite{Angulo:2018}.

\subsubsection{Constructing the differential in \texorpdfstring{$\hat{\fD},\hat{\tilde{\fD}}$}{D}}
To begin, we first endow a grading where the $\Gamma$-charges have degree-0 and the $\alpha$-charges have degree-1. The idea is to perform a pushforward at the differential forms factor $\Omega^\bullet$ of $\hat{\tilde{\fD}}$ (or $\hat\fD$). This is accomplished by viewing the THF $\Phi$ as inducing a fibration \cite{Scrdua2017OnTH}
\begin{equation*}
    \bR_{\mathrm{d}x_3} \hookrightarrow Y\rightarrow M
\end{equation*}
over a (compact) complex manifold $M$, where $\mathbb{R}_{\mathrm{d}x_3}$ is the 1-dimensional fibre along the covector direction $\mathrm{d}x_3$ of the THF (which may differ from the chirality $\mathrm{d}\ell$). For simplicity, we compactify the fibre from $\mathbb{R}_{\mathrm{d}x_3}$ to a circle $S^1_{\mathrm{d}x_3}$ such that $Y$ defines a {Seifert fibration} $p:Y\rightarrow M$ \cite{Aganagic:2017tvx}, then we can perform a fibrewise integration $\pi_*:\Omega^{\bullet}(Y)\rightarrow \Omega^{\bullet-1}(M)$ such that,\footnote{This gives rise to the well-known {\it Gysin homomorphism} $\pi^!:H^\bullet(Y)\rightarrow H^{\bullet-\operatorname{rank}E}(M)$ \cite{book-charclass} on cohomology, where $E$ is the compact fibre of $Y\rightarrow M$.} on 1-forms, 
\begin{equation}\label{integrate}
    (\pi_*\omega)_{(w,\bar w)} = \int_{S^1_{\mathrm{d}x_3} = p^{-1}(w,\bar w)}\omega,\qquad \omega\in\Omega^1(Y).
\end{equation}
We make use of this map to define the differential $\hat\mu_1:\hat\fD_1\rightarrow \hat\fD_0$ as a pullback by $\pi_*\otimes\mu_1$,
\begin{equation*}
    (\hat\mu_1\tilde Q)_\Gamma = \tilde Q_{(\pi_*\otimes \mu_1)\Gamma} = \langle \mu_1(\pi_*\Gamma),H_\perp\rangle,
\end{equation*}
assigning a $\Gamma$-charge to an $\alpha$-charge. We construct the same differential for the dual charges $\hat\fD$.

\medskip

We are finally to prove the main theorem of this section.
\begin{theorem}\label{affine2alg}
    The tuple $(\hat{\tilde{\fD}},\hat\mu_1,\hat\mu_2=[-,-])$ with $[-,-]$ given by \eqref{bracket1} is a Lie 2-algebra fitting into the central extension sequence
    \begin{equation*}
        \underline{\mathbb{K}} \rightarrow \hat{\tilde\fD}\rightarrow \tilde\fD,
    \end{equation*} 
    iff \eqref{central} defines a Lie 2-algebra 2-cocycle associated to this extension, where $\underline{\mathbb{K}} = \mathbb{K}\xrightarrow{0}\mathbb{K}$.
\end{theorem}
\begin{proof}
    Recall $\Gamma_\ell=0$. We will first perform some general computations before we specialize $(\alpha,\Gamma)$ to the space $\mathring{\cL}_{3d}^R$ parameterizing the charges $\hat{\tilde{\fD}}$. From the central term \eqref{central}, we shall define
    \begin{equation}
        s_1(\Gamma,\Gamma') = \langle \Gamma,\mu_1\Gamma'\rangle,\qquad s_2(\Gamma,\alpha) = \langle \operatorname{tor}\Gamma,\alpha\rangle,\qquad s_3=0\label{2cocycle}.
    \end{equation}
    It is useful to extend $s_2$ to a skew-symmetric map by an integration by parts,
    \begin{equation*}
       s_2(\alpha,\Gamma) = \langle \partial \alpha,\Gamma\rangle =- \langle \operatorname{tor}\Gamma,\alpha\rangle = -s_2(\Gamma,\alpha).
    \end{equation*}
    As $\phi=0,s_3=0$, and since $\pi_*$ acts trivially on the $\alpha$-charges, we see that \eqref{cond2}, \eqref{jac2} are trivially satisfied.
    
    Let us now check the graded Leibniz rule,
    \begin{equation*}
        \hat\mu_1\circ[-,-] = [-,-]\circ (\hat\mu_1\otimes 1 +(-1)^\text{deg} 1\otimes \hat\mu_1).
    \end{equation*}
    From \eqref{bracket}, the left-hand side gives 
    \begin{align*}
        \hat\mu_1[Q_{(\alpha,\Gamma)},Q_{(\alpha',\Gamma')}] &= -\hat\mu_1(Q_{([\alpha',\alpha],\mu_2(\alpha',\Gamma) - \mu_2(\alpha,\Gamma'))} + \langle (\alpha,\Gamma),\hat d(\alpha',\Gamma')\rangle )\\
        &= - Q_{[\alpha',(\pi_*\otimes \mu_1)\Gamma] - [\alpha,(\pi_*\otimes \mu_1)\Gamma']}\\
        &= -Q_{[\alpha',\mu_1(\pi_*\Gamma)]} - Q_{[\alpha,\mu_1(\pi_*\Gamma')]},
    \end{align*}
    while the right-hand side gives
    \begin{align*}
        [(\hat\mu_1Q)_\Gamma,Q_{(\alpha',\Gamma')}] - [Q_{(\alpha,\Gamma)},(\hat\mu_1Q)_{\Gamma'}] &= -Q_{([\alpha',(\pi_*\otimes\mu_1)\Gamma],-\mu_2((\pi_*\otimes\mu_1)\Gamma,\Gamma'))} \\
        &\qquad + Q_{[(\pi_*\otimes\mu_1)\Gamma',\alpha], \mu_2((\pi_*\otimes\mu_1)\Gamma',\Gamma)} \\
        &\qquad +\langle \mu_1(\pi_*\Gamma),\operatorname{tor}\Gamma'\rangle - \langle \Gamma,\partial \mu_1(\pi_*\Gamma')\rangle \\
        &= -Q_{[\alpha',\mu_1(\pi_*\Gamma)]} + Q_{[\mu_1(\pi_*\Gamma'),\alpha]}  \\
       &\qquad + \langle \mu_1(\pi_*\Gamma),\operatorname{tor}\Gamma'\rangle + \langle \operatorname{tor}\Gamma,\mu_1(\pi_*\Gamma')\rangle\\
       &\qquad - (Q_{\mu_2(\mu_1(\pi_*\Gamma'),\Gamma)}+ Q_{\mu_2(\mu_1(\pi_*\Gamma),\Gamma')}).
    \end{align*}
    Now given the parameters $\Gamma$ in $\mathring{\cL}^R_{3d}$ must be valued in the maximal Abelian $\ft\subset\fh$, the final terms involving $\mu_2(\mu_1(\pi_*\Gamma'),\Gamma) = [\pi_*\Gamma',\Gamma] =0$ vanish. The two sides then coincide iff $s_2$ \eqref{2cocycle} satisfies \eqref{cond1}.

    We now check the graded Jacobi identities. Due to the form of the bracket \eqref{bracket}, the graded Jacobi identities for the non-central extension terms follow directly from those of $\fG$. The remaining terms read
    \begin{align}
        \langle [\alpha,\alpha'],\operatorname{tor}\Gamma\rangle - \langle \alpha',\operatorname{tor}\mu_2(\alpha,\Gamma)\rangle + \langle \alpha,\operatorname{tor}\mu_2(\alpha',\Gamma)\rangle &=0 \label{proofcond},
        \\
        \langle \mu_1\Gamma,\mu_2(\alpha,\Gamma')\rangle + \langle \alpha,[\Gamma',\Gamma]\rangle - \langle \mu_1\Gamma',\mu_2(\alpha,\Gamma')\rangle &=0 \nonumber,
    \end{align}
    which are precisely the conditions \eqref{jac}, \eqref{jac1} for $s$ \eqref{2cocycle}. However, due to the lack of a $\langle\Gamma,\mu_1\Gamma'\rangle$ contribution in the central term of the bracket \eqref{bracket1} for $\hat{\tilde{\fD}}$, \eqref{jac1} is trivially satisfied for $\hat{\tilde\fD}$. And finally, since $\Gamma$ is Abelian, \eqref{jac3} is also trivially satisfied. This completes the proof.
Hence $\check\mu_1$ indeed preserves the bracket.
\end{proof}

From \eqref{bracket}, we see that the above general computations also applies to the charges $\hat\fD$. Whence, under mild assumptions,\footnote{For $\hat\fD$, we must keep quadratic terms $[\Gamma,\Gamma'],\langle\Gamma,\mu_1\Gamma'\rangle$ in our computations. Assuming the quantity $[\pi_*\Gamma,\Gamma'] = \llbracket \Gamma,\Gamma'\rrbracket$ defines a Lie bracket $Q_{\llbracket \Gamma,\Gamma'\rrbracket} = [Q_{\Gamma'},Q_\Gamma]$ on the charges, \eqref{jac3} follows from \eqref{proofcond} given $\hat\mu_1$ preserves this bracket. One can check that this is indeed the case:
\begin{equation*}
        (\pi_*\otimes\mu_1)(\llbracket\Gamma,\Gamma'\rrbracket) = \int_{S^1_{\mathrm{d}x_3}} \mu_1[(\pi_*\Gamma)_{(w,\bar w)},\Gamma'_{(x_3,w',\bar w')}] = [\mu_1(\pi_*\Gamma_{(w,\bar w)}),\mu_1(\pi_*\Gamma_{(w',\bar w')})],
    \end{equation*}
    where $(w,\bar w),(w',\bar w')\in M$ are local coordinates on the base space $M$ and $x_3\in S^1_\Phi$ is along the fibre.} one can prove that it also forms a centrally extended Lie 2-algebra. We call $\hat\fD,\hat{\tilde{\fD}}$ the \textbf{homotopy affine Lie algebras of planar currents}. 

\begin{remark}\label{alignedtrvi}
    Recall that $\Gamma_\ell=0$. In the affine Lie 2-algebras $\hat\fD,\hat{\tilde\fD}$, the pushforward $\pi_*=0$ \eqref{integrate} is trivial if the THF $\Phi$ aligns with $\mathrm{d}\ell$. This renders the charge algebra $\hat\fD$ both strict and skeletal, and hence lacks a degree-3 {Postnikov class} $\kappa=0$ \cite{Baez:2003fs,chen:2022,Wagemann+2021}; see also \S \ref{classification}. As one can construct gauge-invariant local perturbative observables from such a class (strictly speaking one does this from the cohomology $H(\hat\fD)$ \cite{Arvanitakis:2020rrk}, which contains the Postnikov class as a $k$-invariant), this suggests that $S_{3d}$ hosts no perturbative anomalies on 3-manifolds. This is consistent with our result found in \S \ref{bordinv} that, in the aligned case, the surface holonomies are bordism invariants.
\end{remark}

All in all, the conserved Noether charges in our theory $S_{3d}$ form centrally-extended Lie 2-algebras $\hat\fD,\hat{\tilde\fD}$. These algebras are parameterized by the symmetries $\cL_\text{sym}\subset C^\infty(Y)\oplus \Omega^1(Y)$ of $S_{3d}$, which are function spaces of infinite-dimension over $\mathbb{K}$. The infinite dimensionality of the symmetry algebra of our three-dimensional theory thus leads to integrability. This is of course in direct analogy with the affine Lie algebra underlying the charges in the 2d WZW model \cite{KNIZHNIK198483}. 

We conclude our paper by briefly mentioning that one of the authors (HC) has developed a notion of {\it 2-Kac-Moody algebra} in \cite{Chen:2023integrable}, which governs the global symmetries of the model \eqref{covariant3daction}. This is in analogy to the role that the affine Kac-Moody algebra $\widehat{\Omega_k\fg}$ plays in the 2d WZW model \cite{book-integrable,KNIZHNIK198483}, whose corresponding Kac-Moody group $\widehat{\Omega_k G}$ \cite{Baez:2005sn} is crucial for quantizing the WZW model. It is known that the extension class $k$ (i.e. the level) defining the Kac-Moody group $\widehat{\Omega_kG}$ comes from the $S^1$-transgression map\footnote{Geometrically, the transgression map is a way of associating bundle gerbes --- a higher notion of bundles --- on a space to a line bundle on its loop space; see \cite{Willerton:2008gyk}.} \cite{Carey_1997,Sharpe:2015mja,Waldorf:2012,Waldorf2015TransgressiveLG}
\begin{equation*}
    H^3(G,\bC^\times)\rightarrow H^2(\Omega G,\bC^\times)\,.
\end{equation*}
Therefore, one expects that our model \eqref{covariant3daction} may admit an analogous description in terms of a "surface transgression" from a certain Lie 2-group cohomology class (cf. \cite{Ginot:2009pil,Angulo:2018}). This shall be a crucial step in quantizing our 3d model, and we leave this for a future endeavour.

\section{Outlook}\label{outlook}
In this paper, we have constructed a 3d integrable field theory from a higher-gauge theory by following through with the Costello-Yamazaki localization procedure \cite{Costello:2019tri}. We also studied in detail the properties of this 3d field theory, and proved several very interesting facts about its properties. We believe that there are many more open questions to be answered in regards to this theory, and the connection between higher-gauge theories and higher-dimensional integrable field theories in general. Here, we list a few of them in this section.

\paragraph{Higher Monodromies.} Recall that the conserved 2-monodromy matrices \eqref{conservedcharges} in our 3d field theory come labelled by the categorical characters $\mathcal{X}_k$ of the Lie 2-group $\mathbb{G}$ and homotopy classes $[\Sigma]$ of surfaces in $Y$. We are lead to two very interesting problems to tackle when considering the properties of these higher monodromies.
\begin{enumerate}
    \item \textbf{The label $[\Sigma]$: differential gerbes and 2-tangle invariants.} In 3d TQFTs, it is known that the algebra of the Wilson lines define invariants of framed links \cite{Witten:1988hc}. The usual singular homology is not enough to capture such data: indeed, framed link invariants satisfy certain knot and satellite relations that are seen only by differential --- possibly Dolbeault, when the chirality and foliation are misaligned --- cohomology \cite{FreedMonopole} (eg. the \textit{Deligne-Beilinson} cohomology \cite{Guadagnini2008DeligneBeilinsonCA} in Abelian Chern-Simons theory). As such, the label $[\Sigma]$ should be treated as differential non-Abelian gerbes \cite{Nikolaus2011FOUREV} which classify invariants of framed 2-tangles \cite{Baez:1995xq,BAEZ1998145} in $Y$ (or its thickening $Y\times [0,1]$).
    
    \item \textbf{The label $k$: orthogonality of categorical characters.} In regular representation theory, one can prove that there are countably many orthogonal irreducible characters of a compact Lie group that span the representation ring \cite{Woronowicz1988}. Correspondingly, the label $k$ should run over all indecomposable $\mathbb{G}$-module categories, but categorical character theory is not yet powerful enough to prove this (see eg. \S 9.6 in \cite{Bartlett:2009PhD} and \textbf{ Proposition 1.2.19} in \cite{Douglas:2018}). Moreover, while an inner product is known to exist for categorical characters \cite{Huang:2024}, there is no satisfactory notion of "orthogonality" for them at the moment. One of the authors (HC) of this paper is currently investigating this problem.
\end{enumerate}
Resolving these two issues are of major interest in recent literature. For instance, higher-tangle invariants would help define 4d non-semisimple TQFTs (which have a chance of definition exotic 4-manifold invariants \cite{Reutter:2020bav}), and categorical character theory would have very explicit applications in the construction of 4d gapped phases in condensed matter theory \cite{Bullivant:2019tbp,Delcamp:2023kew}.

\paragraph{Quantization of $S_{3d}$: the 2-Kac-Moody group.} An obvious next step to take would be to try and quantize the theory $S_{3d}$. Towards this, one of the authors (HC) has devised in \cite{Chen:2023integrable} a notion of "2-Kac-Moody algebra" $\widehat{\Sigma_s\fG}$ which underlies the algebraic properties of the global Noether {charges} of the theory, where $s$ is the extension 2-cocycle \eqref{2cocycle}. The higher monodromy matrices \eqref{conservedcharges} would then be controlled by the integrated 2-Kac-Moody group $\widehat{\Sigma_s\mathbb{G}}$. As $S_{3d}$ is a topological-holomorphic field theory, it would be interesting to understand the structure of this object $\widehat{\Sigma_s\mathbb{G}}$ and its relation to the raviolo vertex operator algebra \cite{Garner:2023zqn,Alfonsi:2024qdr}.

The quantum Hilbert space associated to $S_{3d}$ are then given by unitary (categorical) representations $\operatorname{UMod}_{\mathsf{2Vect}}(\widehat{\Sigma_s\mathbb{G}})$ of this 2-Kac-Moody group. In analogy with the modules of vertex operator algebras in general \cite{KazhdanLusztig:1994}, this 2-category should have equipped higher modular and braided tensor structures, which are of major interest in both pure categorical algebra \cite{Johnson_Freyd_2023} as well as the construction of higher-skein "lasagna modules" in 4d TQFTs \cite{Morrison2019InvariantsO4,Manolescu2022SkeinLM}.

\paragraph{Other choices of disorder operators} In this article we have considered a unique choice of disorder operator $\omega = z^{-1}\dr z$, which led to a rich family of three-dimensional theories satisfying properties which are higher homotopical analogues of those satisfied by the WZW model in two dimensions. Notably, in \cite{Schenkel:2024dcd} they considered a variety of disorder operators, except for the one hereby considered. Their construction led to different $3$d theories which have a Lax connection with spectral parameter. It would be very interesting to perform a deep analysis of the holonomies, focusing on the invariance under homotopies relative boundary for the Lax connections constructed there. Morever, it would be appealing to analyze the symmetries of the theories obtained in \cite{Schenkel:2024dcd}, and construct the corresponding current algebras.

\section{Appendix}

\subsection{Appendix A: Classification of Lie 2-groups and Lie 2-algebras}\label{classification}

Let us begin with a brief overview of the notion of 2-groups, Lie 2-algebras and 2-gauge theory. Recall a 2-group $\mathbb{G}$ is, by definition, a monoidal groupoid $\Gamma\rightrightarrows G$ in which all objects are invertible. There are many different but equivalent characterizations of a 2-group, namely as a pointed 2-groupoid, as a group object internal to the category of groups, or as a 2-term crossed-complex of groups as in \textbf{Definition \ref{2grpdef}}. 

The fundamental theorem of Lie 2-groups state that there is a one-to-one correspondence between connected, simply-connected Lie 2-groups and Lie 2-algebras. \textbf{Definition \ref{lie2alg}} gives Lie 2-algebras as a differential crossed-complex \cite{Baez:2003fs}, which is equivalent to the $L_2$-algebra definition by simply identifying $\mu_1=t$ and $\mu_2 = (\rhd,[-,-])$. The homotopy Jacobi identities are equivalent to \eqref{pfeif1} and the 2-Jacobi identities. 

The goal of this section is to briefly recall some classification results for Lie 2-groups and Lie 2-algebras. If $\mathbb{G},\mathbb{G}'$ are two Lie 2-groups, a \emph{Lie 2-group homomorphism} $(\Phi,\Psi): \mathbb{G}\rightarrow\mathbb{G}'$ is a tuple of Lie group maps $\Phi:G\rightarrow G',\Psi:\mathsf{H}\rightarrow\mathsf{H}'$ such that
\begin{enumerate}
    \item $\bar\mu_1'\circ \Phi = \Psi\circ\bar\mu_1$, and
    \item $\Phi(x\rhd y) = (\Psi x)\rhd' (\Phi y)$ for all $x\in G,y\in\mathsf{H}$.
\end{enumerate}
Similarly, If $\mathfrak{G},\mathfrak{G}'$ are two Lie 2-algebras, then a \emph{Lie 2-algebra homomorphism} $(\phi,\psi):\mathfrak{G}\rightarrow\mathfrak{G}'$ is a tuple of Lie algebra maps $\phi: \fh\rightarrow \fh',\psi: \fg\rightarrow \fg'$ such that
\begin{enumerate}
    \item $\mu_1'\circ \phi = \psi \circ \mu_1$, and
    \item $\phi(\sfx\rhd \sfy) = (\psi \sfx)\rhd'(\phi \sfy)$ for each $\sfx\in \fg,\sfy\in \fh$.
\end{enumerate}
As in the Lie 1-algebra case, this notion transports to maps between principal 2-bundles $\mathcal{P}\rightarrow\mathcal{P}$ \cite{Baez:2004in,Chen:2022hct}.

In contrast to Lie groups, the notion of "isomorphism" for Lie groupoids is categorical equivalence \cite{maclane:71}. A bit more explicitly, we say $\mathbb{G}\simeq\mathbb{G}'$ are equivalent iff there exist Lie 2-group maps $(\Phi,\Psi):\mathbb{G}\rightarrow\mathbb{G}'$ and $(\Phi,\Psi)^{-1}:\mathbb{G}'\rightarrow\mathbb{G}$ such that there are invertible natural transformations 
\begin{equation*}
    \mathcal{H}:(\Phi,\Psi)^{-1}\circ (\Phi,\Psi)\Rightarrow\mathrm{id}_{\mathbb{G}}, \qquad \mathcal{E}: (\Phi,\Psi)\circ (\Phi,\Psi)^{-1}\Rightarrow \mathrm{id}_{\mathbb{G}'}.
\end{equation*}
The following is a classic result first proven by Ho{\`a}ng Xu{\^ a}n S{\'i}nh \cite{Nguyen2014CROSSEDMA,Ang2018,Kapustin:2013uxa,baez2023ho}, who is a Vietnamese mathematician taught by Alexander Grothendieck.
\begin{theorem}
    2-groups $\mathbb{G}$ are classified by the following data: a group $\Pi_1=\operatorname{coker}\bar\mu_1$, a $\Pi_1$-module $\Pi_2=\operatorname{ker}\bar\mu_1$, and a degree-3 group cohomology class $\tau\in H^3(\Pi_1,\Pi_0)$ called the \textbf{Postnikov class}.
\end{theorem}
\noindent Note $\Pi_2\subset\mathsf{H}$ must be an Abelian group by the Peiffer identity \eqref{pfeif1}. The tuple $(\Pi_1,\Pi_2,\tau)$ is often called the \textit{Ho{\`a}ng data} of $\mathbb{G}$.

An analogous result holds for Lie 2-algebras. Here, one can state equivalence in the context of chain complexes. We say $\mathfrak{G}\simeq\mathfrak{G}'$ are equivalent iff there exist Lie 2-algebra maps $(\phi,\psi):\mathfrak{G}\rightarrow \mathfrak{G}',(\phi,\psi)^{-1}:\mathfrak{G}'\rightarrow \mathfrak{G}$ and invertible chain homotopies
\begin{equation*}
    \eta: (\phi,\psi)^{-1}\circ(\phi,\psi)\Rightarrow \mathrm{id}_\mathfrak{G},\qquad \varepsilon:(\phi,\psi)\circ (\phi,\psi)^{-1}\Rightarrow\mathrm{id}_{\mathfrak{G}'}.
\end{equation*}
The following is also a classic result, which was claimed to be first proven by Gerstenhaber by Wagemann in \cite{Wagemann+2021}.
\begin{theorem}
Lie 2-algebras are classified up to equivalence by the following data: a Lie algebra $\mathfrak{n}=\operatorname{coker}\mu_1$, a $\mathfrak{n}$-module $V=\operatorname{ker}\mu_1$ and a degree-3 Lie algebra cohomology class $\kappa\in H^3(\mathfrak{n},V)$, called the \textbf{Postnikov class}.
\end{theorem}
\noindent One can show that, if two principal 2-bundles $\mathcal{P},\mathcal{P}'\rightarrow X$ have structure 2-groups $\mathbb{G},\mathbb{G}'$ that have distinct Postnikov classes, then they cannot be isomorphic 2-bundles over $X$.

\subsection{Appendix B: Homotopy transfer}\label{sec:homtransf}
The BV-BRST formalism is a way to organize the algebraic and geometric data in a field theory with gauge symmetry into cochain complexes, which are designed such that their cohomologies describes precisely the gauge-invariant on-shell quantities. The mathematical structure that arises out of this construction is an \emph{$L_\infty$-algebra}, which can be described as a $\mathbb{Z}$-graded differential complex 
$$\mathfrak{L}\cong \dots \rightarrow \mathfrak{l}_{-2}\xrightarrow{\ell_1}\mathfrak{l}_{-1}\xrightarrow{\mu_1}\mathfrak{l}_0\xrightarrow{\ell_1}\mathfrak{l}_1\xrightarrow{\ell_1}\mathfrak{l}_2\rightarrow \dots$$ equipped with higher-brackets $\ell_n\in \operatorname{Hom}^{2-n}(\mathfrak{L}^{n\otimes},\mathfrak{L})$ for $n\geq 2$ satisfying the Koszul-Jacobi identities. In gauge field theories with structure (2-)group $\mathbb{G}$, such as the homotopy Maurer-Cartan theories \cite{Jurco:2018sby}, for instance, this complex is given by $\mathfrak{L} = \Omega^\bullet(X)\otimes \mathfrak{G}$ where $\mathfrak{G}=\operatorname{Lie}\mathbb{G}$.

Now from a purely mathematical perspective, it is known that $L_\infty$-algebras retracts onto its cohomology \cite{stasheff2018,Alfonsi_2023}.\footnote{$L_\infty$-algebras are objects in the derived category of Lie algebras. More generally in any derived Abelian category, equivalences are precisely quasi-isomorphisms, ie. derived chain maps that induce isomorphisms in cohomologies.} In other words, given a $L_\infty$-algebra $\mathfrak{L}$, the inclusion $\iota:H(\mathfrak{L})\hookrightarrow \mathfrak{L}$ treating $H^\bullet(\mathfrak{L})\cong\operatorname{ker}\ell_1/\operatorname{im}\ell_1$ as a subcomplex of $\mathfrak{L}$ admits a homotopy inverse (ie. an adjunction) given by the projection $p:\mathfrak{L}\rightarrow H(\mathfrak{L})$ --- there exists {chain homotopies} $\eta: \iota\circ p\Rightarrow \mathrm{id}_{\mathfrak{G}}$, $\varepsilon: p\circ \iota\Rightarrow \mathrm{id}_{H(\mathfrak{G})}$ that "witness" this retraction. To be a chain homotopy, the map $\eta=(\eta_n)$ --- where $\eta_n: \mathfrak{L}_n\rightarrow \mathfrak{L}_{n-1}$ --- for instance must satisfy
\begin{equation*}
    \iota \circ p - 1 = \eta \circ \ell_1 + \ell_1\circ \eta 
\end{equation*}
as linear maps on $\mathfrak{L}$.

In this context, it can be shown that a $L_\infty$-algebra structure, ie. the higher brackets $\tilde\ell=(\tilde\ell_n)_{n\geq 2}$, on $H(\mathfrak{L})$ can be induced from that $\ell=(\ell_n)_{n\geq 2}$ on $\mathfrak{L}$ by using the projector $p$; notice, by construction, the cohomology $H(\mathfrak{L})$ has trivial differential. This is given by the formula
\begin{equation}\label{homotopytransfer}
    \tilde\ell = p\circ (\ell + \ell\circ \eta\circ \ell + \ell \circ \eta\circ \ell\circ \eta\circ \ell + \dots ) \circ \iota,
\end{equation}
and it can be proven that $\tilde\ell$ indeed satisfies the Koszul-Jacobi identities on $H(\mathfrak{L})$ \cite{stasheff2018,Arvanitakis:2020rrk}. This is the celebrated \textbf{homotopy transfer theorem} for $L_\infty$-algebras: the $L_\infty$-algebra structure on $\mathfrak{L}$ is transferred to a $L_\infty$-algebra structure on its cohomology $H(\mathfrak{L})$ through the projection quasi-isomorphism $p$.

An interesting case occurs when $\mathfrak{L}$ is only a dgla, such that $\ell_n=0$ for all $n>2$. The formula \eqref{homotopytransfer} implies that homotopy transfer in fact induces higher homotopy brackets on $H(\mathfrak{L})$. Indeed, at degree-3, we see that we have a generally non-trivial contribution given for each $\sfx_1,\sfx_2,\sfx_3\in\mathfrak{L}$ by
\begin{equation*}
    \tilde\ell_3([\sfx_1],[\sfx_2],[\sfx_3]) = ~\circlearrowright  [\ell_2(\eta(\ell_2([\sfx_1],[\sfx_2])),\eta([\sfx_3]))],
\end{equation*}
where $\circlearrowright$ denotes a cyclic sum and we have denoted the cohomology class of an element $\sfx\in\mathfrak{L}$ by $p(\sfx)=[\sfx]\in H(\mathfrak{L})$. This map $\tilde\ell_3$ is usually called the \textbf{homotopy map} in the literature \cite{Chen:2012gz}, and can be seen to measure the failure of the chain homotopy $\eta$ to respect the graded Jacobi identity. This is relevant for us, as given a strict Lie 2-algebra $\mathfrak{G}$, the associated BV-BRST $L_\infty$-algebra $\mathfrak{L} = \Omega^\bullet(X)\otimes \mathfrak{G}$ is indeed a dgla with $\ell_1=d - \mu_1$. However, upon performing a homotopy transfer, we see that a degree-3 homotopy map is induced on the cohomology $H(\mathfrak{L})$. 

This idea can be used to immediately to prove Gerstenhaber's theorem in \S \ref{classification}. The cohomology $H(\mathfrak{G})$ of a Lie 2-algebra $\mathfrak{G}$ consist of $M\xrightarrow{0}\mathfrak{n}$, and the homotopy map $\tilde\ell_3$ is a 3-cocycle representing the Postnikov class $\kappa\in H^3(\mathfrak{n},M)$.

\chapter{Conclusion}

This thesis investigates the connection between Chern-Simons theories and integrable field theories. Building on the correspondence introduced in the seminal works of Costello, Witten, and Yamazaki \cite{Costello:2017dso, Costello:2018gyb, Costello:2019tri}, it examines both the original construction and several of its generalizations. The articles comprising this thesis have contributed to the development of the field across all of its major directions and, moreover, have initiated a new and independent line of research within this framework. Below, we summarize the main results of each article included in this thesis, highlighting their contributions to the advancement of the field.

Four-dimensional Chern-Simons theory is essentially defined by two key pieces of input data that determine the resulting two-dimensional field theory: a meromorphic 1-form \( \omega \), and the boundary conditions imposed on the gauge field at the poles of \( \omega \). Following the original proposal, a number of works have explored different choices for \( \omega \) and boundary conditions, and investigated the corresponding two-dimensional integrable field theories that result from them. Based on the homotopical framework developed in \cite{Benini:2020skc}, it was shown in \cite{Lacroix:2021iit} that for a meromorphic 1-form with a double pole at infinity, and otherwise arbitrary, the corresponding two-dimensional integrable field theories were \emph{non-degenerate} $\mathcal{E}$-models. These are a general class of 2d theories  introduced by Klim\v{c}\'\i{}k and \v{S}evera in \cite{Klimcik:1995dy, Klimcik:1995ux, Klimcik:1996np} as $\sigma$-models providing a natural setting for describing a non-abelian generalisation of $T$-duality, known as Poisson-Lie $T$-duality. The work presented in Chapter~\ref{chap:4dCS} considers the most general choice of the meromorphic 1-form \( \omega \), lifting the restriction that it must have a double pole at infinity, and thereby extending the results of \cite{Lacroix:2021iit}. We show that this more general setting gives rise to \emph{degenerate} \( \mathcal{E} \)-models, which constitute a further generalization of the non-degenerate models previously studied.
This work therefore offers a comprehensive answer to the question of what constitutes the most general class of two-dimensional field theories derivable from four-dimensional Chern-Simons theory.

Shortly after the original correspondence was introduced, the framework was extended to relate six-dimensional Chern-Simons theory on twistor space to the anti-self-dual Yang–Mills (ASDYM) equations \cite{Kcostello, Bittleston:2020hfv, Penna:2020uky}. The work of Bittleston and Skinner \cite{Bittleston:2020hfv} considered a specific meromorphic $(3,0)$-form, which enabled them to explore this generalized setting and derive the four-dimensional WZW model \cite{Losev:1995cr} from six-dimensional Chern-Simons theory. However, in contrast to the extensive developments in the four-dimensional case, this six-dimensional generalization had received relatively little attention—likely due to the technical challenges associated with twistor theory. Building on the insights gained from four-dimensional Chern-Simons theory, in Chapter~\ref{sec:intdeformations} we explored a different choice of meromorphic \((3,0)\)-form within the framework of six-dimensional Chern-Simons theory. While the use of twistor space introduced several technical challenges in extending the ideas from the 4d setting, these obstacles were successfully overcome. This led to the construction of a novel four-dimensional integrable field theory—an integrable deformation of the WZW model—closely analogous to the two-dimensional \( \lambda \)-deformation of the WZW model \cite{Sfetsos:2013wia}. The model is integrable in the sense that its equations of motion can be reformulated as the anti-self-dual Yang--Mills equations for a suitable gauge connection. In particular it provides a new example of the Penrose-Ward correspondence. 

Within the same framework, Chapter~\ref{sec:gaugingthediamond} presents a different generalization, this time aimed at constructing four-dimensional gauged WZW models. To achieve this, we extended the ideas of Stedman \cite{Stedman:2021wrw}, which involve introducing an additional gauge field and a new term in the six-dimensional action. This approach contrasts with the standard method of generating different four-dimensional theories by varying the choice of meromorphic \((3,0)\)-form. Our construction successfully led to a four-dimensional generalization of the gauged WZW model, which was shown to be integrable. This provides yet another example of the Penrose--Ward correspondence. In this context, our work on six-dimensional Chern-Simons theory on twistor space focused on constructing two distinct generalizations of the original models developed in \cite{Bittleston:2020hfv, Penna:2020uky}. These constructions not only led to previously unknown integrable field theories in four dimensions, but also provided valuable insight into the structural features of the correspondence.

Finally, Chapter~\ref{chap:5d2CS} introduces a new correspondence between five-dimensional 2-Chern-Simons theory and three-dimensional integrable field theories. Unlike the other developments in this thesis, this work does not extend an existing construction but instead establishes a fundamentally new framework. In this sense, it represents the most original contribution of the thesis. The central idea was to apply the "categorical ladder = dimensional ladder" principle to construct a higher-dimensional analogue of four-dimensional Chern-Simons theory, which then serves as a parent theory for integrable models in three dimensions. This entirely novel approach was successfully carried out in \cite{Chen:2024axr}. The three-dimensional field theories constructed in this work are integrable in a somewhat non-standard sense. On the one hand, they refine earlier ideas on higher-dimensional integrability \cite{Alvarez:1997ma}, demonstrating that the language of higher category theory is naturally suited to formalize and extend these concepts. In particular, it provides a systematic method for constructing theories that are integrable in this broader sense. On the other hand, one of the models introduced in \cite{Chen:2024axr} exhibits an infinite-dimensional symmetry algebra, which can be precisely described using the Raviolo formalism developed in \cite{Garner:2023zqn}. This ongoing work has the potential to serve as a foundation for developing a bootstrap program for topological-holomorphic three-dimensional quantum field theories.

\printbibliography

\end{document}